**UNIVERSIDAD NACIONAL AGRARIA**

**LA MOLINA**

**ESCUELA DE POST GRADO**

**MAESTRÍA EN NUTRICIÓN**

# EVALUACIÓN DE UN PRODUCTO A BASE DE ACEITE ESENCIAL DE ORÉGANO SOBRE LA INTEGRIDAD INTESTINAL, LA CAPACIDAD DE ABSORCIÓN DE NUTRIENTES Y EL COMPORTAMIENTO PRODUCTIVO DE POLLOS DE CARNE

**Presentada por:**

**DIEGO ANDRÉS MARTÍNEZ PATIÑO-PATRONI**

**TESIS PARA OPTAR EL GRADO DE**

**MAGISTER SCIENTIAE EN NUTRICIÓN**

**UNIVERSIDAD NACIONAL AGRARIA LA MOLINA**
**Escuela de Post-Grado**
Telf. 614-7800 Anexo 209 Telefax 614-7142
Apartado 12-056 La Molina
**LIMA - PERU**

# ACTA DE SUSTENTACIÓN

Los Miembros del Jurado que suscriben, reunidos para evaluar la sustentación de tesis presentada por el alumno **DIEGO ANDRÉS MARTÍNEZ PATIÑO-PATRONI**, denominada: "EVALUACIÓN DE UN PRODUCTO A BASE DE ACEITE ESENCIAL DE ORÉGANO SOBRE LA INTEGRIDAD INTESTINAL, LA CAPACIDAD DE ABSORCIÓN DE NUTRIENTES Y EL COMPORTAMIENTO PRODUCTIVO DE POLLOS DE CARNE", para cumplir con uno de los requisitos para optar el grado académico de *Magister Scientiae* en la Especialidad de **NUTRICIÓN**.

Teniendo en consideración los méritos del referido trabajo así como los conocimientos demostrados por el sustentante, declaramos la tesis como:

APROBADA

con el calificativo (*) de EXCELENTE

UNIVERSIDAD NACIONAL AGRARIA - LA MOLINA
Escuela de Post Grado

En consecuencia, queda en condición de ser considerado APTO por el Consejo Universitario y recibir el grado académico de *Magister Scientiae*, de conformidad con lo estipulado en el Artículo 41° del Reglamento de la Escuela de Post Grado.

La Molina, 5 de noviembre del 2012

Dra. María Elena Villanueva Espinoza
PRESIDENTE

Dr. Carlos Vílchez Perales
PATROCINADOR

Dr. Carlos Gómez Bravo
MIEMBRO

Dr. Víctor Guevara Carrasco
MIEMBRO

(*) De acuerdo con el Artículo 17° del Reglamento de Tesis, éstas deberán ser calificadas con términos de: EXCELENTE, MUY BUENO, BUENO o REGULAR.

# RESUMEN

Se llevaron a cabo ocho experimentos en diferentes condiciones observadas en la industria para evaluar el efecto de un producto a base de aceite esencial de orégano (PRO) sobre el comportamiento productivo, integridad intestinal y capacidad de absorción de nutrientes de pollos de carne: En los tres primeros, hasta los 14 días de edad y sin exposición a desafíos entéricos específicos, se evaluaron los efectos sobre las características histológicas del intestino, el crecimiento corporal y la integridad esquelética. En otros dos experimentos, hasta el día 28 de edad y con pollos desafiados con el Síndrome de Tránsito Rápido (STR), se evaluaron los efectos sobre el estado antioxidante y el comportamiento productivo, y luego ésto fue contrastado frente a un desafío de campo. Finalmente, en otros dos experimentos, hasta el día 21 de edad y con pollos desafiados con coccidia, se evaluaron los efectos sobre la coccidiosis y la digestibilidad y eficiencia nutricional. Se emplearon dietas a base de maíz, soya y harina de pescado. En ausencia de desafío se observaron respuestas significativas, por efecto del PRO, en la altura y área superficial de la vellosidad intestinal, en la densidad e índice de estructura de la mucosa, en las relaciones entre la vellosidad y la cripta, en la densidad de las células caliciformes y la producción de mucina, en la ganancia de peso y conversión alimentaria ($P<0.04$), y en el rendimiento de carcasa y densidad ósea ($P<0.08$). En las aves afectadas por el STR se observaron respuestas significativas ($P<0.05$) en la actividad de la superóxido dismutasa, ganancia de peso, conversión alimentaria, frecuencia de alteraciones en heces, diámetro de la bursa e incidencia de disbacteriosis. En las aves afectadas por coccidiosis se observaron respuestas significativas en el porcentaje de aves positivas y score microscópico de coccidia, recuento de ooquistes, diámetro de la bursa, ganancia de peso, conversión alimentaria, digestibilidad aparente de la materia seca y proteína, y eficiencias energética y proteica ($P<0.01$); así como en la retención aparente de nitrógeno ($P<0.07$). Los resultados indican que el PRO añade valor a la dieta favoreciendo la eficiencia del alimento y el comportamiento productivo.

Palabras clave: pollos de carne, aceite esencial, orégano, histología, digestibilidad, absorción, coccidia, antioxidante.

# ÍNDICE GENERAL

# ÍNDICE DE CUADROS Y GRÁFICOS

**Gráfico** **Página**

# ÍNDICE DE IMÁGENES

# ÍNDICE DE ANEXOS

## ABREVATURAS

| | |
|---|---|
| AAM | Área absoluta de mucina |
| AE | Altura del enterocito |
| AEO | Aceite esencial de orégano |
| AEV | Área de enterocitos en la vellosidad |
| ACC | Área de la célula caliciforme |
| ARM | Área relativa de mucina |
| ASC | Área superficial de la cripta |
| ASV | Área superficial de la vellosidad |
| ATTD | Digestibilidad aparente en el tracto total |
| AV | Altura de la vellosidad |
| Bu | Bursa de Fabricio |
| Ba | Bazo |
| CA | Consumo de alimento |
| CCA | Coeficiente de crecimiento alométrico |
| CE | Contenido de energía metabolizable en la dieta |
| CMA | Contenido del marcador en el alimento |
| CMS | Contenido de materia seca en el alimento |
| CMO | Contenido mineral en el hueso |
| CMH | Contenido del marcador en las heces |
| CNA | Contenido del nutriente en el alimento |
| CNH | Contenido del nutriente en las heces |
| CPD | Contenido de proteína en la dieta |
| CPH | Contenido de proteína en las heces |
| DACC | Densidad absoluta de células caliciformes |
| DCC | Diámetro cráneo-caudal |
| DEXA | Absorciometría de rayos-X de energía dual |
| Dimar | Daño intestinal medido a través de residuos |
| DLL | Diámetro latero-lateral |
| DMO | Densidad mineral ósea |
| DP | Diámetro promedio de la diáfisis |
| DRCC | Densidad relativa de células caliciformes |
| DM | Densidad de la mucosa |
| DRE | Densidad relativa de enterocitos |
| DV | Densidad de las vellosidades |
| EER | Relación de eficiencia energética |
| EN | Excreción de nitrógeno |
| GAV | Grosor apical de la vellosidad |
| GALP | Grosor apical de la lámina propia |
| GBV | Grosor basal de la vellosidad |
| GBLP | Grosor basal de la lámina propia |
| GC | Grosor de la cripta |
| GCC | Grosor de la célula caliciforme |
| GLP | Grosor de la lámina propia |
| GV | Grosor de la vellosidad |
| HE | Hematoxilina-eosina |
| IEM | Índice de estructura de la mucosa |
| IEV | Índice de estructura de la vellosidad |
| IFV | Índice de forma de la vellosidad |
| IM | Índice morfométrico |
| IMS | Ingesta de material seca |
| IN | Ingesta de nitrógeno |
| KOH | Hidróxido de potasio |
| LC | Largo de la cripta |
| LCC | Largo de la célula caliciforme |
| MAS | Síndrome de Mala Absorción |
| MCN | Material de cama nuevo |
| MCR | Material de cama reusado |
| MIT | Índice de mitosis en la cripta |
| PC | Peso corporal |
| PCm | Peso corporal medio |
| PCMm | Peso corporal metabolizable medio |
| PER | Relación de eficiencia proteica |
| PO | Peso del órgano o estructura |
| PRO | Producto a base de aceite esencial de orégano |
| Rba | Índice morfométrico del bazo |
| Rbu | Índice morfométrico de la bursa |
| RN | Retención aparente de nitrógeno |
| RSS | Síndrome de Enanismo y Retraso |
| Rti | Índice morfométrico del timo |
| SOD | Superóxido dismutasa |
| STR | Síndrome de Tránsito Rápido |
| Ti | Timo |
| X | Aumentos (cuando está después de un número) |

# I. INTRODUCCIÓN

La producción de pollos de carne demanda eficiencia debido al reducido margen que existe entre el costo de producción y el precio de venta del pollo. Por ello, y considerando la elevada participación del costo de alimentación en la estructura de costos, resulta imperativo maximizar la eficiencia de las aves para convertir el alimento en proteína corporal así como diseñar estrategias costo-efectivas para controlar las condiciones que vulneran la eficiencia productiva de las parvadas.

Uno de los principales aspectos que influyen sobre la eficiencia nutricional es la integridad intestinal pues de ella depende la adecuada absorción de nutrientes. La integridad de la mucosa intestinal puede verse afectada por factores relacionados a la calidad de la dieta y de los insumos del alimento, por agentes infecciosos, entre otros. Muchos de estos factores, además de influir sobre la integridad intestinal ejercen influencia sobre el normal desarrollo de la mucosa intestinal y consecuentemente la capacidad de absorción de nutrientes en los primeros días de vida del ave.

Actualmente, la industria cuenta con recursos y estrategias para controlar los factores antes mencionados y favorecer el óptimo desarrollo e integridad intestinal. Sin embargo, debido al constante incremento en el costo del alimento y a los nuevos desafíos de campo, es necesario disponer de estrategias innovadoras que permitan el control de la salud intestinal de manera integral y sean rentables para el productor.

El aceite esencial de orégano (AEO) reúne algunos atributos de interés para la salud intestinal de las aves; sin embargo, sus cualidades deben ser estandarizadas y validadas pues es un recurso con características variables. Si bien se han realizado diversos estudios empleando AEO en pollos de carne, muchas de estas investigaciones han sido llevadas a cabo en laboratorios en condiciones *in vitro* y solo algunas en situaciones comerciales o que se asemejen a ellas. Por otro lado, se han realizado pruebas de campo, incluyendo algunas en granjas comerciales de empresas locales con resultados que favorables (Martínez, 2010); sin embargo, en algunos casos estas evaluaciones no han contado con el rigor metodológico necesario que respalde la calidad de la información obtenida.

El objetivo del presente estudio fue evaluar el efecto de un producto a base de AEO sobre el comportamiento productivo, la integridad intestinal y la capacidad de absorción de nutrientes de pollos de carne. Para lograrlo se llevó a cabo una serie de experimentos en que las aves fueron sometidas a diferentes condiciones de crianza, semejantes a aquellas observadas en la industria, que influyen sobre la integridad intestinal y la capacidad de absorción de nutrientes, tal como se indica a continuación.

Los tres primeros experimentos que se presentan en este trabajo (Experimentos 1, 2 y 3) fueron conducidos durante las 2 primeras semanas de vida de las aves y en condiciones normales de crianza; es decir, sin exponer las aves a factores específicos que vulneren la integridad intestinal. En estos experimentos se evaluó el efecto del producto sobre la morfometría histológica de la mucosa intestinal, la capacidad de absorción de nutrientes (Experimento 1) y algunos indicadores de la absorción de nutrientes, como el comportamiento productivo, el crecimiento alométrico de órganos y estructuras corporales (Experimento 2) y la mineralización ósea e integridad esquelética (Experimento 3).

Posteriormente, para evaluar el producto en condiciones de desafío entérico, fue necesario desarrollar la metodología necesaria que permitieran inducir, de manera controlada, el Síndrome de Tránsito Rápido (STR) y la coccidiosis bajo las formas típicamente observadas en las explotaciones comerciales (Anexo 1).

Empleando dicha metodología se evaluó el efecto del producto, en presencia del STR, sobre el estado antioxidante de las aves (Experimento 4) y el comportamiento productivo (Experimento 5) en condiciones experimentales. Posteriormente, para corroborar los hallazgos obtenidos, se realizó una evaluación en una granja comercial empleando el producto en el control del STR (Experimento 6). Finalmente, la metodología de desafío desarrollada fue empleada para inducir un cuadro de coccidiosis típico de campo y evaluar el efecto del producto sobre la propia coccidiosis (Experimento 7) y sobre la digestibilidad y eficiencia nutricional (Experimento 8).

# II. REVISIÓN DE LITERATURA

## 2.1 Desarrollo del tejido óseo

El hueso está compuesto aproximadamente en un 70% por minerales, 20% materia orgánica y 10% agua (Rath *et al*, 2000). La matriz mineral está compuesta en un 95% por calcio y fósforo en la forma de hidroxiapatita, que constituye de 60 a 70% del peso del hueso (Rath *et al*, 1999; 2000). Aproximadamente 80 a 90% de la matriz orgánica está constituida por colágeno tipo I, que con su característica fibrosa y su estructura de triple hélice forma el andamiaje primario de los tejidos esqueléticos y provee la base para el proceso de mineralización (Rath *et al*, 2000).

La rigidez del hueso radica en la deposición de calcio y fósforo en la forma de hidroxiapatita durante el proceso de mineralización (Rath *et al*, 1999, 2000; Almeida Paz y Bruno, 2006), siendo el contenido de ceniza del hueso proporcional a su resistencia a la compresión (Bonser and Casinos, 2003). La flexibilidad del hueso depende del componente orgánico (Velleman, 2000). La triple hélice del colágeno tipo I confiere resistencia a la tensión mecánica, mientras que los enlaces cruzados en él mantienen las hélices aseguradas. Si la cantidad de enlaces cruzados es reducida las hélices se pueden separar, mientras que si son muy abundantes disminuye la habilidad del hueso para absorber energía (Seeman and Delmas, 2006).

Si bien la resistencia del hueso a la rotura es resultado de su mineralización (Mendes *et al*, 2006), el balance existente entre los componentes orgánico y mineral es también importante (Rath *et al*, 1999). Así, cuando se incrementa el contenido relativo de mineral aumenta la fortaleza del hueso, pero se pierde flexibilidad (Seeman and Delmas, 2006). Al respecto, en un experimento realizado con gallinas Wight Leghorn, Zhang y Coon (1997) encontraron que la resistencia del hueso a la rotura no sólo está en función al porcentaje de ceniza sino también a su volumen.

Las características mecánicas de los huesos dependen de su función, siendo los huesos largos palancas diseñadas para la carga y movimiento, en que se favorece la rigidez sobre la flexibilidad (Seeman and Delmas, 2006). De esta forma, la madurez ósea se alcanza cuando se ha completado el desarrollo estructural básico y la mineralización, y se ha alcanzado la fortaleza mecánica óptima (Rath *et al*, 2000). Así, los huesos de pollos de 3 semanas de edad son significativamente más rígidos y menos elásticos que los de 2 semanas de edad, debido al mayor contenido relativo de mineral y menor grosor del hueso (Kim *et al*, 2011).

Desde el punto de vista histológico, el hueso es un tejido complejo compuesto por diferentes tipos de células y realiza un continuo proceso de renovación y reparación. Los dos principales tipos celulares responsables de la remodelación son los osteoclastos, que resorben el hueso, y los osteoblastos, que forman nuevo hueso (van't Hof and Ralston, 2001; Jamal, 2011). La coordinación que existe entre las actividades de resorción y formación del hueso se conoce como “ciclo de remodelación ósea” (Bain and Watkins, 1993).

Si bien se ha establecido la influencia que ejerce el balance hormonal entre andrógenos y estrógenos sobre el desarrollo óseo, este factor resulta poco relevante en las líneas actuales de pollos típicamente precoces (Leeson and Summers, 2005). Se ha observado también en pollos de líneas genéticas con mayor velocidad de crecimiento tibias más fuertes en edades iguales; sin embargo, estas diferencias no se observan cuando los pesos corporales son iguales (McDevitt *et al*, 2006). Por lo tanto, la fortaleza ósea no está en función de la velocidad de crecimiento, sino del peso corporal. No obstante, la incidencia de problemas de patas en pollos de líneas comerciales con mayor velocidad de crecimiento es mayor (Kestin *et al*, 1992).

Dado que el hueso se adapta a la carga física, la remodelación debe coincidir con la ganancia de peso. Si no hay coincidencia entre la ganancia de peso y la modelación y remodelación del hueso esto puede conllevar al desgaste y desgarro, causando la separación de la cabeza articular en pollos de carne (Durairaj, 2008).

## 2.2 Síndromes de tránsito rápido y de mala absorción

Desde la década de los setentas en todo el mundo se han registrado síndromes entéricos con mayor o menor magnitud en la presencia de signos tales como alta morbilidad y mortalidad variable. Si bien los agentes causales no se han determinado claramente estas condiciones son usualmente llamadas “enteritis viral”. Estos procesos se han identificado con diferentes nombres: Síndrome de Tránsito Rápido (STR), Síndrome de Mala Absorción (MAS por su nombre en inglés: Malabsorption Syndrome), Síndrome de Enanismo y Retraso (RSS por sus siglas en inglés: Runting-Stunting Syndrome), Síndrome del Ave Pálida (PS por sus siglas en inglés: Pale Syndrome), Enfermedad del “Pollo Helicóptero”, Síndrome Infeccioso de Falta de Crecimiento, Síndrome de Retraso del Pavo, Mortalidad en Pico (Celis, 2000; Songserm *et al*, 2002), Enanismo Infeccioso, Necrosis de la Cabeza de Fémur, Enfermedad del Hueso Quebradizo, Osteoporosis (Bustamante y Chávez, 2010), entre otros.

En las parvadas afectadas se observa disparidad en cuanto a peso y eventualmente pigmentación (Martínez, 2010). Las heces presentan gránulos de alimento parcialmente sin digerir, pueden ser acuosas, presentar descamaciones mucosas (Celis, 2000) y permanecer adheridas en la cloaca y en la región perianal en algunas aves produciendo escaldaduras y prolapsos (Smith, 2006).

Los casos que se observan de estos síndromes difieren en cuanto a epizootiología, signos y lesiones que se observan; sin embargo, todos se caracterizan porque el agente etiológico no ha sido definido con precisión y las mayores pérdidas económicas son producidas por el bajo crecimiento, elevada conversión alimentaria, infecciones secundarias y desórdenes metabólicos (Celis, 2000), así como incremento en los descartes, menor viabilidad, y tiempo para alcanzar el peso de mercado (Martínez, 2010). Los picos de ocurrencia son variables pudiendo presentarse alrededor de los 10 o 12 días de edad en el caso del RSS (Zavala and Sellers, 2005; Payne, 2008) o entre el día 21 y 28 de edad en el caso del MAS (Bustamante y Chávez, 2010) y del STR (Martínez, 2010). Puede observarse palidez en tarsos, desorientación y en algunos casos presencia de “plumas helicóptero”.

En la necropsia frecuentemente se observa fragilidad ósea, desprendimiento y necrosis de la epífisis superior del fémur (Márquez, 2007; citado por Bustamante y Chávez, 2010). En casos leves de síndromes virales es posible observar inflamación del gastrocnemio y del tendón flexor, mientras en estados avanzados se presenta ruptura del gastrocnemio, pancreatitis y atrofia del páncreas (Van der Heide, 1983; citado por Bustamante y Chávez, 2010). En diversos síndromes es frecuente observar erosión de molleja (Angel, 2008; Hoerr, 2010) vesículas biliares y proventrículos agrandados, enteritis, intestinos pálidos con paredes muy delgadas y translúcidas, conteniendo abundante líquido y alimento sin digerir (Zavala and Sellers, 2005; Smith, 2006); lesiones que impiden la correcta absorción de vitaminas, calcio, fósforo y aminoácidos esenciales (Van der Heide, 1980; Bustamante y Chávez, 2010) y que afectan de manera inequívoca la absorción de lípidos y la utilización de la energía dietaria, incluso en las aves aparentemente sanas (Lilburn *et al*, 1982). En algunos síndromes virales en ocasiones se observa también atrofia del timo y de la bursa de Fabricio (Nili *et al*, 2007).

A nivel histológico, y especialmente en los síndromes virales, puede observarse quistes envolviendo las criptas de Lieberkühn catalogado como “enteropatía cística” para luego de evolucionar hacia una lesión inflamatoria convertirse en “enteritis cística” y producir acortamiento y atrofia de las vellosidades intestinales (Songserm *et al*, 2002; Zavala and Sellers, 2005), siendo el yeyuno la porción más afectada (Songserm *et al*, 2002). La enteropatía cística característica del RSS ha sido reproducida inoculando a aves sanas contenidos intestinales provenientes de aves afectadas y luego de ser filtrados contra bacterias (Zavala and Sellers, 2005) así como colocando a las aves sobre materiales de cama re-utilizados provenientes de parvadas afectadas (Smith, 2008; Perry *et al*, 1991) lo que confirmaría su característica viral; sin embargo, también ha sido producida consistentemente induciendo deficiencias de las vitaminas del complejo B (Zavala and Sellers, 2005).

Los síndromes entéricos pueden afectar la actividad enzimática en el intestino delgado, presentándose disturbios en la función exocrina del páncreas (Sinclair *et al*, 1984; Sályi and Glávits, 1999), reduciéndose la actividad enzimática de la lipasa, tripsina, glutamil transferasa, amilasa, quimotripsina (Mazurkiewicz *et al*, 1993), así como reduciéndose la actividad de las disacaridasas intestinales (Sályi and Glávits, 1999) como maltasa y sucrasa, y finalmente disminuyendo la retención de proteína, grasa, energía bruta, materia seca y cenizas (Angel *et al*, 1990).

Algunas características hematológicas y bioquímicas de las aves afectadas por síndromes entéricos son menores recuentos de hemoglobina, eritrocitos y proteína sérica total, así como mayores niveles de amilasa, fosfatasa alcalina, y alanino y asparto aminotransferasas séricas, consistentes con la menor digestión y absorción de proteína y el daño hepático e intestinal (Prameela Rani *et al*, 2011).

Por otro lado, en pollos sometidos al MAS, asociado al menor peso corporal, enteritis con diarrea, deshidratación y elevada mortalidad, se ha observado también menor concentración plasmática de 25-hidroxi vitamina D3, calcio (Perry *et al*, 1991a) y fósforo (Perry *et al*, 1991A; Sályi and Glávits, 1999), así como menor mineralización de la tibia (Perry *et al*, 1991a; Sályi and Glávits, 1999) y mayor frecuencia de deformaciones en la tibia (Perry *et al*, 1991b).

Estos síndromes son usualmente referidos como trastornos multifactoriales (Angel, 2008) y han sido correlacionados con diferentes factores derivados de la calidad del alimento (Celis, 2000) como micotoxinas (Raghavan, 1997) y antinutrientes de la soya (Campabadal, 2008, 2009), con agentes bacterianos como *Clostridium perfringens* (Rebel *et al*, 2006; Dustan and Jones, 2008; Hoerr, 2010), virales como reovirus, enterovirus, adenovirus, parvovirus, bacteriófagos (Songserm *et al*, 2002), genéticos, de manejo, ambientales (Zavala and Sellers, 2005), entre otros. Si bien se ha descubierto factores virales asociados a síndromes como el MAS y el RSS, el consenso general es que es poco probable que estos sean los únicos responsables e incluso que puedan ser elementos comunes en todos los casos de campo (Zavala and Sellers, 2005) y se considera que las bacterias cumplen un rol importante al menos como agentes secundarios (Cervantes, 2007; Martínez, 2010).

Estos síndromes no siempre están asociados a factores virales. Ello se evidencia en los reportes que indican que el empleo de las vacunas contra reovirus no siempre reduce la frecuencia del síndrome (Bustamante y Chávez, 2010), observándose el síndrome en parvadas negativas a virus. Al respecto, Dustan and Jones (2008) indican que la vacunación de pollos contra reovirus es una estrategia de prevención efectiva sólo en el 50% de los casos. Asimismo, existen reportes que indican además la presencia de coccidia asociada a estos síndromes (Rodríguez *et al*, 1988).

Si bien los síndromes de tipo infeccioso y en particular los virales son los mejor estudiados, existen también una serie de agentes de tipo no infeccioso que pueden conducir a la presentación de cuadros similares a los que con frecuencia se refieren como "tránsito rápido" (STR), "mala absorción" o "enteritis no específica" (Cervantes, 2007).

En la actualidad, las medidas de control empleadas son en su mayoría sintomáticas, dependiendo de las manifestaciones clínicas y la intensidad del proceso, y tienen como objetivo mejorar la condición clínica de las aves (Bustamante y Chávez, 2010), tal como sucede con la suplementación de altas cantidades de vitamina D3 sobre la ganancia de peso en aves afectadas por el RSS (Sályi and Glávits, 1999).

Asimismo, investigaciones realizadas en la Universidad de Georgia determinaron que las aves afectadas por el MAS y que recibieron suplementación de selenio y vitamina E ganan más peso que las aves control (Colnago *et al*, 1982). Un efecto similar se observa cuando las reproductoras reciben niveles superiores de vitamina E y selenio, permitiendo que sus pollos sean capaces de recuperarse más rápidamente del MAS inducido experimentalmente (Rebel *et al*, 2004). La suplementación de vitamina A en pollos afectados por STR es una práctica muchas veces común para favorecer la regeneración de la mucosa intestinal (Martínez, 2010); sin embargo, otros estudios revelan que su administración a aves afectadas por el MAS tiene un efecto exacerbativo, elevando la conversión alimentaria y reduciendo la ganancia de peso y el porcentaje de ceniza ósea, concluyendo que un antagonismo nutricional entre cantidades excesivas de vitamina A y vitaminas E o D pudiera ser responsable de los efectos exacerbativos (Veltmann *et al*, 1985).

Se ha observado que sustituir la soya por harina de pescado y pasta de girasol en pavos afectados por el síndrome de retraso, previene completamente o al menos mitiga los efectos del síndrome sobre la eficiencia productiva (Angel *et al*, 1992). Asimismo, para el control del MAS existen disponibles en el mercado vacunas de reovirus, a pesar que su rol en esta condición parece incierto (McNulty *et al*, 2008).

En el caso del RSS el consenso general es que el agente primario es viral y que los agentes bacterianos son secundarios (Smith, 2006), cuyos efectos secundarios pueden ser atenuados con la administración de antibióticos, observándose un mejor peso y conversión alimentaria en pavos afectados tras la adición de 22 ppm de virginiamicina en la dieta (Al-Batshan *et al*, 1992), así como mayor peso con la adición de 220 ppm de bacitracina (Trampel and Sell, 1994). En casos del STR se observan resultados satisfactorios en condiciones de campo con el uso de 150 ppm de neomicina en la dieta cuando el cuadro se asocia a algunos problemas de calidad del alimento o a *C. perfringens*, así como con el uso de sulfas cuando el cuadro se asocia a procesos en que coccidia participa de manera sub-clínica (Martínez, 2010).

## 2.3 Modelos no virales de desafío entérico

Los esfuerzos realizados para reproducir las lesiones generadas por procesos entéricos han producido resultados extremamente variables y existe un consenso general en que no son fácilmente reproducibles bajo condiciones experimentales. Sin embargo, los primeros trabajos fueron más exitosos en producir lesiones de Enteritis Necrótica (Al-Sheikhly and Truscott, 1977a,b; Truscott and Al-Sheikhly, 1977; Prescott *et al*, 1978; Hamdy *et al*, 1983). Los esfuerzos más recientes, sin embargo, han resultado menos exitosos. Tal es el caso de Cowen *et al* (1987) quienes lograron reproducir sólo una pequeña incidencia de Enteritis Necrótica en pollos desafiados con *C. perfringens*, pero no lograron inducir los signos en otras aves a pesar que se encontró un elevado recuento de dicho microorganismo en el tracto intestinal; el caso de Craven (2000) quien no observó incremento alguno en la mortalidad en pollos alimentados con dietas suplementadas con centeno, a pesar de los altos niveles de *C. perfringens* en el contenido intestinal; o el caso de Pedersen *et al* (2003) quienes tampoco lograron inducir mortalidad o signos clínicos de Enteritis Necrótica a pesar de inocular altas dosis de patógenos. En este contexto, es posible que en comparación a las aves entre 1960 y 1980, los actuales genotipos sean más resistentes y que los patrones de respuesta a los desafíos por patógenos hayan cambiado.

### 2.3.1 Modelos sin el uso de coccidia como factor predisponente

Kulkarni *et al* (2007) evaluaron cinco modelos de desafío con diferentes grados de severidad, empleando niveles variados de los siguientes factores: número de días en que se administró el inóculo (3 a 5), y la presencia del inóculo en el alimento de forma continua o discontinua. Los investigadores observaron una correlación directa entre el score de lesiones intestinales y los factores evaluados.

Por su parte Chalmers *et al* (2007) demostraron que la administración de un inóculo de *C. perfringens* es capaz de reproducir un cuadro de enteritis; sin embargo, la patogenicidad de las cepas es variable y, en consecuencia, también su capacidad para producir daño a nivel entérico. En sus evaluaciones estudiaron también el efecto de cinco cepas de *C. perfringens* obtenidas de parvadas comerciales criadas sin antibióticos y emplearon un modelo caracterizado por la administración del inóculo vía alimento durante 24 horas, tras 8 horas de ayuno, en el día 13 de edad, empleando dosis entre $10^9$ y $10^{11}$ UFC por pollo. Mantuvieron a las aves alojadas en corrales en piso con material de cama nuevo y administraron una dieta a base de maíz y soya con 25% de trigo para fomentar la enteritis. Los investigadores observaron que de las cinco cepas evaluadas sólo una produjo efectos evidentes en mortalidad (35.8%) y lesiones intestinales, mientras que las otras cuatro produjeron mortalidades entre 3.8 y 7.7% y lesiones reducidas.

Años atrás, Kaldhusdal *et al* (1999) ya habían llevado a cabo dos experimentos para evaluar el efecto de las variables implicadas en los modelos de desafío empleando aves en corrales en piso con material de cama nuevo. En el primer experimento administraron 5 x $10^5$ UFC de *C. perfringens* por pollo en un alimento conteniendo trigo, avena, anticoccidial, pero no promotor de crecimiento. En el segundo experimento administraron por pollo 4 x $10^9$ UFC de *C. perfringens* de dos cepas con diferentes tasas de producción de la toxina alfa en un alimento conteniendo trigo, sorgo y avena, y sin anticoccidial y promotor de crecimiento. Los investigadores observaron que en el primer experimento sólo las aves que recibieron el inóculo presentaron mortalidad elevada (17%) demostrando el efecto del inóculo administrado. Sin embargo, el mayor recuento de *C. perfringens* en las aves no desafiadas y la presencia indistinta de lesiones intestinales incluso en las aves no desafiadas, refleja la contaminación cruzada que se produjo en ambos experimentos.

Estudios posteriores como el reportado por Jia *et al* (2009) en que se administró $10^8$ UFC de *C. perfringens* en el día 13 de edad a pollos alimentados con dietas a base de maíz o trigo y suplementadas o no con carbohidrasas, demostraron el efecto predisponente de los polisacáridos no almidonosos presentes en granos como el trigo, ya que se observaron que sólo las aves desafiadas, alimentadas con dietas conteniendo trigo pero sin carbohidrasas fueron las únicas afectadas negativamente en términos de comportamiento productivo, recuento intestinal de *C. perfringens* y score de lesiones en la mucosa.

Otra combinación de factores de desafío es la que fue empleada por Nikpiran *et al* (2008), quienes lograron reproducir signos clínicos, lesiones intestinales y mortalidad por Enteritis Necrótica empleando un modelo de desafío que consistió en la administración de un inóculo 2 veces por día, durante 5 días continuos, de manera simultánea en el alimento y en el pico de las aves. Para la administración oral se aplicó directamente en el pico de las aves 9 x $10^8$ UFC de *C. perfringens*. La cepa de *C. perfringens* fue obtenida de parvadas afectadas por brotes agudos o severos de Enteritis Necrótica. Para predisponer la enfermedad se empleó los siguientes factores predisponentes: mayor densidad de alojamiento de las aves, altos niveles de trigo y harina de pescado en la dieta, y la inducción de inmunodepresión relativa.

Sin embargo, existen en la literatura reportes de experiencias como aquella de Siragusa *et al* (2008), quienes no lograron producir lesiones intestinales a pesar de administrar $10^7$ UFC por pollo de una cepa de *C. perfringens* aislada de un brote de campo de Enteritis Necrótica, mediante un inóculo administrado en los días 15 y 16 de edad, manteniendo a las aves alojadas en jaulas dispuestas en baterías con piso de malla y alimentadas con una dieta a base de maíz y soya pero sin promotor de crecimiento.

No obstante, la enteritis producida por *C. perfringens* tiene importancia económica no solo cuando se presenta en forma clínica, sino también cuando el cuadro presenta una manifestación subclínica. Una muestra de ello son las evaluaciones reportadas por Olkowski *et al* (2006) quienes aislaron una cepa de *C. perfringens* de un brote de campo de Enteritis Necrótica fulminante y la administraron vía oral aplicando modelos de corto y largo plazo. En el modelo de corto plazo se administró 1 ml conteniendo 3 x $10^{10}$ UFC por pollo como dosis única. En el modelo de largo plazo se administró diariamente 1 ml conteniendo 1 o 2 x $10^8$ UFC por pollo entre los días 14 y 21 de edad. En ambos casos las aves recibieron una dieta conteniendo 40% de proteína y harina de pescado como principal fuente proteica. A pesar que ninguno de los modelos evaluados produjo signos clínicos de Enteritis Necrótica ni mortalidad elevada, fue posible reproducir lesiones sub-clínicas macro y microscópicas distintivamente pronunciadas en la mucosa intestinal. Los investigadores consideran que las lesiones sub-clínicas producidas son de indudable significancia pato-fisiológica, ya que desplegar una respuesta inflamatoria en el ave es demandante en términos metabólicos; además, la naturaleza de las lesiones y los disturbios producidos en la interface entre los enterocitos y la lámina propia indudablemente afecta la fisiología digestiva. Por lo tanto, los efectos potenciales de estos cambios sobre el comportamiento productivo de las aves no deben ser subestimados. En términos prácticos, estos procesos pueden ser de considerable significancia para la industria y, por la naturaleza sub-clínica de las lesiones, el origen del problema puede no ser fácilmente identificable en las parvadas afectadas. Los investigadores consideran que los protocolos de evaluación deben incluir exámenes histológicos con criterios de clasificación claramente definidos para verificar o descartar el diagnóstico macroscópico de la Enteritis Necrótica.

### 2.3.2 Modelos que utilizan coccidia como factor predisponente

Diferentes estudios realizados han demostrado que la inclusión de coccidia como parte de los modelos de desafío mejora la efectividad de los mismos. Tal es el caso del experimento realizado por Lillehoj *et al* (2007) quienes reportaron que asociando un inóculo de *Eimeria acervulina* a uno de *C. perfringens* se logra desarrollar una patología intestinal más severa que cuando se emplea sólo uno de dichos inóculos, y que juntos desarrollan un efecto sinérgico para desencadenar un cuadro más severo que conlleva a una alteración en la respuesta inmunológica del ave. Esto fue corroborado por Shelton *et al* (2007) quienes administraron un inóculo de coccidia en el día 9 y de *C. perfringens* en los días 14 a 16 a pollos alojados en corrales en piso encontrando mayor número de lesiones intestinales en las aves desafiadas.

Por su parte Jerzsele *et al* (2011) lograron reproducir lesiones de Enteritis Necrótica empleando un modelo de desafío mixto que consistió en la administración de un inóculo de *C. perfringens* tipo A, emplear dietas sin promotor de crecimiento, con 25% de harina de pescado, aplicar una vacuna contra la Enfermedad de Gumboro antes del inóculo y una contra coccidia durante la administración del inóculo bacteriano.

Otros estudios lograron reproducir cuadros de enteritis más severos, tal es el caso de McDougald *et al* (2008) quienes evaluaron dos modelos de desafío, consistentes en la administración de un inóculo de coccidia conteniendo 25,000 ooquistes de *E. acervulina* y 15,000 de *E. maxima* en el día 14 de edad y 2 x $10^7$ o 3.2 x $10^8$ UFC de *C. perfringens* en los días 19 a 21 de edad. Las aves que recibieron la menor y mayor concentración del inóculo fueron muestreadas en el día 28 y 22 de edad, respectivamente. Las aves estuvieron alojadas en jaulas en baterías y fueron alimentadas con dietas conteniendo o no antimicrobianos promotores del crecimiento. La mortalidad producida en las aves que recibieron la menor dosis del inóculo y fueron alimentadas con dietas conteniendo o no promotor de crecimiento fue 39 y 48%, respectivamente; mientras que las que recibieron la mayor dosis del inóculo y fueron alimentadas con dietas conteniendo o no promotor de crecimiento mostraron 42 y 67% de mortalidad. Las lesiones fueron encontradas sólo en las aves que recibieron el alimento sin promotor de crecimiento. Los investigadores concluyen que *C. perfringens* puede causar alta mortalidad, pero baja incidencia de lesiones en aves aparentemente sanas, lo que implica una alta variabilidad en el estado histológico de las aves.

Asimismo, McReynolds *et al* (2004) evaluaron el efecto de diferentes factores en la capacidad del modelo para reproducir un cuadro entérico. Los factores empleados en los modelos evaluados fueron la administración o no de cada uno de los siguientes: 24 dosis por pollo de una vacuna contra coccidia en el día 14 de edad para comprometer el sistema inmune; 10 dosis por pollo de una vacuna contra la Enfermedad de Gumboro vía ocular en el día 14 de edad para producir inmunodepresión; y 3 x $10^7$ UFC de *C. perfringens* administrado 2 veces por día, vía oral, en los días 17 a 19 de edad. En todos los casos las aves estuvieron alojadas en corrales con material de cama nueva y recibieron una dieta a base de maíz y soya con 50% de trigo. Los investigadores observaron que la administración del inóculo de *C. perfringens* incrementa el índice de lesiones, y que su administración asociada a las vacunas, en dosis altas, contra la Enfermedad de Gumboro y contra coccidia produce un mayor score de lesiones. Estos modelos de desafío se caracterizaron por producir alta mortalidad (26 a 28%) hasta el día 25 de edad y se encontraron evidencias de contaminación cruzada en las aves no desafiadas.

Por otro lado, modelos de desafío que incluyen la administración de coccidia también han sido exitosos para reproducir cuadros sub-clínicos de enteritis. Tal es el caso de Pedersen *et al* (2008) quienes desarrollaron un modelo para generar enteritis sub-clínica en pollos de carne sin producir mortalidad ni un efecto evidente en el consumo de alimento y ganancia de peso. Dichos investigadores administraron un inóculo de *C. perfringens* conteniendo los productos extracelulares vía alimento por cuatro días continuos y 10 dosis de vacuna contra coccidia para generar un ligero daño en la mucosa que promueva el desarrollo de enteritis. Para predisponer el cuadro incluyeron además 50% de trigo en una dieta basal en forma de harina y sin promotor de crecimiento ni anticoccidial. Adicionalmente, de manera simultánea a la administración del inóculo de *C. perfringens*, suministraron una ración constituida por 75% de la dieta basal y 25% de harina de pescado. Posteriormente, Gholamiandehkordi *et al* (2007) realizaron una descripción detallada de los cambios patológicos e histopatológicos causados por este modelo.

## 2.4 Inmunocompetencia, características de los órganos linfoides y salud intestinal

En las aves la inmunodepresión es un síndrome clínico que consiste en la disminución de células linfoides, disfunción del sistema inmune que afecta la capacidad del ave de sintetizar anticuerpos y sustancias humorales alterando, en consecuencia, la capacidad de ofrecer resistencia a las enfermedades, comprometiendo el comportamiento productivo de la parvada (Perozo-Marín *et al*, 2004).

Esta condición puede ser inducida por diferentes factores no infecciosos como drogas, antibióticos, temperatura ambiental, estrés en general, micotoxinas, estado nutricional, o factores infecciosos como virus o toxinas bacterianas. Se ha observado que las líneas genéticas de aves con órganos linfoides de mayor tamaño responden mejor a los desafíos que las líneas de aves con órganos linfoides de menor tamaño (Li *et al*, 2001). Asimismo, se ha determinado que el estado nutricional guarda una estrecha relación con la inmunocompetencia de las aves. Al respecto, un experimento diseñado para determinar la influencia de la concentración de arginina en la dieta sobre el desarrollo de los tejidos linfoides, demostró que las dietas pobres en este aminoácido afectan el desarrollo de los órganos linfoides, desencadenando estados inmunodepresivos en las aves (Kwak *et al*, 1999).

Los criterios de evaluación de la capacidad de defensa del sistema inmune son complementarios entre sí. La evaluación clínica indicando la severidad y frecuencia de los procesos infecciosos, es un buen indicador del estado inmunológico de la parvada, así como la evaluación macroscópica, que consiste en determinar el tamaño, peso, y apariencia de los órganos linfoides. Otros indicadores de gran relevancia para conocer la inmunocompetencia de las aves son la evaluación microscópica de los órganos linfoides y la respuesta serológica a las vacunaciones (Perozo-Marín *et al*, 2004).

Los métodos macroscópicos más utilizados son: 1) evaluación del timo, bursa y médula ósea en distintas edades críticas, 2) comparación entre el tamaño de la bursa y el bazo, siendo el tamaño de la bursa mayor que el del bazo hasta el día 30 o 35 de edad, 3) relación entre el peso de los órganos linfoides y el peso corporal, y 4) relación entre los pesos de los órganos linfoides (Pulido *et al*, 2001; Perozo-Marín *et al*, 2004).

Existe una alta correlación entre las variables relacionadas a los órganos linfoides, tal como se muestra en el Anexo 56. Los mayores índices de correlación se presentan entre el peso corporal y los pesos del timo y el bazo (0.87 a 0.99 y 0.89 a 0.99, respectivamente), y estos órganos mantienen constante su crecimiento en relación a la evolución del peso del ave. Similar correlación se observa entre el peso del timo y el peso del bazo (0.77 a 0.98), y entre el peso de la bursa y el diámetro de la bursa (0.93). Algunos autores reportan menores pero aún significativos índices de correlación entre el peso vivo y el peso de la bursa (0.64 a 0.92) o entre el peso de la bursa y los pesos del timo y el bazo (0.62 a 0.87 y 0.61 a 0.83, respectivamente), debido principalmente al menor desarrollo de la bursa a partir de la cuarta semana de edad (Hernández, 1998; Ulloa *et al*, 1999; Perozo-Marín *et al*, 2004). Dichas correlaciones son particularmente evidentes en animales jóvenes en los cuales el sistema inmunológico se halla en plena etapa de desarrollo y maduración (Giambrone and Diener, 1990). Posteriormente, cuanto más tiempo el ave conserve la bolsa de Fabricio intacta, la inmunosupresión será menor (Siegel, 1990; citado por Sandoval *et al*, 2002).

La alta correlación entre el peso de los órganos linfoides y el peso corporal permite emplear índices relativos y estimar con mayor precisión la incidencia de aquellos factores que actúan sobre el tejido linfoide (Sandoval *et al*, 2002). Se ha establecido que valores muy bajos del índice morfométrico bursal pueden indicar atrofia por regresión, e inmunosupresión (Kuney *et al*, 1981; Sandoval *et al,* 2002).

En las aves sometidas a estrés se produce una liberación sistémica de glucocorticoides y catecolaminas, sustancias que afectan la respuesta inmune generando la involución de los tejidos linfoides, evidente en la respuesta inmune humoral y celular. Desde este punto de vista, se ha establecido que la pérdida de peso asociada a la atrofia y regresión de los órganos linfoides representa un indicador sensible del estrés en las aves (Dohms and Saif, 1984), siendo el tamaño del timo un indicador del estado de salud del ave y de la respuesta aguda o crónica a situaciones de estrés (Mostl and Palme, 2002). Puvadolpirod and Thaxton (2000) observaron que el peso relativo no sólo del timo sino además del bazo y bursa disminuyó como consecuencia de la exposición de las aves a factores estresantes.

Pulido *et al* (2001) consideran que la relación entre el tamaño de la bursa y del bazo (bursa/bazo) representa una lectura física de la capacidad de respuesta inmune en pollos de engorde, y esta variable como indicador de la inmunocompetencia de aves vacunadas contra la Enfermedad de Gumboro, en una zona caracterizada por la condición endémica de la enfermedad, reportando que índices superiores a 2 pueden ser considerados como propios de una adecuada inmunocompetencia.

Por otro lado, las características de los órganos linfoides guardan relación también con el nivel de desafío y/o estado sanitario de las aves. Al respecto, Deshmukh *et al* (2007) reportaron que cuando las aves son sometidas a un desafío por *Salmonella gallinarum* se produce hiperplasia del bazo.

## 2.5 Calidad de la torta de soya y salud intestinal

La torta de soya es un co-producto obtenido de la industria aceitera para consumo humano (Caprita and Caprita, 2010) y es empleando en alimentación animal por su alto contenido de proteína, mismo que fluctúa entre 44 y 49%. La torta de soya posee una alta digestibilidad y un adecuado balance de aminoácidos para la nutrición de aves con excepción de los aminoácidos azufrados, lo que puede verificarse contrastando el contenido de aminoácidos azufrados (digestibilidad verdadera), que en la harina de soya con 48% de proteína es 0.45 veces el contenido de lisina (Rostagno *et al*, 2011), con el requerimiento de aminoácidos azufrados, que es 0.72 a 0.78 veces el requerimiento de lisina (Leclercq, 1998).

La principal fracción de proteínas de la soya son globulinas, albúminas y en menor grado glutelinas, las que son solubles en soluciones salinas diluidas, soluciones alcalinas o soluciones ácidas, respectivamente. Por lo anterior, al someter a la torta de soya a un tratamiento con solución salina u otras bases se incrementa su solubilidad. Por esta razón la solubilidad de la proteína en sustancias álcalis es usada como un indicador del procesamiento de la soya (Celis, 2000).

La semilla de soya contiene varios componentes perjudiciales para las aves, que incluyen inhibidores de proteasas, generalmente llamados inhibidores de tripsina, lectinas o hemaglutininas, estrógenos, entre otros (Vohra and Kratzer, 2001; Caprita and Caprita, 2010). Estos factores antinutricionales se desnaturalizan por medio de calor, logrando un efecto favorable en su digestibilidad, debido a la destrucción de los inhibidores de proteasas, a la desnaturalización de la proteína exponiéndola a la acción enzimática y al rompimiento de las células ricas en grasa, lo que incrementa su disponibilidad (Celis, 2000).

Sin embargo, cuando la soya recibe un tratamiento de temperatura excesiva, algunos aminoácidos, en especial la lisina y otros como cistina, histidina y ácido aspártico, reaccionan con azúcares formando enlaces resistentes a la acción enzimática o bien forman compuestos indisponibles, reduciéndose la solubilidad y digestibilidad de la proteína (Celis, 2000). Dicho efecto se corrobora por la prueba de solubilidad de la proteína con hidróxido de potasio (KOH). Al respecto, Araba and Dale (1990) establecieron, tras realizar un estudio en pollos, que la solubilidad de la proteína en KOH (0.2%; 0.036 N) es un indicador confiable de la calidad de la torta de soya.

Por lo expuesto, resulta importante determinar si la soya ha recibido un tratamiento térmico adecuado para inactivar los factores antinutricionales, pero también para asegurar la digestibilidad de la proteína. Entre las pruebas de mayor utilidad para el control de calidad de la torta se encuentra, además de la determinación de la solubilidad de la proteína en KOH, la determinación de la actividad ureásica. La prueba de actividad ureásica indica el grado de destrucción de la enzima ureasa al tratar la soya con calor. La técnica mide la actividad de una enzima termolábil presente en la semilla de soya cruda, la ureasa. Este grado de actividad se mide determinando el pH de la muestra al agregarle urea. La determinación de la actividad ureásica resulta adecuada para detectar soyas con un procesamiento térmico insuficiente, pero no detecta los efectos del sobrecalentamiento, por lo que es necesario considerar además la prueba de solubilidad de la proteína en KOH para valorar adecuadamente la calidad de la soya (Celis, 2000; Vohra and Kratzer, 2001; Caprita *et al*, 2010).

USSEC (2006) considera aceptable una actividad ureásica alrededor de 0.30 puntos de variación en pH y como mínimo 0.01. Campabadal (2009) considera que una actividad ureásica mayor de 0.5 es propia de una soya cruda, 0.5 a 0.3 cuando la cocción ha sido insuficiente, y 0.3 a 0.05 cuando ha sido adecuadamente procesada.

Se ha establecido rangos referenciales para la solubilidad de la proteína de soya en KOH. Así, para lograr una aceptable digestibilidad de la lisina se considera que la solubilidad debe ser al menos 74% (Caprita and Caprita, 2010; Parsons *et al*, 1991). Al respecto, USSEC (2006) considera aceptable un rango de solubilidad entre 73 y 88%, mientras que Campabadal (2009) considera que una solubilidad de la proteína entre 75 y 85% es indicativo de un adecuado procesamiento térmico, por debajo de 75% se considera sobre cocinada y por encima de 85% cruda. Sin embargo, es mejor no relacionar la solubilidad con soyas crudas, pues existen nuevos métodos de procesamiento de alta tecnología que reportan valores sobre 85% de solubilidad con un adecuado procesamiento (Campabadal, 2008).

## 2.6 Salud intestinal, estado antioxidante y comportamiento productivo

Los radicales libres son átomos o moléculas que contienen uno o más electrones libres, se derivan del oxígeno o nitrógeno, y son muy inestables, reactivos y capaces de dañar moléculas como proteínas, ADN, lípidos y carbohidratos. El daño que los radicales libres causan a las proteínas produce modificaciones en las funciones de receptores y de transportadores de iones, así como la alteración de la actividad de diversas enzimas. Por otro lado, la oxidación de los ácidos grasos poliinsaturados altera la composición, estructura, fluidez y permeabilidad de las membranas y la actividad de enzimas ligadas a ellas. Finalmente, el daño a las moléculas biológicas compromete el crecimiento, desarrollo, inmunocompetencia y reproducción (Surai *et al*, 2003).

Existe un gran número de factores que influyen negativamente sobre el estado antioxidante de las aves. Adicionalmente a los radicales libres formados como consecuencia natural del metabolismo, los componentes celulares del sistema inmune producen una gran variedad de moléculas destructivas del entorno, tales como radicales libres, que son producidos como mecanismo para la destrucción de patógenos (Marikovsky *et al*, 2003; Costantini and Dell'Omo, 2006; Davies, 2011). Se ha determinado que el estrés oxidativo juega un papel principal en muchas patologías degenerativas y se considera que la formación y metabolismo de radicales libres es un mecanismo patobioquímico involucrado en la fase de iniciación o progresión de varias enfermedades (Surai *et al*, 2003).

Las deficiencias de minerales que son cofactores de enzimas con actividad antioxidante pueden causar estrés oxidativo, tal como sucede con el selenio, que en la forma de selenocisteína se encuentra en el sitio activo de la glutatión peroxidasa y la tioredoxina reductasa, con el zinc, cobre y manganeso, que son parte integral de las superóxido dismutasas, o el hierro, que es parte esencial de la catalasa (Surai *et al*, 2003). En condiciones de estrés oxidativo la producción de radicales libre se incrementa dramáticamente (Afzal and Armstrong, 2002); entonces, sin la suplementación dietaria de antioxidantes resulta difícil prevenir el daño a la mayoría de órganos y sistemas. Sin embargo, el desafío para el nutricionista es determinar la oportunidad y magnitud necesaria para esta suplementación y su justificación económica (Surai *et al*, 2003).

Koinarski *et al* (2005) evaluaron el efecto de *E. acervulina* sobre el estado antioxidante de pollos, administrando $3x10^5$ ooquistes esporulados de *E. acervulina* provenientes de un desafío de campo en los días 12, 14 y 16 de edad. Los investigadores observaron en las aves desafiadas menor ganancia de peso y eficiencia en la conversión alimentaria ($P<0.05$), un score de lesiones alrededor de 3 en la escala de Johnson y Reid (1970), 30% menor actividad enzimática de la superóxido dismutasa ($P<0.01$) y menor concentración de caroteno y de las vitaminas A y C.

Koinarski *et al* (2006a) evaluaron el efecto de la inclusión de un complejo glicitano de zinc ($2Gly.ZnCl_2.2H2O$) sobre el estado antioxidante de pollos de carne sometidos a un desafío por *E. acervulina*, encontrando un efecto positivo traducido en un menor nivel de malondialdehído y mayor ganancia de peso ($P<0.01$). Gabrashanska *et al* (2008) evaluaron el efecto del mismo compuesto en ratas infectadas experimentalmente con *Fasciola hepática* para inducir estrés oxidativo. Los investigadores confirmaron que la infección parasitaria indujo el estrés oxidativo y encontraron una respuesta favorable al tratamiento con glicinato de zinc evidente en la mayor actividad de las enzimas superóxido dismutasa y glutatión peroxidasa, la menor concentración de malondialdehído y mayores niveles de las vitaminas E y A y de zinc y selenio. Los investigadores, a pesar de no observar variación alguna en la carga parasitaria, observaron un incremento en la ganancia de peso en los animales afectados y tratados, por lo que se asocia que el mejor estado antioxidante influyó directamente en la ganancia de peso.

## 2.7 Aceite esencial de orégano y salud intestinal

### 2.7.1 Descripción del aceite esencial de orégano

El nombre común "orégano" reúne más de veinte diferentes especies de plantas, muchas de ellas empleadas con fines culinarios o para la elaboración de cosméticos, fármacos y licores, y con un olor característico. Una de las especies más explotadas es el *Origanum vulgare* (Arcila-Lozano *et al*, 2004).

Se conoce comúnmente como aceite esencial de orégano (AEO) al producto líquido, aromático y con características hidrofóbicas que se extrae del orégano y que está conformado por la mezcla natural de metabolitos secundarios de esta planta (Parry, 1922). Si bien se ha documentado la presencia de entre 30 y 50 metabolitos secundarios en el AEO, algunos de los principales son los fenoles carvacrol y timol, y los hidrocarburos monoterpenoides ρ-cimeno y γ-terpineno (Ruso *et al*, 1998). Estos metabolitos son a su vez indicadores del rendimiento de la planta, por cuanto a mayor relación (carvacrol + timol) / (ρ-cimeno + γ-terpineno) mayor es el rendimiento de aceite esencial (Baser, 2002)

Las sustancias contenidas en el AEO están clasificadas como Generalmente Reconocidas como Seguras (GRAS, Generally Recognized As Safe) por la Administración de Alimentos y Fármacos de los Estados Unidos (FDA, 2010).

Se ha demostrado que el AEO posee diferentes propiedades de interés para la avicultura; sin embargo, su actividad biológica depende del contenido y proporción de metabolitos secundarios, ya que cada una de estas sustancias posee un diferente grado de actividad biológica en cada mecanismo de acción que desarrolla el AEO (Martínez, 2010).

### 2.7.2 Propiedades del aceite esencial de orégano

#### 2.7.2.1 Acción sobre la mucosa intestinal

Greathead and Kamel (2006) evaluaron el efecto del carvacrol y timol, ambos metabolitos secundarios del orégano, sobre la estructura de la mucosa intestinal de pollos de 35 días de edad desafiados con *E. acervulina*, observando que las aves alimentadas con dietas conteniendo estos metabolitos del orégano presentaron una altura de la vellosidad intestinal 13% mayor, una profundidad de cripta 35% menor y una mejor relación entre la altura de la vellosidad y la profundidad de la cripta.

García *et al* (2007) evaluaron el efecto de los aceites esenciales de salvia, tomillo y romero en la alimentación de pollos observando a los 42 días de edad vellosidades intestinales 10% más largas y 28% mayor área superficial de absorción.

Se ha postulado que este efecto del AEO se basa en el incremento de la tasa de renovación celular de la mucosa intestinal (Bruerton, 2002; citado por Ferket, 2003) producido en parte por lo fenoles naturales contenidos en el orégano, manteniendo una población más saludable de enterocitos.

#### 2.7.2.2 Acción sobre la actividad de las enzimas digestivas

Se ha documentado para varios derivados fitogénicos, y entre ellos el AEO, el efecto estimulante de la actividad enzimática a nivel del sistema digestivo. Lee *et al* (2004) reportan que al evaluar el efecto del timol, metabolito secundario del orégano, en la alimentación de pollos de 21 días sobre la actividad de las enzimas pancreáticas, se observaron incrementos del 14%, 17% y 28% en la actividad enzimática de la quimotripsina, tripsina y lipasa, respectivamente. Jamroz *et al* (2005) por su parte observaron un incremento del 1.2% en la proporción del peso del músculo de la pechuga asociado a un incremento en la actividad lipasa a nivel pancreático e intestinal en pollos suplementados con una combinación de carvacrol, cinamaldehído y capsaicina. Asimismo, Jang *et al* (2007) observaron que la inclusión de una combinación de aceites esenciales de hierbas en la alimentación de pollos hasta los 35 días de edad incrementó la actividad enzimática de tripsina, α-amilasa pancreática y maltasa intestinal. Recientemente, Basmacioglu Malayoglu *et al* (2010) observaron que la inclusión del AEO en dietas a base de trigo y soya para pollos de 1 a 21 días de edad, incrementa significativamente la actividad de la quimotripsina en el sistema digestivo y la consecuente digestibilidad de la proteína dietaria.

#### 2.7.2.3 Acción antioxidante

El efecto antioxidante del AEO se debe en gran medida a la presencia de grupos hidroxilo en los compuestos fenólicos que actúan como donadores de hidrógeno para los radicales peróxido producidos durante la primera etapa en la oxidación de los lípidos, retardando por tanto la formación de peróxido de hidrógeno (Lee *et al*, 2004). Mediante métodos de captura del peróxido de hidrógeno y la prueba de rancidez, se ha logrado establecer que la actividad antioxidante del AEO es similar a la del α-Tocoferol y BHT en términos de su capacidad para inhibir la peroxidación de lípidos, protegiendo al ADN del daño oxidativo (Martínez-Tomé *et al*, 2001;

Faleiro *et al*, 2005). Si bien se ha reportado que la actividad anti-radical es atribuible principalmente al Carvacrol y Timol (Deighton *et al*, 1993), se ha establecido que al menos 14 sustancias presentes en el AEO poseen actividad antioxidante (Máthé, 2009; Applegate *et al*, 2010) entre las que destacan carvacrol, timol, ρ-cimeno, γ-terpineno, sabineno, linalol y β-cariofileno (Máthé, 2009).

#### 2.7.2.4 Acción antibacteriana

La acción antibacteriana del AEO ha sido documentada ampliamente. Son al menos 19 las sustancias presentes en el AEO que poseen actividad bactericida (Máthé, 2009). Al respecto, carvacrol y timol actúan desnaturalizando y coagulando de las proteínas de la membrana celular de la bacteria, incrementando su permeabilidad y produciendo pérdida de material celular y muerte de la bacteria (Lee *et al*, 2004). Se ha reportado la acción antibacteriana del AEO contra *C. perfringens, Salmonella entérica, Escherichia coli, Staphylococcus aureus*, entre otras (Martínez, 2010).

De los metabolitos secundarios contenidos en el AEO, particularmente los terpenoides, pueden alcanzar el espacio intracelular de las bacterias Gram-negativas a través de las porinas de la membrana celular (Helander *et al*, 1998) por su carácter lipofílico y coadyuvar en la acción antibacteriana (Lee et, 2004).

Los metabolitos secundarios contenidos en el AEO actúan afectando la captación bacteriana de nutrientes, la síntesis de ácido nucleico, (Lambert *et al*, 2001), la actividad respiratoria (Conner and Beuchat, 1984), la gradiente de pH, afectando en este último caso la bomba de protones $H^+$-APTasa y produciendo la pérdida de iones, ATP, ácidos nucleicos y aminoácidos como glutamato (Lambert *et al*, 2001).

#### 2.7.2.5 Acción prebiótica

Se considera como prebiótico a aquel "ingrediente del alimento no digestibles que afectan benéficamente al hospedero a través de la estimulación selectiva del crecimiento y/o actividad de de una o un limitado número de bacterias en el colon que puede mejorar la salud del hospedero" (Gibson and Roberfroid, 1995).

En este sentido, el AEO modifica la microflora intestinal (Ferket, 2003) debido a que su efecto antibacteriano es mayor contra especies patógenas que contra benéficas (Martínez, 2010). Diversos estudios *in vitro* demuestran que el AEO presenta una menor Concentración Inhibitoria Mínima para especies bacterianas patógenas como *Escherichia coli*, *Clostridium sp.*, y *Salmonella sp.*, en relación a especies benéficas como *Enterococcus sp.*, *Bifidobacterium sp.*, o *Lactobacillus sp.* (Hammer *et al*, 1999; Lee *et al*, 2004; Betancourt *et al*, 2011). Esta característica se evidencia también en cultivos bacterianos, en que el AEO produce halos de inhibición mayores contra *Salmonella sp.*, en comparación a *Enterococcus sp.* (Dorman and Deans, 2000). Esta actividad antibacteriana de manera diferenciada desarrolla un efecto de exclusión competitiva a nivel entérico favoreciendo el crecimiento relativo de la flora benéfica y, en consecuencia, la salud intestinal (Martínez, 2010). Zheng *et al* (2010) observaron en peces que la suplementación dietaria de AEO reduce significativamente el recuento de bacterias intestinales patógenas como *Enterobacteriaceae sp.*, *Aeromonas sp.* y *Bacteroides sp.* e incrementa el número de bacterias benéficas como *Bifidobacterium sp.*, *Lactobacillus sp.* y *Bacillus sp.* Asimismo, Betancourt *et al* (2011) evaluaron el efecto de la suplementación de AEO en la dieta de pollos de carne y observaron mayor proporción del género *Clostridium sp.* en las aves control y mayor proporción del género *Lactobacillus sp.* en las aves suplementadas con AEO. Finalmente, Yew (2008) reporta mayores recuentos de bacterias acido lácticas como *Lactobacillus sp.* y menores de patógenas como *Aeromonas sp.* en el tracto gastrointestinal de peces alimentados con AEO.

#### 2.7.2.6 Acción anticoccidial

El AEO incrementa la tasa de renovación celular de la mucosa intestinal (Levkut *et al*, 2011) y previene el ataque de coccidia manteniendo una población más saludable de enterocitos (Bruerton, 2002; citado por Ferket, 2003). El AEO acelera el proceso natural de descamación (Fabian *et al*, 2006), eliminando las células infectadas con esporozoitos antes de alcanzar la fase de merozoito, creando condiciones adversas para que la coccidia complete su ciclo de vida. Este mecanismo permite que una fracción pequeña de ooquistes sea eliminada al lumen intestinal, lo que favorece el desarrollo de la inmunidad contra la coccidia en el ave (Lillehoj and Trout, 1996). Se ha establecido que los fenoles tienen acción oocisticida (Williams, 1997); sin embargo, el principal efecto del AEO sobre el control de la coccidiosis parece estar mediado por su acción sobre la mucosa intestinal. A estos efectos se suma la acción antimicrobiana del AEO contra *C. perfringens* (Lee *et al*, 2004), patógeno ubicuo que en ciertas condiciones exacerba y predispone la coccidiosis en las aves (McDougald *et al*, 2008).

### 2.7.3 Calidad del aceite esencial de orégano

Es importante establecer que los criterios para evaluar la calidad del AEO dependerán del uso objetivo. En el caso de la industria avícola resulta de particular interés la composición de metabolitos activos del orégano en el AEO, la ausencia de sustancias contaminantes, la estabilidad del AEO en el agua de bebida y/o en el alimento balanceado y la actividad biológica en las aves destino (Martínez, 2010).

Si bien el AEO puede ser obtenido empleando métodos químicos o físicos, es recomendable que el proceso de extracción se realice por medio de arrastre de vapor, ya que así se evitará la presencia de contaminantes que pudieran afectar la actividad de los metabolitos secundarios del orégano o acarrear efectos no deseados en el producto final. Por otro lado, los metabolitos secundarios del orégano contenidos en el AEO son en su mayoría sustancias altamente volátiles e insolubles en agua; por ello, es necesario que el AEO sea incorporado en preparaciones que controlen la volatilidad de sus componentes y aseguren su estabilidad tras el proceso de peletización que se aplica al alimento balanceado, y en el agua de bebida, según sea el caso (Martínez, 2010).

En base a su composición química el orégano puede tener diferentes quimotipos (Máthé, 2009). El contenido y proporción de los metabolitos secundarios en esta planta depende de muchos factores, como son los factores climáticos, las prácticas agronómicas, la floración, la variedad genética, la altitud, época de cosecha y estado de desarrollo de la planta (Arcila-Lozano *et al*, 2004; Ariza, 2011). Se ha determinado rangos de variación común del contenido de estos metabolitos en el aceite esencial tan variables como de 10 a 90% para carvacrol, de 10 a 50% para timol y de 3 a 10% para ρ-cimeno o γ-terpineno (Ruso *et al*, 1998; Daferera *et al*, 2000; Plaus *et al*, 2001; Chorianopoulos *et al*, 2004; Muñoz Acevedo *et al*, 2007).

Adicionalmente, en términos de su actividad biológica y efectos, cada constituyente químico individual tiene sus propias cualidades características (Máthé, 2009), y se ha determinado que cada metabolito secundario del orégano posee un diferente grado de actividad en cada uno de los mecanismos que desarrolla el AEO (Adam *et al*, 1998). Esto implica que la actividad biológica final del AEO dependerá, en gran medida, del contenido de cada principal metabolito secundario y de la proporción que exista entre ellos (Janssen *et al*, 1987; Silva *et al*, 2010). Es entonces evidente la necesidad imperativa de emplear AEOs o productos derivados con características estandarizadas (Martínez, 2010).

# III. CARACTERÍSTICAS DEL PRODUCTO EMPLEADO

Como fuente de aceite esencial de orégano (AEO) se empleó el producto Orevitol®, desarrollado por la empresa CKM S.A.C. (Lima, Perú) a base de AEO. El producto se encuentra en dos presentaciones:

1) Orevitol®-M, en forma de polvo dispersable, diseñado para ser administrado en el alimento balanceado y contener AEO en cantidad suficiente para aportar 48 g de carvacrol por kg de producto (CKM, 2010b).

2) Orevitol®-L, en forma de solución acuosa, diseñado para ser administrado en el agua de bebida y contener AEO en cantidad suficiente para aportar 40 g de carvacrol por litro de producto (CKM, 2010a).

Evaluaciones realizadas han evidenciado las propiedades antimicrobianas del producto (UNMSM, 2009a; UNMSM, 2009b) y su estabilidad a la temperatura del proceso de peletización del alimento (UNMSM, 2010).

En adelante el producto será referido en todo los casos como PRO.

# IV. SECCIÓN EXPERIMENTAL

# EXPERIMENTO N° 1

# EFECTO DE UN PRODUCTO A BASE DE ACEITE ESENCIAL DE ORÉGANO SOBRE LA HISTOMORFOMETRÍA INTESTINAL Y LA CAPACIDAD DE ABSORCIÓN DE NUTRIENTES DE POLLOS DE CARNE

## 1. Introducción

El incremento de los costos de alimentación de las aves comerciales obliga a maximizar la eficiencia con que el ave utiliza los nutrientes dietarios. Para lograr este óptimo aprovechamiento del alimento es necesario asegurar una adecuada salud intestinal y un óptimo desarrollo de la capacidad de absorción de nutrientes.

Actualmente, para favorecer la calidad de la mucosa intestinal, la industria cuenta con diferentes estrategias, como es el uso de dietas a base de ingredientes altamente digestibles y/o con propiedades funcionales derivadas del uso de probióticos, acidificantes, prebióticos, entre otros.

Una nueva estrategia es el uso de aceites esenciales con contenidos específicos de metabolitos secundarios de plantas aromáticas para mejorar el desarrollo intestinal, entre los que destaca el aceite esencial de orégano (AEO). El objetivo de este experimento fue determinar el efecto que ejerce un producto a base de AEO (referido en adelante como PRO) sobre las características morfométricas de la mucosa intestinal a nivel histológico y la capacidad de absorción de nutrientes.

## 2. Materiales y métodos

### 2.1. Lugar y fecha

La crianza de las aves se llevó a cabo en las instalaciones del Programa de Aves de la Facultad de Zootecnia de la Universidad Nacional Agraria La Molina en Lima-Perú a mediados del 2010. El sacrificio de las aves y la toma de muestras intestinales se llevó a cabo en las instalaciones del Laboratorio de Patología Aviar de la Facultad de Medicina Veterinaria de la Universidad Nacional Mayor de San Marcos (FMV-UNMSM) en Lima, Perú. El procesamiento de las muestras histológicas se llevó a cabo en el Laboratorio de Procedimientos Histológicos del Departamento de Patología Humana del Hospital María Auxiliadora, perteneciente a la red de hospitales del Ministerio de Salud, Perú. La lectura de las láminas histológicas se realizó en el Laboratorio de Histología, Embriología y Patología Veterinaria de la FMV-UNMSM.

### 2.2. Periodo de crianza

La evaluación se llevó a cabo desde la recepción de las aves en las instalaciones de crianza hasta el día 14 de edad y el periodo requerido para el procesamiento de las muestras recabadas.

### 2.3. Instalaciones, equipos y materiales

Las aves estuvieron alojadas sobre material de cama reutilizado (Anexo 6) a razón de 21.4 pollos/$m^2$ (0.047 $m^2$/pollo). Durante los primeros 3 días de vida se colocó papel periódico para evitar el acceso directo de los pollos BB al material de cama, mientras que el alimento fue suministrado en bandejas plásticas y el agua de bebida en bebederos tipo tongo. A partir del cuarto día de edad el alimento y agua de bebida fue provisto en comederos y bebederos lineales, respectivamente.

La calefacción fue provista por un sistema eléctrico con resistencias y termostatos, y controlada de acuerdo a las recomendaciones de la línea genética (Cobb-Vantress, 2008b) empleando un termo-higrómetro digital con una aproximación de 0.1 °C para temperatura y 1 % para humedad. El agua de bebida fue potabilizada empleando 1 ml de hipoclorito de sodio al 4.5% por cada 10 L de agua.

Para el pesaje de los ingredientes mayores del alimento se utilizó una balanza digital con capacidad de 150 kg y aproximación de 0.02 kg, mientras que para el pesaje de los ingredientes de la premezcla se utilizó una balanza electrónica con capacidad de 6 kg y aproximación de 1 g. Se utilizó mezcladoras horizontales de cintas de 400 y 30 kg de capacidad para la mezcla de los ingredientes durante la preparación del alimento y para las premezclas, respectivamente.

Para la necropsia de las aves se empleó bisturí y tijeras y guantes quirúrgicos. Para el muestreo de las secciones intestinales se empleó guantes quirúrgicos, hojas de afeitar, frascos plásticos para muestras individuales y una solución de formaldehído al 4% bufferada con PBS pH 7.0 (solución salina de buffer fosfato), con la siguiente composición por L: 4 g de fosfato sódico monobásico, 6.5 g de fosfato sódico dibásico, 100 ml de formaldehído al 37 o 40% y agua destilada en c.s.p. 1 L (INS, 1997; Lu *et al*, 2008).

Las láminas histológicas fueron preparadas empleando alcohol etílico en diferentes concentraciones, parafina, xilol, albumina de Mayer (glicerina y clara de huevo en partes iguales), hematoxilina (C.I. 75290), eosina amarilla (C.I. 45380), sulfato doble de aluminio y potasio, agua corriente, óxido rojo de mercurio, ácido clorhídrico, amoniaco, acido acético glacial, bálsamo de Canadá, cápsulas histológicas plásticas (cassettes) y láminas porta y cubreobjetos.

La deshidratación de las muestras e infiltración de parafina se realizó empleando un procesador automático de tejidos marca Leica TP1020. Los bloques de parafina fueron preparados empleando un centro de inclusión marca Leica EG1150H. Para la solidificación de la parafina se empleó un centro de enfriamiento marca Leica EG1150C. Para el pesaje de los colorantes se empleó una balanza de precisión marca Cobos. La parafina fue calentada empleando una estufa marca Memmert. Para el corte de las muestras incluidas se empleó un micrótomo de rotación tipo Minott marca Leica RM2125RT. Para la colocación de cortes sobre portaobjetos se empleó un baño de flotación marca Medax Nagel KG Kiel. Para la verificación de las láminas se empleó un microscopio óptico de campo claro marca Zeiss con aumentos efectivos de 100X y 400X. La lectura de las láminas histológicas se realizó mediante un microscopio óptico marca Zeiss Axiostar.

### 2.4. Animales experimentales

Se empleó 128 pollos machos de la línea Cobb 500 (características en el Anexo 4). Las aves fueron asignadas de manera aleatoria a los tratamientos.

## 2.5. Tratamientos

Se evaluaron 6 tratamientos, de acuerdo al siguiente cuadro:

| Tratamiento | Sección intestinal | Alimento |
|---|---|---|
| 1 | Duodeno | Dieta basal |
| 2 | Duodeno | Dieta basal + 500 ppm de Orevitol® |
| 3 | Yeyuno | Dieta basal |
| 4 | Yeyuno | Dieta basal + 500 ppm de Orevitol® |
| 5 | Íleon | Dieta basal |
| 6 | Íleon | Dieta basal + 500 ppm de Orevitol® |

El producto indicado será referido, en adelante, como PRO.

## 2.6. Alimentación

Se suministró una dieta basal a base de maíz, soya y harina de pescado, complementada con aceite de pescado y aminoácidos sintéticos, y suplementada con una premezcla de vitaminas y minerales. La alimentación fue *ad libitum*. Las características de la dieta basal se presentan en el Cuadro 1. El aditivo indicado en los tratamientos dietarios fue suministrado de 1 a 14 días de edad a expensas del maíz.

## 2.7. Toma y preparación de muestras histológicas

Al final de la segunda semana de vida se sacrificó 8 pollos por tratamiento cortando las arterias carótidas con la subsecuente exanguinación (Mutus *et al*, 2006), de acuerdo al protocolo del LPA. Se tomó muestras intestinales dentro de los primeros 5 minutos después de la muerte para prevenir la autolisis de los ápices de las vellosidades (Hoerr, 2010) cortando una sección de 2 cm largo de cada sección del intestino, no removiéndose el contenido intestinal (Hoerr, 2010).

**Cuadro 1. Características de la dieta empleada en el Experimento 1.**

| Ingredientes | % |
|---|---|
| Maíz amarillo | 52.517 |
| Torta de soya | 26.699 |
| Harina de pescado | 14.940 |
| Aceite semirefinado de pescado | 2.024 |
| DL-Metionina | 0.190 |
| L-Lisina | 0.116 |
| Cloruro de colina | 0.085 |
| Fosfato dicálcico | 1.609 |
| Carbonato de calcio | 0.978 |
| Sal común | 0.361 |
| Marcador inerte [1] | 0.300 |
| Premezcla [2] | 0.085 |
| Antifúngico [2] | 0.085 |
| Antioxidante [2] | 0.013 |

| Nutriente | Aporte calculado |
|---|---|
| Energía metabolizable, Kcal/kg | 3028 |
| Proteína cruda, % | 26.72 |
| Lisina, % | 1.72 |
| Metionina + Cistina, % | 1.10 |
| Treonina, % | 1.05 |
| Triptófano, % | 0.29 |
| Calcio, % | 1.54 |
| Fósforo disponible, % | 0.67 |
| Sodio, % | 0.34 |
| Grasa total, % | 5.97 |
| Fibra cruda, % | 2.70 |

| Componente | Composición proximal [3], % |
|---|---|
| Humedad | 11.89 |
| Proteína total (N x 6.25) | 26.06 |
| Extracto Etéreo | 4.87 |
| Fibra Cruda | 2.12 |
| Cenizas | 7.13 |
| ELN [4] | 47.94 |

[1] Óxido crómico como marcador inerte.

[2] Premezcla de vitaminas y minerales Proapak 2A®. Composición: Retinol: 12’000,000 UI; Colecalciferol: 2’500,000 UI; DL α-Tocoferol Acetato: 30,000 UI; Riboflavina: 5.5 g; Piridoxina: 3 g; Cianocobalamina: 0.015 g; Menadiona: 3 g; Ácido Fólico: 1 g; Niacina: 30 g; Ácido Pantoténico: 11 g; Biotina: 0.15 g; Zn: 45 g; Fe: 80 g; Mn: 65 g; Cu: 8 g; I: 1 g; Se: 0.15 g; Excipientes c.s.p. 1,000 g. Antifúngico: Mold Zap®; antioxidante: Danox®

[3] Informes de ensayo 1235/2010 LENA y 1236/2010 LENA, Universidad Nacional Agraria La Molina.

[4] ELN = Extracto Libre de Nitrógeno (calculado).

Las muestras de cada sección del intestino delgado se tomaron aproximadamente en el centro de la distancia de cada sección (Smirnov *et al*, 2005), considerando: duodeno, desde el final de la molleja hasta el final del conducto pancreático y biliar; yeyuno, desde el final del conducto pancreático y biliar hasta el divertículo de Meckel; e íleon, desde el divertículo de Meckel hasta la división de los ciegos.

Para evitar la autolisis de los tejidos de la mucosa intestinal y preservar sus características morfométricas, las muestras fueron fijadas durante 24 horas en una solución bufferada de formaldehido al 4% en una proporción fijador-muestra de al menos 40 a 1 (Hib, 2001) para asegurar la calidad del proceso de fijación. Luego de transcurridas 24 horas se transfirió las muestras a una solución de alcohol metílico al 70% para evitar la sobrefijación por acción del formaldehído y la consecuente menor inmunopositividad (INS, 1997; Webster *et al*, 2009), para ser empleadas posteriormente en estudios de inmunohistoquímica en óptimas condiciones. Las muestras fueron conservadas en esta solución hasta ser procesadas en el Laboratorio de Procedimientos Histológicos.

En el laboratorio las muestras fueron impregnadas con parafina para lograr una consistencia adecuada para realizar los cortes. Debido a que la parafina es una sustancia hidrofóbica, las muestras fueron previamente deshidratadas empleando soluciones de alcohol etílico de concentraciones progresivamente crecientes desde 70% hasta 100% en un periodo de 14 horas. Luego fueron sumergidas en xilol, como solvente intermediario (agente desalcoholizante), y colocadas en una estufa con parafina líquida entre 56 y 60 °C. Luego de impregnar las muestras, se dejó solidificar la parafina con las muestras de tejidos incluidas, obteniéndose los tacos histológicos adheridos finalmente a cápsulas histológicas plásticas (cassettes) empleando una espátula caliente.

Con un micrótomo se cortó láminas de aproximadamente 4 µm de grosor, suficientemente delgadas para permitir que puedan sean atravesarlas por la luz. Los cortes fueron extendidos en agua tibia y colocadas sobre portaobjetos previamente cubiertos con albúmina de Mayer. Finalmente, las láminas fueron coloreadas empleando el método de coloración de Harris empleando hematoxilina y eosina (HE) (INS, 1997). Luego de la coloración se realizó el montaje respectivo de las muestras empleando el bálsamo de Canadá.

## 2.8. Mediciones

Al término de la segunda semana de crianza se tomó al azar 8 pollos por tratamiento, en cada ave se muestreó las tres secciones intestinales y de cada una de ellas se obtuvo una muestra histológica con la que se preparó una lámina conteniendo 4 cortes seriados. En cada muestra y sección intestinal se realizó 10 mediciones por cada una de las 11 variables directas evaluadas, realizando en total 5280 mediciones, con lo que fue posible estimar otras variables derivadas. Se empleó aumentos efectivos de 40X para localizar los puntos de medición y las lecturas fueron realizadas con aumentos efectivos de 400X. Las variables en estudio fueron las siguientes:

### 2.8.1. Morfometría de la vellosidad intestinal

- Altura de vellosidad (AV):

  Fue medida, empleando vellosidades íntegras y perpendiculares a la pared intestinal, desde la base de la vellosidad (antes de la zona glandular) hasta el ápice de los enterocitos (µm).

- Grosor de la vellosidad (GV):

  El grosor basal de la vellosidad (GBV; µm) se determinó en la base de esta, y el grosor apical de la vellosidad (GAV; µm) antes de llegar a la zona cóncava del ápice. El grosor medio de la vellosidad o simplemente grosor de la vellosidad (GV) se calculó como el promedio de los dos anteriores (µm).

- Índice de forma (IFV):

  Fue calculado empleando la fórmula presentada a continuación, donde IFV: índice de forma de la vellosidad, AV: altura de la vellosidad (µm), GV: grosor de la vellosidad (µm).

$$IFV = \frac{AV}{GV}$$

- Grosor de la lámina propia (GLP):

  Fue calculado empleando la fórmula presentada a continuación, donde GLP: grosor de la lámina propia (µm), GALP: grosor apical de la lámina propia (µm) medido aproximadamente en el 90% de la longitud de la vellosidad, GBLP: grosor basal de la lámina propia (µm) medido en la base de la vellosidad.

$$\text{GLP} = \frac{\text{GALP} + \text{GBLP}}{2}$$

- Área superficial de la vellosidad (ASV):

  Fue estimada empleando una aproximación cilíndrica (Gookin *et al*, 2003, Solis de los Santos *et al*, 2007; Law *et al*, 2007) (pero modificada para no considerar las áreas de las bases) con la fórmula presentada a continuación, donde ASV: área superficial de la vellosidad ($mm^2$), AV: altura de la vellosidad (µm), GV: grosor de la vellosidad (µm).

$$\text{ASV} = \frac{3.1416 \times \text{AV} \times \text{GV}}{1000 \times 1000}$$

### 2.8.2. Características de la capa de enterocitos

- Altura del enterocito (AE):

  Se calculó empleando la fórmula siguiente, donde AE: altura del enterocito (µm), GV: grosor de la vellosidad (µm), GLP: grosor de la lámina propia (µm).

$$\text{AE} = \frac{\text{GV} - \text{GLP}}{2}$$

- Área de enterocitos en la vellosidad (AEV):

  Es el área de la capa de enterocitos en el corte longitudinal de la vellosidad intestinal. Se calculó empleando la fórmula siguiente, donde AEV: área de enterocitos en la sección de corte longitudinal de la vellosidad ($mm^2$/vellosidad), GV: grosor de la vellosidad (µm), GLP: grosor de la lámina propia (µm), AV: altura de la vellosidad (µm), GV: grosor de la vellosidad (µm).

$$\text{AEV} = \frac{\text{GV} - \text{GLP}}{2000} \times \frac{(2 \times \text{AV}) + \text{GV}}{1000}$$

- Densidad relativa de enterocitos (DRE):

  Es el porcentaje de área que la capa de enterocitos representa del área total de la sección de corte longitudinal de la vellosidad. Se calculó empleando la fórmula presentada a continuación (Anexo 12), donde DRE: densidad relativa de enterocitos (%), GV: grosor de la vellosidad (µm), GLP: grosor de la lámina propia (µm), AV: altura de la vellosidad (µm), GV: grosor de la vellosidad (µm).

$$\text{DRE} = (\text{GV} - \text{GLP}) \times \frac{(2 \times \text{AV}) + \text{GV}}{\text{AV} \times \text{GV}} \times 50$$

### 2.8.3. Estructura de la mucosa intestinal

- Densidad de las vellosidades (DV):

  Se estimó empleando la fórmula siguiente, donde DV: densidad de las vellosidades (N°/mm lineal de la muscularis) en la sección de corte transversal del intestino, GV: grosor de la vellosidad (µm), GC: grosor de la cripta (µm).

$$\text{DV} = \frac{1000}{\text{GV} + \text{GC}}$$

- Densidad de la mucosa (DM):

  Se define como el área ($mm^2$) de mucosa en la sección de corte longitudinal de la vellosidad por cada mm lineal de muscularis. Se estimó empleando la fórmula presentada a continuación (Anexo 12), donde DM: densidad de la mucosa ($mm^2$ de mucosa / mm lineal de muscularis), GV: grosor de la vellosidad (µm), AV: altura de la vellosidad (µm), GC: grosor de la cripta (µm) medido en la mitad de su longitud.

$$\text{DM} = \frac{\text{GV} \times \text{AV}}{1000 \times (\text{GV} + \text{GC})}$$

- Índice de estructura de la vellosidad (IEV):

  Se define como el área superficial de mucosa existente por unidad de área de muscularis en la base de la vellosidad. Se estimó empleando la fórmula presentada a continuación (Anexo 12), donde IEV: índice de estructura de la vellosidad, AV: altura de la vellosidad (µm), GV: grosor de la vellosidad (µm).

$$\mathrm{IEV} = \frac{4 \times \mathrm{AV}}{\mathrm{GV}}$$

- Índice de estructura de la mucosa (IEM):

  Se define como el área superficial de mucosa existente en las vellosidades por unidad de área de muscularis en el intestino. Se estimó empleando la fórmula presentada a continuación (Anexo 12), donde IEM: índice de estructura de la mucosa, GV: grosor de la vellosidad (µm), AV: altura de la vellosidad (µm), GC: grosor de la cripta (µm) medido en la mitad de la longitud.

$$\mathrm{IEM} = \frac{3.1416 \times \mathrm{GV} \times \mathrm{AV}}{(\mathrm{GV} + \mathrm{GC})^2}$$

### 2.8.4. Indicadores de proliferación celular

- Largo de la cripta (LC):

  Fue medido desde la abertura de la cripta en el inicio de la zona glandular (en la base del epitelio luminal) hasta la muscularis mucosae (µm).

- Grosor de la cripta (GC):

  Se realizó en el 50% de la longitud de la cripta.

- Área superficial de la cripta (ASC):

  Se calculó empleando la fórmula presentada a continuación, donde ASC: área superficial de la cripta ($mm^2$), LC: largo de la cripta (µm), GC: grosor de la cripta (µm).

$$\mathrm{ASC} = \frac{3.1416 \times \mathrm{LC} \times \mathrm{GC}}{1000 \times 1000}$$

- Relación altura de vellosidad / largo de cripta (AV/LC):

  Se calculó empleando la fórmula presentada a continuación, donde AV/LC: relación entre la altura de la vellosidad y el largo de la cripta, AV: altura de la vellosidad (µm), LC: largo de la cripta (µm).

$$\mathrm{AV/LC} = \frac{\mathrm{AV}}{\mathrm{LC}}$$

- Relación entre las áreas superficiales de la vellosidad y la cripta (ASV/ASC):

  Se determinó empleando la fórmula presentada a continuación, donde ASV/ASC: relación entre las áreas superficiales de la vellosidad y de la cripta, AV: altura de vellosidad (µm), GV: grosor de la vellosidad (µm), LC: largo de la cripta (µm), GC: grosor de la cripta (µm)

$$\mathrm{ASV/ASC} = \frac{\mathrm{AV} \times \mathrm{GV}}{\mathrm{LC} \times \mathrm{GC}}$$

- Índice de mitosis en la cripta (MIT):

  Se calculó empleando la fórmula presentada a continuación, donde MIT: índice de mitosis en la cripta (N° de mitosis / $mm^2$ de área superficial de la cripta), NM: número de mitosis por cripta, LC: largo de la cripta (µm), GC: grosor de la cripta (µm).

$$\mathrm{MIT} = \frac{\mathrm{NM} \times 1000 \times 1000}{3.1416 \times \mathrm{LC} \times \mathrm{GC}}$$

### 2.8.5. Dinámica de la mucina intestinal

- Densidad de células caliciformes (DACC, DRCC):

  La densidad absoluta de células caliciformes (DACC) se determinó por conteo directo y se expresa en número de células caliciformes por vellosidad (N°/vellosidad). La densidad relativa de células caliciformes (DRCC; número de células presentes por unidad de área de la sección de corte longitudinal de la vellosidad) se calculó empleando la fórmula presentada a continuación, donde

DRCC: densidad relativa de células caliciformes (N°/mm$^2$), AV: altura de vellosidad, GV: grosor de la vellosidad.

$$\mathrm{DRCC} = \frac{\mathrm{DACC} \times 1000 \times 1000}{\mathrm{AV} \times \mathrm{GV}}$$

- Largo de la célula caliciforme (LCC):

  Se consideró como tal el diámetro mayor de la célula caliciforme (µm) medido desde la abertura luminal de la célula hasta el lugar de constricción en dirección a la lámina propia.

- Grosor de la célula caliciforme (GCC):

  Se consideró como tal el diámetro menor de la célula caliciforme (µm) medido en el punto medio del diámetro mayor.

- Área de la célula caliciforme (ACC):

  Se refiere al área de la célula caliciforme en la sección de corte longitudinal de la vellosidad. Se calculó empleando una aproximación elíptica (Smirnov *et al*, 2004) con la fórmula siguiente, donde ACC: área de la célula caliciforme (µm$^2$), LCC: largo de la célula caliciforme (µm), GCC: grosor de la célula caliciforme (µm).

$$\mathrm{ACC} = 0.7854 \times \mathrm{LCC} \times \mathrm{GCC}$$

- Área absoluta de mucina (AAM):

  Es el área total de mucina que se encuentra en la sección de corte longitudinal de la vellosidad. Se calculó con la fórmula presentada a continuación, donde AAM: área absoluta de mucina (µm$^2$/vellosidad), ACC: área de la célula caliciforme (µm$^2$), DACC: densidad absoluta de las células caliciformes (N°/vellosidad).

$$\mathrm{AAM} = \mathrm{ACC} \times \mathrm{DACC}$$

- Área relativa de mucina (ARM):

  El área relativa de mucina (ARM) se refiere al porcentaje de área que representan las células caliciformes en la sección de corte longitudinal de la vellosidad y se calculó empleando la fórmula presentada a continuación, donde ARM: área relativa de mucina (%), AAM: área absoluta de mucina (µm$^2$/vellosidad), AV: altura de vellosidad (µm), GV: grosor de la vellosidad (µm).

$$\text{ARM} = \frac{\text{AAM} \times 100}{\text{AV} \times \text{GV}}$$

### 2.9. Diseño estadístico

Se empleó el Diseño Completamente al Azar con arreglo factorial 3 (sección intestinal) x 2 (nivel de inclusión del aditivo) con 8 repeticiones por tratamiento aplicando el procedimiento GLM del programa Statistical Analysis System SAS 9.0 (SAS Institute, 2009). Las diferencias entre medias se evaluaron mediante la prueba de Duncan (1955). El Modelo Aditivo Lineal General aplicado a las variables evaluadas fue el siguiente:

Y ijk = U + Ei + Aj + E*A (i, j) + Eijk

i = 1, 2, 3, j = 1, 2 k=1, 2, 3, 4, 5, 6, 7, 8

Donde:

| | |
|---|---|
| Yijk | Observación en la i-ésima sección intestinal recibiendo el j-ésimo nivel de inclusión del aditivo, en la k-ésima repetición |
| U | Media general |
| Ei | Efecto de la i-ésima sección intestinal |
| Aj | Efecto del j-ésimo nivel de inclusión del aditivo |
| E*A (i,j) | Efecto de la interacción de la i-ésima sección intestinal por el j-ésimo nivel de inclusión del aditivo |
| Eijk | Error experimental |

Siendo reducido el tamaño de muestra por la existencia de unidades perdidas (Anexo 13) se consideró significativos los valores con $P < 0.10$ (Crespo and Esteve-Garcia, 2001; Zhu *et al*, 2003, Kilburn and Edwards, 2004; van Nevel *et al*, 2005; Tahir *et al*, 2008; Teeter *et al*, 2008) o menores de 0.05, según se indica en cada caso.

Para una mayor amplitud del análisis se comparó de los resultados del uso o no del PRO independientemente en cada sección intestinal, empleando en todos los casos, e independientemente de la existencia o no de homogeneidad de varianzas, la prueba-t por el método Satterthwaite (Satterthwaite, 1946), mediante el procedimiento TTEST con aproximación SATTERTHWAITE del programa SAS (SAS Institute, 2009), que no asume homogeneidad en las varianzas y permite un mejor control del error Tipo I en comparación a la determinación inicial de la homogeneidad de varianzas y posterior elección de la prueba-t sólo en caso ésta exista (Ruxton, 2006).

## 3. Resultados y discusión

La metodología propuesta para la evaluación histológica de la mucosa intestinal brinda mayor amplitud en su análisis, permitiendo delinear los probables mecanismos involucrados en la respuesta observada.

### 3.1. Morfometría de la vellosidad intestinal

Los resultados de este conjunto de variables se presentan en el Cuadro 2 y Anexo 13. La altura de la vellosidad intestinal es significativamente diferente en las tres secciones del intestino (P<0.01), siendo mayor en el duodeno y menor en el íleon. Según los resultados observados, la altura de la vellosidad intestinal decrece linealmente en dirección al íleon, observándose reducciones de alrededor del 30% en cada sección respecto a la anterior, entre duodeno, yeyuno e íleon (Cuadro 2). Esto coincide con lo reportado por Awad *et al* (2006), quienes observaron vellosidades 36% más cortas en el yeyuno que en el duodeno, y con lo reportado por Pelicano *et al* (2005) y Khambualai *et al* (2010) quienes observaron reducciones progresivas en dirección al íleon del 21% y 35%, respectivamente, en relación a la sección inmediata anterior.

Por otro lado, se observa mayor altura de vellosidad (P<0.01) por efecto del PRO, independientemente de la sección intestinal, incrementándola en 15% respecto a las aves control (Cuadro 2). Este hallazgo es consistente con lo reportado por García *et al* (2007) quienes observaron un incremento del 10% en la altura de la vellosidad

**Cuadro 2. Efecto de un producto a base de aceite esencial de orégano sobre la morfometría de la vellosidad intestinal.**

| | | | Morfometría de la vellosidad intestinal | | | | | | |
|---|---|---|---|---|---|---|---|---|---|
| | | | Altura de vellosidad (µm) | Grosor de la vellosidad: Apical (µm) | Grosor de la vellosidad: Basal (µm) | Grosor de la vellosidad: Medio (µm) | Índice de forma | Grosor de la lámina propia (µm) | Área superficial de la vellosidad ($mm^2$) |
| -------------- Factores en estudio --------------- | | | AV | GAV | GBV | GV | IFV | GLP | ASV |
| Tratamiento | Sección | Aditivo | | | | | | | |
| 1 | Duodeno | Control | 1209 | 108 | 149 | 129 | 9.50 | 51.4 | 0.4938 £ |
| 2 | | PRO | 1420 | 120 | 181 | 150 | 9.00 | 57.1 | 0.6950 $ |
| 3 | Yeyuno | Control | 808 | 115 | 142 | 129 | 6.80 | 57.2 | 0.3300 €# |
| 4 | | PRO | 1033 | 112 | 135 | 121 | 8.52 | 47.8 | 0.3963 £€ |
| 5 | Íleon | Control | 568 | 110 | 116 | 113 | 5.24 | 44.2 | 0.2050 # |
| 6 | | PRO | 647 | 101 | 112 | 106 | 6.40 | 44.5 | 0.2112 # |
| P (sección*aditivo) | | | 0.3649 | 0.6136 | 0.2560 | 0.3632 | 0.1904 | 0.3122 | 0.0918 |
| Efecto de la sección intestinal | | Duodeno | 1286 a | 113 | 162 a | 138 a | 9.29 a | 53.8 | 0.5670 a |
| | | Yeyuno | 901 b | 114 | 139 a | 125 ab | 7.52 b | 53.3 | 0.3576 b |
| | | Íleon | 602 c | 106 | 114 b | 110 b | 5.74 c | 44.3 | 0.2077 c |
| P (sección) | | | < .0001 | 0.7038 | 0.0010 | 0.0408 | <.0001 | 0.1052 | <.0001 |
| Efecto del aditivo (1) | | Control | 848 b | 111 | 135 | 123 | 7.09 | 50.9 | 0.3367 b |
| | | PRO | 982 a | 110 | 141 | 125 | 7.88 | 49.5 | 0.4019 a |
| P (aditivo) | | | 0.0011 | 0.9972 | 0.5090 | 0.7928 | 0.1176 | 0.7745 | 0.0148 |

(1) Aditivo: Control: sin aditivo; PRO: 500 ppm de Orevitol®.

a,b,c,$,£,€,# Promedios significativamente diferentes no comparten la misma letra (a,b,c; P<0.05) o el mismo símbolo ($,£,€,#; 0.05<P<0.10)

intestinal en aves que reciben una combinación de aceites esenciales de orégano, canela y tomillo, así como con los resultados de Roldán (2010) quien observó incremento del 16%, 15% y 10% cuando suministraron a las aves aceites esenciales de albahaca, romero y tomillo, respectivamente, siendo timol el principal componente de éste último, el cual se encuentra también en el AEO.

De acuerdo a los resultados observados el PRO incrementa la altura de la vellosidad en 17, 28 y 14% en el duodeno, yeyuno e íleon, respectivamente (Cuadro 2). Este incremento en la altura de la vellosidad intestinal, con mayor énfasis en el yeyuno brinda una capacidad de absorción de nutrientes aún mayor de aquellos que son principalmente absorbidos en esta sección intestinal, como aminoácidos (Moe *et al*, 1987), péptidos (Michalski and Weinberg, 1999), lípidos (Gurr *et al*, 1989), minerales como el zinc (Suchithra *et al*, 2007; Yu *et al*, 2008), fosfatos (Gaynor and Cornejo, 2002) y bicarbonato (Kimchi *et al*, 2011), y vitaminas como carotenoides, vitamina A, ácido fólico, colina (Hoffmann-La Roche, 1989) y biotina (Bowman and Rosemberg, 1987).

Asimismo, el incremento en la longitud de las vellosidades intestinales favorece directamente la capacidad de absorción de nutrientes ya que este proceso se produce principalmente en el tercio apical de la vellosidad (Stange and Dietschy, 1983; Pappenheimer and Michel, 2003).

La longitud de la vellosidad es resultado de la tasa de proliferación celular (Cera, 1988) y la tasa de descamación celular (Martínez, 2010). Son diferentes los mecanismos que ejerce el AEO y que favorecen el incremento de la longitud de la vellosidad intestinal. El rol fundamental que cumple el equilibrio de la flora microbiana sobre la salud intestinal ha sido ampliamente documentado (Ferket *et al*, 2005; Yegani and Korver, 2008) y se ha establecido que el control de la flora patógena reduce la presencia de toxinas asociadas con cambios en la morfometría intestinal como vellosidades de menor longitud (Xu *et al*, 2003). Al respecto, diferentes estudios han demostrado la actividad antimicrobiana del AEO (Conner and Beuchat, 1984; Helander *et al*, 1998; Lambert *et al*, 2001; Lee *et al*, 2004; Máthé, 2009) así como su efecto prebiótico favoreciendo el crecimiento relativo de la flora intestinal benéfica (Ferket, 2003; Jamroz *et al*, 2005) por medio de un mayor control antimicrobiano de la flora intestinal patógena (Hammer *et al*, 1999; Dorman and Deans, 2000; Lee *et al*, 2004; Yew, 2008; Zheng *et al*, 2010). Por otro lado, se ha demostrado en pollos que fitonutrientes como carvacrol son potentes moduladores del control transcripcional de linfocitos intraepiteliales a nivel intestinal, ejerciendo influencia sobre la inmunidad, metabolismo y fisiología del ave por alteración de la expresión de genes asociados con la resistencia del huésped contra patógenos (Kim *et al*, 2010). Se ha establecido, además, que el carvacrol regula la expresión génica favoreciendo la expresión de 26 genes y reduciendo la expresión de otros 48, algunos de ellos relacionados con el desarrollo de tejidos (Lillehoj *et al*, 2011), lo que explicaría su influencia en el incremento de la longitud de la vellosidad intestinal. Por otro lado, se ha postulado que el efecto del AEO se basa en el incremento de la tasa de renovación celular de la mucosa intestinal (Bruerton, 2002; citado por Ferket, 2003) producido en parte por los fenoles naturales contenidos en el orégano, manteniendo una población más saludable de enterocitos.

El grosor apical de la vellosidad intestinal no muestra diferencias por efecto de su localización en el intestino; sin embargo, se observa diferencias significativas en el grosor de la vellosidad cuando este es medido a nivel basal ($P<0.01$), siendo éste último mayor en el duodeno y yeyuno respecto al íleon, así como el grosor medio ($P<0.05$), siendo mayor en el duodeno en relación al íleon. En consecuencia, la vellosidad tiene una forma más rectangular y menos triangular cuanto más se aproxima al íleon. No se observa diferencias estadísticamente significativas en ninguna de las mediciones del grosor de la vellosidad por efecto del PRO.

Respecto al índice de forma de la vellosidad se observa que éste es significativamente diferente ($P<0.01$) en las tres secciones intestinales, siendo mayor en el duodeno y menor en el íleon; sin embargo, no se observan diferencias significativas por efecto del PRO. Esto indica que la vellosidad es proporcionalmente más ancha en el íleon (Cuadro 2); es decir, que mientras las vellosidades reducen su longitud en 53% de duodeno a íleon, el grosor se reduce sólo en 20%. Estos resultados son consistentes con los reportados por Awad *et al* (2006) quienes observaron valores de altura y grosor de la vellosidad correspondientes a un menor índice de forma en yeyuno que en el íleon en pollos de 42 días de edad. Al respecto, Khempaka *et al* (2011), administraron 1% de glutamina en la dieta a pollos de carne y observaron un incremento significativo en el grosor de la vellosidad pero no en la digestibilidad de nutrientes, por lo que debe considerase que un incremento en el grosor de la vellosidad no necesariamente conlleva a un incremento en la absorción de nutrientes.

Los resultados indican diferencias significativas en el área superficial de la vellosidad intestinal entre las tres secciones del intestino ($P<0.01$), siendo mayor en el duodeno y menor en el íleon; así como por efecto del PRO ($P<0.05$), siendo mayor cuando éste es administrado (Cuadro 2). Estos resultados concuerdan con los observados respecto a la longitud de la vellosidad intestinal. Las aves que reciben el PRO presentan vellosidades intestinales con 19% mayor área de superficie y en consecuencia mayor capacidad de absorción de nutrientes.

Los resultados de este conjunto de variables indican que la interacción entre los factores localización y PRO tiene un efecto significativo ($P<0.10$) solo en el área superficial de la vellosidad, encontrándose mayor área superficial principalmente en el las vellosidades del duodeno (Cuadro 2). Así, en el duodeno se observa un incremento del 40% en el área superficial de la vellosidad, lo que fue verificado tras realizar una comparación de las medias de las aves con y sin el PRO en cada sección intestinal de manera independiente mediante la prueba-t por el método Satterthwaite (Satterthwaite, 1946; SAS Institute, 2009) encontrando diferencias significativas ($P=0.0213$) en el área superficial de la vellosidad entre las aves con y sin el PRO sólo en el duodeno. Al respecto, Roldán (2010) observó mayor área superficial de la vellosidad intestinal sólo en el yeyuno en aves suplementadas con aceite esencial de romero y albahaca.

En el presente estudio no se observó variaciones estadísticamente significativas en el grosor de la lámina propia por efecto de la localización ni del PRO (Cuadro 2). Al respecto, Silva *et al* (2009) observaron un incremento significativo en el grosor de la lámina propia en pollos desafiados con coccidia, pero no en aquellas aves que además recibieron AEO y que presentaron la menor tasa de eliminación de ooquistes, concluyendo que el grosor de la lámina propia es un indicador indirecto de estado sanitario del ave y del desafío de patógenos, viéndose incrementado a consecuencia de estos. La ausencia de diferencias en el grosor de la lámina propia en nuestro experimento por efecto del PRO demuestra, entonces, que las aves no estuvieron sometidas a condiciones de desafío entérico, información que resulta importante para el análisis de los indicadores de proliferación celular como se verá más delante.

### 3.2. Características de la capa de enterocitos

Las características de la capa de enterocitos se presentan en el Cuadro 3 y Anexo 13. Se observan diferencias significativas en la altura del enterocito sólo por efecto de la localización en el intestino ($P<0.05$) siendo ésta 37 mayor en el duodeno que en el íleon (Cuadro 3). Esto concuerda con lo reportado por Geyra *et al* (2001), quienes encontraron aproximadamente 80% mayor altura de enterocitos en el duodeno que en íleon. Sin embargo, esta altura no presenta variación por efecto del PRO.

El área de enterocitos en la sección de corte de la vellosidad presenta diferencias significativas entre las 3 localizaciones del intestino ($P<0.01$), siendo mayor en el duodeno y menor en el íleon, así como por efecto del PRO ($P<0.05$), siendo mayor en presencia de éste. De acuerdo a los resultados observados, las vellosidades duodenales presentan tres veces más área de enterocitos en el duodeno que en el íleon; asimismo, la suplementación del PRO determina 27% mayor área de enterocitos en las vellosidades intestinales (Cuadro 3). Sin embargo, parece ser que estas diferencias, tanto por efecto de su localización como por efecto del PRO se explican por las variaciones en la altura de la vellosidad, ya que la densidad relativa de enterocitos no presenta diferencias significativas por efecto de la localización de la vellosidad ni por efecto del PRO (Cuadro 3). Finalmente, no se observa diferencias significativas por efecto de la interacción del PRO y la localización en este conjunto de variables evaluadas.

### 3.3. Estructura de la mucosa intestinal

Las características estructurales de la mucosa intestinal se presentan en el Cuadro 4 y Anexo 13. La densidad de las vellosidades intestinales presenta diferencias significativas sólo por efecto de su localización ($P<0.05$), siendo 20% mayor en el íleon que en el duodeno; mientras que la densidad de la mucosa presenta diferencias significativas entre las tres secciones del intestino ($P<0.01$), pero es mayor en el

**Cuadro 3. Efecto de un producto a base de aceite esencial de orégano sobre las características de la capa de enterocitos**

| | | | Características de la capa de enterocitos | | |
|---|---|---|---|---|---|
| | | | Altura del enterocito (µm) | Área de enterocitos en la sección de corte de la vellosidad ($mm^2$/vellosidad) | Densidad relativa de enterocitos (% del área de corte de la vellosidad) |
| ------------------ Factores en estudio ------------------- | | | AE | AEV | DRE |
| Tratamiento | Sección | Aditivo | | | |
| 1 | Duodeno | Control | 38.49 | 0.0999 | 62.9 |
| 2 | | PRO | 41.23 | 0.1494 | 64.7 |
| 3 | Yeyuno | Control | 35.71 | 0.0637 | 60.0 |
| 4 | | PRO | 36.42 | 0.0815 | 63.4 |
| 5 | Íleon | Control | 30.50 | 0.0384 | 60.7 |
| 6 | | PRO | 30.83 | 0.0420 | 62.2 |
| P (sección*aditivo) | | | 0.5340 | 0.1206 | 0.9505 |
| Efecto de la sección intestinal | | Duodeno | 41.90 a | 0.1180 a | 63.7 |
| | | Yeyuno | 36.01 ab | 0.0712 b | 61.4 |
| | | Íleon | 30.64 b | 0.0399 c | 61.3 |
| P (sección) | | | 0.0145 | <.0001 | 0.7317 |
| Efecto del aditivo (1) | | Control | 34.70 | 0.0661 b | 61.2 |
| | | PRO | 37.53 | 0.0838 a | 63.3 |
| P (aditivo) | | | 0.3382 | 0.0124 | 0.4164 |

(1) Aditivo: Control: sin aditivo; PRO: 500 ppm de Orevitol®.
a, b, c Promedios significativamente diferentes no comparten la misma letra (P<0.05).

**Cuadro 4. Efecto de un producto a base de aceite esencial de orégano sobre la estructura de la mucosa intestinal.**

| | | | Estructura de la mucosa intestinal | | | |
|---|---|---|---|---|---|---|
| | | | Densidad | | Índices de estructura | |
| | | | De las vellosidades (n° / mm de muscularis) | De la mucosa ($mm^2$ / mm de muscularis) | De la vellosidad | De la mucosa |
| ------------------- Factores en estudio -------------------- | | | DV | DM | IEV | IEM |
| Tratamiento | Sección | Aditivo | | | | |
| 1 | Duodeno | Control | 5.46 | 0.836 | 38.04 | 14.11 |
| 2 | | PRO | 4.74 | 1.016 | 36.97 | 14.11 |
| 3 | Yeyuno | Control | 5.81 | 0.557 | 27.17 | 9.86 |
| 4 | | PRO | 5.76 | 0.718 | 34.10 | 12.89 |
| 5 | Íleon | Control | 6.25 | 0.384 | 20.94 | 7.38 |
| 6 | | PRO | 6.35 | 0.423 | 25.65 | 8.47 |
| P (sección*aditivo) | | | 0.5812 | 0.2653 | 0.2996 | 0.1597 |
| Efecto de la sección intestinal | | Duodeno | 5.16 b | 0.902 a | 37.65 a | 14.11 a |
| | | Yeyuno | 5.79 ab | 0.624 b | 30.06 b | 11.12 b |
| | | Íleon | 6.29 a | 0.401 c | 22.96 c | 7.85 c |
| P (sección) | | | 0.0210 | <0.0001 | <0.0001 | <0.0001 |
| Efecto del aditivo (1) | | Control | 5.86 | 0.583 b | 28.37 £ | 10.31 b |
| | | PRO | 5.66 | 0.679 a | 31.49 $ | 11.61 a |
| P (aditivo) | | | 0.5106 | 0.0026 | 0.0947 | 0.0350 |

(1) Aditivo: Control: sin aditivo; PRO: 500 ppm de Orevitol®.

a,b,c,$,£,€,# Promedios significativamente diferentes no comparten la misma letra (a,b,c; P<0.05) o el mismo símbolo ($,£,€,#; 0.05<P<0.10

duodeno y menor en el íleon. A pesar de la mayor densidad con que las vellosidades se presentan en el íleon, la mayor longitud que estas tienen en el duodeno determina que la densidad de la mucosa sea mayor cuanto más proximal es su localización en el intestino (Cuadro 4). Estos resultados coinciden con lo reportado por Fischer da Silva *et al* (2007) y Pelicano *et al* (2007) quienes observaron mayor densidad de vellosidades cuanto más distal la sección intestinal, así como con los resultados de Marchini *et al* (2011) quienes encontraron mayor área de mucosa en el duodeno que en el yeyuno en pollos de 14 días de edad.

Se observa que la densidad de la mucosa es mayor ($P<0.01$) con el uso del PRO. Dado que el PRO no ejerce una influencia significativa sobre la densidad de las vellosidades (Cuadro 4), la mayor densidad de la mucosa observada en las aves con el PRO se debe principalmente a la mayor altura de sus vellosidades (Cuadro 2).

Por otro lado, el índice de estructura de la vellosidad (IEV) se define como el área superficial de mucosa que existe en la vellosidad por unidad de área de muscularis en la base de la misma vellosidad, mientras que el índice de estructura de la mucosa (IEM) se define como el área superficial de mucosa existente en las vellosidades por unidad de área de muscularis a través del intestino. Ambos índices presentan diferencias ($P<0.01$) entre las tres secciones intestinales, siendo mayores en el duodeno y menores en el íleon. De igual forma, los índices de estructura de la vellosidad y de la mucosa son mayores en las aves suplementadas con el PRO ($P<0.10$ y $<0.05$, respectivamente). Estos resultados evidencian estructuras más complejas cuanto más proximal es la sección intestinal, así como por efecto del PRO. Así, las aves suplementadas con el PRO presentan 3 veces más área superficial de mucosa por cada unidad de área en la base de la vellosidad.

Los resultados del índice de estructura de la mucosa muestran la misma tendencia observada en el índice de estructura de la vellosidad; sin embargo, los valores son menores debido a que se cuantifica también las criptas, que aunque no aportan área superficial de mucosa a este cálculo, sí la correspondiente área de muscularis. El IEV es una variable de gran utilidad para conocer la estructura de la mucosa por efecto de la propia vellosidad y está influenciada no sólo por la altura de la vellosidad sino también por su grosor y densidad. El IEM, por el contrario, permite conocer la estructura de la mucosa incluyendo el efecto de la cripta y estimar el área total de la mucosa en el intestino del ave.

Los resultados observados en los indicadores de estructura de la mucosa intestinal así como en el área superficial de la vellosidad evidencian el estímulo que ejerce el PRO sobre la disponibilidad de nutrientes, ya que la capacidad de absorción de nutrientes depende principalmente del área superficial de la mucosa (Mitchel and Moretó, 2006). Al respecto, Noy y Sklan (1995) indican que el área superficial representa el potencial de absorción, mientras que la toma de nutrientes es dependiente de la concentración de sustrato, la disponibilidad de transportadores y la tasa de recambio.

### 3.4. Indicadores de proliferación celular

Los resultados observados en las variables relativas a la proliferación celular se presentan en el Cuadro 5 y Anexo 13. Al respecto, no se observa interacción entre la sección intestinal y el uso del PRO. Se observan diferencias significativas en el largo ($P<0.10$) y grosor ($P=0.05$) de la cripta sólo por efecto de la localización, siendo mayores en el duodeno que en el íleon. Por otro lado, no se observa diferencias en el área superficial de la cripta por efecto del PRO, pero sí por efecto de su localización ($P<0.05$), siendo mayor en el duodeno que en íleon.

Las relaciones entre la altura de la vellosidad y el largo de la cripta, y entre las áreas superficiales de ambas estructuras presentan diferencias significativas entre las secciones del intestino ($P<0.01$), siendo mayores en el duodeno y menores en el íleon. Ambas relaciones son mayores ($P<0.05$) en aves suplementadas con el PRO. Finalmente, el índice de mitosis en la cripta presenta diferencias significativas ($P<0.10$) por efecto de la localización, siendo mayor en el íleon que en el duodeno. No se observan diferencias estadísticamente significativas por efecto del PRO. En la Imagen 1 se presenta una muestra fotográfica de mitosis en las criptas intestinales.

Teniendo presente los resultados observados, es importante considerar que el balance entre la proliferación celular por mitosis y la pérdida celular en el ápice de la vellosidad determinan el recambio constante y el mantenimiento de la longitud de la vellosidad (Maiorka et al, 2003) y por lo tanto, el crecimiento de la vellosidad ocurre

**Cuadro 5. Efecto de un producto a base de aceite esencial de orégano sobre los indicadores de proliferación celular**

| | | | Indicadores de proliferación celular | | | | | |
|---|---|---|---|---|---|---|---|---|
| | | | Largo de la cripta (µm) | Grosor de la cripta (µm) | Área en la superficie de la cripta ($mm^2$) | Relación entre altura de vellosidad y largo de cripta | Relación entre área superficial de vellosidad y cripta | Índice de mitosis en la cripta (n°/$mm^2$ de superficie de la cripta) |
| Factores en estudio | | | LC | GC | ASC | AV/LC | ASV/ASC | MIT |
| Tratamiento | Sección | Aditivo | | | | | | |
| 1 | Duodeno | Control | 186 | 57.3 | 0.03235 | 6.50 | 15.87 | 150 |
| 2 | Duodeno | PRO | 202 | 62.1 | 0.03926 | 7.43 | 18.90 | 149 |
| 3 | Yeyuno | Control | 183 | 55.6 | 0.03187 | 4.55 | 10.27 | 175 |
| 4 | Yeyuno | PRO | 173 | 53.5 | 0.02940 | 6.19 | 15.53 | 179 |
| 5 | Íleon | Control | 157 | 52.9 | 0.02598 | 3.77 | 6.73 | 212 |
| 6 | Íleon | PRO | 165 | 53.9 | 0.02797 | 4.19 | 8.50 | 203 |
| P (sección*aditivo) | | | 0.6695 | 0.4375 | 0.3038 | 0.5035 | 0.5953 | 0.9692 |
| Efecto de la sección intestinal | | Duodeno | 193 $ | 59.3 $ | 0.03549 a | 6.87 a | 17.08 a | 150 £ |
| | | Yeyuno | 179 $£ | 54.7 $£ | 0.03084 ab | 5.22 b | 12.18 b | 177 $£ |
| | | Íleon | 160 £ | 53.4 £ | 0.02683 b | 3.95 c | 7.55 c | 208 $ |
| P (sección) | | | 0.0978 | 0.0501 | 0.0160 | <.0001 | <.0001 | 0.0703 |
| Efecto del aditivo (1) | | Control | 174 | 55.2 | 0.02976 | 4.82 b | 10.71 b | 181 |
| | | PRO | 179 | 56.3 | 0.03194 | 5.71 a | 13.48 a | 179 |
| P (aditivo) | | | 0.7347 | 0.5717 | 0.3709 | 0.0291 | 0.0264 | 0.9108 |

(1) Aditivo: Control: sin aditivo; PRO: 500 ppm de Orevitol®.

a,b,c,$,£ Promedios significativamente diferentes no comparten la misma letra (a, b, c; P<0.05) o el mismo símbolo ($, £; 0.05<P<0.10).

**Imagen 1. Proliferación celular en la cripta.** Se observan las células en mitosis (flechas) en el epitelio de la cripta. Microscopía a 400 X. Coloración HE.

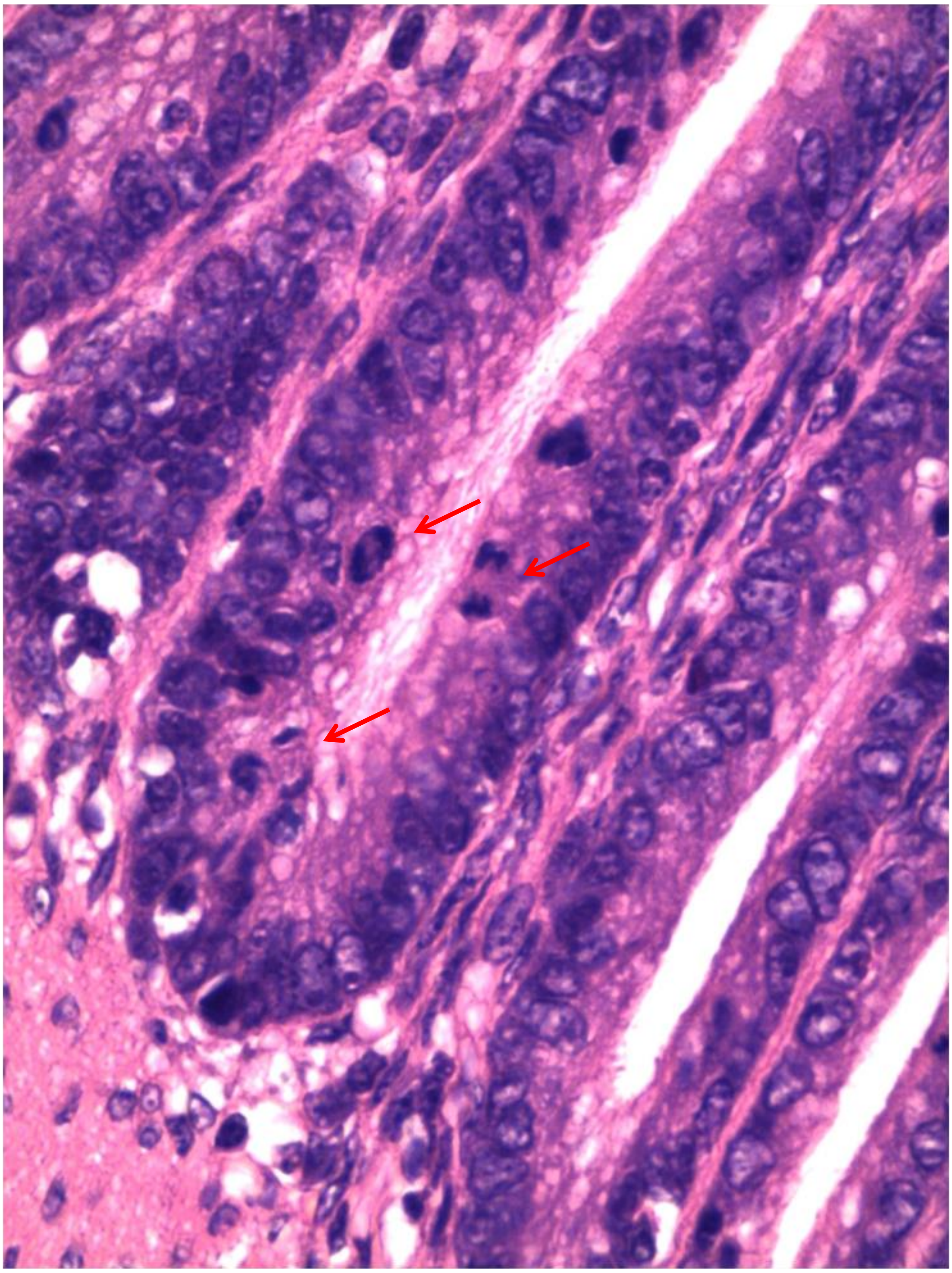

cuando la tasa de mitosis es mayor que la tasa de descamación de enterocitos, o cuando la tasa de descamación se reduce en relación a la de mitosis (Riccardi et al, 2011). Al respecto, resulta importante considerar que en las aves, a diferencia de los mamíferos, la proliferación celular se produce no sólo en la cripta sino también en la propia vellosidad intestinal (Uni et al, 1998; Uni, 2006; Levkut et al, 2011), tal como ocurre en los peces y reptiles (Ariza et al, 2011). Así, las divisiones mitóticas de las criptas son responsables de aproximadamente un 55% de la proliferación celular, la región media de las vellosidades de un 32% y la zona apical de un 8% (Applegate et al, 1994; citados por Roldán, 2010).

Para delinear los principales factores que influyen sobre la longitud de la vellosidad intestinal es importante determinar a su vez cuáles son los principales factores que influyen por un lado sobre la tasa de descamación y apoptosis celular y por otro sobre la proliferación celular. Al respecto, se ha establecido que la alta tasa de proliferación celular en el epitelio intestinal está regulada por la disponibilidad de nutrientes, la gastrina, la hormona del crecimiento, la actividad neuro-regulatoria, entre otros factores (Wilmore, 1997). Asimismo, se ha establecido que la microflora intestinal en aves convencionales promueve una mayor proliferación y recambio celular en comparación a las aves libres de microorganismos (Rolls et al, 1978); sin embargo, esta condición normal en el epitelio intestinal de las aves convencionales ha sido descrita como una "inflamación fisiológica" (Sprinz, 1962; citado por Rolls et al, 1978). En el presente estudio las aves estuvieron alojadas sobre material de cama y sometidas a condiciones normales de crianza, observándose una escasa frecuencia de vellosidades descamadas, por lo tanto se considera que la tasa de proliferación celular es el principal factor que habría influido sobre la longitud de la vellosidad (Maiorka et al, 2006). Por esta razón, en el presente estudio se considera como principales indicadores de la proliferación celular las características de la cripta, las relaciones existentes entre la vellosidad intestinal y la cripta, y finalmente la tasa de mitosis en la cripta para verificar la proliferación que al menos en esta región se estaría produciendo. El análisis conjunto de estas variables permite dilucidar aspectos importantes de la dinámica de la mucosa intestinal.

Una cripta de mayor profundad indica una mayor tasa de proliferación celular (Cannon, 2009) en respuesta a factores no sólo patológicos sino también fisiológicos.

Cuando se incrementa la tasa de descamación de enterocitos o la apoptosis celular se incrementa también la tasa de proliferación celular para su restitución (Zempleni, 2003), lo que incrementa la demanda de nutrientes y puede disminuir la eficiencia productiva del animal (Roldán, 2010) ya que la renovación de enterocitos representa una gran proporción de la tasa metabólica basal (Ferket, 2004), siendo la mucosa del tracto gastrointestinal el tejido que muestra la mayor tasa de recambio en el organismo (Podolsky *et al*, 1993). En aquellos casos en que el incremento del largo de la cripta esté asociado al incremento de la longitud de la vellosidad, el que la mayor capacidad potencial de absorción de nutrientes sea efectiva dependerá del estado de la propia vellosidad. Asimismo, dependerá de la relación entre la altura de la vellosidad y el largo de la cripta si la mayor capacidad de absorción de nutrientes supera o no la mayor demanda nutricional por efecto del incremento en la tasa de proliferación. Cuanto mayor sea, en términos proporcionales, el incremento en el largo de la vellosidad respecto al incremento en el largo de la cripta, mayor será también la probabilidad de tener una variación positiva en el balance energético del ave (Zempleni, 2003). Por el contrario, una vellosidad corta y una cripta profunda, pueden conducir a una menor absorción de nutrientes, a un incremento de la secreción en el tracto intestinal, a un mayor contenido de humedad en las heces, a la disbacteriosis, y exacerbación de coccidia y/o *Clostridium perfringens*, produciendo enteritis de diferentes tipos y menor eficiencia productiva.

En el presente estudio, las diferencias observadas en el largo y grosor de la cripta entre las secciones intestinales guardan relación con la altura de la vellosidad en estas secciones; sin embargo, dicha relación no es directamente proporcional, ya que la relación entre altura de vellosidad y largo de cripta (AV/LC) así como la relación entre las áreas superficiales de la vellosidad y la cripta (ASV/ASC) son diferentes en las tres secciones el intestino (Cuadro 5). Es posible entonces presumir que las mayores relaciones AV/LC y ASV/ASC observadas en el duodeno que en el íleon se deben a una mayor tasa de proliferación celular en el duodeno, ya que, como evidencian los resultados obtenidos, cuanto más proximal su localización, la proliferación debe ser tal que permita mantener una mayor longitud de vellosidad. Sin embargo, esta mayor proliferación celular requerida en el duodeno no se explica por la tasa de mitosis en la cripta ya que la actividad mitótica por $mm^2$ de área superficial registrada en la cripta es 38% mayor en el íleon que en el duodeno (Cuadro 5). Estos resultados coinciden con Marchini *et al* (2011) quienes observaron, en pollos de carne, que el porcentaje de células de la cripta en mitosis es 16% mayor en el íleon que en el duodeno.

Por otro lado, se observa que las aves que reciben el PRO presentan una mayor relación entre las longitudes (AV/LC) y las áreas (ASV/ASC) de la vellosidad y la cripta; sin embargo, no presentan diferencias en cuanto a las características de la cripta ni en cuanto a la tasa de mitosis en la cripta, lo que indicaría que en este experimento un factor importante para determinar la mayor longitud de la vellosidad en las aves que reciben el PRO es la mayor tasa de proliferación celular en la propia vellosidad. Al respecto, los resultados obtenidos por Levkut *et al* (2011) respaldan esta afirmación ya que dichos investigadores, tras administrar AEO a pollos de carne durante 42 días, determinaron 40% mayor proliferación celular en la propia vellosidad ($P<0.01$) mediante la técnica de determinación de PCNA (antígeno nuclear de células en proliferación). Esta mayor proliferación puede ser resultado de la regulación de la expresión génica (Zempleni, 2003) que ejercen sustancias contenidas en el AEO como el carvacrol (Lillehoj *et al*, 2011), en particular sobre el ciclo celular (Kim *et al*, 2010), o del leve incremento en la tasa de apoptosis celular de enterocitos observada con el AEO y el carvacrol (Fabian *et al*, 2006) y que también puede estar, al menos en parte, mediada por la influencia de este metabolito secundario sobre la expresión génica (Kim *et al*, 2010).

En relación a lo anterior, un incremento en la tasa de restitución de los enterocitos puede incrementar el costo nutricional y afectar el comportamiento productivo de las aves criadas sin condiciones particulares de desafío (Ferket, 2004). En el presente experimento, sin embargo, las aves fueron criadas en condiciones normales observándose 14% mayor peso y 10% menor conversión alimentaria en el día 14 de edad en las aves suplementadas con el PRO. Esto corrobora el carácter leve del incremento ejercido por el AEO sobre la apoptosis celular reportada por Fabian *et al* (2006), influyendo positivamente sobre la proliferación celular y la longitud de la vellosidad intestinal, y aumentando el área de absorción de nutrientes, lo que permitiría compensar un eventual incremento en el requerimiento de mantenimiento del ave por efecto del propio incremento en la apoptosis celular; sin embargo, es también posible que la mayor tasa de proliferación celular sea resultado no sólo de un incremento en la apoptosis celular inducida por el AEO, sino además consecuencia de su influencia sobre la regulación de la expresión génica.

Otros factores del AEO que coadyuvan al mantenimiento de una mayor relación de longitudes y áreas entre vellosidad y cripta son su acción antimicrobiana y prebiótica, así como su acción antioxidante. En relación a ello, se ha establecido la influencia de la flora patógena sobre la apoptosis celular (Xia and Talley, 2001) y que la disminución en la actividad microbiana en la digesta o a nivel del borde de cepillo de los enterocitos puede reducir el daño a estas células y la necesidad de renovación celular en el intestino (Hughes, 2003), así como que las especies reactivas de oxígeno son uno de los factores que influyen sobre la apoptosis celular en las vellosidades intestinales (Rosenfeld, 1998; Godlewski *et al*, 2005).

De manera complementaria, el análisis conjunto de los resultados observados en las áreas superficiales de la vellosidad (ASV) y de la cripta (ASC) (Cuadro 5), indican que la localización determina cambios estructurales en la mucosa intestinal tanto a nivel de vellosidad como de la cripta; mientras que los cambios ejercidos por el PRO se producen principalmente a nivel de la vellosidad intestinal. No obstante, tras realizar una comparación de las medias de las aves control y PRO (con y sin PRO, respectivamente) en cada sección intestinal de manera independiente, mediante la prueba-t y por el método Satterthwaite (Satterthwaite, 1946; SAS Institute, 2009), se verificó la existencia de una mayor área superficial en las criptas del duodeno en las aves suplementadas con el PRO (P=0.0784) lo que indica una mayor proliferación celular en las criptas del duodeno por efecto del PRO (Cuadro 5).

### 3.5. Dinámica de la mucina intestinal

La dinámica de la mucina intestinal se presenta en el Cuadro 6 y Anexo 13. Se observan diferencias significativas por efecto de la localización en la densidad absoluta ($P<0.05$) y relativa ($P<0.01$) de células caliciformes, siendo mayor la densidad absoluta en el duodeno que en el íleon, tal como se ilustra en la Imagen 2, mientras que la densidad relativa es diferente en las tres secciones y mayor en el íleon que en el duodeno. Se observan también diferencias significativas por efecto

**Cuadro 6. Efecto de un producto a base de aceite esencial de orégano sobre la dinámica de la mucina intestinal**

| | | | Dinámica de la mucina | | | | | | |
|---|---|---|---|---|---|---|---|---|---|
| | | | Densidad de células caliciformes | | Características de la célula caliciforme | | | Área de mucina | |
| | | | Absoluta (N° por vellosidad) | Relativa (N°/ mm² de vellosidad) | Largo (µm) | Grosor (µm) | Área (µm²) | Absoluta (µm² por vellosidad) | Relativa (% del área de la vellosidad) |
| ----------- Factores en estudio ------------- | | | DACC | DRCC | LCC | GCC | ACC | AAM | ARM |
| Tratamiento | Sección | Aditivo | | | | | | | |
| 1 | Duodeno | Control | 63.9 | 419 | 9.39 a | 9.03 | 67.14 | 4433 | 2.15 |
| 2 | | PRO | 116.8 | 521 | 8.58 ab | 8.72 | 58.80 | 6739 | 2.88 |
| 3 | Yeyuno | Control | 61.7 | 640 | 8.06 b | 8.06 | 51.57 | 3259 | 3.32 |
| 4 | | PRO | 103.0 | 864 | 9.54 a | 8.78 | 65.80 | 6836 | 5.66 |
| 5 | Íleon | Control | 52.6 | 836 | 7.63 b | 7.56 | 45.50 | 2484 | 3.78 |
| 6 | | PRO | 69.7 | 1061 | 7.85 b | 8.02 | 50.50 | 3856 | 5.48 |
| P (sección*aditivo) | | | 0.2093 | 0.7978 | 0.0337 | 0.3050 | 0.1029 | 0.2917 | 0.4860 |
| Efecto de la sección intestinal | | Duodeno | 83.1 a | 456 c | 9.05 a | 8.90 a | 63.67 a | 5208 a | 2.48 b |
| | | Yeyuno | 79.0 ab | 733 b | 8.68 a | 8.36 ab | 57.50 ab | 4749 a | 4.29 a |
| | | Íleon | 59.9 b | 934 a | 7.72 b | 7.58 b | 47.64 b | 2986 b | 4.51 a |
| P (sección) | | | 0.0189 | 0.0002 | 0.0071 | 0.0096 | 0.0132 | 0.0066 | 0.0044 |
| Efecto del aditivo (1) | | Control | 59.1 b | 642 b | 8.33 | 8.10 | 54.32 | 3319 b | 3.16 b |
| | | PRO | 93.3 a | 851 a | 8.61 | 8.48 | 57.88 | 5538 a | 4.73 a |
| P (aditivo) | | | 0.0001 | 0.0337 | 0.3743 | 0.2138 | 0.3813 | 0.0008 | 0.0049 |

(1) Aditivo: Control: sin aditivo; PRO: 500 ppm de Orevitol®.
a, b, c Promedios significativamente diferentes no comparten la misma letra (P>0.05).

**Imagen 2. Densidad de las células caliciformes.** Se observa una mayor densidad de células caliciformes (flechas) en el yeyuno (Y) que en el duodeno (D), y aun mayor en el íleon (I). Las tres son microscopías a 400 X. Coloración HE.

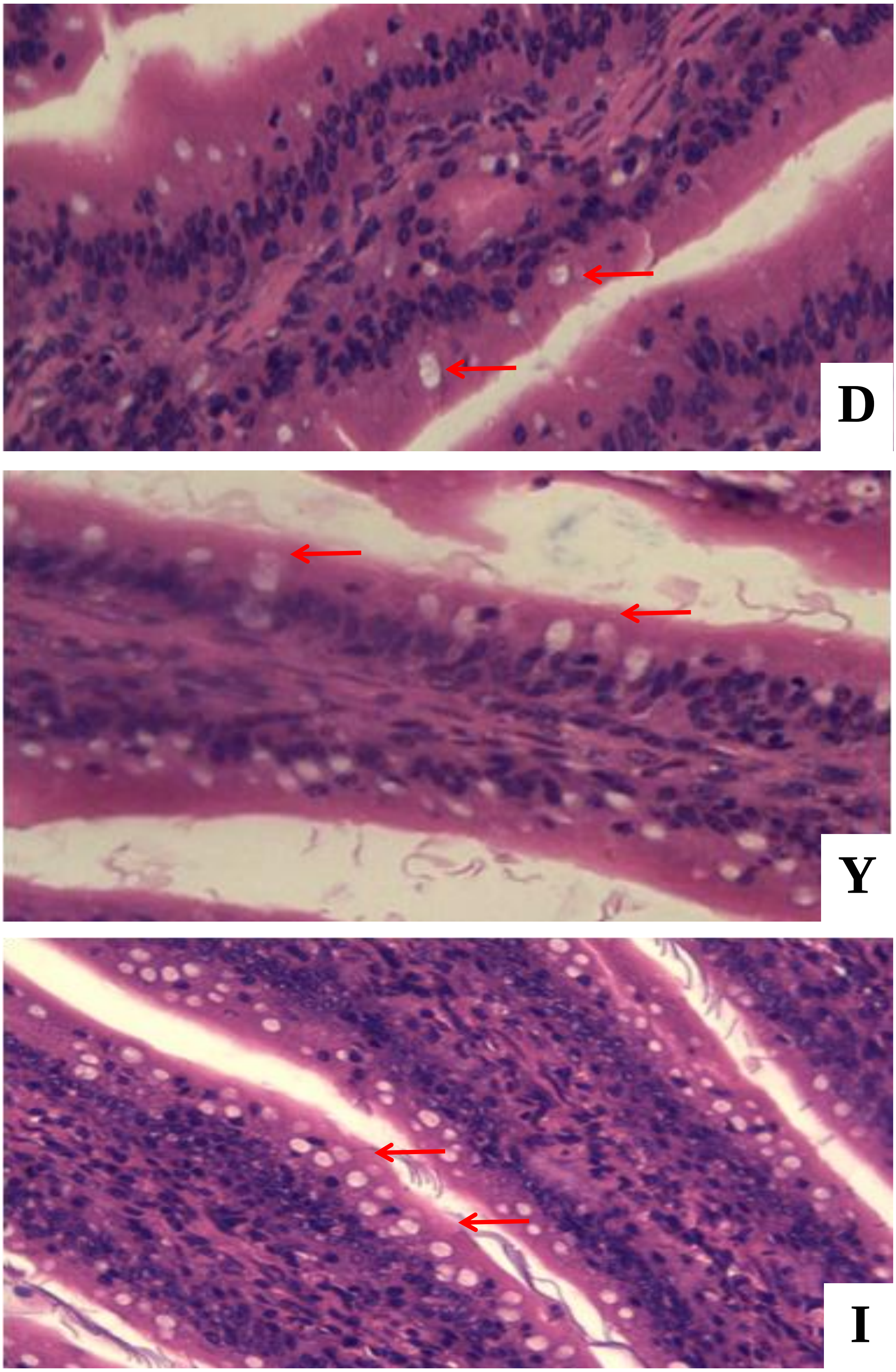

del PRO en las densidades absoluta ($P<0.01$) y relativa ($P<0.05$) de las células caliciformes, siendo mayor en ambos casos en presencia del PRO.

Las células caliciformes presentan diferencias significativas en su longitud por efecto de la localización ($P<0.01$), siendo mayor en el duodeno y yeyuno que en el íleon, pero no por efecto del PRO. Se observan, además, diferencias por efecto de la interacción de ambos factores ($P<0.05$), encontrándose mayor longitud de estas células en presencia del PRO sólo en el yeyuno, así como menor longitud de estas células en el yeyuno que en el duodeno sólo en ausencia del PRO. Por otro lado, se observan diferencias significativas en el grosor ($P<0.01$) y el área ($P<0.05$) de la célula caliciforme sólo por efecto de su localización en el intestino; siendo mayor, en ambos casos, en el duodeno que en el íleon, tal como se observa en la Imagen 3.

Finalmente, se observan diferencias significativas ($P<0.01$) en las áreas absoluta y relativa de mucina por efecto de la sección intestinal como por efecto del PRO. El área absoluta de mucina es mayor en el duodeno y yeyuno respecto al íleon, mientras que el área relativa de mucina es mayor en el yeyuno e íleon que en el duodeno.

Estos resultados indican que si bien en el duodeno existe mayor número de células caliciformes por vellosidad (DACC), esto se debe a la mayor longitud que las vellosidades intestinales tienen en esta sección del intestino (Cuadro 6). Además, existen diferencias importantes en la cantidad por $mm^2$ de vellosidad (DRCC); por lo cual, para tener una clara aproximación respecto a la producción de mucina, resulta conveniente determinar directamente el espesor de la capa adherente de mucina (Smirnov *et al*, 2004; Smirnov *et al*, 2005; Major *et al*, 2011) o emplear variables correlacionadas con la tasa de producción de mucina. Al respecto, en el presente estudio se consideró como principal indicador el área relativa de mucina (ARM), que indica el porcentaje del área de la sección de corte longitudinal de la vellosidad que corresponde al contenido de las células caliciformes, refleja las reservas de mucina y está correlacionada con el espesor de la capa de mucina en la superficie de la vellosidad (Smirnov *et al*, 2004). Otras variables como la densidad de células caliciformes fueron empleadas para establecer el origen de las diferencias observadas (Tanabe *et al*, 2005).

**Imagen 3. Tamaño de las células caliciformes.** Se observa que las células caliciformes (flechas) son de mayor tamaño en el duodeno (D) que en el íleon (I). Ambas son microscopías a 400 X. Coloración HE.

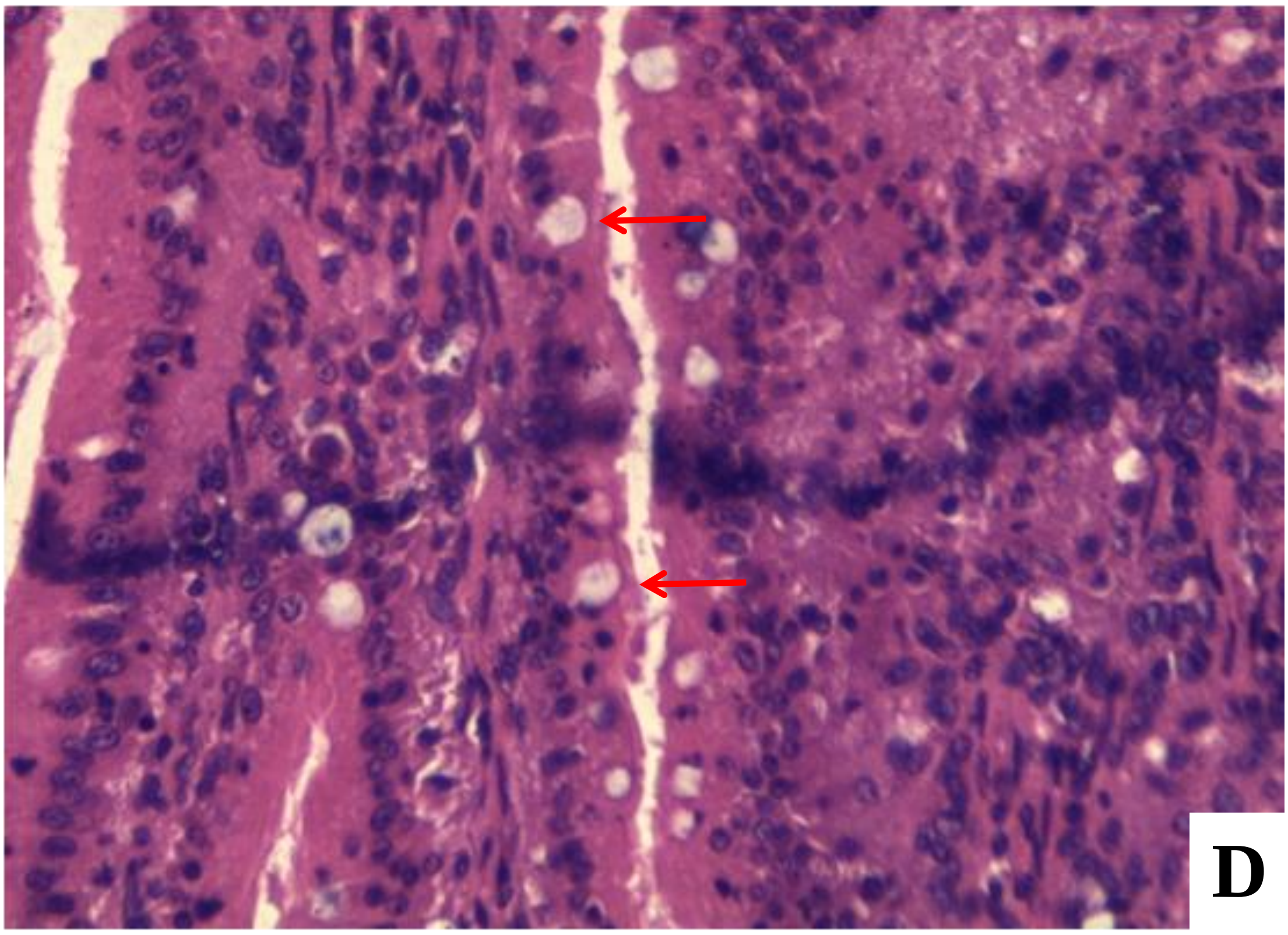

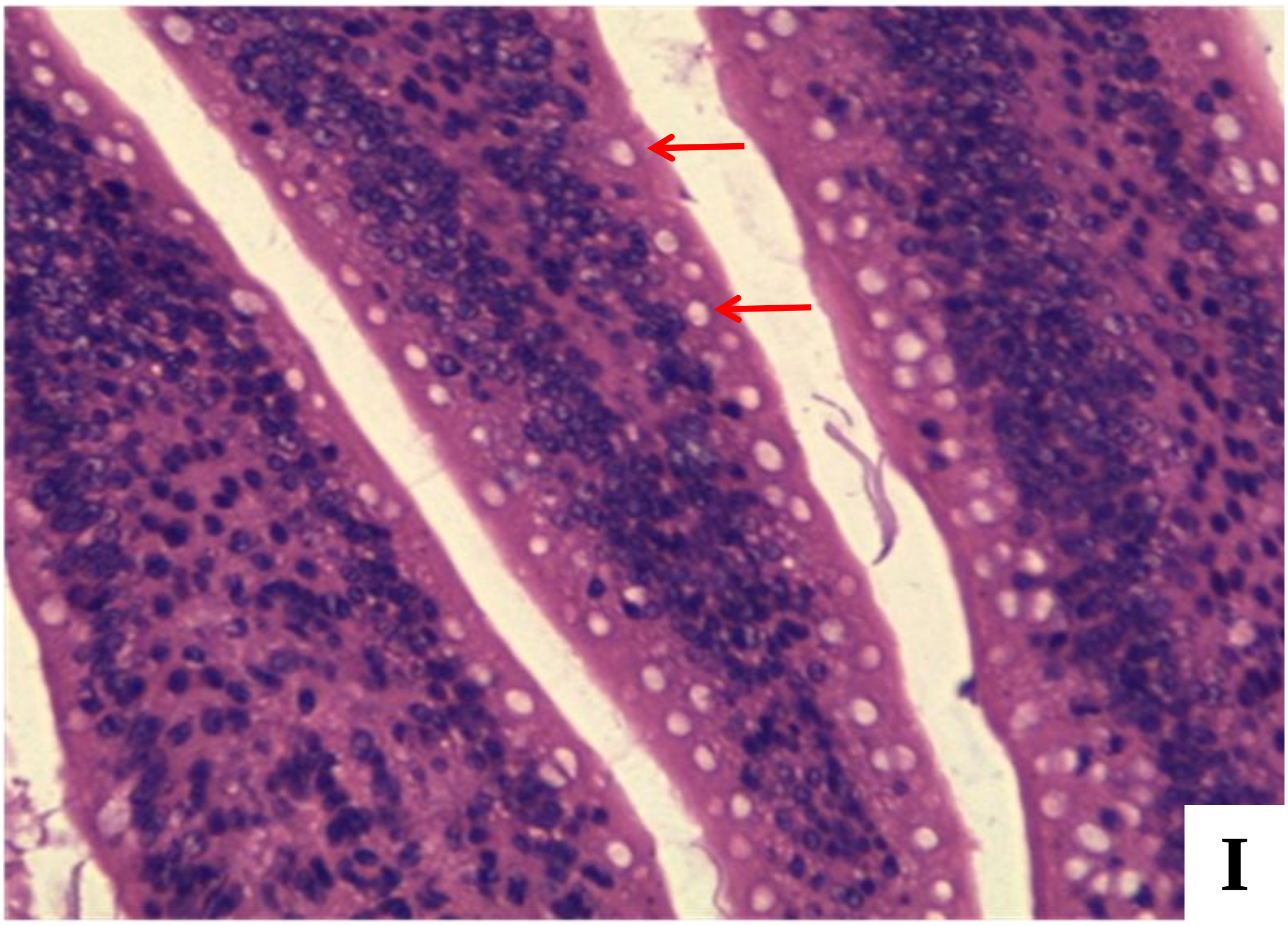

En el presente experimento se encontró que el número de células caliciformes por vellosidad (DACC) es mayor en el duodeno que en el íleon; sin embargo, el área relativa de mucina en el yeyuno e íleon es 73% y 82% mayor que en el duodeno, respectivamente (Cuadro 6). Esto es consistente con reportes previos que indican mayor cantidad de células caliciformes por vellosidad en el duodeno que en el íleon (Smirnova *et al*, 2003), pero mayores reservas y potencial de producción de mucina cuanto más distal su localización en el intestino delgado (Uni *et al*, 2003; Smirnov *et al*, 2004). Esta aparente contradicción se explica porque la vellosidad intestinal es más larga en el duodeno que en el íleon, por lo que la mayor cantidad de tejido en la vellosidad posibilita una mayor cantidad de células caliciformes, encontrándose 60% y 100% mayor número de células caliciformes por $mm^2$ de vellosidad (DRCC) en el yeyuno e íleon que en el duodeno, respectivamente (Cuadro 6). Estos hallazgos concuerdan con los reportes previos (Corcuera Pindado *et al*, 2005; Brady, 2011) y con el mayor índice de mitosis observado en las criptas del íleon.

La forma de la célula caliciforme no presenta diferencias, siendo constante la relación entre largo y ancho; sin embargo, su tamaño fue mayor en el duodeno que en el íleon (34% mayor área de la sección de corte de la célula caliciforme en el duodeno que en el íleon) (Cuadro 6). Ello concuerda con Smirnov *et al* (2005) quienes encontraron células caliciformes de mayor tamaño cuanto más proximal su localización en el intestino. Es decir, que la mayor producción potencial de mucina en el íleon que en el duodeno se debe a la mayor cantidad de células caliciformes por unidad de área de corte de la vellosidad, a pesar que en el íleon son más pequeñas.

Respecto al efecto del PRO (Cuadro 6), se observa un incremento significativo en las reservas de mucina en el intestino de las aves suplementadas con el AEO, del 50% y 67% al ser medida a través del área relativa (ARM) y absoluta (AAM) de mucina, respectivamente. Esto se explica por la presencia de 58% más células caliciformes por vellosidad (DACC) en las aves suplementadas con el PRO, ciertamente influenciado por la mayor longitud de la vellosidad en estas aves, así como por el incremento del 33% en el número de células caliciformes por unidad de área en el corte longitudinal de la vellosidad (DRCC). Estos resultados son consistentes con lo reportado por Reisinger *et al* (2011), quienes observaron que pollos suplementados con una combinación de aceites esenciales conteniendo AEO presentaron 30% mayor número de células caliciformes por vellosidad (DACC) y 23% mayor densidad relativa de células caliciformes (DRCC), y con lo reportado por Jamroz *et al* (2006) quienes administraron una combinación de carvacrol, cinamaldehído y capsaicina a pollos de carne hasta el día 42, observando un incremento significativo de mucus en el proventrículo y yeyuno, logrando un efecto protector que redujo la probabilidad de adhesión de *Escherichia coli* y *C. perfringens* al epitelio intestinal. Asimismo, el incremento en las reservas de mucina ha sido también observado con los antimicrobianos promotores de crecimiento tradicionales (Smirnov *et al*, 2005).

Es importante establecer que la secreción mucosa de las células caliciformes está compuesta por glicoproteínas mucínicas, inmunoglobulina A, agua y electrolitos; y si bien cumple una función lubricante y de protección contra agentes luminales como microorganismos, también constituye una reserva de otros productos secretorios de las células caliciformes que pueden contribuir al mantenimiento de la integridad de la mucosa incluyendo factores de naturaleza peptídica que regulan la tasa de recambio de la mucosa y la integridad estructural (Podolsky *et al*, 1993).

En el presente experimento el incremento en las reservas de mucina por efecto del PRO (Cuadro 6) está directamente correlacionado con la mayor proliferación celular evidenciada a través de los indicadores empleados (Cuadro 5). Asimismo, la mayor producción de mucina en el íleon se correlaciona directamente con el mayor índice de mitosis en esta sección. Esto hace pensar que los factores peptídicos referidos por Podolsky *et al* (1993) coadyuvan en el incremento observado en la proliferación celular por efecto del PRO.

Por otro lado, se ha establecido que las glucoproteínas de la mucina se unen a cadenas heterogéneas de polisacáridos de la pared bacteriana previniendo la adhesión de bacterias a las células epiteliales y la subsecuente invasión del tejido (Deplancke and Gaskins, 2001). Además, la capa de mucus adherente se une a las proteínas receptoras ligantes de mucina en las células epiteliales, previniendo el acceso de las bacterias a algunos receptores epiteliales (Slomiany *et al*, 2001). Incluso se ha reportado que la mucina posee factores protectivos ya que la capa mucosa acumula compuestos bactericidas y bacteriostáticos (Thompson and Applegate, 2005). Sin embargo, el gel mucoso también provee de un sustrato altamente nutritivo para las poblaciones microbianas (Deplancke and Gaskins, 2001).

Se ha establecido que el grosor de la capa adherente de mucina depende del balance entre la tasa de secreción y su degradación a través de digestión enzimática o remoción mecánica (Smirnov *et al*, 2004). Al respecto, se ha reportado que las bacterias y/o sus productos metabólicos pueden alterar la producción de mucina a través de cambios en la proliferación o forma las de células caliciformes, así como en la expresión génica de la mucina (Meslin *et al*, 1999), y se ha postulado que algunas cepas bacterianas probióticas actúan sobre la síntesis y secreción de mucina a través de la expresión del mRNA de la mucina (Smirnov *et al*, 2005) y de la producción de prostaglandina (Belley and Chadee, 1999; Shibolet et al, 2002). Sin embargo, es escasa la información respecto a si dichos cambios en la producción de mucina afectan también la propia estructura de los carbohidratos contenidos en ella (Smirnova *et al*, 2003), y si, en consecuencia, estos cambios son favorables al epitelio intestinal previniendo la colonización de patógenos, o si favorecen su instalación y la utilización de la mucina (Forder *et al*, 2005).

Por otro lado, se ha establecido que la respuesta inflamatoria mediada por las células T (Smirnov *et al*, 2004; Collier *et al*, 2008), como aquella desencadenada por el daño intestinal causado por una infección de coccidia, así como la propia toxina alfa producida por *C. perfringens* (Hofacre, 2007) estimulan la mucogénesis. Además, el AEO ejerce un efecto irritante sobre el epitelio intestinal por su contenido de fenoles, exacerbando la erosión en el ápice de la vellosidad (Schwartz, 2010), lo que podría implicar que el incremento en la producción de mucina en las aves que reciben el PRO (Cuadro 6) sea, al menos en parte, consecuencia de una respuesta inflamatoria.

Al respecto, Major *et al* (2011) indican que el incremento el grosor de la capa adherente de mucina observado en aves asociado a la suplementación de AEO puede deberse a un incremento en las tasas de producción de mucina y de recambio de la mucosa intestinal, y que el incremento en el grosor de la capa mucosa adherente tiene un efecto protector contra coccidia; sin embargo, se ha establecido también que la mayor producción de mucina favorece el crecimiento de bacterias mucolíticas y de *C. perfringens*, incrementando el riesgo de Enteritis Necrótica (Collier *et al*, 2008:

Hofacre, 23007), así como que dicho incremento en la producción de mucina podría actuar como una barrera luminar evitando el pasaje de algunos nutrientes a través de la pared intestinal (Esmail, 2011).

En el presente estudio, el efecto irritante del AEO puede ser una de las razones que influyan en la mayor proliferación celular; sin embargo, su efecto neto no resulta perjudicial para el epitelio intestinal ya que se observa una mejor estructura en la mucosa intestinal de las aves suplementadas con el PRO (Cuadro 4). Asimismo, el efecto neto del incremento observado en las reservas de mucina por efecto del PRO (Cuadro 6) resulta favorable al epitelio intestinal ya que la acción antibacteriana y prebiótica del AEO (Jamroz *et al*, 2005) reduce los riesgos asociados a patógenos bacterianos (Major *et al*, 2011), permitiendo que esta mayor cantidad de mucina determine una menor adhesión de patógenos (Reisinger *et al*, 2011).

## 4. Conclusiones

Los resultados obtenidos bajo las condiciones del presente estudio permiten llegar a las siguientes conclusiones:

- El PRO incrementa en 15% la longitud de las vellosidades intestinales y en 19% el área de su superficie, favorece la estructura de la vellosidad y la mucosa intestinal, la mayor densidad de la mucosa y la capacidad de absorción de nutrientes. Estos efectos se producen principalmente, aunque no de forma exclusiva, en el duodeno.

- Los cambios ejercidos por el PRO se manifiestan en las vellosidades intestinales a lo largo de todo el intestino delgado y en las criptas del duodeno.

- La mayor proliferación celular en las aves suplementadas con el PRO es responsable, al menos en parte, del incremento en la longitud de las vellosidades intestinales.

- El PRO incrementa en 33% la densidad de células caliciformes en las vellosidades intestinales y en 50% el potencial de producción de mucina.

# EXPERIMENTO N° 2

# EFECTO DE UN PRODUCTO A BASE DE ACEITE ESENCIAL DE ORÉGANO SOBRE EL CRECIMIENTO ALOMÉTRICO Y COMPORTAMIENTO PRODUCTIVO DE POLLOS DE CARNE EN LA ETAPA DE INICIO

## 1. Introducción

En la actualidad, la industria avícola está expuesta al continuo incremento en los costos del maíz y la soya y a la volatilidad en los precios de venta; ambos son factores que afectan la rentabilidad de las operaciones. Se estima que el periodo de crianza se reduce a una tasa aproximada de un día por año (Panda *et al*, 2009) y que cada año se incrementa en 3% la eficiencia productiva (Ponce, 2011) a consecuencia de la continua mejora genética de las líneas comerciales; sin embargo, para lograrlo con índices de rentabilidad aceptables es necesario maximizar la eficiencia del costo del alimento. Esto implica que la industria debe optimizar sus procesos de adquisición de granos y desarrollar tecnologías que permitan maximizar la eficiencia productiva de las aves.

Para establecer estrategias idóneas es importante considerar que si bien el pollo de carne consume sólo 8% de su ingesta acumulada de alimento en los primeros 14 días (Cobb-Vantress, 2008a), existe una alta correlación entre los pesos en las primeras semanas y el comportamiento productivo al final de la campaña (Ponce de León, 2011). Es conveniente, entonces, implementar estrategias costo-efectivas en el inicio de la crianza, que permitan maximizar la ganancia de peso en las primeras semanas y confieran a las parvadas mejores condiciones para lograr un óptimo desempeño en el resto de la campaña.

Se ha descrito el efecto que ejerce el AEO sobre las características histológicas del intestino delgado. En el Experimento 1, asociado al presente (Peebles *et al,* 1999a; 1999b), se han reportado mejoras en las características morfológicas de la mucosa intestinal y en la capacidad de absorción de nutrientes en las aves suplementadas con un producto a base de AEO (referido en adelante como PRO). El objetivo de este experimento fue evaluar el efecto del PRO sobre el crecimiento alométrico y el comportamiento productivo de pollos de carne en la etapa inicial de crianza.

## 2. Materiales y métodos

### 2.1. Lugar, fecha y duración

La crianza de las aves se llevó a cabo en las instalaciones del Programa de Aves de la Facultad de Zootecnia de la Universidad Nacional Agraria La Molina en Lima-Perú a inicios del segundo semestre del 2010. El sacrificio de las aves y mediciones posteriores se realizó en las instalaciones del Laboratorio de Patología Aviar de la Facultad de Medicina Veterinaria de la Universidad Nacional Mayor de San Marcos en Lima-Perú. La evaluación se llevó a cabo desde la recepción de las aves en las instalaciones de crianza y el periodo de evaluación fue el comprendido de 0 a 14 días de edad.

### 2.2. Instalaciones, equipos y materiales

Las aves estuvieron alojadas sobre material de cama a razón de 21.4 pollos/$m^2$ (0.047 $m^2$/pollo). Cada unidad experimental contó con un comedero y un bebedero lineales. Durante los primeros 3 días de vida se colocó papel periódico para evitar el acceso directo de los pollos BB al material de cama, mientras que el alimento fue suministrado en bandejas plásticas y el agua de bebida en bebederos tipo tongo. La calefacción fue provista por un sistema eléctrico con resistencias y termostatos, y controlada de acuerdo a las recomendaciones de la línea genética (Cobb-Vantress, 2008b) empleando un termo-higrómetro digital con una aproximación de 0.1 °C para temperatura y 1 % para humedad.

Para el pesaje de los ingredientes mayores del alimento se utilizó una balanza digital con capacidad de 150 kg y aproximación de 0.02 kg, mientras que para el pesaje de los ingredientes de la premezcla se utilizó una balanza electrónica con capacidad de 6 kg y aproximación de 1 g. Se utilizó mezcladoras horizontales de cintas de 400 y 30 kg de capacidad para la mezcla de los ingredientes durante la preparación del alimento y para las premezclas, respectivamente.

Para la identificación individual de las aves se empleó el método presentado en el Anexo 3. Para la medición de los pesos vivos se utilizó una balanza electrónica con capacidad para 200 g y con aproximación de 10 mg y otra con capacidad para 15 kg y con aproximación de 1 g. El agua de bebida fue potabilizada empleando 1 ml de hipoclorito de sodio al 4.5% por cada 10 L de agua. Para la necropsia de las aves se empleó bisturí, tijeras y guantes quirúrgicos.

### 2.3. Animales experimentales

Se empleó 192 pollos machos de la línea Cobb 500 (ver Anexo 4). Las aves fueron asignadas de manera aleatoria a cada tratamiento.

### 2.4. Tratamientos

Se evaluaron 3 tratamientos, definidos a continuación:

Tratamiento 1 : Dieta basal
Tratamiento 2 : Dieta basal + 500 ppm de Orevitol®
Tratamiento 3 : Dieta basal + 150 ppm de neomicina

El producto Orevitol® será referido, en adelante, como PRO. El tratamiento 3 fue incluído como un control positivo por ser el sulfato de neomicina frecuentemente empleado como promotor de crecimiento en condiciones comerciales.

### 2.5. Alimentación

Se suministró una dieta basal a base de maíz, soya y harina de pescado, complementada con aceite de pescado y aminoácidos sintéticos, y suplementada con una premezcla de vitaminas y minerales. La alimentación fue *ad libitum*. Las características de la dieta basal se presentan en el Cuadro 7. Los aditivos indicados en los tratamientos dietarios fueron suministrados de 1 a 14 días de edad a expensas del maíz.

**Cuadro 7. Características de la dieta empleada en el Experimento 2.**

| Ingredientes | % |
|---|---|
| Maíz amarillo | 52.517 |
| Torta de soya | 26.699 |
| Harina de pescado | 14.940 |
| Aceite semirefinado de pescado | 2.024 |
| DL-Metionina | 0.190 |
| L-Lisina | 0.116 |
| Cloruro de colina | 0.085 |
| Fosfato dicálcico | 1.609 |
| Carbonato de calcio | 0.978 |
| Sal común | 0.361 |
| Marcador inerte [1] | 0.300 |
| Premezcla [2] | 0.085 |
| Antifúngico [2] | 0.085 |
| Antioxidante [2] | 0.013 |

| Nutriente | Aporte calculado |
|---|---|
| Energía metabolizable, Kcal/kg | 3028 |
| Proteína cruda, % | 26.72 |
| Lisina, % | 1.72 |
| Metionina + Cistina, % | 1.10 |
| Treonina, % | 1.05 |
| Triptófano, % | 0.29 |
| Calcio, % | 1.54 |
| Fósforo disponible, % | 0.67 |
| Sodio, % | 0.34 |
| Grasa total, % | 5.97 |
| Fibra cruda, % | 2.70 |

| Componente | Composición proximal [3], % |
|---|---|
| Humedad | 11.89 |
| Proteína total (N x 6.25) | 26.06 |
| Extracto Etéreo | 4.87 |
| Fibra Cruda | 2.12 |
| Cenizas | 7.13 |
| ELN [4] | 47.94 |

[1] Óxido crómico como marcador inerte.

[2] Premezcla de vitaminas y minerales Proapak 2A®. Composición: Retinol: 12’000,000 UI; Colecalciferol: 2’500,000 UI; DL α-Tocoferol Acetato: 30,000 UI; Riboflavina: 5.5 g; Piridoxina: 3 g; Cianocobalamina: 0.015 g; Menadiona: 3 g; Ácido Fólico: 1 g; Niacina: 30 g; Ácido Pantoténico: 11 g; Biotina: 0.15 g; Zn: 45 g; Fe: 80 g; Mn: 65 g; Cu: 8 g; I: 1 g; Se: 0.15 g; Excipientes c.s.p. 1,000 g. Antifúngico: Mold Zap®; antioxidante: Danox®

[3] Informes de ensayo 1235/2010 LENA y 1236/2010 LENA, Universidad Nacional Agraria La Molina.

[4] ELN = Extracto Libre de Nitrógeno (calculado).

## 2.6. Mediciones

### 2.6.1. Crecimiento alométrico

Para la determinación de los coeficientes de crecimiento alométrico se realizó las mediciones en una muestra de 15 pollos BB durante la fase pre-experimental y las mismas mediciones se repitieron al término de la primera y segunda semana en 8 aves muestreadas por tratamiento (una por cada repetición). Se pesó la carcasa, la pechuga, el intestino, el hígado y el páncreas. Las mediciones fueron realizadas de forma inmediata tras el sacrificio para evitar variaciones en el peso del órgano por efecto del tiempo transcurrido (Bowes and Julian, 1998). El peso de la carcasa se determinó luego de retirar las vísceras (excepto pulmones y riñones) y la piel. El peso de la pechuga se determinó luego de retirar la piel. Para la evaluación del crecimiento alométrico se determinó los coeficientes respectivos empleando la siguiente fórmula (Fisher, 1984; citado por Cuervo *et al*, 2002):

$$CCA = (PO_b / PO_a) / (PC_b / PC_a)$$

Donde: CCA: coeficiente de crecimiento alométrico, PO: peso del órgano o estructura a muestrear, PC: peso corporal, a: día de nacimiento, b: días tras eclosión.

Cuando el órgano crece en la misma proporción al peso corporal, el CCA es de uno, si el crecimiento del órgano es menor al del peso corporal el CCA es menor a uno y cuando hay un crecimiento rápido en relación a la ganancia de peso corporal el CCA es mayor a uno (Cuervo *et al*, 2002).

### 2.6.2. Comportamiento productivo

Se midió a través de las siguientes variables:

- Peso vivo: Al iniciar el experimento y al término de cada semana se pesó individualmente a las aves.
- Ganancia de peso: Se calculó valores semanales de forma individual.
- Consumo de alimento: Se calculó el consumo acumulado al día 14 de edad.
- Conversión alimentaria: Se calculó en función a los valores acumulados de consumo de alimento y ganancia de peso al día 14.

- Rendimiento de carcasa: Se determinó semanalmente como la relación entre los pesos de la carcasa y del ave viva (%).

### 2.7. Diseño estadístico

Se empleó el Diseño Completo al Azar con tres tratamientos y ocho repeticiones. El análisis de varianza se realizó aplicando el procedimiento ANOVA del programa Statistical Analysis System SAS 9.0 (SAS Institute, 2009) y la diferencia de medias empleando la prueba de Duncan (1955). El Modelo Aditivo Lineal General aplicado fue el siguiente:

Yij = U + Ti + Eij

Donde:

| | | |
|---|---|---|
| Yij | = | variable respuesta |
| U | = | media general |
| Ti | = | i-ésimo tratamiento ( i = 1, 2, 3 ) |
| Eij | = | Error experimental |

Se consideró significativos valores con P < 0.10 (Crespo and Esteve-Garcia, 2001; Zhu *et al*, 2003, Kilburn and Edwards, 2004; van Nevel *et al*, 2005; Tahir *et al*, 2008; Teeter *et al*, 2008) o menores de 0.05, según se indica en cada caso.

## 3. Resultados y discusión

Los resultados observados en el presente estudio (Cuadro 8 y Anexo 14) indican un coeficiente de crecimiento alométrico (CCA) de la carcasa significativamente mayor (P<0.10) durante la primera semana de edad y de forma acumulada al día 14 de edad en las aves suplementadas con el PRO. Tanto en el día 7 como en el 14 de edad se observa un rendimiento de carcasa significativamente mayor (P<0.10) en las aves suplementadas con el PRO (Cuadro 8). Por el contrario, los CCA de la pechuga, intestino, hígado y páncreas no presentaron diferencias significativas (P>0.1) a lo largo del presente estudio.

**Cuadro 8. Efecto de un producto a base de aceite esencial de orégano sobre el crecimiento alométrico y comportamiento productivo de pollos en la etapa de inicio.**

| Variable | Tratamientos [1] | | | | | | P |
|---|---|---|---|---|---|---|---|
| | 1 | | 2 | | 3 | | |
| **Coeficientes de crecimiento alométrico [3]** | | | | | | | |
| Día 7 de edad | | | | | | | |
| Carcasa | 1.09 | £ | 1.12 | $ | 1.10 | $£ | 0.0962 |
| Pechuga de la carcasa | 4.44 | | 4.64 | | 4.28 | | 0.1469 |
| Resto de la carcasa | 0.91 | | 0.93 | | 0.93 | | 0.1435 |
| Intestino | 2.52 | | 2.41 | | 2.48 | | 0.6120 |
| Hígado | 1.81 | | 1.79 | | 1.81 | | 0.9726 |
| Páncreas | 2.84 | | 2.82 | | 2.78 | | 0.9735 |
| Día 14 de edad | | | | | | | |
| Carcasa | 1.13 | £ | 1.21 | $ | 1.19 | $£ | 0.0611 |
| Pechuga de la carcasa | 5.95 | | 6.22 | | 6.23 | | 0.3250 |
| Resto de la carcasa | 0.87 | | 0.94 | | 0.92 | | 0.2191 |
| Intestino | 1.89 | | 1.79 | | 1.72 | | 0.1711 |
| Hígado | 1.51 | | 1.46 | | 1.49 | | 0.4913 |
| Páncreas | 2.45 | | 2.37 | | 2.61 | | 0.8117 |
| **Comportamiento productivo** | | | | | | | |
| Día 7 de edad | | | | | | | |
| Peso corporal día 0, g [2] | 47.9 | | 48.4 | | 48.1 | | 0.7607 |
| Peso corporal día 7, g [2] | 193.3 | b | 201.6 | a | 202.7 | a | 0.0474 |
| Ganancia de peso 0 a 7 días, g [2] | 145.4 | b | 153.2 | ab | 154.6 | a | 0.0478 |
| Rendimiento de carcasa día 7, % [3] | 65.78 | £ | 67.55 | $ | 66.50 | $£ | 0.0962 |
| Día 14 de edad | | | | | | | |
| Peso corporal día 14, g [2] | 430.2 | c | 488.9 | a | 462.4 | b | <0.0001 |
| Ganancia de peso 8 a 14 días, g [2] | 237.2 | c | 286.9 | a | 259.6 | b | <0.0001 |
| Ganancia de peso 0 a 14 días, g [2] | 382.0 | c | 440.4 | a | 414.3 | b | <0.0001 |
| Consumo de alimento 0 a 14 días, g [3] | 555.4 | | 578.9 | | 549.7 | | 0.3367 |
| Conversión alimentaria 0 a 14 días, g [3] | 1.46 | a | 1.31 | b | 1.33 | b | 0.0022 |
| Rendimiento de carcasa día 14, % [3] | 67.96 | £ | 72.56 | $ | 71.82 | $£ | 0.0611 |

[1] Tratamientos: 1: aves control; 2: 500 ppm de Orevitol®; 3: 150 ppm de neomicina.
[2] Los valores son promedios de 64 observaciones por tratamiento.
[3] Los valores son promedios de 8 observaciones por tratamiento.
a,b,c,$,£ Promedios significativamente diferentes no comparten la misma letra (a,b,c; P<0.05) o el mismo símbolo ($,£; 0.05<P<0.10).

Estos resultados indican que la pechuga es la porción de la carcasa con la mayor velocidad de crecimiento. Este rápido crecimiento se explica por la presión genética de las casas comerciales (Cobb-Vantress, 2008b) y coincide con Clemente (2005) quien observó un incremento en el porcentaje de pechuga desde 2% al nacimiento, hasta 4% en el día 7 de edad y 7% en el día 14; valores que sin embargo corresponden a CCA para la pechuga menores a los del presente estudio debido probablemente a los cambios en la conformación corporal a través de los años, o a diferencias en el procedimiento de extracción de la carcasa. Por ello, si bien se ha demostrado la validez de las relaciones alométricas para caracterizar y modelar el crecimiento corporal empleando funciones de Gompertz (Hruby *et al*, 1994), se recomienda su uso solo en comparaciones intra-experimentales.

Los CCA observados para el intestino, hígado y páncreas de 2.47, 1.80 y 2.81 en la primera semana y de 1.80, 1.49 y 2.48 en las 2 primeras semanas, respectivamente, concuerdan con reportes previos de 2.70, 2.50 y 3.00 para el intestino, hígado y páncreas, respectivamente, en el día 7 de edad (García, 2000), y de 1.50 y 2.70 para el hígado y páncreas, respectivamente, en las 2 primeras semanas de edad (Cuervo *et al*, 2002). Los CCA mayores a 1 en estos órganos se explica por su elevada actividad metabólica, necesaria para proveer energía al resto del organismo (García, 2000). Así, entre los días 4 y 14 de edad la longitud del intestino se incrementa entre 20 a 30% y el diámetro entre 5 a 10%, lo que supone 50 y 60% mayor volumen intestinal; además aumenta el largo de las vellosidades y criptas intestinales, y el número de enterocitos por vellosidad (Uni *et al*, 1995), por lo que se espera que el CCA del intestino en estas etapas sea mayor a la unidad (Sell *et al*, 1991).

Los resultados indican que la suplementación del PRO incrementa la tasa de crecimiento alométrico de la carcasa en 3% durante la primera semana y en 7% durante los primeros 14 días de vida del ave. Al respecto, se ha documentado que la disponibilidad de aminoácidos como lisina tiene un efecto específico sobre la composición de la carcasa, el porcentaje de pechuga (Schutte and Pack, 1995) y la retención de proteína en la carcasa (Sibbald and Wolynetz, 1986); y que aportes de metionina (Ahmed and Abbas, 2011) o lisina (Si *et al*, 2004) dietaria superiores a los niveles recomendados por NRC (1994) incrementan el porcentaje de pechuga. El incremento del 19% en el área superficial de las vellosidades intestinales reportado en el Experimento 1 por efecto del PRO explica la mayor disponibilidad de nutrientes que habría determinado la mayor velocidad de crecimiento de la carcasa en las aves suplementadas con el PRO. Los resultados en la tasa de crecimiento de la en las aves suplementadas con el PRO concuerdan con reportes previos que indican mayor rendimiento de carcasa en aves suplementadas con AEO (Alcicek el al, 2003).

En el presente estudio, se observa hasta el día 14 de edad un crecimiento de la pechuga 5% mayor en las aves del tratamiento 2; sin embargo, no se ha podido atribuir dicha diferencia al efecto de la suplementación del PRO. Al respecto, se ha determinado en condiciones similares a las del presente estudio que pollos de 48 días de edad suplementados con una combinación de carvacrol, cinamaldehído y capsaicina incrementan significativamente el porcentaje de pechuga (Jamroz and Kamel, 2002), por lo que en las condiciones del presente estudio, un mayor periodo de evaluación permitiría determinar de forma significativa la influencia del PRO sobre la mayor velocidad de crecimiento de la pechuga observada en estas aves.

Por otro lado, al término de la primera semana de edad se observa diferencias significativas ($P<0.05$) en la ganancia de peso de las aves suplementadas con neomicina, así como en los pesos corporales de estas mismas aves y de aquellas suplementadas con el PRO respecto a las aves control (Cuadro 8 y Anexos 15 y 16). Asimismo, se hallan diferencias ($P<0.0001$) en el peso corporal al término de la segunda semana de edad, y en la ganancia de peso durante la segunda semana de vida y en el periodo acumulado durante los 14 días de evaluación, observándose los mayores valores en las aves suplementadas con el PRO y los menores en las aves control. Sin embargo, no se observa diferencias significativas ($P>0.1$) en el consumo de alimento en el periodo experimental. Por esta razón las aves suplementadas con el PRO o neomicina presentan una conversión alimentaria significativamente menor ($P<0.05$) que las aves control (Cuadro 8 y Anexo 17). Finalmente, se observa un rendimiento de carcasa significativamente mayor ($P<0.10$) en las aves suplementadas con el PRO.

De acuerdo a estos resultados, al final de la primera semana de vida el efecto neto de la suplementación del PRO o neomicina es similar en cuanto a peso corporal, observándose pesos 4.3% y 4.8% mayores que el control en las aves suplementadas con el PRO y neomicina, respectivamente. Sin embargo, al final de la segunda semana las diferencias en el peso en las aves suplementadas con el PRO y neomicina respecto al control es de 13.6% y 7.5%, respectivamente. El incremento en el peso de las aves suplementadas con el PRO se explica por el efecto que éste ejerce sobre la estructura de la mucosa intestinal y que fue reportado en el Experimento 1, ya que se ha demostrado que la altura de vellosidad intestinal está positivamente correlacionada con la ganancia de peso corporal vacío y la ingesta de material seca (Pluske *et al*, 1997). Además, se ha reportado el incremento en la concentración de acetato y butirato en el contenido cecal de pollos de 43 días de edad suplementados con una combinación de timol, cinamaldehído y capsaicina. Esto es importante ya que el ácido butírico es la mayor fuente de energía para las células de la membrana mucosa del colon, promoviendo la multiplicación celular, además reduce el pH intestinal favoreciendo la resistencia a enfermedades y controlando bacterias patógenas como *Escherichia coli* (Cao *et al*, 2010).

El comportamiento productivo de las aves en el presente estudio indica que la mayor ganancia de peso y pesos finales observados en las aves suplementadas con el PRO o neomicina se debe principalmente a un mayor aprovechamiento del alimento, observándose conversiones alimentarias 10% y 9% menores respecto al control en las aves suplementadas con el PRO y neomicina, respectivamente. Esta mayor eficiencia en la utilización del alimento por efecto del PRO se corrobora con el mayor rendimiento de carcasa observado en las aves suplementadas con el PRO al final de la primera y segunda semana, y concuerdan con lo reportado por Jamroz *et al* (2005) quienes observaron un incremento de 1.2% en la proporción del peso del músculo de la pechuga asociado a un incremento en la actividad lipasa a nivel pancreático e intestinal en pollos suplementados con una combinación de carvacrol, cinamaldehído y capsaicina. Al respecto, Fotea *et al* (2010) observaron 5% mayor peso corporal y una reducción en la conversión alimentaria de 1.93 a 1.85 en pollos de 42 días suplementados con 1% de AEO en la dieta, y Hernández *et al* (2004) observaron una mayor digestibilidad iliaca de la materia seca y del almidón en las aves que recibieron un suplemento a base de extractos de orégano, canela y pimienta.

En el presente experimento, la menor conversión alimentaria observada en las aves suplementadas con el PRO indica una mayor disponibilidad de nutrientes para ganancia de peso como resultado del balance entre la demanda nutricional para satisfacer los requerimientos de mantenimiento del ave y los nutrientes absorbidos.

Las principales fuentes de variación entre los tratamientos del presente estudio que influyen sobre los requerimientos de mantenimiento de las aves están referidas a la dinámica de la restitución de la mucosa intestinal, incluyendo la proliferación y las tasas de descamación y apoptosis celular, así como al nivel de actividad inmunológica para resguardar al ave de los efectos patogénicos de la flora intestinal. En relación a ellos, se ha demostrado que el AEO incrementa la apoptosis celular (Fabian *et al*, 2006), pero también la proliferación celular (Levkut *et al*, 2011) y se ha observado la los metabolitos secundarios contenidos en el AEO incrementan la longitud de la vellosidad intestinal (García *et al*, 2007; Roldán, 2010), por lo que es posible establecer que el balance entre apoptosis/descamación y proliferación es favorable a esta última. Por otro lado, el control que ejerce el AEO sobre la flora intestinal patógena por su acción antimicrobiana (Conner and Beuchat, 1984; Helander *et al*, 1998; Lambert *et al*, 2001; Lee *et al*, 2004; Máthé, 2009) y la regulación de la flora comensal por su acción prebiótica (Hammer *et al*, 1999; Dorman and Deans, 2000; Ferket, 2003; Lee *et al*, 2004; Jamroz *et al*, 2005; Zheng *et al*, 2010), así como el incremento en la producción de mucina por efecto de los metabolitos secundarios contenidos en el AEO (ver Experimento 1; Jamroz *et al*, 2006; Reisinger *et al*, 2011) que reduce la adhesión de patógenos (Deplancke and Gaskins, 2001) implica que la demanda de actividad inmunológica por parte del ave será menor, reduciendo el gasto energético por este concepto (Wolowczuk *et al*, 2008; Jiang *et al*, 2010). La mayor diversidad de efectos del AEO contenido en el PRO respecto a la actividad antimicrobiana de la neomicina explican la mayor disponibilidad de nutrientes requerida para el crecimiento de tejidos observado en el presente estudio, en términos generales, como peso corporal.

Por otro lado, en el presente estudio los principales factores que influyen sobre la cantidad de nutrientes absorbidos están referidos a la capacidad de la mucosa intestinal para absorber nutrientes y a la cantidad de nutrientes disponibles en el lumen intestinal tras la digestión por la propia ave y la parcial utilización por la microflora intestinal. Al respecto, los resultados reportados en el Experimento 1 indican que el PRO incrementa la altura y el área superficial de la vellosidad intestinal y la densidad de la mucosa lo que brinda al ave una mayor capacidad para la absorción de nutrientes (Roldán, 2010). Asimismo, se ha documentado que el AEO y metabolitos secundarios como el timol, contenidos en él, incrementan la actividad de enzimas digestivas como la quimotripsina (Lee *et al*, 2004; Basmacioglu Malayoglu *et al*, 2010), tripsina (Lee *et al*, 2004) y lipasa (Lee *et al*, 2004; Jamroz *et al*, 2005) lo que favorece la disponibilidad de nutrientes para ser absorbidos.

## 4. Conclusiones

Los resultados obtenidos bajo las condiciones del presente estudio permiten llegar a las siguientes conclusiones:

- La suplementación del PRO influye positivamente en el crecimiento alométrico de la carcasa, respecto a las aves no suplementadas.

- La suplementación del PRO, al igual que de la neomicina, incrementa el peso de los pollos al final de la primera y segunda semana de vida y reduce la conversión alimentaria acumulada durante los primeros 14 días de vida.

- Al término de la segunda semana de crianza, el efecto sobre el peso corporal es mayor con la suplementación del PRO que con la neomicina.

# EXPERIMENTO N° 3

# EFECTO DE UN PRODUCTO A BASE DE ACEITE ESENCIAL DE ORÉGANO SOBRE LA MORFOMETRÍA Y MINERALIZACIÓN ÓSEA E INTEGRIDAD ESQUELÉTICA DE POLLOS DE CARNE

## 1. Introducción

La industria avícola está sometida al constante incremento del costo del alimento; por lo tanto, para lograr operaciones rentables, es necesario optimizar los procesos de producción y la eficiencia del alimento, así como incrementar la viabilidad y calidad del producto final, sean estos procesados o aves en pie.

Uno de los aspectos que compromete la calidad del pollo es el porcentaje de aves afectadas por problemas relacionados con la integridad esquelética y la incidencia de lesiones y patologías como la epifisiólisis femoral proximal agravada por la manipulación durante la saca (Mitchell and Boon, 1986), la presencia de fracturas de fémur y tibia, y salpicaduras de sangre en la carne (Applegate and Angel, 2008). Los descartes de aves por problemas locomotores así como mortalidades de entre 2 y 8% asociadas a problemas esqueléticos afectan negativamente las ganancias del avicultor (Thorp, 1994; Barreiro *et al*, 2010). La integridad esquelética se ve influenciada de forma negativa en gran medida por la rápida velocidad de crecimiento que presentan las líneas genéticas modernas (Mitchell and Boon, 1986; Deeb and Lamont, 2002; Zhou *et al*, 2005) y en los últimos se está convirtiendo en el principal factor que limita la mejora genética para reducir el periodo de crianza (Leeson, 2012).

Si bien existen diferentes estrategias para reducir los problemas asociados a la integridad esquelética de las aves, es conveniente determinar la influencia que ejerce, entre otros, la calidad nutricional de la dieta, su inocuidad, la disponibilidad de nutrientes y el uso de aditivos no nutricionales.

Uno de los aditivos empleados recientemente para reemplazar a los antimicrobianos promotores del crecimiento y/o mejorar el comportamiento productivo de las aves son los aceites esenciales de plantas. Diversos estudios se han realizado con resultados positivos sobre la digestibilidad y disponibilidad de nutrientes con el uso de aceites esenciales de tomillo, anís estrella (Amad *et al*, 2011), orégano, canela y pimienta (Hernández *et al*, 2004), con metabolitos secundarios propios de estas plantas, como el carvacrol, cinamaldehído y capsaicina (García *et al*, 2007), incluyendo la disponibilidad de la ceniza dietaria medida en las heces (Jamroz and Kamel, 2002); sin embargo, son escasos los estudios realizados sobre la morfometría y mineralización ósea, y en términos generales sobre la integridad esquelética.

Finalmente, las características óseas también han sido utilizadas para estimar los requerimientos nutricionales (Dhandu and Angel, 2003; Applegate and Angel, 2008) y se han propuesto diversas metodologías para estimar la mineralización ósea, como el contenido de ceniza, la resistencia a la rotura, el peso, el volumen, e incluso el ultrasonido (Mutus *et al*, 2006). El objetivo del presente estudio fue determinar la influencia de un producto a base de AEO (referido en adelante como PRO) sobre las características morfométricas, indicadores de mineralización y alteraciones óseas, que en conjunto reflejan la integridad esquelética de los pollos de carne.

## 2. Materiales y métodos

### 2.1. Lugar, fecha y duración

La crianza de las aves se condujo en el Programa de Aves de la Facultad de Zootecnia de la Universidad Nacional Agraria La Molina en Lima-Perú a inicios del segundo semestre del 2010. El sacrificio y procesamiento de muestras se realizó en las instalaciones del Laboratorio de Patología Aviar de la Facultad de Medicina Veterinaria de la Universidad Nacional Mayor de San Marcos en Lima-Perú. La evaluación se llevó a cabo desde la recepción de las aves en las instalaciones de crianza y el periodo de evaluación fue el comprendido de 0 a 14 días de edad.

### 2.2. Instalaciones, equipos y materiales

Las aves estuvieron alojadas sobre material de cama a razón de 21.4 pollos/$m^2$ (0.047 $m^2$/pollo). Cada unidad contó con un comedero y un bebedero lineales. Durante los primeros 3 días de vida se colocó papel periódico para evitar el acceso directo de los pollos BB al material de cama, mientras que el alimento fue suministrado en bandejas plásticas y el agua de bebida en bebederos tipo tongo. La calefacción fue provista por un sistema eléctrico con resistencias y termostatos, y controlada de acuerdo a las recomendaciones de la línea genética (Cobb-Vantress, 2008b) empleando un termo-higrómetro digital con una aproximación de 0.1 °C para temperatura y 1% para humedad.

Para el pesaje de los ingredientes mayores del alimento se utilizó una balanza digital con capacidad de 150 kg y aproximación de 0.02 kg, mientras que para el pesaje de los ingredientes de la premezcla se utilizó una balanza electrónica con capacidad de 6 kg y aproximación de 1 g. Se utilizó mezcladoras horizontales de cintas de 400 y 30 kg de capacidad para la mezcla de los ingredientes durante la preparación del alimento y para las premezclas, respectivamente.

Para la identificación individual de las aves se empleó el método presentado en el Anexo 3. El agua de bebida fue potabilizada empleando 1 ml de hipoclorito de sodio al 4.5% por cada 10 L de agua. Para la necropsia de las aves se empleó bisturí, tijeras y guantes quirúrgicos.

Los huesos muestreados fueron pesados empleando una balanza electrónica de precisión (Kocabagli, 2001) con capacidad para 200 g y con aproximación de 10 mg. Para la determinación del volumen de los huesos muestreados se empleó probetas graduadas con capacidad para 10 y 25 $cm^3$ y con aproximación de 0.2 $cm^3$. Los huesos fuero medidos empleando un vernier (Kocabagli, 2001) marca Uyustools Professional de con capacidad para 15 cm y aproximación de 0.05 mm. Para realizar el corte de las tibias se empleó tijeras especiales de cocina con cuchillas gruesas y mangos reforzados para vencer la resistencia del hueso al corte.

### 2.3. Animales experimentales

Se empleó 128 pollos machos de la línea Cobb 500, cuyas características se presentan en el Anexo 4. De manera aleatoria las aves fueron asignadas a cada tratamiento.

## 2.4. Tratamientos

Se evaluaron 6 tratamientos, definidos a continuación:

| Tratamiento | Hueso | Alimento |
|---|---|---|
| 1 | Fémur | Dieta basal |
| 2 | Fémur | Dieta basal + 500 ppm de Orevitol® |
| 3 | Tibia | Dieta basal |
| 4 | Tibia | Dieta basal + 500 ppm de Orevitol® |
| 5 | Tarso | Dieta basal |
| 6 | Tarso | Dieta basal + 500 ppm de Orevitol® |

El producto indicado será referido, en adelante, como PRO.

## 2.5. Alimentación

Se suministró una dieta basal a base de maíz, soya y harina de pescado, complementada con aceite de pescado y aminoácidos sintéticos, y suplementada con una premezcla de vitaminas y minerales. La alimentación fue *ad libitum*. Las características de la dieta basal se presentan en el Cuadro 9. El aditivo indicado en la sección tratamientos fue suministrado de 1 a 14 días de edad a expensas del maíz.

## 2.6. Mediciones

En las aves *in vivo*, así como en los huesos extraídos de éstas tras su sacrificio, se evaluó un conjunto de variables relacionadas con la integridad esquelética de las aves como son la morfometría y mineralización ósea, y algunas mediciones complementarias indicativas de alteraciones.

En el día 14 de edad se sacrificó al azar 16 aves de cada tratamientos, se realizó la necropsia y se muestreó los muslos, piernas y patas, las que fueron conservadas en congelación hasta su procesamiento (Dhandu and Angel, 2003; Mendes *et al*, 2006) Las muestras fueron descongeladas a temperatura ambiente y luego colocadas en agua hirviente durante 15 minutos (Kocabagli, 2001; Applegate and Lilburn, 2002;

**Cuadro 9. Características de la dieta empleada en el Experimento 3.**

| Ingredientes | % |
|---|---|
| Maíz amarillo | 52.517 |
| Torta de soya | 26.699 |
| Harina de pescado | 14.940 |
| Aceite semirefinado de pescado | 2.024 |
| DL-Metionina | 0.190 |
| L-Lisina | 0.116 |
| Cloruro de colina | 0.085 |
| Fosfato dicálcico | 1.609 |
| Carbonato de calcio | 0.978 |
| Sal común | 0.361 |
| Marcador inerte [1] | 0.300 |
| Premezcla [2] | 0.085 |
| Antifúngico [2] | 0.085 |
| Antioxidante [2] | 0.013 |

| Nutriente | Aporte calculado |
|---|---|
| Energía metabolizable, Kcal/kg | 3028 |
| Proteína cruda, % | 26.72 |
| Lisina, % | 1.72 |
| Metionina + Cistina, % | 1.10 |
| Treonina, % | 1.05 |
| Triptófano, % | 0.29 |
| Calcio, % | 1.54 |
| Fósforo disponible, % | 0.67 |
| Sodio, % | 0.34 |
| Grasa total, % | 5.97 |
| Fibra cruda, % | 2.70 |

| Componente | Composición proximal [3], % |
|---|---|
| Humedad | 11.89 |
| Proteína total (N x 6.25) | 26.06 |
| Extracto Etéreo | 4.87 |
| Fibra Cruda | 2.12 |
| Cenizas | 7.13 |
| ELN [4] | 47.94 |

[1] Óxido crómico como marcador inerte.

[2] Premezcla de vitaminas y minerales Proapak 2A®. Composición: Retinol: 12'000,000 UI; Colecalciferol: 2'500,000 UI; DL α-Tocoferol Acetato: 30,000 UI; Riboflavina: 5.5 g; Piridoxina: 3 g; Cianocobalamina: 0.015 g; Menadiona: 3 g; Ácido Fólico: 1 g; Niacina: 30 g; Ácido Pantoténico: 11 g; Biotina: 0.15 g; Zn: 45 g; Fe: 80 g; Mn: 65 g; Cu: 8 g; I: 1 g; Se: 0.15 g; Excipientes c.s.p. 1,000 g. Antifúngico: Mold Zap®; antioxidante: Danox®

[3] Informes de ensayo 1235/2010 LENA y 1236/2010 LENA, Universidad Nacional Agraria La Molina.

[4] ELN = Extracto Libre de Nitrógeno (calculado).

Moraes, 2006), proceso que no afecta el contenido mineral, la densidad del hueso, ni la resistencia a la rotura (Orban *et al*, 1993), pero que puede extraer alrededor del 80% de la grasa (Almeida Paz *et al*, 2008).

Posteriormente se retiró manualmente los restos de tejidos y obtener los correspondientes fémures (fémur), tibiotarsos (tibias) y tarsometatarsos (tarsos) (Baumel and Witmer, 1993). Se puso particular énfasis en retirar completa y cuidadosamente los cartílagos en las superficies articulares, el periostio y otros tejidos no óseos en los extremos de cada hueso, de los surcos intercondilares del fémur y tibia, y de la fosa infracotilar y de las incisuras intratrocleares del tarso (ver Imagen 4). Luego de limpiadas, las piezas fueron enjuagadas y secadas con papel toalla. Luego de terminar la limpieza de las muestras se dejó secar los huesos durante 24 horas a temperatura ambiente en un lugar ventilado (Kocabagli, 2001; Mutus *et al*, 2006).

**Imagen 4. Huesos muestreados**

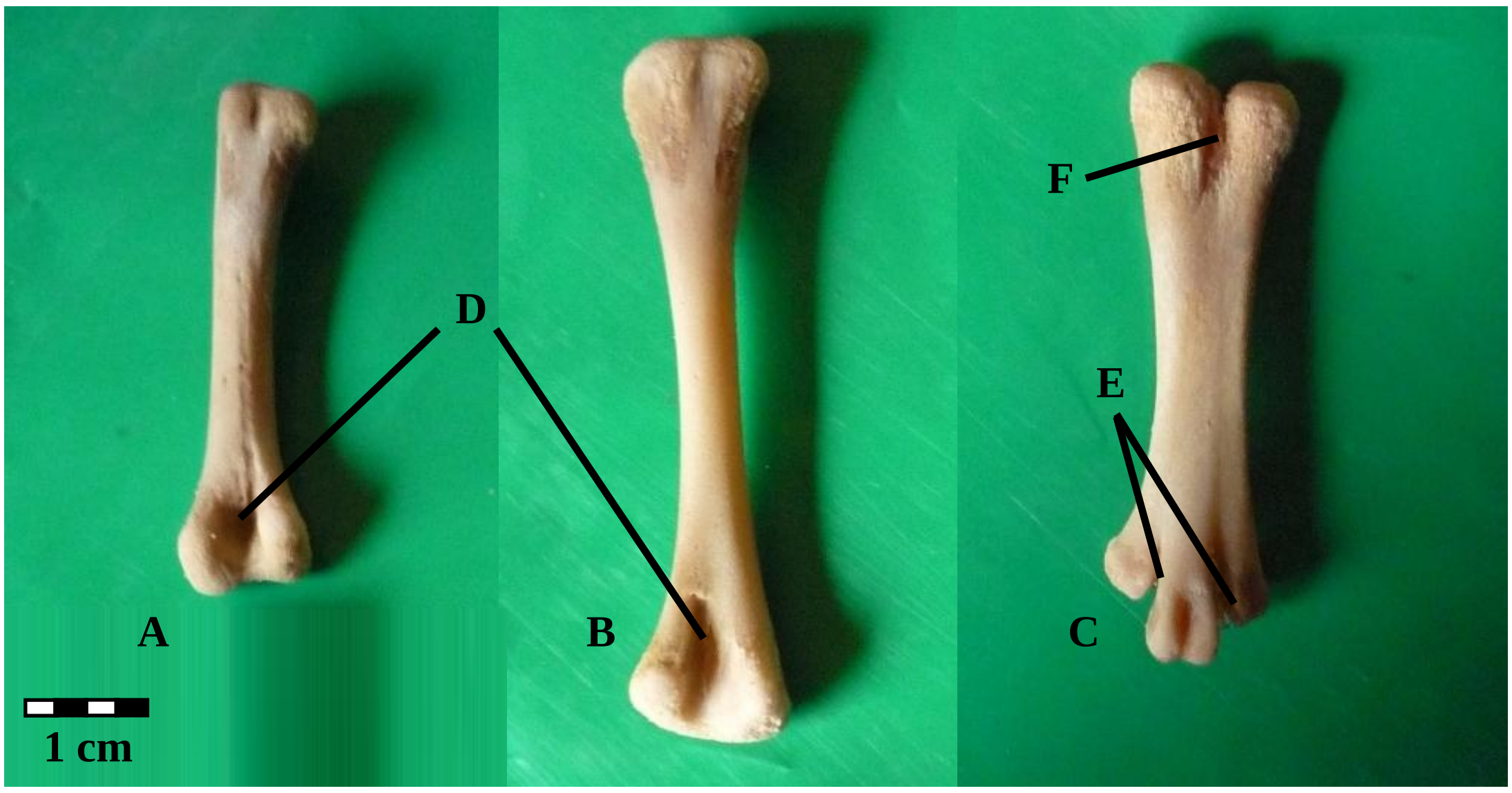

Vista ventral del fémur (A) y tibia (B) derechos, y del tarso izquierdo (C). Se observa los surcos intercondilares (D), las incisuras intratrocleares (E) y la fosa infracotilar (F).

En cada hueso muestreado se determinó las variables presentadas a continuación, obteniéndose valores individuales y luego promedio por ave:

### 2.6.1. Características morfométricas

- Largo:

  Se determinó considerando la longitud mayor de extremo a extremo del hueso, manteniendo el eje longitudinal del hueso paralelo al brazo principal del vernier. Los valores se presentan en milímetros (mm).

- Diámetro de la diáfisis:

  Se determinó los diámetros latero-lateral (DLL) y cráneo-caudal (DCC) de la diáfisis en la mitad de la longitud del hueso (Kocabagli, 2001; Quarantelli *et al*, 2007). El diámetro promedio de la diáfisis (DP) se expresa en milímetros (mm) y se calculó empleando la siguiente fórmula:

$$\mathrm{DP} = \frac{\mathrm{DLL} + \mathrm{DCC}}{2}$$

- Volumen:

  El volumen de cada hueso ($cm^3$) se determinó midiendo el desplazamiento de agua en probetas graduadas al sumergir completamente cada hueso (Sato, 1995; Zhang and Coon, 1997; Quarantelli *et al*, 2007).

- Índice de forma:

  Se determinó dividiendo el largo del hueso entre el diámetro de la diáfisis. Este índice indica cuántas veces está contenido el diámetro de la diáfisis del hueso en el largo del mismo.

### 2.6.2. Indicadores de mineralización ósea

- Peso del hueso:

  Los huesos fueron pesados de forma individual. Los valores se presentan en miligramos (mg).

- Densidad:

  La densidad ósea se determinó mediante la volumetría del hueso (Quarantelli *et al*, 2007); de esta forma, se consideró como densidad ósea a la masa de material orgánico e inorgánico en el hueso por unidad de volumen del mismo (Rath *et al*, 2000) y se calculó empleando la siguiente fórmula:

$$\text{Densidad (mg/cm}^3\text{)} = \frac{\text{Peso, mg}}{\text{Volumen, cm}^3}$$

- Índice modificado de Seedor:

  Este índice fue inicialmente propuesto por Seedor *et al* (1991) como indicador de la densidad y calidad ósea (Barreiro *et al*, 2010; Souza da Silva, 2010), para ser calculado en base al peso de las cenizas; sin embargo, desde entonces en diferentes estudios se ha empleado una modificación de este índice considerando el peso del hueso entero y no sólo de la ceniza (Monteagudo *et al*, 1997; Kocabagli, 2001; Mendes *et al*, 2006; Moraes, 2006; Mutus *et al*, 2006; Potenca, 2008; Candido, 2009; Barreiro *et al*, 2010; Souza da Silva, 2010). El Índice modificado de Seedor se basa en el concepto de que es la fracción mineral del hueso la que tiene la mayor densidad específica. Así, cuando mayor es este índice mayor es la densidad del hueso. Se calculó empleando la siguiente fórmula:

$$\text{I. modificado de Seedor} = \frac{\text{Peso, mg}}{\text{Largo, mm}}$$

- Índice Quetelet:

  Este índice, también llamado índice de masa corporal, fue propuesto por Adolphe Quetelet alrededor de 1740, es ampliamente utilizado en medicina humana para determinar la obesidad y ha sido empleado en estudios con pollos de carne (Rutten *et al*, 2002). Si bien se expresa de manera estándar en $kg/m^2$, los valores reportados en $mg/mm^2$ son numéricamente idénticos. Cuanto menor es el índice Quetelet, el hueso es relativamente más liviano pero más largo, mientras que cuanto mayor es el índice, el hueso es relativamente más pesado pero más corto. Se calculó empleando la siguiente fórmula:

$$\text{I. Quetelet } (\text{mg/mm}^2) = \frac{\text{Peso, mg}}{(\text{Largo, mm})^2}$$

- Índice de robusticidad:

  Este índice fue propuesto por Alphonse Riesenfeld en 1972 y ha sido empleado en estudios con pollos de carne (Kocabagli, 2001; Mutus *et al*, 2006; Somkuwar *et al*, 2010). Cuanto menor es este índice se considera que la estructura del hueso es más fuerte. Se calculó empleando la siguiente fórmula:

$$\text{I. de robusticidad} = \frac{\text{Largo, cm}}{\sqrt[3]{\text{Peso, g}}}$$

### 2.6.3. Mediciones complementarias

En las aves muestreadas se registró las siguientes variables:

- Capacidad para caminar:

  Antes del sacrificio de las aves se observó la presencia de problemas locomotores y se midió empleando una versión modificada del score de la capacidad para caminar propuesto previamente (Kestin *et al*, 1992), que tiene una escala de 1 a 6, donde:

  - 6 Normal
  - 5 Defectos leves
  - 4 Anormalidad definida para caminar
  - 3 Lesión evidente, por ejemplo cojera o inestabilidad. En este caso las aves suelen estar postradas.
  - 2 Grave daño y dificultad para caminar
  - 1 Incapacidad para sostenerse sobre sus patas, para su traslado requieren de las alas, se apoyan sobre los corvejones.

  Se considera probable que en los scores 3 a 1 exista dolor crónico (López *et al*, 2011). Se determinó el score correspondiente para cada ave muestreada.

- Degeneración femoral:

  Durante la necropsia de cada ave se realizó la dislocación de las dos cabezas femorales del acetábulo y se determinó el nivel de degeneración observada empleando el Índice de Degeneración Femoral, desarrollado por la empresa Hybro® y presentado por Almeida Paz *et al* (2008), con una escala de 1 a 5, donde:

  - 1 Hueso sin lesión
  - 2 El cartílago está ausente en la cabeza femoral y el hueso está intacto
  - 3 La cabeza femoral no tiene el cartílago y está parcialmente rota
  - 4 La cabeza femoral está considerablemente dañada pero su contorno está aún visible
  - 5 La cabeza del fémur está completamente rota y no es posible reconocer su contorno (lesión clínicamente conocida como epifisiólisis femoral proximal).

  Se determinó el índice de cada hueso evaluado y luego el promedio por ave.

- Discondroplasia tibial:

  Luego de realizar las mediciones necesarias para determinar las características morfométricas y los indicadores de mineralización ósea, se hizo un corte sagital en la diáfisis proximal de cada tibia muestreada. El grado de lesión se midió visualmente empleando una modificación del score propuesto por Thorp *et al* (1997) y Zhang *et al* (1997), y posteriormente ilustrado por Almeida Paz *et al* (2005), con una escala de 1 a 4, siendo 1 ausente, 2 leve, 3 moderado y 4 severo, de acuerdo al tamaño de la masa cartilaginosa no mineralizada, medida desde la placa en dirección distal hacia la metáfisis. Se determinó el score para cada tibia muestreada y se obtuvo un valor promedio por ave.

## 2.7. Diseño estadístico

Se empleó el Diseño Completamente al Azar con arreglo factorial 3 (hueso muestreado) x 2 (nivel de inclusión del aditivo) con 16 repeticiones por tratamiento, aplicando el procedimiento GLM del programa Statistical Analysis System SAS 9.0 (SAS Institute, 2009). Las diferencias entre medias se evaluaron mediante la prueba de Duncan (1955). El Modelo Aditivo Lineal General aplicado fue el siguiente:

Y ijk = U + Ei + Aj + E*A (i, j) + Eijk

Donde:

| | |
|---|---|
| Yijk | Observación en la i-ésima sección intestinal recibiendo el j-ésimo nivel de inclusión del aditivo, en la k-ésima repetición |
| U | Media general |
| Ei | Efecto del i-ésimo hueso |
| Aj | Efecto del j-ésimo nivel de inclusión del aditivo |
| E*A (i,j) | Efecto de la interacción del i-ésimo hueso por el j-ésimo nivel de inclusión del aditivo |
| Eijk | Error experimental |

i = 1, 2, 3,  j = 1, 2  k=1, 2, 3, 4, 5, 6, 7, 8

Para los índices de capacidad para caminar, degeneración femoral y discondroplasia tibial se empleó un Diseño Completo al Azar en su forma simple, con 2 tratamientos (control y PRO; con y sin PRO, respectivamente) y 16 repeticiones para cada uno; ello debido a que para estar variables se empleó las propias aves o solo algún hueso.

Se consideró significativos aquellos valores con P menores de 0.10 (Crespo and Esteve-Garcia, 2001; Zhu *et al*, 2003, Kilburn and Edwards, 2004; van Nevel *et al*, 2005; Tahir *et al*, 2008; Teeter *et al*, 2008) o 0.05, según se indica en cada caso.

## 3. Resultados y discusión

En los últimos años el mantenimiento de la integridad esquelética de las aves se ha convertido en uno de los principales factores que limitan la mejora genética para reducir el periodo de crianza (Leeson, 2012) debido a la presión de selección ejercida a través de los años para lograr líneas genéticas de aves con mayor velocidad de crecimiento (Mitchell and Boon, 1986; Deeb and Lamont, 2002; Zhou *et al*, 2005). Al respecto, la mineralización ósea guarda relación directa con la calidad del pollo de carne (Barreiro *et al*, 2010); en consecuencia, este aspecto de la integridad esquelética, así como la incidencia de problemas locomotores y anomalías óseas, debe ser constantemente evaluada.

Las características morfométricas e indicadores de mineralización ósea observados en el presente estudio se muestran en los Cuadros 10 y 11 y Anexos 19 a 21. En el presente estudio se observan diferencias significativas (P<0.01), entre los tres huesos evaluados, respecto a sus características morfométricas e índices modificado de Seedor, Quetelet y de robusticidad (Cuadro 11).

Los resultados corroboran que el fémur, tibia y tarso son todos distintos entre sí desde el punto de vista de sus características morfométricas (P<0.01). Así, la tibia es el hueso más pesado, largo y de mayor volumen, mientras que el tarso es el que presenta los menores valores en estas tres variables (Cuadro 10). El mayor diámetro latero-lateral de la diáfisis se observa en el tarso, y el menor en la tibia, mientras que el mayor diámetro cráneo-caudal de la diáfisis se observa en el fémur, y el menor en la tibia. El fémur presenta, entonces, un diámetro promedio de la diáfisis significativamente mayor que la tibia y el tarso. Finalmente, el mayor índice de forma se observa en la tibia y el menor en el fémur.

Por otro lado, los tres huesos evaluados presentan también valores significativamente diferentes (P<0.01) en cuanto su peso e índices modificado de Seedor, de robusticidad y Quetelet. Así, el tarso presenta un menor índice modificado de Seedor que el fémur y la tibia, los cuales son similares, mientras que el menor índice Quetelet se presenta en la tibia y el mayor en el fémur. Finalmente, la tibia presenta un mayor índice de robusticidad que el fémur y tarso, siendo estos similares. A pesar de las diferencias observadas en las variables mencionadas, no se encuentran diferencias significativas (P>0.10) entre los huesos en términos de su densidad. Al respecto, si bien los índices modificado de Seedor, de robusticidad y Quetelet han sido empleados para evaluar la mineralización ósea, los resultados del presente estudio indican que, a diferencia de la densidad ósea, estas variables no deben ser empleados para comparar diferentes huesos.

Respecto al efecto del PRO, no se observan diferencias significativas en las variables morfométricas ni en los índices modificado de Seedor, de robusticidad y Quetelet; sin embargo, sí se observan diferencias significativas (P<0.10) en la densidad ósea, siendo esta mayor en los huesos de las aves suplementadas con el PRO.

**Cuadro 10. Efecto de un producto a base de aceite esencial de orégano sobre la integridad esquelética: Morfometría ósea.**

| Efectos evaluados | Tratamientos | | Largo del hueso (mm) | Diámetro de la diáfisis (mm) | | | Volumen del hueso ($cm^3$) | Índice de forma |
|---|---|---|---|---|---|---|---|---|
| | | | | Latero-Lateral | Cráneo-Caudal | Promedio | | |
| Aditivo | | Control | 42.72 | 5.333 | 4.615 | 4.974 | 1.446 | 8.872 |
| | | PRO | 42.89 | 5.402 | 4.718 | 5.060 | 1.411 | 8.795 |
| | P (aditivo) | | 0.7859 | 0.6327 | 0.6915 | 0.6524 | 0.2882 | 0.584 |
| Hueso | | Fémur | 38.76 b | 5.431 b | 5.959 a | 5.695 a | 1.322 b | 7.088 c |
| | | Tibia | 51.38 a | 4.573 c | 4.370 c | 4.472 b | 1.759 a | 11.535 a |
| | | Tarso | 38.13 b | 6.120 a | 3.636 b | 4.878 b | 1.197 c | 7.847 b |
| | P (hueso) | | <0.0001 | <0.0001 | <0.0001 | <0.0001 | <0.0001 | <0.0001 |
| Interacción | Fémur | Control | 38.72 | 5.375 | 5.847 | 5.611 | 1.338 | 7.176 |
| | | PRO | 38.80 | 5.488 | 6.072 | 5.780 | 1.306 | 7.001 |
| | Tibia | Control | 51.34 | 4.569 | 4.350 | 4.459 | 1.788 | 11.553 |
| | | PRO | 51.43 | 4.578 | 4.391 | 4.484 | 1.731 | 11.517 |
| | Tarso | Control | 38.11 | 6.056 | 3.647 | 4.852 | 1.213 | 7.887 |
| | | PRO | 38.16 | 6.190 | 3.623 | 4.907 | 1.180 | 7.805 |
| | P (aditivo*hueso) | | 0.9981 | 0.9542 | 0.8738 | 0.9443 | 0.9536 | 0.9483 |

a,b,c Promedios significativamente diferentes no comparten la misma letra (a,b,c; P<0.05).

**Cuadro 11. Efecto de un producto a base de aceite esencial de orégano sobre integridad esquelética: Indicadores de mineralización ósea.**

| Efectos evaluados | Tratamientos | | Peso del hueso (mg) | Densidad ósea (mg/cm$^3$) | Índice Modificado de Seedor | Índice Quetelet | Índice de robusticidad |
|---|---|---|---|---|---|---|---|
| Aditivo | | Control | 701.7 | 486.72 £ | 16.1028 | 0.3890 | 4.809 |
| | | PRO | 703.8 | 501.80 $ | 16.5864 | 0.3867 | 4.826 |
| | P (aditivo) | | 0.9910 | 0.0713 | 0.1352 | 0.7815 | 0.7030 |
| Hueso | | Fémur | 650.0 b | 493.44 | 16.7374 a | 0.4315 a | 4.486 b |
| | | Tibia | 861.6 a | 491.89 | 16.7523 a | 0.3259 c | 5.409 a |
| | | Tarso | 593.2 c | 497.80 | 15.5262 b | 0.4068 b | 4.549 b |
| | P (hueso) | | <0.0001 | 0.8500 | 0.0037 | <0.0001 | <0.0001 |
| Interacción | Fémur | Control | 650.6 | 485.46 | 16.6479 | 0.4330 | 4.479 |
| | | PRO | 649.4 | 501.41 | 16.8262 | 0.4300 | 4.493 |
| | Tibia | Control | 858.8 | 486.55 | 16.5069 | 0.3255 | 5.410 |
| | | PRO | 864.4 | 497.23 | 16.9977 | 0.3264 | 5.408 |
| | Tarso | Control | 595.6 | 488.24 | 15.0904 | 0.4086 | 4.540 |
| | | PRO | 590.7 | 506.75 | 15.9348 | 0.4048 | 4.560 |
| | P (aditivo*hueso) | | 0.9682 | 0.9249 | 0.7214 | 0.9586 | 0.9503 |

a,b,c,$,£ Promedios significativamente diferentes no comparten la misma letra (a,b,c; P<0.05) o el mismo símbolo ($,£; 0.05<P<0.10).

Durante la necropsia, en una de las muestras control se observó una lesión femoral clasificada con un score de 5 (Almeida Paz *et al*, 2008). En esta misma ave se observó discondroplasia tibial moderada, pero en ninguna otra de las aves muestreadas. Por ello el índice de degeneración femoral y el score de discondroplasia tibial (Cuadro 12 y Anexo 18) fueron, en promedio, menores en las aves suplementadas con el PRO; sin embargo, estas diferencias no son significativas (P>0.10). A pesar de ello, no se observó problemas locomotores, presentando todas las aves plena capacidad para caminar (Kestin *et al*, 1992).

De acuerdo a los resultados del presente estudio, de las variables evaluadas, sólo la densidad ósea se puede emplear para comparaciones entre diferentes huesos, ya que no se observan diferencias significativas entre la densidad del fémur, tibia y tarso.

Se ha observado que si bien las aves con mayor velocidad de crecimiento presentan mayor densidad ósea, también presentan mayor probabilidad de ocurrencia de problemas de patas (Eusebio, 2010). En el presente estudio, si bien no se observó problemas locomotores clínicos en las aves experimentales, sí se encontró la rotura de una cabeza de fémur en un ave del grupo control. Al respecto, existen antecedentes del efecto de la manipulación de las aves sobre la manifestación de anomalías compatibles con la lesión observada en la cabeza femoral proximal en una muestra de las aves control (Mitchell and Boon, 1986); sin embargo, es poco probable que dicha lesión haya sido producida durante la manipulación de las aves, ya que éstas fueron manipuladas sosteniéndolas lateralmente por el cuerpo y no de las patas, lo que minimiza el riesgo de lesiones (Chen and Moran, 1995).

**Cuadro 12. Efecto de un producto a base de aceite esencial de orégano sobre integridad esquelética: Mediciones complementarias.**

| Variable | Efecto del aditivo | | P |
|---|---|---|---|
| | Control | PRO | |
| Capacidad para caminar [1] | 6.00 | 6.00 | 1.0000 |
| Degeneración femoral [2] | 1.25 | 1.00 | 0.3253 |
| Discondroplasia tibial [3] | 1.13 | 1.00 | 0.3253 |

[1,2,3] Medidos en escalas de 1 a 6, de 1 a 5, y de 1 a 4, respectivamente.

En relación a la morfometría ósea, el valor promedio observado en el presente estudio para el largo de la tibia de 51.4 (Cuadro 10) es menor que el valor de 57 mm para largo reportado por Skinner and Waldroup (1995) en pollos de la misma edad y línea genética, mientras que el valor promedio de 4.5 mm observado en el presente estudio para el grosor de la tibia es mayor que el valor de 3.4 mm reportado por dichos investigadores. Por otro lado, el valor promedio observado en el presente estudio para el largo del fémur (38.8 mm) es menor que el valor reportado por Barreiro *et al* (2010) de aproximadamente 45 mm en pollos de la misma edad y línea genética, mientras que el valor promedio de 5.7 mm observado en el presente estudio para el grosor del fémur es mayor que el valor de aproximadamente 4 mm reportado por dichos investigadores. Las diferencias observadas en el presente estudio respecto a los valores reportados por Skinner and Waldroup (1995) y Barreiro *et al* (2010) guardan relación con el cambio en la conformación corporal producido a través de los años, y se verifica por el mayor peso corporal promedio de la aves del presente estudio (460 g) respecto a los 307 g reportado por Skinner and Waldroup (1995).

### 3.1. Densidad ósea

La densidad del hueso es uno de los indicadores más importantes para medir la integridad y calidad del hueso (Almeida Paz *et al*, 2008) y puede ser medida de forma directa, como en el presente estudio, o con el uso de métodos que permiten su estimación, como son la absorciometría de rayos-X de energía dual (DEXA; conocida en medicina humana como densitometría ósea; Muñóz-Torres *et al*, 2005) (Mitchell *et al*, 1997; Angel *et al*, 2006), empleando un lixiscopio (McKay *et al*, 2000), mediante la composición mineral, la resistencia a la rotura, el índice Seedor, entre otros (Almeida Paz y Bruno, 2006). La mayor densidad ósea observada en el presente estudio en las aves suplementadas con el PRO (Cuadro 11) refleja su mayor contenido mineral y es considerada como un factor atenuante del riesgo de fracturas (Rath *et al*, 2000), debido a que el contenido de ceniza en el hueso y su densidad son indicadores de la fortaleza y del estado de salud del hueso, ya que la mineralización provee a los huesos de resistencia a la compresión (Rath *et al*, 2000).

Los resultados del presente estudio indican que la menor densidad ósea en los pollos no suplementados con el PRO es significativa pero moderada. Al respecto, Angel *et* *al* (2006) reportaron que grados de mineralización marginalmente menores pueden no incrementar la tasa de decomisos y pérdidas durante el procesamiento. Sin embargo, las características del proceso de saca y faenado, así como la experticia del personal involucrado pueden incrementar las lesiones en las patas y alas. Por lo tanto, el incremento observado en la densidad ósea, si bien es moderado, favorece la reducción del riesgo de pérdidas durante la crianza y procesamiento de las aves.

Se ha observado, además, un incremento en la densidad de los huesos de las aves suplementadas con el PRO (Cuadro 11). Sin embargo, no se observó diferencias en largo o grosor de los huesos estudiados, a pesar que las aves suplementadas con el PRO presentaron un mayor peso corporal que las aves control (489 vs 430 g; data no presentada). Al respecto, se ha establecido que, manteniendo constantes las demás características del hueso, su mayor densidad se correlaciona con una mayor fortaleza. Se ha determinado también que la densidad del hueso puede estar influenciada por factores relacionados a las características químicas de su matriz orgánica (Knott and Bailey, 1998), debido a modificaciones en los enlaces cruzados de colágeno intermoleculares, la interacción entre proteoglicanaos y proteínas no colágenas con colágeno, y otros cambios glico-oxidativos en el colágeno y los proteoglicanos que alteran la mineralización y las propiedades bioquímicas (Oxlund *et al*, 1995).

A diferencia de lo observado en el presente estudio, Williams *et al* (2000) reportan que los animales que crecen más rápido presentan huesos más largos, con mayores diámetros y mayor espesor en la región cortical. Shim *et al* (2011b) por su parte reportan que en una misma población, los pollos que presentan mayor tasa de crecimiento tienen huesos más largos, anchos, pesados y fuertes que aquellos que presentan una menor velocidad de crecimiento (Shim *et al*, 2011b), debido probablemente a que la hormona del crecimiento estimula la síntesis de ILGF-1 a nivel del disco epifisiario, lo que induce la mitogénesis, resultando en el crecimiento del hueso (Isaksson *et al*, 1987); efecto que ha sido documentado también por Kocamis *et al* (2000), quienes tras administrar ILGF-1 *in ovo* observaron mayor peso corporal, longitud del fémur y tibia, y concentración de hidroxiprolina en los huesos.

Se ha establecido, que la modelación y remodelación del hueso modifican su tamaño y contornos externos, así como su arquitectura interna mediante la deposición o remoción de material de la superficie del hueso. Además, si la estructura de un hueso determina la carga que puede tolerar, también la carga determina su estructura (Seeman and Delmas, 2006), ya que el modelaje es un proceso adaptativo (Bain and Watkins, 1993; Rath *et al*, 2000; Brickett *et al*, 2007). Así, el peso que el hueso soporta estimula su crecimiento e incrementa su densidad (Knowles and Broom, 1990), observándose que en aves de mayor peso corporal se puede incrementar el peso del hueso y reducir su crecimiento longitudinal, debido a la mayor presión sobre la placa de crecimiento (Rutten *et al*, 2002), mientras que aves en corrales abiertos presentan mayor densidad ósea que aquellas en baterías (Fleming *et al*, 1994). Así, se ha establecido que la tibia puede remodelarse en cualquier punto de su estructura para adaptarse al estrés mecánico (Bimener *et al*, 1986).

En el presente estudio, la ausencia de diferencias en el largo y ancho de los huesos (Cuadro 10) a pesar del mayor peso corporal puede ser consecuencia del reducido periodo de evaluación. Al respecto, Applegate and Liburn (2002) observaron en pollos que en el día 15 de edad el fémur y la tibia han alcanzado, respectivamente, el 50% y 45% de la longitud y grosor que logran en el día 43 de edad; sin embargo, las mismas aves al día 15 de edad sólo han alcanzado el 15% del peso corporal que logran al día 43. Esto demuestra que, en los pollos, el desarrollo óseo se produce con una menor velocidad que el incremento del peso corporal, y explica la ausencia de diferencias en el largo y grosor de los huesos de las aves en el presente estudio.

Por lo anteriormente expuesto es posible establecer que la mayor densidad ósea observada en las aves suplementadas con el PRO pueda ser resultado de dicho factor evaluado y no del peso corporal, ya que no se observa diferencias en el tamaño de los huesos (Cuadro 10). Los resultados de Kim *et al* (2011) concuerdan con esta hipótesis, reportando que pollos suplementados con 250 µg/kg de Vitamina D en la dieta presentan sólo un incremento en la densidad ósea pero no el tamaño del hueso.

Resultan consistentes también los hallazgos de Liu *et al* (2004), quienes evaluaron el efecto de diferentes fuentes de lípidos dietarios sobre el desarrollo óseo de codornices, observando diferencias significativas entre los tratamientos evaluados en el contenido porcentual de ceniza en la tibia, pero no en el peso, largo y diámetro de dicho hueso, ni en el peso corporal, por lo que los investigadores concluyen que es necesario que la duración del factor en estudio sea lo suficientemente prolongado para causar un efecto significativo en las variables en estudio. Por su parte, Gallinger *et al* (2004) evaluaron el efecto de la inclusión de salvado de arroz sobre el desarrollo y mineralización óseo y observaron diferencias significativas entre los niveles de inclusión de 0 y 20% en la dieta, en el contenido porcentual de ceniza en la tibia, pero no en el largo ni en el peso del hueso, ni en el peso corporal de las aves.

Al respecto, Moraes (2006) evaluó el efecto de diferentes fuentes de grasa dietaria sobre la mineralización del fémur de pollos de carne, observando en el día 14 de edad, diferencias estadísticamente significativas entre tratamientos en el contenido porcentual de cenizas en el hueso, pero no en su peso, largo, diámetro e índice Seedor, como tampoco en el peso corporal de las aves, sugiriendo que el periodo necesario para observar diferencias en estas últimas variables puede ser mayor. Esto fue verificado por dichos investigadores al evaluar los mismos indicadores en el día 42 de edad, en que observaron diferencias significativas en el contenido porcentual de cenizas en el hueso, pero también en el índice Seedor.

La mayor densidad ósea observada por efecto del PRO (Cuadro 11) puede ser aun más favorable a la integridad esquelética en pollos de carne de la línea Cobb, ya que presentan mayor velocidad de crecimiento que los de la línea Ross (Marcato *et al*, 2008) y, consecuentemente, mayor frecuencia de problemas de patas (Kestin *et al*, 1992), aunado a la posible menor capacidad para utilizar los minerales de la dieta en las líneas genéticas actuales (McDevitt *et al*, 2006).

El efecto observado en la mineralización de las aves por efecto de la suplementación del PRO en el presente experimento, resulta de interés para otras líneas de aves que suelen ver afectada su integridad esquelética. Tal es el caso de los pavos machos, los cuales suelen presentar fracturas de fémur a partir de la semana 15 (Applegate and Angel, 2008), y las gallinas ponedoras, ya que la integridad esquelética de prácticamente todas ellas resulta afectada por el desarrollo de osteoporosis a pesar de contar con una adecuada suplementación mineral (Thorp, 1994), observándose que la fragilidad de sus huesos puede conducir a fracturas en un 30% de las aves al final del ciclo productivo (Gregory and Wilkins, 1989).

## 3.2. Mecanismos del AEO que influyen sobre la mineralización ósea

### 3.2.1. Disponibilidad de nutrientes y mineralización ósea

Se ha establecido que el requerimiento de nitrógeno de los pollos de carne para el óptimo desarrollo de la matriz orgánica del hueso es usualmente mayor que el requerimiento aparente para crecimiento corporal (Leeson and Summers, 2005). Al respecto, se ha reportado que la inclusión de AEO en la dieta de pollos de carne incrementa el coeficiente de metabolización aparente de nitrógeno (Hernández *et al*, 2004; Díaz, 2011), la actividad de la quimotripsina y consecuentemente la digestibilidad de la proteína dietaria (Basmacioglu Malayoglu *et al*, 2010), y que el timol, metabolito secundario del orégano, incrementa la actividad enzimática de la quimotripsina y tripsina en 14% y 17%, respectivamente (Lee *et al*, 2004).

Bittar (2011) reportan que las lesiones entéricas disminuyen la absorción de todos los nutrientes, en especial de la vitamina D3, y que en general las aves que presentan disturbios intestinales tienen mayor incidencia de problemas locomotores. Por esta razón, toda mejora en la capacidad intestinal para la absorción de nutrientes puede favorecer la mineralización ósea. Al respecto, se ha documentado el efecto favorable que ejerce el AEO sobre la digestibilidad de los nutrientes dietarios, por lo que resulta esperable que también se produzca un incremento en la disponibilidad de minerales, tal como ha sido reportado con el uso de otras plantas (*Cissus quadrangularis, Zingiber officinale, Lepidium sativum, Terminalia arjuna, Cestrum diurnum* y *Uraria picta*) demostrándose un efecto positivo sobre la absorción de calcio y fósforo, sobre la formación de colágeno y sobre la mineralización ósea (Deka *et al*, 1994; Somkuwar *et al*, 2010).

La mayor densidad ósea observada en las aves que reciben el PRO (Cuadro 11) es consistente con el incremento en la disponibilidad de nutrientes a nivel intestinal, y con los hallazgos de Eusebio (2010), quien observó en aves con una misma velocidad de crecimiento, que dietas a base de trigo redujeron la integridad esquelética incrementando la probabilidad de ocurrencia de problemas de patas por deformaciones respecto a las dietas a base de maíz, debido a su alto contenido de polisacáridos no almidonosos, que reduce la digestibilidad de nutrientes, incluyendo minerales y otros factores involucrados en el crecimiento y desarrollo de los huesos.

### 3.2.2. Microflora intestinal y mineralización ósea

Se ha establecido que el AEO favorece el crecimiento de la flora intestinal benéfica (Lee *et al*, 2004; Hammer *et al*, 1999; Dorman and Deans, 2000; Zheng *et al*, 2010; Betancourt *et al*, 2011). Al respecto, Plavnik and Scott (1980) observaron que la adición de 2.5 a 5% de levadura de cervecería en la dieta de pollos reduce la debilidad de las patas y la incidencia de discondroplasia tibial, e incrementa la fortaleza del hueso. Por su parte, Mutus *et al* (2006) adicionaron probióticos a la dieta de pollos y observaron incrementos significativos en el grosor del canal medular y de la pared lateral de la tibia, mayor índice tibiotarsal –entendido como la fracción ósea no medular en el diámetro de la diáfisis–, y mayor contenido de fósforo y ceniza. Asimismo, Nahashon *et al* (1994) reportaron una respuesta favorable en la retención de calcio y fósforo tras la suplementación dietaria de *Lactobcillus sp.*

Se ha observado que los prebióticos y probióticos favorecen la mineralización y desarrollo óseo mediante mecanismos que incluyen: el incremento de la solubilidad de materiales debido a la mayor producción bacteriana de ácidos grasos de cadena corta; el incremento en la superficie de absorción debido a la proliferación de enterocitos mediado por productos de la fermentación bacteriana, como lactato y butirato; incremento en la expresión de proteínas ligantes de calcio; la mejora de la salud intestinal; degradación del ácido fítico que forma complejos minerales no disponibles; liberación de factores moduladores del hueso como fitoestrógenos de la dieta; estabilización de la flora, ecología y mucina intestinal; y el impacto en factores moduladores del crecimiento como las poliaminas (Scholz-Ahrens *et al*, 2007).

Esta respuesta positiva en el balance mineral de las aves suplementadas con probióticos y prebióticos, así como el efecto prebiótico que desarrolla el AEO (Hammer *et al*, 1999; Lee *et al*, 2004; Dorman and Deans, 2000; Yew, 2008; Zheng *et al*, 2010; Betancourt *et al*, 2011) permiten correlacionar la acción del AEO sobre la microflora intestinal con la mayor mineralización ósea e integridad esquelética de las aves.

### 3.2.3. Antioxidantes, expresión génica y mineralización ósea.

Se ha determinado que los radicales libres pueden incrementar la resorción mineral de los huesos a través de la activación del factor nuclear κB (NF-κB) (Schenck, 2011), que es un complejo proteico que regula la expresión de genes que codifican proteínas que participan en los procesos inflamatorios, en la activación del sistema inmune y en la transducción de la señal antiapoptótica (Paur *et al*, 2010).

Al respecto, Paur *et al* (2010) evaluaron el efecto de una preparación a partir de extractos de orégano, tomillo, café, clavo y nueces sobre la actividad transcipcional del NF-κB *in vitro* e *in vivo* empleando ratones, y observaron una drástica y significativa reducción, dependiente de la dosis, en la actividad del NF-κB, lo que podría reducir la resorción mineral de los huesos.

Por otro lado, se ha establecido que el óxido nítrico influye sobre la remodelación del hueso a través de la modulación que ejerce sobre la actividad de los osteoblastos y osteoclastos (Schenck, 2011), observándose que la alta concentración de óxido nítrico inhibe la diferenciación y actividad de los osteoblastos (Schenck, 2011), pero en mayor magnitud de los osteoclastos (van't Hof and Ralston, 2001; Jamal, 2011). Se ha determinado también que cuando los osteoblastos y osteocitos son sometidos a estrés mecánico, producen óxido nítrico (van't Hof and Ralston, 2001), que inhibe la resorción mineral de hueso mediada por prostaglandinas (Schenck, 2011).

Al respecto, se ha demostrado que el anión superóxido, que es la especie reactiva de oxígeno más abuntante en el organismo, puede transformar el óxido nítrico en anión peroxinitrito, que es altamente reactivo. Esto reduce la concentración de óxido nítrico y afecta el balance osteoblastos-osteoclastos, produciendo disturbios en el metabolismo del hueso (Schenck, 2011).

En consecuencia, el efecto antioxidante que ha sido documentado en el AEO (Máthé, 2009; Applegate *et al*, 2010) explica, al menos en parte, la mayor densidad ósea en las aves suplementadas con el PRO, al capturar radicales superóxido y probablemente favorecer la disponibilidad de óxido nítrico a nivel local. Este efecto es consistente con reportes previos (Jamal, 2011), en que se ha observado el incremento en la densidad ósea y en los indicadores de fortaleza del hueso tras la administración farmacológica de sustancias donadoras del óxido nítrico como la nitroglicerina, que incrementan la disponibilidad de óxido nítrico.

### 3.3. La mineralización no es un proceso homogéneo en el hueso

Los índices modificado de Seedor, de robusticidad y Quetelet empleados en el presente estudio fueron calculados a partir del peso y longitud del hueso, mientras que el índice de forma fue calculado a partir del largo y ancho del hueso, y en ninguna de estas variables se observó diferencias significativas, mientras que sí en la densidad del hueso (Cuadro 11). Es posible que los índices mencionados carezcan de la sensibilidad para evidenciar cambios marginales en la mineralización del hueso debido a que la menor mineralización es más evidente en las epífisis que en la diáfisis, donde fue medido el grosor del hueso.

Al respecto, se ha establecido que, en los huesos largos, la mineralización no se produce de forma homogénea a lo largo del hueso, observándose que la porción proximal del hueso es más propensa a tener menor mineralización, tiene menor contenido de calcio, fósforo y ceniza, y aquí se produce una mayor incidencia de problemas de patas que en el extremo distal (Durairaj, 2008).

En relación a lo anterior, Almeida Paz *et al* (2008) observaron que los pollos machos tienden a presentar menor densidad en la epífisis del fémur, a pesar de su mayor resistencia a la rotura en la diáfisis, asociado a una mayor suceptibilidad a los síndromes metabólicos que afectan la calidad ósea e integridad esquelética produciendo lesiones en la epífisis proximal del fémur con un grado correlacionado inversamente con la densidad en esta región del hueso, demostrando que la menor mineralización del hueso resulta más evidente en la epífisis que en la diáfisis.

Se ha determinado que los desórdenes que ocurren durante el crecimiento de los huesos son frecuentemente resultado de patologías en la placa de crecimiento o anormalidades en el modelaje del hueso (Thorp, 1994), tal como sucede en la discondroplasia tibial, en que se observa una anormal masa cartilaginosa blanca, opaca, no mineralizada ni vascularizada en el extremo proximal de la tibia (Shim *et* *al*, 2011a), desde la placa de crecimiento de la epífisis hacia la metáfisis (Farquharson *et al*, 1992), lo que corrobora que una menor mineralización tiende a ser más acentuada en la epífisis del hueso.

### 3.4. Sensibilidad de los huesos evaluados

Los resultados del presente estudio indican que si bien el incremento en la densidad ósea en las aves suplementadas con el PRO se produce tanto en el fémur como en la tibia y el tarso, el incremento en el fémur es 50% mayor que en la tibia.

Al respecto, la tibia es el hueso con mayor velocidad de crecimiento en el ave, ha sido considerado por muchos años como el más sensible a las deficiencias de calcio y fósforo (McLean and Urist, 1961; citados por Yan *et al*, 2005), y presenta mayor sensibilidad en cuanto al contenido de ceniza que la ganancia de peso ante variaciones dietarias de fósforo (Skinner *et al*, 1992; Waldroup *et al*, 2000; Yan *et al*, 2001; Dhandu and Angel, 2003), especialmente en la zona de proliferación en aves jóvenes (Nelson and Walker, 1964). Sin embargo, se ha determinado que otras estructuras pueden ser empleadas para determinar la mineralización ósea en pollos de carne con igual confiabilidad que en la tibia, como por ejemplo la pata (García and Dale, 2006) o los dedos de las patas (Yan *et al*, 2005), se haya o no extraído la grasa de las muestras empleadas, así como el fémur (Moraes, 2006). Al respecto, Dalmagro *et al* (2011) reportan que el contenido de ceniza en la pata y la densidad ósea guardan una relación directa con los niveles dietarios de calcio y fósforo.

Más aun, se ha determinado que el fémur presenta mayor sensibilidad que la tibia a cambios nutricionales o a otros factores que influyen sobre la mineralización ósea. Al respecto, se ha demostrado en estudios sobre el recambio del calcio y la relación calcio-fósforo en la dieta, que el fémur es más sensible que la tibia a los cambios dietarios (Itoh and Hatano, 1964; Dilworth and Day, 1965). Por ello, a pesar que el contenido de ceniza en la tibia ha sido el principal criterio para determinar los requerimientos nutricionales de calcio y fósforo en la mayoría de especies avícolas (Skinner and Waldroup, 1995), el estado general del desarrollo esquelético en pollos de carne no debe establecerse sólo en base a las mediciones en la tibia (Applegate and Lilburn, 2002) ya que emplearla como base para la evaluación de la integridad esquelética en pollos de carne parece no ser apropiada (Chen and Moran, 1995) debido a que diferentes huesos presentan diferentes características de desarrollo (Applegate and Lilburn, 2002), siendo el fémur más sensible que la tibia al daño mecánico durante el deshuesado y a cambios en la dieta (Moran and Todd, 1994).

Se ha observado incluso que los problemas femorales de los pollos al beneficio son más frecuentes que aquellos relacionados con la tibia (Mitchell and De Boom, 1986; Duff and Randall, 1987; Gregory and Austin, 1992), mientras que durante el aturdimiento y desplume son más frecuentes las roturas en la clavícula, coracoide, escápula e isquion (Gregory and Wilkins, 1992; citados por Chen and Moran, 1995).

Ello se debe en parte a que todos los huesos no se desarrollan de la misma forma. Chen and Moran (1995) evaluaron el efecto de 0.40% y 0.25% de fósforo disponible, mediante la suplementación de diferentes niveles de fosfato dicálcico, en dietas a base de maíz y soya sobre las características óseas de pollos de carne, observando a las seis semanas de edad diferencias significativas en el contenido porcentual de ceniza total en el hueso así como en la densidad en la diáfisis, tanto en el fémur como en la tibia. Sin embargo, si bien se observó diferencias significativas en el contenido relativo de mineral en las epífisis proximal y distal en el fémur, no se observó diferencias atribuibles al tratamiento dietario en la tibia. Esto evidencia que el fémur y en particular las regiones de las epífisis presentan una mayor sensibilidad a los niveles bajos de fósforo disponible en la dieta que la tibia, y que resultados experimentales de densidad ósea por densitometría pueden reflejar falsos negativos si se evalúan las epífisis de la tibia. Los investigadores observaron además, que en condiciones de restricción de fósforo dietario en pollos en crecimiento existe una deposición mineral preferente en la epífisis proximal de la tibia en relación al fémur, pudiéndose observar deficiencias en la mineralización en las epífisis del fémur pero no de la tibia; es decir, que el fémur es más sensible a las deficiencias dietarias de fósforo que la tibia. Los investigadores consideran que las clavículas, las costillas y el isquion son igualmente, sino más sensibles que el fémur.

Esto ha sido corroborado por Applegate and Liburn (2002) quienes observaron que la mineralización en la diáfisis, medida a través del contenido de ceniza, es comparativamente menor en el fémur que en la tibia, mientras que en la epífisis se produce un ligero, pero no significativo, efecto opuesto. Ambas observaciones explican la mayor frecuencia de problemas óseos en el fémur que en la tibia (Applegate and Liburn, 2002).

Para estudiar la mineralización ósea en pollos de carne, Angel *et al* (2006) evaluaron el efecto de la suplementación dietaria de diferentes niveles de fósforo disponible y de una fitasa sobre la densidad ósea y el contenido mineral del hueso, medidos ambos por DEXA, así como sobre el peso y contenido porcentual de ceniza en el hueso en el día 49 de edad. Los investigadores observaron que si bien la tibia presenta mayor sensibilidad que el esqueleto completo ante diferentes tratamientos dietarios, el fémur presenta una sensibilidad aun mayor que tibia. Al respecto, se ha reportado que el contenido de ceniza en el fémur en las últimas etapas de crecimiento es un mejor indicador de estado de mineralización del hueso (Moran and Todd, 1994; Chen and Moran, 1995; Dhandu and Angel, 2003).

Finalmente, empleando gallinas reproductoras pesadas, Almeida Paz *et al* (2006) y Mendes *et al* (2006) determinaron que la correlación entre la calidad ósea en el fémur y la tibia es reducida. Así, al evaluar la correlación entre un mismo indicador en la tibia y el fémur observaron que, para la densidad por DEXA fue de 0.67, para el contenido de ceniza fue de 0.72, no encontrando una correlación significativa entre la resistencia a la rotura del fémur y la tibia.

### 3.5. Sensibilidad de las variables en estudio

Se ha reportado que existe una alta variabilidad en los datos referentes al desarrollo óseo en aves en función a las diferentes metodologías empleadas (Oviedo-Rondón and Ferket, 2005).

Así, empleando gallinas reproductoras pesadas, Almeida Paz *et al* (2006) y Mendes *et al* (2006) determinaron una limitada consistencia entre los diferentes indicadores de mineralización ósea. Así, al evaluar la correlación del mismo indicador entre tibia y fémur observaron que, si bien para el índice modificado de Seedor fue de 0.95, para el contenido de ceniza fue de 0.72, para la densidad medida por DEXA fue de 0.67, y no encontraron una correlación significativa entre la resistencia a la rotura en el fémur y en la tibia. Asimismo, observaron correlaciones negativas entre el índice modificado de Seedor y el contenido de ceniza (-0.47 en la tibia y -0.76 en el fémur), así como entre la densidad medida mediante DEXA y el contenido de ceniza (-0.38 en la tibia y -0.46 en el fémur), mientras que no encontraron correlaciones significativas entre el índice modificado de Seedor y la resistencia a la rotura en la tibia.

Otros reportes sobre la mineralización y desarrollo óseo corroboran las diferencias en sensibilidad de los diferentes indicadores empleados y respaldan la validez de la densidad ósea. Al respecto, Adebiyi *et al* (2009) evaluaron el efecto de la suplementación de 2, 4 y 6% de diatomita en la dieta de gallos sobre la mineralización de la tibia, observando mejoras significativas en la ganancia de peso y conversión alimentaria, e incrementos significativos en el índice modificado de Seedor y en los contenidos porcentuales de Ca, P y ceniza; sin embargo, no observaron diferencias significativas en el peso, largo e índice de robusticidad de la tibia.

Por su parte, Kocabagli (2001) evaluó el efecto de una fitasa en la dieta de pollos de carne sobre algunos indicadores de mineralización de la tibia, observando un incremento en el peso corporal, en el contenido de ceniza en la tibia y en los índices de robusticidad y modificado de Seedor, pero no en el peso y largo del hueso, ni en el diámetro de la diáfisis. Monteagudo *et al* (1997) comparó el índice modificado de Seedor y el índice de robusticidad con la densidad ósea, determinada por DEXA, y con el contenido de mineral en el hueso de ratas. Los investigadores determinaron que el índice modificado de Seedor y el índice de robusticidad tienen una mayor correlación con el contenido mineral en el hueso que con la densidad del mismo. Asimismo, observaron que el índice modificado de Seedor tiene mayor correlación que el índice de robusticidad con el contenido de mineral en el hueso; sin embargo, en ratas overiectomizadas –como modelo de estudio de la osteoporosis (Seedor *et al*, 1991; Sato, 1995; Arjmandi *et al*, 1996)– ambos índices fueron significativamente diferentes de la densidad ósea, por lo que no resultan apropiados en estas condiciones.

Dhandu and Angel (2003) evaluaron el efecto de cuatro niveles dietarios de fósforo disponible que variaron entre 0.15% y 0.31% en pollos de carne de 42 días de edad, observando diferencias significativas en el contenido absoluto y porcentual de ceniza en la tibia, pero no en las variables directas de fortaleza del hueso, como la fuerza requerida para romper el hueso (kg), o el módulo de elasticidad ($kg/cm^2$) entendido como la tasa de deformación del hueso por unidad de fuerza ejercida sobre él, por lo que el contenido de contenido de ceniza en una variable que brinda mayor sensibilidad.

Angel *et al* (2006) evaluaron el efecto de la suplementación dietaria de diferentes niveles de fósforo disponible y de una fitasa sobre la densidad mineral ósea (DMO) y el contenido mineral del hueso (CMO), medidos ambos por DEXA, así como sobre el peso y contenido porcentual de ceniza en el hueso en el día 49 de edad. Los investigadores observaron que el CMO es más sensible que la DMO, y que el CMO es más sensible que el peso de la ceniza del hueso. Observaron además que las medidas logradas del contenido de ceniza en el hueso coinciden con las mediciones mediante DEXA, lo que también ha sido reportado en otros estudios (Akpe *et al*, 1987; Mitchell *et al*, 1997) que correlacionan ambas mediciones.

En comparación a la determinación de la mineralización ósea mediante el contenido de ceniza o mediante técnicas avanzadas como DEXA que son más rápidas (Akpe *et al*, 1987; Angel *et al*, 2006), el método aplicado en el presente estudio para la determinación de la densidad del hueso no requiere que las muestras sean enviadas a un laboratorio especializado. Por el contrario puede realizarse en los propios centros de producción avícola, con costos reducidos, de manera simple y con resultados confiables, como se corrobora de los resultados del presente estudio, pudiendo ser realizado incluso por los encargados de granja.

### 3.6. Ventajas del procesamiento aplicado a las muestras

En el presente estudio las muestras óseas fueron hervidas de acuerdo a experiencias previas (Orban *et al*, 1993; Kocabagli, 2001; Applegate and Lilburn, 2002; Moraes, 2006; Almeida Paz *et al*, 2008). Si bien la obtención de los huesos por extracción directa puede tomar menos tiempo que hirviendo las muestras (Orban *et al*, 1993), hervir las muestras permite retirar completamente los cartílagos y demás tejidos no óseos sin dañar el hueso en caso éste presentara lesiones o deficiencias de mineralización, favoreciendo la sensibilidad del método.

Si bien el contenido de mineral, la densidad y la resistencia a la rotura, son influenciados por el proceso de secado de las muestras (Orban *et al*, 1993), la parcial extracción que se logra del contenido de grasa durante este proceso térmico (Almeida Paz *et al*, 2008), así como la reducción en el contenido de humedad tras el secado a temperatura ambiente, incrementan la sensibilidad a las variaciones en la densidad del hueso por efecto de su contenido mineral. Al respecto, Orban *et al* (1993) observaron una reducción del 25 a 28% en la densidad de la tibia luego de ser sometida a secado a temperatura ambiente durante 3 a 7 días. Esta metodología podría evitar la extracción de la grasa con solventes orgánicos, así como el empleo de estufas para el proceso de secado de las muestras (Murakami *et al*, 2009), representando una ventaja operativa respecto a la extracción de grasa recomendada por la AOAC que es usualmente un paso limitante en la evaluación de la mineralización ósea (Yan *et al*, 2005).

## 4. Conclusiones

Los resultados obtenidos bajo las condiciones del presente estudio permiten llegar a las siguientes conclusiones:

- La suplementación del PRO no produce cambios en la morfometría ósea de pollos de carne.

- La suplementación del PRO favorece la integridad esquelética incrementando la mineralización ósea, determinada mediante la densidad del hueso.

- El incremento en la mineralización ósea como resultado de la suplementación del PRO se produce tanto en el fémur y tarso como en la tibia.

# EXPERIMENTO N° 4

# EFECTO DE UN PRODUCTO A BASE DE ACEITE ESENCIAL DE ORÉGANO SOBRE EL ESTADO ANTIOXIDANTE DE POLLOS DE CARNE

## 1. Introducción

En los sistemas industriales de producción las aves se encuentran sometidas a diversos factores que limitan su eficiencia productiva como las micotoxinas, las coccidias, las enteritis bacterianas, la calidad sub-óptima de los insumos dietarios, la producción de amoniaco en los ambientes de crianza, los desafíos infecciosos y el estrés por calor. Los efectos de estos factores en las aves incluyen depresión y disminución en la ingesta de alimento, como sucede en los procesos respiratorios infecciosos, el daño a la mucosa intestinal y la disminución en la eficiencia alimentaria y ganancia de peso, como sucede los cuadros de coccidia y/o *Clostridium perfringens*, o el daño hepático producido por micotoxinas u otros factores dietarios.

Un aspecto común en la mayoría de estos procesos es el impacto negativo que tienen sobre el estado antioxidante del ave, incrementando el gasto energético de mantenimiento y reduciendo la eficiencia de conversión alimentaria. Considerando que los requerimientos de mantenimiento son mucho más altos que los de crecimiento, es necesario entonces concentrarse en controlar los factores que incrementan el gasto de mantenimiento y en lo posible reducirlo (Clements, 2011).

Se ha reportado que el aceite esencial de orégano (AEO) tiene actividad antioxidante reduciendo la peroxidación de lípidos (Faleiro *et al*, 2005); sin embargo, es escasa la información disponible sobre su efecto a nivel sistémico. El objetivo de este experimento fue determinar el efecto de un producto a base de AEO (referido en adelante como PRO) sobre el estado antioxidante de pollos de carne.

# 2. Materiales y métodos

## 2.1. Lugar, fecha y duración

La crianza de las aves se llevó a cabo en las instalaciones del Programa de Aves de la Facultad de Zootecnia de la Universidad Nacional Agraria La Molina (UNALM) a mediados del segundo semestre del 2010. Los inóculos de *C. perfringens* y coccidia, empleados para desafiar a las aves, fueron preparados en el Laboratorio de Biología y Genética Molecular de la Facultad de Medicina Veterinaria de la Universidad Nacional Mayor de San Marcos (UNMSM) en Lima-Perú y en el Programa de Aves de la UNALM, respectivamente. Las muestras de sangre de las aves fueron colectadas en las instalaciones del Laboratorio de Patología Aviar de la UNMSM y fueron analizadas por el Laboratorio Clínico ROE. La evaluación se llevó a cabo desde la recepción de las aves en las instalaciones de crianza y el periodo de evaluación fue el comprendido de 0 a 28 días de edad.

## 2.2. Instalaciones, equipos y materiales

### 2.2.1. Instalaciones de crianza

Las aves estuvieron alojadas en 12 jaulas distribuidas en baterías con comederos y bebederos lineales y las características que se detallan en el Anexo 2. Las aves fueron alojadas a razón de 21.4 pollos/$m^2$ sobre material de cama reutilizado (0.047 $m^2$/pollo). Durante los primeros 3 días de vida se colocó papel periódico para evitar el acceso directo de los pollos BB al material de cama, el alimento fue suministrado en bandejas plásticas y el agua de bebida en bebederos tipo tongo.

Dos semanas antes de la recepción de las aves se inició un programa de control de vectores, empleando cebaderos, rodenticidas y mosquicidas comerciales. Antes y después del periodo experimental se realizó la limpieza y desinfección de las instalaciones.

### 2.2.2. Calefacción, control de la temperatura y ventilación

La calefacción fue provista por un sistema eléctrico con resistencias y termostatos, y fue controlada de acuerdo a las recomendaciones de la línea genética (Cobb-Vantress, 2008b). La ventilación fue controlada con cortinas de polipropileno instaladas en el perímetro del área de crianza. La humedad ambiental en el área de crianza se mantuvo alrededor de 40%. Para la medición de la temperatura y humedad ambiental se empleó un termo-higrómetro electrónico digital con una aproximación de 0.1 °C para temperatura y 1% para humedad.

### 2.2.3. Preparación del alimento

Para el pesaje de los ingredientes mayores del alimento se utilizó una balanza digital con capacidad de 150 kg y aproximación de 0.02 kg, mientras que para el pesaje de los ingredientes de la premezcla se utilizó una balanza electrónica con capacidad de 6 kg y aproximación de 1 g.

Se utilizó mezcladoras horizontales de cintas de 400 y 30 kg de capacidad para la mezcla de los ingredientes durante la preparación del alimento y para las premezclas, respectivamente. Después de la preparación de cada dieta se realizó el flushing y limpieza de los equipos (FAO, 2004; Ratcliff, 2009). El alimento fue envasado en sacos de papel y polietileno laminado, y almacenado en cilindros metálicos.

### 2.2.4. Manejo y actividades especiales

Para la identificación individual de las aves se empleó el método que se presenta en el Anexo 3. Los pesos vivos se obtuvieron utilizando una balanza electrónica con capacidad para 200 g y con aproximación de 10 mg y otra con capacidad para 15 kg y con aproximación de 1 g. Para la medición del alimento consumido, se utilizó una balanza electrónica con capacidad para 15 kg y con aproximación de 1 g.

En el día 10 de edad las aves fueron vacunadas, por vía ocular, contra la Enfermedad de Newcastle, empleando una vacuna viva liofilizada con una cepa VG/GA (Avinew®, Merial Limited).

Como parte del modelo entérico empleado para inducir el desafío al estado antioxidante de las aves se empleó un inóculo de coccidia que fue preparado a partir de una vacuna comercial contra coccidia (Immucox® for Chickens II, Vetech Laboratories) y un inóculo de *C. perfringens* proveniente de un aislamiento de un brote sub-clínico de campo de enteritis en pollos de 3 semanas de edad. Las características de los inóculos se presentan en los Anexos 9 y 10.

El agua de bebida fue potabilizada empleando 1 ml de hipoclorito de sodio al 4.5% por cada 10 L de agua. Para evitar que la presencia de cloro interfiera con la viabilidad de los inóculos la clorinación del agua se realizó tres días antes de proveerla a las aves, y diariamente, al momento de suministrarla a las aves se verificó la ausencia de cloro empleando un kit comercial de evaluación.

Para la toma de muestras de sangre se empleó guantes quirúrgicos y tubos especiales provistos por el Laboratorio Clínico ROE.

### 2.3. Animales experimentales

Se empleó 96 pollos machos de la línea Cobb 500, cuyas características se presentan en el Anexo 4. Los pollos fueron alojados al azar en 12 jaulas.

### 2.4. Modelo de desafío entérico

Para inducir el desafío al estado antioxidante se empleó un modelo entérico para inducir el STR desarrollado previamente (ver Anexo 1, modelo de desafío B).

### 2.5. Tratamientos

Se evaluaron 2 tratamientos, que fueron definidos de la siguiente manera:

Tratamiento 1: Dieta basal (grupo control)
Tratamiento 2: Dieta basal + 500 ppm de Orevitol®

Todas las aves estuvieron sometidas a un modelo entérico para inducir el desafío al estado antioxidante del ave.  El producto será referido, en adelante, como PRO.

## 2.6. Alimentación

En el periodo de 1 a 14 días de edad se suministró una dieta basal a base de maíz, soya y harina de pescado, y a partir del día 15 de edad se suministró una dieta basal a base de maíz y soya. Ambas dietas fueron complementadas con aceite de pescado y aminoácidos sintéticos, y suplementadas con una premezcla de vitaminas y minerales. La alimentación fue *ad libitum*. Las características de las dietas basales empleadas se presentan en el Cuadro 13. El aditivo indicado en los tratamientos dietarios fue incorporado en la dieta a expensas del maíz de 1 a 28 días de edad.

## 2.7. Mediciones

### 2.7.1. Estado antioxidante

Se determinó la actividad enzimática de la superóxido dismutasa (SOD) en sangre (Georgieva *et al*, 2006) y el contenido de caroteno sérico (Koinarski *et al*, 2005), muestreando seis aves por tratamiento en el día 28 de edad.

Las muestras para determinar SOD y caroteno sérico fueron tomadas colectando sangre directamente del ave en tubos de 4 ml de capacidad conteniendo 68 unidades USP de heparina litio como anticoagulante (BD Vacutainer® Lithium Heparin, Ref. 367884; BD Diagnostics, Preanalytical Systems) y de 10 ml de capacidad conteniendo un agente activador de la coagulación (BD Vacutainer® Serum, Ref. 367820; BD Diagnostics, Preanalytical Systems), respectivamente (Ison *et al*, 2005).

Luego de verificar la completa coagulación de las muestras para la determinación de caroteno, todos los tubos fueron remitidos inmediatamente al laboratorio de análisis. Para la determinación de SOD y caroteno fue necesario colectar por muestra no menos de 1 ml de sangre o suero, respectivamente.

La actividad de la SOD fue determinada por el método desarrollado por McCord y Fridovich en 1969 (Abiaka *et al*, 2000), empleando un kit comercial (RANSOD®, Ref. SD125; Randox Laboratories). Este emplea xantina y xantin oxidasa para generar radicales superóxido, los cuales reaccionan con cloruro de 2-(4-yodofenil)-

**Cuadro 13. Características de las dietas empleadas en el Experimento 4.**

| Ingredientes | % | |
|---|---|---|
| | De 1 a 14 días | De 15 a 28 días |
| Maíz amarillo | 52.517 | 61.775 |
| Torta de soya | 26.699 | 31.405 |
| Harina de pescado | 14.940 | 0.000 |
| Aceite semirefinado de pescado | 2.024 | 2.380 |
| DL-Metionina | 0.190 | 0.224 |
| L-Lisina | 0.116 | 0.135 |
| Cloruro de colina | 0.085 | 0.100 |
| Fosfato dicálcico | 1.609 | 1.892 |
| Carbonato de calcio | 0.978 | 1.150 |
| Sal común | 0.361 | 0.424 |
| Marcador inerte [1] | 0.300 | 0.300 |
| Premezcla [1] | 0.085 | 0.100 |
| Antifúngico [1] | 0.085 | 0.100 |
| Antioxidante [1] | 0.013 | 0.015 |

| Nutriente | Aporte nutricional | |
|---|---|---|
| | De 1 a 14 días | De 15 a 28 días |
| EM [2], Kcal/kg | 3028 | 3008 |
| Proteína cruda, % | 26.72 | 20.04 |
| Lisina, % | 1.72 | 1.16 |
| Metionina + Cistina, % | 1.10 | 0.87 |
| Treonina, % | 1.05 | 0.76 |
| Triptófano, % | 0.29 | 0.23 |
| Calcio, % | 1.54 | 1.17 |
| Fósforo disponible, % | 0.67 | 0.44 |
| Sodio, % | 0.34 | 0.19 |
| Grasa total, % | 5.97 | 5.11 |
| Fibra, % | 2.70 | 3.17 |

| Componente | Composición proximal [3], % | |
|---|---|---|
| | De 1 a 14 días [4] | De 15 a 28 días [5] |
| Humedad | 11.89 | 12.35 |
| Proteína total [2] | 26.06 | 19.66 |
| Extracto Etéreo | 4.87 | 3.83 |
| Fibra Cruda | 2.12 | 2.49 |
| Cenizas | 7.13 | 5.88 |
| ELN [2] | 47.94 | 55.79 |

[1] Marcador inerte: óxido crómico. Premezcla de vitaminas y minerales Proapak 2A®. Composición: Retinol: 12'000,000 UI; Colecalciferol: 2'500,000 UI; DL α-Tocoferol Acetato: 30,000 UI; Riboflavina: 5.5 g; Piridoxina: 3 g; Cianocobalamina: 0.015 g; Menadiona: 3 g; Ácido Fólico: 1 g; Niacina: 30 g; Ácido Pantoténico: 11 g; Biotina: 0.15 g; Zn: 45 g; Fe: 80 g; Mn: 65 g; Cu: 8 g; I: 1 g; Se: 0.15 g; Excipientes c.s.p. 1,000 g. Antifúngico: Mold Zap®; Antioxidante: Danox®

[2] EM: Energía metabolizable; Proteína total: N x 6.25; ELN: Extracto Libre de Nitrógeno (calculado).

[3] Informes de ensayo 1235/2010 LENA y 1236/2010 LENA, Universidad Nacional Agraria La Molina.

[4] Calculado a partir de los análisis proximales de la dieta empleada de 15 a 28 días de edad (85%) y de la harina de pescado (15%) empleadas en su producción.

[5] Informe de ensayo 1235/2010 LENA, Universidad Nacional Agraria La Molina.

3-(4-nitrofenol)-5-fenil tetrazolio (INT) para formar un colorante formazán rojo, cuya intensidad puede medirse por espectrofotometría (a 505 nm). La actividad de la SOD se mide por el grado de inhibición de esta reacción. Una unidad de SOD es la que produce un 50% de inhibición del valor de reducción de INT (Fathi *et al*, 2011), ya que la presencia de SOD reduce la formación de dicho colorante de manera proporcional a su concentración (Zanón, 2008). La actividad enzimática de la SOD se expresa en unidades SOD por ml de sangre entera (U/ml). La determinación de caroteno sérico se llevó a cabo por espectrofotometría cuantitativa (Costantini and Dell'Omo, 2006), y los resultados se expresan en µg/dL.

#### 2.7.2. Comportamiento productivo

Se midió empleando las siguientes variables:

- Peso vivo: Se registró los pesos individuales al recibir los pollos BB y en el día 28 de edad.
- Ganancia de peso: Se calculó valores acumulados al día 28 de edad.
- Consumo de alimento: Se calculó el consumo acumulado al día 28 de edad.
- Conversión alimentaria: Se calculó en función a los valores acumulados de consumo de alimento y ganancia de peso al día 28.

### 2.8. Diseño estadístico

Se empleó el Diseño Completo al Azar con 2 tratamientos y 6 repeticiones. El análisis de varianza se realizó con el programa Statistical Analysis System SAS 9.0 (SAS Institute, 2009) y la diferencia de medias empleando la prueba de Duncan (1955). El Modelo Aditivo Lineal General aplicado fue el siguiente:

$$Yij = U + Ti + Eij$$

Donde:

| | | |
|---|---|---|
| Yij | = | variable respuesta |
| U | = | media general |
| Ti | = | i-ésimo tratamiento ( i = 1, 2 ) |
| Eij | = | Error experimental |

## 3. Resultados y discusión

En el presente estudio se observa un efecto favorable en el estado antioxidante y en el comportamiento productivo de las aves suplementadas con el PRO en el alimento. En el día 28 de edad las aves que reciben el PRO presentan una actividad de la superóxido dismutasa 15% mayor que las aves control ($P<0.05$), tienen un peso corporal 10% mayor (136 g más por ave; $P<0.01$) y una mayor ganancia de peso ($P<0.01$), así como una menor conversión alimenticia ($P<0.05$), equivalente a 160 g menos alimento consumido por cada kg de peso ganado. Sin embargo, no se observan diferencias en los niveles de caroteno sérico (Cuadro 14 y Anexos 46 y 47).

Al respecto, se ha documentado que uno de los principales componentes del AEO, carvacrol, reduce la expresión génica de marcadores del estrés oxidativo como el CYP1A1 (citocromo P450, familia 1, subfamilia A, polipéptido 1) reflejando un mejor estatus antioxidante (Vinothini and Nagini; citados por Lillehoj *et al*, 2011).

**Cuadro 14. Efecto de un producto a base de aceite esencial de orégano sobre el estado antioxidante y comportamiento productivo de pollos de carne.**

| Variable | Tratamientos [1] | | P |
|---|---|---|---|
| | 1 | 2 | |
| **Estado antioxidante** | | | |
| Superóxido dismutasa, U/ml | 66.5 b | 76.7 a | 0.0417 |
| Caroteno sérico, µg/dL | 2.0 | 2.0 | 1.0000 |
| **Comportamiento productivo** | | | |
| Peso inicial (día 0 de edad), g | 47.7 | 48.6 | 0.2028 |
| Peso final (día 28 de edad), g | 1336.7 b | 1472.9 a | 0.0047 |
| Ganancia de peso, g | 1288.8 b | 1423.9 a | 0.0049 |
| Consumo de alimento, g | 1917.3 | 1886.8 | 0.4733 |
| Conversión alimentaria | 1.49 a | 1.33 b | 0.0180 |

[1] Tratamientos: 1: dieta basal; 2: dieta basal + 500 ppm de Orevitol®

a,b: Promedios en una misma fila con la misma letra no son significativamente diferentes.

Young *et al* (2003) evaluaron el efecto de la inclusión de 1000 ppm de orégano o 200 ppm α-tocoferol en el alimento de pollos sobre el estado antioxidante de las aves y no observaron efectos significativos; sin embargo, la crianza de las aves fue realizada en instalaciones experimentales y sin desafíos aparentes que influyan sobre el estado antioxidante de las aves, lo que explicaría la ausencia de efecto.

Por su parte, Botsoglou *et al* (2008) emplearon tetracloruro de carbono como factor inductor del estrés oxidativo en un modelo animal para evaluar durante seis semanas el efecto de la inclusión de 1% de orégano en la dieta de ratas sobre el estado antioxidante. El efecto negativo en el estado antioxidante a consecuencia del modelo empleado fue evidente en la mayor activad sérica de enzimas marcadoras de la función hepática, como la aspartato transaminasa, alanina transaminasa y fosfatasa alcalina, así como en los mayores niveles de malondialdehído como indicador de la peroxidación de lípidos en tejidos, y en la menor actividad captadora de los radicales ABTS•+ (2,2´azinobis-(3-etilbenzotiazolin 6-ácido sulfónico)), hidroxilo, superóxido e hidrógeno en tejidos. Los investigadores observaron que la inclusión de 1% de orégano en la dieta de las ratas en algunos casos atenuó y en otros revirtió completamente el efecto del tetracloruro de carbono sobre el estado antioxidante, demostrando la capacidad antioxidante del orégano.

En el presente estudio, el estrés oxidativo fue inducido empleando un modelo de desafío que incluyó, entre otros factores, la administración de un inóculo de coccidia (Anexo 1, modelo de desafío B). Los resultados encontrados en el presente estudio son consistentes con lo reportado por Koinarski *et al* (2005) en cuanto a que la actividad de la superóxido dismutasa es un indicador del estado antioxidante sensible a cambios en los estados fisiológico y de salud del ave, y con los numerosos reportes disponibles sobre la actividad antioxidante del AEO (Deighton *et al*, 1993; Martínez-Tomé *et al*, 2001; Lee *et al*, 2004; Faleiro *et al*, 2005). Asimismo, el incremento del 15% en la actividad de la superóxido dismutasa en el actual experimento y las reducciones en la actividad de esta misma enzima de entre 20% (Koinarski *et al*, 2006a; Koinarski *et al*; 2006b) a 30% (Koinarski *et al*, 2005) en aves sometidas a desafíos entéricos en estudios previos, implica que el PRO atenúa el impacto negativo del desafío entérico sobre el estado antioxidante del ave.

Se ha estimado que en las mitocondrias humanas alrededor de 1 a 3% del oxígeno consumido puede ser derivado de la cadena transportadora de electrones para formar radicales superóxido y que un adulto de 70 kg en descanso produce alrededor de 1.72 kg de radicales superóxido por año (Surai *et al*, 2003). A partir de estos valores es posible estimar que una parvada de 20 mil pollos de carne durante su periodo de crianza (45 días y 2.5 kg de peso corporal), y sólo a consecuencia de su metabolismo basal, produce alrededor de 150 kg de radicales superóxido. A esta cantidad se suma la formación de radicales por actividad física, procesos de digestión, entre otros. En condiciones de estrés o enfermedad esta situación se agrava aun más (Marikovsky *et al*, 2003; Davies, 2011). Estas estimaciones muestran que la producción de radicales libres en el organismo es sustancial y evidencia el hecho que miles de moléculas biológicas en el ave pueden ser fácilmente dañadas si no son protegidas. Al respecto, Surai *et al* (2003) consideran que en condiciones de estrés oxidativo, sin la suplementación dietaria de antioxidantes resulta difícil prevenir el daño a la mayoría de órganos y sistemas.

En el presente estudio, el mejor comportamiento productivo de las aves suplementadas con el PRO, asociada al mejor estado antioxidante, concuerda con el menor daño que los radicales libres estarían causando a las moléculas biológicas reduciendo el gasto excesivo para mantenimiento y el consecuente compromiso sobre el crecimiento de las aves (Surai *et al*, 2003). Asimismo, concuerdan con los hallazgos de Georgieva *et al* (2011a) quienes observaron un incremento en la ganancia de peso asociado a un incremento en el estatus antioxidante de aves sometidas a un desafío por *Eimeria acervulina* y suplementadas con selenito de sodio.

A la fecha, se han reportado incrementos en la actividad de la SOD en experimentos con animales suplementados con aceites esenciales, extractos de plantas o metabolitos secundarios de estas obtenidos por síntesis. Tal es el caso del AEO en peces (Zheng *et al*, 2010); del cinamaldehído (Gowder and Devaraj, 2006), carvacrol (Aristatile *et al*, 2009; Jayakumar *et al*, 2012), mejorana (*Origanum majorana*; Shati, 2011) y extracto de orégano francés (*Plectranthus amboinicus*; Chiu *et al*, 2012) en ratas; y del extracto de kion (raíz de jengibre) en pollos (Zhang *et al*, 2009). Sin embargo, los resultados del presente estudio son los primeros en demostrar el efecto antioxidante del AEO a través del incremento de la actividad de la SOD en pollos de carne.

Finalmente, dado que el daño oxidativo ocurre cuando los efectos de las especies reactivas de oxígeno son mayores que los mecanismos antioxidantes en los sistemas biológicos (Schenck, 2011), la mayor actividad antioxidante observada en las aves suplementadas indica que el AEO favorece la estabilidad de los sistemas biológicos a nivel celular.

## 4. Conclusiones

Los resultados obtenidos bajo las condiciones del presente estudio permiten llegar a las siguientes conclusiones:

- La inclusión del PRO en la dieta de pollos de carne favorece el estado antioxidante, a través del incremento en la actividad de la enzima superóxido dismutasa, favoreciendo la estabilidad de los sistemas biológicos a nivel celular.

- La administración del PRO en el alimento de pollos de carne favorece la ganancia de peso y la eficiencia de utilización del alimento balanceado.

# EXPERIMENTO N° 5

# EFECTO DE UN PRODUCTO A BASE DE ACEITE ESENCIAL DE ORÉGANO SOBRE EL COMPORTAMIENTO PRODUCTIVO DE POLLOS DE CARNE SOMETIDOS A UN MODELO DE DESAFÍO ENTÉRICO

## 1. Introducción

Uno de los factores que afectan la eficiencia productiva de las aves en las explotaciones comerciales es el Síndrome de Tránsito Rápido (STR). Si bien esta condición cursa con procesos entéricos sub-clínicos, impacta negativamente en el comportamiento productivo, afectando la ganancia de peso, la conversión alimentaria y la eficiencia productiva. Una de las estrategias más comúnmente empleadas para controlar este síndrome es la implementación de programas a base de antimicrobianos. Si bien ésta es una medida válida y efectiva para controlar los factores bacterianos subyacentes durante el proceso, el daño causado a la mucosa intestinal permanece durante al menos 4 días que es el periodo que le toma al ave restituir naturalmente su integridad intestinal (Maiorka y Rocha, 2009).

Considerando que el mayor impacto económico durante el STR se debe a la pérdida en eficiencia productiva y que ésta es consecuencia directa del estado afectado de la mucosa intestinal, resulta necesario plantear estrategias de control que favorezcan la restitución de la integridad intestinal. El objetivo de este experimento fue evaluar el efecto de un producto a base de AEO (referido en adelante como PRO) sobre el comportamiento productivo de pollos de carne desafiados por el STR.

## 2. Materiales y métodos

### 2.1. Lugar, fecha y duración

La crianza de las aves se llevó a cabo en las instalaciones del Programa de Aves de la Facultad de Zootecnia de la Universidad Nacional Agraria La Molina (UNALM) en Lima-Perú durante el segundo semestre del 2010. Los inóculos de *Clostridium* *perfringens* y de coccidia fueron preparados en el Laboratorio de Biología y Genética Molecular de la Facultad de Medicina Veterinaria de la Universidad Nacional Mayor de San Marcos (UNMSM) en Lima-Perú y en el Programa de Aves de la UNALM, respectivamente. La necropsia de las aves muestreadas y las evaluaciones posteriores se realizó en las instalaciones del Laboratorio de Patología Aviar de la UNMSM. La evaluación se llevó a cabo desde la recepción de las aves en las instalaciones de crianza y el periodo de evaluación fue el comprendido de 0 a 28 días de edad.

## 2.2. Instalaciones, equipos y materiales

### 2.2.1. Instalaciones de crianza

Las aves estuvieron alojadas en 24 jaulas con piso de malla metálica distribuidas en baterías de 4 pisos. Cada jaula contó con un comedero y un bebedero lineales. Las aves fueron alojadas a razón de 21.4 pollos/$m^2$ (0.047 $m^2$/pollo) sobre material de cama reutilizado cuyas características se detallan en el Anexo 6. Durante los primeros 3 días de vida se colocó papel periódico para evitar el acceso directo de los pollos BB al material de cama, el alimento fue suministrado en bandejas plásticas y el agua de bebida en bebederos tipo tongo. Para poder contener material de cama fue necesario acondicionar las jaulas empleando malla sintética, láminas y precintos de polietileno, y silicona, de manera tal que permitiera su ventilación y la retuviera dentro de la jaula. Las adaptaciones realizadas se observan en el Anexo 2. Antes de la recepción de las aves se inició un programa de control de vectores, empleando cebaderos, rodenticidas y mosquicidas comerciales. Antes y después del periodo experimental se realizó la limpieza y desinfección de las instalaciones.

### 2.2.2. Calefacción, control de la temperatura y ventilación

La calefacción fue provista por un sistema eléctrico con resistencias y termostatos, y fue controlada de acuerdo a las recomendaciones de la línea genética (Cobb-Vantress, 2008b). Para reducir la variabilidad en la temperatura ambiental a consecuencia de las diferentes alturas entre los pisos de las baterías se adaptó resistencias adicionales en la base de cada batería. La ventilación fue controlada con cortinas de polipropileno instaladas en el perímetro del área de crianza. La humedad ambiental en el área de crianza se mantuvo alrededor de 40%. Para la medición de la temperatura y humedad ambiental se empleó un termo-higrómetro electrónico digital con una aproximación de 0.1 °C para temperatura y 1% para humedad.

### 2.2.3. Preparación del alimento

Para el pesaje de los ingredientes mayores se utilizó una balanza digital con capacidad de 150 kg y aproximación de 0.02 kg, y para el pesaje de los ingredientes de la premezcla se empleó otra con capacidad de 6 kg y aproximación de 1 g. Se utilizó mezcladoras horizontales de cintas de 400 y 30 kg de capacidad para la mezcla de los ingredientes. Después de la preparación de cada dieta se realizó el flushing y limpieza de equipos (FAO, 2004; Ratcliff, 2009). Se envasó el alimento en sacos de papel y polietileno laminado, y se almacenó en cilindros metálicos.

### 2.2.4. Manejo y actividades especiales

Para la identificación individual de las aves se empleó el método que se presenta en el Anexo 3. Para la medición de los pesos vivos se utilizó una balanza electrónica con capacidad para 200 y con aproximación de 10 mg y otra con capacidad para 15 kg y con aproximación de 1 g. Para la medición del alimento consumido, se utilizó una balanza electrónica con capacidad para 15 kg y con aproximación de 1 g.

Para remover el material de cama se empleó espátulas de polietileno. Para reducir la contaminación cruzada entre tratamientos se empleó guantes quirúrgicos desechables, alcohol metílico y desinfectante a base de sales cuaternarias de amonio y aldehídos (CKM-DESIN®, Laboratorio CKM S.A.C.).

En el día 10 de edad las aves fueron vacunadas, vía ocular, contra la Enfermedad de Newcastle, empleando una vacuna viva liofilizada con una cepa VG/GA (Avinew®, Merial Limited). Para el desafío de las aves se empleó un inóculo de coccidia que fue preparado a partir de una vacuna comercial contra coccidia (Immucox® for Chickens II, Vetech Laboratories) y un inóculo de *C. perfringens*, proveniente de un aislamiento de un brote sub-clínico de campo de enteritis en pollos de 3 semanas de edad. Las características de los inóculos se presentan en los Anexos 9 y 10.

El agua de bebida fue potabilizada empleando 1 ml de hipoclorito de sodio al 4.5% por cada 10 L de agua. Para evitar que la presencia de cloro interfiera con la viabilidad de los inóculos la clorinación del agua se realizó tres días antes de proveerla a las aves, y diariamente, al momento de suministrarla a las aves se verificó la ausencia de cloro empleando un kit comercial de evaluación.

#### 2.2.5. Necropsias y las mediciones posteriores

Para la medición de los pesos se utilizó una balanza electrónica con capacidad para 15 kg y con aproximación de 1 g. Para la necropsia de las aves se empleó bisturí, tijeras y guantes quirúrgicos. Para la medición de los pesos de las vísceras se empleó una balanza electrónica con capacidad para 200 g y con aproximación de 10 mg. Para la medición del diámetro de la Bursa de Fabricio se empleó un Bursómetro comercial con perforaciones crecientes desde 1/8 hasta 8/8 de pulgada (Fort Dodge, 2001).

### 2.3. Animales experimentales

Se empleó 192 pollos machos Cobb 500, cuyas características se presentan en el Anexo 4. Los pollos fueron asignados aleatoriamente a los tratamientos.

### 2.4. Tratamientos

Se evaluaron 3 tratamientos, que fueron definidos de la siguiente manera:

Tratamiento 1: Dieta basal (control negativo)
Tratamiento 2: Dieta basal + 500 ppm de Orevitol®
Tratamiento 3: Dieta basal + 150 ppm de neomicina (control positivo)

Todas las aves estuvieron sometidas a un modelo de desafío desarrollado previamente para inducir el STR (Anexo 1, modelo B). El producto Orevitol® será referido en adelante como PRO. El tratamiento 3 fue incluido como control positivo ya que el sulfato de neomicina, al igual que el PRO, ejerce acción antibacteriana; y es empleado en condiciones comerciales para el control de procesos entéricos como el STR que son causados o cursan con exacerbación de *Clostridium perfringens*.

### 2.5. Alimentación

En el periodo de 1 a 14 días de edad se suministró una dieta basal a base de maíz, soya y harina de pescado, y a partir del día 15 de edad una dieta basal a base de maíz y soya. Ambas dietas fueron complementadas con aceite de pescado y aminoácidos sintéticos, y suplementadas con una premezcla de vitaminas y minerales. La alimentación fue *ad libitum*. Las características de las dietas empleadas se presentan en el Cuadro 15. Durante todo el periodo experimental, los aditivos indicados en los tratamientos dietarios fueron incorporados en la dieta a expensas del maíz.

### 2.6. Mediciones

#### 2.6.1. Evaluación de heces

En la tercera y cuarta semana de edad, dos veces por semana y tres veces en cada fecha, se registró y calculó lo siguiente:

- Alteraciones en heces:
  En las heces muestreadas se determinó lo siguiente:
    - Heces acuosas, %
    - Heces con alimento sin digerir, %
    - Heces con descamaciones mucosas, %
    - Heces hemorrágicas, %

- Índice Dimar:
  El índice Dimar (daño intestinal medido a través de residuos) es propuesto como un indicador del daño a la mucosa intestinal, considerando la relación de cada alteración en las heces con el comportamiento productivo de las parvadas en condiciones comerciales. Se mide en una escala de 0 a 100, donde 0 y 100 son el menor y mayor grado de daño intestinal, respectivamente, medibles mediante este método. Para el cálculo, a cada excreta se asignó un score de acuerdo a la siguiente escala:

**Cuadro 15. Características de las dietas empleadas en el Experimento 5.**

| Ingredientes | % | |
|---|---|---|
| | De 1 a 14 días | De 15 a 28 días |
| Maíz amarillo | 52.517 | 61.775 |
| Torta de soya | 26.699 | 31.405 |
| Harina de pescado | 14.940 | 0.000 |
| Aceite semirefinado de pescado | 2.024 | 2.380 |
| DL-Metionina | 0.190 | 0.224 |
| L-Lisina | 0.116 | 0.135 |
| Cloruro de colina | 0.085 | 0.100 |
| Fosfato dicálcico | 1.609 | 1.892 |
| Carbonato de calcio | 0.978 | 1.150 |
| Sal común | 0.361 | 0.424 |
| Marcador inerte [1] | 0.300 | 0.300 |
| Premezcla [1] | 0.085 | 0.100 |
| Antifúngico [1] | 0.085 | 0.100 |
| Antioxidante [1] | 0.013 | 0.015 |

| Nutriente | Aporte nutricional | |
|---|---|---|
| | De 1 a 14 días | De 15 a 28 días |
| EM [2], Kcal/kg | 3028 | 3008 |
| Proteína cruda, % | 26.72 | 20.04 |
| Lisina, % | 1.72 | 1.16 |
| Metionina + Cistina, % | 1.10 | 0.87 |
| Treonina, % | 1.05 | 0.76 |
| Triptófano, % | 0.29 | 0.23 |
| Calcio, % | 1.54 | 1.17 |
| Fósforo disponible, % | 0.67 | 0.44 |
| Sodio, % | 0.34 | 0.19 |
| Grasa total, % | 5.97 | 5.11 |
| Fibra cruda, % | 2.70 | 3.17 |

| Componente | Composición proximal [3], % | |
|---|---|---|
| | De 1 a 14 días [4] | De 15 a 28 días [5] |
| Humedad | 11.89 | 12.35 |
| Proteína total [2] | 26.06 | 19.66 |
| Extracto Etéreo | 4.87 | 3.83 |
| Fibra Cruda | 2.12 | 2.49 |
| Cenizas | 7.13 | 5.88 |
| ELN [2] | 47.94 | 55.79 |

[1] Marcador inerte: óxido crómico. Premezcla de vitaminas y minerales Proapak 2A®. Composición: Retinol: 12’000,000 UI; Colecalciferol: 2’500,000 UI; DL α-Tocoferol Acetato: 30,000 UI; Riboflavina: 5.5 g; Piridoxina: 3 g; Cianocobalamina: 0.015 g; Menadiona: 3 g; Ácido Fólico: 1 g; Niacina: 30 g; Ácido Pantoténico: 11 g; Biotina: 0.15 g; Zn: 45 g; Fe: 80 g; Mn: 65 g; Cu: 8 g; I: 1 g; Se: 0.15 g; Excipientes c.s.p. 1,000 g. Antifúngico: Mold Zap®; Antioxidante: Danox®

[2] EM: Energía metabolizable; Proteína total: N x 6.25; ELN: Extracto Libre de Nitrógeno (calculado).

[3] Informes de ensayo 1235/2010 LENA y 1236/2010 LENA, Universidad Nacional Agraria La Molina.

[4] Calculado a partir de los análisis proximales de la dieta empleada de 15 a 28 días de edad (85%) y de la harina de pescado (15%) empleadas en su producción.

[5] Informe de ensayo 1235/2010 LENA, Universidad Nacional Agraria La Molina.

0: si es normal (seca, bien formada y no contiene alimento sin digerir)
1: si la excreta es acuosa
2: si contiene alimento sin digerir
3: si presenta descamaciones de mucosa
4: si es hemorrágica

Y se aplicó la siguiente fórmula, donde SPHM: score promedio de las heces.

$$\text{Índice Dimar} = SPHM \times 25$$

### 2.6.2. Morfometría de los órganos linfoides

Al final del periodo experimental se tomó al azar 8 pollos por tratamiento para ser pesados y sacrificados cortando las arterias carótidas con la subsecuente exanguinación (Mutus *et al*, 2006), de acuerdo al protocolo del Laboratorio de Patología Aviar de la UNMSM. La necropsia se realizó de acuerdo a los protocolos de Bermúdez y Stewart-Brown (2003) y Colas *et al* (2010). Se colectó la bursa y el timo y, previa eliminación del tejido graso, se determinó las siguientes variables:

- Diámetro de la bursa de Fabricio: La medida cualitativa del diámetro mayor de la bursa se determinó mediante un bursómetro comercial (Fort Dodge, 2001).

- Índice morfométrico de la bursa (Rbu): Se calculó empleando la fórmula presentada a continuación (Anexo 1), donde Rbu: índice morfométrico de la bursa, PB: peso de la bursa (g), PC: peso corporal (g).

$$\text{Rbu} = \frac{\text{PB}}{\text{PC}} \times 1000$$

- Relación bursa/timo (Bu-Ti): Se calculó empleando la fórmula (Anexo 1) presentada a continuación, donde Bu-Ti: Relación bursa/timo, PB: peso de la bursa (g), PT: peso del timo (g).

$$\text{Bu-Ti} = \text{PB}/\text{PB}$$

#### 2.6.3. Comportamiento productivo

Se midió empleando las siguientes variables:

- Peso vivo: Se registró los pesos individuales al recibir los pollos BB y al final del periodo experimental (día 28).
- Ganancia de peso: Se calculó valores acumulados.
- Consumo de alimento: Al final de cada semana se pesó el residuo de alimento y se calculó el consumo acumulado al día 28 de edad.
- Conversión alimentaria: Se calculó empleando valores acumulados de consumo de alimento y ganancia de peso.

### 2.7. Diseño estadístico

Se empleó el Diseño Completo al Azar con 3 tratamientos y 8 repeticiones. El análisis de varianza se realizó con el programa Statistical Analysis System SAS 9.0 (SAS Institute, 2009) y la diferencia de medias empleando la prueba de Duncan (1955). El Modelo Aditivo Lineal General aplicado fue el siguiente:

Yij = U + Ti + Eij Donde:

| | | |
|---|---|---|
| Yij | = | variable respuesta |
| U | = | media general |
| Ti | = | i-ésimo tratamiento ( i = 1, 2, 3 ) |
| Eij | = | Error experimental |

Se consideró significativos valores con P menores de 0.10 (Crespo and Esteve-Garcia, 2001; Zhu *et al*, 2003, Kilburn and Edwards, 2004; van Nevel *et al*, 2005; Tahir *et al*, 2008; Teeter *et al*, 2008) o de 0.05, según se indica en cada caso.

## 3. Resultados y discusión

### 3.1. Características de las heces

Desde el día 14 de edad las aves con el PRO o neomicina presentan una menor ($P<0.01$) frecuencia de heces con alteraciones (Cuadro 16, Imágenes 5 a 7 y Anexo 48). Hasta el día 21, el porcentaje de heces con descamaciones mucosas (Imagen 6) es bajo en todos los tratamientos (≤ 3%) y luego se incrementa sólo en las aves control (7.25%), manteniendo los tratamientos con el PRO o neomicina niveles bajos (< 2.5%) y significativamente menores (P<0.01) que las aves control. Hasta el día 21 las aves suplementadas con el PRO o neomicina no muestran diferencias significativas en cuanto a porcentaje de heces acuosas (Imagen 5) y heces normales (Imagen 8); sin embargo, a partir de esta edad ambas variables presentan resultados favorables en las aves suplementadas con el PRO. Los resultados observados indican que las aves suplementadas con el PRO presentan sostenidamente la menor frecuencia de heces con alimento sin digerir (Imagen 7) y menores índices Dimar.

La presencia de heces húmedas (Imagen 5) en la cama es un indicador de disbacteriosis en la parvada (Mortimer, 2002) y en el presente estudio es resultado del modelo de desafío empleado. La menor frecuencia de heces húmedas en las aves que reciben el PRO o neomicina, es consistente con la acción antimicrobiana que ambos desarrollan y que ha sido documentada para el AEO (Conner and Beuchat, 1984; Helander *et al*, 1998; Lambert *et al*, 2001; Lee *et al*, 2004) y para el PRO (UNMSM, 2009a). Se ha reportado, además, que el AEO tiene un efecto prebiótico y por su acción antimicrobiana selectiva y mayor contra bacterias patógenas (Hammer *et al*, 1999; Dorman and Deans, 2000; Lee *et al*, 2004), promoviendo el crecimiento de la flora benéfica (Ferket, 2003; Yew, 2008; Zheng *et al*, 2010). La menor frecuencia de heces húmedas en las aves suplementadas con el PRO respecto a aquellas con neomicina durante la cuarta semana de vida, es entonces consistente con el balance más eficiente que el PRO que promueve en la flora intestinal.

A consecuencia del modelo de desafío empleado las aves presentan mayor porcentaje de heces con descamaciones mucosas (Imagen 6) a partir del día 21 de edad, tal como se verifica en el grupo control. En el presente experimento las descamaciones mucosas guardan relación, entre otros factores, con la enteritis por *C. perfringens* reportada en el Anexo 1, a consecuencia del modelo de desafío empleado en el presente estudio (modelo B); por ello, la acción antibacteriana del PRO y la neomicina reduce la frecuencia de descamaciones mucosas en las heces, efecto que resulta significativo y más evidente durante la cuarta semana de crianza (Cuadro 16).

**Cuadro 16. Efecto de un producto a base de aceite esencial de orégano en la alimentación de pollos sometidos al Síndrome de Tránsito Rápido sobre las características fecales y linfoideas, y comportamiento productivo.**

| Variable | Tratamientos [1] 1 | 2 | 3 | P |
|---|---|---|---|---|
| **Características de las heces** | | | | |
| De 15 a 21 días de edad | | | | |
| Normales, % | 24.41 b | 52.39 a | 49.01 a | <0.0001 |
| Acuosas, % | 75.59 a | 47.61 b | 50.99 b | <0.0001 |
| Alimento sin digerir, % | 64.99 a | 35.59 c | 40.88 b | <0.0001 |
| Descamaciones mucosas, % | 3.05 | 1.14 | 1.82 | 0.2915 |
| Hemorrágicas, % | 0.00 | 0.00 | 0.00 | ND |
| Índice Dimar | 35.91 a | 21.09 c | 23.42 b | <0.0001 |
| De 22 a 28 días de edad | | | | |
| Normales, % | 24.07 c | 58.36 a | 44.72 b | <0.0001 |
| Acuosas, % | 75.93 a | 41.64 c | 55.28 b | <0.0001 |
| Alimento sin digerir, % | 55.09 a | 24.88 c | 35.21 b | <0.0001 |
| Descamaciones mucosas, % | 7.25 a | 1.85 b | 2.45 b | 0.0002 |
| Hemorrágicas, % | 0.00 | 0.00 | 0.00 | ND |
| Índice Dimar | 34.57 a | 17.09 c | 23.24 b | <0.0001 |
| **Morfometría de órganos linfoides** | | | | |
| Diámetro de la bursa [2] | 7.25 a | 6.25 b | 7.13 a | 0.0070 |
| IM de la bursa (Rbu) | 2.91 a | 1.97 b | 2.76 a | 0.0122 |
| Relación bursa/timo (Bu-Ti) | 0.52 $£ | 0.39 £ | 0.56 $ | 0.0552 |
| **Comportamiento productivo** | | | | |
| Peso inicial (día 0), g | 47.9 | 48.4 | 48.1 | 0.7077 |
| Peso final (día 28), g | 1295.3 c | 1459.5 a | 1378.3 b | 0.0007 |
| Ganancia de peso, g/pollo | 1247.0 c | 1410.6 a | 1329.9 b | 0.0008 |
| Consumo de alimento, g/pollo | 1897.4 | 1875.8 | 1925.7 | 0.5417 |
| Conversión alimentaria | 1.53 a | 1.33 b | 1.45 a | 0.0023 |

[1] Tratamientos: 1: aves desafiadas control; 2: aves desafiadas + 500 ppm de Orevitol®; 3: aves desafiadas + 150 ppm de neomicina.
[2] Medido con un bursómetro en una escala de 1/8 a 8/8 de pulgada.
a,b,c,$,£ Promedios significativamente diferentes no comparten la misma letra (a,b; P<0.05) o el mismo símbolo ($,£; 0.05<P<0.10).
ND No determinado.

**Imagen 5.** **Heces acuosas.** Predominantes en el grupo control.

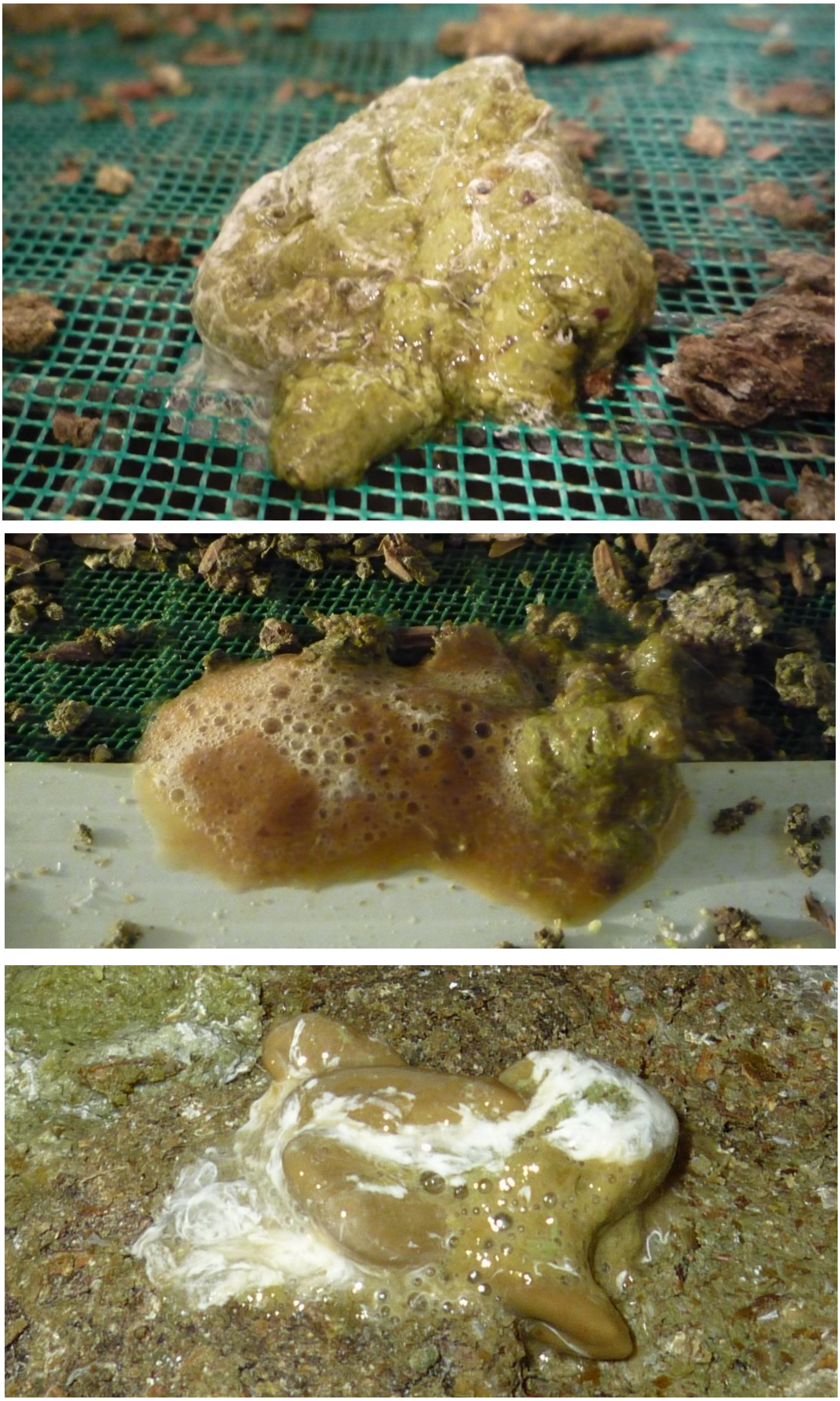

**Imagen 6.** **Heces con descamaciones de la mucosa intestinal.** Predominantes en el grupo control.

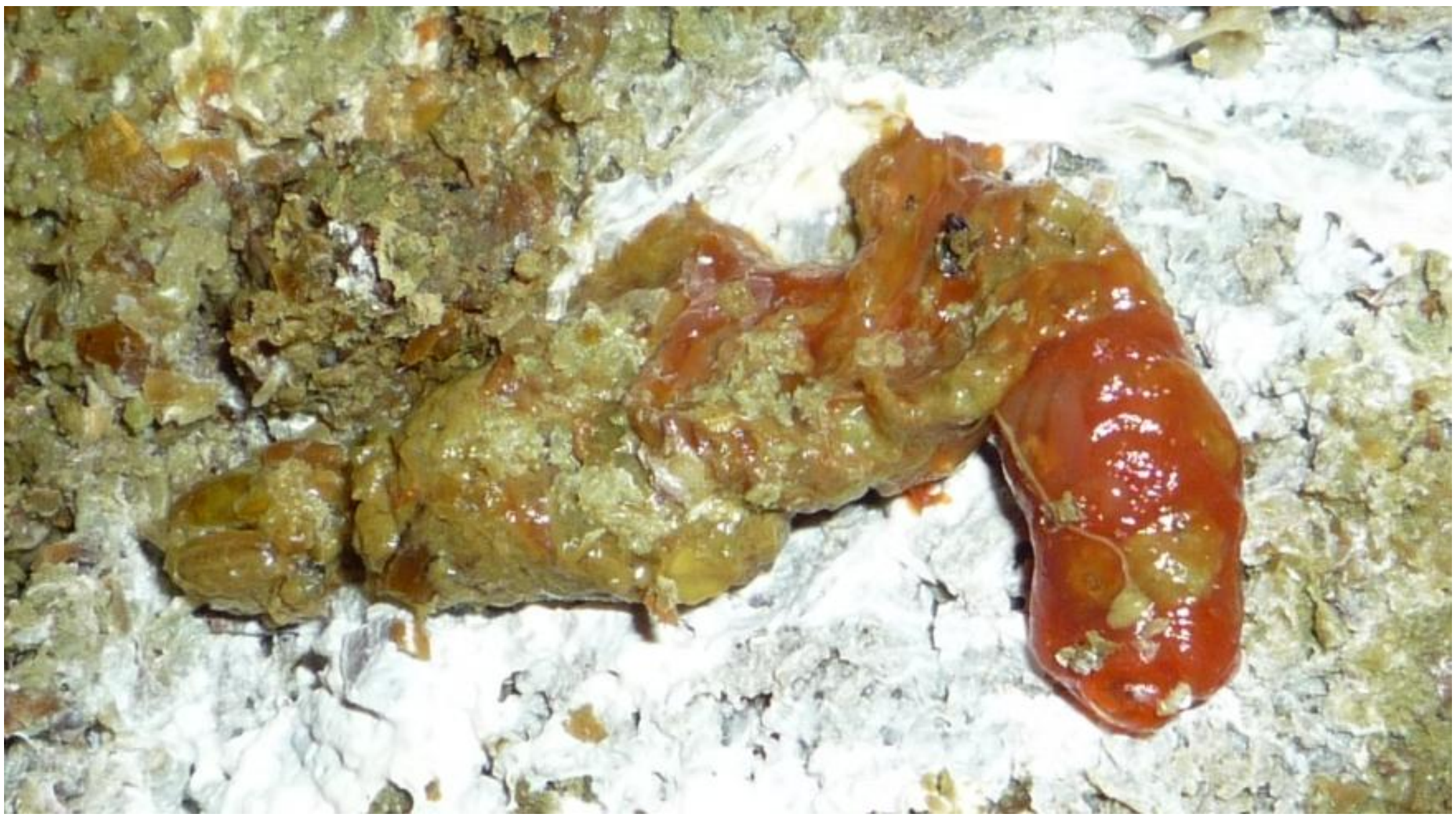

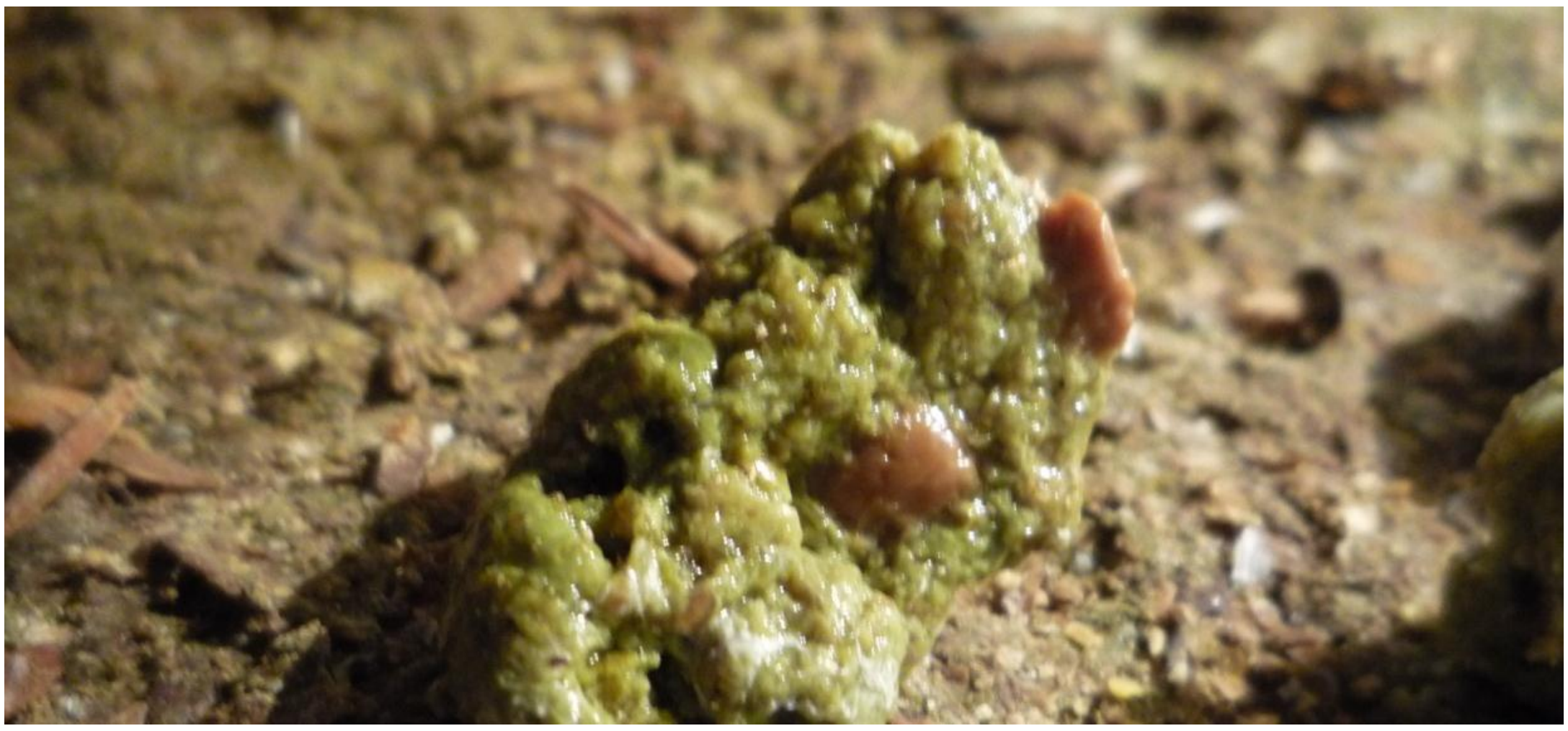

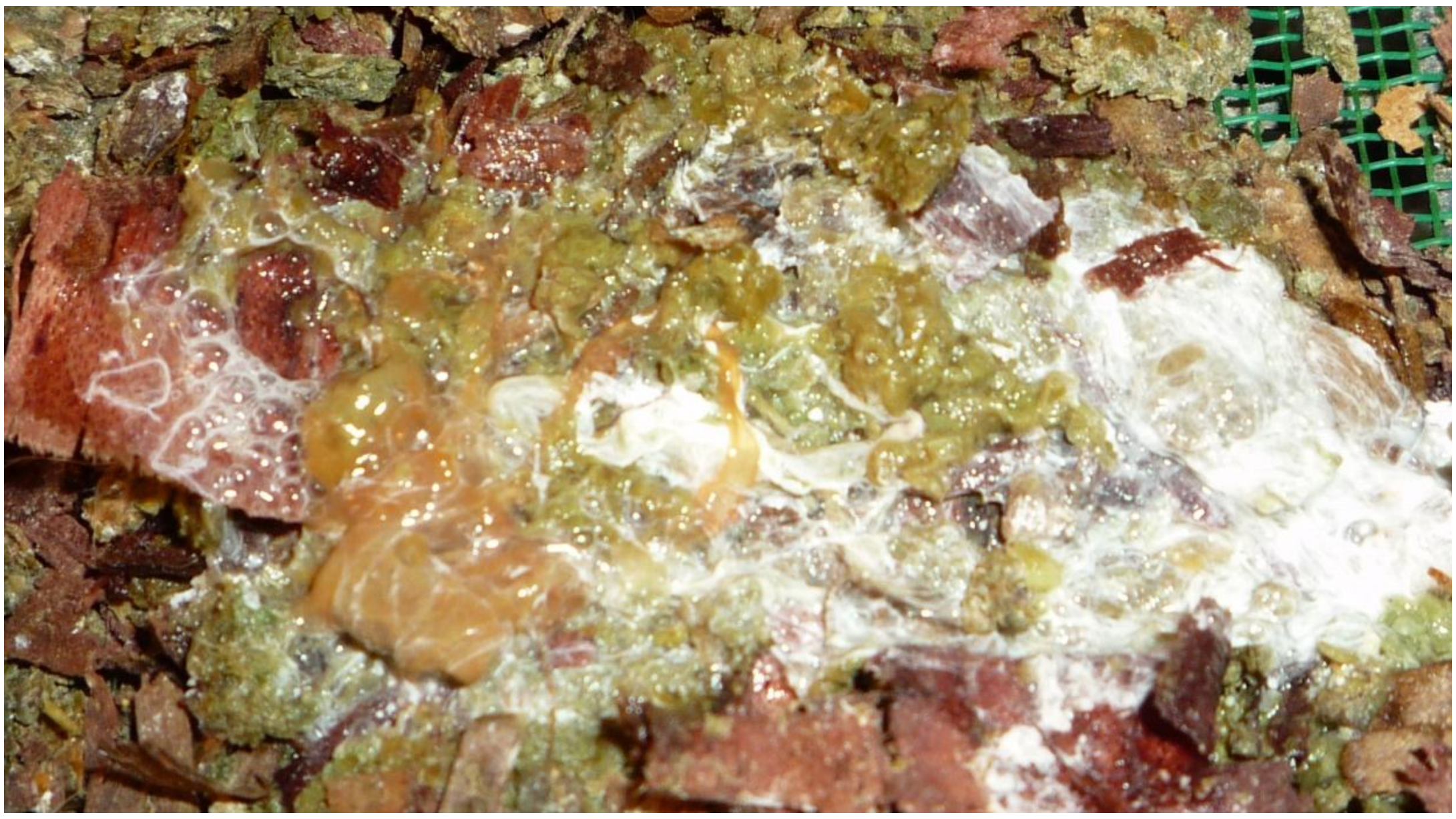

**Imagen 7.** **Heces con alimento sin digerir.** Predominantes en el grupo control.

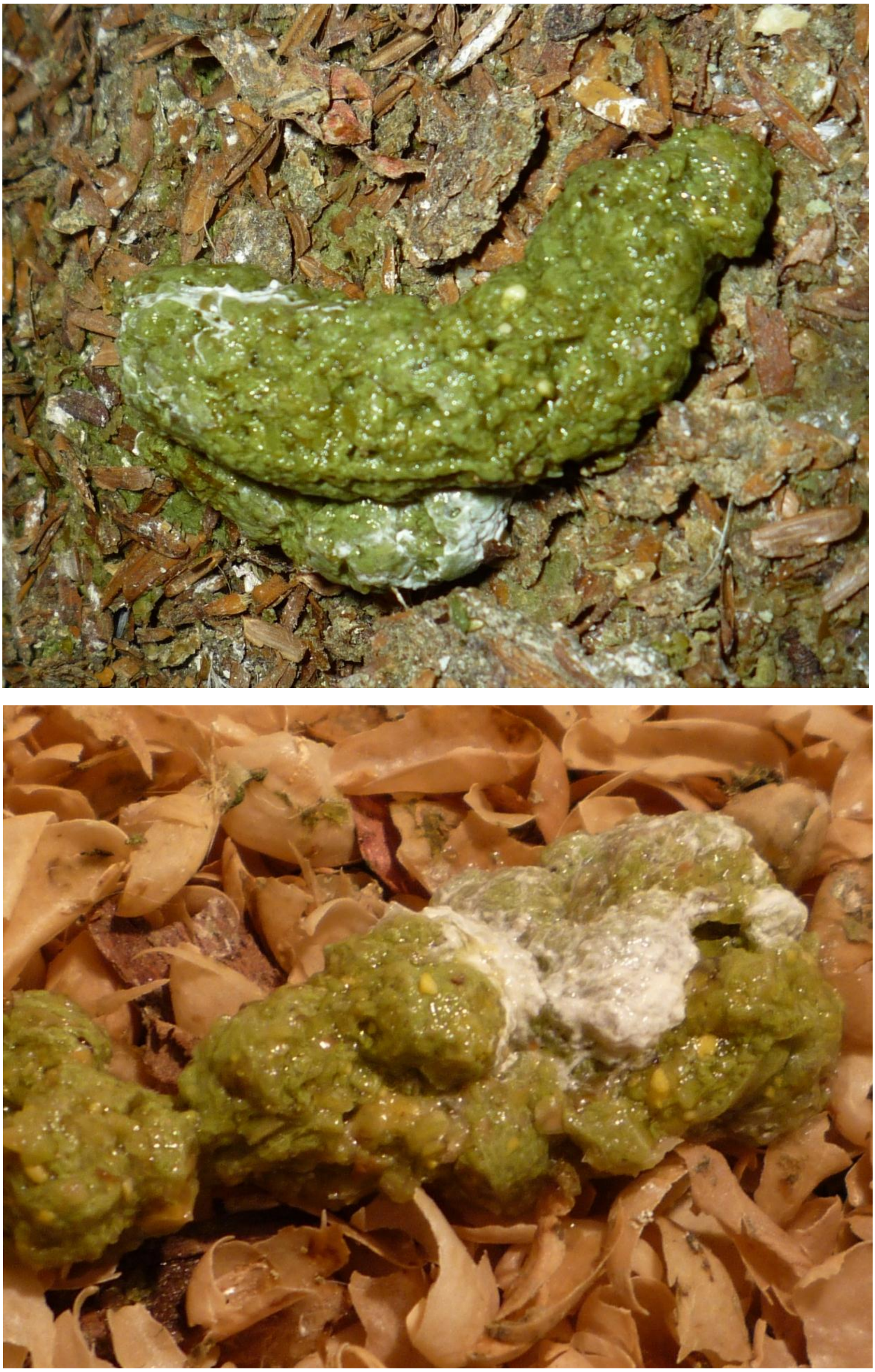

**Imagen 8.** **Heces normales.** Predominantes en las aves suplementadas con el producto a base de aceite esencial de orégano.

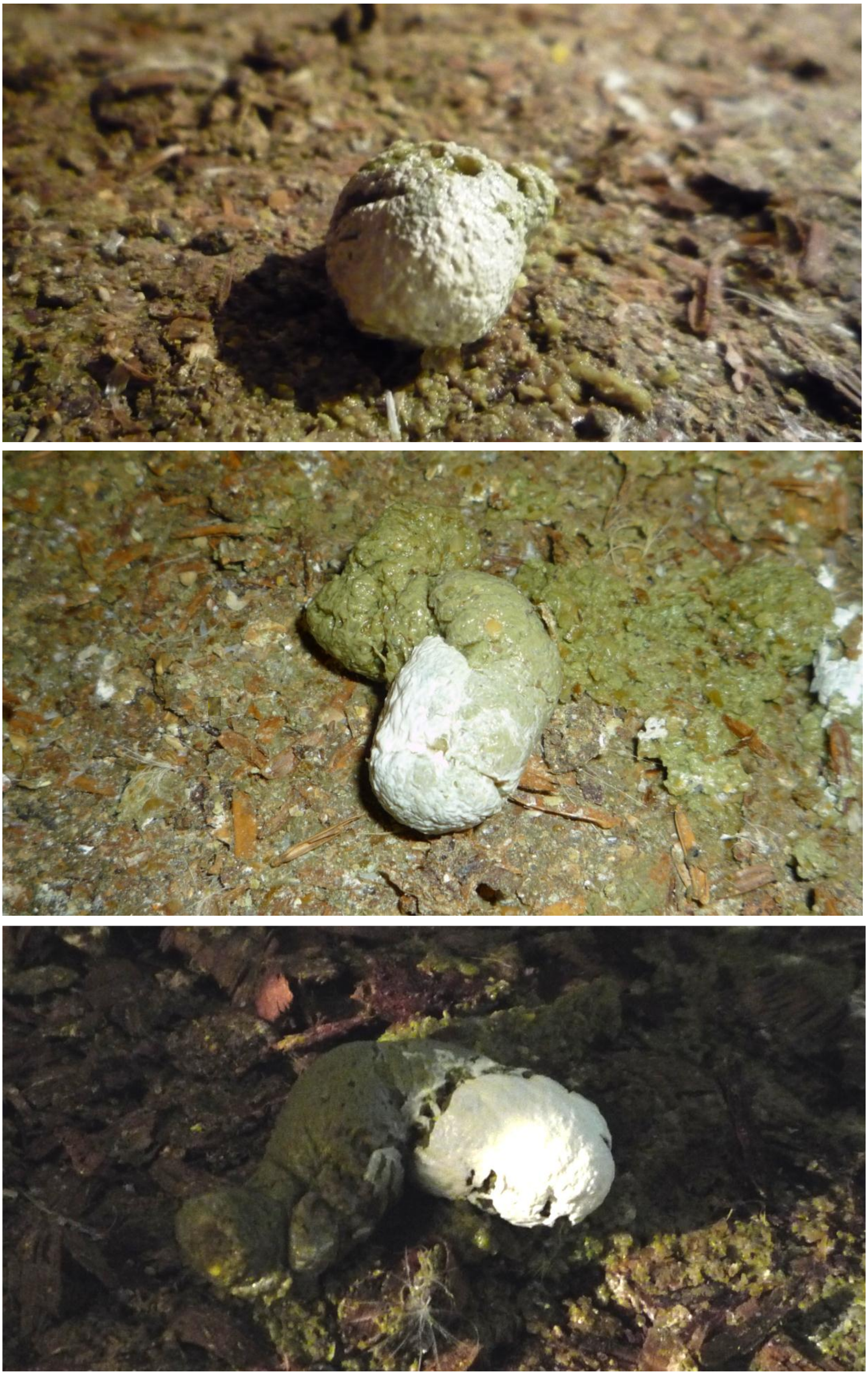

Por otro lado, se ha demostrado que los metabolitos secundarios del AEO incrementan la actividad de la tripsina (Lee *et al*, 2004). Se ha establecido también que, entre otras enzimas, ésta inactiva la toxina α del *C. perfringens* tipo A y la toxina β del *C. perfringens* tipo C (Niilo, 1965; Baba *et al*, 1992). Por lo tanto, es posible presumir que la acción indirecta del AEO sobre esta toxina, reduzca aun en mayor medida el impacto del desafío entérico y explica, al menos en parte, el mejor comportamiento productivo de las aves suplementadas con el PRO respecto a las suplementadas con neomicina (Mitsch *et al*, 2004).

La menor frecuencia de heces con alimento sin digerir en las aves con neomicina en relación al grupo testigo se explica por el control de la disbacteriosis y el menor daño concomitante a la mucosa intestinal, tal como reportan Smits *et al* (1999) quienes indican, además, que la disbacteriosis y el daño a la mucosa afectan la capacidad de absorción y digestibilidad de nutrientes e incrementan la presencia de alimento en las heces. La frecuencia de heces con alimento sin digerir en las aves con el PRO aun menor en relación al tratamiento con neomicina es consistente con la capacidad del PRO para promover la proliferación celular en la mucosa intestinal (ver Experimento 1), logrando restituir la integridad de la mucosa intestinal afectada en menor tiempo, tal como ha sido reportado previamente con el uso del AEO (Bruerton, 2002; citado por Ferket, 2003), favoreciendo la elongación de las vellosidades intestinales (Greathead and Kamel, 2006) e incrementando el área superficial de absorción de nutrientes (García *et al*, 2007), lo que también fue verificado en el Experimento 1.
El menor índice Dimar en las aves suplementadas con el PRO (Cuadro 16) corrobora los hallazgos de alteraciones específicas en las heces y demuestra su mayor capacidad para atenuar los efectos del STR en comparación a la neomicina.

### 3.2. Morfometría de los órganos linfoides

En el presente estudio se observa que las aves control y aquellas suplementadas con neomicina muestran los mayores tamaños ($P<0.01$) e índices morfométricos de la bursa ($P<0.05$) (Cuadro 16 y Anexo 48). Se observa, además, que la relación bursa/timo en las aves suplementadas con neomicina es significativamente mayor ($P<0.10$) que en aquellas suplementadas con el PRO. El menor tamaño, diámetro e índice morfométrico de la bursa, así como la menor relación bursa/timo, en las aves que reciben el PRO es consistente con el mayor control que el PRO ejerce sobre los efectos del STR y el concomitante desafío inmunológico. Concuerda también con los hallazgos de Deshmukh *et al* (2007) quienes reportan que aves desafiadas con *Salmonella gallinarum* presentan hiperplasia del bazo; y con las observaciones reportadas en el Anexo 1, en que las aves sometidas a un desafío entérico intenso (modelo B), mismo que se empleó en este experimento para inducir el STR, presentaron bursas más grandes.

### 3.3. Comportamiento productivo

Las aves suplementadas con el PRO y neomicina presentan pesos corporales significativamente mayores que las aves control; sin embargo, el peso de las aves suplementadas con el PRO es aun mayor que el de aquellas suplementadas con neomicina (P<0.01) (Cuadro 16 y Anexo 50). La ganancia acumulada de peso muestra el mismo comportamiento que el peso vivo. No se observa diferencias estadísticas significativas en el consumo de alimento entre los tres tratamientos (P>0.10). Al final del periodo experimental, luego de transcurridos los 28 días de edad, las aves suplementadas con el PRO presentan una conversión alimentaria significativamente menor que las aves control (P<0.01). Los resultados observados en el peso vivo indican que tanto el PRO como la neomicina favorecen la ganancia de peso de las aves afectadas por el STR. Sin embargo, debido a que el consumo de alimento no presenta variaciones por efecto de los tratamientos; la mayor ganancia de peso, incluso en el tratamiento con neomicina, se debe principalmente a la mejor conversión alimentaria.

Las observaciones del Anexo 1, tras evaluar dos modelos de desafío para inducir el STR, reflejaron un desafío entérico por el STR más intenso alrededor del día 28 de edad. En el presente experimento, las aves de 28 días de edad suplementadas con el PRO presentan una conversión alimentaria aun mejor que las aves con neomicina, respecto a las aves control, lo que influye del mismo modo en la ganancia de peso.

En relación a lo anterior, tanto la neomicina como el PRO poseen actividad antimicrobiana como se ha referido anteriormente; sin embargo, se ha documentado que el AEO además de esta acción promueve el adecuado balance de la flora intestinal por su mayor actividad antimicrobiana contra especies patógenas (Hammer *et al*, 1999b; Dorman and Deans, 2000; Lee *et al*, 2004), tiene actividad antioxidante (Deighton *et al*, 1993; Lee *et al*, 2004; Faleiro *et al*, 2005), favorece la proliferación celular en la mucosa intestinal (ver Experimento 1; Levkut *et al*, 2011) y consecuentemente su restitución (Bruerton, 2002; citado por Ferket, 2003; Greathead and Kamel, 2006; García *et al*, 2007), y tiene un efecto favorable sobre la actividad de las enzimas digestivas (Jang *et al*, 2007; Basmacioglu Malayoglu *et al*, 2010). Estos mecanismos de acción en su conjunto explican la mejor utilización de los nutrientes en las aves que reciben el PRO respecto a aquellas con neomicina.

## 4. Conclusiones

Los resultados obtenidos bajo las condiciones del presente estudio permiten llegar a las siguientes conclusiones:

- La suplementación del PRO en el alimento de pollos de carne reduce la frecuencia de alteraciones en las heces y favorece la ganancia de peso de las aves sometidas al STR.

- En las aves afectadas por el STR, la administración del PRO en el alimento balanceado reduce la frecuencia de heces acuosas, con alimento sin digerir y el índice Dimar. Este efecto favorable en las características de las heces es incluso mayor que con el uso de neomicina.

- La administración del PRO en el alimento de pollos de carne sometidos al STR favorece la ganancia de peso y conversión alimentaria, logrando una mayor eficiencia en el uso de los nutrientes dietarios y un mejor comportamiento productivo. Esta mejora en el comportamiento productivo es incluso mayor que con el uso de neomicina.

# EXPERIMENTO N° 6

# EFECTO DE UN PRODUCTO A BASE DE ACEITE ESENCIAL DE ORÉGANO EN EL CONTROL DEL SÍNDROME DE TRÁNSITO RÁPIDO EN PARVADAS COMERCIALES DE POLLOS DE CARNE

## 1. Introducción

Los trastornos entéricos afectan la eficiencia productiva de las aves, incluso en aquellas condiciones en que no se manifiestan signos clínicos. Es común observar en condiciones de campo índices de conversión alimentaria 5% menos eficientes en aves sometidas a procesos entéricos sub-clínicos; sin embargo, muchas veces estos resultados se asocian a factores aleatorios de diferentes tipos.

Existen indicadores que permiten presumir la existencia de un trastorno entérico sub-clínico, entre los que destacan la presencia de lesiones intestinales, la desuniformidad en el peso y/o pigmentación de las aves y la presencia de alteraciones en las heces. Sin embargo, la existencia del trastorno muchas veces es sub-estimado. Experiencias previas demuestran que el aceite esencial de orégano (AEO) tiene cualidades de interés para la salud intestinal. El objetivo de este experimento fue determinar el efecto de un producto a base de AEO (referido en adelante como PRO) en el control del STR en parvadas comerciales de pollos de carne afectadas por un desafío de campo.

## 2. Materiales y métodos

### 2.1. Lugar, fecha y duración

La evaluación se llevó a cabo en un plantel comercial de pollos de carne localizado en el distrito de Ventanilla, Lima - Perú, en Diciembre de 2010. La evaluación tuvo una duración de siete días, contados desde el inicio del tratamiento al detectarse el trastorno entérico en las parvadas.

### 2.2. Instalaciones, equipos y materiales

Las aves se encontraron alojadas a razón de 9.5 aves por $m^2$ en galpones comerciales no automáticos. El alimento y agua de bebida fueron provistos empleando comederos tipo tolva y bebederos lineales de canaleta, respectivamente.

Se empleó una balanza tipo reloj con capacidad para 10 kg y con aproximación de 0.1 kg para el pesaje de las aves. Para la necropsia de las aves se empleó tijeras y guantes quirúrgicos. Las mediciones en heces para cuantificar la frecuencia de alteraciones se realizaron empleando papel absorbente Watman® y manta arpillera.

### 2.3. Animales experimentales

Se empleó 18,000 pollos de carne de la línea Cobb 500 alojados en dos galpones. 11,000 de ellos, 50% machos y 50% hembras, de 33 días de edad, y los 7,000 restantes, 50% machos y 50% hembras, de 23 días de edad.

### 2.4. Caracterización del estado inicial de las parvadas

Las parvadas empleadas se caracterizaron por presentar una evidente desuniformidad en cuanto a peso vivo y pigmentación de patas (Imagen 9).

Para determinar las características de las heces se realizó un muestreo colocando durante 1 hora 8 pliegos (de aproximadamente 1 $m^2$ cada uno) en cada galpón y corral. Se consideró como húmeda aquella excreta que produjo un halo de humedad con al menos del doble del diámetro que el de la misma excreta (Panneman and van der Stroom-Kruyswijk, 2002) (Imagen 10). Mediante este método se verificó la presencia de más de 30% de las heces húmedas (Imagen 10). Se observó además, más del 50% de las heces con presencia de alimento sin digerir (Imagen 11) y menos del 5% de las heces con descamaciones mucosas (Imagen 12).

**Imagen 9. Desuniformidad característica de las parvadas del Experimento 6.**

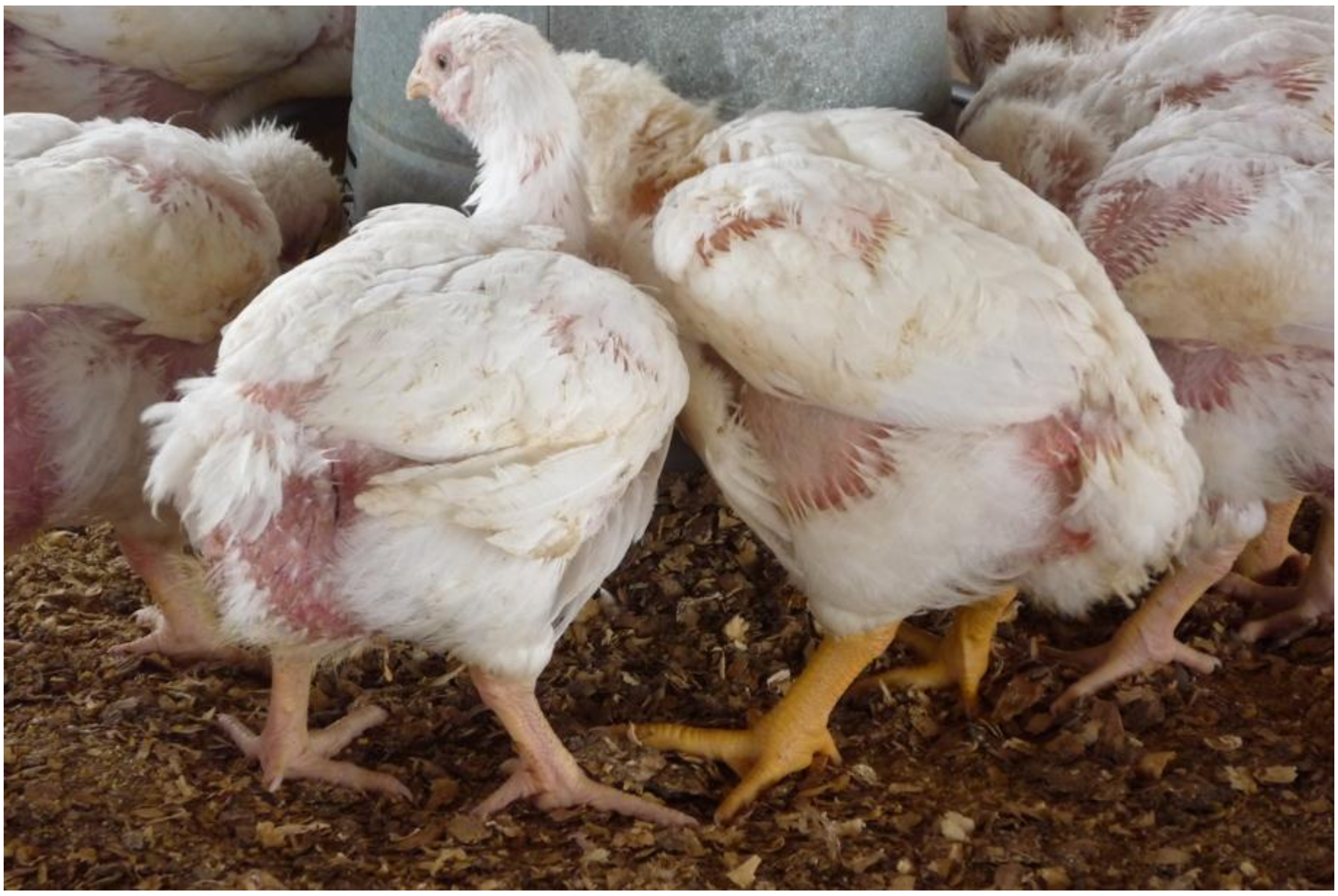

**Imagen 10. Heces húmedas muestreadas en el Experimento 6 sobre papel absorbente en el piso del galpón.**

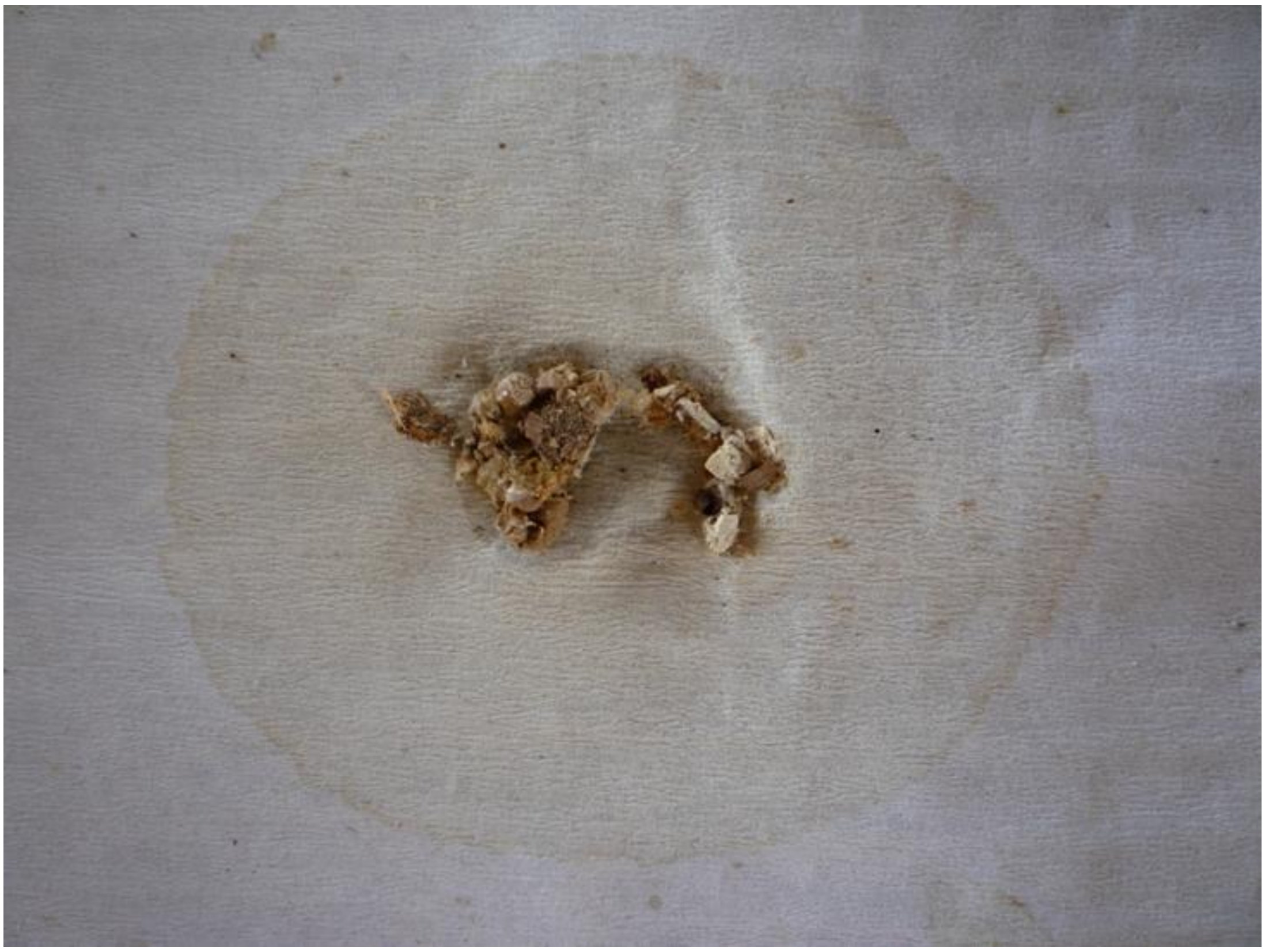

**Imagen 11. Alimento sin digerir en las heces de las aves del Experimento 6.**

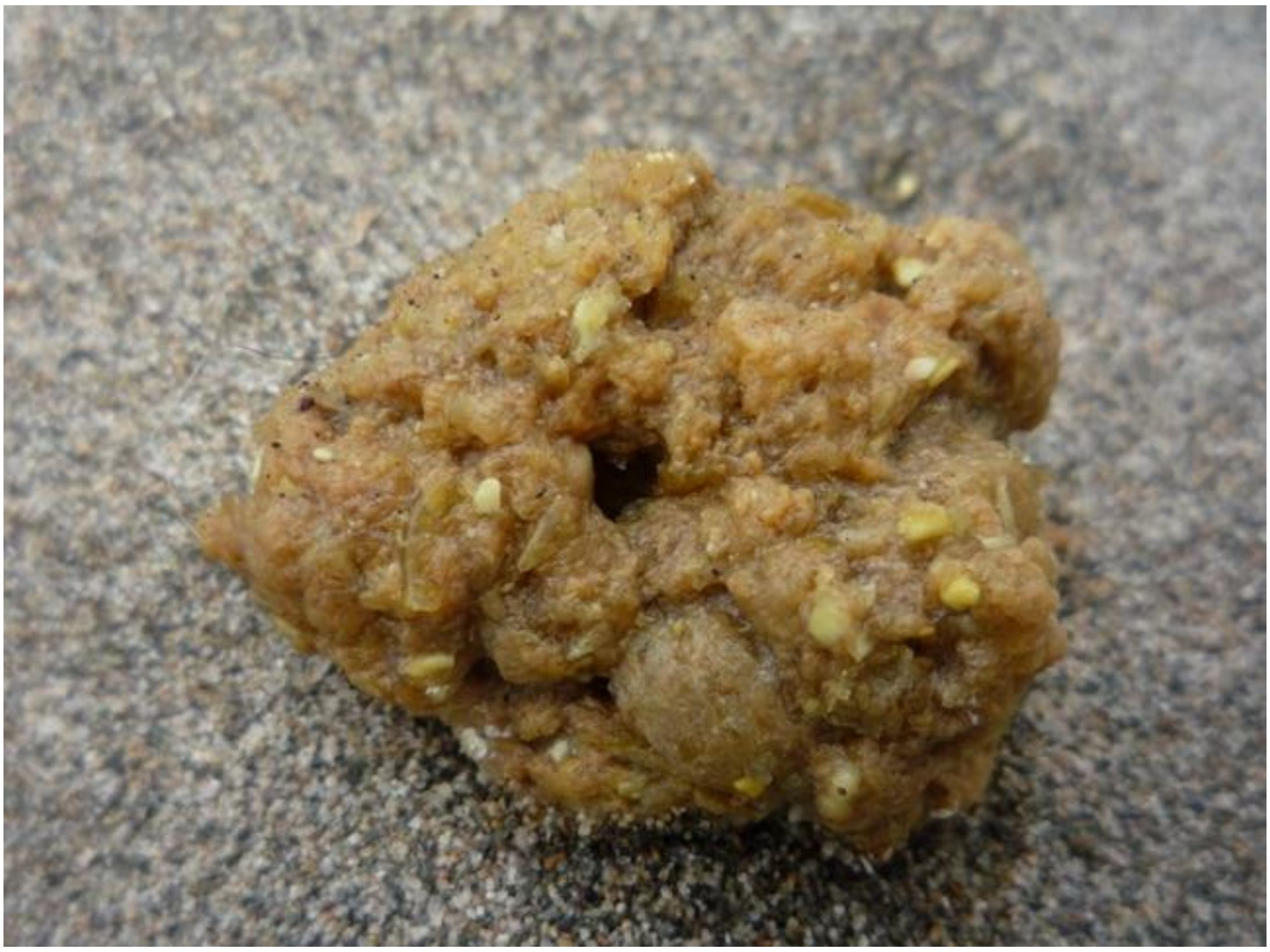

**Imagen 12. Descamaciones mucosas en las heces de las aves del Experimento 6.**

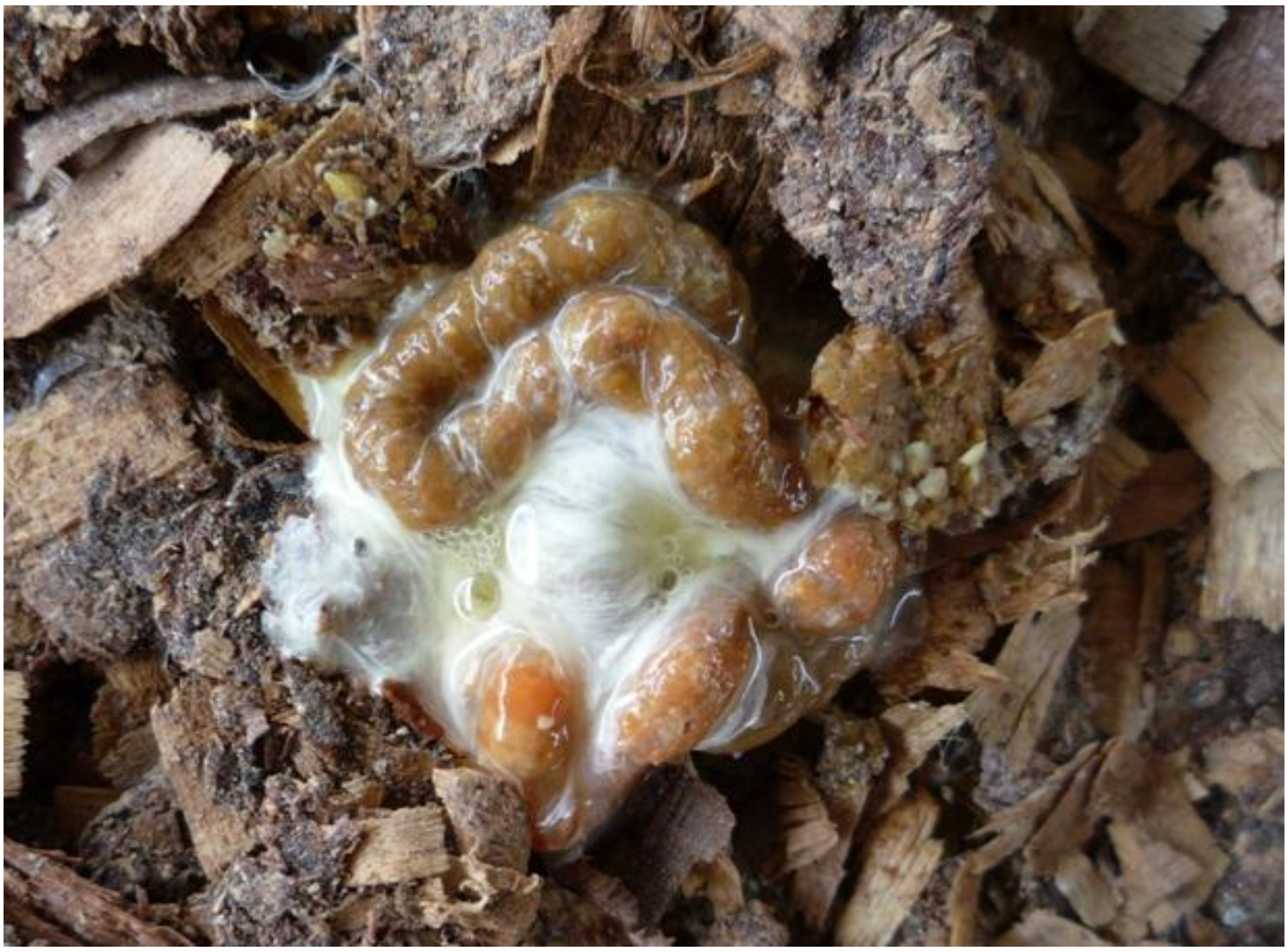

Asimismo, se tomó al azar una muestra de aves aparentemente sanas y se realizó la necropsia de acuerdo a los protocolos propuestos por Bermúdez y Stewart-Brown (2003) y Colas *et al* (2010), encontrándose enteritis duodenal en 10% de las aves (Imagen 13) y alimento sin digerir en el íleon en 90% de las aves (Imagen 14). No se encontró descamaciones de mucosa ni lesiones por coccidia en las aves muestreadas.

Finalmente, se verificó la presencia de disbacteriosis en las parvadas empleando un método validado mediante cultivos microbiológicos semi-cuantitativos (Panneman and van der Stroom-Kruyswijk, 2002). Para ello, se consideró positivo (+) a disbacteriosis cuando la parvada presentó al menos 30% de heces clasificadas como húmedas de acuerdo al criterio antes descrito.

## 2.5. Tratamientos

Se evaluaron 2 tratamientos, que fueron definidos de la siguiente manera:

- T1: Aves control
- T2: Aves tratadas con Orevitol® a razón de 300 ml por cada $m^3$ de agua de bebida, durante 5 días

El producto indicado será referido, en adelante, como PRO.

## 2.6. Alimentación

Las aves fueron alimentadas con la dieta empleada por la empresa, formulada a base de maíz y soya, siguiendo las recomendaciones nutricionales de la línea genética (Cobb-Vantress, 2008a), cuyo aporte nutricional fue el siguiente: Energía metabolizable 3170 Kcal/kg, proteína cruda 18%, lisina 1.05%, metionina + cistina 0.82%, treonina 0.72%, triptófano 0.19%, calcio 0.9%, fósforo disponible 0.45% y sodio 0.19%. Todas las aves recibieron la misma dieta.

**Imagen 13. Duodenos normales (abajo) y con enteritis (arriba) en las aves del Experimento 6.**

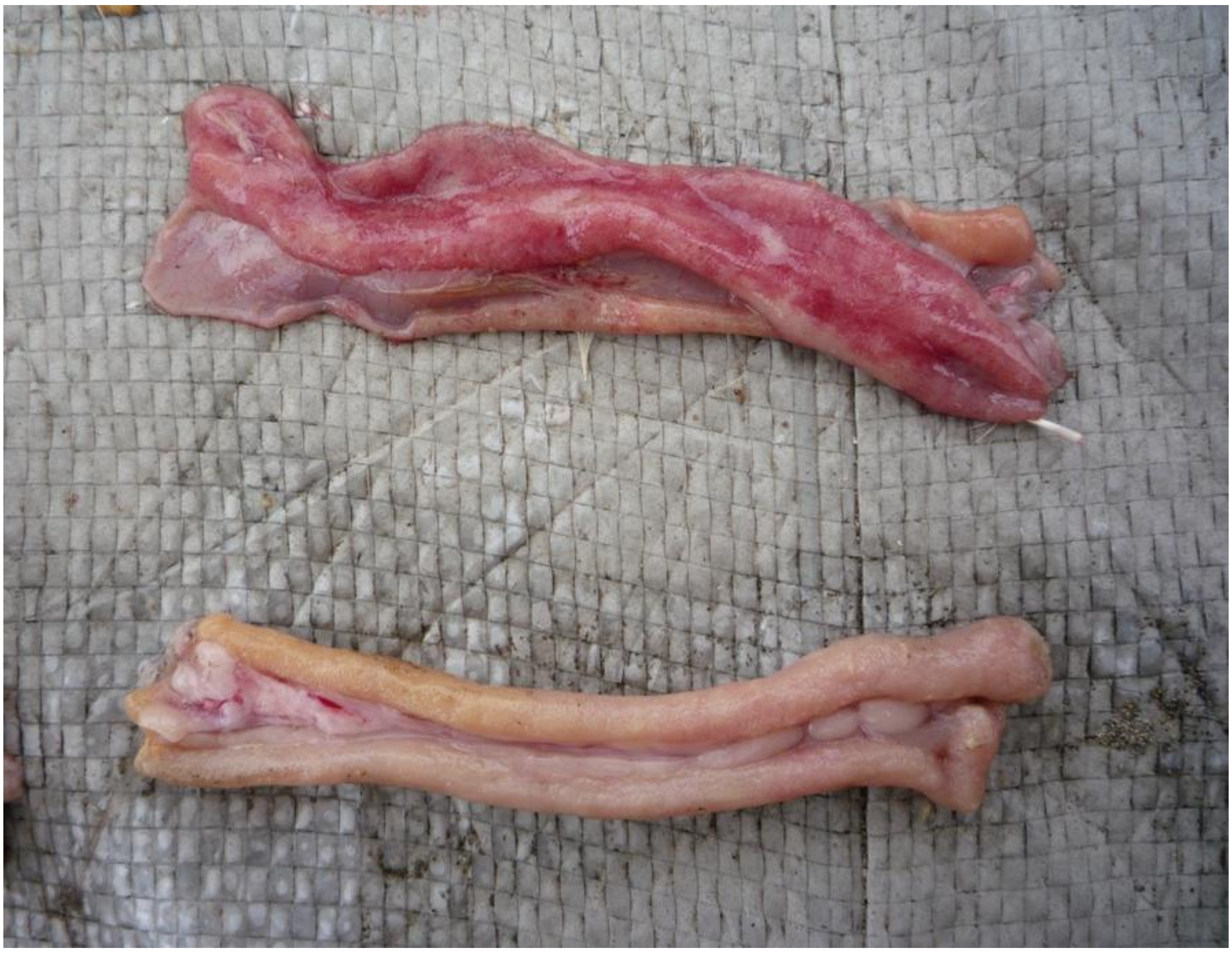

**Imagen 14. Alimento sin digerir en el íleon de las aves del Experimento 6.**

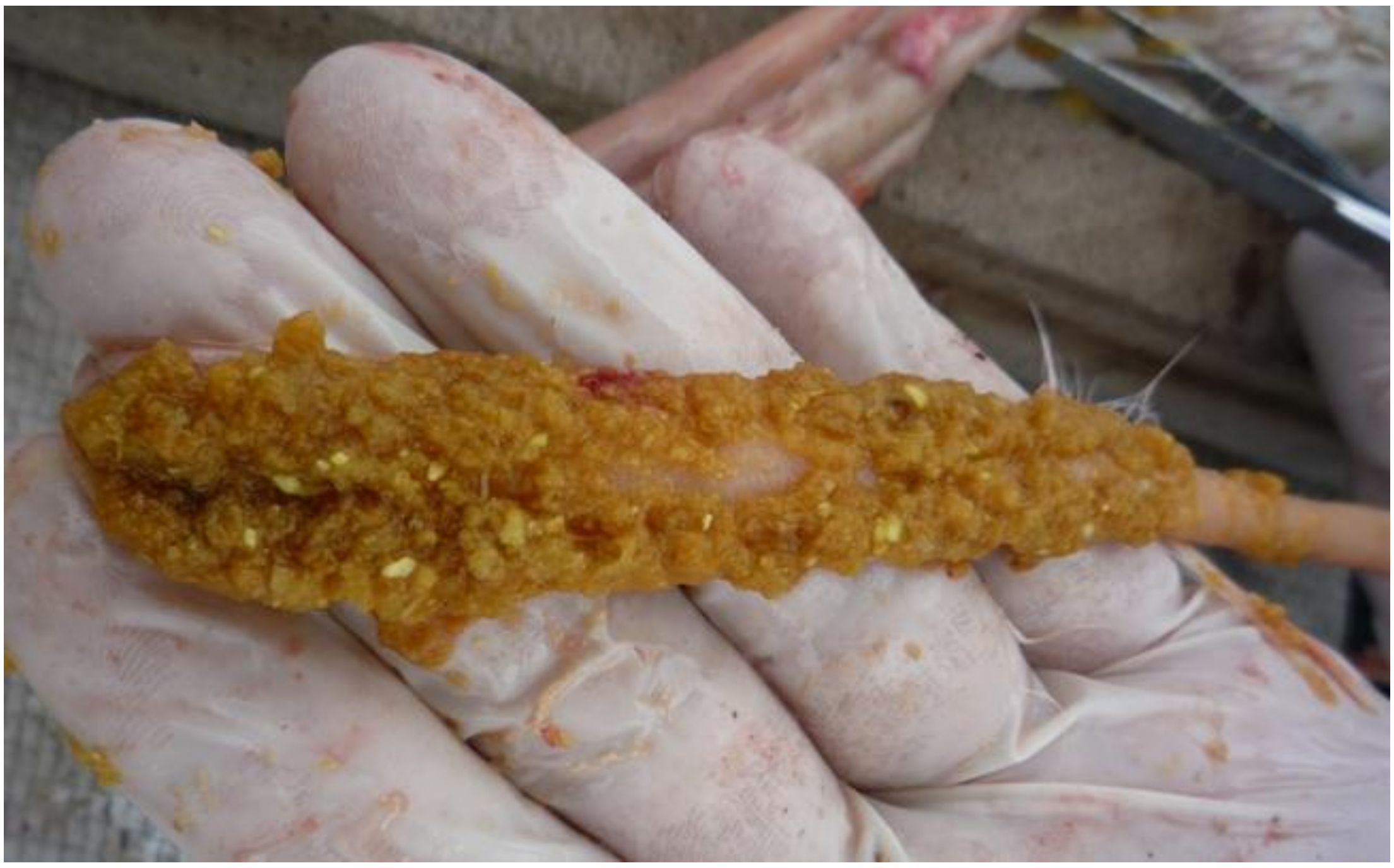

## 2.7. Mediciones

### 2.7.1. Disbacteriosis

Al final del periodo experimental, la presencia o ausencia clínica de disbacteriosis en cada unidad experimental se determinó empleando un método de Panneman and van der Stroom-Kruyswijk (2002) antes presentado. Para ello, se consideró positiva (+) a disbacteriosis aquellas unidades experimentales con al menos 30% de heces clasificadas como húmedas y negativa (-) aquella con un porcentaje menor. Para determinar el grado de humedad de las heces de cada parvada se colocó en el piso del galpón papel absorbente (8 pliegos de aproximadamente 1 $m^2$ cada uno) durante 1 hora, considerándose como húmeda aquella excreta que produjo un halo de humedad con al menos del doble del diámetro que el de la misma excreta (Imagen 10).

### 2.7.2. Evaluación de las heces

En los días 5 y 7 de la evaluación se registró las siguientes variables:

- Porcentaje de heces acuosas.
- Porcentaje de heces con alimento sin digerir
- Porcentaje de heces con descamaciones mucosas
- Porcentaje de heces hemorrágicas
- Índice Dimar:

  A cada excreta muestreada se asignó un score de acuerdo a la siguiente escala:

  0: si es normal (seca, bien formada y no contiene alimento sin digerir)

  1: si la excreta es acuosa

  2: si contiene alimento sin digerir

  3: si presenta descamaciones de mucosa

  4: si es hemorrágica

  Luego se aplicó la fórmula presentada a continuación, donde SPHM: score promedio de las heces muestreadas.

$$\text{Índice Dimar} = \text{SPHM} \times 25$$

#### 2.7.3. Crecimiento corporal

Se pesó a las aves antes de iniciar el tratamiento (día 0), al final del tratamiento (día 5) y dos días después de concluido el tratamiento (día 7). En cada fecha se tomó al azar una muestra representativa de aves y se pesó de forma grupal. El peso de las aves se obtuvo dividiendo las mediciones realizadas entre el número de aves empleadas, el cual fue siempre mayor o igual a 100 aves.

### 2.8. Diseño estadístico

Se utilizó el Diseño de Bloques Completo al Azar con dos tratamientos y cuatro repeticiones debido a que se empleó aves de dos edades distintas y en cada caso de ambos sexos. Cada combinación posible de edad y sexo constituyó un bloque. El análisis de varianza se llevó a cabo usando el programa Statistical Analysis System SAS 9.0 (SAS Institute, 2009) y la diferencia de medias usando la prueba de Duncan (1955). El Modelo Aditivo Lineal General aplicado fue el siguiente:

$$Y_{ij} = U + T_i + B_j + E_{ij}$$

Donde:

| | | |
|---|---|---|
| Yij | = | variable respuesta |
| U | = | media general |
| Ti | = | i-ésimo tratamiento ( i = 1, 2 ) |
| Bj | = | j-ésimo bloque ( j = 1, 2, 3, 4 ) |
| Eij | = | Error experimental |

Se consideró significativos aquellos valores con P menores de 0.10 (Crespo and Esteve-Garcia, 2001; Zhu *et al*, 2003, Kilburn and Edwards, 2004; van Nevel *et al*, 2005; Tahir *et al*, 2008; Teeter *et al*, 2008) o menores de 0.05, según se indica en cada caso.

## 3. Resultados y discusión

El presente experimento fue realizado en parvadas de aves en que se reportó retraso en el crecimiento y desuniformidad como principales características anómalas. Tras una inspección fue posible observar, además, alteraciones en las heces compatibles con un trastorno entérico. Al respecto, la desuniformidad en pollos de carne ha sido asociada con parvadas inmunodeprimidas (McIlroy *et al*, 1992; Lukert and Saif, 2003) y con daño a la mucosa intestinal (McDougald, 2003), entre otros factores.

Durante el presente estudio, la principal alteración observada en las heces fue la presencia de alimento sin digerir (Cuadro 17 y Anexo 51); observándose, en ambos tratamientos y fechas de muestreo, mayor prevalencia de esta alteración que heces acuosas; y no se presentaron heces hemorrágicas. Esto coincide con observaciones de campo respecto a la presencia de excretas secas, aparentemente normales, bien formadas, pero que al ser inspeccionadas se observa alimento sin digerir en su contenido. Esta forma del STR resulta menos evidente por la menor presencia de heces acuosas, lo que hace que sea difícilmente perceptible a simple vista; sin embargo, su impacto sobre el comportamiento productivo de la parvada es igualmente importante (Martínez, 2010).

En el presente estudio, tras el periodo de evaluación, las aves que recibieron el PRO muestran menor frecuencia de heces con alimento sin digerir ($P < 0.04$), menos heces acuosas ($P < 0.05$), mayor porcentaje de heces normales ($P < 0.04$) y un menor índice Dimar ($P < 0.03$) que las aves control. El índice Dimar es la variable que, en promedio, muestra la menor probabilidad de error estadístico ($P<0.0213$) entre aquellas que reflejan el estado de las heces.

Al respecto, si bien la literatura científica no reporta un tratamiento efectivo y formalmente establecido contra el STR (Merck, 2011) precisamente por su carácter multifactorial, es frecuente en condiciones comerciales emplear estrategias terapéuticas definidas generalmente por la principal característica anómala observada (Martínez, 2010). Así, cuando lo más evidente es la presencia de heces acuosas, se emplea caolín, pectina y/o sustancias que atenúan el desbalance osmótico en el lumen intestinal; cuando se trata de enteritis, lesiones por *Clostridium sp.* o presencia

**Cuadro 17. Efecto de un producto a base de aceite esencial de orégano vía agua de bebida en el control del Síndrome de Tránsito Rápido.**

| Variable | Tratamientos [1] 1 | Tratamientos [1] 2 | P |
|---|---|---|---|
| **Evaluación de heces [2]** | | | |
| 5 días después de iniciado el tratamiento | | | |
| Heces normales, % | 34.6 b | 61.9 a | 0.0266 |
| Heces acuosas, % | 56.1 a | 25.8 b | 0.0118 |
| Heces con alimento sin digerir, % | 65.4 a | 38.1 b | 0.0226 |
| Heces con descamaciones mucosas, % | 2.1 | 1.0 | 0.4802 |
| Heces hemorrágicas, % | 0.0 | 0.0 | ND |
| Índice Dimar | 33.2 a | 19.3 b | 0.0204 |
| 7 días después de iniciado el tratamiento | | | |
| Heces normales, % | 47.6 b | 78.6 a | 0.0389 |
| Heces acuosas, % | 44.1 a | 17.4 b | 0.0411 |
| Heces con alimento sin digerir, % | 52.4 a | 21.4 b | 0.0389 |
| Heces con descamaciones mucosas, % | 0.0 | 0.0 | ND |
| Heces hemorrágicas, % | 0.0 | 0.0 | ND |
| Índice Dimar | 28.2 a | 12.6 b | 0.0222 |
| **Disbacteriosis [3]** | | | |
| Al iniciar el tratamiento | + | + | ND |
| 5 días del inicio del tratamiento | + | – | ND |
| 7 días del inicio del tratamiento | + | – | ND |
| **Crecimiento corporal** | | | |
| Peso día 0 (inicio de la medicación), g | 1189 | 1189 | ND |
| Peso día 5 (fin de la medicación), g | 1543 | 1560 | 0.7174 |
| Peso día 7, g | 1758 £ | 1822 $ | 0.0594 |

1 Tratamientos: 1: aves control sin tratamiento; 2: aves tratadas con Orevitol® vía agua de bebida a razón de 300 ml/$m^3$ por cinco días.

2 En estas variables se verificó la ausencia de efecto del factor bloque, por ello se realizó el análisis estadístico empleando el Diseño Completo al Azar para incrementar los grados de libertad del error y mejorar la calidad del análisis.

3 Los resultados se expresan como positivo (+) o negativo (–) a disbacteriosis.

a,b,$,£ Promedios significativamente diferentes no comparten la misma letra (a,b; P<0.05) o el mismo símbolo ($,£; 0.05<P<0.10).

ND No determinado

de alimento sin digerir en las heces se suelen emplear antimicrobianos como la neomicina o colistina; si se observan lesiones por *Eimeria sp.* se administran coccidicidas específicos o sulfas (Martínez, 2010). Sin embargo, en la mayoría de los casos se trata de estrategias empíricas que principalmente controlan la manifestación del cuadro, pero que difícilmente logran atenuar el impacto negativo que este tiene sobre el comportamiento productivo del ave.

El efecto menos evidente pero económicamente más importante que un cuadro entérico produce sobre la eficiencia productiva de la parada, independientemente su naturaleza, es la pérdida de eficiencia en la utilización del alimento debido en gran medida al daño producido sobre la mucosa intestinal, tal como sucede con el STR cuando este es predispuesto por la inadecuada calidad de la soya (Celis, 2000) o por el uso de grasas rancias (Hoerr, 1998; Dibner *et al*, 1996), así como en casos de micotoxicosis (Leeson and Summers, 2005; Brown *et al*, 2008), Enteritis Necrótica (Williams, 2005), o coccidiosis (McDougald, 2003). Las estrategias de control antes mencionadas tienen escaso efecto sobre este aspecto; por ello, a pesar que la principal manifestación o incluso el agente sean controlados, el daño a la integridad intestinal permanecerá durante el tiempo que naturalmente le toma al ave restituir la estructura de su mucosa intestinal. Al respecto, se ha establecido que el recambio celular de los enterocitos en condiciones fisiológicas normales se produce en un periodo de 72 horas en pollos de 4 días de edad y en 96 horas en aves mayores (Uni *et al*, 1998; Geyra *et al*, 2001; Maiorka y Rocha, 2009). Entonces, considerando que el periodo de crianza es de 42 días, este proceso de restitución representa 9.5% del ciclo de producción; es decir, que durante el 10% del periodo de crianza las aves verán afectadas su capacidad de absorción de nutrientes y, en consecuencia, su comportamiento productivo.

En el presente experimento las aves presentaron de forma no inducida un cuadro entérico que cursó con disbacteriosis, presencia de alimento sin digerir en las heces, enteritis, desuniformidad y retraso en el crecimiento (Cuadro 17 y Anexos 51 y 52). Por las características observadas fue posible descartar, por un lado, la participación de coccidia como origen del cuadro o como un importante agente secundario involucrado ya que no se observaron lesiones intestinales características (Conway and McKenzie, 2007). También fue posible descartar factores relacionados al alimento, ya que otras parvadas del mismo plantel alimentadas con las mismas dietas no estuvieron afectadas.

La significativa disminución observada en la frecuencia de heces con alteraciones, el control de la disbacteriosis en las aves tratadas con el PRO, así como la enteritis característica observada al inicio de la evaluación, indican que el cuadro observado inicialmente en las aves fue causado, predispuesto o influenciado por *Clostridium perfringens* y/u otros patógenos bacterianos.

Los resultados obtenidos en las aves tratadas con el PRO son consistentes con la acción antimicrobiana que ha sido documentada para el AEO (Conner and Beuchat, 1984; Helander *et al*, 1998; Lambert *et al*, 2001; Lee *et al*, 2004) y para el mismo PRO (UNMSM, 2009a). Los resultados también son consistentes con el efecto prebiótico del AEO (Hammer *et al*, 1999b; Dorman and Deans, 2000; Ferket, 2003; Lee *et al*, 2004; Yew, 2008; Zheng *et al*, 2010).

En el presente experimento se observó que las aves tratadas con el PRO presentaron un peso vivo que no fue significativamente mayor que las aves control al final del tratamiento (quinto día; +17 g; P>0.10; Cuadro 17 y Anexo 52) pero sí al final del periodo de evaluación (dos días después; +64 g; P<0.06). Al respecto, se ha observado un incremento en la proliferación celular en la mucosa intestinal en las aves suplementadas con AEO (ver Experimento 1; Levkut *et al*, 2011) aumentando la longitud de las vellosidades intestinales y la capacidad para la absorción de nutrientes. En condiciones de desafío y/o daño a la mucosa intestinal como las observadas en el presente estudio, la mayor proliferación celular por efecto del PRO se traduce en la restitución de la mucosa intestinal en menor tiempo (Bruerton, 2002; citado por Ferket, 2003), favoreciendo la elongación de las vellosidades intestinales descamadas a consecuencia de la enteritis (Greathead and Kamel, 2006) y restituyendo el área superficial para la absorción de nutrientes (García *et al*, 2007). En consecuencia, es probable que los primero cinco días de tratamiento favorecieran la restitución del epitelio intestinal, y recién a partir de ese momento se produjera una mayor absorción de nutrientes, traduciéndose en un mayor peso corporal en los días siguientes.

## 4. Conclusiones

Los resultados obtenidos bajo las condiciones del presente estudio permiten llegar a las siguientes conclusiones:

- La administración del PRO vía agua de bebida reduce la frecuencia de heces acuosas y la presencia de alimento sin digerir en las heces.

- En parvadas afectadas por el STR, la administración del PRO controla la disbacteriosis.

- La administración del PRO atenúa el efecto negativo del STR sobre el crecimiento de las aves.

# EXPERIMENTO N° 7

# EFECTO DE UN PRODUCTO A BASE DE ACEITE ESENCIAL DE ORÉGANO SOBRE LA COCCIDIOSIS EN POLLOS DE CARNE

## 1. Introducción

La coccidiosis es uno de los principales factores que afectan la eficiencia productiva de los pollos de carne, su curso es particularmente importante por cuanto modifica afecta la salud intestinal y el comportamiento productivo de las aves. Si bien sólo en ciertos casos este proceso cursa con incremento en la mortalidad, es común que predisponga la ocurrencia de procesos y síndromes entéricos muchas veces caracterizados por un efecto productivo poco aparente, pero económicamente dramático. Asimismo, en todos los casos, la conversión alimentaria se ve afectada, generando a la industria avícola mundial costos por US$ 2,400 millones anuales (Shirley *et al*, 2005; Peek, 2010).

Para su prevención y control la industria cuenta con estrategias bien definidas, y emplea productos y sustancias de diferentes clases. Sin embargo, cada vez son más frecuentes las resistencias a ciertas sustancias anticoccidiales y las restricciones de los mercados internacionales al uso de sustancias potencialmente perjudiciales para el consumidor.

En los últimos años se han realizado diversas investigaciones respecto al uso de sustancias y productos fitogénicos para el control de la coccidiosis. El objetivo de este experimento fue determinar el efecto de un producto a base de AEO (referido en adelante como PRO) en la alimentación de aves desafiadas con coccidia sobre los indicadores empleados en condiciones comerciales para el control de esta enfermedad.

# 2. Materiales y métodos

## 2.1. Lugar, fecha y duración

La crianza de las aves y la preparación del inóculo de coccidia se llevaron a cabo en las instalaciones del Programa de Aves de la Facultad de Zootecnia de la Universidad Nacional Agraria La Molina en Lima-Perú a mediados del segundo semestre del 2010. La necropsia de las aves y toma de muestras intestinales se realizó en las instalaciones del Laboratorio de Patología Aviar de la Facultad de Medicina Veterinaria de la Universidad Nacional Mayor de San Marcos (UNMSM) en Lima-Perú. El procesamiento de las muestras histológicas se llevó a cabo en el Laboratorio de Procedimientos Histológicos del Departamento de Patología Humana del Hospital María Auxiliadora, perteneciente a la red de hospitales del Ministerio de Salud, Perú. La lectura de las láminas histológicas se realizó en el Laboratorio de Histología, Embriología y Patología Veterinaria de la UNMSM. La evaluación se llevó a cabo desde la recepción de las aves en las instalaciones de crianza y el periodo de evaluación fue el comprendido de 0 a 21 días de edad.

## 2.2. Instalaciones, equipos y materiales

### 2.2.1. Instalaciones de crianza

Las aves estuvieron alojadas sobre material de cama reutilizado (Anexo 6) a razón de 21.4 pollos/$m^2$ (0.047 $m^2$/pollo). Durante los primeros 3 días de vida se colocó papel periódico para evitar el acceso directo de los pollos BB al material de cama, mientras que el alimento fue suministrado en bandejas plásticas y el agua de bebida en bebederos tipo tongo. A partir del cuarto día de edad el alimento y agua de bebida fue provisto en comederos y bebederos lineales, respectivamente.

Dos semanas antes del ingreso de las aves se inició un programa de control de vectores, empleando rodenticidas y mosquicidas comerciales. Antes y después del periodo experimental se realizó la limpieza y desinfección de las instalaciones.

### 2.2.2. Calefacción, control de la temperatura y ventilación

La calefacción fue provista por un sistema eléctrico con resistencias y termostatos, y fue controlada de acuerdo a las recomendaciones de la línea genética (Cobb-Vantress, 2008b). Para reducir la variabilidad en la temperatura ambiental a consecuencia de las diferentes alturas entre los pisos de las baterías se adaptó resistencias adicionales en la base de cada batería. La ventilación fue controlada con cortinas de polipropileno instaladas en el perímetro del área de crianza. La humedad ambiental en el área de crianza se mantuvo alrededor de 40%. Para la medición de la temperatura y humedad ambiental se empleó un termo-higrómetro electrónico digital con una aproximación de 0.1 °C para temperatura y 1% para humedad.

### 2.2.3. Preparación del alimento

Para el pesaje de los ingredientes mayores del alimento se utilizó una balanza digital con capacidad de 150 kg y aproximación de 0.02 kg, mientras que para el pesaje de los ingredientes de la premezcla se utilizó una balanza electrónica con capacidad de 6 kg y aproximación de 1 g.

Se utilizó mezcladoras horizontales de cintas de 400 y 30 kg de capacidad para la mezcla de los ingredientes durante la preparación del alimento y para las premezclas, respectivamente. Después de la preparación de cada dieta se realizó el flushing y limpieza de los equipos (FAO, 2004; Ratcliff, 2009). El alimento fue envasado en sacos de papel y polietileno laminado, y almacenado en cilindros metálicos.

### 2.2.4. Manejo y actividades especiales

Para la identificación individual de las aves se empleó el método que se presenta en el Anexo 3. Para la medición de los pesos vivos se utilizó una balanza electrónica con capacidad para 200 g y con aproximación de 10 mg y otra con capacidad para 15 kg y con aproximación de 1 g. Para la medición del alimento consumido, se utilizó una balanza electrónica con capacidad para 15 kg y con aproximación de 1 g.

Para remover el material de cama se empleó espátulas de polietileno. Se empleó guantes quirúrgicos desechables, alcohol metílico y desinfectante a base de sales cuaternarias de amonio y aldehídos (CKM-DESIN®, Laboratorio CKM S.A.C.).

En el día 10 de edad las aves fueron vacunadas, por vía ocular, contra la Enfermedad de Newcastle, empleando una vacuna viva liofilizada con una cepa VG/GA (Avinew®, Merial Limited).

Para el desafío de las aves se empleó un inóculo de coccidia preparado a partir de una vacuna comercial contra coccidia (Immucox® for Chickens II, Vetech Laboratories; ver Anexo 10).

El agua de bebida fue potabilizada empleando 1 ml de hipoclorito de sodio al 4.5% por cada 10 L de agua. Para evitar que la presencia de cloro interfiera con la viabilidad del inóculo, la clorinación del agua se realizó tres días antes de proveerla a las aves, y diariamente, al momento de suministrarla a las aves, se verificó la ausencia de cloro empleando un kit comercial de evaluación.

#### 2.2.5. Necropsias, mediciones y procesamiento de muestras histológicas

Para la medición de los pesos vivos se utilizó una balanza electrónica con capacidad para 15 kg y con aproximación de 1 g. Para la necropsia de las aves se empleó bisturí, tijeras y guantes quirúrgicos. Para las mediciones ulteriores se utilizó un microscopio óptico de luz artificial binocular LeicaDM500®, láminas porta y cubre objetos, papel secante y bolsas plásticas de cerrado hermético. Para la medición de los pesos de las vísceras se empleó una balanza electrónica con capacidad para 200 g y con aproximación de 10 mg. Para la medición del diámetro de la bursa de Fabricio se empleó un bursómetro comercial con perforaciones crecientes desde 1/8 hasta 8/8 de pulgada (Fort Dodge, 2001).

Para el muestreo de las secciones intestinales se empleó guantes quirúrgicos, hojas de afeitar, frascos plásticos para muestras individuales y una solución de formaldehído al 4% bufferada con PBS pH 7.0 (solución salina de buffer fosfato), con la siguiente composición por L: 4 g de fosfato sódico monobásico, 6.5 g de fosfato sódico dibásico, 100 ml de formaldehído al 37 o 40% y agua destilada en c.s.p. 1 L (INS, 1997; Lu *et al*, 2008).

Para la preparación de las láminas histológicas se empleó la coloración HE (hematoxilina-eosina) empleando hematoxilina (C.I. 75290), eosina amarilla (C.I. 45380), alcohol etílico en diferentes concentraciones, parafina, xilol, albumina de Mayer (glicerina y clara de huevo en partes iguales), sulfato doble de aluminio y potasio, agua corriente, óxido rojo de mercurio, ácido clorhídrico, amoniaco, acido acético glacial, bálsamo de Canadá, cápsulas histológicas plásticas (cassettes) y láminas porta y cubreobjetos.

Para la deshidratación de las muestras e infiltración de parafina se empleó un procesador automático de tejidos marca Leica TP1020. Para la preparación de los bloques de parafina se empleó un centro de inclusión marca Leica EG1150H. Para la solidificación de la parafina se empleó un centro de enfriamiento marca Leica EG1150C. Para el pesaje de los colorantes se empleó una balanza de precisión marca Cobos. Para el calentamiento de la parafina se empleó una estufa marca Memmert. Para el corte de las muestras incluidas se empleó un micrótomo de rotación tipo Minott marca Leica RM2125RT. Para la colocación de cortes sobre portaobjetos se empleó un baño de flotación marca Medax Nagel KG Kiel. Para la verificación de las láminas se empleó un microscopio óptico de campo claro marca Zeiss con aumentos de 100X y 400X. Para la lectura de las láminas histológicas se empleó un microscopio óptico marca Zeiss Axiostar.

## 2.3. Animales experimentales

Se empleó 128 pollos machos de la línea Cobb 500 con un peso promedio de 48.2 g (Anexo 4). Los pollos fueron alojados al azar en 16 unidades experimentales. Al final de la evaluación se muestreó al azar un pollo por unidad experimental para los controles respectivos.

### 2.4. Tratamientos

Se evaluaron 2 tratamientos, que fueron definidos de la siguiente manera:

Tratamiento 1: Dieta basal (grupo control)
Tratamiento 2: Dieta basal + 500 ppm de Orevitol®

Las aves de ambos tratamientos fueron desafiadas con coccidia. El producto indicado será referido, en adelante, como PRO.

### 2.5. Modelo de desafío

En el día 14 de edad, por vía oral y empleando cánulas oro-ingluviales (Christaki *et al*, 2004; Conway and McKenzie, 2007; Elmusharaf *et al*, 2010; Georgieva *et al*, 2011b), se administró una dosis de un inóculo de coccidia preparado para contener oocistos no atenuados provenientes de aislamientos de campo (Conway and McKenzie, 2007) de *Eimeria acervulina*, *E. maxima*, *E. tenella*, *E. necatrix* y *E. brunetti* con una concentración total no menor a 24 x $10^5$ oocistos vivos por dosis, equivalente a 10 dosis de una vacuna comercial (Immucox® for Chickens II, Vetech Laboratories).

### 2.6. Alimentación

En el periodo de 1 a 14 días de edad se suministró una dieta basal a base de maíz, soya y harina de pescado, y a partir del día 15 de edad se suministró una dieta basal a base de maíz y soya. Ambas dietas fueron complementadas con aceite de pescado y aminoácidos sintéticos, y suplementadas con una premezcla de vitaminas y minerales. La alimentación fue *ad libitum*. Las características de las dietas empleadas se presentan en el Cuadro 18. El aditivo indicado en los tratamientos dietarios fue incorporado en la dieta de 1 a 21 días a expensas del maíz.

**Cuadro 18. Características de las dietas empleadas en el Experimento 7.**

| Ingredientes | % | |
|---|---|---|
| | De 1 a 14 días | De 15 a 21 días |
| Maíz amarillo | 52.517 | 61.775 |
| Torta de soya | 26.699 | 31.405 |
| Harina de pescado | 14.940 | 0.000 |
| Aceite semirefinado de pescado | 2.024 | 2.380 |
| DL-Metionina | 0.190 | 0.224 |
| L-Lisina | 0.116 | 0.135 |
| Cloruro de colina | 0.085 | 0.100 |
| Fosfato dicálcico | 1.609 | 1.892 |
| Carbonato de calcio | 0.978 | 1.150 |
| Sal común | 0.361 | 0.424 |
| Marcador inerte [1] | 0.300 | 0.300 |
| Premezcla [2] | 0.085 | 0.100 |
| Antifúngico [2] | 0.085 | 0.100 |
| Antioxidante [2] | 0.013 | 0.015 |

| Nutriente | Aporte nutricional | |
|---|---|---|
| | De 1 a 14 días | De 15 a 21 días |
| EM [3], Kcal/kg | 3028 | 3008 |
| Proteína cruda, % | 26.72 | 20.04 |
| Lisina, % | 1.72 | 1.16 |
| Metionina + Cistina, % | 1.10 | 0.87 |
| Treonina, % | 1.05 | 0.76 |
| Triptófano, % | 0.29 | 0.23 |
| Calcio, % | 1.54 | 1.17 |
| Fósforo disponible, % | 0.67 | 0.44 |
| Sodio, % | 0.34 | 0.19 |
| Grasa total, % | 5.97 | 5.11 |
| Fibra cruda, % | 2.70 | 3.17 |

| Componente | Composición proximal [4], % | |
|---|---|---|
| | De 1 a 14 días [5] | De 15 a 21 días [6] |
| Humedad | 11.89 | 12.35 |
| Proteína total [3] | 26.06 | 19.66 |
| Extracto Etéreo | 4.87 | 3.83 |
| Fibra Cruda | 2.12 | 2.49 |
| Cenizas | 7.13 | 5.88 |
| ELN [3] | 47.94 | 55.79 |

[1] Óxido crómico como marcador inerte.
[2] Premezcla de vitaminas y minerales Proapak 2A®. Composición: Retinol: 12’000,000 UI; Colecalciferol: 2’500,000 UI; DL α-Tocoferol Acetato: 30,000 UI; Riboflavina: 5.5 g; Piridoxina: 3 g; Cianocobalamina: 0.015 g; Menadiona: 3 g; Ácido Fólico: 1 g; Niacina: 30 g; Ácido Pantoténico: 11 g; Biotina: 0.15 g; Zn: 45 g; Fe: 80 g; Mn: 65 g; Cu: 8 g; I: 1 g; Se: 0.15 g; Excipientes c.s.p. 1,000 g. Antifúngico: Mold Zap®; Antioxidante: Danox®
[3] EM: Energía metabolizable; Proteína total: N x 6.25; ELN: Extracto Libre de Nitrógeno (calculado).
[4] Informes de ensayo 1235/2010 LENA y 1236/2010 LENA, Universidad Nacional Agraria La Molina.
[5] Calculado a partir de los análisis proximales de la dieta empleada de 15 a 28 días de edad (85%) y de la harina de pescado (15%) empleadas en su producción.
[6] Informe de ensayo 1235/2010 LENA, Universidad Nacional Agraria La Molina.

## 2.7. Mediciones

### 2.7.1. Indicadores de coccidia

Al término del periodo experimental se tomó al azar un pollo de cada unidad experimental, para luego ser pesados y sacrificados cortando las arterias carótidas con la subsecuente exanguinación (Mutus *et al*, 2006), de acuerdo al protocolo del Laboratorio de Patología Aviar de la UNMSM. Se realizó la necropsia de acuerdo a los protocolos propuestos por Bermúdez y Stewart-Brown (2003) y Colas *et al* (2010), registrándose las siguientes variables:

- Aves positivas a coccidia:
  Se realizó el raspado de la mucosa intestinal, se preparó un frotis y se observó en el microscopio (Cardona, 2005). Se determinó la presencia o ausencia de ooquistes en el raspado intestinal. Se expresa en porcentaje.

- Score microscópico de coccidia:
  Se realizó el raspado de la mucosa intestinal, se preparó un frotis y se observó en el microscopio a un aumento total de 100X para el rastreo de los ooquistes y 400X para la lectura. Se empleó el siguiente score en base a la presencia y cantidad de ooquistes encontrados: 0: ausentes; 1: escasos; 2: cantidad moderada; 3: abundantes.

- Score de lesiones de coccidia:
  Se empleó los scores descritos por Johnson and Reid (1970), e ilustrados por Conway and McKenzie (2007), en una escala de 0 a 4.

- Pigmentación de tarsos:
  Se determinó en cada ave muestreada al final del periodo experimental empleando el abanico colorimétrico de Roche (Beardsworth and Hernández, 2004; Galobart *et al*, 2004).

- Recuento de ooquistes en cama:
  Se determinó al final del periodo experimental tomando muestras directamente de material de cama. Los métodos empleados para el muestreo del material de cama y para el recuento de ooquistes se presentan en el Anexo 11.

### 2.7.2. Morfometría de los órganos linfoides

Al final del periodo experimental, en las aves muestreadas se colectó los órganos linfoides y, previa eliminación del tejido graso, se determinó las siguientes variables:

- Diámetro de la bursa de Fabricio:
  La medida cualitativa del diámetro mayor de la bursa se determinó mediante un bursómetro comercial (Fort Dodge, 2001).

- Índices morfométricos:
  Se calculó empleando la fórmula (Anexo 1) presentada a continuación, donde IM: índice morfométrico, PO: peso del órgano (g), PC: peso corporal (g).

$$IM = (PO/PC) \times 1000$$

- Relación ente órganos:
  Se determinó las relaciones bursa/bazo (Bu-Ba), bursa/timo (Bu-Ti) y timo/bazo (Ti-Ba), empleando la fórmula (Anexo 1) presentada a continuación, donde $RO_{A-B}$: relación entre los órganos A y B, $PO_A$: peso del órgano A (g), $PO_B$: peso del órgano B (g).

$$RO_{A-B} = \frac{PO_A}{PO_B}$$

### 2.7.3. Comportamiento productivo

Se midió empleando las siguientes variables:

- Peso vivo: Se registró los pesos individuales al recibir los pollos BB y al final del periodo experimental (día 21).

- Ganancia de peso: Se calculó la ganancia total de peso (g) durante el periodo experimental.

- Consumo de alimento: Al término de cada semana se pesó los residuos de alimento y se calculó el consumo acumulado de alimento (g) durante el periodo experimental.

- Conversión alimentaria: Se calculó dividiendo el consumo acumulado de alimento entre el peso corporal al término del periodo experimental, de acuerdo al método empleado en la industria (Ponce de León, 2011).

- Mortalidad: Se calculó valores acumulados al día 21 de edad (%).

## 2.8. Diseño estadístico

Se utilizó el Diseño Completo al Azar con dos tratamientos y ocho repeticiones. La prueba de bondad de ajuste a la distribución normal de los datos obtenidos a partir de las variable evaluadas se realizó empleando la prueba de Chi-Cuadrado (Calzada, 1982).

El análisis de varianza de los datos distribuidos normalmente se llevó a cabo aplicando el procedimiento paramétrico ANOVA del programa Statistical Analysis System SAS 9.0 (SAS Institute, 2009) y la diferencia de medias se realizó usando la prueba de Duncan (1955).

Los datos referidos a lesiones intestinales y aves positivas a coccidia se analizaron empleando la prueba no paramétrica de Kruskal-Wallis mediante el procedimiento NPAR1WAY con restricción WILCOXON del programa SAS (Schlotzhauer and Littell, 1997; McDonald, 2009), por no estar distribuidos normalmente. El Modelo Aditivo Lineal General fue:

Yij = U + Ti + Eij;  Donde:

| | | |
|---|---|---|
| Yij | = | variable respuesta |
| U | = | media general |
| Ti | = | i-ésimo tratamiento ( i = 1, 2 ) |
| Eij | = | Error experimental |

Se consideró significativos aquellos valores con P menores de 0.05.

## 3. Resultados y discusión

### 3.1. Diseño de la evaluación

El objetivo de esta evaluación fue verificar la acción contra coccidia que algunos autores han postulado para el AEO (Lillehoj and Trout 1996; Bruerton, 2002; citado por Ferket, 2003). Para lograr dicho objetivo se empleó parcialmente el modelo de desafío B presentado en el Anexo 2 (ver sección Modelo de Desafío del presente experimento), que incluyó como principal fuente de contaminación el uso de un inóculo de coccidia. Al respecto, se ha determinado que la infección con coccidia puede lograrse empleando material de cama contaminada; sin embargo, la inoculación de ooquistes vía oral es un medio más efectivo para establecer la infección y produce resultados similares a la inoculación en cama (Elmusharaf *et al*, 2010). Además, para reproducir condiciones de crianza que se asemejen más a las comerciales en que los desafíos de campo suelen estar constituidos por un conjunto de especies de coccidia (Alcaíno *et al*, 2002), el inóculo administrado en el presente estudio contuvo una dosis mixta de oocistos vivos no atenuados constituida por cinco especies del agente.

La evaluación se realizó empleando un sistema de jaulas elevadas para reducir el riesgo de contaminación cruzada observada en estudios previos con coccidia (McReynolds *et al*, 2004) cuando se emplean corrales en piso, debido a la forma en que se realiza el manejo de las aves. Sin embargo, debido a las características propias del ciclo de vida de la coccidia es necesario que en las pruebas diseñadas para evaluar la actividad de sustancias contra este agente, las aves estén alojadas sobre material de cama pues ello permite la replicación, el reciclaje y la reinfección de la coccidia (Conway and McKenzie, 2007). Para suplir esta necesidad se diseñó un sistema de alojamiento en que las jaulas contuvieron material de cama (Anexo 2), lo que a su vez permitió reproducir las condiciones de campo en cuanto a la concentración de amoniaco en el ambiente.

En el presente estudio se consideró como periodo de evaluación crítico la edad de 21 días, 7 días después de la administración del inóculo de coccidia, de acuerdo a los reportes de campo y, entre otros, a las recomendaciones de la Administración de Alimentos y Fármacos de los Estados Unidos (CVM, 1992) y a los resultados de Christaki *et al* (2004) quienes desafiaron pollos de carne con *E. tenella* en el día 14 de edad, observando un efecto negativo en el peso corporal en el día 21 y una respuesta favorable en las aves suplementadas con un extracto fitogénico vía alimento. En el presente estudio, el periodo de 7 días comprendido desde la inoculación de las aves y las evaluaciones fue suficiente para lograr la reinfección necesaria con los oocistos vacunales y desencadenar el cuadro deseado de coccidiosis subclínica.

Como resultado del modelo de desafío aplicado en el presente estudio, se logró desarrollar un cuadro sub-clínico caracterizado por la presencia de aves positivas a coccidia (63%), encontrándose lesiones intestinales compatibles con *E. acervulina* (63% de las aves; score promedio de 1.4) y *E. maxima* (25% de las afectadas; score promedio de 0.4) (Cuadro 19 y Anexos 53 y 54). Las lesiones fueron clasificadas como compatibles con coccidia de acuerdo a su localización en el intestino, a las características de las lesiones macroscópicas, de acuerdo a lo descrito por Conway and McKenzie (2007), y a su localización vertical en la mucosa intestinal, sea en el epitelio de las vellosidades intestinales (*E. acervulina, E. maxima, E. tenella, E. necatrix, E. brunetti, E. mivati, E. praecox* y *E. mitis*), en la lámina propia (*E. necatrix* y *E. tenella*), en las criptas de Lieberkühn (*E. acervulina, E. tenella y E. praecox*) o en la muscularis (*E. necatrix*) (Quiroz, 2005). Tras desafiar a las aves del presente estudio, no se incrementó la mortalidad (μ = 3.1%), ni se observaron aves postradas o decaídas; es decir, fue un cuadro sub-clínico. Esto concuerda con

**Cuadro 19. Efecto de un producto a base de aceite esencial de orégano sobre la coccidiosis en pollos de carne.**

| Variable | Tratamientos (1) 1 | | Tratamientos (1) 2 | | P |
|---|---|---|---|---|---|
| **Indicadores de coccidia** | | | | | |
| Aves positivas a coccidia (2), % | 62.5 | a | 12.5 | b | 0.0455* |
| Score microscópico de coccidia (3) | 0.875 | a | 0.125 | b | 0.0381* |
| Score de lesiones por *E. acervulina* (4) | 1.375 | | 1.000 | | 0.4395* |
| Score de lesiones por *E. maxima* (4) | 0.375 | | 0.125 | | 0.4875* |
| Recuento de ooquistes, N°/g cama | 1629 | a | 1359 | b | 0.0006 |
| Pigmentación de tarsos (5) | 2.13 | | 2.50 | | 0.4091 |
| **Morfometría de órganos linfoides** | | | | | |
| Diámetro de la bursa (6) | 6.00 | a | 5.25 | b | 0.0313 |
| Índice morfométrico de la bursa (Rbu) | 2.53 | | 2.22 | | 0.4339 |
| Índice morfométrico del bazo (Rba) | 1.18 | | 1.24 | | 0.8217 |
| Índice morfométrico del timo (Rti) | 5.67 | | 5.41 | | 0.6884 |
| Relación bursa/bazo (Bu-Ba) | 2.22 | | 2.30 | | 0.8882 |
| Relación bursa/timo (Bu-Ti) | 0.45 | | 0.43 | | 0.7765 |
| Relación timo/bazo (Ti-Ba) | 5.08 | | 5.88 | | 0.6071 |
| **Comportamiento productivo** | | | | | |
| Peso final (día 21 de edad), g | 848.4 | b | 925.9 | a | 0.0024 |
| Ganancia de peso, g | 800.4 | b | 877.5 | a | 0.0026 |
| Consumo de alimento, g | 1130.1 | | 1127.0 | | 0.8945 |
| Conversión alimentaria | 1.33 | a | 1.22 | b | 0.0012 |
| Mortalidad, % | 3.1 | | 3.1 | | 1.0000 |

(1) Tratamientos: 1: aves desafiadas control; 2: aves desafiadas + 500 ppm de Orevitol®.
(2) El diagnóstico fue realizado por microscopía. Todos los casos fueron sub-clínicos.
(3) Score microscópico de coccidia. 0: ausente; 1: escasos; 2: cantidad moderada; 3: abundantes.
(4) Score de lesiones de Johnson and Reid (1970) en una escala de 0 a 4.
(5,6) Empleando el abanico colorimétrico Roche y bursómetro (1/8" a 8/8"), respectivamente.
a,b Promedios significativamente diferentes no comparten la misma letra (a,b; P<0.05).
* Datos analizados empleando la prueba de Kruskal-Wallis.

recientes observaciones de campo en nuestro medio, en que la mayor proporción de casos asociados a coccidia son compatibles principalmente con *E. acervulina* y en menor proporción con *E. maxima* u otras especies (Martínez, 2010).

De manera complementaria se presenta un registro fotográfico de duodenos afectados por coccidiosis en el presente estudio que presentan lesiones compatibles con *E. acervulina* con scores clasificados como 1 (Imágenes 15 a 18), 2 (Imágenes 19 a 22) y 3 (Imágenes 23 a 26). Esto permite correlacionar las observaciones macroscópicas del score de Johnson y Reid (1970) para lesiones compatibles con *E. acervulina*, que son las más frecuentemente encontradas en las explotaciones comerciales, con las respectivas observaciones histológicas, verificándose el significativo daño causado a la integridad de la mucosa intestinal.

## 3.2. Indicadores de coccidia

### 3.2.1. Lesiones intestinales y microscopía diagnóstica

En el presente estudio las aves fueron consideradas positivas a coccidia desde dos puntos de vista: el macroscópico (presencia de lesiones intestinales compatibles) y el microscópico (presencia de formas de coccidia en el raspado de la mucosa intestinal; McDougald, 2003). Así, al término del periodo experimental la evaluación macroscópica de lesiones intestinales muestra menores scores de lesiones compatibles con *E. acervulina* y *E. maxima* en las aves que reciben el PRO; sin embargo, no ha sido posible determinar que estas diferencias sean atribuibles al aditivo empleado (P>0.10). En la evaluación microscópica se observa un menor porcentaje de aves positivas a coccidia (13% vs 63%; P<0.05) y un menor score microscópico de coccidia (0.13 vs 0.88; P<0.05) en el tratamiento con el PRO en comparación al grupo control (Cuadro 19 y Anexos 53 y 54).

Al respecto, es importante precisar que si bien el porcentaje de aves positivas a coccidia en el tratamiento 1 fue el mismo en las evaluaciones macroscópica y microscópica (62.5%), en el tratamiento 2 el 75% de las aves fueron positivas en la evaluación macroscópica, pero sólo el 12.5% lo fue en la evaluación microscópica (Anexo 54); es decir, que no se encontró ooquistes en la mayor parte de las aves que

**Imagen 15. Duodeno de un pollo positivo a *E. acervulina* que presenta lesiones con score 1.** Se observan las lesiones características en forma de placas blanquecinas (flechas) orientadas transversalmente al eje del intestino en la superficie mucosa del asa duodenal.

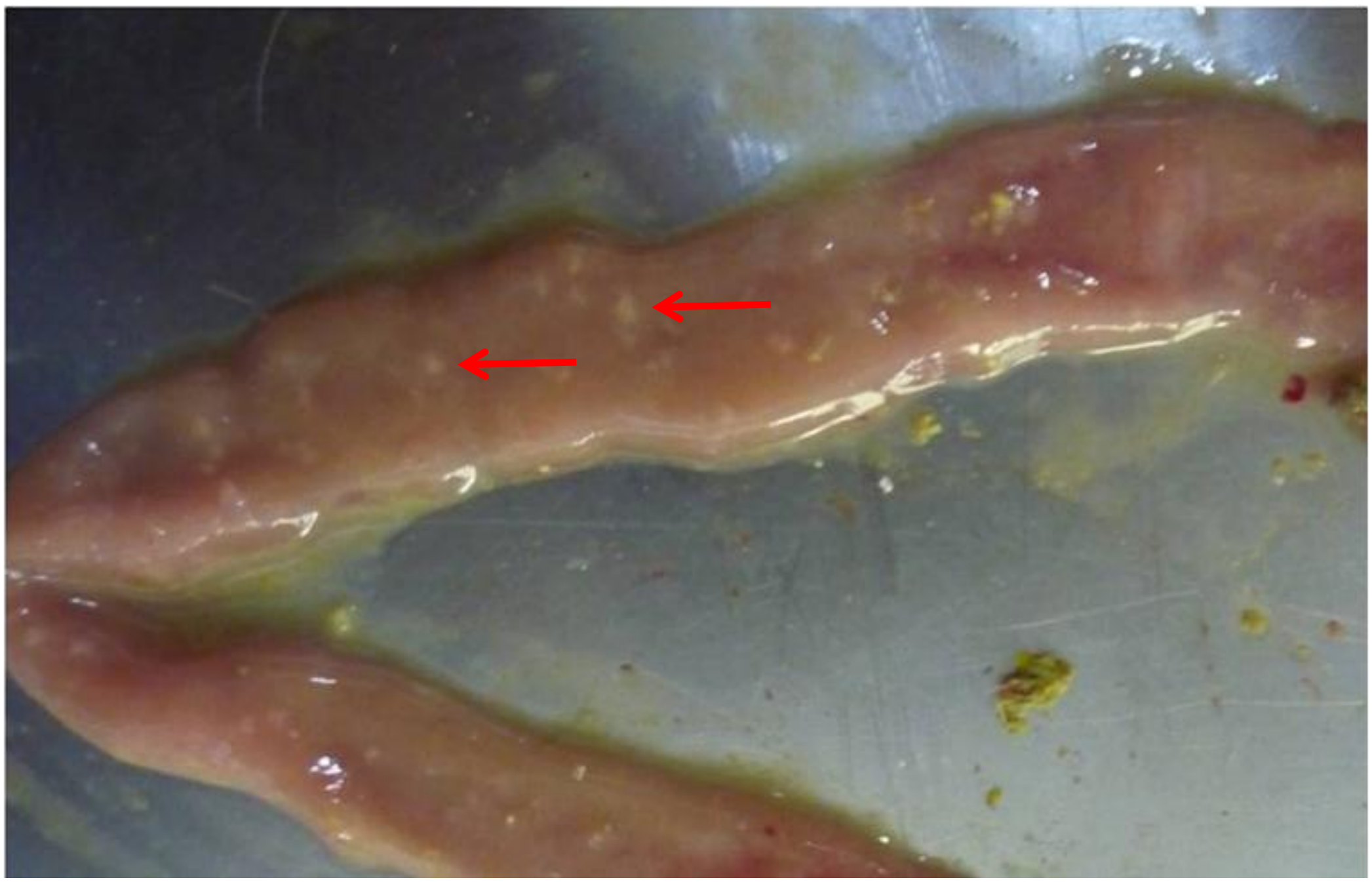

**Imagen 16. Sección transversal del duodeno de un pollo positivo a *E. acervulina* que presentó lesiones con score 1 en la evaluación macroscópica.** Se observan vellosidades descamadas (flechas). Microscopía a 40 X con coloración HE. Las zonas marcadas en recuadros corresponden a las Imágenes 17 y 18.

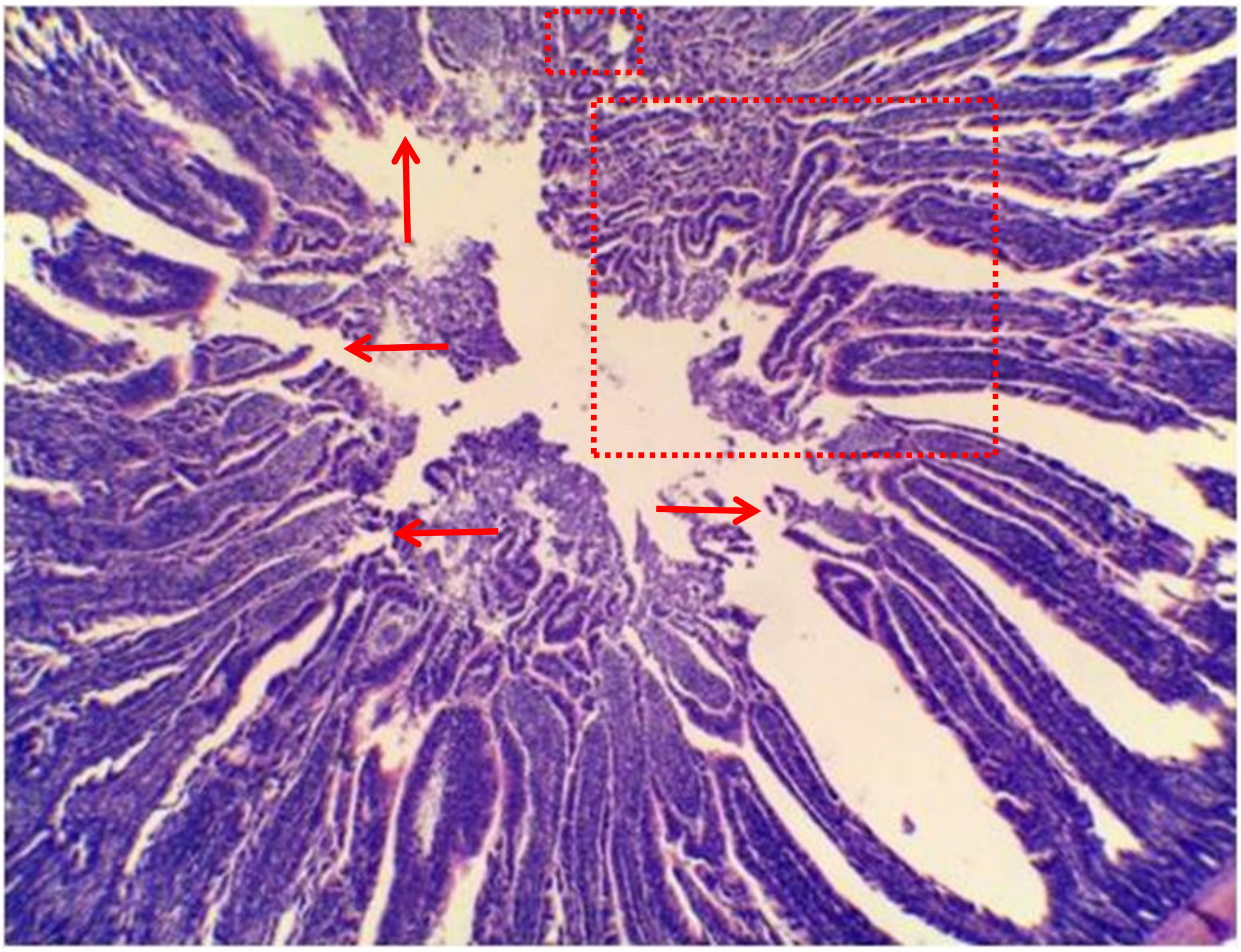

**Imagen 17. Vellosidades intestinales del duodeno de un pollo positivo a *E. acervulina* que presentó lesiones con score 1 en la evaluación macroscópica.** Se observa la descamación de enterocitos en la luz y la infiltración de linfocitos en la lámina propia. Microscopía a 100 X con coloración HE.

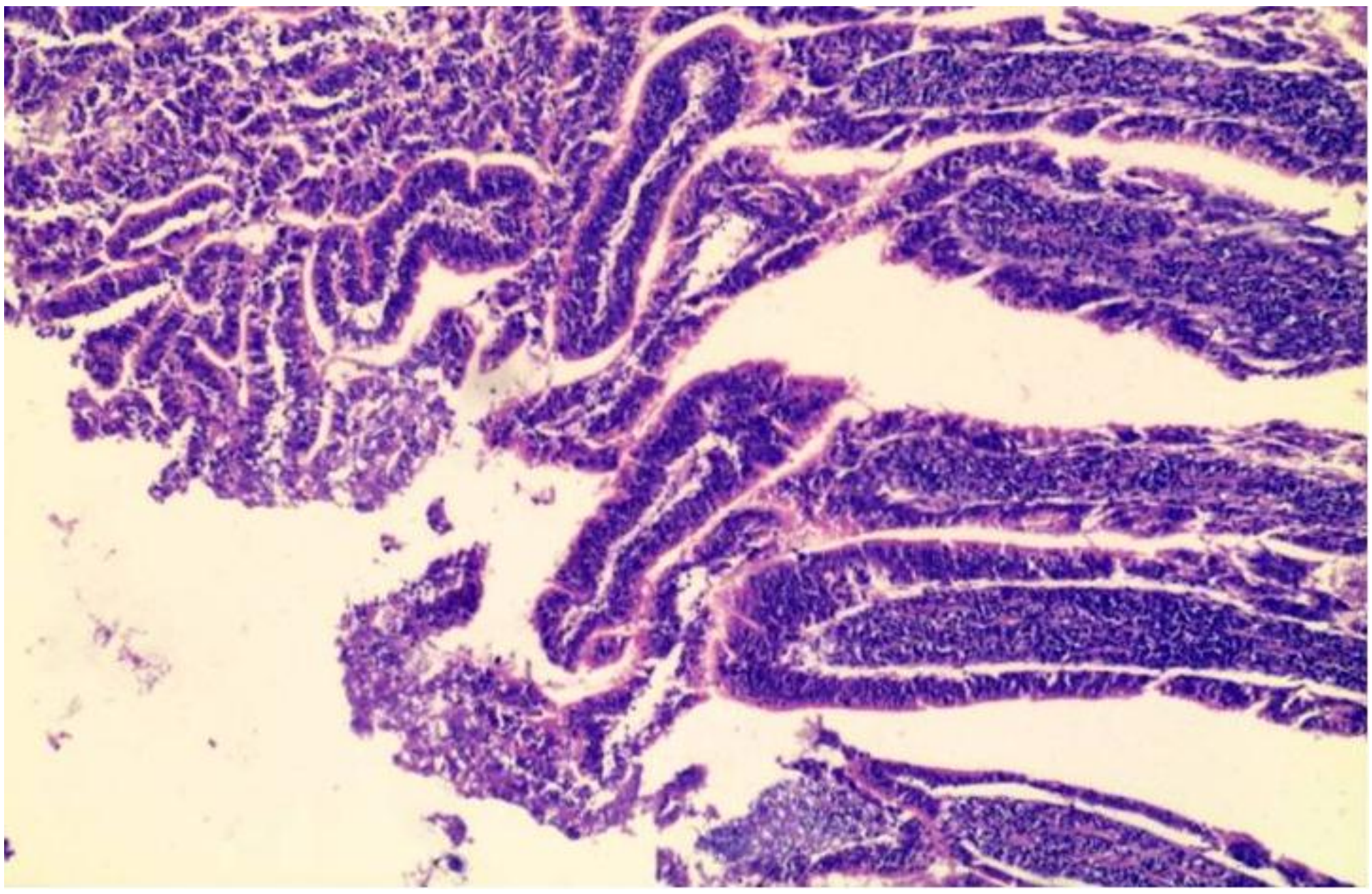

**Imagen 18. Vellosidad intestinal del duodeno de un pollo positivo a *E. acervulina* que presentó lesiones con score 1 en la evaluación macroscópica.** Se observa algunas formas inmaduras de coccidias en el epitelio (flechas largas), y heterófilos (flecha mediana) y abundantes linfocitos (flechas cortas) en la lámina propia. Microscopía a 400 X con coloración HE.

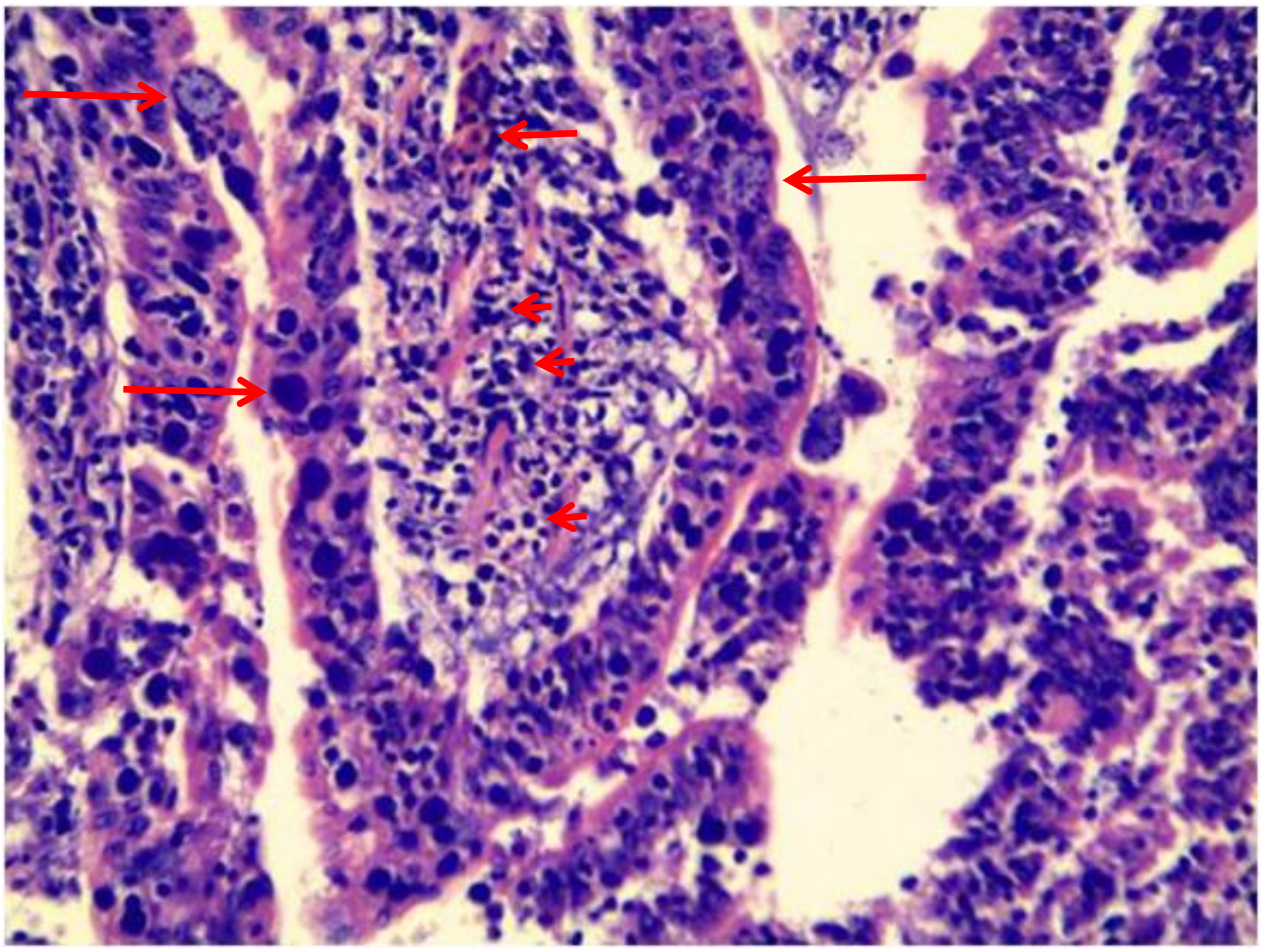

**Imagen 19. Duodeno de un pollo positivo a *E. acervulina* que presenta lesiones con score 2.** Se observa que las lesiones blanquecinas características (flechas) se presentan en mayor densidad que en el score 1 (Imagen 15).

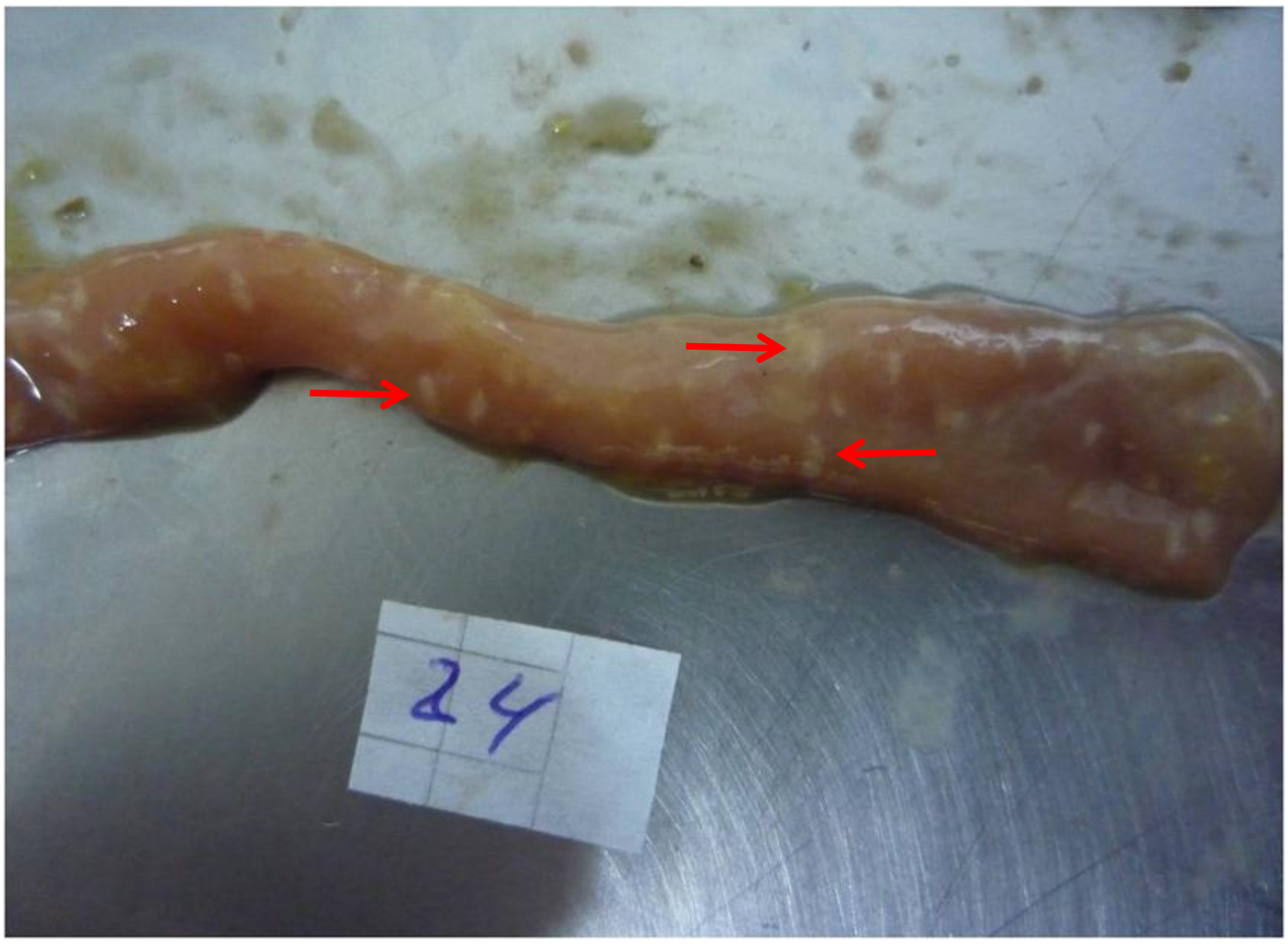

**Imagen 20. Sección transversal del duodeno de un pollo positivo a *E. acervulina* que presentó lesiones con score 2 en la evaluación macroscópica.** Se observan vellosidades descamadas (flechas). Microscopía a 40 X con coloración HE. Las zonas marcadas en recuadros corresponden a las Imágenes 21 y 22.

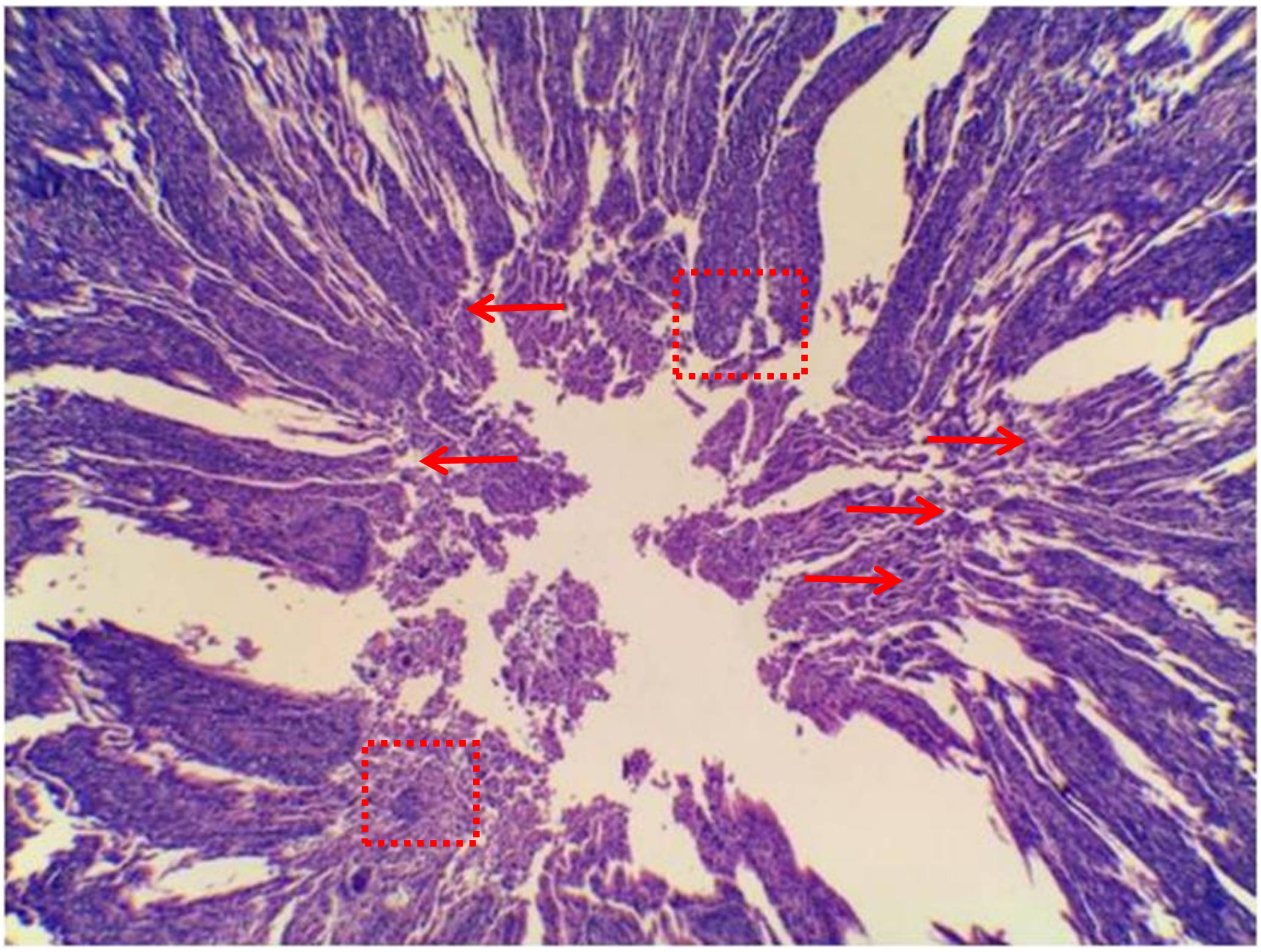

**Imagen 21. Contenido intestinal del duodeno de un pollo positivo a *E. acervulina* que presentó lesiones con score 2 en la evaluación macroscópica.** Entre los restos de la mucosa descamada se observa formas inmaduras de coccidias (flechas largas) y células caliciformes (flechas medianas), así como restos de enterocitos, linfocitos y material celular (flechas cortas). Microscopía a 400 X con coloración HE.

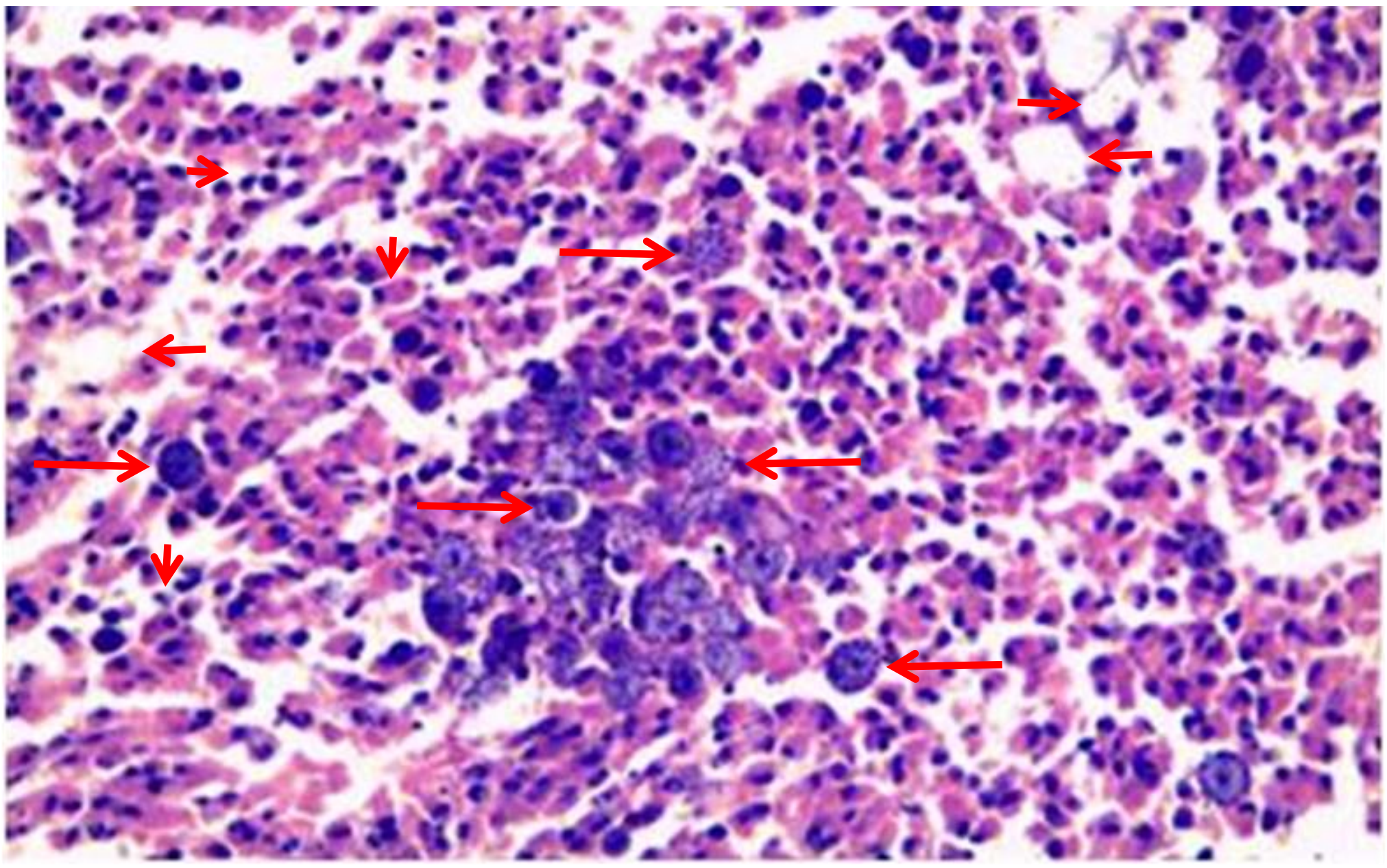

**Imagen 22. Vellosidad intestinal del duodeno de un pollo positivo a *E. acervulina* que presentó lesiones con score 2 en la evaluación macroscópica.** En algunas zonas del epitelio se observa la abundancia de formas maduras (flechas largas) e inmaduras (flechas cortas) de coccidias, que no permite la visualización de los enterocitos. Microscopía a 400 X con coloración HE.

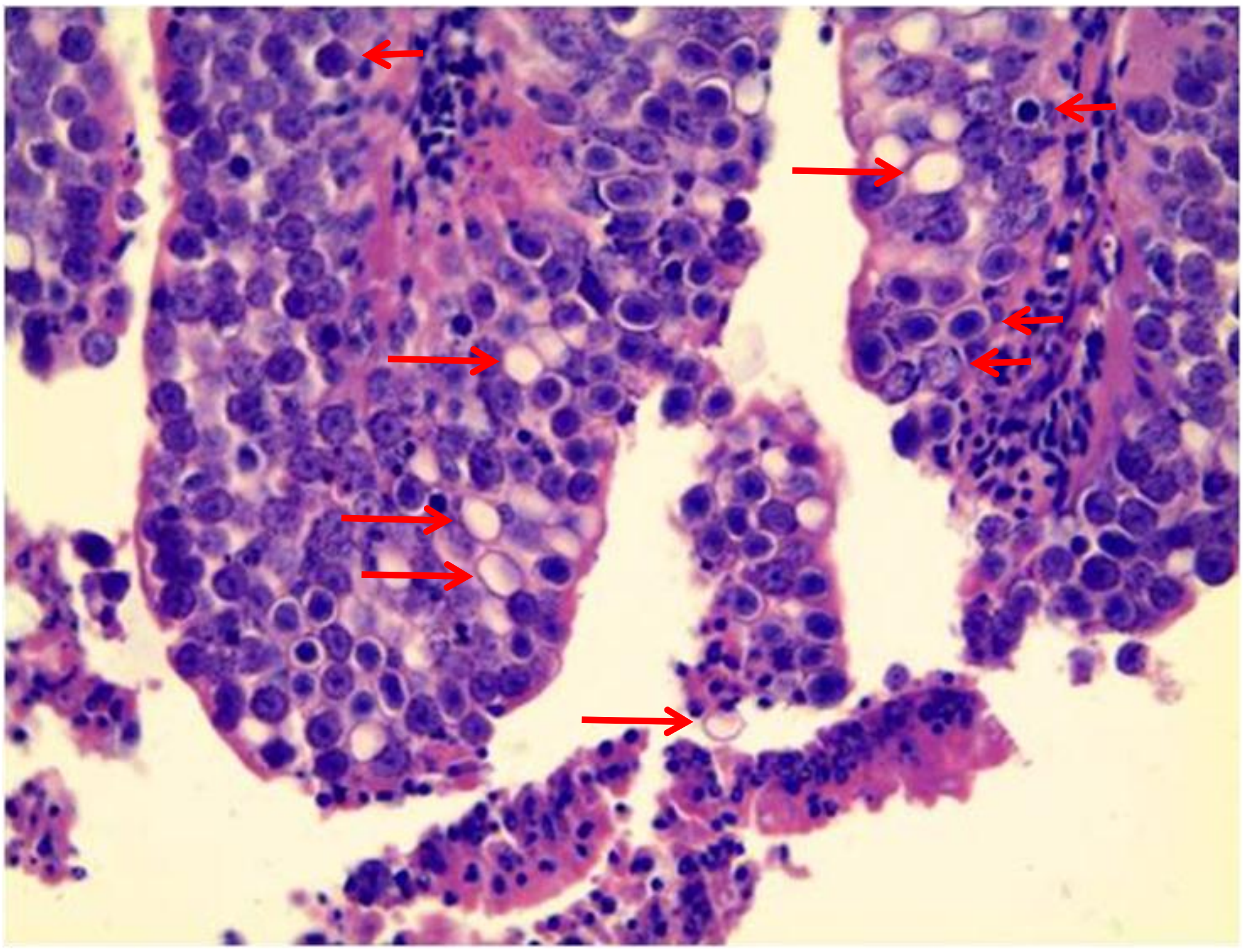

**Imagen 23. Duodeno de un pollo positivo a *E. acervulina* que presenta lesiones con score 3.** Se observa que las lesiones blanquecinas características se presentan en mayor densidad que en el score 2 (Imagen 19) y que algunas están en coalescencia (círculos).

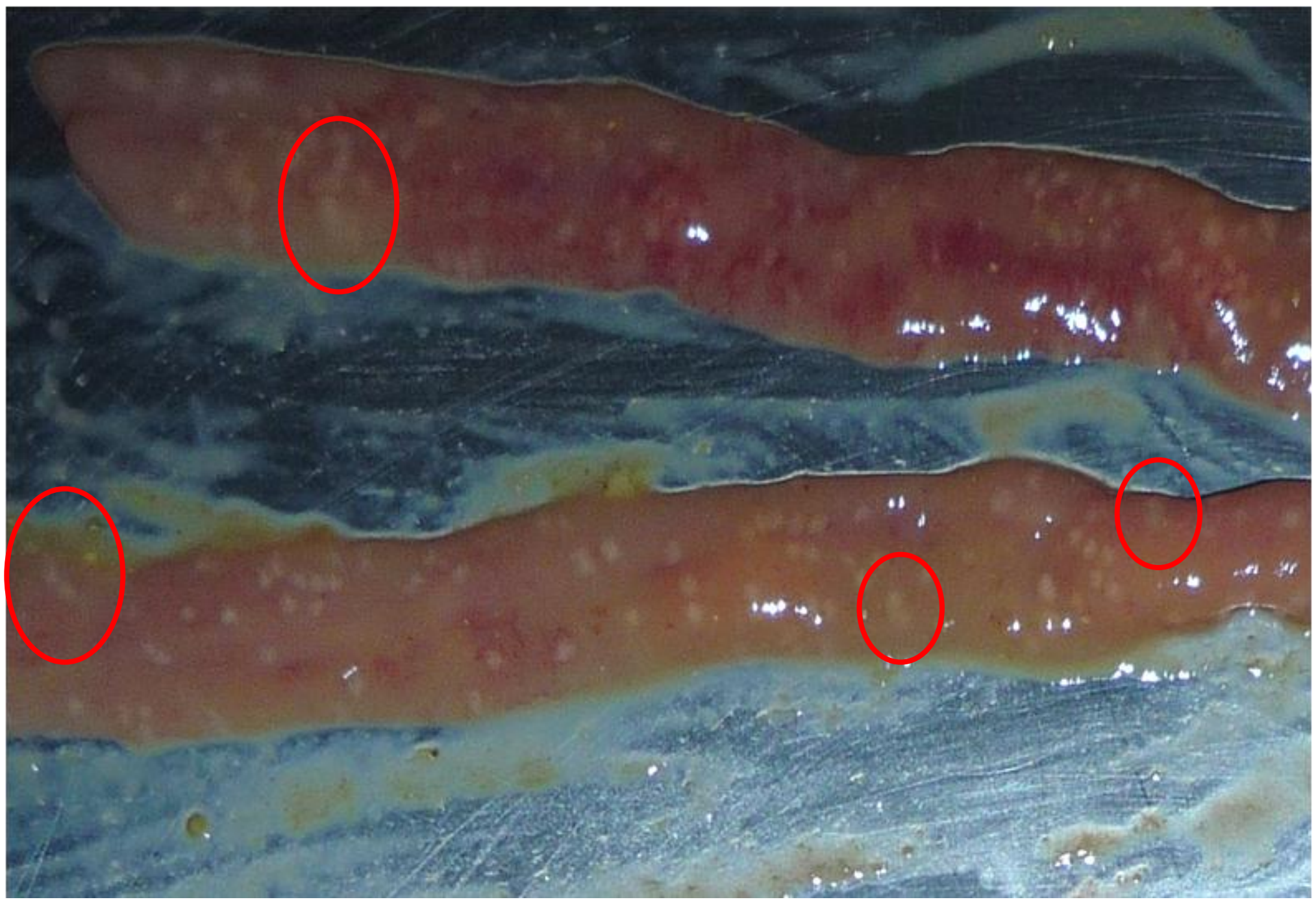

**Imagen 24. Sección transversal del duodeno de un pollo positivo a *E. acervulina* que presentó lesiones con score 3 en la evaluación macroscópica.** Se observa vellosidades descamadas (flechas). Microscopía a 40 X con coloración HE. Las zonas marcadas en recuadros corresponden a las Imágenes 25 y 26.

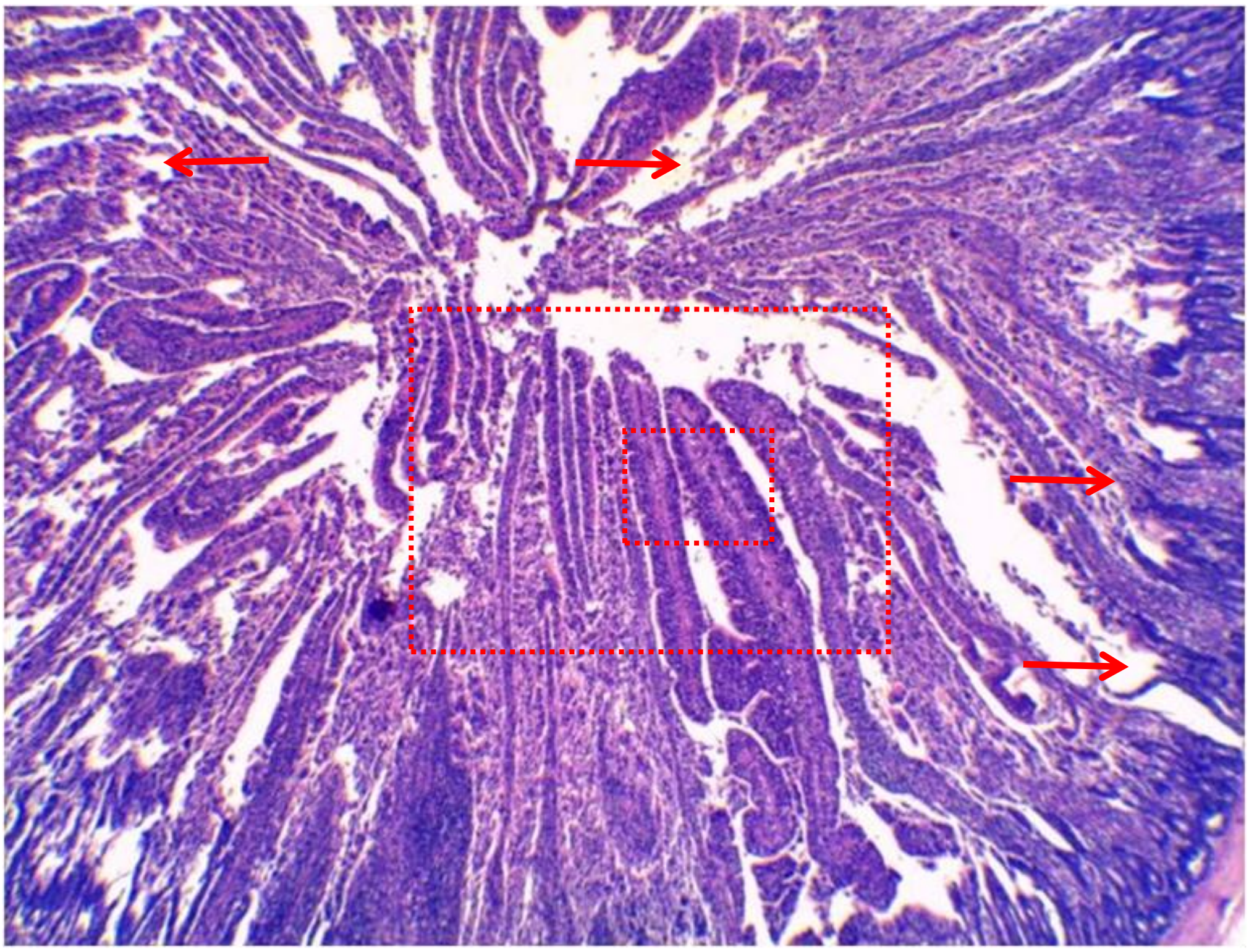

**Imagen 25. Vellosidades intestinales del duodeno de un pollo positivo a *E. acervulina* que presentó lesiones con score 3 en la evaluación macroscópica.** Las coccidias son abundantes y están presentes en todas las vellosidades y en la luz intestinal. Microscopía a 100 X con coloración HE.

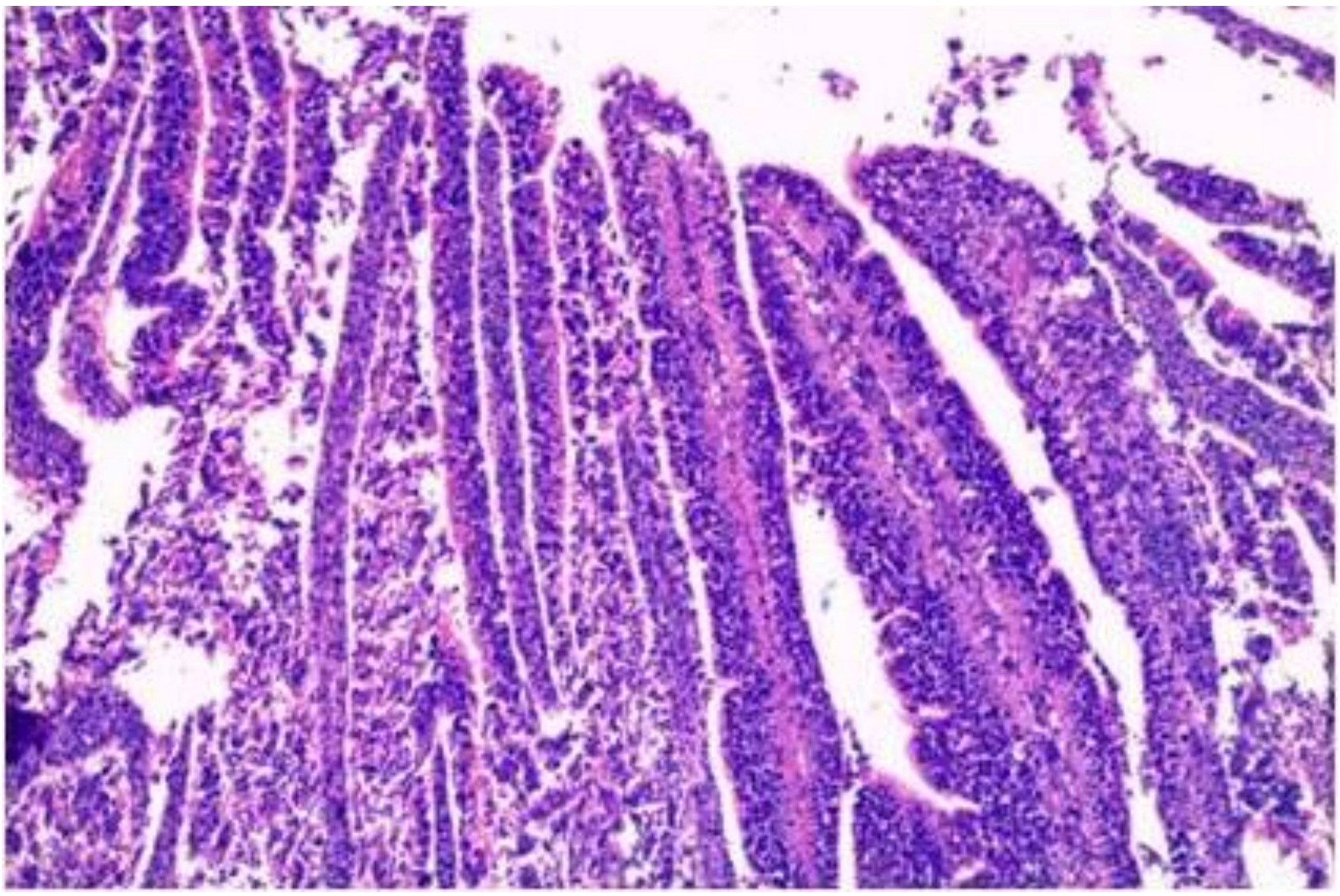

**Imagen 26. Distribución de las coccidias en la vellosidad intestinal del duodeno de un pollo positivo a *E. acervulina* que presentó lesiones con score 3 en la evaluación macroscópica.** Se observa sólo formas inmaduras de coccidia debido a la replicación del ciclo asexual; sin embargo, el grado de invasión es mayor que en el score 2. Se observa escasa reacción inflamatoria no supurativa en la lámina propia. Microscopía a 400 X con coloración HE.

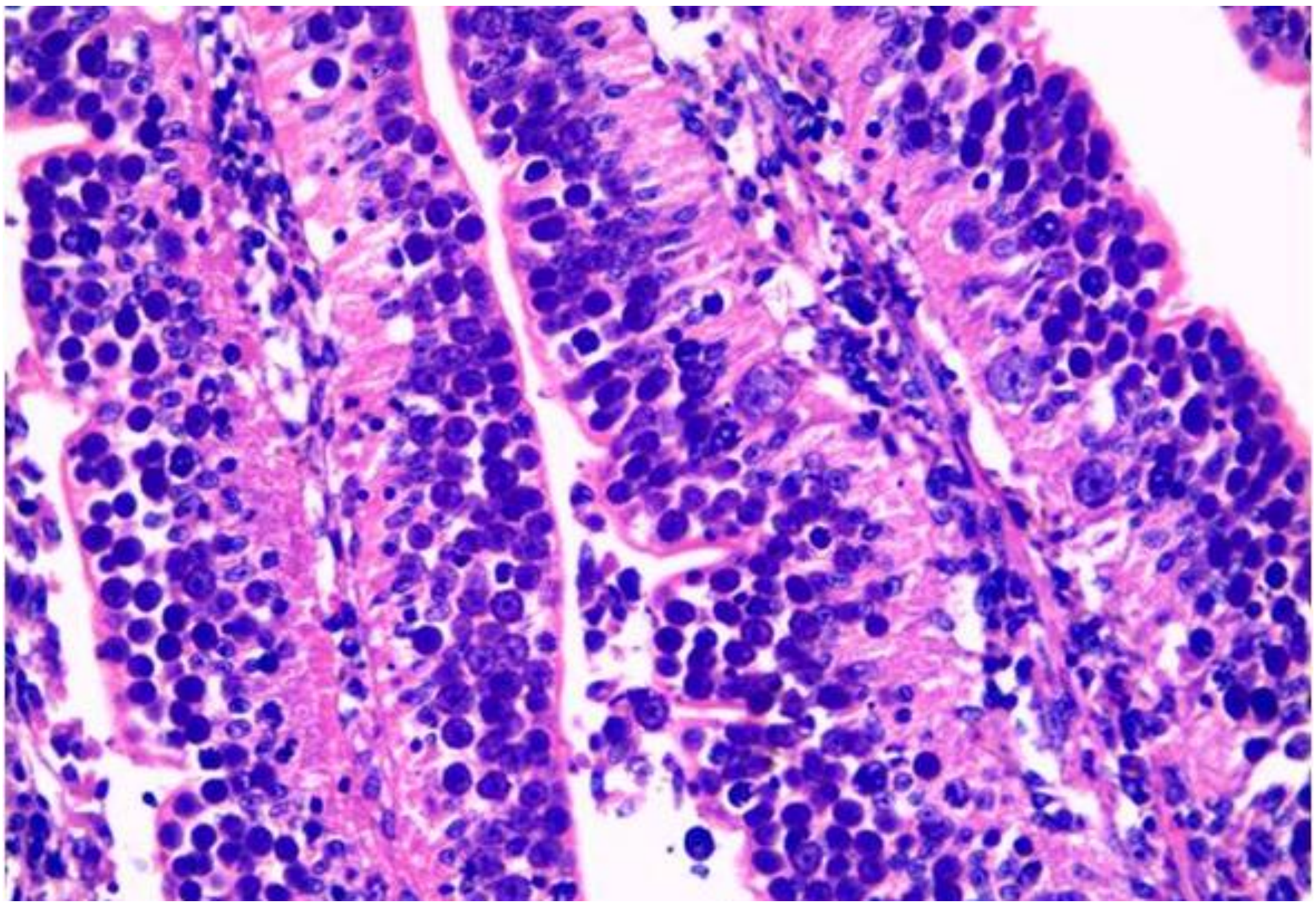

presentaron lesiones duodenales compatibles con *E. acervulina*. La frecuencia de pollos con lesiones intestinales, pero sin formas de coccidia fue del 83% en las aves suplementadas con el PRO, lo que guarda relación con el proceso de cicatrización que ocurre en los sitios de las lesiones tras la eliminación de las coccidias (Quiroz, 2005).

Estos resultados indican que si bien la evaluación de la coccidiosis mediante el score de lesiones de Johnson y Reid (1970) tiene una alta utilidad práctica, el diagnostico a partir de este indicador es presuntivo y debe ser confirmado. Al respecto, Conway and McKenzie (2007) sostienen que la identificación de especies específicas de coccidia en cada ave es generalmente difícil y no es recomendable cuando se califican infecciones mixtas.

En el presente estudio, el menor score microscópico de coccidia observado en las aves que reciben el PRO y los resultados del score macroscópico evidencian que la suplementación del PRO reduce la presencia de coccidias asociadas a las lesiones macroscópicas.

La existencia de lesiones intestinales compatibles con coccidia, pero sin presencia de oocistos ha sido documentada previamente. Williams (2003) observó tras un desafío infectivo, que las aves vacunadas tienen escasos oocistos asociados a las lesiones intestinales, mientras que las aves no vacunadas y no protegidas tienen grandes números de parásitos endógenos asociados a las lesiones. Bafundo and Donovan (1988), citados por Conway and McKenzie (2007), observaron que aves con un score de lesiones de 2.7 no presentaron depresión sobre la ganancia de peso, mientras que con un score promedio de 3.2 presentaron sólo un moderado efecto sobre la ganancia de peso.

Si bien el score de lesiones puede no predecir el impacto productivo, otros reportes (Teeter *et al*, 2008) indican que incluso scores de lesiones de 1.0 de acuerdo a la escala de Johnson and Reid (1970) tienen un efecto negativo en el metabolismo animal y su comportamiento productivo, observando un impacto en el apetito del ave, en la ganancia de peso, en el costo energético de mantenimiento y en la retención de energía.

En el presente experimento se observó scores de coccidia 27% y 67% menores para *E. acervulina* y *E. maxima*, respectivamente, en las aves suplementadas con el PRO; sin embargo, no se observa significancia estadística en estas diferencias (P>0.10), lo que no las descarta, ya que la elevada variabilidad en los resultados de las unidades experimentales en estas variables (Anexo 54) redujo el poder estadístico de la prueba; es decir, la probabilidad de que el efecto del tratamiento sea detectado si dicho efecto existe (Berndtson, 1991; Aaron and Hays, 2004).

Los resultados del presente experimento indican que la administración del PRO controla las coccidias, y que si bien se observa un efecto positivo en el score de lesiones, esta variable presenta una menor sensibilidad al tratamiento. Asimismo, los resultados del score macroscópico de coccidia en las aves suplementadas con el PRO coinciden con lo observado por Conway *et al* (1990) quienes reportan que elevados scores de lesiones de coccidia en aves medicadas se asocian a menores cambios en la ganancia de peso en comparación con las aves no medicadas. Al respecto, los trabajos de Conway *et al* (1999) sobre la relación entre la dosis del inóculo de coccidia y los scores de lesiones demostraron que en aves no medicadas las dosis relativamente bajas de ooquistes pueden producir scores de lesiones de hasta 2.0 y que para producir un mismo score de lesiones en aves medicadas se requiere una dosis mayor. Asimismo, diferentes estudios realizados demuestran que la relación entre la dosis administrada de ooquistes y el score de lesiones es no lineal, observándose un incremento en el score de lesiones en una tasa decreciente conforme se incrementa la dosis de ooquistes (McKenzie *et al*, 1989). Más aun, dosis bajas de inóculo pueden producir un alto score de lesiones (Conway *et al*, 1993).

En base a lo anteriormente expuesto, en el presente experimento se considera los scores macroscópicos de lesiones de manera referencial, y se asigna mayor relevancia a las evaluaciones microscópicas realizadas mediante raspados intestinales, ya que muestra directamente la cantidad de coccidias que replicarán el proceso infectivo. McDougald (2003) coincide con este criterio indicando que el score macroscópico de lesiones es útil para evaluar la severidad de las infecciones pero puede no correlacionarse con la calificación microscópica.

En el presente estudio, la administración del PRO resulta en una reducción de 80% en la proporción de aves positivas a coccidia y se observa menor cantidad de ooquistes en el intestino, tal como se verifica en el score microscópico de coccidia. Al respecto, evaluaciones *in vitro* e *in vivo* han mostrado que los fenoles pueden ser usados específicamente como oocisticida contra *E. tenella* (Williams, 1997; citado por Christaki *et al*, 2004) y que el AEO tiene un alto contenido en fenoles naturales (Ruso *et al*, 1998).

Los resultados del presente estudio respecto al control que ejerce el PRO sobre la coccidia es consistente con reportes previos. Al respecto, Giannenas *et al* (2003) demostraron la acción anticoccidial del AEO administrando 300 mg/kg de AEO en la dieta de pollos experimentalmente infectados con 5 x $10^4$ oocistos esporulados de *E. tenella* en el día 14 de edad, observando una reducción significativa en el score de lesiones intestinales, pero con un efecto en magnitud menor a 75 mg/kg de lasalocido. Saini *et al* (2003) desafiaron pollos de carne con $10^5$ oocistos esporulados de *Eimeria sp*. en el día 15 de edad, y encontraron una reducción en el score de lesiones intestinales de 4.33 a 3.22 en el día 19 en las aves desafiadas y suplementadas con AEO respecto a aquellas sin AEO. Tsinas *et al* (2011) implementaron un modelo de desafío semejante, pero administraron un inóculo conteniendo $10^5$ oocistos esporulados de *E. acervulina* y *E. maxima*, encontrando una reducción similar a la reportada por Saini *et al* (2003) en el score de lesiones intestinales asociado a una mejora en el comportamiento productivo.

### 3.2.2. Recuento de ooquistes

La técnica del recuento de ooquistes presenta algunos inconvenientes para su interpretación como son la inhibición de la motilidad intestinal por efecto de *E. necatrix*, retrasando la aparición de ooquistes en las heces, y la ausencia de una relación lineal entre las concentraciones de ooquistes en las excretas y en el inóculo (CVM, 1992). Adicionalmente, las especies más prolíficas de coccidia son menos patógenas, y se pueden encontrar recuentos más altos sin un cuadro clínico severo (McDougald, 2003). Finalmente, cuando el recuento de ooquistes se realiza directamente en las heces o en el contenido cecal (Jordan *et al*, 2011), los valores obtenidos pueden ser comparados entre diferentes estudios, lo que no es posible cuando el recuento se realiza en la cama de aves ya que se desconoce la proporción de material de cama y la de heces en la cama de aves que se muestrea. A pesar de estas restricciones, el recuento de ooquistes en muestras en cama de aves es una técnica usualmente empleada en condiciones comerciales para monitorear el desafío de coccidia a que las aves están sometidas y permite establecer comparaciones dentro de la misma empresa ya que otras fuentes de variación, como son el tipo de material de cama, la cantidad de material empleado, las prácticas de manejo de la cama, el tipo de equipos, programas anticoccidiales, entre otras, son controladas.

Los valores encontrados en el recuento de ooquistes en el presente estudio se consideran, en términos generales, bajos respecto a los valores esperados en presencia de un cuadro subclínico de coccidia como el reproducido en el presente estudio. Por el contrario, coinciden con valores de 1133 +/- 373 ooquistes/g de cama encontrados en planteles comerciales en condiciones normales no correlacionadas con cuadros de coccidia (Caiña *et al*, 2001; Salinas *et al*, 2001) y en trabajos experimentales (Condemarín, 2002; Vejarano, 2005; Vejarano *et al*, 2008) en Lima–Perú. Al respecto, en condiciones comerciales, se puede considerar que valores alrededor de los 1,000 ooquistes por gramo de cama de pollos de carne no guardan relación con un desafío relevante a la integridad intestinal ni con cuadros subclínicos de coccidia detectables mediante lesiones intestinales, mientras que valores por encima de 2,500 ooquistes por gramo de cama pueden tomarse como un indicador de cuadros de coccidia que, aunque pueden ser subclínicos, deben ser controlados (L.A. Quevedo, comunicación personal) debido a la correlación positiva existente entre el recuento de ooquistes en cama y las lesiones macroscópicas y microscópicas causadas por coccidia (Salinas *et al*, 2001), así como entre el recuento de ooquistes y la enfermedad subclínica (CVM, 1992).

Se ha reportado que las coccidias vacunales provenientes de vacunas vivas atenuadas presentan una limitada capacidad de replicación (Schnitzler and Shirley, 1999; Allen and Fetterer, 2002; Dalloul and Lillehoj, 2005), requiriendo incluso altos contenidos de humedad en la cama para su esporulación (L.A.Quevedo, comunicación personal). Sin embargo, la vacuna contra coccidia empleada en el presente estudio como material inoculante contuvo cepas vivas no atenuadas; en consecuencia, los bajos recuentos de ooquistes encontrados en la cama de ambos tratamientos no son atribuibles a la capacidad de replicación de las coccidias vacunales. Es posible que dichos recuentos en la cama se deban al manejo aplicado a este material durante la crianza o a los procedimientos de muestro y análisis posteriores.

Se ha observado que al final del periodo experimental los grupos control presentan un recuento de ooquistes en el material de cama que es sólo ligeramente mayor que lo observado en el material de cama reusado en el presente estudio antes del ingreso de las aves (1441 ooquistes/g de cama; Anexos 6 y 54). Es posible que los ooquistes encontrados en la cama de las aves al inicio del periodo experimental no necesariamente fueran viables e infectivos a las aves, ya que como parte del proceso de acondicionamiento del material de cama re-utilizado se realiza el flameado de la cama y otros procedimientos que reducen la viabilidad de la coccidia (Reyna *et al*, 1983). Al respecto, en una evaluación realizada a escala comercial, Caiña *et al* (2001) observaron que no existe una relación directa entre el recuento de ooquistes al inicio de la campaña y el recuento en los días 21 y 42 de edad.

En el presente estudio, al final del periodo experimental, en la cama de las aves suplementadas con el PRO se observa un recuento de ooquistes 17% menor que las aves control ($P<0.01$). Esta observación coincide con lo reportado por Tsinas *et al* (2011) quienes evaluaron el efecto comparativo del AEO y la salinomicina en aves desafiadas experimentalmente con $10^5$ oocistos esporulados de *E. acervulina* e igual cantidad de *E. maxima*, observando en las aves que recibieron el AEO un recuento de ooquistes en las heces 63% menor que en las aves control entre los días 20 a 28 de edad, pero 77% mayor que en las aves suplementadas con salinomicina. Es importante resaltar que cada dosis administrada a las aves en nuestro experimento contuvo una cantidad total de oocistos 12 veces mayor que la dosis administrada por Tsinas *et al* (2011) y además dicha dosis se administró por triplicado, en comparación a una única dosis administrada por los investigadores referidos. Si bien no es posible comparar los recuentos reportados por dichos investigadores con los encontrados en el presente estudio ya que ellos muestrearon heces, mientras que en nuestro estudio se muestreó cama, que está conformada por heces, material de cama, restos de alimento, plumas, etc., los hallazgos del presente estudio concuerdan con lo reportado por los referidos investigadores respecto al efecto ejercido por el AEO sobre la coccidia y que éste puede ser corroborado mediante recuentos de ooquistes.

Silva *et al* (2009) administraron 5 x $10^4$ oocistos esporulados de *E. tenella* por pollo en el día 19 de edad y siete días tras la inoculación observaron un recuento de ooquistes en la cama significativamente menor en las aves suplementadas con 500 o 1000 ppm de AEO respecto a las aves no tratadas, y similar a las aves suplementadas con ionóforos (125 ppm de nicarbazina de 0 a 21 días de edad y 66 ppm de salinomicina desde entonces). Major *et al* (2011) inocularon 25 x $10^3$ oocistos esporulados de *E. acervulina* por pollo y, luego de transcurridos siete días desde la inoculación, observaron recuentos de ooquistes en promedio 60% menores en las heces de las aves desafiadas y suplementadas con AEO respecto a aquellas desafiadas pero no tratadas.

Se ha establecido que el AEO incrementa la apoptosis celular en el ápice de la vellosidad intestinal (Fabian *et al*, 2006) y la proliferación celular en la mucosa intestinal (Levkut *et al*, 2011), lo que incrementa la tasa de recambio de los enterocitos (Bruerton, 2002; citado por Ferket, 2003). Se ha postulado también que este mecanismo puede ser el responsable de la acción anticoccidial del AEO, y que la más rápida eliminación de los enterocitos crea un medio hostil para que la coccidia complete su ciclo, eliminando las células infectadas con esporozoitos son antes que se desarrolle el estado de merozoito, que causa los principales signos clínicos en los brotes de coccidiosis (McDougald, 2003). Este mecanismo permite que una cantidad reducida de oocistos completen su ciclo favoreciendo el desarrollo de la inmunidad contra la coccidia en el ave (Joyner and Norton, 1976; Williams, 2002; McDougald, 2003; Yuño y Gogorza, 2008; Major *et al*, 2011).

Al respecto, Newman and Mathis (2008) observaron los mayores recuentos de ooquistes en granjas limpias y con un descanso sanitario de al menos dos semanas, mientras que observaron los recuentos menores y más tempranos en parvadas con altas densidades de crianza y alojadas sobre cama reusada durante cinco años. Los investigadores concluyeron que la limpieza y la baja densidad de crianza retrasan el desarrollo de la inmunidad en aves alimentadas con dietas conteniendo anticoccidiales, pudiendo desarrollarse altas cargas de coccidiosis subclínica antes del beneficio.

### 3.2.3. Pigmentación de tarsos

El consumidor peruano, como el de muchos otros mercados, es susceptible a las variaciones en el color de los productos avícolas como huevos y carne de aves (Beardsworth and Hernandez, 2004) pues éste se asocia al estado de salud del ave. En aves alimentadas con dietas maíz-soya, la pobre pigmentación de piel y patas está relacionada a una menor absorción de carotenoides a nivel de duodeno y yeyuno, y puede deberse a procesos tóxicos alimentarios, a la coccidiosis aviar o a procesos infecciosos bacterianos como Coriza Infecciosa y Cólera Aviar, o virales como la Enfermedad de Newcastle (Lake, 1977). Si bien los programas con anticoccidiales en el alimento generalmente reducen la probabilidad de ocurrencia de coccidiosis clínicas o sub-clínicas, frecuentemente se presentan en campo casos de coccidiosis clínicas que si bien no causan mortalidad, si afectan las variables productivas, la pigmentación y la uniformidad de las aves (Caiña *et al*, 2001).

Al respecto, las evaluaciones realizadas por Caiña *et al* (2001) indican que no existe una correlación positiva entre la pigmentación de los tarsos y el recuento de ooquistes / g de cama. Esto puede deberse a que si bien el recuento de ooquistes permite tener una impresión general respecto a la coccidiosis, no identifica especies de *Eimeria* ni su viabilidad. Estos aspectos resultan críticos ya que el efecto negativo sobre la absorción intestinal está relacionado principalmente con *E. acervulina* y *E. maxima*, que afectan el duodeno, provocando una menor absorción de carotenoides e induciendo una menor pigmentación, siendo *E. maxima* la especie de coccidia que tiene el mayor efecto negativo sobre la absorción de pigmentos aun cuando no se observen lesiones intestinales graves (McDougald, 2003).

Los resultados obtenidos en el presente estudio no permiten verificar una correlación positiva entre la pigmentación de de los tarsos y el recuento de ooquistes en la cama. Si bien las aves suplementadas con el PRO muestran índices de pigmentación 17% mayores que las aves control, la elevada variabilidad no permite determinar si esta diferencia se debe a la suplementación del PRO (P>0.10). Los valores de pigmentación obtenidos en el presente estudio coinciden con los reportados por Caiña *et al* (2001), quienes observaron una pigmentación típica de 2.30 +/- 0.60 al evaluar planteles comerciales situados en las zonas de Huacho, Huaral y Ventanilla, en el departamento de Lima-Perú. Tanto las aves del presente estudio como las evaluadas por dichos investigadores fueron alimentadas con dietas a base de maíz y soya. Si bien las aves del presente estudio fueron desafiadas por coccidia y por ello se podría haber esperado valores menores de pigmentación, la época en que Caiña *et al* (2001) realizaron el trabajo de campo fue entre marzo y junio, meses en que, por la menor temperatura ambiental, es usual reducir la ventilación en los galpones para conservar la temperatura ambiental dentro del galpón, lo que incrementa la frecuencia de trastornos entéricos incluyendo procesos sub-clínicos por coccidia (Martínez, 2010). Al respecto, Condemarín (2002) realizó una evaluación en corrales en piso con pollos alimentados también con dietas maíz y soya; sin embargo, observaron valores de pigmentación de de 4.80 +/- 0.90, notablemente mayores a los obtenidos en nuestro experimento, que pueden ser explicados por las condiciones óptimas de crianza y la ausencia de factores de desafío intestinal.

Las observaciones de campo, realizadas de manera subjetiva, indican mejor pigmentación en machos que en hembras (Caiña *et al*, 2001). Al respecto, se ha propuesto que de producirse una mayor pigmentación en machos ésta podría deberse al mayor consumo diario de alimento que los machos tienen en relación a las hembras por unidad de peso corporal, lo que implicaría una mayor ingesta de pigmentos (Caiña *et al*, 2001), o a las diferencias en la capacidad de engrase entre hembras y machos (Piraces y Cortez, 1994; citados por Caiña *et al*, 2001). Sin embargo, se ha demostrado que en igualdad de condiciones las hembras presentan una mayor tasa de deposición lipídica (Shahin and Abd El Azeem, 2006) lo que a su vez podría favorecer la mayor pigmentación en hembras. Finalmente, no se ha podido demostrar la influencia del sexo en la pigmentación de patas en trabajos controlados en condiciones de campo (Caiña *et al*, 2001). Por esta razón, a pesar de haber empleado exclusivamente pollos machos en el presente estudio, los resultados encontrados pueden ser considerados como válidos para poblaciones comerciales mixtas.

### 3.3. Morfometría de órganos linfoides

La relación bursa/bazo mayor a 2 observada en ambos tratamientos refleja la adecuada inmunocompetencia de las aves al final del periodo experimental, tal como indican Pulido *et al* (2001), quienes reportaron que valores superiores a 2 en este índice pueden ser considerados como propios de una adecuada inmunocompetencia. Sin embargo, se encontraron bursas de un tamaño significativamente mayor (P<0.05) en las aves del grupo control en comparación al tratamiento con el PRO (6.0 y 5.3 cm, respectivamente), pero no se observaron diferencias significativas en otras variables morfométricas (P>0.10). Al respecto, es importante considerar que los resultados observados en los indicadores de coccidia evidencian que el PRO ejerce un efecto activo contra la coccidia, reduciendo el desafío sanitario en estas aves. El mayor tamaño de la bursa observado en las aves control guarda relación, entonces, con dicha situación y es consistente con el mayor desafío al que estas aves se encontraron sometidas respecto a las aves suplementadas con el PRO. Ello fue verificado mediante la evaluación presentada en el Anexo 1 (modelo B), en que aves sometidas a un modelo de desafío similar al empleado en el presente estudio presentaron un mayor tamaño de bursa en los días 21 y 28 de edad así como mayores índices morfométricos del bazo y del timo en el día 28. Este comportamiento de los órganos linfoides ante un desafío sanitario también ha sido documentado por otros autores. Tal es el caso de Tambini *et al* (2010) quienes observaron que los pollos alojados en material de cama reutilizado por cinco campañas presentaron un mayor índice morfométrico de la bursa respecto a las aves alojadas sobre material de cama nuevo. Asimismo, Deshmukh *et al* (2007) reportaron que aves sometidas a un desafío entérico con *Salmonella gallinarum* presentaron hiperplasia del bazo.

### 3.4. Comportamiento productivo

En el presente estudio, al término del periodo experimental las aves que recibieron el PRO presentaron un peso corporal 9% mayor (P<0.01), una ganancia de peso 10% mayor (P<0.01) y una menor conversión alimentaria (1.22 vs 1.33; P<0.01). Este menor impacto negativo del desafío de coccidia sobre el comportamiento productivo de las aves que reciben el PRO coincide con lo observado por Conway *et al* (1990) quienes reportan que, independientemente del score de lesiones de coccidia, las aves medicadas presentan mayor ganancia de peso en comparación con las aves no medicadas. Long *et al* (1980) observaron un efecto similar en aves parcialmente inmunes a *E. tenella*, las cuales desarrollaron lesiones cecales severas y similares a las no inmunizadas, pero sin una reducción comparable en la ganancia de peso o hematocrito. Por su parte, Conway *et al* (1999) reportan que las mejoras en ganancia de peso observadas en las aves medicadas en relación a aquellas sin medicar se deben a menores niveles de infección que el score de lesiones puede no lograr medir con la misma precisión, y consideran que se debe poner particular énfasis en el comportamiento productivo del ave, indicando que la ganancia de peso es la variable más sensible y con mayor capacidad para informar sobre la eficacia anticoccidial de una sustancia. Existe un consenso general respecto al comportamiento productivo como principal indicador de respuesta de una estrategia anticoccidial (CVM, 1992; McDougald, 2003; Williams, 2003; Conway and McKenzie, 2007).

El efecto del PRO sobre la ganancia de peso de pollos desafiados con coccidia observado en el presente estudio es consistente con reportes previos sobre la suplementación del PRO en aves desafiadas por este agente. Al respecto, Jamroz and Kamel (2002) observaron mayor peso corporal en aves de 48 días de edad suplementadas con una combinación de carvacrol, cinamaldehído y capsaicina, mientras que Lillehoj *et al* (2011) desafiaron pollos de carne con 2 x $10^4$ oocistos esporulados de *E. acervulina* en el día 14 de edad y administraron una combinación de carvacrol, cinamaldehído y capsaicina, observando un incremento significativo en la ganancia de peso en el día 24 de edad de 650 g a 720 g respecto a las aves control.

En dos estudios realizados para evaluar el efecto del 300 ppm de AEO y 2.5, 5.0, 7.5 y 10 g/kg de orégano molido y 75 ppm de lasalócido en pollos desafiados con 5 x $10^4$ oocistos esporulados de *E. tenella* en el día 14 de edad (Giannenas *et al*, 2003; 2004) se observó que el peso corporal en las aves suplementadas con AEO fue similar al de las aves control sin desafío, mayor al de las aves control con desafío, y menor al de las aves suplementadas con lasalócido. Asimismo, se observó que la suplementación de 5.0 y 7.5 g/kg de orégano molido y lasalócido produjeron mayor ganancia de peso que 2.5 y 10.0 g/kg de orégano molido, pero que todos los niveles de orégano molido determinan pesos superiores en comparación a las aves control con desafío.

De forma similar, Batungbacal *et al* (2007) compararon la eficacia anticoccidial del extracto de orégano respecto al amprolio. Para ello administraron un extracto de orégano o amprolio vía agua de bebida a pollos desafiados artificialmente con coccidia, observando que si bien el tratamiento con amprolio o extracto de orégano redujo el score de lesiones y la eliminación de ooquistes en las heces respecto a las aves desafiadas y sin tratamiento, sólo las aves tratadas con el extracto de orégano presentaron una ganancia de peso similar al grupo control no desafiado, e incluso mayor que el de las aves desafiadas sin tratamiento o tratadas con amprolio.

En el presente estudio, el mayor peso final y ganancia de peso observados en las aves suplementadas con el PRO se explica por la mayor eficiencia de estas aves en la utilización del alimento, debido a que no se observan diferencias en el consumo de alimento, tal como fue reportado en otros experimentos con el uso del PRO (Experimentos 2, 5 y 6). Si bien se ha reportado que aves desafiadas con coccidia presentan una reducción en el consumo de alimento siete días después de la administración del inóculo (Christaki *et al*, 2004), los resultados del presente estudio no permiten determinar si la similitud en el consumo de alimento de las aves de ambos tratamientos se debe a la reversión del efecto reductor de la coccidiosis sobre el consumo de alimento, o si este efecto realmente se ha producido.

Se ha establecido que la reducción en la ganancia de peso en las aves afectadas por coccidiosis es proporcional a la dosis infectiva de coccidia y que *E. acervulina* se caracteriza por el incremento en la conversión alimentaria (McDougald, 2003). Por esta razón, la mayor ganancia de peso y menor conversión alimentaria en las aves suplementadas con el PRO es indicativo del control ejercido sobre la coccidia. Esto se corrobora con los resultados de Giannenas *et al* (2003) quienes observaron no sólo una mayor ganancia de peso, sino también una mejor conversión alimentaria en las aves desafiadas por coccidia y suplementadas con AEO.

Es importante mencionar que, de forma complementaria al control que ejerce el PRO sobre la coccidiosis, el estímulo que produce sobre la proliferación celular en la mucosa intestinal (ver Experimento 1; Levkut *et al*, 2011) promueve la restitución de las vellosidades afectadas por el proceso en menor tiempo. Esto tiene gran relevancia debido a que, si bien el recambio celular de los enterocitos en condiciones fisiológicas normales se produce en un periodo de 72 horas en pollos de 4 días de edad, en aves mayores el tiempo que toma el recambio celular se extiende a 96 horas (Uni *et al*, 1998; Geyra *et al*, 2001; Maiorka y Rocha, 2009), representando 9.5 del periodo de crianza, ya que éste es de 42 días; es decir, que durante el 10% del periodo de crianza las aves verán afectadas su capacidad de absorción de nutrientes.

## 4. Conclusiones

Los resultados obtenidos bajo las condiciones del presente estudio permiten llegar a las siguientes conclusiones:

- El PRO ejerce un control significativo sobre la coccidiosis, reduciendo el porcentaje de aves positivas a coccidia y la producción de ooquistes.

- El PRO atenúa el impacto negativo de la coccidiosis sobre el comportamiento productivo, incrementando la ganancia de peso y reduciendo la conversión alimentaria.

# EXPERIMENTO N° 8

# EFECTO DE UN PRODUCTO A BASE DE ACEITE ESENCIAL DE ORÉGANO SOBRE LA DIGESTIBILIDAD Y EFICIENCIA NUTRICIONAL EN POLLOS DE CARNE

## 1. Introducción

Debido al constante incremento en los costos de producción, la industria avícola busca mejorar cada vez más la eficiencia de utilización del alimento de las aves para la producción de carne destinada al consumo humano. Los factores que limitan la óptima utilización de los nutrientes dietarios por las aves son diversos e incluyen la calidad del alimento, las buenas prácticas de alimentación, la salud intestinal de las aves y las condiciones de crianza que incrementan el gasto nutricional para soportar funciones homeostáticas, inmunológicas, de restitución de tejidos, entre otras.

La industria dispone de diferentes estrategias para reducir el efecto negativo de estos factores sobre el comportamiento productivo de las aves. Uno de ellos es el uso de antimicrobianos promotores de crecimiento, que si bien tienen un efecto importante sobre la eficiencia productiva de las aves, los mercados internacionales imponen restricciones al uso de ciertas sustancias por producir condiciones potencialmente perjudiciales para el consumidor.

Una de las tecnologías más estudiadas en los últimos años para sustituir los antimicrobianos promotores de crecimiento son los derivados fitogénicos como extractos, aceites esenciales, y sus metabolitos secundarios. Entre ellos destaca el AEO que ha mostrado efectos significativos en el comportamiento productivo de las aves.

El objetivo de este experimento fue determinar el efecto de un producto a base de AEO (referido en adelante como PRO) sobre la digestibilidad del alimento y la utilización de los nutrientes en pollos de carne.

# 2. Materiales y métodos

## 2.1. Lugar, fecha y duración

La crianza de las aves se llevó a cabo en las instalaciones del Programa de Aves de la Facultad de Zootecnia de la Universidad Nacional Agraria La Molina (UNALM) en Lima-Perú a mediados del segundo semestre del 2010 y los análisis de químicos se llevaron a cabo en el Laboratorio de Evaluación Nutricional de Alimentos de la UNALM en Lima-Perú. La evaluación se llevó a cabo desde la recepción de las aves en las instalaciones de crianza y el periodo de evaluación fue el comprendido de 0 a 21 días de edad.

## 2.2. Instalaciones, equipos y materiales

### 2.2.1. Instalaciones de crianza

Las aves estuvieron alojadas sobre material de cama reutilizado (Anexo 6) a razón de 21.4 pollos/$m^2$ (0.047 $m^2$/pollo). Durante los primeros 3 días de vida se colocó papel periódico para evitar el acceso directo de los pollos BB al material de cama, mientras que el alimento fue suministrado en bandejas plásticas y el agua de bebida en bebederos tipo tongo. A partir del cuarto día de edad el alimento y agua de bebida fue provisto en comederos y bebederos lineales, respectivamente. Dos semanas antes del ingreso de las aves se inició un programa de control de vectores, empleando rodenticidas y mosquicidas comerciales. Antes y después del periodo experimental se realizó la limpieza y desinfección de las instalaciones.

### 2.2.2. Calefacción, control de la temperatura y ventilación

La calefacción fue provista por un sistema eléctrico con resistencias y termostatos, y fue controlada de acuerdo a las recomendaciones de la línea genética (Cobb-Vantress, 2008b). Para reducir la variabilidad en la temperatura ambiental a consecuencia de las diferentes alturas entre los pisos de las baterías se adaptó resistencias adicionales en la base de cada batería. La ventilación fue controlada con cortinas de polipropileno instaladas en el perímetro del área de crianza. La humedad ambiental en el área de crianza se mantuvo alrededor de 40%. Para la medición de la temperatura y humedad ambiental se empleó un termo-higrómetro electrónico digital con una aproximación de 0.1 °C para temperatura y 1% para humedad.

### 2.2.3. Preparación del alimento

Para el pesaje de los ingredientes mayores del alimento se utilizó una balanza digital con capacidad de 150 kg y aproximación de 0.02 kg, mientras que para el pesaje de los ingredientes de la premezcla se utilizó una balanza electrónica con capacidad de 6 kg y aproximación de 1 g.

Se utilizó mezcladoras horizontales de cintas de 400 y 30 kg de capacidad para la mezcla de los ingredientes durante la preparación del alimento y para las premezclas, respectivamente. Después de la preparación de cada dieta se realizó el flushing y limpieza de los equipos (FAO, 2004; Ratcliff, 2009). El alimento fue envasado en sacos de papel y polietileno laminado, y almacenado en cilindros metálicos.

### 2.2.4. Manejo y actividades especiales

Para la medición de los pesos vivos se utilizó una balanza electrónica con capacidad para 200 g y con aproximación de 10 mg y otra con capacidad para 15 kg y con aproximación de 1 g. Para la medición del alimento consumido, se utilizó una balanza electrónica con capacidad para 15 kg y con aproximación de 1 g.

En el día 10 de edad las aves fueron vacunadas, por vía ocular, contra la Enfermedad de Newcastle, empleando una vacuna viva liofilizada con una cepa VG/GA (Avinew®, Merial Limited). Para desafiar a las aves se empleó un inóculo de coccidia preparado a partir de una vacuna comercial contra coccidia (Immucox® for Chickens II, Vetech Laboratories; ver Anexo 10) para inducir condiciones de desafío.

El agua de bebida fue potabilizada con 1 ml de hipoclorito de sodio al 4.5% por cada 10 L de agua. Para evitar que la presencia de cloro interfiera con la viabilidad del inóculo, la clorinación del agua se realizó tres días antes de proveerla a las aves, y diariamente, al momento de suministrarla a las aves, se verificó la ausencia de cloro empleando un kit comercial de evaluación. Para lo toma de muestras de heces se empleó espátulas metálicas, bolsas de polietileno y plumón indeleble.

### 2.3. Animales experimentales

Se empleó 128 pollos machos de la línea Cobb 500 (Anexo 4), distribuidos al azar en 16 unidades experimentales.

### 2.4. Tratamientos

Se evaluaron 2 tratamientos, que fueron definidos de la siguiente manera:

Tratamiento 1: Dieta basal (grupo control)
Tratamiento 2: Dieta basal + 500 ppm de Orevitol®

Las aves de ambos tratamientos fueron desafiadas con coccidia. El producto indicado será referido, en adelante, como PRO.

### 2.5. Modelo de desafío

En el día 14 de edad, por vía oral y empleando cánulas oro-ingluviales (Christaki *et al*, 2004; Conway and McKenzie, 2007; Elmusharaf *et al*, 2010; Georgieva *et al*, 2011b), se administró una dosis de un inóculo de coccidia preparado para contener oocistos vivos no atenuadas de aislamientos de campo (Conway and McKenzie, 2007) de *Eimeria acervulina, E. maxima, E. tenella, E. necatrix* y *E. brunetti* con una concentración total no menor a 24 x $10^5$ oocistos vivos por dosis, equivalente a 10 dosis de una vacuna comercial (Immucox® for Chickens II, Vetech Laboratories).

### 2.6. Alimentación

En el periodo de 1 a 14 días de edad se suministró una dieta basal a base de maíz, soya y harina de pescado, y a partir del día 15 de edad se suministró una dieta basal a base de maíz y soya. Ambas dietas fueron complementadas con aceite de pescado y aminoácidos sintéticos, y suplementadas con una premezcla de vitaminas y minerales. La alimentación fue *ad libitum*. Las características de las dietas empleadas se presentan en el Cuadro 20. El aditivo indicado en los tratamientos dietarios fue incorporado en la dieta de 1 a 21 días de edad a expensas del maíz.

## 2.7. Mediciones

### 2.7.1. Digestibilidad, balance nitrogenado y eficiencia de nutrientes

En los días 19, 20 y 21 de edad se colectaron excretas limpias y frescas de la cama de cada unidad experimental. Se determinó en estas muestras de excretas y en la dieta basal los contenidos de nitrógeno, proteína cruda y ceniza insoluble en ácido. El contenido total de ceniza insoluble en ácido fue empleado como marcador (Furuichi and Takahashi, 1981; Lemme *et al*, 2004; Sales and Janssens, 2006); para ello se incorporó en la dieta 0.3% de un marcador inerte como fuente de ceniza insoluble. Con esta información y el comportamiento productivo de las aves durante la tercera semana de vida (Anexo 56) se determinó las variables correspondientes al periodo de 15 a 21 días de edad que se presentan a continuación.

- Digestibilidad aparente:

  En diversos estudios en aves se ha empleado el análisis de excretas para determinar la digestibilidad de la proteína u otros nutrientes ("Digestibilidad Aparente en el Tracto Total" o ATTD, por su sigla en inglés) (Rotter *et al*, 1990; Langhout, 1998; Ravindran *et al*, 1999; Danicke *et al*, 2000; Hernández *et al*, 2004; Dotas *et al*, 2010; Gourdouvelis *et al*, 2010; Schøyen *et al*, 2007; Onimisi *et al*, 2008; Chiang *et al*, 2010; Lee *et al*, Roldán, 2010; 2010; Botha, 2011; Baek *et al*, 2012). Si bien la ATTD en aves no mide la digestibilidad de la proteína propiamente dicha de acuerdo a la definición clásica, ya que las heces y la orina son eliminadas juntas, sino su metabolicidad (Ravindran *et al*, 1999; Lemme *et al*, 2004; Sakomura y Rostagno, 2007; Makiyama *et al*, 2012), este último no es un término comúnmente empleado (Sakomura y Rostagno, 2007).

**Cuadro 20. Características de las dietas empleadas en el Experimento 8.**

| Ingredientes | %, De 1 a 14 días | %, De 15 a 21 días |
|---|---|---|
| Maíz amarillo | 52.517 | 61.775 |
| Torta de soya | 26.699 | 31.405 |
| Harina de pescado | 14.940 | 0.000 |
| Aceite semirefinado de pescado | 2.024 | 2.380 |
| DL-Metionina | 0.190 | 0.224 |
| L-Lisina | 0.116 | 0.135 |
| Cloruro de colina | 0.085 | 0.100 |
| Fosfato dicálcico | 1.609 | 1.892 |
| Carbonato de calcio | 0.978 | 1.150 |
| Sal común | 0.361 | 0.424 |
| Marcador inerte [1] | 0.300 | 0.300 |
| Premezcla [2] | 0.085 | 0.100 |
| Antifúngico [2] | 0.085 | 0.100 |
| Antioxidante [2] | 0.013 | 0.015 |

| Nutriente | Aporte nutricional, De 1 a 14 días | Aporte nutricional, De 15 a 21 días |
|---|---|---|
| EM [3], Kcal/kg | 3028 | 3008 |
| Proteína cruda, % | 26.72 | 20.04 |
| Lisina, % | 1.72 | 1.16 |
| Metionina + Cistina, % | 1.10 | 0.87 |
| Treonina, % | 1.05 | 0.76 |
| Triptófano, % | 0.29 | 0.23 |
| Calcio, % | 1.54 | 1.17 |
| Fósforo disponible, % | 0.67 | 0.44 |
| Sodio, % | 0.34 | 0.19 |
| Grasa total, % | 5.97 | 5.11 |
| Fibra cruda, % | 2.70 | 3.17 |

| Componente | Composición proximal [4], %, De 1 a 14 días [5] | Composición proximal [4], %, De 15 a 21 días [6] |
|---|---|---|
| Humedad | 11.89 | 12.35 |
| Proteína total [3] | 26.06 | 19.66 |
| Extracto Etéreo | 4.87 | 3.83 |
| Fibra Cruda | 2.12 | 2.49 |
| Cenizas | 7.13 | 5.88 |
| ELN [3] | 47.94 | 55.79 |

1 Óxido crómico como fuente de ceniza insoluble en ácido.

2 Premezcla de vitaminas y minerales Proapak 2A®. Composición: Retinol: 12’000,000 UI; Colecalciferol: 2’500,000 UI; DL α-Tocoferol Acetato: 30,000 UI; Riboflavina: 5.5 g; Piridoxina: 3 g; Cianocobalamina: 0.015 g; Menadiona: 3 g; Ácido Fólico: 1 g; Niacina: 30 g; Ácido Pantoténico: 11 g; Biotina: 0.15 g; Zn: 45 g; Fe: 80 g; Mn: 65 g; Cu: 8 g; I: 1 g; Se: 0.15 g; Excipientes c.s.p. 1,000 g. Antifúngico: Mold Zap®; Antioxidante: Danox®

3 EM: Energía metabolizable; Proteína total: N x 6.25; ELN: Extracto Libre de Nitrógeno (calculado).

4 Informes de ensayo 1235/2010 LENA y 1236/2010 LENA, Universidad Nacional Agraria La Molina.

5 Calculado a partir de los análisis proximales de la dieta empleada de 15 a 28 días de edad (85%) y de la harina de pescado (15%) empleadas en su producción.

6 Informe de ensayo 1235/2010 LENA, Universidad Nacional Agraria La Molina.

Si bien el efecto de las pérdidas endógenas, fermentación cecal y excreciones urinarias se puede corregir muestreando el contenido ilíaco (“Digestibilidad Aparente Ilíaca” o AID, por su sigla en inglés) (Yap *et al*, 1997), la cantidad de muestra disponible a nivel del íleon es reducida y, en consecuencia, menos representativa que las muestras obtenidas de excretas, conduciendo a una mayor variabilidad en los resultados analíticos y, por lo tanto menor poder estadístico aún cuando las diferencias entre tratamientos observadas en las heces sean menores que las encontradas en el contenido ilíaco (Schøyen *et al*, 2007). Por esta razón, se considera que, además de laboriosa, la obtención de una sola muestra ilíaca válida demanda utilizar varias aves, y se recomienda emplearla solo en aves maduras (Lemme *et al*, 2004).

Por otro lado, se ha reportado que la diferencia entre los valores obtenidos en aves mediante ATTD y AID es mínima y no significativa (Schøyen *et al*, 2007; Onimisi *et al*, 2008), que es leve cuando se trata de oleaginosas e incluso ínfima cuando se trata de cereales (Williams, 1995), y que si bien se obtienen valores diferentes cuando se trata de trigo y algunas harinas proteicas de origen animal, los valores obtenidos son similares cuando se trata de sorgo, maíz, soya y harina de pescado (Ravindran *et al*, 1999).

En consecuencia, para realizar una comparación entre ambos tratamientos dietarios, resulta conveniente emplear la Digestibilidad Aparente en el Tracto Total (ATTD) como un indicador válido de la digestibilidad aparente de la proteína y materia seca (Dotas *et al*, 2010).

En el presente estudio se calculó la digestibilidad aparente de la proteína y de la materia seca empleando la fórmula presentada a continuación (Chiang *et al*, 2010), donde ATTD: digestibilidad aparente del nutriente en el tracto total (%), CMA: contenido del marcador en el alimento (%), CMH: contenido del marcador en las heces (%), CNH: contenido del nutriente en las heces (%), CNA: contenido del nutriente en el alimento (%). CMA, CMH, CNH y CNA son expresados en base 100% seca.

$$\text{ATTD} = 100 \times \left(1 - \frac{\text{CMA}}{\text{CMH}} \times \frac{\text{CNH}}{\text{CNA}}\right)$$

- Ingesta de nitrógeno:

  Se calculó empleando las fórmulas presentadas a continuación, donde IN: ingesta de nitrógeno (g/ave/día), IMS: ingesta de material seca (g/día), CPD: contenido de proteína en la dieta expresado en base seca (%), CA: consumo de alimento (g/día), CMS: contenido de materia seca en el alimento (%).

$$\text{IN} = \frac{\text{IMS} \times \text{CPD}}{625} \qquad \text{IMS} = \frac{\text{CA} \times \text{CMS}}{100}$$

- Excreción de nitrógeno:

  Se consideró como tal la cantidad de nitrógeno total excretada en las heces y orina (Kerr and Kidd, 1999), y no se determinó la excreción de acido úrico en las excretas (Lopez and Leeson, 2005). La excreción de nitrógeno se calculó empleando las fórmulas presentadas a continuación, donde EN: excreción de nitrógeno (g/ave/día), CA: consumo de alimento (g/día), CMS: contenido de materia seca en el alimento (%), IMS: ingesta de material seca (g/día), CMA: contenido del marcador en el alimento (%), CPH: contenido de proteína en las heces (%), CMH: contenido del marcador en las heces (%). CMA, CPH y CMH son expresados en base 100% seca.

$$\text{IMS} = \frac{\text{CA} \times \text{CMS}}{100} \qquad \text{EN} = \frac{\text{IMS} \times \text{CMA} \times \text{CPH}}{\text{CMH} \times 625}$$

- Retención aparente de nitrógeno:

  Se calculó empleando la fórmula presentada a continuación, donde RAN: retención aparente de nitrógeno (g/ave/día), IN: ingesta de nitrógeno (g/ave/día), EN: excreción de nitrógeno (g/ave/día).

$$\text{RAN} = \text{IN} - \text{EN}$$

- Balance nitrogenado:

  Refleja la cantidad de nitrógeno retenido por unidad de peso corporal. Se calculó de dos formas, para ser expresado por unidad de peso absoluto (kg) (Bonato, 2010) o metabolizable ($kg^{0.75}$) (Rabello *et al*, 2002), empleando las fórmulas presentadas a continuación, donde BN: balance nitrogenado, IN: ingesta de nitrógeno (g/ave/día), EN: excreción de nitrógeno (g/ave/día), PCm: peso corporal medio (kg) equivalente al promedio de los pesos corporales al inicio y fin del periodo de evaluación, PCMm: peso corporal metabolizable medio (kg) equivalente al promedio de los pesos metabolizables ($kg^{0.75}$) al inicio y fin del periodo de evaluación.

$$\text{BN (g/kg/día)} = \frac{\text{IN} - \text{EN}}{\text{PCm}}$$

$$\text{BN (g/kg}^{0.75}\text{/día)} = \frac{\text{IN} - \text{EN}}{\text{PCMm}}$$

- Relación de Eficiencia Energética:

  Se calculó empleando la fórmula presentada a continuación (Hosseini-Vashan *et al*, 2010), donde EER: relación de eficiencia energética (%), GP3: ganancia de peso durante la tercera semana (g/día), CA3: consumo de alimento durante la tercera semana (kg/día), CE: contenido de energía metabolizable en la dieta (kcal/kg).

$$\text{EER} = \frac{\text{GP3} \times 100}{\text{CA3} \times \text{CE}}$$

- Relación de Eficiencia Proteica:

  Se calculó empleando la fórmula presentada a continuación (Hosseini-Vashan *et al*, 2010), donde PER: relación de eficiencia proteica (g/g), GP3: ganancia de peso durante la tercera semana (g/día), CA3: consumo de alimento durante la tercera semana (g/día), CPD: contenido de proteína en la dieta (%).

$$\text{PER} = \frac{\text{GP3} \times 100}{\text{CA3} \times \text{CPD}}$$

### 2.7.2. Comportamiento productivo

Se midió empleando las siguientes variables:

- Peso vivo: Se registró los pesos individuales al recibir los pollos BB y al final del periodo experimental (día 21).
- Ganancia de peso: Se calculó la ganancia promedio por vía (a/ave/día), dividiendo la ganancia total de peso por ave entre el número de días de evaluación.
- Consumo de alimento: Al término de cada semana se pesó los residuos de alimento y se calculó el consumo diario promedio por ave durante las tres primeras semanas de vida.
- Conversión alimentaria: Se calculó con los valores acumulados de consumo de alimento y ganancia de peso al día 21 de edad.

## 2.8. Diseño estadístico

Se empleó el Diseño Completo al Azar con dos tratamientos y ocho repeticiones. El análisis de varianza se llevó a cabo aplicando el procedimiento paramétrico ANOVA del programa Statistical Analysis System SAS 9.0 (SAS Institute, 2009) y la diferencia de medias se realizó usando la prueba de Duncan (1955). El Modelo Aditivo Lineal General aplicado a las variables evaluadas fue el siguiente:

$$Yij = U + Ti + Eij$$

Donde:

Yij = variable respuesta

U = media general

Ti = i-ésimo tratamiento ( i = 1, 2 )

Eij = Error experimental

Se consideró significativos aquellos valores con P menores de 0.10 (Crespo and Esteve-Garcia, 2001; Zhu *et al*, 2003, Kilburn and Edwards, 2004; van Nevel *et al*, 2005; Tahir *et al*, 2008; Teeter *et al*, 2008) o menores de 0.05, según se indica en cada caso.

## 3. Resultados y discusión

En el presente estudio, para asemejar las condiciones de campo, se desafió a las aves con un inóculo de coccidia induciendo un cuadro subclínico con lesiones intestinales compatibles con *E. acervulina* (score 1.4) y E. maxima (score 0.4), reducida mortalidad (alrededor del 3%) y un adecuado nivel de actividad de las aves, no observándose aves clínicamente afectadas. Las características del cuadro de coccidia inducido están ilustradas en el Experimento 7 y los resultados del presente experimento en el Cuadro 21 y Anexos 55 y 56.

### 3.1. Digestibilidad de nutrientes

En las aves suplementadas con el PRO se observa una digestibilidad aparente de la materia seca y proteína cruda en el tracto total significativamente mayor ($P<0.01$) que las aves control en 7.7 y 9.9 puntos porcentuales, respectivamente. Al respecto, es importante mencionar que el efecto de los aceites esenciales sobre la promoción del crecimiento de las aves ha sido atribuido en parte a un incremento en la digestibilidad de nutrientes (Basmacioglu Malayoglu *et al*, 2010; Hernández *et al*, 2004). Así, los incrementos observados en la digestibilidad aparente de la materia seca y la proteína por efecto del PRO en el presente estudio son consistentes con reportes previos en pollos de carne suplementados con aceites esenciales, incluyendo el AEO, extractos de plantas y metabolitos secundarios de éstas.

Hernández *et al* (2004) observaron una mayor digestibilidad iliaca de la materia seca y del almidón en las aves que recibieron un suplemento a base de extractos de orégano, canela y pimienta. García *et al* (2007) suplementaron la dieta de pollos de carne con avilamicina o con una combinación de aceites esenciales de orégano, canela y pimienta con un alto contenido de cinamaldehído, carvacrol y capsaicina, y observaron en estas aves un incremento en la digestibilidad de la proteína cruda de 61% a 71% respecto a las aves control sin suplementación; incremento que fue estadísticamente similar al observado en las aves suplementadas con avilamicina. En ambos tratamientos el incremento en la digestibilidad estuvo asociado en un incremento en la conversión alimentaria pero no se observó diferencias en la longitud de las vellosidades intestinales, por lo que los investigadores atribuyen las

**Cuadro 21. Efecto de un producto a base de aceite esencial de orégano sobre la utilización de nutrientes y el comportamiento productivo.**

| Variable | Tratamientos [(1)] | | | | P |
|---|---|---|---|---|---|
| | 1 | | 2 | | |
| **Digestibilidad, balance nitrogenado y eficiencia de nutrientes** | | | | | |
| Digestibilidad aparente de la materia seca [(2)], % | 69.77 | b | 77.51 | a | 0.0097 |
| Digestibilidad aparente de la proteína [(2)], % | 63.96 | b | 73.83 | a | 0.0083 |
| Ingesta de nitrógeno, g/ave/día | 2.582 | a | 2.464 | b | 0.0165 |
| Excreción de nitrógeno, g/ave/día | 0.931 | a | 0.647 | b | 0.0048 |
| Retención aparente de nitrógeno, g/ave/día | 1.651 | £ | 1.826 | $ | 0.0642 |
| Balance nitrogenado, g/kg$^{0.75}$/día | 2.323 | | 2.389 | | 0.5767 |
| Balance nitrogenado, g/kg/día | 2.569 | | 2.583 | | 0.9191 |
| Relación de eficiencia energética, % | 23.71 | b | 26.11 | a | 0.0024 |
| Relación de eficiencia proteica, g/g | 3.64 | b | 4.01 | a | 0.0024 |
| **Comportamiento productivo** | | | | | |
| Peso del pollo BB, g | 47.94 | | 48.40 | | 0.4344 |
| Ganancia de peso, g/día | 38.10 | b | 41.78 | a | 0.0025 |
| Consumo de alimento, g/día | 53.82 | | 53.67 | | 0.8920 |
| Conversión alimentaria | 1.41 | a | 1.29 | b | 0.0012 |

(1) Tratamientos: 1: aves desafiadas control; 2: aves desafiadas + 500 ppm de Orevitol®.

(2) Digestibilidad aparente en el tracto total (ATTD).

a,b,$,£ Promedios significativamente diferentes no comparten la misma letra (a,b; P<0.05) o el mismo símbolo ($,£; 0.05<P<0.10).

diferencias observadas en la digestibilidad a la acción antimicrobiana de ambas sustancias y además a la capacidad de los aceites esenciales de mejorar la digestión.

Otros investigadores han reportado incrementos en la digestibilidad de nutrientes a nivel ilíaco y en el tracto total en pollos (Hernandez *et al*, 2004; Jamroz *et al*, 2005). En aves de 21 días suplementados con una combinación de carvacrol, cinamaldehído y capsaicina observaron un incremento en la digestibilidad de la fibra y la grasa, y una mayor disponibilidad de la ceniza y del nitrógeno dietarios (Jamroz and Kamel, 2002). Amad *et al* (2011) administraron diferentes niveles de un producto conteniendo aceites esenciales de tomillo y anís estrella a pollos de carne de 1 a 42 días de edad y observaron mejoras en la conversión alimentaria e incrementos en la digestibilidad de la ceniza, proteína cruda y grasa total, así como en la disponibilidad de calcio y fósforo, en todos los casos lineales en función a la dosis suplementada.

### 3.2. Metabolismo nitrogenado y disponibilidad de la energía

En el presente estudio, se observa que la ingesta y excreción de nitrógeno son mayores en las aves control ($P < 0.05$ y $< 0.01$, respectivamente), mientras que la retención aparente de nitrógeno es mayor en las aves suplementadas con el PRO ($P < 0.10$). Por otro lado, las aves suplementadas con el PRO presentaron relaciones de eficiencia proteica y energética significativamente mayores ($P<0.01$) en las aves suplementadas con el PRO. Estas mejoras observadas en el metabolismo nitrogenado y en la eficiencia de la proteína y energía dietarias por efecto del PRO son consistentes con reportes previos. Al respecto, Hernández *et al* (2004) observaron que la inclusión de una mezcla de aceites esenciales conteniendo AEO en la dieta de pollos incrementó los coeficientes de metabolización aparente de la materia seca y de nitrógeno, mientras que Díaz (2011) administró a las aves 900 ppm de AEO en el alimento y también encontró un mayor coeficiente de metabolización aparente de nitrógeno.

Bravo *et al* (2009) observaron un incremento en la energía metabolizable aparente corregida por nitrógeno cuando las dietas fueron suplementadas con una mezcla de metabolitos secundarios de plantas, conteniendo carvacrol.

Finalmente, se ha reportado también que el carvacrol, metabolito secundario del AEO, influye sobre la expresión de genes relacionados con señales de transducción, metabolismo de proteínas, el movimiento intracelular de proteínas, metabolismo de carbohidratos, metabolismo de lípidos, ácidos grasos y sustancias esteroideas (Kim *et al*, 2010) como andrógenos y estrógenos (Lillehoj *et al*, 2011).

### 3.3. Influencia de la coccidiosis

Las aves del presente estudio fueron sometidas a un desafío por coccidia que indujo un cuadro subclínico compatible principalmente con *E. acervulina* y con las características típicas observadas en condiciones de campo. Esta especie de coccidia puede causar pérdida de peso, debido a la anorexia causada por la enfermedad y a la severa depresión en la absorción de nutrientes en el duodeno (Martínez *et al*, 1995). Al respecto, se ha reportado que la coccidiosis afecta la digestibilidad de los nutrientes dietarios. Esto fue observado por Adams *et al* (1996) quienes administraron 6 x $10^5$ oocistos esporulados de *E. acervulina* por pollo en el día 18 de edad y observando una disminución en la digestibilidad de la grasa en el tracto total de 86 a 22% en el día 21 de edad. Los investigadores determinaron que este resultado es atribuible a la disminución en la secreción de sales biliares, a la reducción en la actividad de la lipasa pancreática y el daño a la mucosa intestinal.

Se ha sugerido además que la coccidiosis afecta de digestión de proteínas (Turk, 1972); sin embargo, no se ha determinado el mecanismo directo.

Diversos autores han documentado la influencia de la coccidiosis sobre la reducción de la actividad de enzimas digestivas vinculadas a la digestión de carbohidratos. Russell and Ruff (1978) observaron una reducción en la actividad amilolítica en el páncreas y en el peso relativo de este órgano, además observaron que la actividad amilolítica ligada a la superficie de la mucosa disminuyó conforme se incrementó la severidad del cuadro medida a través de score de lesiones. Major and Ruff (1978) observaron una disminución en la actividad de la sacarasa y maltasa en la fracción celular del epitelio intestinal, pero sólo en la región intestinal afectada por la coccidia. Enigk and Dey-Hazra (1976) observaron una reducción en la actividad ezimática de la maltasa y sucarasa proporcional al grado de la infección.

El efecto negativo sobre la absorción intestinal está relacionado principalmente con *E. acervulina* y *E. maxima*, provocando una menor absorción de carotenoides y afectando la pigmentación, siendo *E. maxima* la especie de coccidia que tiene el mayor efecto negativo sobre la absorción de pigmentos aun cuando no se observen lesiones intestinales graves (McDougald, 2003).

Al respecto, Teeter *et al* (2008) indican que incluso scores de lesiones de 1.0 en la escala de Johnson and Reid (1970) tienen un efecto negativo en el metabolismo animal y su comportamiento productivo, observando un impacto en el apetito del ave, en la ganancia de peso, en el costo energético de mantenimiento y en la retención de energía.

### 3.4. Mecanismos involucrados en la acción del AEO

Debido al efecto anticoccidial que ejerce el AEO (Giannenas *et al*, 2003; Saini *et al*, 2003; Tsinas *et al*, 2011) es posible atribuir al propio control de la coccidiosis cierta influencia en el incremento en la digestibilidad de nutrientes; sin embargo, otros mecanismos del AEO guardan relación directa con la mayor digestibilidad. Tal es el caso de los cambios morfométricos inducidos por el AEO en la mucosa intestinal, que han sido reportados previamente (Jamroz *et al*, 2006) y corroborados en el Experimento 1, incrementando la longitud de las vellosidades intestinales, el área superficial de la vellosidad y, por lo tanto, la capacidad para la absorción de nutrientes.

Por otro lado, el incremento en la actividad de enzimas digestivas juega también un rol importante. Al respecto, Basmacioglu Malayoglu *et al* (2010) observaron que la inclusión de 250 o 500 ppm de AEO en la dieta de pollos de carne incrementó la actividad de la quimotripsina y consecuentemente la digestibilidad de la proteína dietaria. Roldán (2010), por su parte, observó incrementos en las digestibilidades aparentes de la proteína y de la grasa en el tracto total en aves suplementadas con aceites esenciales de albahaca y tomillo, atribuyéndolo al incremento en la actividad de las enzimas digestivas a consecuencia de la suplementación de aceites esenciales, de acuerdo a lo reportado en estudios previos (Lee *et al*, 2004; Jamroz *et al*, 2005;

Jang *et al*, 2007; Basmacioglu Malayoglu *et al*, 2010). Al respecto, se ha establecido que el carvacrol induce la actividad lipasa en el páncreas y en la pared intestinal (Jamroz *et al*, 2005), lo que puede ser resultado de la acción que, entre otras posibles sustancias contenidas en el AEO, ejerce el carvacrol, ya que se ha documentado su influencia en la regulación de la expresión génica, favoreciendo la expresión de 26 genes y reduciendo la expresión de otros 48, algunos de ellos relacionados con el metabolismo lipídico (Lillehoj *et al*, 2011).

Adicionalmente, debe tenerse presente el efecto regulador del AEO sobre la flora intestinal. Al respecto, se ha establecido que la disponibilidad de nutrientes para el hospedero puede aumentarse mediante la alteración de la microflora intestinal, favoreciendo la absorción de nutrientes (Dias, 2011). Así, Levkut *et al* (2011) observaron incrementos en la disponibilidad de calcio y magnesio en aves suplementadas con AEO, atribuyéndolo a la menor viabilidad de la microflora intestinal por su efecto antimicrobiano. En otro estudio reciente se administró una combinación de timol y cinamaldehído a pollos de carne y se observó de 11 a 14 días de edad incrementos significativos en la retención aparente de la materia seca, energía y proteína cruda en 3.6, 4.1 y 7.5 unidades porcentuales, respectivamente y consecuentemente un incremento en la energía metabolizable aparente en 191 Kcal/k (Cao *et al*, 2010). En ese mismo estudio los investigadores encontraron además, en el día 43 de edad, mayor concentración de acetato y butirato en el contenido cecal de estas aves, y asociaron este efecto con la regulación que ejerce el AEO sobre la proporción de especies y número de bacterias debido a su efecto antibacteriano, a su acción prebiótica y a su efecto sobre la promoción de la exclusión competitiva de patógenos (Ferket, 2003; Zheng *et al*, 2010). Dicho incremento en la concentración de butirato no sólo promueve la multiplicación de las células de la membrana mucosa del colon, por ser el ácido butírico la mayor fuente de energía para estas células, sino que además reduce el pH intestinal favoreciendo la resistencia a enfermedades, reprimiendo bacterias patógenas como *Escherichia coli* (Cao *et al*, 2010).

Los cuadros de coccidiosis, e incluso las terapias implementadas en su control (Neumann *et al*, 2011), pueden desencadenar la disbacteriosis, y se ha determinado que la disbacteriosis y el consecuente daño a la mucosa intestinal afectan la digestibilidad aparente de los nutrientes e incrementan la presencia de alimento sin digerir en las heces (Smits *et al*, 1999).

El control de la coccidia por el PRO en el presente estudio, el control de la disbacteriosis (Experimento 6) concomitante a la coccidiosis, y la estabilización de la microflora bacteriana intestinal por la acción prebiótica del AEO (Hammer *et al*, 1999; Ferket, 2003; Lee *et al*, 2004; Zheng *et al*, 2010; Betancourt *et al*, 2011), reducen además la necesidad del ave de desarrollar mecanismos de respuesta inmunológica (Hashemi and Davoodi, 2010), reduciendo el gasto de nutrientes para suplir dicha función (Yaqoob and Calder, 2003; Clements, 2011).

Dicho efecto adverso se verifica en el presente estudio. La retención aparente de nitrógeno del 73.84% observada en las aves del presente estudio suplementadas con el PRO y alimentadas con una dieta a base de maíz y soya conteniendo 19.66% de proteína cruda coincide con la retención de 74.69% reportada por Kerr and Kidd (1999) en pollos alimentados también con una dieta similar (base maíz-soya; 18.2% de proteína cruda). Por el contrario, la retención aparente de nitrógeno del 63.96% observada en las aves control del presente estudio, desafiadas con coccidia y no suplementadas con el PRO, coincide con la retención del 16.7% reportada por estos mismos autores en pollos alimentados con una dieta conteniendo sólo 16.7% de proteína cruda. Al respecto, Humphrey *et al* (2002) reportaron el costo de lisina, expresado en µmol del aminoácido por kg de peso corporal por día, para algunas funciones inmunológicas de aves desafiadas con lipopolisacáridos respecto a aves normales, mostrando incrementos de 45.5 a 90.9 µmol/kg para leucopoyesis, de 65.6 a 69.7 µmol/kg para síntesis de inmunoglobulinas, de 0 a 386 µmol/kg para síntesis de proteínas de fase aguda, y de 111 a 547 µmol/kg para la inmunocompetencia total. Ello implica que el efecto del AEO sobre el control de la coccidiosis y la regulación de la microflora intestinal es nutricionalmente significativo.

### 3.5. Comportamiento productivo

En el presente estudio, las aves que fueron suplementadas con el PRO presentaron una ganancia de peso 10% mayor ($P<0.01$) y una menor conversión alimentaria (1.29 vs 1.41; $P<0.01$), mientras que el consumo de alimento fue similar en ambos tratamientos (P>0.10). Este mayor peso final y ganancia de peso observados en las aves suplementadas con el PRO se explica por la mayor eficiencia de estas aves en la utilización del alimento, debido a que no se observan diferencias en el consumo de alimento, tal como fue reportado en otros experimentos con el uso del PRO (Experimentos 2, 5 y 6). Esta mayor eficiencia en la utilización del alimento en las aves suplementadas con el PRO se verifica en la mayor digestibilidad aparente de nutrientes, mayor retención de nitrógeno y mayores eficiencias de la proteína y energía dietarias. Al respecto, el incremento observado en la digestibilidad aparente de la proteína dietaria contribuye a un mejor balance de de aminoácidos, menor perdida de energía a través del proceso de desaminación y consecuentemente mayor disponibilidad de energía para el crecimiento de tejidos (Olukosi *et al*, 2008).

Es importante mencionar que el estímulo que ejerce el AEO sobre la proliferación celular en la mucosa intestinal (ver Experimento 1; Levkut *et al*, 2011) favorece el mayor crecimiento de la vellosidad intestinal y, en consecuencia, mayor capacidad para la absorción de nutrientes (Noy y Sklan, 1995; Mitchel and Moretó, 2006).

## 4. Conclusiones

Los resultados obtenidos bajo las condiciones del presente estudio permiten llegar a las siguientes conclusiones:

- La suplementación del PRO favorece la absorción de nutrientes, incrementa la digestibilidad aparente de la materia seca y proteína dietarias, y repercute positivamente en el comportamiento productivo de las aves.

- La inclusión del PRO en la dieta de pollos de carne agrega valor a la dieta, incrementando la eficiencia de la proteína y energía para la ganancia de peso.

- La suplementación del PRO favorece la conservación del ambiente, incrementando la retención de nitrógeno y reduciendo su eliminación.

# V. DISCUSIÓN GENERAL

La utilización de antibióticos en el alimento animal ha sido vinculada con la aparición de resistencias bacterianas, lo que ha motivado la progresiva prohibición del uso de antibióticos promotores de crecimiento en diferentes países en todo el mundo. Estas restricciones han afectado las ganancias de los productores avícolas por la menor tasa de crecimiento y eficiencia en la utilización del alimento, así como por el incremento en la incidencia de problemas sanitarios como la Enteritis Necrótica.

Para mejorar el retorno económico de la producción avícola y asegurar el abastecimiento de alimentos saludables y seguros destinados al consumo humano, se han desarrollado tecnologías tendientes a suplir la función de los antimicrobianos promotores de crecimiento y atenuar las consecuencias derivadas de su restricción. Tal es el caso de los probióticos, prebióticos, simbióticos, enzimas, acidificantes, derivados fitogénicos, entre otros. Particularmente, los extractos de plantas y aceites esenciales han sido ampliamente estudiados (Jamroz and Kamel, 2002; Botsoglou el al, 2002; Alcicek el al, 2003; Mitsch el al, 2004; Botsoglou *et al*, 2005) y se ha documentado en los últimos años cualidades de interés para el productor avícola.

Diferentes investigaciones *in vitro* han demostrado, entre otros, los efectos antimicrobiano (Helander *et al*, 1998; Faleiro *et al*, 2005) y antioxidante (Deighton *et al*, 1993; Martínez-Tomé *et al*, 2001; Faleiro *et al*, 2005) del AEO. Sin embargo, algunos de los estudios realizados *in vivo* (Fukayama, 2004; Fukayama *et al*, 2005) no han logrado evidenciar los efectos del AEO. Al respecto, se ha postulado que la ausencia de diferencias significativas puede deberse a la ausencia de un desafío sanitario mayor para que la actividad antimicrobiana de los extractos vegetales y sus constituyentes pueda ser evidenciada (Fukayama *et al*, 2005; Jesus, 2007; Toledo *et al*, 2007; Barreto *et al*, 2008), lo que concuerda con los resultados de algunas evaluaciones realizadas en empresas avícolas de nuestro medio, en granjas experimentales en condiciones sanitarias óptimas (observación personal). Asimismo, Allen *et al* (1997) reportan que los aceites esenciales pueden incrementar el peso de las aves sólo cuando éstas se encuentran sometidas a un desafío intenso por coccidia.

Un comportamiento similar, respecto a los aceites esenciales, se ha observado no sólo en aves sino también en otras especies animales, y aun con el uso de otras tecnologías. Tal es el caso de los antimicrobianos promotores de crecimiento, en que la respuesta sobre la ganancia de peso es menor cuando los animales están en una instalación de crianza nueva (Holden *et al*, 1998) o la contaminación por patógenos es reducida (Hays and Speer, 1960) o cuando el estado de salud del animal, la ganancia de peso o eficiencia del alimento es mayor. Por otro lado, diversos estudios han reportado que los aceites esenciales permiten un desempeño igual o mayor al observado en animales que reciben antimicrobianos convencionales (Alcicek *et al*, 2003; Ertas *et al*, 2005; Oetting *et al*, 2006; Franco *et al*, 2007; Silva *et al*, 2009).

Finalmente, se ha observado que el contenido y proporción de los metabolitos secundarios del orégano en el aceite esencial es altamente variable (Ruso *et al*, 1998; Daferera *et al*, 2000; Plaus *et al*, 2001; Chorianopoulos *et al*, 2004; Muñoz Acevedo *et al*, 2007), así como que cada metabolito secundario presenta un diferente grado de actividad biológica en cada uno de los mecanismos de acción que ejerce el aceite esencial (Janssen *et al*, 1987; Silva *et al*, 2010), por lo que es posible considerar que los resultados de la administración del aceite esencial dependa también de sus propias características.

En los estudios realizados como parte del presente trabajo, se ha evaluado el efecto de la inclusión del PRO sobre diferentes aspectos que influyen en el comportamiento productivo del ave bajo diferentes condiciones sanitarias. En el Anexo 1 se ha presentado la evaluación de dos modelos de desafío intestinal para las aves, logrando reproducir condiciones similares a las observadas en campo. En estas condiciones el PRO ha mostrado efectos significativos en aves sometidas al STR o coccidiosis, tal como ha sido presentado en los Experimentos 5 a 7, mejorando el estado antioxidante, controlando el daño intestinal, incrementando la absorción y utilización de los nutrientes dietarios y mejorando el comportamiento productivo de las aves.

Sin embargo, los hallazgos reportados en los Experimentos 1 a 3, indican que los efectos del PRO también pueden ser evidenciados en aves en condiciones normales, no sometidas a desafíos específicos por coccidia o patógenos bacterianos como

*Clostridium perfringens*, observándose un mayor desarrollo de la estructura de la mucosa intestinal, mejores condiciones para enfrentar la microflora intestinal patógena, mayor capacidad de absorción de nutrientes, mayor mineralización ósea e integridad esquelética, y mejor comportamiento productivo.

Para entender de manera integral los beneficios del AEO sobre el comportamiento productivo de las aves, es necesario considerar las múltiples interacciones existentes entre sus mecanismos de acción. En el Gráfico 1 se presenta una ilustración de ello. Como se observa, el AEO posee diferentes propiedades que confieren beneficios a la salud y al comportamiento productivo de las aves. Al respecto, la Unión Europea clasifica los aditivos para el alimento en cinco categorías: tecnológicos, sensoriales, nutricionales, zootécnicos y coccidiostatos e histomonastatos; sin embargo, la complejidad y multifuncionalidad del AEO, lo localiza en todas las categorías descritas (Afanador *et al*, 2011).

Algunos mecanismos, como la acción anticoccidial, pueden tener menor relevancia en aves no sometidas a condiciones de desafío. Sin embargo, la acción que ejerce el PRO sobre la promoción de la proliferación celular en el epitelio intestinal, la consecuente elongación de las vellosidades intestinales e incremento en la capacidad de absorción de nutrientes, así como la acción antioxidante documentadas en el presente trabajo, son importantes en cualesquiera sean las condiciones de crianza, ya que el desarrollo de la mucosa intestinal y la capacidad digestiva del ave es un factor crítico que puede limitar el comportamiento productivo de las aves, particularmente en las primeras semanas de vida (Uni, 2006; Mitchell and Moretó, 2006).

Particularmente en el caso de las aves, es necesario que el nutricionista diseñe alimentos y estrategias de alimentación que permitan suministrar a las aves el nivel adecuado de nutrientes, de forma que puedan ser digeridos y absorbidos de manera eficiente, pero también de forma segura y libre de patógenos. Para ello es necesario enfocarse en el desarrollo y calidad de la mucosa intestinal, modular la microflora del tracto gastrointestinal para controlar trastornos intestinales, proteger a las aves de los efectos negativos de la oxidación, atenuar el desarrollo de enfermedades no infecciosas, reducir los factores de riesgo para las enfermedades infecciosas subclínicas y favorecer la inmuno-competencia de las aves.

**Gráfico 1. Principales interacciones involucradas en los mecanismos de acción del aceite esencial de orégano sobre la mejora del comportamiento productivo.**

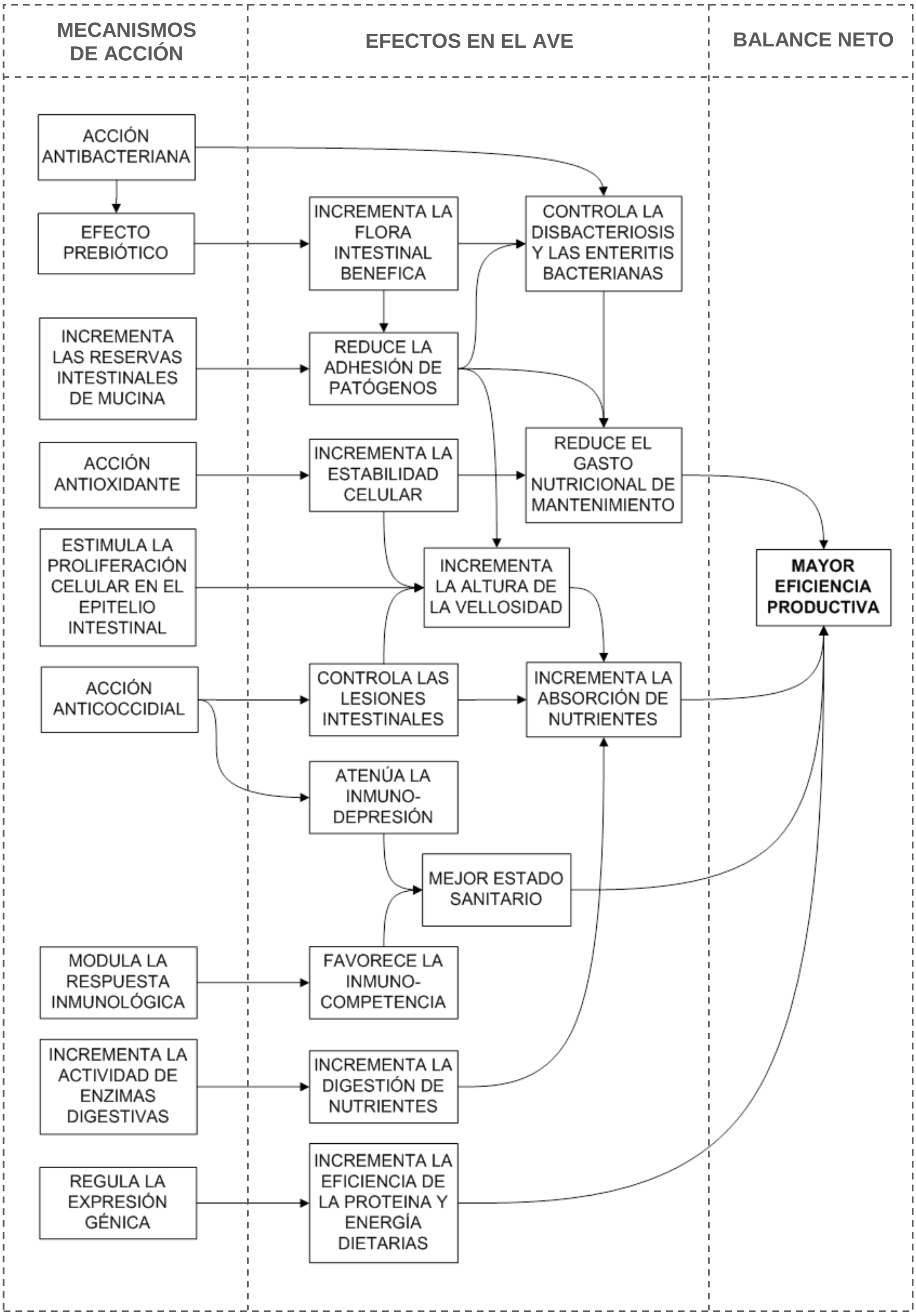

Fuente: Elaboración propia

# VI. CONCLUSIONES GENERALES

- El PRO incrementa la proliferación celular en el epitelio intestinal, lo que se traduce en el incremento de la longitud de las vellosidades intestinales, del área superficial para la absorción de nutrientes, y consecuentemente de la digestibilidad de los nutrientes dietarios y de la mineralización ósea e integridad esquelética.

- El balance neto del incremento en la proliferación celular en el epitelio intestinal, de la acción antibacteriana y de la acción antioxidante del PRO, resulta positivo para el balance nutricional del ave, produciendo una mayor eficiencia en la conversión alimentaria.

- El PRO ejerce un control activo de la coccidiosis y atenúa el impacto negativo que sobre el comportamiento productivo tienen los trastornos y procesos entéricos que cursan con el tránsito rápido del alimento, exacerbación de coccidia y enteritis como aquella producida por *Clostridium perfringens*.

- El PRO favorece la restitución del estado antioxidante en las aves sometidas a condiciones de desafío intestinal, favoreciendo el comportamiento productivo.

- Las mayores reservas de mucina intestinal por efecto del PRO observadas en el presente trabajo, así como su acción antibacteriana, favorecen la menor adhesión de bacterias patógenas, lo que brinda mayor salud intestinal y mejores condiciones para la eficiente utilización del alimento.

- La suplementación de PRO en el alimento de pollos agrega valor a las dietas, incrementando la eficiencia de utilización de la proteína y energía dietarias.

# VII. RECOMENDACIONES

- Establecer programas de suplementación del PRO costo-efectivos en pollos de carne para favorecer el desarrollo del epitelio intestinal y la capacidad de absorción de nutrientes durante las primeras semanas de vida, así como para mejorar la salud intestinal, reducir la probabilidad de ocurrencia de trastornos y procesos entéricos en los periodos de mayor desafío y consecuentemente preservar la capacidad productiva de las parvadas.

- Evaluar el uso del PRO en pavos, gallinas ponedoras y gallinas reproductoras.

- Incorporar la evaluación de las heces mediante los índices propuestos en los programas de monitoreo de la salud intestinal de las aves en parvadas comerciales.

- Incluir los muestreos de fémur en los estudios relativos al desarrollo e integridad ósea.

- Implementar políticas y sistemas para la gestión de la calidad del valor nutricional y de la inocuidad de las dietas de las aves comerciales.

# VIII. REFERENCIAS

# IX. ANEXOS

# ANEXO 1. EVALUACIÓN DE DOS MODELOS DE DESAFÍO PARA INDUCIR EL SÍNDROME DE TRÁNSITO RÁPIDO EN POLLOS DE CARNE

**Contenido** Página

**Cuadros** Página

**Gráficos** Página

# 1. Introducción

El Síndrome de Tránsito Rápido (STR) es una condición recurrente y de gran importancia económica en las explotaciones avícolas. Las parvadas afectadas por este síndrome presentan disbacteriosis, heces acuosas, con alimento sin digerir y/o descamaciones mucosas. Esta condición impacta negativamente sobre los resultados productivos de la campaña retrasando el crecimiento de las aves, incrementando la conversión alimentaria y reduciendo la eficiencia productiva; sin embargo, puede pasar inadvertida atribuyéndose la menor eficiencia productiva a la variabilidad que normalmente ocurre entre parvadas o entre planteles de la misma empresa.

Las estrategias dirigidas a mejorar la eficiencia productiva de las parvadas deben ser evaluadas en condiciones similares a las observadas en las explotaciones comerciales, y para lograrlo es necesario contar con una metodología que permita reproducir las condiciones de campo. El objetivo de este experimento fue establecer un modelo de desafío entérico en pollos de carne que permita reproducir, de manera controlada, el STR y las condiciones sub-clínicas observadas en las explotaciones avícolas comerciales.

# 2. Materiales y métodos

## 2.1. Lugar, fecha y duración

La crianza de las aves se llevó a cabo en las instalaciones del Programa de Aves de la Facultad de Zootecnia de la Universidad Nacional Agraria La Molina (UNALM) en Lima-Perú a inicios del segundo semestre del 2010. Los inóculos de *Clostridium perfringens* y coccidia fueron preparados en el Laboratorio de Biología y Genética Molecular de la Facultad de Medicina Veterinaria de la Universidad Nacional Mayor de San Marcos (UNMSM) en Lima-Perú y en el Programa de Aves de la UNALM, respectivamente. La necropsia de las aves muestreadas y las mediciones ulteriores se llevaron a cabo en las instalaciones del Laboratorio de Patología Aviar de la UNMSM. La evaluación se llevó a cabo desde la recepción de las aves en las instalaciones de crianza y el periodo de evaluación fue el comprendido de 0 a 28 días de edad.

## 2.2. Instalaciones, equipos y materiales

### 2.2.1. Instalaciones de crianza

Las aves estuvieron alojadas sobre material de cama a razón de 21.4 pollos/$m^2$ (0.047 $m^2$/pollo) en 16 jaulas con piso de malla metálica distribuidas en baterías de 4 pisos. Cada jaula contó con un comedero y un bebedero lineales y las características que se detallan en la sección "Modelos de desafío" de este experimento. Durante los primeros 3 días de vida se colocó papel periódico para evitar el acceso directo de los pollos BB al material de cama, el alimento fue suministrado en bandejas plásticas y el agua de bebida en bebederos tipo tongo.

Para poder contener material de cama fue necesario acondicionar las jaulas empleando malla sintética, láminas y precintos de polietileno, y silicona, de manera tal que permitiera su ventilación y la retuviera dentro de la jaula. Las adaptaciones realizadas se observan en el Anexo 2.

Dos semanas antes de la recepción de las aves se inició un programa de control de vectores, empleando cebaderos, rodenticidas y mosquicidas comerciales. Antes y después del periodo experimental se realizó la limpieza y desinfección de las instalaciones.

### 2.2.2. Calefacción y ventilación

La calefacción fue provista por un sistema eléctrico con resistencias y termostatos, y fue controlada de acuerdo a las recomendaciones de la línea genética. Para reducir la variabilidad en la temperatura ambiental a consecuencia de las diferentes alturas entre los pisos de las baterías se adaptó resistencias adicionales en la base de cada batería. La ventilación fue controlada con cortinas de polipropileno instaladas en el perímetro del área de crianza. La humedad ambiental en el área de crianza se mantuvo alrededor de 40%. Para la medición de la temperatura y humedad ambiental se empleó un termo-higrómetro electrónico digital con una aproximación de 0.1 °C para temperatura y 1% para humedad.

### 2.2.3. Preparación del alimento

Para el pesaje de los ingredientes mayores del alimento se utilizó una balanza digital con capacidad de 150 kg y aproximación de 0.02 kg, mientras que para el pesaje de los ingredientes de la premezcla se utilizó una balanza electrónica con capacidad de 6 kg y aproximación de 1 g.

Se utilizó mezcladoras horizontales de cintas de 400 y 30 kg de capacidad para la mezcla de los ingredientes durante la preparación del alimento y para las premezclas, respectivamente. El alimento fue envasado en sacos de papel y polietileno laminado, y almacenado en cilindros metálicos.

### 2.2.4. Manejo y actividades especiales

Para la identificación individual de las aves se desarrolló un método dual en base a precintos de colores y tinta indeleble (Anexo 3), a partir de métodos previos (Pinilla, 2000; SEO, 2000; Palanca, 2004).

Para la medición de los pesos vivos se utilizó una balanza electrónica con capacidad para 200 g y con aproximación de 10 mg y otra con capacidad para 15 kg y con aproximación de 1 g. Para la medición del alimento consumido, se utilizó una balanza electrónica con capacidad para 15 kg y con aproximación de 1 g.

Para remover el material de cama se empleó espátulas de polietileno. Para reducir la contaminación cruzada entre tratamientos se empleó guantes quirúrgicos desechables, alcohol metílico y desinfectante a base de sales cuaternarias de amonio y aldehídos (CKM-Desin®, Laboratorio CKM S.A.C.).

En el día 10 de edad las aves fueron vacunadas, por vía ocular, contra la Enfermedad de Newcastle, empleando una vacuna viva liofilizada con una cepa VG/GA (Avinew®, Merial Limited)

Para el desafío de las aves se empleó un inóculo de coccidia que fue preparado a partir de una vacuna comercial contra coccidia (Immucox® for Chickens II, Vetech Laboratories) y un inóculo de *C. perfringens* proveniente de un aislamiento de un brote sub-clínico de campo de enteritis en pollos de 3 semanas de edad. Las características de los inóculos se presentan en la sección "Modelos de desafío" del presente estudio.

El agua de bebida fue potabilizada empleando 1 ml de hipoclorito de sodio al 4.5% por cada 10 L de agua. Para evitar que la presencia de cloro interfiera con la viabilidad de los inóculos la clorinación del agua se realizó tres días antes de proveerla a las aves, y diariamente, al momento de suministrarla a las aves se verificó la ausencia de cloro empleando un kit comercial de evaluación.

### 2.2.5. Necropsias y las mediciones posteriores

Para la medición de los pesos vivos se utilizó una balanza electrónica con capacidad para 15 kg y con aproximación de 1 g. Para la necropsia de las aves se empleó bisturí, tijeras y guantes quirúrgicos. Para las mediciones ulteriores se utilizó un microscopio óptico de luz artificial binocular LeicaDM500®, láminas porta y cubre objetos, papel secante y bolsas plásticas de cerrado hermético. Para la medición de los pesos de las vísceras se empleó una balanza electrónica con capacidad para 200 g y con aproximación de 10 mg. Para la medición del diámetro de la Bursa de Fabricio se empleó un Bursómetro comercial, que es una regla con 8 perforaciones circulares con diámetros crecientes desde 1/8 hasta 8/8 de pulgada (Fort Dodge, 2001).

### 2.3. Animales experimentales

Se empleó 128 pollos machos de la línea Cobb 500, cuyas características se presentan en el Anexo 4. Los pollos fueron alojados al azar en 16 jaulas. Al término de cada semana se muestreó al azar 1 pollo por jaula para los controles respectivos.

### 2.4. Tratamientos

Se evaluaron 2 tratamientos, que fueron definidos de la siguiente manera:

Tratamiento 1 : Aves sometidas al modelo de desafío A
Tratamiento 2 : Aves sometidas al modelo de desafío B

En el Cuadro A se presenta un resumen de los factores empleados las características generales de los modelos de desafío empleados. El detalle de dichos modelos se presenta en la sección "Modelos de desafío" de este experimento.

Para minimizar el riesgo de contaminación cruzada entre tratamientos se implementó el protocolo que indica en el Anexo 5.

**Cuadro A. Características generales de los modelos de desafío empleados.**

| Factores | Niveles de cada factor [1] | |
|---|---|---|
| | Modelo A | Modelo B |
| Baja digestibilidad de la proteína dietaria | + + | + |
| Exceso de proteína dietaria | - | + |
| Inóculo de coccidia | - | + |
| Inóculo de *Clostridium pefringens* | - | + |
| Material de cama re-utilizado | - | + |

[1] (-) Factor ausente; (+) factor presente; (++) factor presente y con mayor intensidad

## 2.5. Factores de desafío empleados

Los modelos de desafío empleados consistieron en combinaciones de los siguientes factores:

### 2.5.1. Baja digestibilidad de la proteína dietaria

La menor digestibilidad de la proteína dietaria se logró empleando una dieta basal conteniendo, como principal fuente proteica, una torta de soya sobre-cocida. Las aves del tratamiento 1 (modelo de desafío A) estuvieron sujetas a esta condición durante todo el periodo experimental; sin embargo, las aves del tratamiento 2 (modelo de desafío B) sólo hasta el día 14 de edad. Durante las 2 primeras semanas las aves del tratamiento 1 recibieron una ración con mayor digestibilidad proteica equivalente a 85% la dieta basal + 15% de harina de pescado. El detalle se presenta en la sección “Alimentación” de este experimento.

### 2.5.2. Exceso de proteína dietaria

Para proveer a las aves proteína dietaria en exceso las aves del tratamiento 2 (modelo de desafío B) recibieron de 1 a 14 días de edad una ración equivalente a 85% la dieta basal + 15% de harina de pescado. El detalle se presenta en la sección “Alimentación” de este experimento.

### 2.5.3. Inóculo de coccidia

En el día 14 de edad las aves del tratamiento 2 (modelo de desafío B) recibieron un inóculo de coccidia preparado para contener oocistos no atenuadas de aislamientos de campo (Conway and McKenzie, 2007) de *Eimeria acervulina, E. maxima, E. tenella, E. necatrix* y *E. brunetti* con una concentración total no menor a 24 x $10^5$ oocistos vivos por dosis, equivalente a 10 dosis de una vacuna comercial (Immucox® for Chickens II, Vetech Laboratories). Las aves del tratamiento 1 recibieron, como placebo, un volumen similar de agua destilada, según lo recomendado por Conway and McKenzie (2007). El inóculo fue administrado a cada pollo vía oro-ingluvial (Christaki *et al*, 2004; Conway and McKenzie, 2007; Elmusharaf *et al*, 2010;

Georgieva *et al*, 2011b) empleando cánulas adheridas a jeringas hipodérmicas desechables de 5 ml sin aguja.

#### 2.5.4. Inóculo de *Clostridium perfringens*

Las aves del tratamiento 2 (modelo de desafío B) recibieron en los días 18, 19 y 20 de edad un inóculo conteniendo $10^8$ UFC de *C. perfringens* por dosis en un volumen de 3 ml. Las aves del tratamiento 1 recibieron, como placebo, 3 ml del medio de cultivo empleado en la preparación del inóculo de *C. perfringens*. Las características del inóculo se presentan en el Anexo 9. El inóculo se administró de manera individual, vía oro-ingluvial, empleando cánulas adheridas a jeringas hipodérmicas desechables de 5 ml sin aguja.

#### 2.5.5. Material de cama reusado

Las aves del tratamiento 2 (modelo de desafío B) fueron alojadas sobre material de cama reusada. Las características de los materiales de cama empleados se presentan en el Anexo 6.

### 2.6. Alimentación

La alimentación fue *ad libitum* y se utilizó dos dietas conformadas por la combinación de una dieta basal y harina de pescado de manera suplementaria, tal como se indica a continuación: Las aves del tratamiento 1 (modelo de desafío A) recibieron 100% la dieta basal de 1 a 28 días de edad, mientras que las aves del tratamiento 2 (modelo de desafío B) recibieron una dieta equivalente a 85% la dieta basal y 15% harina de pescado de 1 a 14 días de edad, y 100% la dieta basal de 15 a 28 días de edad; ello con la finalidad de proveer a las aves del tratamiento 2 proteína en exceso en relación al requerimiento de la línea genética para estimular la instalación del inóculo de *C. perfringens*. Las características de las dietas empleadas se presentan en el Cuadro B, incluyendo la dieta basal designada como Dieta 1.

**Cuadro B. Características de las dietas empleadas.**

| Ingredientes | Composición, % | |
|---|---|---|
| | Dieta 1 | Dieta 2 [1] |
| Maíz amarillo | 61.775 | 52.508 |
| Torta de soya | 31.405 | 26.695 |
| Harina de pescado | 0.000 | 14.938 |
| Aceite de pescado | 2.380 | 2.024 |
| DL-Metionina | 0.224 | 0.190 |
| L-Lisina | 0.135 | 0.116 |
| Cloruro de colina | 0.100 | 0.085 |
| Fosfato dicálcico | 1.892 | 1.608 |
| Carbonato de calcio | 1.150 | 0.978 |
| Sal común | 0.424 | 0.360 |
| Marcador inerte [2] | 0.300 | 0.300 |
| Premezcla [2] | 0.100 | 0.085 |
| Antifúngico [3] | 0.100 | 0.085 |
| Antioxidante [3] | 0.015 | 0.013 |

| Nutrientes | Aporte calculado | |
|---|---|---|
| | Dieta 1 | Dieta 2 [1] |
| EM [3], Kcal/kg | 3008 | 3028 |
| Proteína cruda, % | 20.04 | 26.72 |
| Lisina, % | 1.16 | 1.72 |
| Metionina+Cistina, % | 0.87 | 1.10 |
| Treonina, % | 0.76 | 1.05 |
| Triptófano, % | 0.23 | 0.29 |
| Calcio, % | 1.17 | 1.54 |
| Fósforo disponible, % | 0.44 | 0.67 |
| Sodio, % | 0.19 | 0.34 |
| Grasa total, % | 5.11 | 5.97 |
| Fibra cruda, % | 3.17 | 2.70 |

| Componentes proximales | Composición [4], % | |
|---|---|---|
| | Dieta 1 | Dieta 2 [1] |
| Humedad | 12.35 | 11.89 |
| Proteína total [3] | 19.66 | 26.06 |
| Extracto Etéreo | 3.83 | 4.87 |
| Fibra Cruda | 2.49 | 2.12 |
| Cenizas | 5.88 | 7.13 |
| ELN [3] | 55.79 | 47.94 |

| Carga microbiológica | Recuento | |
|---|---|---|
| | Dieta 1 | Dieta 2 [1] |
| *C. perfringens* [5], UFC/g | < 50 | 75 a 118 |
| Coliformes [6], NMP/g | 23 | 20 |
| Aerobios mesófilos [6], UFC/g | $53x10^3$ | $74x10^3$ |

(1) La dieta 2 es equivalente a la combinación de 85% de la dieta basal (dieta 1) y 15% de harina de pescado.
(2) Marcador inerte: óxido crómico. Premezcla de vitaminas y minerales Proapak 2A®. Composición: Retinol: 12’000,000 UI; Colecalciferol: 2’500,000 UI; DL α-Tocoferol Acetato: 30,000 UI; Riboflavina: 5.5 g; Piridoxina: 3 g; Cianocobalamina: 0.015 g; Menadiona: 3 g; Ácido Fólico: 1 g; Niacina: 30 g; Ácido Pantoténico: 11 g; Biotina: 0.15 g; Zn: 45 g; Fe: 80 g; Mn: 65 g; Cu: 8 g; I: 1 g; Se: 0.15 g; Excipientes c.s.p. 1,000 g
(3) Antifúngico: Mold Zap®; Antioxidante: Danox®; EM: Energía metabolizable; Proteína total: N x 6.25; ELN: Extracto Libre de Nitrógeno (calculado).
(4) Dieta 1: Informe 1235/2010 LENA. Dieta 2: Calculado a partir de análisis de dieta empleada de 15 a 28 días de edad (85%) y de harina de pescado (15%).
(5) Dieta 1: Informe 19/Jul/2010 LBGM. Dieta 2: A partir de recuentos en harina de pescado y en dieta basal (Informes 19/Julio/2010 LBGM).
(6) Dieta 1: Informe 3-08632/10 CERPER. Dieta 2: Calculado a partir de recuentos en harina de pescado (Informe 3-08633/10 CERPER) y en dieta basal.

La dieta basal fue elaborada en base a maíz y soya, como principales ingredientes energético y proteico, respectivamente. Asimismo, fue complementada con aceite semirefinado de pescado y aminoácidos sintéticos, y suplementada con una premezcla de vitaminas y minerales.

Las características de las harinas de soya y pescado empleadas en la preparación de las dietas se presentan en el Anexo 7 y el criterio empleado para su asignación a los tratamientos en el Cuadro C.

**Cuadro C. Criterio para la asignación de dietas por tratamiento.**

| Periodo | Tratamientos [1] | |
|---|---|---|
| | 1 | 2 |
| De 1 a 14 días | Dieta 1 | Dieta 2 |
| De 15 a 28 días | Dieta 1 | Dieta 1 |

[1] Tratamientos: 1: aves sometidas al modelo de desafío A; 2: aves sometidas al modelo de desafío B

### 2.7. Mediciones

#### 2.7.1. Comportamiento productivo

Se midió empleando las siguientes variables:

- Peso vivo: Cada semana se pesó individualmente a las aves.
- Ganancia de peso: Se calculó valores individuales y promedio.
- Consumo de alimento: Al término de cada semana se pesó los residuos de alimento y se calculó el consumo de alimento.
- Conversión alimentaria: Se calculó semanalmente en función al consumo de alimento y la ganancia de peso en cada periodo.
- Mortalidad: Se realizó la necropsia, registrando fecha, hora, posible causa de muerte y el residuo de alimento.

### 2.7.2. Evaluación de heces

En la tercera y cuarta semana de edad, dos veces por semana y tres veces en cada fecha, se registró y calculó lo siguiente:

- Alteraciones en heces:

  En las heces muestreadas se determinó las siguientes variables:
  - Heces acuosas, %
  - Heces con alimento sin digerir, %
  - Heces con descamaciones mucosas, %
  - Heces hemorrágicas, %

- Índice Dimar:

  El índice Dimar (daño intestinal medido a través de residuos) es propuesto como un indicador práctico del daño a la mucosa intestinal, ya que considera la relación que cada alteración en las heces tiene con el comportamiento productivo de las parvadas en condiciones comerciales, y reúne en una sola variable las alteraciones más frecuentemente observadas en campo. Se mide en una escala de 0 a 100, donde 0 y 100 son el menor y mayor grado de daño intestinal, respectivamente, medibles mediante este método. Para el cálculo, a cada excreta muestreada se asignó un score de acuerdo a la siguiente escala:

  0: si es normal (seca, bien formada y no contiene alimento sin digerir)
  1: si la excreta es acuosa
  2: si contiene alimento sin digerir
  3: si presenta descamaciones de mucosa
  4: si es hemorrágica

  Luego se aplicó la fórmula presentada a continuación, donde SPHM: score promedio de las heces muestreadas.

$$\text{Índice Dimar} = \text{SPHM} \times 25$$

### 2.7.3. Evidencias del daño intestinal

Al término de cada semana se tomó al azar un pollo de cada jaula, para luego ser pesados y sacrificados cortando las arterias carótidas con la subsecuente exanguinación (Mutus *et al*, 2006), de acuerdo al protocolo del Laboratorio de Patología Aviar de la UNMSM. Se realizó la necropsia de acuerdo a los protocolos propuestos por Bermúdez y Stewart-Brown (2003) y Colas *et al* (2010). Se cortó longitudinalmente toda la extensión del intestino para exponer su interior y realizar su inspección, y se registró las siguientes variables:

- Enteritis: Se empleó el siguiente score para la evaluación macroscópica: 0: no presenta; 1: presenta enteritis.

- Presencia de lesiones: Se empleó el siguiente score para la evaluación macroscópica: 0: sin lesiones; 1: presenta lesiones.

- Densidad de lesiones: Se determinó por medición directa y los resultados se expresan en número de lesiones por $cm^2$.

- Score de lesiones por *C. perfringens*: Se determinó empleando el siguiente score: 0: no presenta; 1: intestino delgado con la pared adelgazada o friable; 2: necrosis focal o ulceración; 3: placas más grandes de necrosis; 4: necrosis severa y extensa, Enteritis Necrótica clásica (Prescott *et al*, 1978).

- Score de lesiones por coccidia: Se determinó empleando el scores de coccidia descrito por Johnson and Reid (1970), e ilustrado por Conway and McKenzie (2007), en una escala de 0 a 4.

- Score microscópico de coccidia: En cada ave y sección intestinal en que se observó lesiones se realizó el raspado de la mucosa, se preparó un frotis (McDougald, 2003) y se observó en el microscopio a un aumento total de 100X para el rastreo de los ooquistes y 400X para la lectura. Se empleó el siguiente score en base a la presencia y cantidad de ooquistes encontrados: 0: ausentes; 1: se encuentran pero son escasos; 2: cantidad moderada; 3: abundantes.

- Aves positivas a coccidia:

  A nivel macroscópico y microscópico se determinó como la presencia de lesiones intestinales o como presencia de ooquistes en el raspado intestinal, respectivamente. Se expresa en porcentaje.

#### 2.7.4. Morfometría de los órganos linfoides

Al realizar las necropsias se colectaron la bursa y el timo y, previa eliminación del tejido graso, se determinó las siguientes variables:

- Diámetro de la bursa de Fabricio:

  La medida cualitativa del diámetro mayor de la bursa se determinó mediante un bursómetro comercial (Fort Dodge, 2001).

- Índices morfométricos:

  Se determinó los índices morfométricos de la bursa (Rbu), del bazo (Rba) y del timo (Rti), empleando la fórmula utilizada por Rosales *et al* (1989), Ismail *et al* (1990), Grieve (1991), Alamsyah *et al* (1993), Salazar (1997), Ulloa (1999), Sandoval *et al* (2002) y Perozo-Marín *et al* (2004) que se presenta a continuación, donde IM: índice morfométrico, PO: peso del órgano (g), PC: peso corporal (g).

$$IM = (PO/PC) \times 1000$$

- Relación ente órganos:

  Se determinó las relaciones bursa/bazo (Bu-Ba), bursa/timo (Bu-Ti) y timo/bazo (Ti-Ba), empleando la fórmula propuesta y utilizada por Rosales *et al* (1989), Ismail (1990), Grieve (1991), Alamsyah *et al* (1993), Salazar (1997), Ulloa (1999), Sandoval *et al* (2002) y Perozo-Marín *et al* (2004) que se presenta a continuación, donde $RO_{A-B}$: relación entre los órganos A y B, $PO_A$: peso del órgano A (g), $PO_B$: peso del órgano B (g).

$$RO_{A-B} = \frac{PO_A}{PO_B}$$

#### 2.7.5. Características de la carcasa y biometría de órganos

Semanalmente se determinó las siguientes variables:

- Rendimientos de carcasa y de pechuga
- Porcentaje de pechuga en la carcasa
- Relación carcasa/pechuga
- Pesos relativos de intestino, hígado y páncreas
- Relación páncreas/hígado

Los rendimientos de carcasa y pechuga fueron definidos como la relación porcentual entre el peso de la porción indicada y el peso vivo. El peso de la carcasa se determinó luego de retirar las vísceras (excepto pulmones y riñones) y la piel. El peso de la pechuga se determinó luego de retirar la piel. El porcentaje de pechuga se calculó en relación al peso de la carcasa. La relación entre el peso de dos estructuras u órganos fue definida como la razón entre ambos. Los pesos relativos fueron definidos como la relación porcentual entre el peso del órgano indicado y el peso vivo.

### 2.8. Diseño estadístico

Se utilizó el Diseño Completo al Azar con dos tratamientos y ocho repeticiones (Calzada, 1982). La prueba de bondad de ajuste a la distribución normal de los datos obtenidos a partir de las variable evaluadas se realizó empleando la prueba de Chi-Cuadrado (Calzada, 1982). El análisis de varianza de los datos distribuidos normalmente se llevó a cabo aplicando el procedimiento paramétrico ANOVA del programa Statistical Analysis System SAS 9.0 (SAS Institute, 2009) y la diferencia de medias se realizó usando la prueba de Duncan (1955). Los datos de las variables relacionadas a las lesiones intestinales no están distribuidos normalmente, por esta razón el análisis de este conjunto de variables se realizó empleando la prueba no paramétrica de Kruskal-Wallis mediante el procedimiento NPAR1WAY con restricción WILCOXON del programa SAS (Schlotzhauer and Littell, 1997; McDonald, 2009). Debido a la existencia de sólo dos tratamientos no fue necesario realizar las pruebas de Dunn o LSD para determinar diferencias entre medias (Schlotzhauer and Littell, 1997).

El Modelo Aditivo Lineal General aplicado a las variables evaluadas fue el siguiente:

$$Yij = U + Ti + Eij$$

Donde:

| | | |
|---|---|---|
| Yij | = | variable respuesta |
| U | = | media general |
| Ti | = | i-ésimo tratamiento ( $i = 1, 2$ ) |
| Eij | = | Error experimental |

Se consideró significativos aquellos valores con $P < 0.10$ (Crespo and Esteve-Garcia, 2001; Zhu *et al*, 2003, Kilburn and Edwards, 2004; van Nevel *et al*, 2005; Tahir *et al*, 2008; Teeter *et al*, 2008) o menores de 0.05, según se indica en cada caso.

## 3. Resultados

### 3.1. Comportamiento productivo

El comportamiento productivo de las aves se presenta en el Cuadro D y Anexos 21 a 27 y 29 a 32). Las aves sometidas a los modelos de desafío evaluados mostraron comportamientos productivos significativamente diferentes. Durante las cuatro semanas de evaluación las aves del tratamiento 2 mostraron los mayores pesos. Hasta el día 14 de edad, a pesar que el consumo de alimento fue similar en ambos tratamientos, la conversión alimentaria fue menor en el tratamiento 2. Por el contrario, a partir de esta edad, el consumo de alimento pero también la conversión alimentaria fueron mayores en las aves del tratamiento 2. La mortalidad fue reducida (6%) y no mostró diferencias entre tratamientos. Todas las aves se mostraron clínicamente sanas.

**Cuadro D. Comportamiento productivo de pollos sometidos a dos modelos de desafío entérico.**

| Variable | Tratamientos [1] 1 | | Tratamientos [1] 2 | | P |
|---|---|---|---|---|---|
| **Resultados acumulados** | | | | | |
| De 0 a 7 días de edad | | | | | |
| Peso inicial (en la recepción), g | 48.2 | | 47.9 | | 0.5797 |
| Peso final (día 7 de edad), g | 178.0 | b | 193.3 | a | 0.0093 |
| Ganancia de peso, g | 129.7 | b | 145.4 | a | 0.0096 |
| Consumo de alimento, g/pollo | 194.2 | | 196.8 | | 0.2709 |
| Conversión alimentaria | 1.50 | a | 1.36 | b | 0.0364 |
| De 0 a 14 días de edad | | | | | |
| Peso final (día 14 de edad), g | 379.6 | b | 430.2 | a | 0.0016 |
| Ganancia de peso, g | 331.2 | b | 382.0 | a | 0.0018 |
| Consumo de alimento, g/pollo | 532.3 | | 555.4 | | 0.2038 |
| Conversión alimentaria | 1.61 | a | 1.46 | b | 0.0134 |
| De 0 a 21 días de edad | | | | | |
| Peso final (día 21 de edad), g | 737.1 | b | 848.4 | a | <0.0001 |
| Ganancia de peso, g | 688.5 | b | 800.0 | a | <0.0001 |
| Consumo de alimento, g/pollo | 963.8 | b | 1130.1 | a | <0.0001 |
| Conversión alimentaria | 1.40 | | 1.41 | | 0.6235 |
| De 0 a 28 días de edad | | | | | |
| Peso final (día 28 de edad), g | 1175.7 | b | 1295.3 | a | 0.0106 |
| Ganancia de peso, g | 1127.4 | b | 1247.0 | a | 0.0107 |
| Consumo de alimento, g/pollo | 1562.9 | b | 1897.4 | a | 0.0002 |
| Conversión alimentaria | 1.38 | b | 1.53 | a | 0.0228 |
| Mortalidad, % | 7.81 | | 4.69 | | 0.5125 |
| **Resultados parciales** | | | | | |
| 2° semana de edad | | | | | |
| Ganancia de peso, g | 200.2 | b | 237.2 | a | 0.0039 |
| Consumo de alimento, g/pollo | 338.1 | | 358.6 | | 0.2460 |
| Conversión alimentaria | 1.70 | a | 1.51 | b | 0.0200 |
| 3° semana de edad | | | | | |
| Ganancia de peso, g | 350.0 | b | 410.9 | a | 0.0002 |
| Consumo de alimento, g/pollo | 431.5 | b | 574.6 | a | <0.0001 |
| Conversión alimentaria | 1.24 | b | 1.40 | a | 0.0095 |
| 4° semana de edad | | | | | |
| Ganancia de peso, g | 408.0 | | 418.7 | | 0.6983 |
| Consumo de alimento, g/pollo | 599.1 | b | 767.3 | a | 0.0055 |
| Conversión alimentaria | 1.46 | b | 1.87 | a | 0.0122 |

[1] Tratamientos 1 y 2: aves sometidas al modelo de desafío A o B, respectivamente.

a, b: Promedios en una misma fila con la misma letra no son significativamente diferentes.

### 3.2. Evaluación de heces

Las heces de las aves sometidas a ambos modelos de desafío mostraron alteraciones propias de los trastornos entéricos: de 64 a 76% de las heces fueron acuosas, 40 a 65% contuvieron alimento sin digerir, y hasta 7% presentaron descamaciones mucosas (Cuadro E y Anexos 33 y 34). La proporción de heces acuosas fue significativamente mayor en la primera y segunda semana de edad en el tratamiento 1, mientras que en la cuarta semana de edad fue mayor en el tratamiento 2 (Gráfico A). La proporción de heces con alimento sin digerir mostró la misma tendencia, siendo significativamente mayor durante las primeras 2 semanas en el tratamiento 1 y durante el periodo subsecuente en el tratamiento 2. La presencia de descamaciones mucosas en las heces fue escasa en las 2 primeras semanas de vida; sin embargo, se observó con mayor frecuencia a partir de los 14 días de edad, encontrándose una proporción significativamente mayor en el tratamiento 2 en la cuarta semana. No se observó heces hemorrágicas. El índice Dimar refleja una incidencia de alteraciones significativamente mayor en el tratamiento 1 durante las dos primeras semanas de edad y en el tratamiento 2 en las semanas subsecuentes. Esta variable mostró, en todos los muestreos, la menor probabilidad de error estadístico ($P<0.0001$).

### 3.3. Evidencias directas de daño intestinal

En las dos primeras semanas de vida se observó enteritis entre 13 y 25% de las aves como única característica patológica; por el contrario, en la tercera y cuarta semanas la incidencia de enteritis se incrementó afectando entre 63 y 75% de las aves, no encontrándose diferencias significativas entre tratamientos (Cuadro F y Anexos 35 a 38). En las semanas 3 y 4 de edad se observó además lesiones compatibles con *C. perfringens* y dos especies de coccidia en ambos tratamientos (Gráfico B), con scores entre 0.25 y 0.75 para *C. perfringens*, hasta 0.50 para *E. maxima*, y entre 0.13 y 1.50 para *E. acervulina*, siendo mayor en el tratamiento 2 en ambas semanas. En la tercera y cuarta semana de edad tanto la evaluación macroscópica como microscópica indicaron un mayor porcentaje de aves positivas a coccidia en el tratamiento 2.

**Cuadro E. Evaluación de heces en pollos sometidos a dos modelos de desafío entérico.**

| Variable | Tratamientos [1] 1 | | Tratamientos [1] 2 | | P |
|---|---|---|---|---|---|
| De 0 a 7 días de edad | | | | | |
| Heces normales, % | 24.97 | £ | 29.08 | $ | 0.0586 |
| Heces acuosas, % | 75.03 | $ | 70.92 | £ | 0.0586 |
| Heces con alimento sin digerir, % | 65.64 | $ | 60.70 | £ | 0.0581 |
| Heces con descamaciones mucosas, % | 0.00 | | 0.00 | | - |
| Heces hemorrágicas, % | 0.00 | | 0.00 | | - |
| Índice Dimar | 35.17 | a | 32.90 | b | 0.0054 |
| De 8 a 14 días de edad | | | | | |
| Heces normales, % | 24.79 | b | 35.22 | a | 0.0003 |
| Heces acuosas, % | 75.21 | a | 64.78 | b | 0.0003 |
| Heces con alimento sin digerir, % | 59.63 | a | 49.83 | b | 0.0001 |
| Heces con descamaciones mucosas, % | 0.60 | | 0.66 | | 0.9448 |
| Heces hemorrágicas, % | 0.00 | | 0.00 | | - |
| Índice Dimar | 33.86 | a | 28.82 | b | <0.0001 |
| De 15 a 21 días de edad | | | | | |
| Heces normales, % | 25.14 | | 24.41 | | 0.7575 |
| Heces acuosas, % | 74.86 | | 75.59 | | 0.7575 |
| Heces con alimento sin digerir, % | 50.73 | b | 64.99 | a | <0.0001 |
| Heces con descamaciones mucosas, % | 2.43 | | 3.05 | | 0.6390 |
| Heces hemorrágicas, % | 0.00 | | 0.00 | | - |
| Índice Dimar | 32.01 | b | 35.91 | a | <0.0001 |
| De 22 a 28 días de edad | | | | | |
| Heces normales, % | 29.89 | a | 24.07 | b | 0.0332 |
| Heces acuosas, % | 70.11 | b | 75.93 | a | 0.0332 |
| Heces con alimento sin digerir, % | 40.17 | b | 55.09 | a | <0.0001 |
| Heces con descamaciones mucosas, % | 5.41 | b | 7.25 | a | 0.0256 |
| Heces hemorrágicas, % | 0.00 | | 0.00 | | - |
| Índice Dimar | 28.92 | b | 34.57 | a | <0.0001 |

[1] Tratamientos: 1: aves sometidas al modelo de desafío A; 2: aves sometidas al modelo de desafío B.

a,b,$,£ Promedios significativamente diferentes no comparten la misma letra (a,b; $P<0.05$) o el mismo símbolo ($,£; $0.05<P<0.10$).

**Gráfico A.** **Alteraciones en heces de pollos sometidos a dos modelos de desafío entérico.**

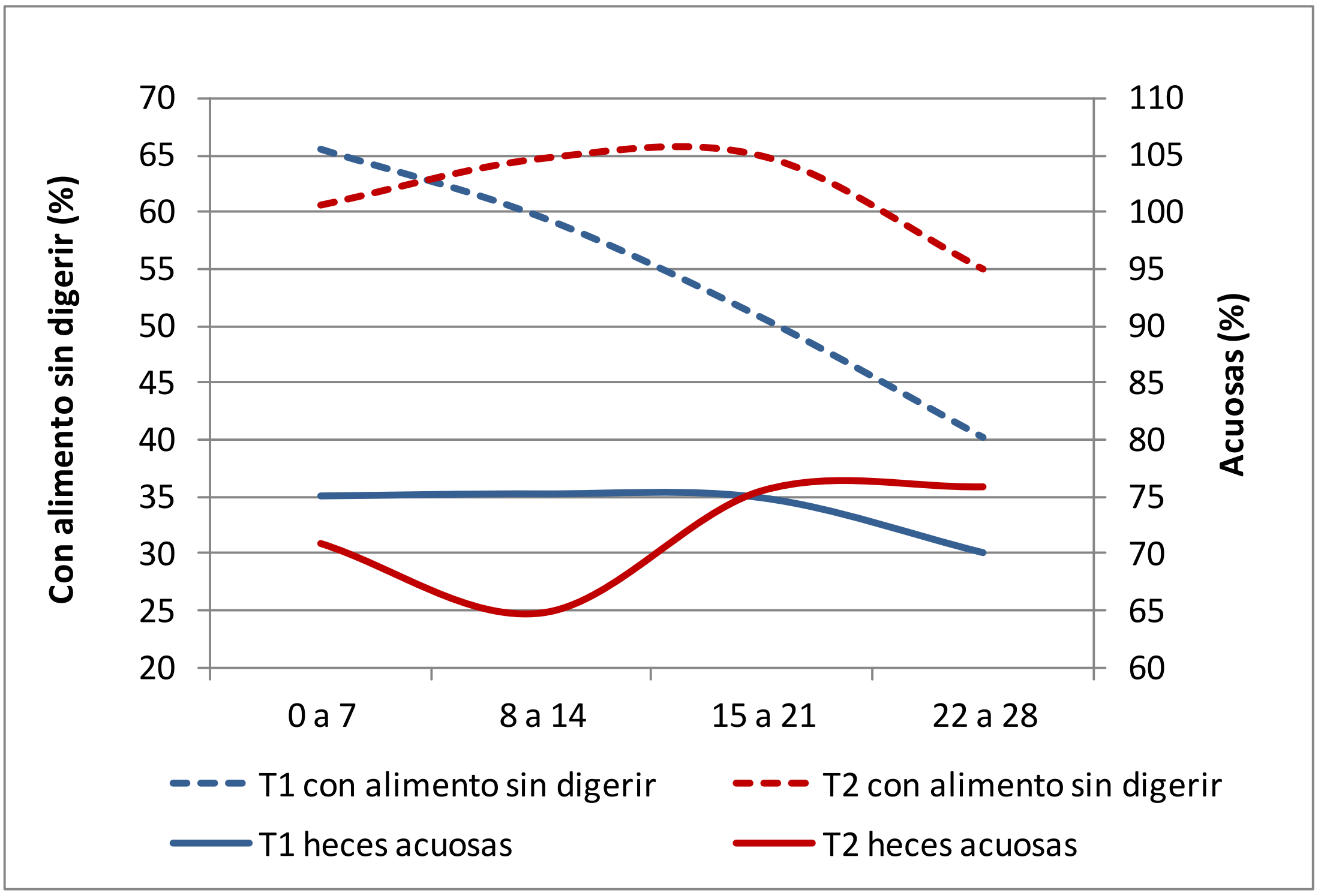

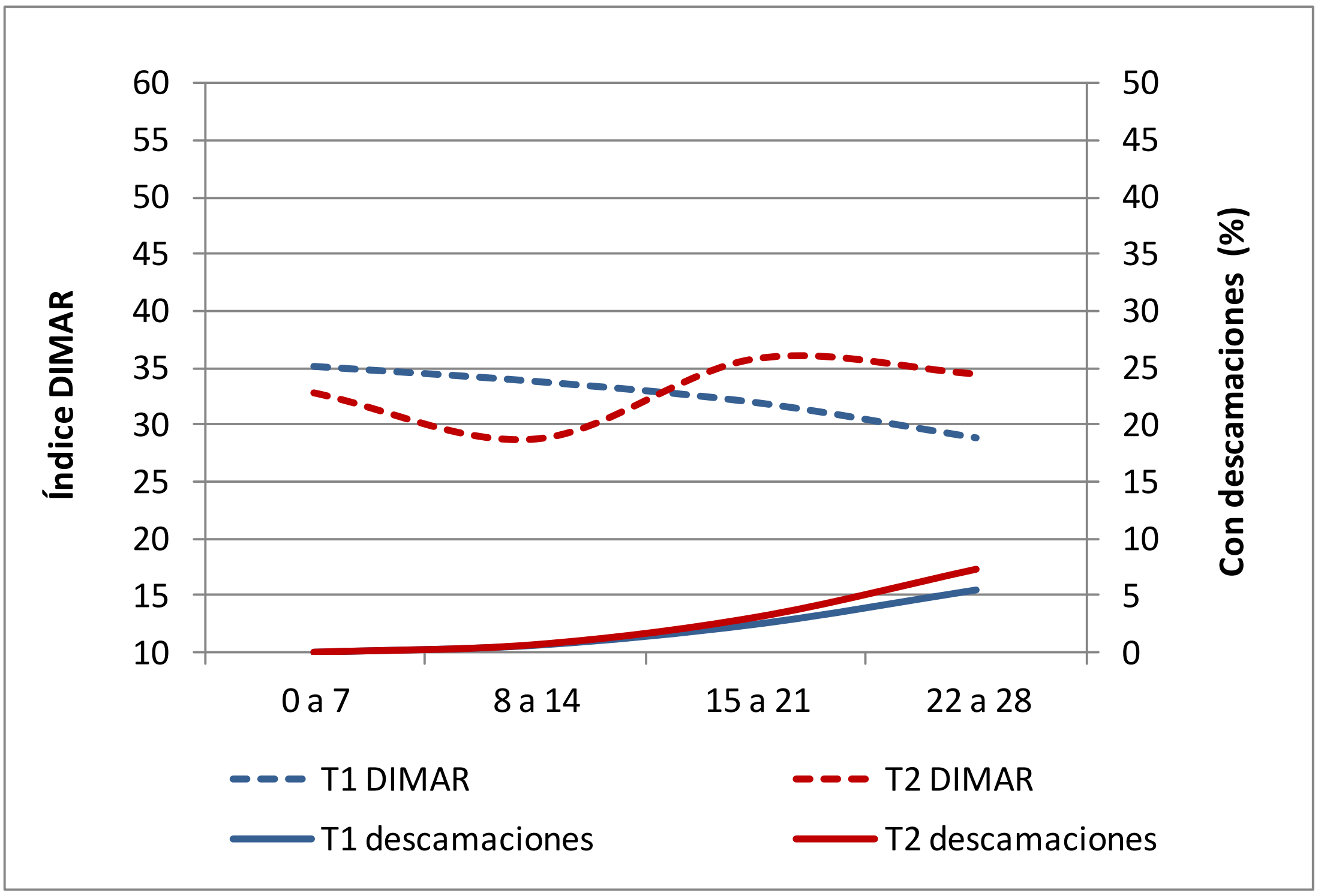

**Cuadro F. Evidencias de daño intestinal en pollos sometidos a dos modelos de desafío entérico**

| Variable | Tratamientos [1] 1 | | Tratamientos [1] 2 | | P |
|---|---|---|---|---|---|
| Día 21 de edad | | | | | |
| Enteritis [2] | 0.63 | | 0.63 | | 1.0000* |
| Presencia de lesiones intestinales [3] | 0.38 | | 0.75 | | 0.1432* |
| Densidad de las lesiones, N° /cm2 | 0.50 | b | 6.50 | a | 0.0128* |
| Score de lesiones por *C. perfringens* [4] | 0.25 | | 0.38 | | 0.6015* |
| Score de lesiones por *E. acervulina* [5] | 0.13 | b | 1.38 | a | 0.0242* |
| Score de lesiones por *E. maxima* [5] | 0.50 | | 0.38 | | 0.8897* |
| Aves positivas a *E. acervulina* (macroscópico) | 13% | b | 63% | a | 0.0455* |
| Aves positivas a *E. maxima* (macroscópico) | 25% | | 25% | | 1.0000* |
| Aves positivas a coccidia (macroscópico) [6] | 25% | £ | 75% | $ | 0.0528* |
| Score microscópico de coccidia [7] | 0.38 | | 0.88 | | 0.1777* |
| Aves positivas a coccidia (microscópico) [8] | 25% | | 63% | | - |
| Día 28 de edad | | | | | |
| Enteritis [2] | 0.75 | | 0.63 | | 0.6015* |
| Presencia de lesiones intestinales [3] | 0.75 | | 0.88 | | 0.5351* |
| Densidad de las lesiones, N° /cm2 | 4.00 | b | 14.38 | a | 0.0105* |
| Score de lesiones por *C. perfringens* [4] | 0.75 | | 0.50 | | 0.3173* |
| Score de lesiones por *E. acervulina* [5] | 0.63 | £ | 1.50 | $ | 0.0615* |
| Score de lesiones por *E. maxima* [5] | 0.25 | | 0.00 | | 0.3173* |
| Aves positivas a *E. acervulina* (macroscópico) | 38% | b | 88% | a | 0.0455* |
| Aves positivas a *E. maxima* (macroscópico) | 13% | | 0% | | 0.3173* |
| Aves positivas a coccidia (macroscópico) [6] | 38% | b | 88% | a | 0.0455* |
| Score microscópico de coccidia [7] | 2.13 | | 2.50 | | 0.3858* |
| Aves positivas a coccidia (microscópico) [8] | 88% | | 100% | | - |

1 Tratamientos 1 y 2: aves sometidas a los modelos de desafío A o B, respectivamente.
2 Score: 0: no se observa; 1: se observa congestión en alguna sección del intestino.
3 Presencia de lesiones intestinales. Score: 0: no presenta; 1, presenta.
4 Score: 0: no presenta; 1: intestino delgado con la pared adelgazada o friable; 2: necrosis focal o ulceración; 3: placas más grandes de necrosis; 4: necrosis severa y extensa.
5 Score de lesiones de Johnson and Reid (1970) en una escala de 0 a 4.
6 Porcentaje de aves que presentan lesiones macroscópicas por el género *Eimeria sp.*
7 Score: 0: ausente; 1: escasos; 2: cantidad moderada; 3: abundantes.
8 Medido como presencia o ausencia de ooquistes en el raspado de la mucosa intestinal.
a,b,$,£ Promedios significativamente diferentes no comparten la misma letra (a,b; $P<0.05$) o el mismo símbolo ($,£; $0.05<P<0.10$).
* Datos analizados empleando la prueba de Kruskal-Wallis.

**Gráfico B.** **Evidencias de daño intestinal en pollos sometidos a dos modelos de desafío entérico.**

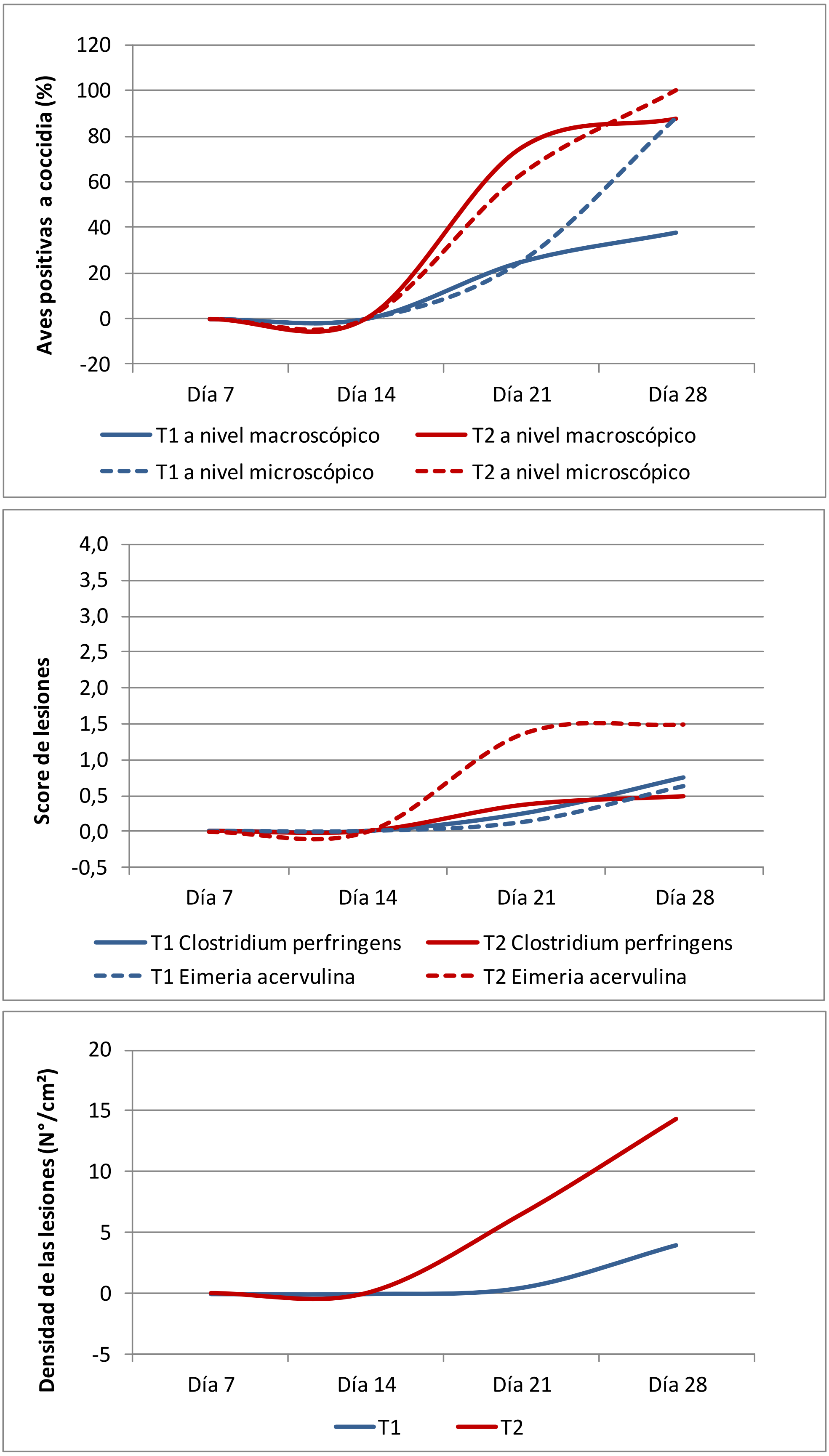

### 3.4. Morfometría de los órganos linfoides

En términos generales los órganos linfoides de las aves de ambos tratamientos mostraron características morfométricas similares en las tres primeras semanas de vida (Cuadro G y Anexos 39 a 41); con excepción del índice morfométrico de la bursa y del diámetro de la misma, que en el día 7 de edad fueron significativamente menores en las aves del tratamiento 1. En el día 21 de edad se observa mayor diámetro de la bursa en el tratamiento 2 pero mayor índice morfométrico del bazo en el tratamiento 1, mientras que en el día 28 de edad se observa diferencias estadísticas significativas en el diámetro de la bursa y en los índices morfométricos del bazo y el timo, siendo todos mayores en las aves del tratamiento 2.

### 3.5. Características de la carcasa y biometría de órganos viscerales

Las aves del tratamiento 2 mostraron un mayor rendimiento de carcasa (Cuadro H y Anexos 42 a 44) en los días 7 y 28 de edad y el mayor rendimiento de pechuga y porcentaje de pechuga en la carcasa en los días 7 y 21 de edad. Las aves del tratamiento 1 mostraron mayor peso relativo del páncreas en el día 14 de edad, mayores pesos relativos del hígado en los días 21 y 28 de edad y mayor peso relativo del intestino en el día 21 de edad, mientras que se observa menor relación páncreas/hígado en estas aves en el día 28 de edad.

## 4. Discusión

Los modelos evaluados tienen como objetivo reproducir el STR bajo los parámetros de ocurrencia en campo; es decir, de manera sub-clínica, en que el comportamiento productivo muestra un impacto poco evidente pero económicamente importante. Para ello se empleó una combinación de factores predisponentes: contenido de proteína dietaria, digestibilidad de esta proteína, inoculación de *C. perfringens*, inoculación de coccidia, y uso de material de cama reutilizado (Cuadro A), empleando pollos machos por la mayor predisposición de este sexo debido a su mayor velocidad de crecimiento (Kouwenhoven *et al*, 1978; Zavala and Barbosa, 2006).

**Cuadro G. Morfometría de los órganos linfoides en pollos sometidos a dos modelos de desafío entérico.**

| Variable | Tratamientos [1] 1 | Tratamientos [1] 2 | P |
|---|---|---|---|
| Día 7 de edad | | | |
| Diámetro de la bursa [2] | 3.00 £ | 3.50 $ | 0.0824 |
| Índice morfométrico de la bursa (Rbu) | 1.85 b | 2.37 a | 0.0376 |
| Índice morfométrico del bazo (Rba) | 0.92 | 1.32 | 0.3040 |
| Índice morfométrico del timo (Rti) | 5.65 | 5.83 | 0.7614 |
| Relación bursa/bazo (Bu-Ba) | 2.09 | 2.47 | 0.5007 |
| Relación bursa/timo (Bu-Ti) | 0.34 | 0.42 | 0.1621 |
| Relación timo/bazo (Ti-Ba) | 6.26 | 5.82 | 0.7036 |
| Día 14 de edad | | | |
| Diámetro de la bursa [2] | 4.50 | 4.50 | 1.0000 |
| Índice morfométrico de la bursa (Rbu) | 2.35 | 2.55 | 0.6484 |
| Índice morfométrico del bazo (Rba) | 1.20 | 1.06 | 0.3646 |
| Índice morfométrico del timo (Rti) | 6.71 | 6.21 | 0.4946 |
| Relación bursa/bazo (Bu-Ba) | 2.03 | 2.44 | 0.2509 |
| Relación bursa/timo (Bu-Ti) | 0.36 | 0.42 | 0.3688 |
| Relación timo/bazo (Ti-Ba) | 6.00 | 6.28 | 0.8212 |
| Día 21 de edad | | | |
| Diámetro de la bursa [2] | 5.38 £ | 6.00 $ | 0.0742 |
| Índice morfométrico de la bursa (Rbu) | 2.68 | 2.53 | 0.7046 |
| Índice morfométrico del bazo (Rba) | 1.58 $ | 1.18 £ | 0.0912 |
| Índice morfométrico del timo (Rti) | 5.40 | 5.67 | 0.7183 |
| Relación bursa/bazo (Bu-Ba) | 1.89 | 2.22 | 0.4368 |
| Relación bursa/timo (Bu-Ti) | 0.55 | 0.45 | 0.2797 |
| Relación timo/bazo (Ti-Ba) | 4.24 | 5.08 | 0.4981 |
| Día 28 de edad | | | |
| Diámetro de la bursa [2] | 6.13 b | 7.25 a | 0.0114 |
| Índice morfométrico de la bursa (Rbu) | 2.32 | 2.91 | 0.1273 |
| Índice morfométrico del bazo (Rba) | 0.98 b | 1.41 a | 0.0462 |
| Índice morfométrico del timo (Rti) | 4.60 £ | 5.74 $ | 0.0791 |
| Relación bursa/bazo (Bu-Ba) | 2.54 | 2.37 | 0.7590 |
| Relación bursa/timo (Bu-Ti) | 0.51 | 0.52 | 0.8880 |
| Relación timo/bazo (Ti-Ba) | 5.20 | 4.64 | 0.6106 |

[1] Tratamientos 1 y 2: aves sometidas a los modelos de desafío A o B, respectivamente.
[2] Medido con un bursómetro en una escala de 1/8 a 8/8 de pulgada.
a,b,$,£ Promedios significativamente diferentes no comparten la misma letra (a,b; P<0.05) o el mismo símbolo ($,£; 0.05<P<0.10).

**Cuadro H. Características de carcasa y biometría de órganos viscerales de pollos sometidos a dos modelos de desafío entérico.**

| Variable | Tratamientos [1] 1 | Tratamientos [1] 2 | P |
|---|---|---|---|
| Día 7 de edad | | | |
| Rendimiento de carcasa, % | 63.89 £ | 65.78 $ | 0.0678 |
| Rendimiento de pechuga, % | 12.32 b | 13.63 a | 0.0031 |
| Porcentaje de pechuga en la carcasa, % | 19.30 b | 20.71 a | 0.0146 |
| Relación carcasa/pechuga | 5.20 a | 4.84 b | 0.0136 |
| Peso relativo del intestino, % | 9.72 | 9.63 | 0.8877 |
| Peso relativo del hígado, % | 4.15 | 3.96 | 0.4641 |
| Peso relativo del páncreas, % | 0.48 | 0.43 | 0.1984 |
| Relación páncreas/hígado | 0.118 | 0.109 | 0.5040 |
| Día 14 de edad | | | |
| Rendimiento de carcasa, % | 71.06 | 67.96 | 0.2651 |
| Rendimiento de pechuga, % | 16.92 | 18.27 | 0.1129 |
| Porcentaje de pechuga en la carcasa, % | 23.85 | 27.28 | 0.1115 |
| Relación carcasa/pechuga | 4.22 $ | 3.77 £ | 0.0818 |
| Peso relativo del intestino, % | 7.66 | 7.24 | 0.4361 |
| Peso relativo del hígado, % | 3.72 | 3.30 | 0.1553 |
| Peso relativo del páncreas, % | 0.43 $ | 0.37 £ | 0.0954 |
| Relación páncreas/hígado | 0.115 | 0.114 | 0.9602 |
| Día 21 de edad | | | |
| Rendimiento de carcasa, % | 72.18 | 72.79 | 0.5042 |
| Rendimiento de pechuga, % | 18.78 b | 20.52 a | 0.0014 |
| Porcentaje de pechuga en la carcasa, % | 26.02 b | 28.21 a | 0.0043 |
| Relación carcasa/pechuga | 3.85 a | 3.55 b | 0.0049 |
| Peso relativo del intestino, % | 6.33 $ | 5.56 £ | 0.0587 |
| Peso relativo del hígado, % | 3.46 a | 2.91 b | 0.0316 |
| Peso relativo del páncreas, % | 0.33 | 0.35 | 0.3422 |
| Relación páncreas/hígado | 0.097 b | 0.123 a | 0.0012 |
| Día 28 de edad | | | |
| Rendimiento de carcasa, % | 69.30 £ | 71.99 $ | 0.0714 |
| Rendimiento de pechuga, % | 19.73 | 21.05 | 0.1142 |
| Porcentaje de pechuga en la carcasa, % | 28.48 | 29.25 | 0.4281 |
| Relación carcasa/pechuga | 3.52 | 3.43 | 0.4147 |
| Peso relativo del intestino, % | 6.05 | 6.15 | 0.8578 |
| Peso relativo del hígado, % | 3.13 a | 2.64 b | 0.0269 |
| Peso relativo del páncreas, % | 0.28 | 0.29 | 0.7271 |
| Relación páncreas/hígado | 0.091 £ | 0.113 $ | 0.0824 |

[1] Tratamientos 1 y 2: aves sometidas a los modelos de desafío A o B, respectivamente.

a,b,$,£ Promedios significativamente diferentes no comparten la misma letra (a,b; $P<0.05$) o el mismo símbolo ($,£; $0.05<P<0.10$).

Los modelos de desafío evaluados determinan resultados diferentes en cuanto a trastornos entéricos (Cuadros E y F) y comportamiento productivo (Cuadro D). La respuesta observada en las aves del tratamiento 2 muestra mayor concordancia con las condiciones de campo en términos comportamiento productivo, ya que si bien la ganancia de peso es mayor que en las aves del tratamiento 1, y se aproxima más a la curva de crecimiento estándar de la línea genética, también es mayor la conversión alimentaria, tal como se muestra en el Gráfico C, determinando una menor eficiencia productiva de la parvada y consecuentemente mayor costo de producción.

**Gráfico C. Comportamiento productivo de pollos bajo dos modelos de desafío.**

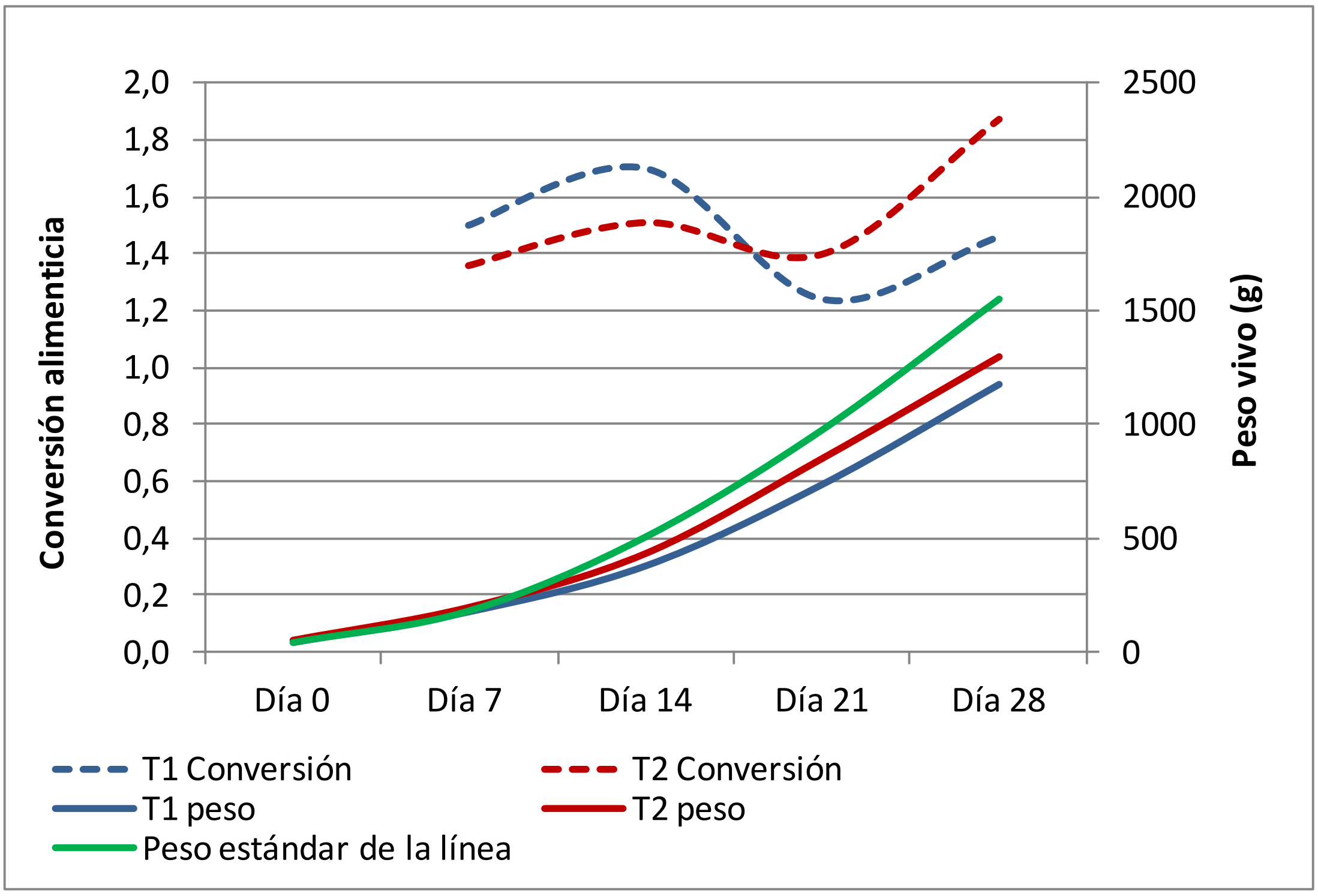

### 4.1. Etapa de 1 a 14 días de edad

Durante las dos primeras semanas de vida, los principales componentes de los modelos de desafío que determinan las diferencias entre ambos tratamientos se relacionan con el tipo de material de cama empleado y con el aporte y digestibilidad de la proteína dietaria.

### 4.1.1. Influencia del material de cama

Los principales factores procedentes del material de cama reusado, que en condiciones de campo influyen negativamente sobre el comportamiento productivo de las aves, son la mayor producción de amoniaco y la mayor carga de microorganismos patógenos. Sin embargo, durante el periodo de evaluación, todas las aves estuvieron en el mismo ambiente y, en consecuencia, las aves de ambos tratamientos fueron expuestas a las mismas concentraciones en el aire del amoniaco procedente del material de cama. Por esta razón se considera que la concentración de amoniaco ambiental no ejerce mayor influencia entre los tratamientos evaluados.

En relación a la carga microbiológica del material de cama, se ha establecido que el reúso de este material puede transmitir patógenos virales, bacterianos, micóticos y parasitarios, entre los que resaltan los virus inmunosupersores como el de la Enfermedad de Gumboro, el de la Anemia Infecciosa Aviar (Tambini *et al*, 2010) y el Reovirus, estando este último vinculado a diversos síndromes entéricos (Dustan and Jones, 2008). Sin embargo, cuando el material es previamente tratado y proviene de crianzas sin procesos infecciosos puede ser utilizado de manera segura (Vejarano, 2005; Vejarano *et al*, 2008). Al respecto, Roll *et al* (2011) observaron que el reúso de cama reduce el recuento de *Salmonella sp.*, mientras que Tambini *et al* (2010) observaron que los pollos alojados en material de cama nuevo o reutilizado por cinco campañas, presentan diferencias en el peso e índice morfométrico de la bursa de Fabricio pero no en el peso corporal de las aves.

Tomando en cuenta lo anterior, en el presente estudio se empleó un material de cama reutilizado procedente de un plantel de pollos de carne sin antecedentes de procesos respiratorios infecciosos atribuibles a virus y con resultados productivos satisfactorios para descartar procesos inmuno-depresivos asociados a la Enfermedad de Gumboro o síndromes entéricos (Anexo 6). Asimismo, el proceso de compostaje al que fue sometido el material de cama puede haber reducido la probabilidad de transmisión de síndromes entéricos infecciosos (Dustan and Jones, 2008) en caso la parvada de la cual provino dicho material hubiera presentado algún síndrome entérico. Los recuentos microbiológicos realizados como parte del presente estudio a los materiales de cama nuevo y reutilizado indican diferencias poco relevantes en cuanto a bacterias aerobias mesófilas, coliformes y *C. perfringens* (Anexo 6). El proceso de acondicionamiento aplicado al material de cama reusado se presenta en el Anexo 6.

Numerosos reportes han vinculado a agentes virales como el reovirus con algunos síndromes intestinales que cursan con tránsito rápido y mala absorción del alimento (Bustamante y Chávez, 2010; Songserm *et al*, 2002; Zavala and Sellers, 2005; Dustan and Jones, 2008). Una característica común a todos estos casos es la presencia de quistes envolviendo las criptas de Lieberkühn afectando principalmente el yeyuno (Songserm *et al*, 2002), anomalía catalogada como “enteropatía cística” para luego de evolucionar hacia una lesión inflamatoria convertirse en “enteritis cística” (Sályi and Glávits, 1999) y producir acortamiento y atrofia de las vellosidades intestinales (Songserm *et al*, 2002; Zavala and Sellers, 2005). Otras características comunes en los cuadros asociados a virus es la presencia de proventriculitis (Songserm *et al,* 2002; Pantin-Jackwood *et al,* 2005; Dormitorio *et al*, 2007; McNulty *et al*, 2008; Kutkat *et al*, 2010) y emplume deficiente (Songserm *et al*, 2002), y en algunos casos osteoporosis, “enfermedad del hueso quebradizo” o “necrosis de cabeza femoral” (van der Heide *et al*, 1981).

En el presente estudio, si bien se observó la rotura de la cabeza de fémur en algunas de las aves de ambos tratamientos, esta se verificó después de intentar dislocar la cabeza femoral del acetábulo, y es posible que esto fuera consecuencia de la menor mineralización ósea causada por la menor absorción de minerales producto del Síndrome de Tránsito Rápido inducido a las aves. Por otro lado, de forma complementaria se realizó una evaluación histológica del intestino de las aves afectadas no observándose enteropatía ni enteritis cística en las criptas intestinales. Las aves no presentaron problemas de emplume y no se observó proventriculitis en las necropsias realizadas al término de cada semana. Por otro lado, las gallinas reproductoras de las cuales procedieron los pollos BB empleados en el presente estudio fueron vacunadas contra reovirus, lo que confiere inmunidad pasiva a la progenie contra este agente (Urdaneta-Vargas *et al*, 2007; Shane, 2008; Lohman Animal Health, 2010). Finalmente, en el Experimento 5 se verificó que aves sometidas al modelo de desafío B presentan una respuesta favorable tras el uso de sustancias con acción antibacteriana, lo que corrobora, junto con las evaluaciones antes mencionadas, que los procesos entéricos inducidos en este experimento no estuvieron asociados a virus.

#### 4.1.2. Influencia del alimento balanceado

En vista de lo anterior, resulta evidente que las principales fuentes de variación en esta etapa fueron las características de las dietas. Al respecto, se confirmó que el balance electrolítico de 257 meq/kg de alimento en la dieta 1 y la relación (K+Cl)/Na de 5.5 (Anexo 8) se encuentran dentro de los parámetros recomendados (Mongin, 1981; Murakami *et al*, 2001; Oviedo-Rondón *et al*, 2001), no influyendo en el desafío entérico. Al respecto, Leeson and Summers (2005) reportan que, debido al alto contenido de K en la harina de soya, cuando no se utilizan fuentes de proteína animal en la dieta, la inclusión de altos niveles de soya puede conllevar a la ocurrencia de enteritis, cama húmeda y lesiones plantares en las patas.

En algunas aves del presente estudio se observó como riñones pálidos e inflamados, lesiones que son compatibles con ocratoxinas (Anexo 28; Hoerr, 2003); sin embargo, no se observan otras lesiones características de esta micotoxina como regresión acelerada de la bursa y atrofia del timo y bazo atribuíbles a inmunodepresión, ni tampoco palidez del hígado ni hipertrofia hepática (Hoerr, 2003; Shane, 2005), ya que el peso relativo del hígado en ambos tratamientos se mantiene dentro de lo esperado a través del periodo de evaluación (Cuadro H; García, 2000), así como tampoco problemas de emplume (Brown *et al*, 2008). No se observaron lesiones compatibles con otras micotoxinas (ver Anexo 28). Por estas razones, en el presente estudio no es posible considerar que las micotoxinas dietarias son un factor relevante de desafío, incluso en los primeros 14 días de edad, lo cual coincide con lo registrado por el Laboratorio de Patología Aviar, quienes no atribuyeron las lesiones encontradas a micotoxinas.

Mientras las aves del tratamiento 1 recibieron una dieta en que el 72% de la proteína provino de una torta de soya sobre-cocida con baja digestibilidad de la proteína (73% de solubilidad en KOH), las aves del tratamiento 2 recibieron una dieta en que 35% de la proteína provino de harina de pescado y sólo el 46% provino de dicha torta de soya. Ello explica la menor ganancia de peso y eficiencia en la conversión del alimento en las aves del tratamiento 1 debido a la menor disponibilidad de proteína digestible. Esto es consistente con reportes que indican que un nivel de solubilidad de la proteína de la torta de soya menor a 75% es indicativo de sobre-cocción (Campabadal, 2009) y que el efecto de este inadecuado procesamiento de la soya es la menor digestibilidad de su proteína (Celis, 2000) y consecuentemente menor ganancia de peso y eficiencia en la conversión alimentaria (Campabadal, 2008). Asimismo, guarda relación con la completa reversión en los efectos del Síndrome de Retraso en la ganancia de peso, observadas por Angel *et al* (1992) en pavos alimentados con una dieta compleja, a base de harina de pescado y girasol como principales insumos proteicos, en comparación a una dieta a base de harina de soya. La torta se soya empleada en el presente estudio fue procedente de Bolivia y su baja solubilidad concuerda con reportes previos que indican que ésta es la que tiene la menor solubilidad de proteína y contenido de lisina digestible (Campabadal, 2008). Al respecto, Morales *et al* (2001) evaluaron el efecto de la composición de la dieta sobre las características del STR y observaron heces con alimento si digerir sólo en los tratamientos conteniendo torta de soya, lo que concuerda con los hallazgos del presente estudio.

En las aves del tratamiento 1 se observa un peso relativo del páncreas 16% mayor que las aves del tratamiento 2 (Cuadro H), lo que guarda consistencia con la menor digestibilidad de la dieta suministrada a estas aves y con la probable hipertrofia temporal en el páncreas como mecanismo adaptativo del ave a una dieta de menor digestibilidad (Applegarth *et al*, 1964). Al respecto, se ha documentado que la inclusión de soya insuficientemente cocida y, en consecuencia, con menor digestibilidad y con altos contenidos de antinutrientes en la dietas de pollos produce menor ganancia de peso y mayor peso relativo del páncreas (Herkelman *et al*, 1993; Mogridge *et al*, 1996; Perilla *et al*, 1997). Sin embargo, en todos estos casos se asocia a la presencia de antinutrientes en la dieta por insuficiente cocción, mientras que en el presente experimento se asocia a una cocción excesiva.

El Gráfico D ilustra el aporte nutricional de las dietas y su relación con los requerimientos nutricionales (Cobb-Vantress, 2008a). El menor aporte de proteína digestible a las aves del tratamiento 1 explica su menor velocidad de crecimiento y mayor conversión alimentaria. Resulta consistente con esta observación el menor rendimiento de carcasa y porcentaje de pechuga en las aves del tratamiento 1 (Cuadro H), corroborando hallazgos previos que indican que la disponibilidad de aminoácidos como lisina tiene un efecto específico sobre la composición de la carcasa, el porcentaje de pechuga (Schutte and Pack, 1995) y la retención de proteína en la carcasa (Sibbald and Wolynetz, 1986); y que aportes de lisina dietaria superiores a lo recomendado por NRC (1994) incrementan el porcentaje de pechuga (Si *et al*, 2004).

**Gráfico D. Aporte nutricional de las dietas empleadas y requerimientos nutricionales de la línea genética.**

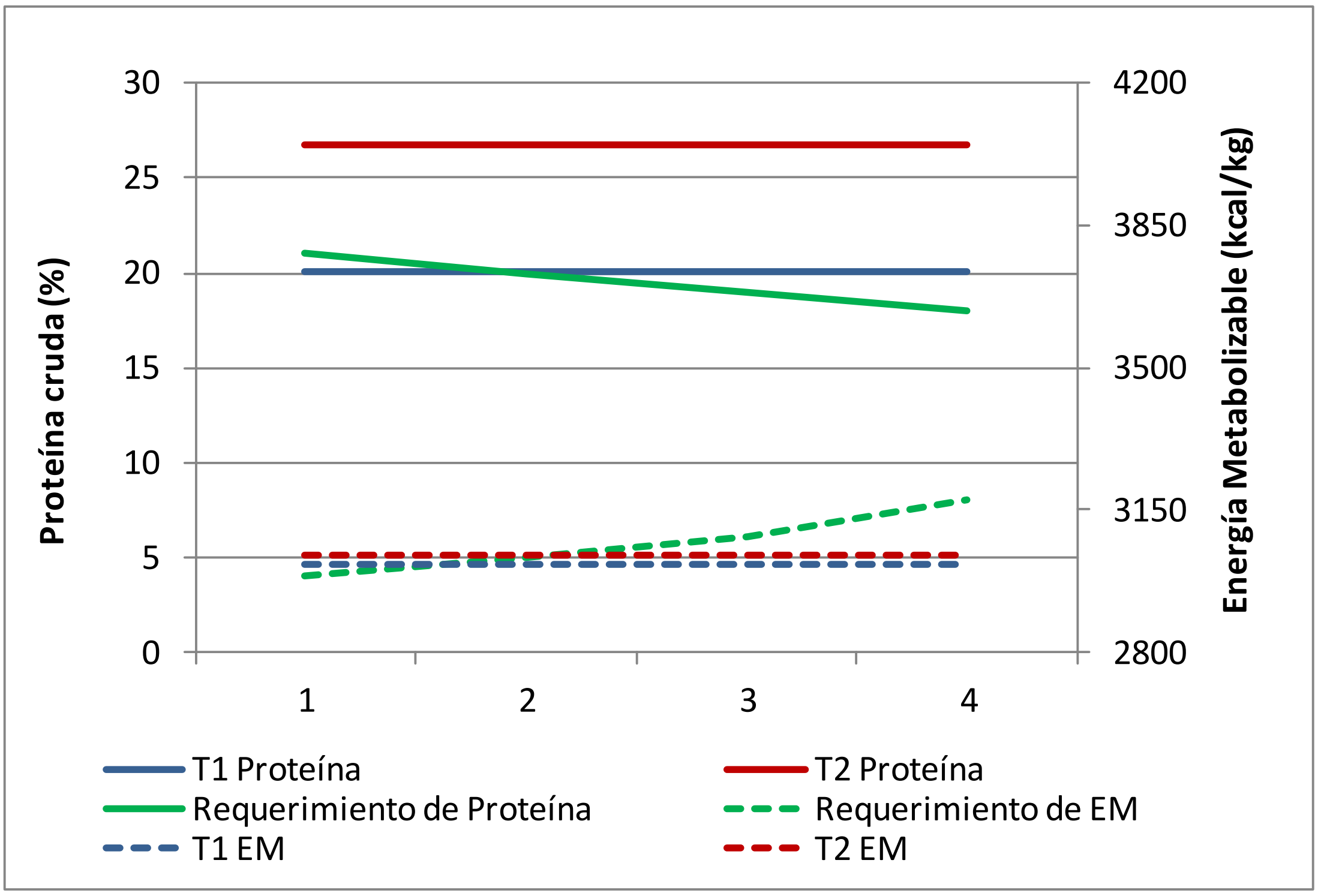

Durante las dos primeras semanas de vida las aves del tratamiento 1, sometidas al modelo de desafío A, presentaron menor ganancia de peso y mayor conversión alimentaria (Cuadro D), lo que concuerda con la esperada menor digestibilidad de la proteína dietaria y la mayor incidencia de heces acuosas y con alimento sin digerir (Cuadro E). Esto concuerda a su vez con hallazgos previos que indican que los factores dietarios que conllevan a una acumulación de nutrientes en el intestino delgado son factores de riesgo de disbacteriosis y que este crecimiento microbiano excesivo en el intestino delgado puede, a su vez, causar una disminución significativa en la digestibilidad aparente de los nutrientes (Smits *et al*, 1999), dando lugar a un círculo vicioso (Martínez, 2010) e incrementándose la presencia de alimento sin digerir en las heces.

En este periodo y sólo en el día 7 de edad se observó un menor desarrollo de la bursa en las aves del tratamiento 1, evidente en un menor tamaño e índice morfométrico (Rbu) (Cuadro G). Se sabe que menores valores del índice morfométrico bursal pueden indicar atrofia por regresión, e inmunosupresión (Kuney *et al*, 1981; Sandoval *et al*, 2002). El transitorio estado inmunodepresivo observado en el presente estudio guarda relación con el menor estado nutricional de estas aves. Al respecto, se ha establecido que si bien la inmunodepresión es un síndrome clínico inducido por diferentes factores, entre ellos se encuentra el estado nutricional (Perozo-Marín *et al*, 2004) y que por ejemplo dietas con niveles bajos de aminoácidos como arginina (Kwak *et al*, 1999) pueden inducir estados inmunodepresivos en aves. Asimismo, se ha demostrado que niveles reducidos de proteína dietaria afectan el comportamiento productivo de pollos vacunados contra coccidia (Lee *et al*, 2011b), hecho que guarda relación con la función inmune de la bursa contra la coccidia, y con la transitoria inmunodepresión en las aves del tratamiento 1, evidente en este órgano linfoide. La aparente inmunodepresión observada en el presente estudio alrededor del día 7 en las aves del tratamiento 1 concuerda con lo reportado por Giambrone and Diener (1990) quienes indican que dichos inmunodepresivos son más evidentes en animales jóvenes en los cuales el sistema inmunológico se halla en pleno desarrollo y maduración.

### 4.2. Etapa de 15 a 28 días de edad

En este periodo el modelo de desafío B, empleado en el tratamiento 2, produjo un daño intestinal significativamente mayor, evidente en un mayor score de lesiones principalmente por *E. acervulina* en el día 21, mayor porcentaje de aves positivas a coccidia en la evaluación macroscópica y mayor porcentaje de aves positivas a *E. acervulina* en la evaluación microscópica, y lesiones intestinales más densas (Cuadro F). En ambos tratamientos la incidencia de enteritis fue superior al 63% pero las lesiones por *C. perfringens* fueron de intensidad reducida (score ≤ 0.75). Al respecto, la vacuna de coccidia empleada como medio inoculante contiene oocistos vivos no atenuados (Conway and McKenzie, 2007) y se ha establecido que este tipo de vacunas predispone la Enteritis Necrótica con mayor severidad (Mathis *et al*, 2007); sin embargo, la capacidad de un modelo de desafío para inducir lesiones de Enteritis Necrótica depende de otros factores como son la dosis del inóculo, la patogenicidad de la cepa empleada (Chalmers *et al*, 2007), el tipo y la presencia del gen de la enterotoxina (Jerzsele *et al*, 2011), la especie animal de procedencia, la presencia de sustrato en el lumen intestinal (Jia *et al*, 2009), factores predisponentes de naturaleza patógena (McReynolds *et al*, 2004; Lillehoj *et al*, 2007), la duración del desafío (Olkowski *et al*, 2006; Nikpiran *et al*, 2008), entre otros.

A pesar del bajo score de lesiones compatibles con *C. perfringens* observado en este experimento (Cuadro F) no puede descartase la relevancia del inóculo bacteriano en los modelos de desafío ni en los resultados del presente estudio, por cuanto está demostrado que coccidia y *C. perfringens* son patógenos que se predisponen mutuamente (Dykstra and Reid, 1977; Al-Sheikhly and Al-Saieg, 1980), pudiendo ser la Enteritis Necrótica resultado del efecto combinado de ambos (McDougald *et al*, 2008; Williams *et al*, 2003). Así, Hoerr (2010) reporta que en Alabama, USA, la mayoría de casos de Enteritis Necrótica están asociados incluso a *E. maxima* moderada a severa. Aun cuando se observen lesiones intestinales leves, el daño que produce la toxina a nivel hepático afecta significativamente el comportamiento productivo de las aves (Martínez, 2010); asimismo, el desencadenamiento de la respuesta inmunológica y la consecuente cascada de reacciones catabólicas liberada por la activación de la respuesta de fase aguda, tiene enormes implicancias en la nutrición del ave, aumentando el gasto de mantenimiento debido al incremento en la degradación del músculo esquelético, el equilibrio negativo de nitrógeno y la redistribución de fuentes de minerales, vitaminas y energía, que aunque es esencial para una respuesta inmunitaria eficaz, resulta económicamente indeseable (Clements, 2011).

El mayor daño intestinal en las aves del tratamiento 2 a partir del día 21 (Cuadro F), 7 días después de la administración del inóculo de coccidia, concuerda con los hallazgos de campo en nuestro medio respecto a la presentación de lesiones por coccidia a partir del día 21 de edad (Martínez, 2010), con los resultados de Christaki *et al* (2004) quienes desafiaron pollos de carne con *E. tenella* en el día 14 de edad, observando un efecto negativo en el peso corporal y conversión alimentaria a partir del día 21, también 7 días después de la administración del inóculo de coccidia, y con el concomitante adelgazamiento de la mucosa intestinal esperado y que se verifica en el menor peso relativo del intestino en las aves del tratamiento 2 en el día 21 de edad (Cuadro H).

Si bien las lesiones intestinales observadas en las aves del tratamiento 1 fueron drásticamente menos densas que en las aves del tratamiento 2 (Cuadro F), su presencia, así como la existencia de algunas lesiones por *C. perfringens* y coccidia demuestra que estas aves también estuvieron sometidas a un desafío por dichos patógenos a pesar que no recibieron los respectivos inóculos, y que se implementó un estricto protocolo de bioseguridad para minimizar el riesgo de contaminación cruzada (Anexo 5), por lo que se descarta que sea la contaminación cruzada el principal factor de origen del desafío por clostridios y coccidia en las aves del tratamiento 1.

Al respecto, estudios previos han reportado la presencia de coccidia en aves no desafiadas e incluso alojadas sobre material de cama nuevo. Biswas *et al* (2001) encontraron 1200 ooquistes por gramo de cama en pollos de 28 días alojados sobre material de cama nuevo y alimentados con dietas comerciales conteniendo promotores de crecimiento y anticoccidiales. Rahmatian *et al* (2010) encontraron 550 y 14,625 oocistos por gramo de cama en los días 21 y 28 de edad, respectivamente, en pollos de carne no desafiados con coccidia y alojados sobre cama nueva de viruta de madera, pero alimentados con dietas no suplementadas con sustancias con acción anticoccidial. Asimismo, se ha establecido que la coccidia puede ser transmitida a las aves susceptibles dándoles de comer escarabajos de la cama (*Alphitobius diaperinus*), moscas, polvo de galpones de aves, entre otros (Reyna *et al*, 1983).

Los recuentos de ooquistes reportados por Rahmatian *et al* (2010) se explican por el carácter ubicuo de la coccidia (Watkins and Opengart, 1997; Elmusharaf, 2007; Batungbacal *et al*, 2007), condición que también comparte *C. perfringens*. Además, es probable que el desafío por estos patógenos en las aves del tratamiento 1 haya sido favorecido por la disbacteriosis precedente y por la menor digestibilidad del alimento, tal como lo manifiestan reportes previos que indican que las enfermedades o trastornos entéricos primarios como la disbacteriosis predisponen a la Enteritis Necrótica (Fukata *et al*, 1991), y que la sensibilización previa de la mucosa influye en la habilidad invasora de los esporozoitos de coccidia (Yuño y Gogorza, 2008).

La mayor intensidad del desafío intestinal en las aves del tratamiento 2 a partir de la tercera semana de edad se verifica en las características morfométricas de los órganos linfoides de estas aves en el día 28 de edad (Cuadro G), en que el diámetro de la bursa y los índices morfométricos del bazo (Rba) y del timo (Rti) mostraron valores significativamente mayores. Esto concuerda con los hallazgos de Deshmukh *et al* (2007) quienes reportan que aves sometidas a un desafío entérico con *Salmonella gallinarum* presentan hiperplasia del bazo. En este periodo sólo el índice morfométrico del bazo es mayor en las aves del tratamiento 1 en el día 21 de edad; sin embargo, ello no guarda relación con lo observado en los otros órganos linfoides.

Por otro lado, la relación bursa/bazo observada en ambos tratamientos y generalmente mayor a 2 (Cuadro G) es un indicador de la adecuada inmunocompetencia que presentan las aves hacia el final del periodo experimental, de acuerdo a los hallazgos de Pulido *et al* (2001), quienes reportan que valores superiores a 2 en este índice pueden ser considerados como propios de una adecuada inmunocompetencia.

El mayor daño intestinal en las aves del tratamiento 2, medido de forma directa a través de las lesiones intestinales (Cuadro F), se correlaciona directamente con las variables empleadas para cuantificar las alteraciones en las heces (Cuadro E). Al respecto, es importante resaltar que en todos los muestreos el índice Dimar es la variable que muestra la menor probabilidad de error estadístico ($P<0.0001$) entre aquellas que reflejan el estado de las heces (Cuadro E). A diferencia de scores propuestos previamente (Siceanu *et al*, 2009), el índice Dimar, empleado en este experimento, integra las diferentes y más frecuentes alteraciones observadas en heces en condiciones comerciales en una sola variable, permitiendo comparar parvadas diferentes empleando un indicador único. Estas observaciones respaldan el uso del índice Dimar para la evaluación general de las alteraciones en las heces y como indicador práctico de la integridad intestinal de las aves.

Las aves del tratamiento 1 presentan, a través del periodo de evaluación, un mayor peso relativo del hígado, siendo estas diferencias significativas en los días 21 y 28 de edad (Cuadro H). Esta diferencia puede explicarse por el menor peso corporal de las aves del tratamiento 1 a través del experimento, ya que los pesos absolutos del hígado son similares entre ambos tratamientos en todas las edades. No existen antecedentes disponibles en la literatura científica que asocien la menor digestibilidad de la proteína de la soya por efecto de sobre-cocción con la hiperplasia del hígado.

En este periodo las aves del tratamiento 2 continúan presentando mayores pesos corporales que las aves del tratamiento 1, debido principalmente al mejor crecimiento durante las dos primeras semanas de vida (Cuadro D). Sin embargo, el daño producido a la salud e integridad intestinal durante la tercera y cuarta semanas de edad por la inoculación de *C. perfringens* y coccidia (Cuadro F) determina en su conjunto la menor capacidad del animal de disponer de los nutrientes de la dieta. Ello se verifica en la menor eficiencia de conversión del alimento en las aves del tratamiento 2, que fueron sometidas al modelo de desafío B.

El modelo empleado en el tratamiento 2 induce un proceso entérico con un claro efecto en la mucosa intestinal (Cuadro F) y disminución en la capacidad de absorción de nutrientes, evidente en la presencia de alimento sin digerir en las heces (Cuadro E). Pedersen *et al* (2008) diseñaron un modelo exitoso para desarrollar un proceso entérico subclínico empleando un inóculo de *C. perfringens*, una vacuna contra coccidia en dosis alta, y una dieta basal con 50% de trigo durante el periodo de administración del inóculo, asociado a harina de pescado en proporción de 3:1. Los investigadores consideran de manera acertada que para evaluar el efecto de diferentes estrategias de intervención en la salud e integridad intestinal es necesario reducir la cantidad de trigo y harina de pescado empleados como parte del modelo. En el presente estudio se empleó, al igual que en el modelo de Pedersen *et al* (2008), un inóculo de *C. perfringens* y una vacuna de coccidia en dosis alta como material inoculante; sin embargo, se logró obviar el uso de trigo y emplear un menor contenido de harina de pescado, logrando una mejor aproximación a una dieta a base de maíz y soya.

En este experimento, el estrés producido por la manipulación de las aves durante la inoculación de *C. perfringens* y coccidia probablemente ha tenido influencia en el desafío logrado, tal como lo demuestran trabajos previos que reportan que el estrés predispone, entre otros, la disbacteriosis (Ferket *et al*, 2005), la inmunodepresión (Mostl and Palme, 2002), y cambios en el comportamiento productivo (Clements, 2011).

En el presente estudio, las aves del tratamiento 2 muestran una ganancia de peso mayor y más cercana al estándar de la línea genética, menor eficiencia en la conversión alimentaria (Gráfico C) y mayor daño intestinal (Cuadro F y Gráfico B). Por otro lado, ambos tratamientos fueron exitosos en reproducir el STR; sin embargo, el cuadro inducido por el modelo de desafío B en el tratamiento 2 se asemeja con mayor precisión a las características del STR en condiciones comerciales, en que este trastorno se presenta de manera sub-clínica con un impacto productivo muchas veces poco evidente.

Diferentes modelos de desafío empleados para inducir cuadros entéricos en pollos muestran resultados exitosos y reproducen los trastornos entéricos (Prescott *et al*, 1978; Kaldhusdal *et al*, 1999; McReynolds *et al*, 2004; Olkowski *et al*, 2006; Kulkarni *et al*, 2007; Shelton *et al*, 2007; McDougald *et al*, 2008; Nikpiran *et al*, 2008; Pedersen *et al*, 2008; Jia *et al*, 2009); sin embargo, la reproductibilidad de algunos métodos propuestos es reducida debido a la dificultad que representa controlar la contaminación cruzada (Kaldhusdal *et al*, 1999; McReynolds *et al*, 2004). La mayoría de estos trabajos fueron realizados en corrales en piso pues es necesario que las aves tengan acceso continuo al material de cama para permitir el ciclo de vida de la coccidia y la reinfección con estos patógenos (Williams, 2005; Elmusharaf *et al*, 2010). En contraste, el uso de jaulas en baterías es limitado, pues reduce el contacto de las aves con las heces. Al respecto, el método empleado en este experimento permitió el acceso continuo de las aves al material de cama y redujo el contacto innecesario del personal con los materiales contaminantes, reduciendo así el riesgo de contaminación cruzada.

## 5. Conclusiones

Los resultados obtenidos bajo las condiciones del presente estudio permiten llegar a las siguientes conclusiones:

- Los modelos de desafío empleados inducen procesos entéricos sub-clínicos de manera controlada con características similares a lo observado en condiciones de campo.
- Empleando modelos a pequeña escala como los propuestos en el presente estudio es posible optimizar los costos de investigación.
- Ambos modelos de desafío reproducen cuadros compatibles con el STR; sin embargo, los hallazgos clínicos y el comportamiento de las aves sometidas al modelo de desafío B se asemejan con mayor precisión a las características del STR observadas en condiciones de campo.
- El uso de torta de soya sobre-cocida con reducida solubilidad y digestibilidad de la proteína es un factor predisponente de los trastornos entéricos en las aves.
- El uso de vacunas contra coccidia, conteniendo oocistos vivos no atenuados, en dosis alta es un medio inoculante válido para inducir cuadros sub-clínicos de coccidiosis en pollos de carne.
- Existe una relación positiva entre la incidencia e intensidad de las lesiones intestinales y la conversión alimentaria.
- El índice Dimar es un indicador válido para evaluar de manera integral el estado de las heces y estimar el daño a la mucosa intestinal.

# ANEXO 2. Adaptaciones realizadas a las jaulas para contener cama

## Adaptaciones realizadas

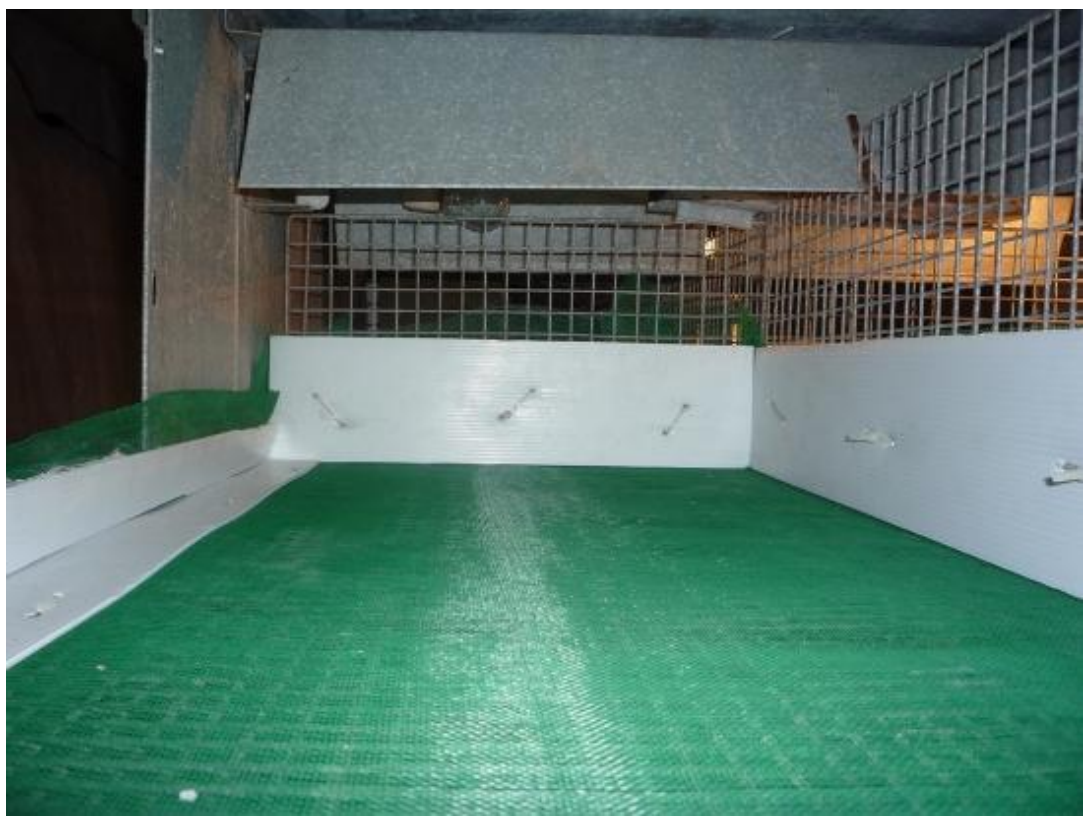

Medidas: 41.5 x 90 cm

## Jaulas conteniendo material de cama

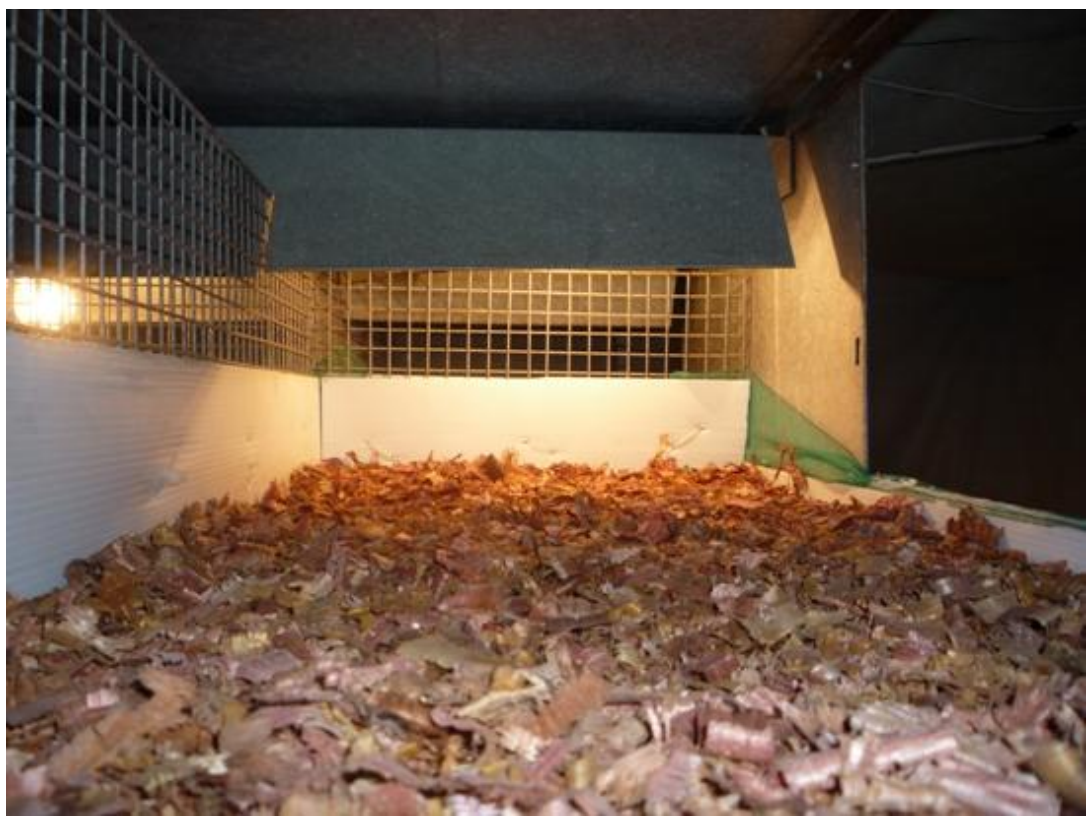

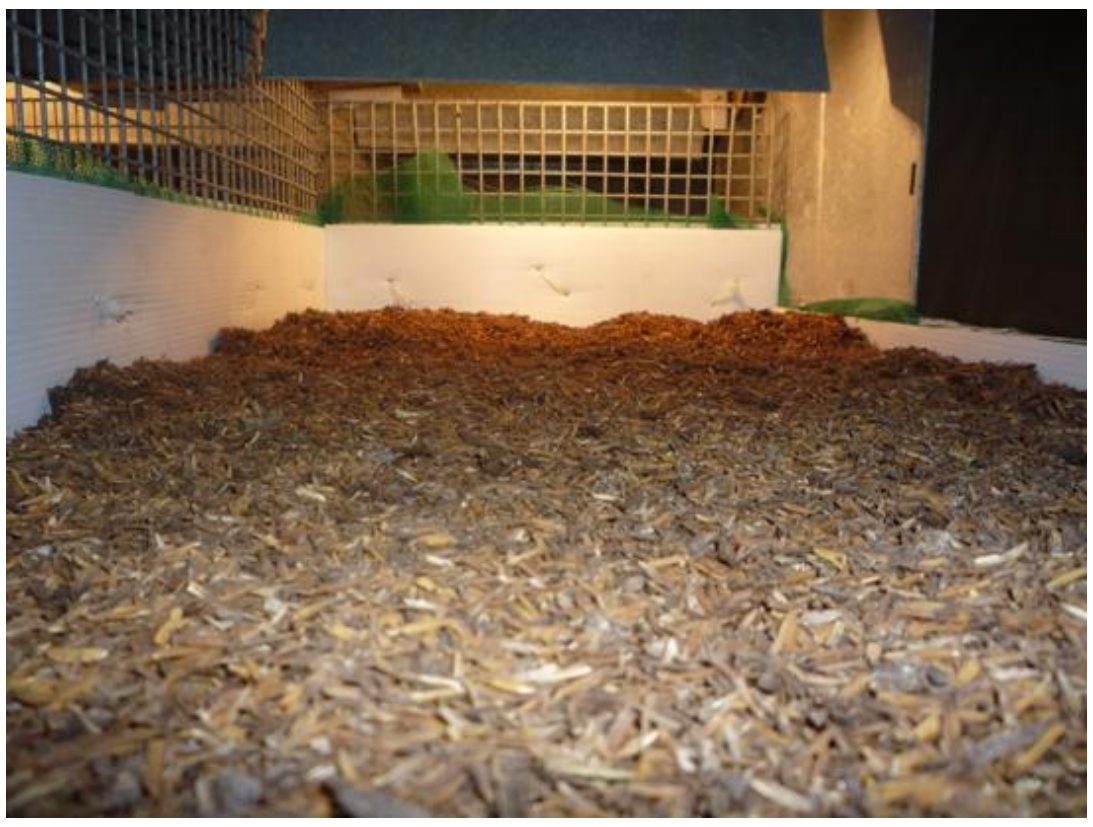

## Aves alojadas sobre material de cama

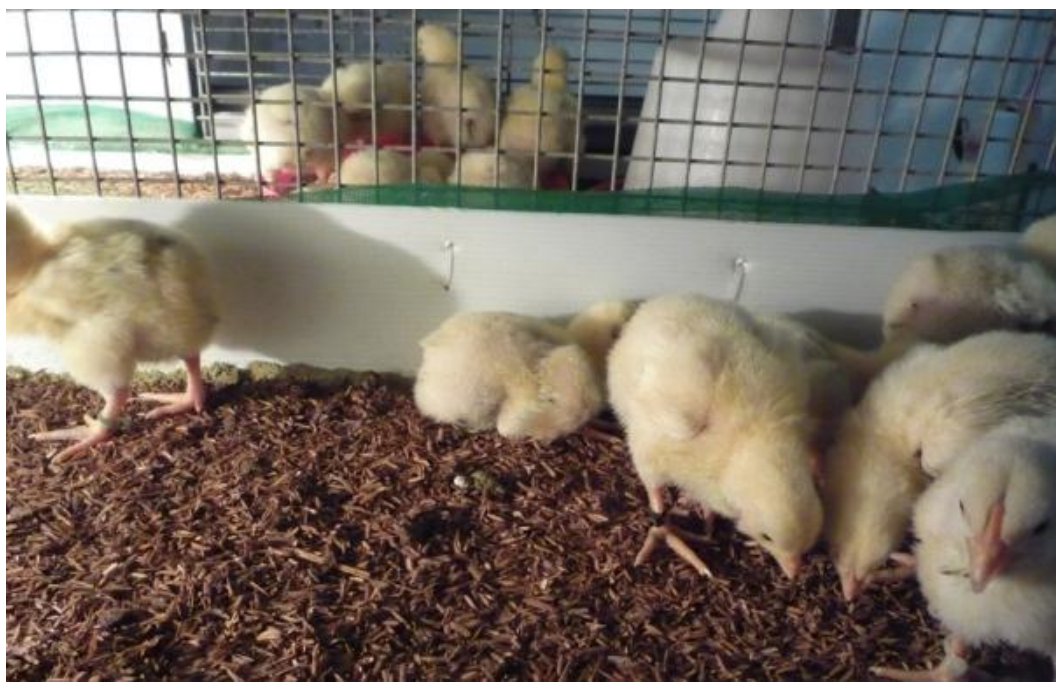

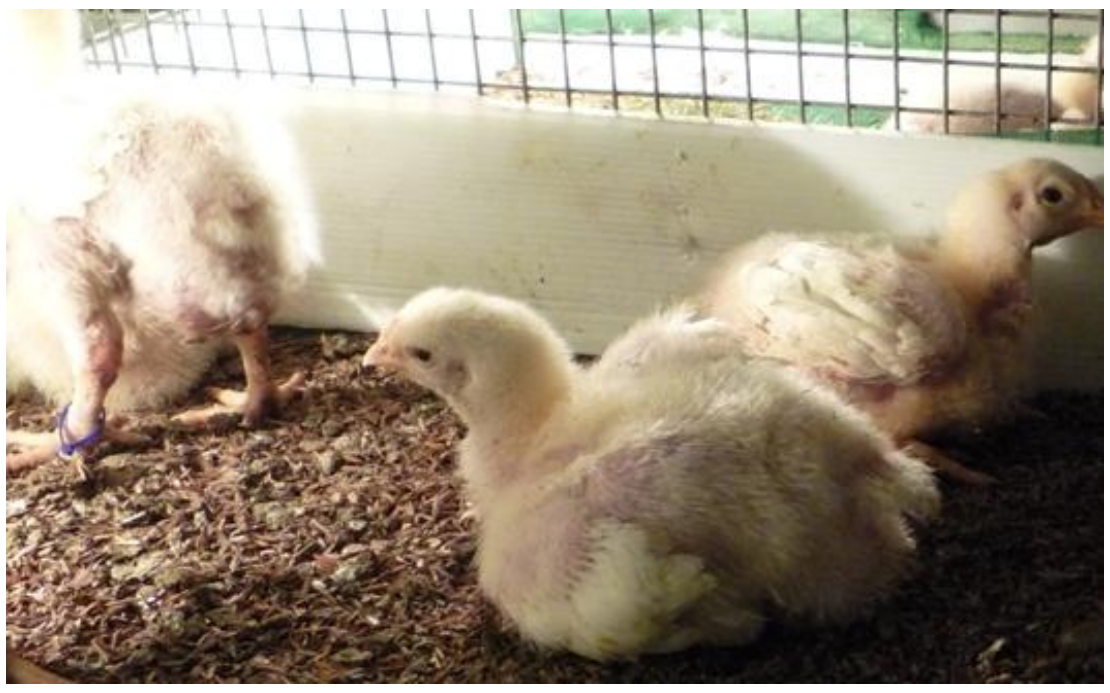

## ANEXO 3. Método desarrollado para la identificación individual de las aves

**Sistema de identificación:** Las aves fueron identificadas empleando un código numérico que asocia el grupo o jaula de procedencia y el número de ave dentro del grupo:

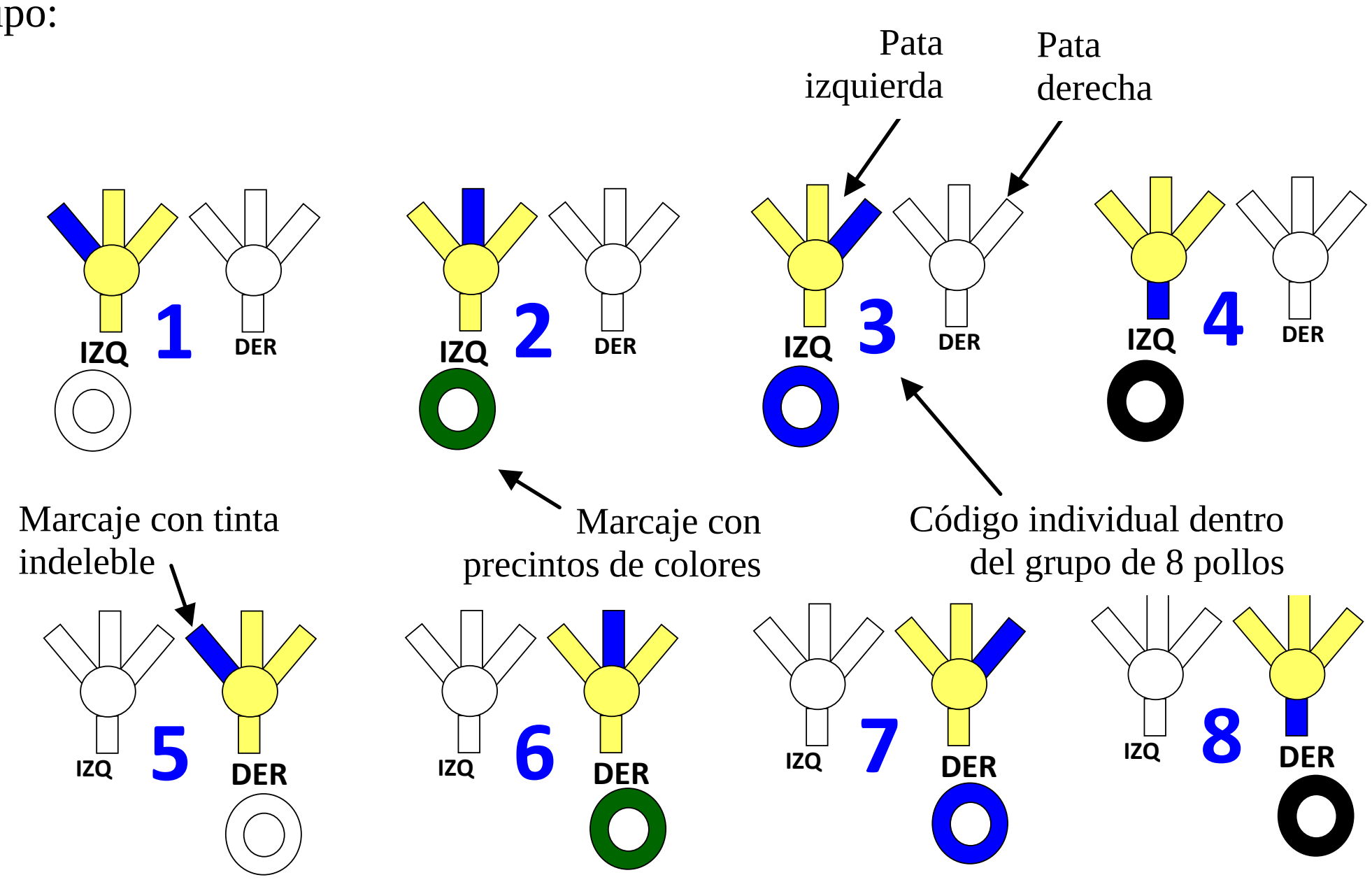

-------------------------------- **Colocación de precintos** --------------------------------------

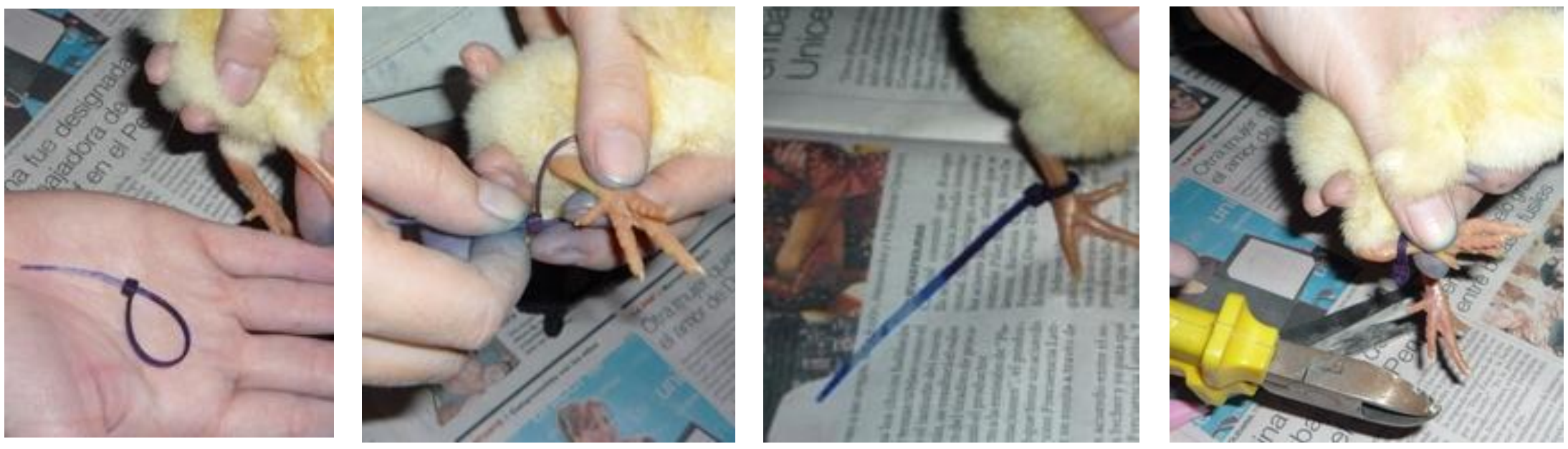

----------------------- **Marcaje de dedos** ----------------------- ---- **Resultado** -----

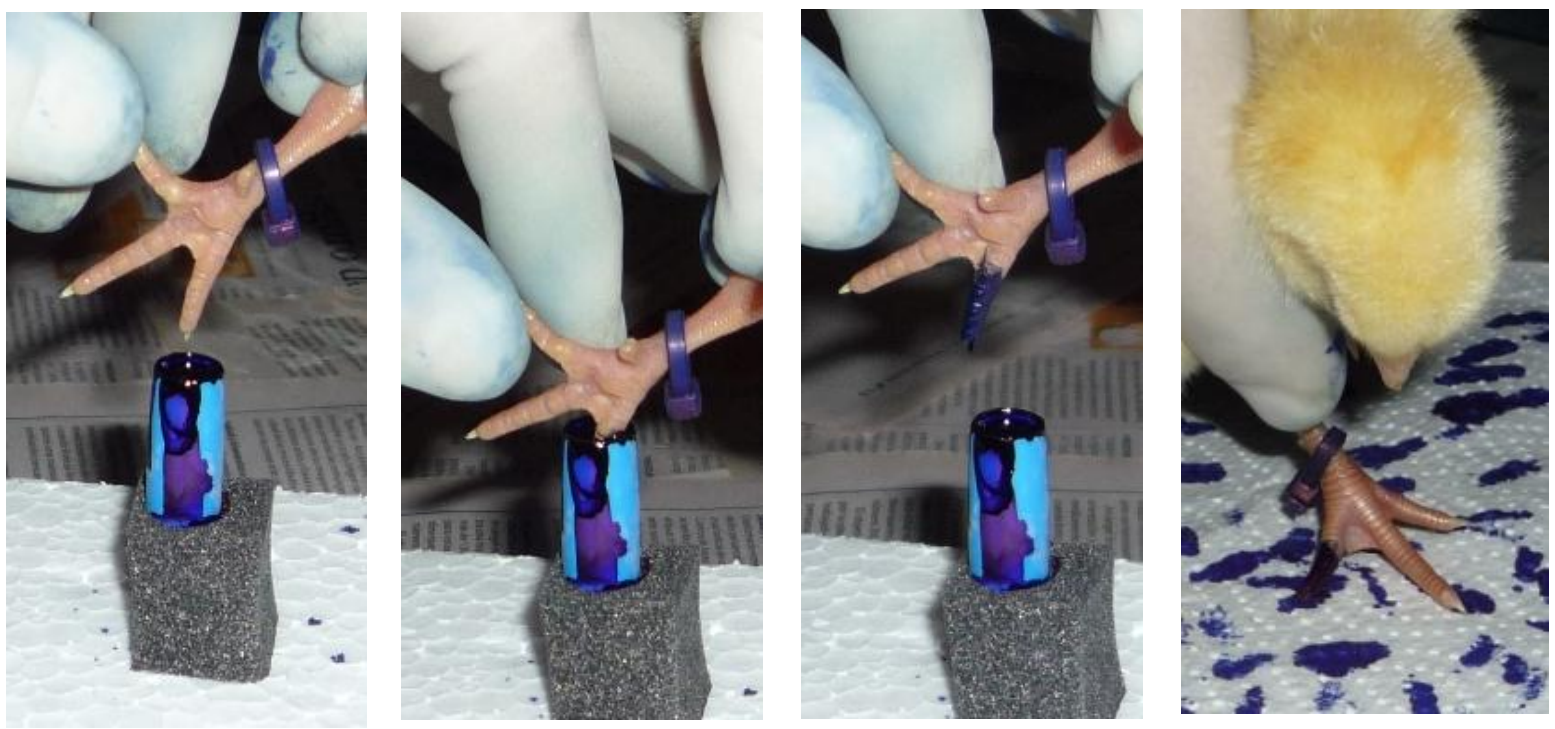

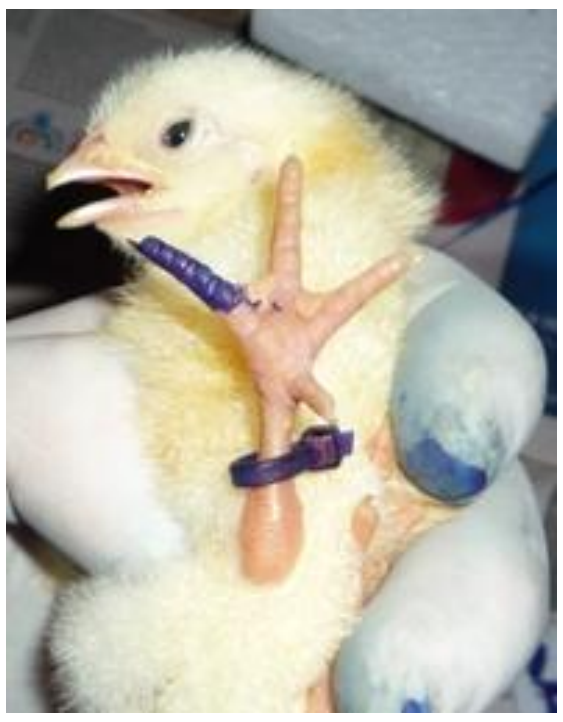

La última imagen muestra un pollo marcado con el número 7. Los precintos fueron cambiados semanalmente y/o en función al crecimiento corporal. Materiales empleados: precintos de colores, alicate de corte, tijeras, tinta indeleble, papel secante y guantes quirúrgicos desechables.

## ANEXO 4. Características de las aves empleadas

| Características de las aves | Empleadas en los experimentos |
|---|---|
| - Línea Cobb 500<br>- Provistos por la empresa Gramobier S.A.C., Lima, Perú.<br>- Procedentes de un lote de reproductoras de 45 semanas de edad (65A).<br>- Vacunas administradas en la planta de incubación:<br>o Recombinante contra la Enfermedad de Marek (HVT) y la Enfermedad de Gumboro (Vaxxitek®, Merial)<br>o Recombinante contra Viruela Aviar y Laringotraqueítis Aviar (Vectormune® FP-LT, Ceva Biomune)<br>o Oleosa contra la Enfermedad de Newcastle cepa La Sota (Cevac® Broiler ND K, Ceva Sante Animale)<br>- Se aplicó a los pollos Ceftiofur sódico (Excenel®, Pfizer Inc.). | 1, 2, 3,<br>5, 6, 8, 9<br>Anexo 1 |

# ANEXO 5. Protocolo de bioseguridad

## Limpieza general, desinfección y control de vectores:

- 2 semanas antes de la recepción de las aves retirar todo resto de alimento.
- Colocar veneno para roedores en cebaderos en lugares estratégicos. 2 veces por verificar y replantear las ubicaciones.
- Colocar cebaderos con veneno para moscas lejos del acceso de las aves. El producto a emplear debe producir muerte por contacto.
- Barrer las instalaciones diariamente.
- En la entrada del ambiente de crianza colocar pediluvios con agua y solución desinfectante. Estos deben ser usados al ingresar y salir del ambiente y la solución desinfectante debe ser renovada diariamente.
- Colocar basureros en cada zona para evitar el traslado de material contaminado de forma expuesta entre ambientes.

## Control de la contaminación cruzada entre tratamientos:

- Inspeccionar al detalle el diseño de las instalaciones y equipos e implementar las modificaciones que fueran necesarias para evitar el contacto entre aves, heces o agua de bebida de tratamientos con diferentes niveles de desafío.
- Evitar o modificar la forma en que se realiza toda práctica que pueda incrementar el contenido de polvo en el ambiente.
- Evitar todo contacto innecesario con el material de cama.
- El contacto con las aves, el material de cama o los equipos de crianza en uso debe hacerse empleando guantes quirúrgicos.
- No emplear el mismo guante para tener contacto con tratamientos diferentes. Desinfectarse las manos con alcohol en cada cambio de guantes quirúrgicos.
- El manejo de la cama debe hacerse empleando diferentes materiales y equipos para cada tratamiento. Al concluir deben ser sanitizados con la solución desinfectante. No intercambiar materiales ni equipos entre tratamientos.
- Cuando sea necesario retirar material de cama en uso, por ejemplo para secarlo, éste debe ser colocado en recipientes independientes para cada tratamiento.
- En general emplear solo equipos y utensilios hechos con materiales no porosos para facilitar su desinfección. En caso sea necesario emplear otros materiales (cajas de cartón, etc.) estos deben ser sustituidos con la mayor frecuencia posible para evitar la acumulación de contaminación.

- Para el pesaje de las aves colocar láminas protectoras sobre la balanza. Reemplazarlas y desinfectar con alcohol al cambiar de tratamiento.
- Almacenar el alimento de cada tratamiento en contenedores independientes.
- Para el reparto de alimento pesarlo en bolsas descartables de polietileno.
- Para pesar los residuos de alimento en los comederos emplear recipientes diferentes para tratamientos con diferentes niveles de desafío.
- Lavar y desinfectar diariamente los bebederos con una solución desinfectante.
- Desinfectar los equipos empleados al final del día y el piso de las instalaciones una vez por semana con una solución desinfectante.

**Control de riesgos asociados al uso de un inóculo bacteriano:**

- Emplear equipo de protección (uniforme, botas, mascarillas y lentes de seguridad) para evitar el contacto accidental del inóculo con las mucosas.
- Todo material empleado o que haya tenido contacto con el inóculo debe ser desinfectado con alcohol antes y después de ser retirado del ambiente.
- Cada operario debe tener consigo alcohol para desinfectar pisos, equipos u otros en caso se produzcan derrames accidentales del inóculo.
- Al concluir la aplicación del inóculo todo equipo de protección empleado debe ser colocado en bolsas plásticas, sellado para su posterior lavado y desinfección.
- Colocar basureros con bolsas de polietileno en el ambiente en que se aplica el inóculo para eliminar el material contaminado y reducir su libre exposición.
- Los residuos sólidos deben ser finalmente incinerados.

La solución desinfectante a que se hace referencia se preparó con CKM-DESIN® (CKM S.A.C.), y se eligió por su efectividad contra *Clostridium perfringens*.

# ANEXO 6. Materiales de cama empleados

## 1) Características generales

Material de cama nuevo tipo 1 (MCN1)

- Material: Viruta de madera sin ningún uso previo.
- Antes de su uso se aplicó un desinfectante (amonio cuaternario y glutaraldehído).

Material de cama re-usado (MCR)

- Material: cascarilla de arroz con un uso previo en un plantel de pollos.
- Procedente de empresa en Huacho (Lima, Perú). Plantel sin antecedentes respiratorios infecciosos atribuibles a virus y con resultado productivo satisfactorio.
- Al final de la campaña anterior se flameó; luego de 7 días se aglomeró en conos, mojó y dejó reposar por 7 días más, alcanzando una temperatura mayor o igual a 60 °C. Se esparció y aplicó un desinfectante (amonio cuaternario y glutaraldehído).

Material de cama nuevo tipo 2 (MCN2)

- Material: cascara de café. Acondicionamiento previo a su uso: Ninguno.
- Empleado como material de recambio.

## 2) Características físico-químicas y microbiológicas

| Variable | MCN1 | MCR |
|---|---|---|
| Recuento de *C. perfringens*, UFC/g | < 50 [1] | 100 [1] |
| Recuento de coliformes, NMP/g | 4 [2] | < 3 [3] |
| Recuento de aerobios mesófilos, UFC/g | $13 \times 10^3$ [2] | $57 \times 10^4$ [3] |
| Recuento de ooquistes, N°/g cama | - | 1441 [4] |

(1) Informes del 19/Julio/2010 Laboratorio de Biología y Genética Molecular, FMV-UNMSM
(2) Informe de ensayo 3-08635/10 CERPER (Certificaciones del Perú S.A.)
(3) Informe de ensayo 3-08634/10 CERPER (Certificaciones del Perú S.A.)
(4) Certificado con Ficha N° 100786 Laboratorio de Patología Aviar, FMV-UNMSM

## 3) Tratamientos en que se empleó los materiales de cama

| Experimento | Tratamientos en que se empleó cada material | | |
|---|---|---|---|
| | MCN1 | MCR | MCN2 |
| 1 | - | 1, 2 | - |
| 2 | - | 1, 2, 3 | - |
| 3 | - | 1, 2 | - |
| Anexo 1 | 1 | 2 | 1, 2 |
| 5 | - | 1, 2 | 1, 2 |
| 6 | - | 1, 2, 3 | 1, 2, 3 |
| 8 | - | 1, 2 | - |
| 9 | - | 1, 2 | - |

## ANEXO 7. Características de los ingredientes usados en las dietas

| Variable | Torta de soya | Harina de pescado |
|---|---|---|
| Humedad, % | 12.26 (2) | 9.28 (3) |
| Proteína total (N x 6.25), % | 46.16 (2) | 62.35 (3) |
| Grasa, % | 1.88 (2) | 10.75 (3) |
| Fibra Cruda, % | 4.31 (2) | 0 (3) |
| Cenizas, % | 5.29 (2) | 14.18 (3) |
| ELN, % (1) | 30.10 (2) | 3.44 (3) |
| Proteína soluble en KOH, % | 73.09 (2) | - |
| Actividad ureásica, ΔpH | 0.03 (2) | - |
| Histamina, ppm | - | 86 (4) |
| Recuento de *Clostridium perfringens*, UFC/g | - | 500 (5) |
| Recuento de coliformes, NMP/g | - | 20 (6) |
| Recuento de aerobios mesófilos, UFC/g | - | $74 \times 10^3$ (6) |
| Procedencia | Bolivia | Perú |

(1) ELN = Extracto Libre de Nitrógeno (calculado).
(2) Informe de ensayo 1237/2010 LENA, Universidad Nacional Agraria La Molina.
(3) Informe de ensayo 1236/2010 LENA, Universidad Nacional Agraria La Molina.
(4) Informe de ensayo 52068 INASSA (International Analytical Service S.A.C.)
(5) Informe del 19/Julio/2010 Laboratorio de Biología y Genética Molecular, UNMSM
(6) Informe de ensayo 3-08633/10 CERPER (Certificaciones del Perú S.A.)

## ANEXO 8. Balance electrolítico de la dieta N° 1 del Anexo 1

| Variable | Valores observados |
|---|---|
| Na, % (1) | 0.33 |
| K, % (1) | 1.17 |
| Cl, % (1) | 0.66 |
| Balance electrolítico (2), meq/kg | 259 |

(1) Informe de análisis del 05/Oct/2011. Método: espectrofotometría de absorción atómica (Laboratorio de Suelos, Facultad de Agronomía, UNALM)

(2) Balance electrolítico (meq/kg) = [Na (mg/kg)]/23 + [K (mg/kg)]/39.1 – [Cl (mg/kg)]/35.5

## ANEXO 9. Inóculo de *Clostridium perfringens* [1]

### 1) Preparación del inóculo

Bacteria: *Clostridium perfringens*
Tipo: *C. perfringens* Tipo A enterotoxigénico (subtipo cpb2- cpe+)
Procedencia: Pollo de 3 – 4 semanas de edad
Síntomas: Retraso en el crecimiento, plumas erizadas, sin diarrea aparente.
Necropsia: Leve congestión en el íleon

El inóculo fue provisto por el Laboratorio de Patología Clínica y Biología Molecular de la Facultad de Medicina Veterinaria de la Universidad Nacional Mayor de San Marcos. Fue aislado de un brote sub-clínico de campo de enteritis en pollos de 3 semanas de edad. Fueron identificadas varias cepas de *C. perfringens* y fue cultivada aquella Tipo A que mostró tener el gen productor de la enterotoxina.

Características fenotípicas:

- Presencia de actividad hemolítica alfa (Total) y beta (Parcial) en crecimiento de Agar TSA con 5% de sangre ovina desfibrinada a 37 °C por 24 horas en anaerobiosis.
- Presencia de actividad lecitinasa en crecimiento de Agar TSA con 5% de yema de huevo a 37 °C por 24 horas en anaerobiosis.
- Reducción del sulfito a sulfato en crecimiento en Agar TSA con 5% de yema de huevo a 37 °C por 24 horas en anaerobiosis.

Características genotípicas:

- Presencia del gen codificante de la toxina alfa por análisis mediante la prueba de Reacción en Cadena de la Polimerasa (PCR).
- Presencia del gen codificante de la toxina enterotoxina por análisis mediante la prueba de Reacción en Cadena de la Polimerasa (PCR).

### 2) Análisis de recuento del inóculo

De dedujo el número de unidades formadoras de colonias (UFC) mediante el crecimiento por 17 horas a 37 °C de un inóculo de 10 ml en 200 ml de caldo Infusión de Cerebro Corazón (BHI).

Número de Unidades Formadoras de Colonias:
El inóculo preparado contuvo $10^8$ UFC / ml. Luego, a partir de inóculo y para asegurar el proceso de administración a las aves se realizó la dilución de 200 ml de este inóculo conteniendo caldo BHI con 1 x $10^8$ UFC / ml con 400 ml de caldo BHI estéril para la obtención de un inóculo final con recuento de 1 x $10^8$ UFC / 3 ml.

[1] Informe del 19 de Julio de 2010 de Laboratorio de Biología y Genética Molecular, UNMSM.

# ANEXO 10. Inóculo de coccidia

## 1) Origen del inóculo

Vacuna comercial Immucox® for Chickens II, Vetech Laboratories. Contiene cepas vivas no atenuadas de aislamientos de campo (Conway and McKenzie, 2007) de *E. acervulina, E. máxima, E. tenella, E. necatrix* y *E. brunetti* con una concentración total no menor a 2.4 x $10^5$ oocistos vivos.

## 2) Preparación

Se diluyó frascos de 1000 dosis en 300 ml de agua destilada por frasco y se añadió a la preparación 3 ml del colorante provisto con la vacuna para facilitar el control durante el proceso de aplicación. La preparación se realizó en las instalaciones del Programa de Aves de la Universidad Nacional Agraria La Molina.

## 3) Administración

En el día 14 de edad a cada pollo se administró 3 ml de la preparación equivalente a 10 dosis, vía oro-ingluvial (Conway and McKenzie, 2007), empleando cánulas adheridas a jeringas hipodérmicas desechables de 5 ml sin aguja.

## ANEXO 11. Muestreo de material de cama y recuento de ooquistes

### 1) Muestreo:

- En el día 21 de edad, y en dos momentos del día, se tomó 4 muestras de material de cama en puntos diferentes de cada unidad experimental.
- Se evitó tomar muestras en zonas cercanas a los bebederos y/o con material de cama empastado para evitar sesgos (Conway and McKenzie, 2007).
- Las 8 muestras de cada unidad experimental fueron mezcladas y remitidas al Laboratorio.

### 2) Recuento:

- Las muestras de cama fueron analizadas por el método McMaster modificado para cama de galpón de aves.
- Se tomó 10 g de la muestra de cama recolectada y se mezcló con agua hasta completar un volumen de 100 ml.
- Se homogenizó la muestra y se dejó reposando hasta el día siguiente (24 horas aproximadamente).
- Luego se filtró la mezcla, y del filtrado se tomó 30 ml que fueron depositados en otro recipiente. Al filtrado se le añadió 15 ml de agua y se dejó sedimentar por 30 minutos.
- Pasado este tiempo se procedió a eliminar el sobrenadante y se agregó al sedimento ClNa en la misma cantidad que fue eliminada. Se homogenizó y se colocó una pequeña cantidad (0.15 ml) en la cámara McMaster esperándose 2 minutos para su lectura.
- El conteo de ooquistes se realizó en un microscopio de luz con un aumento de 100x. El número de ooquistes encontrados en la cámara se multiplicó por 67 (factor utilizado al contar en una sola cámara).
- Se realizaron 4 repeticiones para dar precisión de la lectura.

Referencia: Certificado con Ficha N° 100786, LPA-UNMSM.

# ANEXO 12. Fórmulas desarrolladas para histología

## 1) Densidad relativa de enterocitos

(1) $$\mathrm{DRE} = \frac{\mathrm{AEV}}{\mathrm{ASCLV}} \times 100$$

Donde: DRE: Densidad de enterocitos
AEV: Área de enterocitos en el corte longitudinal de la vellosidad ($mm^2$)
ASCLV: Área de la sección de corte longitudinal de la vellosidad ($mm^2$)

(2) $$\mathrm{AEV} = \mathrm{AE} \times \mathrm{LCE}$$

Donde: AEV: Área de enterocitos en el corte longitudinal de la vellosidad ($mm^2$)
AE: Altura del enterocito (mm)
LCE: Largo de la capa de enterocitos en el corte longitudinal de la vellosidad (mm)

(3) $$\mathrm{AE} = \frac{\mathrm{GV} - \mathrm{GLP}}{2 \times 1000}$$

Donde: AE: Altura del enterocito (mm)
GV: Grosor de la vellosidad (µm)
GLP: Grosor de la lámina propia (µm)

(4) $$\mathrm{LCE} = \frac{(2 \times \mathrm{AV}) + \mathrm{GV}}{1000}$$

Donde: LCE: Largo de la capa de enterocitos en el corte longitudinal de la vellosidad (mm)
AV: Altura de la vellosidad (µm)
GV: Grosor de la vellosidad (µm)

(5) ASCLV fue calculado empleando la siguiente fórmula:

$$\mathrm{ASCLV} = \frac{\mathrm{AV}}{1000} \times \frac{\mathrm{GV}}{1000}$$

Donde ASCLV: Área de la sección de corte longitudinal de la vellosidad ($mm^2$)
AV: Altura de la vellosidad (µm)
GV: Grosor de la vellosidad (µm)

(6) Reemplazando (3) y (4) en (2), y luego (2) y (5) en (1) tenemos:

$$\mathrm{DRE} = \frac{\frac{\mathrm{GV} - \mathrm{GLP}}{2\,\mathrm{X}\,1000} \times \frac{(2 \times \mathrm{AV}) + \mathrm{GV}}{1000}}{\frac{\mathrm{GV}}{1000} \times \frac{\mathrm{AV}}{1000}} \times 100$$

$$\mathbf{DRE} = (\mathbf{GV} - \mathbf{GLP}) \times \frac{(\mathbf{2} \times \mathbf{AV}) + \mathbf{GV}}{\mathbf{AV} \times \mathbf{GV}} \times \mathbf{50}$$

### 2) Densidad de la mucosa

Se refiere al área ($mm^2$) de mucosa que existe en corte longitudinal de la vellosidad por cada unidad lineal (mm) de muscularis. A mayor relación, mayor capacidad de absorción de nutrientes.

(1) $DM = ASCLV \times NVa$

Donde: DM: Densidad de la mucosa
ASCLV: Área de la sección de corte longitudinal de la vellosidad ($mm^2$)
NVa: Número de vellosidades x mm de muscularis

(2) ASCLV fue calculado mediante una aproximación trapezoidal empleando la siguiente fórmula:

$$ASCLV = \frac{GAV + GBV}{2000} \times \frac{AV}{1000}$$

Donde ASCLV: Área de la sección de corte longitudinal de la vellosidad ($mm^2$)
GAV: Grosor apical de la vellosidad (μm)
GBV: Grosor basal de la vellosidad (μm)
AV: Altura de la vellosidad (μm)

(3) NVa se obtuvo aplicando la siguiente regla de tres: Si en cada (GV+GC) μm hay 1 vellosidad, en 1000 μm hay NVa vellosidades.

$$NVa = \frac{1000}{GV + GC}$$

Donde NVa: Número de vellosidades x mm de muscularis
GV: Grosor de la vellosidad (μm)
AV: Altura de la vellosidad (μm)

(4) Reemplazando (2) y (3) en (1):

$$DM = \frac{GAV + GBV}{2000} \times \frac{AV}{1000} \times \frac{1000}{GV + GC} = \frac{GV}{1000} \times \frac{AV}{GV + GC}$$

$$\mathbf{DM = \frac{GV \times AV}{1000\,(GV + GC)}}$$

### 3) Índice de estructura de la vellosidad

Se refiere al área superficial existente en la vellosidad por unidad de área de muscularis. A mayor relación, mayor capacidad de absorción de nutrientes.

(1) $$IEV = \frac{ASV}{ASCTV}$$

Donde: IEV: Estructura de la vellosidad
ASV: Área superficial de la vellosidad ($\mu m^2$)
ASCTV : Área de la sección de corte transversal de la vellosidad ($\mu m^2$)

(2) $ASV = 3.1416 \times GV \times AV$

Donde: ASV: Área superficial de la vellosidad (μm²)
GV: Grosor de la vellosidad (μm)
AV: Altura de la vellosidad (μm)

(3) $$ASCTV = 3.1416 \times \left(\frac{GV}{2}\right)^2$$

Donde: ASCTV: Área de la sección de corte transversal de la vellosidad (μm²)
GV: Grosor de la vellosidad (μm)

(4) Reemplazando (2) y (3) en (1):

$$IEV = \frac{3.1416 \times GV \times AV}{3.1416 \times \left(\frac{GV}{2}\right)^2}$$

$$\mathbf{IEV = \frac{4 \times AV}{GV}}$$

## 4) Índice de estructura de la mucosa

Se refiere al área superficial de mucosa que existe en el intestino por unidad de área de muscularis. A mayor relación, mayor capacidad de absorción de nutrientes.

(1) $IEM = ASV \times NVb$

Donde: IEM: Estructura de la mucosa
ASV: Área superficial de la vellosidad (mm²)
NVb: Número de vellosidades x mm² de muscularis

(2) $$ASV = \frac{3.1416 \times GV \times AV}{1000 \times 1000}$$

Donde: ASV: Área superficial de la vellosidad (mm²)
GV: Grosor de la vellosidad (μm)
AV: Altura de la vellosidad (μm)

(3) $NVb = NVa^2$

Donde: NVb: Número de vellosidades x mm² de muscularis
NVa: Número de vellosidades x mm de muscularis

(4) NVa se obtuvo aplicando la siguiente regla de tres: Si en cada (GV+GC) μm hay 1 vellosidad, en 1000 μm hay NVa vellosidades.

$$NVa = \frac{1000}{GV + GC}$$

Donde NVa: Número de vellosidades x mm de muscularis
GV: Grosor de la vellosidad (μm)
GC: Grosor de la cripta (μm)

(5) Reemplazando (2), (3) y (4) en (1):

$$IEM = \frac{3.1416 \times GV \times AV}{1000 \times 1000} \times \left(\frac{1000}{GV + GC}\right)^2$$

$$\mathbf{IEM = \frac{3.1416 \times GV \times AV}{(GV + GC)^2}}$$

## ANEXO 13. Experimento 1 – variables histológicas, día 14

| T | F1 | F2 | R | AV | LC | AV/LC | GAV | GBV | GV | IFV | GLP | GC | ASV | ASC | ASV / ASC | DV | DM | IEM | IEV | AE | AEV | MIT | DRE | DACC | DRCC | LCC | GCC | ACC | AAM | ARM |
|---|---|---|---|---|---|---|---|---|---|---|---|---|---|---|---|---|---|---|---|---|---|---|---|---|---|---|---|---|---|---|
| 1 | D | – | 1 | 1221 | *302 | **4.04 | 89 | 120 | 105 | 11.7 | 34 | 69.10 | 0.4008 | *0.0656 | **6.1 | 117 | 0.735 | 13.30 | 46.73 | 35.500 | 0.0904 | 47.270 | 70.85 | 117 | 915 | 10.60 | 9.50 | 79 | 9238 | *2.305 |
| 1 | D | – | 2 | 1410 | 193 | 7.32 | 142 | 160 | 151 | 9.3 | 47 | 57.80 | 0.6689 | 0.0350 | 19.1 | 90 | 1.020 | 15.34 | 37.35 | 52.050 | 0.1546 | 140.035 | 72.63 | 90 | 421 | 8.20 | 7.80 | 50 | 4501 | 0.673 |
| 1 | D | – | 3 | 1414 | 194 | 7.28 | 99 | 192 | 146 | 9.7 | 58 | 50.60 | 0.6475 | 0.0309 | 21.0 | 31 | 1.050 | 16.79 | 38.81 | 43.800 | 0.1303 | 191.216 | 63.20 | 31 | 151 | 10.60 | 10.50 | 87 | 2727 | 0.421 |
| 1 | D | – | 4 | 1163 | 184 | 6.31 | 110 | 160 | 135 | 8.6 | 50 | 64.80 | 0.4917 | 0.0375 | 13.1 | 55 | 0.785 | 12.37 | 34.55 | 42.475 | 0.1045 | 112.004 | 66.77 | 55 | 350 | 10.30 | 9.80 | 79 | 4344 | 0.884 |
| 1 | D | – | 5 | 1183 | 169 | 6.99 | 105 | 138 | 122 | 9.7 | 67 | 48.80 | 0.4517 | 0.0260 | 17.4 | 55 | 0.844 | 15.57 | 38.93 | 27.300 | 0.0679 | 254.282 | 47.23 | 55 | 385 | 9.40 | 9.90 | 73 | 4042 | 0.895 |
| 1 | D | – | 6 | 950 | 195 | 4.88 | 85 | 115 | 100 | 9.5 | 43 | 53.10 | 0.2982 | 0.0325 | 9.2 | 21 | 0.620 | 12.73 | 38.01 | 28.275 | 0.0565 | 154.022 | 59.56 | 21 | 222 | 7.30 | 7.70 | 44 | 932 | 0.312 |
| 1 | D | – | 7 | - | - | - | - | - | - | - | - | - | - | - | - | - | - | - | - | - | - | - | - | - | - | - | - | - | - | - |
| 1 | D | – | 8 | 1124 | 180 | 6.24 | 123 | 159 | 141 | 8.0 | 61 | 57.00 | 0.4979 | 0.0322 | 15.4 | 78 | 0.800 | 12.70 | 31.89 | 40.000 | 0.0956 | 152.019 | 60.30 | 78 | 491 | 9.30 | 8.00 | 58 | 4546 | 0.913 |
| 2 | D | + | 1 | 1430 | 157 | 9.10 | 152 | 221 | 186 | 7.7 | 49 | 67.90 | 0.8374 | 0.0335 | 25.0 | 171 | 1.048 | 12.95 | 30.69 | 68.500 | 0.2087 | 143.233 | 78.29 | 171 | 640 | 9.10 | 8.10 | 58 | 9882 | 1.180 |
| 2 | D | + | 2 | 1492 | 203 | 7.35 | 106 | 170 | 138 | 10.8 | 49 | 57.30 | 0.6447 | 0.0366 | 17.6 | 74 | 1.053 | 16.98 | 43.39 | 44.375 | 0.1385 | 243.430 | 67.50 | 74 | 363 | 8.40 | 8.50 | 56 | 4172 | 0.647 |
| 2 | D | + | 3 | 1410 | 181 | 7.81 | 120 | 191 | 155 | 9.1 | 63 | 60.00 | 0.6874 | 0.0340 | 20.2 | 109 | 1.017 | 14.85 | 36.34 | 46.345 | 0.1379 | 141.079 | 63.01 | 109 | 497 | 7.60 | 8.90 | 53 | 5775 | 0.840 |
| 2 | D | + | 4 | - | - | - | - | - | - | - | - | - | - | - | - | - | - | - | - | - | - | - | - | - | - | - | - | - | - | - |
| 2 | D | + | 5 | 1349 | 247 | 5.45 | 120 | 168 | 144 | 9.4 | 65 | 61.40 | 0.6105 | 0.0477 | 12.8 | 113 | 0.946 | 14.46 | 37.46 | 39.615 | 0.1126 | 83.853 | 57.94 | 113 | 583 | 9.00 | 8.90 | 63 | 7128 | 1.168 |
| 2 | D | + | 6 | - | - | - | - | - | - | - | - | - | - | - | - | - | - | - | - | - | - | - | - | - | - | - | - | - | - | - |
| 2 | D | + | 7 | - | - | - | - | - | - | - | - | - | - | - | - | - | - | - | - | - | - | - | - | - | - | - | - | - | - | - |
| 2 | D | + | 8 | *1038 | 222 | **4.69 | 104 | 154 | 129 | 8.0 | 60 | 64.00 | **0.4212 | 0.0445 | **9.5 | 50 | *0.694 | 11.29 | **32.15 | 34.575 | **0.0762 | 132.479 | 56.87 | **50 | **370 | 8.80 | 9.20 | 64 | **3154 | 0.749 |
| 3 | Y | – | 1 | 882 | 198 | 4.46 | 116 | 122 | 119 | 7.4 | 48 | 54.50 | 0.3295 | 0.0339 | 9.7 | 86 | 0.605 | 10.95 | 29.65 | 35.555 | 0.0669 | 165.354 | 63.81 | 86 | 823 | 9.80 | 8.90 | 69 | 5912 | 1.794 |
| 3 | Y | – | 2 | - | - | - | - | - | - | - | - | - | - | - | - | - | - | - | - | - | - | - | - | - | - | - | - | - | - | - |
| 3 | Y | – | 3 | 629 | 206 | 3.05 | 189 | 201 | 195 | 3.2 | 92 | 66.40 | 0.3857 | 0.0430 | 9.0 | 62 | 0.469 | 5.64 | 12.90 | 51.425 | 0.0747 | 148.862 | 60.89 | 62 | 506 | 6.40 | 6.40 | 32 | 1998 | 0.518 |
| 3 | Y | – | 4 | 830 | 247 | 3.36 | 83 | 114 | 99 | 8.4 | 40 | 46.60 | 0.2572 | 0.0361 | 7.1 | 70 | 0.564 | 12.19 | 33.65 | 29.400 | 0.0517 | 146.688 | 63.15 | 70 | 853 | 8.70 | 9.10 | 62 | 4340 | 1.687 |
| 3 | Y | – | 5 | 1122 | 166 | 6.78 | 114 | 161 | 137 | 8.2 | 62 | 61.80 | 0.4840 | 0.0321 | 15.1 | 95 | 0.774 | 12.21 | 32.69 | 37.500 | 0.0893 | 205.403 | 57.97 | 95 | 619 | 7.60 | 8.10 | 48 | 4613 | 0.953 |
| 3 | Y | – | 6 | 651 | 147 | 4.43 | 67 | 74 | 70 | 9.3 | 34 | 45.60 | 0.1437 | 0.0210 | 6.8 | 47 | 0.395 | 10.71 | 37.08 | 18.350 | 0.0252 | 190.074 | 55.06 | 47 | 1036 | 7.70 | 6.60 | 40 | 1892 | 1.316 |
| 3 | Y | – | 7 | 612 | 170 | 3.60 | 85 | 131 | 108 | 5.7 | 52 | 48.50 | 0.2071 | 0.0259 | 8.0 | 21 | 0.422 | 8.48 | 22.72 | 27.750 | 0.0370 | 185.419 | 56.04 | 21 | 317 | 8.30 | 8.30 | 54 | 1131 | 0.546 |
| 3 | Y | – | 8 | 928 | 150 | 6.19 | 152 | 193 | 173 | 5.4 | 73 | 66.00 | 0.5029 | 0.0311 | 16.2 | 52 | 0.671 | 8.84 | 21.52 | 49.975 | 0.1014 | 186.484 | 63.33 | 52 | 324 | 7.95 | 9.03 | 56 | 2926 | 0.582 |
| 4 | Y | + | 1 | 1223 | 172 | 7.11 | 112 | 155 | 133 | 9.2 | 38 | 51.70 | 0.5125 | 0.0280 | 18.3 | 127 | 0.881 | 14.96 | 36.67 | 47.800 | 0.1233 | 254.001 | 75.57 | 127 | 776 | 10.10 | 9.00 | 71 | 9038 | 1.763 |
| 4 | Y | + | 2 | 1006 | 232 | 4.34 | 111 | 133 | 122 | 8.2 | 43 | 59.50 | 0.3853 | 0.0433 | 8.9 | 83 | 0.676 | 11.70 | 32.98 | 39.425 | 0.0841 | 136.167 | 68.58 | 83 | 674 | 8.60 | 7.90 | 53 | 4408 | 1.144 |
| 4 | Y | + | 3 | 969 | 172 | 5.64 | *84 | 138 | 111 | 8.8 | 52 | 58.70 | 0.3369 | 0.0317 | **10.6 | 106 | 0.633 | 11.75 | 35.03 | 29.100 | 0.0596 | 141.954 | 55.60 | 106 | 988 | 9.90 | 9.30 | 72 | 7658 | 2.273 |
| 4 | Y | + | 4 | 1156 | 142 | 8.15 | 113 | 152 | 132 | 8.7 | 57 | 45.50 | 0.4810 | 0.0203 | 23.7 | 89 | 0.860 | 15.19 | 34.91 | 37.850 | 0.0925 | 222.011 | 60.43 | 89 | 580 | 9.80 | 9.50 | 73 | 6493 | 1.350 |
| 4 | Y | + | 5 | 809 | 145 | 5.58 | 112 | 98 | 105 | 7.7 | 49 | 52.00 | 0.2657 | 0.0237 | 11.2 | 110 | 0.540 | 10.83 | 30.92 | 27.925 | 0.0481 | 139.409 | 56.85 | 110 | 1300 | 9.30 | 8.20 | 60 | 6582 | 2.478 |
| 4 | Y | + | 6 | - | - | - | - | - | - | - | - | - | - | - | - | - | - | - | - | - | - | - | - | - | - | - | - | - | - | - |
| 4 | Y | + | 7 | - | - | - | - | - | - | - | - | - | - | - | - | - | - | - | - | - | - | - | - | - | - | - | - | - | - | - |
| 4 | Y | + | 8 | - | - | - | - | - | - | - | - | - | - | - | - | - | - | - | - | - | - | - | - | - | - | - | - | - | - | - |
| 5 | I | – | 1 | 559 | 153 | 3.66 | 106 | 122 | 114 | 4.9 | 50 | 54.70 | 0.2002 | 0.0263 | 7.6 | 67 | 0.378 | 7.03 | 19.61 | 32.050 | 0.0395 | 167.568 | 61.96 | 67 | 1047 | 8.90 | 8.10 | 57 | 3777 | 1.887 |
| 5 | I | – | 2 | 592 | 210 | 2.82 | 83 | 102 | 92 | 6.4 | 31 | 45.10 | 0.1717 | 0.0297 | 5.8 | 58 | 0.398 | 9.09 | 25.63 | 30.500 | 0.0389 | 171.568 | 71.21 | 58 | 1054 | 7.50 | 7.20 | 42 | 2443 | 1.423 |
| 5 | I | – | 3 | 429 | 160 | 2.69 | 97 | 105 | 101 | 4.2 | 34 | 50.50 | 0.1362 | 0.0253 | 5.4 | 48 | 0.286 | 5.93 | 16.98 | 33.500 | 0.0321 | 181.556 | 74.11 | 48 | 1098 | 7.00 | 6.30 | 35 | 1649 | 1.211 |
| 5 | I | – | 4 | 705 | 194 | 3.63 | 112 | 141 | 127 | 5.6 | 63 | 61.30 | 0.2803 | 0.0374 | 7.5 | 63 | 0.475 | 7.95 | 22.31 | 31.625 | 0.0486 | 147.138 | 54.48 | 63 | 709 | 7.70 | 7.10 | 43 | 2718 | 0.970 |
| 5 | I | – | 5 | 577 | 169 | 3.42 | 104 | 103 | 103 | 5.6 | 42 | 48.10 | 0.1875 | 0.0255 | 7.4 | 52 | 0.394 | 8.16 | 22.31 | 30.775 | 0.0387 | 302.051 | 64.83 | 52 | 876 | 7.80 | 7.20 | 44 | 2307 | 1.230 |
| 5 | I | – | 6 | 454 | 144 | 3.16 | 97 | 98 | 97 | 4.7 | 47 | 53.70 | 0.1388 | 0.0242 | 5.7 | 29 | 0.292 | 6.08 | 18.63 | 25.025 | 0.0251 | 210.665 | 56.90 | 29 | 659 | 6.40 | 7.20 | 36 | 1053 | 0.759 |
| 5 | I | – | 7 | 581 | 124 | 4.67 | 92 | 73 | 82 | 7.0 | 42 | 50.00 | 0.1503 | 0.0195 | 7.7 | 32 | 0.361 | 8.57 | 28.17 | 20.363 | 0.0253 | 209.818 | 52.92 | 32 | 660 | 7.80 | 7.30 | 45 | 1413 | 0.940 |
| 5 | I | – | 8 | 644 | 105 | 6.11 | 191 | 180 | 185 | 3.5 | *105 | 60.00 | 0.3748 | 0.0199 | *18.9 | 72 | 0.486 | 6.23 | 13.91 | 40.125 | 0.0591 | 307.034 | 49.55 | 72 | 605 | 7.91 | *10.06 | 62 | 4512 | 1.204 |
| 6 | I | + | 1 | 605 | 102 | 5.93 | 115 | 122 | 119 | 5.1 | 45 | 52.20 | 0.2253 | 0.0167 | 13.5 | 68 | 0.420 | 7.73 | 20.43 | 36.600 | 0.0486 | 256.816 | 67.82 | 68 | 941 | 7.40 | 8.10 | 47 | 3178 | 1.410 |
| 6 | I | + | 2 | 447 | 113 | 3.96 | 138 | 134 | 136 | 3.3 | 47 | 52.90 | 0.1911 | 0.0187 | 10.2 | 61 | 0.322 | 5.34 | 13.12 | 44.430 | 0.0457 | 347.043 | 75.19 | 61 | 1006 | 5.90 | 5.70 | 26 | 1616 | 0.846 |
| 6 | I | + | 3 | 646 | 245 | 2.64 | 72 | 83 | 78 | 8.3 | 39 | 51.30 | 0.1573 | 0.0395 | 4.0 | 69 | 0.389 | 9.48 | 33.31 | 19.250 | 0.0264 | 124.199 | 52.63 | 69 | 1386 | 7.30 | 8.60 | 49 | 3422 | 2.175 |
| 6 | I | + | 4 | 743 | 151 | 4.92 | 90 | 118 | 104 | 7.1 | 48 | 49.00 | 0.2430 | 0.0233 | 10.4 | 75 | 0.505 | 10.37 | 28.57 | 28.150 | 0.0448 | 219.260 | 57.90 | 75 | 974 | 9.40 | 9.20 | 68 | 5114 | 2.105 |
| 6 | I | + | 5 | 758 | 179 | 4.25 | 90 | 88 | 89 | 8.5 | 45 | 59.10 | 0.2121 | 0.0331 | 6.4 | 91 | 0.456 | 9.66 | 34.05 | 22.050 | 0.0354 | 111.641 | 52.43 | 91 | 1353 | 9.50 | 9.20 | 69 | 6267 | 2.955 |
| 6 | I | + | 6 | - | - | - | - | - | - | - | - | - | - | - | - | - | - | - | - | - | - | - | - | - | - | - | - | - | - | - |
| 6 | I | + | 7 | - | - | - | - | - | - | - | - | - | - | - | - | - | - | - | - | - | - | - | - | - | - | - | - | - | - | - |
| 6 | I | + | 8 | 681 | 197 | 3.46 | 99 | 124 | 112 | 6.1 | 43 | 59.00 | 0.2385 | 0.0365 | 6.5 | 54 | 0.445 | 8.21 | 24.43 | 34.500 | 0.0508 | 161.578 | 66.95 | 54 | 707 | 7.60 | 7.30 | 44 | 2340 | 0.981 |

T: tratamiento (definido por la combinación de los diferentes niveles de los 2 factores en estudio). F1: factor sección intestinal (D: duodeno, Y: yeyuno, I: íleon). F2: factor aditivo ( – : sin aditivo, + : con aditivo). R: repetición. *, ** Se aplicó las pruebas de Dixon (Dean and Dixon, 1951) y Grubbs (Grubbs, 1969) a los valores de cada tratamiento y se identificó (P<0.05) valores anómalos (*) y otros (**) calculados a partir de estos; por ello no fueron considerados en el cálculo de promedios.

# ANEXO 14. Experimento 2 – pesos y coeficientes de crecimiento alométrico

| | Pollo BB | TRATAMIENTO 1 | | | | | | | | TRATAMIENTO 2 | | | | | | | | TRATAMIENTO 3 | | | | | | | |
|---|---|---|---|---|---|---|---|---|---|---|---|---|---|---|---|---|---|---|---|---|---|---|---|---|---|
| | | 1 | 2 | 3 | 4 | 5 | 6 | 7 | 8 | 1 | 2 | 3 | 4 | 5 | 6 | 7 | 8 | 1 | 2 | 3 | 4 | 5 | 6 | 7 | 8 |
| | | PESOS - Día 7 | | | | | | | | | | | | | | | | | | | | | | | |
| Peso vivo, g | 47.90 | 200 | 210 | 202 | 221 | 192 | 203 | 194 | 187 | 190 | 213 | 196 | 204 | 213 | 217 | 216 | 219 | 195 | 200 | 210 | 206 | 201 | 211 | 206 | 192 |
| Carcasa, g | 28.836 | 131 | 136 | 138 | 146 | 124 | 138 | 131 | 115 | 129 | 147 | 129 | 138 | 144 | 149 | 144 | 147 | 132 | 131 | 140 | 136 | 130 | 141 | 140 | 128 |
| Pechuga, g | 1.470 | 28.43 | 27.95 | 27.86 | 29.03 | 25.41 | 30.92 | 26.65 | 23.15 | 26.22 | 34.58 | 25.95 | 24.55 | 29.31 | 34.14 | 32.58 | 30.8 | 25.97 | 24.73 | 26.71 | 25.71 | 25.85 | 30.96 | 29.73 | 23.6 |
| Hígado, g | 1.049 | 6.58 | 6.74 | 7.89 | 9.51 | 7.7 | 8.31 | 8.51 | 8.42 | 7.46 | 8.22 | 7.89 | 8.58 | 8.08 | 8.12 | 8.07 | 9.04 | 7.16 | 7.86 | 8.03 | 8.39 | 8.55 | 7.48 | 7.84 | 8.85 |
| Intestino, g | 1.834 | 19.07 | 18.92 | 17.44 | 21.23 | 17.08 | 18.87 | 21.11 | 20.93 | 20.84 | 18.1 | 18.79 | 19.36 | 19.15 | 18.95 | 18.8 | 19.16 | 17.2 | 18.48 | 19.48 | 21.69 | 21.58 | 18.18 | 18.4 | 18.71 |
| Bursa, g | 0.048 | 0.63 | 0.45 | 0.54 | 0.39 | 0.4 | 0.49 | 0.45 | 0.45 | 0.44 | 0.38 | 0.51 | 0.38 | 0.5 | 0.41 | 0.43 | 0.5 | 0.45 | 0.46 | 0.3 | 0.48 | 0.32 | 0.31 | 0.45 | 0.38 |
| Bazo, g | 0.013 | 0.16 | 0.26 | 0.17 | 0.29 | 0.17 | 0.49 | 0.45 | 0.45 | 0.24 | 0.27 | 0.3 | 0.2 | 0.15 | 0.41 | 0.43 | 0.5 | 0.18 | 0.27 | 0.22 | 0.25 | 0.16 | 0.31 | 0.45 | 0.38 |
| Páncreas, g | 0.072 | 0.75 | 0.95 | 0.71 | 1.05 | 0.98 | 1.07 | 0.76 | 0.62 | 0.98 | 0.86 | 0.8 | 0.79 | 0.78 | 1.03 | 1.03 | 0.8 | 0.85 | 1.17 | 0.44 | 0.9 | 0.88 | 0.88 | 0.95 | 0.69 |
| | | PESOS - Día 14 | | | | | | | | | | | | | | | | | | | | | | | |
| Peso vivo, g | | 486 | 511 | 466 | 481 | 458 | 375 | 388 | 471 | 440 | 538 | 472 | 410 | 557 | 506 | 345 | 600 | 388 | 430 | 362 | 345 | 600 | 352 | 431 | 471 |
| Carcasa, g | | 345 | 357 | 329 | 350 | 317 | 258 | 267 | 247.8 | 307 | 391 | 353 | 301 | 401 | 372 | 250 | 431 | 267 | 309 | 261 | 250 | 431 | 245 | 319 | 347.8 |
| Pechuga, g | | 91.14 | 92.32 | 79.01 | 95.9 | 83.69 | 56.46 | 70.61 | 98.55 | 76.39 | 106 | 98 | 72.61 | 106.97 | 92.7 | 67.9 | 120.56 | 70.61 | 76.62 | 64.49 | 67.9 | 120.6 | 65.9 | 84.77 | 98.55 |
| Hígado, g | | 14.74 | 17.7 | 14.69 | 12.03 | 13.99 | 17.62 | 12.59 | 15.28 | 14.67 | 16.32 | 12.42 | 17.48 | 16.45 | 15.16 | 12 | 16.93 | 12.59 | 13.73 | 12.75 | 12 | 16.93 | 13.46 | 12.22 | 15.28 |
| Intestino, g | | 41.27 | 36.26 | 34.77 | 30.74 | 32.18 | 29.68 | 28.99 | 28.43 | 33.47 | 36.32 | 32.52 | 32.45 | 33.53 | 35.08 | 22.31 | 36.52 | 28.99 | 26.13 | 26.92 | 22.31 | 36.52 | 23.4 | 27.22 | 28.43 |
| Bursa, g | | 0.74 | 1.24 | 1.3 | 1.02 | 1.2 | 1.77 | 0.89 | 0.88 | 1.02 | 1.21 | 0.87 | 1.04 | 1.44 | 1.9 | 0.95 | 0.53 | 0.89 | 0.77 | 1.29 | 0.95 | 0.53 | 1.1 | 1.02 | 0.88 |
| Bazo, g | | 0.43 | 0.36 | 0.59 | 0.39 | 0.43 | 0.52 | 0.48 | 0.59 | 0.61 | 0.43 | 0.33 | 0.35 | 0.61 | 0.7 | 0.17 | 0.75 | 0.48 | 0.47 | 0.62 | 0.17 | 0.75 | 0.37 | 0.54 | 0.59 |
| Páncreas, g | | 1.6 | 1.74 | 1.62 | 1.75 | 1.84 | 1.53 | 1.85 | 1.34 | 1.65 | 2.02 | 1.63 | 1.09 | 2.26 | 1.64 | 1.62 | 1.78 | 1.85 | 1.7 | 1.61 | 1.62 | 1.78 | 1.79 | 1.13 | 1.34 |
| | | COEFICIENTES DE CRECIMIENTO ALOMÉTRICO - Día 7 | | | | | | | | | | | | | | | | | | | | | | | |
| Peso vivo | | 1.000 | 1.000 | 1.000 | 1.000 | 1.000 | 1.000 | 1.000 | 1.000 | 1.000 | 1.000 | 1.000 | 1.000 | 1.000 | 1.000 | 1.000 | 1.000 | 1.000 | 1.000 | 1.000 | 1.000 | 1.000 | 1.000 | 1.000 | 1.000 |
| Carcasa | | 1.088 | 1.076 | 1.135 | 1.097 | 1.073 | 1.129 | 1.122 | 1.022 | 1.128 | 1.146 | 1.093 | 1.124 | 1.123 | 1.141 | 1.107 | 1.115 | 1.124 | 1.088 | 1.107 | 1.097 | 1.074 | 1.110 | 1.129 | 1.107 |
| Pechuga | | 4.632 | 4.337 | 4.494 | 4.280 | 4.312 | 4.963 | 4.476 | 4.034 | 4.497 | 5.290 | 4.314 | 3.921 | 4.484 | 5.126 | 4.915 | 4.583 | 4.339 | 4.029 | 4.144 | 4.067 | 4.190 | 4.781 | 4.702 | 4.005 |
| Hígado | | 1.502 | 1.466 | 1.784 | 1.965 | 1.831 | 1.869 | 2.003 | 2.056 | 1.793 | 1.762 | 1.838 | 1.921 | 1.732 | 1.709 | 1.706 | 1.885 | 1.677 | 1.795 | 1.746 | 1.860 | 1.942 | 1.619 | 1.738 | 2.105 |
| Intestino | | 2.491 | 2.354 | 2.255 | 2.510 | 2.324 | 2.428 | 2.843 | 2.924 | 2.865 | 2.220 | 2.504 | 2.479 | 2.349 | 2.281 | 2.274 | 2.286 | 2.304 | 2.414 | 2.423 | 2.751 | 2.805 | 2.251 | 2.333 | 2.546 |
| Bursa | | 3.161 | 2.150 | 2.682 | 1.771 | 2.090 | 2.422 | 2.327 | 2.415 | 2.324 | 1.790 | 2.611 | 1.869 | 2.355 | 1.896 | 1.998 | 2.291 | 2.316 | 2.308 | 1.433 | 2.338 | 1.597 | 1.474 | 2.192 | 1.986 |
| Bazo | | 2.846 | 4.405 | 2.994 | 4.668 | 3.150 | 8.587 | 8.252 | 8.561 | 4.494 | 4.510 | 5.445 | 3.488 | 2.505 | 6.722 | 7.082 | 8.122 | 3.284 | 4.803 | 3.727 | 4.317 | 2.832 | 5.227 | 7.771 | 7.041 |
| Páncreas | | 2.492 | 3.006 | 2.336 | 3.157 | 3.392 | 3.503 | 2.603 | 2.203 | 3.428 | 2.683 | 2.712 | 2.573 | 2.433 | 3.154 | 3.169 | 2.427 | 2.897 | 3.887 | 1.392 | 2.903 | 2.909 | 2.771 | 3.065 | 2.388 |
| | | COEFICIENTES DE CRECIMIENTO ALOMÉTRICO - Día 14 | | | | | | | | | | | | | | | | | | | | | | | |
| Peso vivo | | 1.000 | 1.000 | 1.000 | 1.000 | 1.000 | 1.000 | 1.000 | 1.000 | 1.000 | 1.000 | 1.000 | 1.000 | 1.000 | 1.000 | 1.000 | 1.000 | 1.000 | 1.000 | 1.000 | 1.000 | 1.000 | 1.000 | 1.000 | 1.000 |
| Carcasa | | 1.179 | 1.161 | 1.173 | 1.209 | 1.150 | 1.143 | 1.143 | 0.874 | 1.159 | 1.207 | 1.242 | 1.220 | 1.196 | 1.221 | 1.204 | 1.193 | 1.143 | 1.194 | 1.198 | 1.204 | 1.193 | 1.156 | 1.229 | 1.227 |
| Pechuga | | 6.110 | 5.887 | 5.525 | 6.496 | 5.954 | 4.906 | 5.930 | 6.818 | 5.657 | 6.420 | 6.765 | 5.770 | 6.258 | 5.969 | 6.413 | 6.547 | 5.930 | 5.806 | 5.805 | 6.413 | 6.547 | 6.100 | 6.409 | 6.818 |
| Hígado | | 1.385 | 1.582 | 1.439 | 1.142 | 1.395 | 2.146 | 1.482 | 1.481 | 1.522 | 1.385 | 1.202 | 1.947 | 1.349 | 1.368 | 1.588 | 1.288 | 1.482 | 1.458 | 1.608 | 1.588 | 1.288 | 1.746 | 1.295 | 1.481 |
| Intestino | | 2.218 | 1.854 | 1.949 | 1.670 | 1.835 | 2.068 | 1.952 | 1.577 | 1.987 | 1.764 | 1.800 | 2.068 | 1.573 | 1.811 | 1.689 | 1.590 | 1.952 | 1.587 | 1.943 | 1.689 | 1.590 | 1.737 | 1.650 | 1.577 |
| Bursa | | 1.528 | 2.435 | 2.799 | 2.128 | 2.629 | 4.736 | 2.302 | 1.875 | 2.326 | 2.257 | 1.850 | 2.545 | 2.594 | 3.768 | 2.763 | 0.886 | 2.302 | 1.797 | 3.576 | 2.763 | 0.886 | 3.136 | 2.375 | 1.875 |
| Bazo | | 3.148 | 2.506 | 4.504 | 2.884 | 3.340 | 4.933 | 4.401 | 4.456 | 4.932 | 2.843 | 2.487 | 3.037 | 3.896 | 4.922 | 1.753 | 4.447 | 4.401 | 3.888 | 6.093 | 1.753 | 4.447 | 3.739 | 4.457 | 4.456 |
| Páncreas | | 2.188 | 2.263 | 2.310 | 2.418 | 2.670 | 2.711 | 3.168 | 1.891 | 2.492 | 2.495 | 2.295 | 1.767 | 2.696 | 2.154 | 3.120 | 1.971 | 3.168 | 2.627 | 2.955 | 3.120 | 1.971 | 3.379 | 1.742 | 1.891 |

Tratamiento 1: dieta basal; Tratamiento 2: dieta basal + 500 ppm de Orevitol®; Tratamiento 3: dieta basal + 500 ppm de neomicina.

## ANEXO 15. Experimento 2 - peso vivo individual, gramos

| Día 1 de edad | | | | | | | | Día 7 de edad | | | | | | | | Día 14 de edad | | | | | | | |
|---|---|---|---|---|---|---|---|---|---|---|---|---|---|---|---|---|---|---|---|---|---|---|---|
| N° | Tratamientos [1] | | | N° | Tratamientos [1] | | | N° | Tratamientos [1] | | | N° | Tratamientos [1] | | | N° | Tratamientos [1] | | | N° | Tratamientos [1] | | |
| | 1 | 2 | 3 | | 1 | 2 | 3 | | 1 | 2 | 3 | | 1 | 2 | 3 | | 1 | 2 | 3 | | 1 | 2 | 3 |
| 1 | 44.55 | 46.81 | 46.01 | 33 | 46.86 | 45.86 | 47.80 | 1 | 159 | 224 | 203 | 33 | 219 | 166 | 266 | 1 | 379 | 526* | 456 | 33 | 487 | 452 | 570 |
| 2 | 40.73 | 49.91 | 48.18 | 34 | 50.02 | 42.65 | 47.06 | 2 | 185 | 194 | 187 | 34 | 192 | 144 | 197 | 2 | 439* | 528 | 414 | 34 | 458 | 390 | 412 |
| 3 | 45.77 | 46.40 | 45.09 | 35 | 48.48 | 53.19 | 48.07 | 3 | 194* | 217 | 191 | 35 | 174 | 196 | 136 | 3 | ** | 504 | 411 | 35 | 407* | 531 | 500* |
| 4 | 46.28 | 51.42 | 44.60 | 36 | 45.70 | 48.83 | 48.90 | 4 | 189 | 197* | 219 | 36 | 193 | 221 | 221 | 4 | 356 | ** | 474 | 36 | 447 | 572 | 519 |
| 5 | 46.10 | 46.07 | 43.19 | 37 | 50.67 | 49.52 | 50.15 | 5 | 197 | 157 | 194* | 37 | 186 | 193* | 202 | 5 | 440 | 347 | ** | 37 | 382 | ** | 444 |
| 6 | 48.62 | 50.82 | 47.99 | 38 | 49.41 | 51.22 | 49.37 | 6 | 193 | 196 | 190 | 38 | 186 | 202 | 218 | 6 | 441 | 459 | 418 | 38 | 416 | 540 | 509 |
| 7 | 45.56 | 44.49 | 45.32 | 39 | 47.43 | 56.57 | 44.76 | 7 | 221 | 210 | 186 | 39 | 193* | 282 | 212 | 7 | 508 | 520 | 369 | 39 | ** | 456* | 469 |
| 8 | 52.40 | 55.53 | 48.17 | 40 | 51.15 | 44.87 | 49.64 | 8 | 208 | 206 | 175 | 40 | 206 | 202 | 207* | 8 | 497 | 489 | 400* | 40 | 488 | 547 | ** |
| 9 | 46.93 | 50.93 | 51.42 | 41 | 48.46 | 49.46 | 50.40 | 9 | 217 | 218 | 190 | 41 | 157 | 182 | 216 | 9 | 465* | 516 | 416 | 41 | 310* | 392* | 524 |
| 10 | 47.82 | 43.46 | 48.28 | 42 | 49.95 | 50.49 | 42.51 | 10 | 192 | 231 | 213 | 42 | 220 | 235 | 233 | 10 | 440 | 575 | 459 | 42 | 473 | 514 | 551 |
| 11 | 48.77 | 46.70 | 46.59 | 43 | 49.53 | 43.86 | 49.78 | 11 | 212* | 200* | 221 | 43 | 165 | 178 | 193 | 11 | ** | ** | 459 | 43 | 369 | 442 | 441 |
| 12 | 41.15 | 44.27 | 47.65 | 44 | 51.29 | 54.20 | 54.14 | 12 | 208 | 207 | 225 | 44 | 195 | 224 | 218 | 12 | 463 | 494 | 473* | 44 | 397 | 529 | 500 |
| 13 | 54.35 | 54.73 | 44.58 | 45 | 46.31 | 47.60 | 47.19 | 13 | 208 | 202 | 202 | 45 | 193* | 221 | 208* | 13 | 438 | 483 | 409 | 45 | ** | 524 | ** |
| 14 | 42.41 | 46.59 | 49.92 | 46 | 40.47 | 46.55 | 50.52 | 14 | 190 | 178 | 221 | 46 | 153 | 207* | 200 | 14 | 454 | 472* | 441 | 46 | 340 | ** | 485 |
| 15 | 52.86 | 48.67 | 43.80 | 47 | 48.54 | 42.81 | 46.12 | 15 | 278 | 196 | 196 | 47 | 212 | 204 | 208 | 15 | 392 | 469 | 435 | 47 | 467 | 460 | 499 |
| 16 | 48.71 | 48.26 | 44.82 | 48 | 49.21 | 46.49 | 49.50 | 16 | 223 | 200 | 209* | 48 | 207 | 199 | 209 | 16 | 477 | 500 | ** | 48 | 473 | 465 | 492* |
| 17 | 47.57 | 48.56 | 46.72 | 49 | 55.56 | 51.33 | 45.36 | 17 | 202 | 199 | 205 | 49 | 165 | 208 | 195 | 17 | 485 | 470 | 444 | 49 | 342* | 477 | 473 |
| 18 | 48.93 | 49.23 | 53.61 | 50 | 48.51 | 41.08 | 50.97 | 18 | 219 | 207* | 200* | 50 | 184 | 202* | 221 | 18 | 482 | ** | ** | 50 | 406 | ** | 515 |
| 19 | 42.18 | 45.46 | 47.10 | 51 | 49.67 | 52.41 | 42.46 | 19 | 198* | 196 | 210 | 51 | 186 | 231 | 191 | 19 | ** | 484 | 457 | 51 | 440 | 546 | 427 |
| 20 | 48.78 | 46.39 | 49.62 | 52 | 48.54 | 53.88 | 46.58 | 20 | 210 | 207 | 202 | 52 | 142 | 213 | 224 | 20 | 470 | 475* | 433* | 52 | 388 | 488 | 484 |
| 21 | 48.51 | 44.13 | 50.92 | 53 | 43.67 | 46.69 | 50.83 | 21 | 207 | 158 | 244 | 53 | 188 | 176 | 210* | 21 | 513 | 438 | 518 | 53 | 427 | 424 | ** |
| 22 | 48.56 | 47.62 | 52.61 | 54 | 52.79 | 45.85 | 47.00 | 22 | 157 | 225 | 231 | 54 | 197 | 189 | 170 | 22 | 391 | 531 | 486 | 54 | 468 | 428* | 304* |
| 23 | 46.42 | 47.15 | 45.93 | 55 | 46.25 | 50.01 | 47.53 | 23 | 181 | 224 | 191 | 55 | 180* | 214 | 219 | 23 | 412* | 549 | 391 | 55 | ** | 534 | 493 |
| 24 | 49.99 | 46.72 | 52.76 | 56 | 49.77 | 50.57 | 50.98 | 24 | 194 | 210 | 190 | 56 | 191 | 187 | 214 | 24 | 455 | 548 | 442 | 56 | 409 | 475 | 490 |
| 25 | 50.61 | 56.46 | 47.60 | 57 | 62.18 | 45.93 | 50.03 | 25 | 191 | 190 | 214 | 57 | 191 | 198 | 232 | 25 | 441* | 511 | 527 | 57 | 425 | 478 | 532 |
| 26 | 46.97 | 53.63 | 45.91 | 58 | 42.64 | 47.11 | 46.34 | 26 | 214 | 178 | 167 | 58 | 166 | 157 | 187 | 26 | 494 | 507 | 455 | 58 | 354 | 389 | 439 |
| 27 | 45.65 | 49.29 | 47.95 | 59 | 44.68 | 45.62 | 56.94 | 27 | 213* | 195* | 197 | 59 | 155 | 177 | 232 | 27 | ** | ** | 496 | 59 | 335 | 426 | 524 |
| 28 | 46.21 | 50.54 | 49.58 | 60 | 44.48 | 45.89 | 46.32 | 28 | 230 | 193 | 186* | 60 | 196 | 212 | 216 | 28 | 529 | 471 | ** | 60 | 435* | 509 | 559* |
| 29 | 45.97 | 51.17 | 40.40 | 61 | 46.03 | 47.87 | 40.92 | 29 | 162 | 222 | 178 | 61 | 151 | 195 | 134 | 29 | 391 | 575 | 487 | 61 | 317 | 468 | 369 |
| 30 | 40.21 | 47.86 | 56.72 | 62 | 54.03 | 51.72 | 49.71 | 30 | 181 | 185 | 163 | 62 | 203 | 217 | 203 | 30 | 433 | 523 | 428 | 62 | 404 | 513 | 442 |
| 31 | 52.48 | 46.12 | 53.18 | 63 | 44.35 | 48.39 | 42.98 | 31 | 222 | 287 | 174 | 63 | 168 | 188* | 203* | 31 | 483 | 455* | 386* | 63 | 374 | ** | ** |
| 32 | 52.92 | 51.51 | 43.72 | 64 | 46.36 | 41.91 | 54.54 | 32 | 218 | 194 | 198 | 64 | 186* | 179 | 205 | 32 | 483 | 551 | 452 | 64 | ** | 438* | 492 |

[1] Tratamientos: 1: aves control; 2: 500 ppm de Orevitol®; 3: 150 ppm de neomicina.

* Aves seleccionadas como muestras o muertas en la semana inmediata siguiente. ** Aves que fueron muestreadas o murieron antes de completar el periodo

## ANEXO 16. Experimento 2 – ganancia de peso individual, gramos

**De 0 a 7 días de edad**

| N° | Tratamientos [1] | | | N° | Tratamientos [1] | | |
|---|---|---|---|---|---|---|---|
| | 1 | 2 | 3 | | 1 | 2 | 3 |
| 1 | 114.45 | 177.19 | 156.99 | 33 | 172.14 | 120.14 | 218.20 |
| 2 | 144.27 | 144.09 | 138.82 | 34 | 141.98 | 101.35 | 149.94 |
| 3 | 148.23 | 170.60 | 145.91 | 35 | 125.52 | 142.81 | 87.93 |
| 4 | 142.72 | 145.58 | 174.40 | 36 | 147.30 | 172.17 | 172.10 |
| 5 | 150.90 | 110.93 | 150.81 | 37 | 135.33 | 143.48 | 151.85 |
| 6 | 144.38 | 145.18 | 142.01 | 38 | 136.59 | 150.78 | 168.63 |
| 7 | 175.44 | 165.51 | 140.68 | 39 | 145.57 | 225.43 | 167.24 |
| 8 | 155.60 | 150.47 | 126.83 | 40 | 154.85 | 157.13 | 157.36 |
| 9 | 170.07 | 167.07 | 138.58 | 41 | 108.54 | 132.54 | 165.60 |
| 10 | 144.18 | 187.54 | 164.72 | 42 | 170.05 | 184.51 | 190.49 |
| 11 | 163.23 | 153.30 | 174.41 | 43 | 115.47 | 134.14 | 143.22 |
| 12 | 166.85 | 162.73 | 177.35 | 44 | 143.71 | 169.80 | 163.86 |
| 13 | 153.65 | 147.27 | 157.42 | 45 | 146.69 | 173.40 | 160.81 |
| 14 | 147.59 | 131.41 | 171.08 | 46 | 112.53 | 160.45 | 149.48 |
| 15 | 225.14 | 147.33 | 152.20 | 47 | 163.46 | 161.19 | 161.88 |
| 16 | 174.29 | 151.74 | 164.18 | 48 | 157.79 | 152.51 | 159.50 |
| 17 | 154.43 | 150.44 | 158.28 | 49 | 109.44 | 156.67 | 149.64 |
| 18 | 170.07 | 157.77 | 146.39 | 50 | 135.49 | 160.92 | 170.03 |
| 19 | 155.82 | 150.54 | 162.90 | 51 | 136.33 | 178.59 | 148.54 |
| 20 | 161.22 | 160.61 | 152.38 | 52 | 93.46 | 159.12 | 177.42 |
| 21 | 158.49 | 113.87 | 193.08 | 53 | 144.33 | 129.31 | 159.17 |
| 22 | 108.44 | 177.38 | 178.39 | 54 | 144.21 | 143.15 | 123.00 |
| 23 | 134.58 | 176.85 | 145.07 | 55 | 133.75 | 163.99 | 171.47 |
| 24 | 144.01 | 163.28 | 137.24 | 56 | 141.23 | 136.43 | 163.02 |
| 25 | 140.39 | 133.54 | 166.40 | 57 | 128.82 | 152.07 | 181.97 |
| 26 | 167.03 | 124.37 | 121.09 | 58 | 123.36 | 109.89 | 140.66 |
| 27 | 167.35 | 145.71 | 149.05 | 59 | 110.32 | 131.38 | 175.06 |
| 28 | 183.79 | 142.46 | 136.42 | 60 | 151.52 | 166.11 | 169.68 |
| 29 | 116.03 | 170.83 | 137.60 | 61 | 104.97 | 147.13 | 93.08 |
| 30 | 140.79 | 137.14 | 106.28 | 62 | 148.97 | 165.28 | 153.29 |
| 31 | 169.52 | 240.88 | 120.82 | 63 | 123.65 | 139.61 | 160.02 |
| 32 | 165.08 | 142.49 | 154.28 | 64 | 139.64 | 137.09 | 150.46 |

**De 8 a 14 días de edad**

| N° | Tratamientos [1] | | | N° | Tratamientos [1] | | |
|---|---|---|---|---|---|---|---|
| | 1 | 2 | 3 | | 1 | 2 | 3 |
| 1 | 220.00 | 302.00 | 253.00 | 33 | 268.00 | 286.00 | 304.00 |
| 2 | 254.00 | 334.00 | 227.00 | 34 | 266.00 | 246.00 | 215.00 |
| 3 | - | 287.00 | 220.00 | 35 | 233.00 | 335.00 | 364.00 |
| 4 | 167.00 | - | 255.00 | 36 | 254.00 | 351.00 | 298.00 |
| 5 | 243.00 | 190.00 | - | 37 | 196.00 | - | 242.00 |
| 6 | 248.00 | 263.00 | 228.00 | 38 | 230.00 | 338.00 | 291.00 |
| 7 | 287.00 | 310.00 | 183.00 | 39 | - | 174.00 | 257.00 |
| 8 | 289.00 | 283.00 | 225.00 | 40 | 282.00 | 345.00 | - |
| 9 | 248.00 | 298.00 | 226.00 | 41 | 153.00 | 210.00 | 308.00 |
| 10 | 248.00 | 344.00 | 246.00 | 42 | 253.00 | 279.00 | 318.00 |
| 11 | - | - | 238.00 | 43 | 204.00 | 264.00 | 248.00 |
| 12 | 255.00 | 287.00 | 248.00 | 44 | 202.00 | 305.00 | 282.00 |
| 13 | 230.00 | 281.00 | 207.00 | 45 | - | 303.00 | - |
| 14 | 264.00 | 294.00 | 220.00 | 46 | 187.00 | - | 285.00 |
| 15 | 114.00 | 273.00 | 239.00 | 47 | 255.00 | 256.00 | 291.00 |
| 16 | 254.00 | 300.00 | - | 48 | 266.00 | 266.00 | 283.00 |
| 17 | 283.00 | 271.00 | 239.00 | 49 | 177.00 | 269.00 | 278.00 |
| 18 | 263.00 | - | - | 50 | 222.00 | - | 294.00 |
| 19 | - | 288.00 | 247.00 | 51 | 254.00 | 315.00 | 236.00 |
| 20 | 260.00 | 268.00 | 231.00 | 52 | 246.00 | 275.00 | 260.00 |
| 21 | 306.00 | 280.00 | 274.00 | 53 | 239.00 | 248.00 | - |
| 22 | 234.00 | 306.00 | 255.00 | 54 | 271.00 | 239.00 | 134.00 |
| 23 | 231.00 | 325.00 | 200.00 | 55 | - | 320.00 | 274.00 |
| 24 | 261.00 | 338.00 | 252.00 | 56 | 218.00 | 288.00 | 276.00 |
| 25 | 250.00 | 321.00 | 313.00 | 57 | 234.00 | 280.00 | 300.00 |
| 26 | 280.00 | 329.00 | 288.00 | 58 | 188.00 | 232.00 | 252.00 |
| 27 | - | - | 299.00 | 59 | 180.00 | 249.00 | 292.00 |
| 28 | 299.00 | 278.00 | - | 60 | 239.00 | 297.00 | 343.00 |
| 29 | 229.00 | 353.00 | 309.00 | 61 | 166.00 | 273.00 | 235.00 |
| 30 | 252.00 | 338.00 | 265.00 | 62 | 201.00 | 296.00 | 239.00 |
| 31 | 261.00 | 168.00 | 212.00 | 63 | 206.00 | - | - |
| 32 | 265.00 | 357.00 | 254.00 | 64 | - | 259.00 | 287.00 |

**De 0 a 14 días de edad**

| N° | Tratamientos [1] | | | N° | Tratamientos [1] | | |
|---|---|---|---|---|---|---|---|
| | 1 | 2 | 3 | | 1 | 2 | 3 |
| 1 | 334.45 | 479.19 | 409.99 | 33 | 440.14 | 406.14 | 522.20 |
| 2 | 398.27 | 478.09 | 365.82 | 34 | 407.98 | 347.35 | 364.94 |
| 3 | - | 457.60 | 365.91 | 35 | 358.52 | 477.81 | 451.93 |
| 4 | 309.72 | - | 429.40 | 36 | 401.30 | 523.17 | 470.10 |
| 5 | 393.90 | 300.93 | - | 37 | 331.33 | - | 393.85 |
| 6 | 392.38 | 408.18 | 370.01 | 38 | 366.59 | 488.78 | 459.63 |
| 7 | 462.44 | 475.51 | 323.68 | 39 | - | 399.43 | 424.24 |
| 8 | 444.60 | 433.47 | 351.83 | 40 | 436.85 | 502.13 | - |
| 9 | 418.07 | 465.07 | 364.58 | 41 | 261.54 | 342.54 | 473.60 |
| 10 | 392.18 | 531.54 | 410.72 | 42 | 423.05 | 463.51 | 508.49 |
| 11 | - | - | 412.41 | 43 | 319.47 | 398.14 | 391.22 |
| 12 | 421.85 | 449.73 | 425.35 | 44 | 345.71 | 474.80 | 445.86 |
| 13 | 383.65 | 428.27 | 364.42 | 45 | - | 476.40 | - |
| 14 | 411.59 | 425.41 | 391.08 | 46 | 299.53 | - | 434.48 |
| 15 | 339.14 | 420.33 | 391.20 | 47 | 418.46 | 417.19 | 452.88 |
| 16 | 428.29 | 451.74 | - | 48 | 423.79 | 418.51 | 442.50 |
| 17 | 437.43 | 421.44 | 397.28 | 49 | 286.44 | 425.67 | 427.64 |
| 18 | 433.07 | - | - | 50 | 357.49 | - | 464.03 |
| 19 | - | 438.54 | 409.90 | 51 | 390.33 | 493.59 | 384.54 |
| 20 | 421.22 | 428.61 | 383.38 | 52 | 339.46 | 434.12 | 437.42 |
| 21 | 464.49 | 393.87 | 467.08 | 53 | 383.33 | 377.31 | - |
| 22 | 342.44 | 483.38 | 433.39 | 54 | 415.21 | 382.15 | 257.00 |
| 23 | 365.58 | 501.85 | 345.07 | 55 | - | 483.99 | 445.47 |
| 24 | 405.01 | 501.28 | 389.24 | 56 | 359.23 | 424.43 | 439.02 |
| 25 | 390.39 | 454.54 | 479.40 | 57 | 362.82 | 432.07 | 481.97 |
| 26 | 447.03 | 453.37 | 409.09 | 58 | 311.36 | 341.89 | 392.66 |
| 27 | - | - | 448.05 | 59 | 290.32 | 380.38 | 467.06 |
| 28 | 482.79 | 420.46 | - | 60 | 390.52 | 463.11 | 512.68 |
| 29 | 345.03 | 523.83 | 446.60 | 61 | 270.97 | 420.13 | 328.08 |
| 30 | 392.79 | 475.14 | 371.28 | 62 | 349.97 | 461.28 | 392.29 |
| 31 | 430.52 | 408.88 | 332.82 | 63 | 329.65 | - | - |
| 32 | 430.08 | 499.49 | 408.28 | 64 | - | 396.09 | 437.46 |

[1] Tratamientos: 1: aves control; 2: 500 ppm de Orevitol®; 3: 150 ppm de neomicina.

* Aves seleccionadas como muestras o muertas en la semana inmediata siguiente. ** Aves que fueron muestreadas o murieron antes de completar el periodo.

## ANEXO 17. Experimento 2 – crecimiento corporal

| Jaula | Consumo de alimento, g/pollo | | | Conversión alimentaria | | |
|---|---|---|---|---|---|---|
| | Tratamientos [1] | | | Tratamientos [1] | | |
| | **1** | **2** | **3** | **1** | **2** | **3** |
| 1 | 580.2 | 615.7 | 543.8 | 1.48 | 1.42 | 1.45 |
| 2 | 550.1 | 587.3 | 485.5 | 1.38 | 1.30 | 1.23 |
| 3 | 574.6 | 590.6 | 537.3 | 1.40 | 1.30 | 1.33 |
| 4 | 595.6 | 620.1 | 502.8 | 1.43 | 1.34 | 1.22 |
| 5 | 556.8 | 616.6 | 578.5 | 1.42 | 1.37 | 1.31 |
| 6 | 498.6 | 509.2 | 598.2 | 1.40 | 1.19 | 1.33 |
| 7 | 587.3 | 577.3 | 564.7 | 1.62 | 1.34 | 1.38 |
| 8 | 500.0 | 514.2 | 586.9 | 1.52 | 1.24 | 1.36 |

[1] Tratamientos: 1: aves control; 2: 500 ppm de Orevitol®; 3: 150 ppm de neomicina.

## ANEXO 18. Experimento 3 – Mediciones complementarias

| Muestras | Capacidad para caminar | | Degeneración femoral | | Discondroplasia tibial | |
|---|---|---|---|---|---|---|
| | **Control** | **PRO** | **Control** | **PRO** | **Control** | **PRO** |
| 1 | 1 | 1 | 1 | 1 | 1 | 1 |
| 2 | 1 | 1 | 1 | 1 | 1 | 1 |
| 3 | 1 | 1 | 1 | 1 | 1 | 1 |
| 4 | 1 | 1 | 1 | 1 | 1 | 1 |
| 5 | 1 | 1 | 1 | 1 | 1 | 1 |
| 6 | 1 | 1 | 1 | 1 | 1 | 1 |
| 7 | 1 | 1 | 5 | 1 | 3 | 1 |
| 8 | 1 | 1 | 1 | 1 | 1 | 1 |
| 9 | 1 | 1 | 1 | 1 | 1 | 1 |
| 10 | 1 | 1 | 1 | 1 | 1 | 1 |
| 11 | 1 | 1 | 1 | 1 | 1 | 1 |
| 12 | 1 | 1 | 1 | 1 | 1 | 1 |
| 13 | 1 | 1 | 1 | 1 | 1 | 1 |
| 14 | 1 | 1 | 1 | 1 | 1 | 1 |
| 15 | 1 | 1 | 1 | 1 | 1 | 1 |
| 16 | 1 | 1 | 1 | 1 | 1 | 1 |

## ANEXO 19. Experimento 3 – resultados generales (1 de 3)

| Hueso | Suple-mento | Rep. | Densidad | Índice Seedor | Índice Quetelet | Índice de forma | Índice de robusticidad | Volumen | Peso | Largo | Diámetro latero-lateral | Diámetro cráneo-caudal | Diámetro promedio |
|---|---|---|---|---|---|---|---|---|---|---|---|---|---|
| Fémur | 1 | 1 | 0.4615385 | 15.15152 | 0.382614 | 4.43697 | 4.69510 | 1.3 | 0.60 | 3.960 | 0.865 | 0.920 | 0.8925 |
| Fémur | 1 | 2 | 0.4583333 | 15.58074 | 0.441381 | 7.06000 | 4.30844 | 1.2 | 0.55 | 3.530 | 0.475 | 0.525 | 0.5000 |
| Fémur | 1 | 3 | 0.4285714 | 15.50388 | 0.400617 | 7.23364 | 4.58839 | 1.4 | 0.60 | 3.870 | 0.495 | 0.575 | 0.5350 |
| Fémur | 1 | 4 | 0.5428571 | 19.21618 | 0.485871 | 7.71707 | 4.33386 | 1.4 | 0.76 | 3.955 | 0.475 | 0.550 | 0.5125 |
| Fémur | 1 | 5 | 0.4571429 | 17.02128 | 0.452694 | 7.40887 | 4.36309 | 1.4 | 0.64 | 3.760 | 0.495 | 0.520 | 0.5075 |
| Fémur | 1 | 6 | 0.5214286 | 19.03520 | 0.496355 | 7.33971 | 4.25916 | 1.4 | 0.73 | 3.835 | 0.495 | 0.550 | 0.5225 |
| Fémur | 1 | 7 | 0.5416667 | 16.86122 | 0.437386 | 7.48544 | 4.45027 | 1.2 | 0.65 | 3.855 | 0.490 | 0.540 | 0.5150 |
| Fémur | 1 | 8 | 0.4714286 | 16.29630 | 0.402378 | 8.18182 | 4.65165 | 1.4 | 0.66 | 4.050 | 0.480 | 0.510 | 0.4950 |
| Fémur | 1 | 9 | 0.4916667 | 14.86146 | 0.475537 | 4.22977 | 4.39970 | 1.6 | 0.78 | 4.050 | 0.920 | 0.995 | 0.9575 |
| Fémur | 1 | 10 | 0.4545455 | 13.55014 | 0.395778 | 7.51485 | 4.57706 | 1.3 | 0.57 | 3.795 | 0.460 | 0.550 | 0.5050 |
| Fémur | 1 | 11 | 0.4500000 | 16.36364 | 0.433311 | 7.64767 | 4.39956 | 1.2 | 0.59 | 3.690 | 0.485 | 0.480 | 0.4825 |
| Fémur | 1 | 12 | 0.5214286 | 18.48101 | 0.449784 | 7.23853 | 4.44305 | 1.4 | 0.70 | 3.945 | 0.515 | 0.575 | 0.5450 |
| Fémur | 1 | 13 | 0.5333333 | 17.06667 | 0.374887 | 8.17391 | 4.64618 | 1.1 | 0.53 | 3.760 | 0.460 | 0.460 | 0.4600 |
| Fémur | 1 | 14 | 0.5153846 | 17.49347 | 0.464010 | 7.22467 | 4.45402 | 1.7 | 0.78 | 4.100 | 0.540 | 0.595 | 0.5675 |
| Fémur | 1 | 15 | 0.4714286 | 17.30013 | 0.377447 | 8.00000 | 4.70049 | 1 | 0.58 | 3.920 | 0.470 | 0.510 | 0.4900 |
| Fémur | 1 | 16 | 0.4466667 | 16.58416 | 0.458338 | 7.91837 | 4.39085 | 1.4 | 0.69 | 3.880 | 0.480 | 0.500 | 0.4900 |
| Fémur | 2 | 1 | 0.4875000 | 19.25926 | 0.374344 | 4.47324 | 4.73340 | 1.2 | 0.59 | 3.970 | 0.880 | 0.895 | 0.8875 |
| Fémur | 2 | 2 | 0.4384615 | 15.01976 | 0.367212 | 7.30693 | 4.64911 | 1.1 | 0.50 | 3.690 | 0.495 | 0.515 | 0.5050 |
| Fémur | 2 | 3 | 0.4916667 | 15.98916 | 0.425030 | 7.12963 | 4.49104 | 1.4 | 0.63 | 3.850 | 0.500 | 0.580 | 0.5400 |
| Fémur | 2 | 4 | 0.5000000 | 17.74398 | 0.467874 | 7.59615 | 4.38688 | 1.4 | 0.73 | 3.950 | 0.490 | 0.550 | 0.5200 |
| Fémur | 2 | 5 | 0.4818182 | 14.09574 | 0.455111 | 7.28155 | 4.35149 | 1.2 | 0.64 | 3.750 | 0.510 | 0.520 | 0.5150 |
| Fémur | 2 | 6 | 0.4588235 | 19.02439 | 0.456749 | 5.28276 | 4.37697 | 1.3 | 0.67 | 3.830 | 0.495 | 0.955 | 0.7250 |
| Fémur | 2 | 7 | 0.5800000 | 14.79592 | 0.453477 | 7.40777 | 4.38174 | 1.4 | 0.66 | 3.815 | 0.500 | 0.530 | 0.5150 |
| Fémur | 2 | 8 | 0.4928571 | 17.78351 | 0.410499 | 7.96059 | 4.61696 | 1.5 | 0.67 | 4.040 | 0.495 | 0.520 | 0.5075 |
| Fémur | 2 | 9 | 0.5125000 | 20.24691 | 0.499924 | 4.19689 | 4.32697 | 1.6 | 0.82 | 4.050 | 0.930 | 1.000 | 0.9650 |
| Fémur | 2 | 10 | 0.4071429 | 15.01976 | 0.395778 | 7.44118 | 4.57706 | 1.4 | 0.57 | 3.795 | 0.490 | 0.530 | 0.5100 |
| Fémur | 2 | 11 | 0.5363636 | 16.01085 | 0.434487 | 7.40704 | 4.39360 | 1.1 | 0.59 | 3.685 | 0.500 | 0.495 | 0.4975 |
| Fémur | 2 | 12 | 0.5642857 | 19.70075 | 0.491290 | 7.29091 | 4.33779 | 1.4 | 0.79 | 4.010 | 0.520 | 0.580 | 0.5500 |
| Fémur | 2 | 13 | 0.4909091 | 14.15465 | 0.371026 | 8.07407 | 4.68486 | 1.1 | 0.54 | 3.815 | 0.475 | 0.470 | 0.4725 |
| Fémur | 2 | 14 | 0.5000000 | 18.40491 | 0.451654 | 7.21239 | 4.48512 | 1.5 | 0.75 | 4.075 | 0.550 | 0.580 | 0.5650 |
| Fémur | 2 | 15 | 0.5166667 | 16.06218 | 0.416119 | 8.00000 | 4.52679 | 1.2 | 0.62 | 3.860 | 0.470 | 0.495 | 0.4825 |
| Fémur | 2 | 16 | 0.5636364 | 15.91784 | 0.408674 | 7.94898 | 4.56783 | 1.1 | 0.62 | 3.895 | 0.480 | 0.500 | 0.4900 |

Suplemento: 1: sin suplemento; 2: 500 ppm de Orevitol®.

## ANEXO 20. Experimento 3 – resultados generales (2 de 3)

| Hueso | Suple-mento | Rep. | Densidad | Índice Seedor | Índice Quetelet | Índice de forma | Índice de robusticidad | Volumen | Peso | Largo | Diámetro latero-lateral | Diámetro craneo-caudal | Diámetro promedio |
|---|---|---|---|---|---|---|---|---|---|---|---|---|---|
| Tibia | 1 | 1 | 0.5000000 | 15.65558 | 0.306371 | 12.02353 | 5.50458 | 1.6 | 0.80 | 5.110 | 0.430 | 0.420 | 0.4250 |
| Tibia | 1 | 2 | 0.4562500 | 14.92843 | 0.305285 | 11.37209 | 5.43085 | 1.6 | 0.73 | 4.890 | 0.430 | 0.430 | 0.4300 |
| Tibia | 1 | 3 | 0.4000000 | 17.25490 | 0.338331 | 10.51546 | 5.32201 | 2.2 | 0.88 | 5.100 | 0.490 | 0.480 | 0.4850 |
| Tibia | 1 | 4 | 0.5176471 | 17.56487 | 0.350596 | 12.21951 | 5.22810 | 1.7 | 0.88 | 5.010 | 0.430 | 0.390 | 0.4100 |
| Tibia | 1 | 5 | 0.4812500 | 14.93695 | 0.289757 | 11.98837 | 5.62426 | 1.6 | 0.77 | 5.155 | 0.440 | 0.420 | 0.4300 |
| Tibia | 1 | 6 | 0.5250000 | 16.32653 | 0.317328 | 11.37017 | 5.45288 | 1.6 | 0.84 | 5.145 | 0.450 | 0.455 | 0.4525 |
| Tibia | 1 | 7 | 0.4888889 | 17.32283 | 0.341001 | 11.22652 | 5.30114 | 1.8 | 0.88 | 5.080 | 0.470 | 0.435 | 0.4525 |
| Tibia | 1 | 8 | 0.4736842 | 17.61252 | 0.344668 | 11.54802 | 5.29265 | 1.9 | 0.90 | 5.110 | 0.450 | 0.435 | 0.4425 |
| Tibia | 1 | 9 | 0.5066667 | 14.91658 | 0.377219 | 10.40000 | 5.16579 | 2 | 1.02 | 5.200 | 0.510 | 0.490 | 0.5000 |
| Tibia | 1 | 10 | 0.4812500 | 15.63452 | 0.308000 | 11.29944 | 5.45515 | 1.7 | 0.77 | 5.000 | 0.455 | 0.430 | 0.4425 |
| Tibia | 1 | 11 | 0.4550000 | 17.61859 | 0.337222 | 10.91892 | 5.31038 | 1.6 | 0.86 | 5.050 | 0.495 | 0.430 | 0.4625 |
| Tibia | 1 | 12 | 0.5352941 | 17.58454 | 0.348964 | 11.27957 | 5.31686 | 1.9 | 0.96 | 5.245 | 0.475 | 0.455 | 0.4650 |
| Tibia | 1 | 13 | 0.4937500 | 15.33981 | 0.282230 | 13.30323 | 5.67381 | 1.6 | 0.75 | 5.155 | 0.400 | 0.375 | 0.3875 |
| Tibia | 1 | 14 | 0.5312500 | 16.69941 | 0.342776 | 11.18974 | 5.41911 | 2.2 | 1.02 | 5.455 | 0.495 | 0.480 | 0.4875 |
| Tibia | 1 | 15 | 0.4888889 | 17.18750 | 0.298082 | 12.34682 | 5.63726 | 1.8 | 0.85 | 5.340 | 0.440 | 0.425 | 0.4325 |
| Tibia | 1 | 16 | 0.4500000 | 17.52678 | 0.319735 | 11.84884 | 5.42148 | 1.8 | 0.83 | 5.095 | 0.450 | 0.410 | 0.4300 |
| Tibia | 2 | 1 | 0.5100000 | 19.61538 | 0.292769 | 12.27711 | 5.58307 | 1.5 | 0.76 | 5.095 | 0.430 | 0.400 | 0.4150 |
| Tibia | 2 | 2 | 0.4529412 | 15.40000 | 0.317452 | 11.19318 | 5.37332 | 1.6 | 0.77 | 4.925 | 0.440 | 0.440 | 0.4400 |
| Tibia | 2 | 3 | 0.5375000 | 17.02970 | 0.341115 | 11.16757 | 5.32995 | 2 | 0.91 | 5.165 | 0.475 | 0.450 | 0.4625 |
| Tibia | 2 | 4 | 0.5052632 | 18.30315 | 0.339798 | 12.62195 | 5.34027 | 1.7 | 0.91 | 5.175 | 0.430 | 0.390 | 0.4100 |
| Tibia | 2 | 5 | 0.4687500 | 14.54898 | 0.297860 | 12.04678 | 5.57098 | 1.6 | 0.79 | 5.150 | 0.435 | 0.420 | 0.4275 |
| Tibia | 2 | 6 | 0.4636364 | 18.69844 | 0.328083 | 11.31111 | 5.37335 | 1.6 | 0.85 | 5.090 | 0.455 | 0.445 | 0.4500 |
| Tibia | 2 | 7 | 0.4722222 | 15.91760 | 0.335693 | 10.44898 | 5.34288 | 1.8 | 0.88 | 5.120 | 0.460 | 0.520 | 0.4900 |
| Tibia | 2 | 8 | 0.4611111 | 16.29048 | 0.341320 | 11.53933 | 5.31855 | 2 | 0.90 | 5.135 | 0.450 | 0.440 | 0.4450 |
| Tibia | 2 | 9 | 0.5666667 | 19.61538 | 0.377219 | 10.19608 | 5.16579 | 1.8 | 1.02 | 5.200 | 0.525 | 0.495 | 0.5100 |
| Tibia | 2 | 10 | 0.4866667 | 14.64393 | 0.293760 | 11.01657 | 5.53636 | 1.5 | 0.73 | 4.985 | 0.470 | 0.435 | 0.4525 |
| Tibia | 2 | 11 | 0.5058824 | 17.06349 | 0.338561 | 10.89730 | 5.29986 | 1.7 | 0.86 | 5.040 | 0.490 | 0.435 | 0.4625 |
| Tibia | 2 | 12 | 0.5000000 | 19.01141 | 0.361434 | 11.37297 | 5.26000 | 2 | 1.00 | 5.260 | 0.475 | 0.450 | 0.4625 |
| Tibia | 2 | 13 | 0.4687500 | 14.46480 | 0.278974 | 13.38065 | 5.70683 | 1.6 | 0.75 | 5.185 | 0.400 | 0.375 | 0.3875 |
| Tibia | 2 | 14 | 0.4714286 | 18.11528 | 0.331478 | 11.21026 | 5.48334 | 2.1 | 0.99 | 5.465 | 0.495 | 0.480 | 0.4875 |
| Tibia | 2 | 15 | 0.6071429 | 16.11374 | 0.305474 | 11.92090 | 5.56864 | 1.4 | 0.85 | 5.275 | 0.445 | 0.440 | 0.4425 |
| Tibia | 2 | 16 | 0.4777778 | 17.13147 | 0.341264 | 11.67442 | 5.27883 | 1.8 | 0.86 | 5.020 | 0.450 | 0.410 | 0.4300 |

Suplemento: 1: sin suplemento; 2: 500 ppm de Orevitol®.

## ANEXO 21. Experimento 3 – resultados generales (3 de 3)

| Hueso | Suple-mento | Rep. | Densidad | Índice Seedor | Índice Quetelet | Índice de forma | Índice de robusticidad | Volumen | Peso | Largo | Diámetro latero-lateral | Diámetro craneo-caudal | Diámetro promedio |
|---|---|---|---|---|---|---|---|---|---|---|---|---|---|
| Tarso | 1 | 1 | 0.5500000 | 14.70588 | 0.393205 | 8.59770 | 4.56475 | 1 | 0.55 | 3.740 | 0.535 | 0.335 | 0.4350 |
| Tarso | 1 | 2 | 0.4500000 | 12.67606 | 0.357072 | 7.63441 | 4.63259 | 1 | 0.45 | 3.550 | 0.570 | 0.360 | 0.4650 |
| Tarso | 1 | 3 | 0.4500000 | 16.44909 | 0.429480 | 7.22642 | 4.46771 | 1.4 | 0.63 | 3.830 | 0.665 | 0.395 | 0.5300 |
| Tarso | 1 | 4 | 0.5333333 | 16.93122 | 0.447916 | 7.95789 | 4.38630 | 1.2 | 0.64 | 3.780 | 0.580 | 0.370 | 0.4750 |
| Tarso | 1 | 5 | 0.5700000 | 15.00000 | 0.394737 | 8.73563 | 4.58309 | 1 | 0.57 | 3.800 | 0.540 | 0.330 | 0.4350 |
| Tarso | 1 | 6 | 0.5250000 | 16.11253 | 0.412085 | 8.02051 | 4.56103 | 1.2 | 0.63 | 3.910 | 0.600 | 0.375 | 0.4875 |
| Tarso | 1 | 7 | 0.4357143 | 15.78266 | 0.408348 | 8.05208 | 4.55729 | 1.4 | 0.61 | 3.865 | 0.625 | 0.335 | 0.4800 |
| Tarso | 1 | 8 | 0.4214286 | 15.34460 | 0.399079 | 7.61386 | 4.58436 | 1.4 | 0.59 | 3.845 | 0.635 | 0.375 | 0.5050 |
| Tarso | 1 | 9 | 0.5200000 | 13.86667 | 0.413971 | 7.00452 | 4.53851 | 1.2 | 0.62 | 3.870 | 0.665 | 0.440 | 0.5525 |
| Tarso | 1 | 10 | 0.4800000 | 13.71429 | 0.382938 | 8.09890 | 4.58251 | 1.1 | 0.52 | 3.685 | 0.570 | 0.340 | 0.4550 |
| Tarso | 1 | 11 | 0.4769231 | 16.31579 | 0.422152 | 8.02128 | 4.46983 | 1.2 | 0.60 | 3.770 | 0.600 | 0.340 | 0.4700 |
| Tarso | 1 | 12 | - | - | 0.420851 | 7.66829 | 4.53685 | 1.3 | 0.65 | 3.930 | 0.645 | 0.380 | 0.5125 |
| Tarso | 1 | 13 | 0.4600000 | 12.10526 | 0.360111 | 8.73563 | 4.72552 | 1 | 0.52 | 3.800 | 0.565 | 0.305 | 0.4350 |
| Tarso | 1 | 14 | 0.5000000 | 15.40436 | 0.493827 | 7.39726 | 4.34470 | 1.6 | 0.81 | 4.050 | 0.665 | 0.430 | 0.5475 |
| Tarso | 1 | 15 | 0.5083333 | 15.84416 | 0.386264 | 7.82828 | 4.64653 | 1.2 | 0.58 | 3.875 | 0.630 | 0.360 | 0.4950 |
| Tarso | 1 | 16 | 0.4428571 | 16.10390 | 0.415773 | 7.60622 | 4.45249 | 1.2 | 0.56 | 3.670 | 0.600 | 0.365 | 0.4825 |
| Tarso | 2 | 1 | 0.5166667 | 16.02067 | 0.369778 | 8.52273 | 4.66334 | 1 | 0.52 | 3.750 | 0.550 | 0.330 | 0.4400 |
| Tarso | 2 | 2 | 0.4727273 | 14.11126 | 0.391837 | 7.21649 | 4.47014 | 1 | 0.48 | 3.500 | 0.630 | 0.340 | 0.4850 |
| Tarso | 2 | 3 | 0.5000000 | 15.91512 | 0.429363 | 7.30769 | 4.45642 | 1.3 | 0.62 | 3.800 | 0.645 | 0.395 | 0.5200 |
| Tarso | 2 | 4 | 0.5000000 | 16.53944 | - | - | - | - | - | - | - | - | - |
| Tarso | 2 | 5 | 0.5200000 | 13.68421 | 0.318560 | 8.63636 | 4.92264 | 1 | 0.46 | 3.800 | 0.560 | 0.320 | 0.4400 |
| Tarso | 2 | 6 | 0.5062500 | 20.00000 | 0.395491 | 7.90863 | 4.61803 | 1.2 | 0.60 | 3.895 | 0.620 | 0.365 | 0.4925 |
| Tarso | 2 | 7 | 0.4833333 | 14.96774 | 0.411537 | 7.66169 | 4.53960 | 1.2 | 0.61 | 3.850 | 0.640 | 0.365 | 0.5025 |
| Tarso | 2 | 8 | 0.4666667 | 15.25886 | 0.418283 | 7.54902 | 4.51506 | 1.4 | 0.62 | 3.850 | 0.640 | 0.380 | 0.5100 |
| Tarso | 2 | 9 | 0.5583333 | 16.94058 | 0.428333 | 7.06250 | 4.51982 | 1.2 | 0.67 | 3.955 | 0.680 | 0.440 | 0.5600 |
| Tarso | 2 | 10 | 0.4750000 | 15.40541 | 0.416362 | 8.08743 | 4.46249 | 1.2 | 0.57 | 3.700 | 0.580 | 0.335 | 0.4575 |
| Tarso | 2 | 11 | 0.5700000 | 14.96063 | 0.392667 | 7.97906 | 4.59515 | 1 | 0.57 | 3.810 | 0.605 | 0.350 | 0.4775 |
| Tarso | 2 | 12 | 0.5307692 | 17.92208 | 0.465509 | 7.66169 | 4.35690 | 1.3 | 0.69 | 3.850 | 0.640 | 0.365 | 0.5025 |
| Tarso | 2 | 13 | 0.5500000 | 14.53104 | 0.383911 | 8.45810 | 4.61968 | 1 | 0.55 | 3.785 | 0.585 | 0.310 | 0.4475 |
| Tarso | 2 | 14 | 0.5000000 | 18.40491 | 0.451654 | 7.44292 | 4.48512 | 1.5 | 0.75 | 4.075 | 0.665 | 0.430 | 0.5475 |
| Tarso | 2 | 15 | 0.4500000 | 13.74046 | 0.349630 | 7.93939 | 4.82608 | 1.2 | 0.54 | 3.930 | 0.635 | 0.355 | 0.4950 |
| Tarso | 2 | 16 | 0.5083333 | 16.55360 | 0.449216 | 7.63731 | 4.34504 | 1.2 | 0.61 | 3.685 | 0.610 | 0.355 | 0.4825 |

Suplemento: 1: sin suplemento; 2: 500 ppm de Orevitol®.

## ANEXO 22. Experimento presentado en el Anexo 1 - peso vivo individual, gramos (1 de 3)

| Día 1 de edad | | | | | | | | Día 7 de edad | | | | | | | |
|---|---|---|---|---|---|---|---|---|---|---|---|---|---|---|---|
| Jaula | N° | Tratamientos [1] | | Jaula | N° | Tratamientos [1] | | Jaula | N° | Tratamientos [1] | | Jaula | N° | Tratamientos [1] | |
| | | 1 | 2 | | | 1 | 2 | | | 1 | 2 | | | 1 | 2 |
| 1 | 1 | 46.61 | 44.55 | 5 | 33 | 47.48 | 46.86 | 1 | 1 | 208 | 159 | 5 | 33 | 263 | 219 |
| | 2 | 50.92 | 40.73 | | 34 | 43.20 | 50.02 | | 2 | 166 | 185 | | 34 | 173* | 192 |
| | 3 | 48.50 | 45.77 | | 35 | 49.84 | 48.48 | | 3 | 182 | 194* | | 35 | 213 | 174 |
| | 4 | 49.83 | 46.28 | | 36 | 51.17 | 45.70 | | 4 | 180 | 189 | | 36 | 150 | 193 |
| | 5 | 50.90 | 46.10 | | 37 | 50.48 | 50.67 | | 5 | 173* | 197 | | 37 | 195 | 186 |
| | 6 | 52.99 | 48.62 | | 38 | 47.38 | 49.41 | | 6 | 161 | 193 | | 38 | 196 | 186 |
| | 7 | 49.17 | 45.56 | | 39 | 42.61 | 47.43 | | 7 | 197 | 221 | | 39 | 99 | 193* |
| | 8 | 49.05 | 52.40 | | 40 | 46.34 | 51.15 | | 8 | 172 | 208 | | 40 | 174 | 206 |
| 2 | 9 | 41.48 | 46.93 | 6 | 41 | 47.08 | 48.46 | 2 | 9 | 179 | 217 | 6 | 41 | 185 | 157 |
| | 10 | 47.15 | 47.82 | | 42 | 46.52 | 49.95 | | 10 | 182 | 192 | | 42 | 183 | 220 |
| | 11 | 56.44 | 48.77 | | 43 | 50.05 | 49.53 | | 11 | 204 | 212* | | 43 | 208 | 165 |
| | 12 | 50.76 | 41.15 | | 44 | 44.19 | 51.29 | | 12 | 169 | 208 | | 44 | 181 | 195 |
| | 13 | 47.68 | 54.35 | | 45 | 52.03 | 46.31 | | 13 | 173 | 208 | | 45 | 210 | 193* |
| | 14 | 46.03 | 42.41 | | 46 | 45.68 | 40.47 | | 14 | 189 | 190 | | 46 | 182* | 153 |
| | 15 | 50.82 | 52.86 | | 47 | 44.37 | 48.54 | | 15 | 177* | 278 | | 47 | 179 | 212 |
| | 16 | 51.62 | 48.71 | | 48 | 49.45 | 49.21 | | 16 | 162 | 223 | | 48 | 182 | 207 |
| 3 | 17 | 53.17 | 47.57 | 7 | 49 | 47.56 | 55.56 | 3 | 17 | 182 | 202 | 7 | 49 | 187 | 165 |
| | 18 | 43.02 | 48.93 | | 50 | 47.35 | 48.51 | | 18 | 183 | 219 | | 50 | 180 | 184 |
| | 19 | 46.19 | 42.18 | | 51 | 43.54 | 49.67 | | 19 | 174 | 198* | | 51 | 188 | 186 |
| | 20 | 45.40 | 48.78 | | 52 | 47.60 | 48.54 | | 20 | 170* | 210 | | 52 | 172 | 142 |
| | 21 | 46.58 | 48.51 | | 53 | 52.37 | 43.67 | | 21 | 179 | 207 | | 53 | 169 | 188 |
| | 22 | 44.89 | 48.56 | | 54 | 42.80 | 52.79 | | 22 | 143 | 157 | | 54 | 162 | 197 |
| | 23 | 46.57 | 46.42 | | 55 | 46.27 | 46.25 | | 23 | 159 | 181 | | 55 | 178* | 180* |
| | 24 | 47.92 | 49.99 | | 56 | 51.55 | 49.77 | | 24 | 181 | 194 | | 56 | 179 | 191 |
| 4 | 25 | 52.81 | 50.61 | 8 | 57 | 46.67 | 62.18 | 4 | 25 | 192 | 191 | 8 | 57 | 172 | 191 |
| | 26 | 46.70 | 46.97 | | 58 | 54.87 | 42.64 | | 26 | 120* | 214 | | 58 | 153 | 166 |
| | 27 | 49.15 | 45.65 | | 59 | 47.58 | 44.68 | | 27 | 184 | 213* | | 59 | 170 | 155 |
| | 28 | 44.20 | 46.21 | | 60 | 49.17 | 44.48 | | 28 | 165 | 230 | | 60 | 184 | 196 |
| | 29 | 48.81 | 45.97 | | 61 | 46.13 | 46.03 | | 29 | 187 | 162 | | 61 | 194 | 151 |
| | 30 | 50.30 | 40.21 | | 62 | 51.78 | 54.03 | | 30 | 172 | 181 | | 62 | 180* | 203 |
| | 31 | 48.89 | 52.48 | | 63 | 50.35 | 44.35 | | 31 | 166* | 222 | | 63 | 167 | 168 |
| | 32 | 47.01 | 52.92 | | 64 | 52.29 | 46.36 | | 32 | 147 | 218 | | 64 | 205 | 186* |

[1] Tratamientos: 1: aves sometidas al modelo de desafío A; 2: aves sometidas al modelo de desafío B

* Las aves marcadas con un asterisco (*) fueron seleccionados al azar como muestras para otras evaluaciones o mueren en la semana inmediata siguiente.

## ANEXO 23. Experimento presentado en el Anexo 1- peso vivo individual, gramos (2 de 3)

| Día 14 de edad | | | | | | | | Día 21 de edad | | | | | | | |
|---|---|---|---|---|---|---|---|---|---|---|---|---|---|---|---|
| Jaula | N° | Tratamientos [1] | | Jaula | N° | Tratamientos [1] | | Jaula | N° | Tratamientos [1] | | Jaula | N° | Tratamientos [1] | |
| | | 1 | 2 | | | 1 | 2 | | | 1 | 2 | | | 1 | 2 |
| 1 | 1 | 400 | 379* | 5 | 33 | 348 | 487 | 1 | 1 | 880 | - | 5 | 33 | 536* | 951 |
| | 2 | 331 | 439* | | 34 | - | 458 | | 2 | 690 | - | | 34 | - | 981 |
| | 3 | 388 | - | | 35 | 419 | 407* | | 3 | 840 | - | | 35 | 730 | - |
| | 4 | 310* | 356 | | 36 | 310 | 447 | | 4 | - | 674* | | 36 | 600 | 861 |
| | 5 | - | 440 | | 37 | 417 | 382 | | 5 | - | 901 | | 37 | 810 | 649* |
| | 6 | 335 | 441 | | 38 | 425 | 416 | | 6 | 604* | 811 | | 38 | 920 | 801 |
| | 7 | 336 | 508 | | 39 | 249* | - | | 7 | 790 | 1011 | | 39 | - | - |
| | 8 | 347 | 497 | | 40 | 365 | 488 | | 8 | 770 | 991 | | 40 | - | 921 |
| 2 | 9 | 392 | 465* | 6 | 41 | 384 | 310* | 2 | 9 | 650 | - | 6 | 41 | 720 | - |
| | 10 | 382 | 440 | | 42 | 367 | 473 | | 10 | 790 | 911 | | 42 | 556* | 961 |
| | 11 | 393 | - | | 43 | 466 | 369 | | 11 | 737* | - | | 43 | 820 | 751 |
| | 12 | 396* | 463 | | 44 | 362 | 397 | | 12 | - | 851 | | 44 | 620 | 735* |
| | 13 | 370 | 438 | | 45 | 470 | - | | 13 | 770* | 751* | | 45 | 930 | - |
| | 14 | 457 | 454 | | 46 | - | 340 | | 14 | 900 | 921 | | 46 | - | 711 |
| | 15 | - | 392 | | 47 | 379* | 467 | | 15 | - | 761 | | 47 | - | 921 |
| | 16 | 318 | 477 | | 48 | 405 | 473 | | 16 | 560* | 884* | | 48 | 750 | 951 |
| 3 | 17 | 379 | 485 | 7 | 49 | 425 | 342* | 3 | 17 | 700 | 1011 | 7 | 49 | 718* | - |
| | 18 | 401 | 482 | | 50 | 374* | 406 | | 18 | 730 | 891 | | 50 | - | 716* |
| | 19 | 427 | - | | 51 | 424 | 440 | | 19 | 870 | - | | 51 | 840 | 891 |
| | 20 | - | 470 | | 52 | 403 | 388 | | 20 | - | 881 | | 52 | 830 | 811 |
| | 21 | 449 | 513 | | 53 | 330 | 427 | | 21 | 850 | 1001 | | 53 | 660 | 691 |
| | 22 | 369* | 391* | | 54 | 354 | 468 | | 22 | - | - | | 54 | 680 | 801 |
| | 23 | 358 | 412* | | 55 | - | - | | 23 | 616* | - | | 55 | - | - |
| | 24 | 417 | 455 | | 56 | 391 | 409 | | 24 | 760 | 791* | | 56 | 750 | 771 |
| 4 | 25 | 390 | 441* | 8 | 57 | 339* | 425 | 4 | 25 | 630* | - | 8 | 57 | - | 941 |
| | 26 | - | 494 | | 58 | 347 | 354 | | 26 | - | 891 | | 58 | 690 | 761 |
| | 27 | 399 | - | | 59 | 414 | 335 | | 27 | 850 | - | | 59 | 810 | 771 |
| | 28 | 332 | 529 | | 60 | 420 | 435* | | 28 | 632* | 1061 | | 60 | 800 | - |
| | 29 | 338 | 391 | | 61 | 440 | 317 | | 29 | 660 | 791 | | 61 | 770 | 593* |
| | 30 | 376 | 433 | | 62 | - | 404 | | 30 | 740 | 809* | | 62 | - | 851 |
| | 31 | - | 483 | | 63 | 381 | 374 | | 31 | - | 871 | | 63 | 680 | 791 |
| | 32 | 288* | 483 | | 64 | 416 | - | | 32 | - | 881 | | 64 | 722* | - |

[1] Tratamientos: 1: aves sometidas al modelo de desafío A; 2: aves sometidas al modelo de desafío B

* Las aves marcadas con un asterisco (*) fueron seleccionados al azar como muestras para otras evaluaciones o mueren en la semana inmediata siguiente.

## ANEXO 24. Experimento presentado en el Anexo 1 - peso vivo individual, gramos (3 de 3)

| Día 28 de edad | | | | | | | | |
|---|---|---|---|---|---|---|---|---|
| Jaula | N° | Tratamientos [1] | | Jaula | N° | Tratamientos [1] | | |
| | | 1 | 2 | | | 1 | 2 |
| 1 | 1 | 1400 | - | 5 | 33 | - | 1432 |
| | 2 | 1090 | - | | 34 | - | 1512 |
| | 3 | 1394 | - | | 35 | 1050 | - |
| | 4 | - | - | | 36 | 813 | 1332 |
| | 5 | - | 1242 | | 37 | 1270 | - |
| | 6 | - | 1234 | | 38 | 1420 | 1187 |
| | 7 | 1290 | 1412 | | 39 | - | - |
| | 8 | 1160 | 1462 | | 40 | - | 1352 |
| 2 | 9 | 910 | - | 6 | 41 | 1030 | - |
| | 10 | 1230 | 1402 | | 42 | - | 1332 |
| | 11 | - | - | | 43 | 1300 | 1130 |
| | 12 | - | 1322 | | 44 | 926 | - |
| | 13 | - | - | | 45 | 1540 | - |
| | 14 | 1196 | 1462 | | 46 | - | 1032 |
| | 15 | - | 1128 | | 47 | - | 1322 |
| | 16 | - | - | | 48 | 1130 | 1302 |
| 3 | 17 | 1081 | 1632 | 7 | 49 | - | - |
| | 18 | 1110 | 1392 | | 50 | - | - |
| | 19 | 1420 | - | | 51 | 1390 | 1312 |
| | 20 | - | 1262 | | 52 | 1370 | 1246 |
| | 21 | 1310 | 1562 | | 53 | 1048 | 1032 |
| | 22 | - | - | | 54 | 1130 | 1222 |
| | 23 | - | - | | 55 | - | - |
| | 24 | 1160 | - | | 56 | 1140 | 1122 |
| 4 | 25 | - | - | 8 | 57 | - | 1352 |
| | 26 | - | 1292 | | 58 | 1090 | 1018 |
| | 27 | 1330 | - | | 59 | 1280 | 1122 |
| | 28 | - | 1592 | | 60 | 1180 | - |
| | 29 | 1050 | 1241 | | 61 | 1018 | - |
| | 30 | 1045 | - | | 62 | - | 1142 |
| | 31 | - | 1222 | | 63 | 1080 | 1142 |
| | 32 | - | 1182 | | 64 | - | - |

[1] Tratamientos: 1: aves sometidas al modelo de desafío A; 2: aves sometidas al modelo de desafío B

* Las aves marcadas con un asterisco (*) fueron seleccionados al azar como muestras para otras evaluaciones o mueren en la semana inmediata siguiente.

## ANEXO 25. Experimento presentado en el Anexo 1 – ganancia de peso individual, gramos (1 de 4)

| De 0 a 7 días de edad | | | | | | | | De 0 a 14 días de edad | | | | | | | |
|---|---|---|---|---|---|---|---|---|---|---|---|---|---|---|---|
| Jaula | N° | Tratamientos [1] | | Jaula | N° | Tratamientos [1] | | Jaula | N° | Tratamientos [1] | | Jaula | N° | Tratamientos [1] | |
| | | 1 | 2 | | | 1 | 2 | | | 1 | 2 | | | 1 | 2 |
| 1 | 1 | 161.39 | 114.45 | 5 | 33 | 215.52 | 172.14 | 1 | 1 | 353.39 | 334.45 | 5 | 33 | 300.52 | 440.14 |
| | 2 | 115.08 | 144.27 | | 34 | 129.80 | 141.98 | | 2 | 280.08 | 398.27 | | 34 | - | 407.98 |
| | 3 | 133.50 | 148.23 | | 35 | 163.16 | 125.52 | | 3 | 339.50 | - | | 35 | 369.16 | 358.52 |
| | 4 | 130.17 | 142.72 | | 36 | 98.83 | 147.30 | | 4 | 260.17 | 309.72 | | 36 | 258.83 | 401.30 |
| | 5 | 122.10 | 150.90 | | 37 | 144.52 | 135.33 | | 5 | - | 393.90 | | 37 | 366.52 | 331.33 |
| | 6 | 108.01 | 144.38 | | 38 | 148.62 | 136.59 | | 6 | 282.01 | 392.38 | | 38 | 377.62 | 366.59 |
| | 7 | 147.83 | 175.44 | | 39 | 56.39 | 145.57 | | 7 | 286.83 | 462.44 | | 39 | 206.39 | - |
| | 8 | 122.95 | 155.60 | | 40 | 127.66 | 154.85 | | 8 | 297.95 | 444.60 | | 40 | 318.66 | 436.85 |
| 2 | 9 | 137.52 | 170.07 | 6 | 41 | 137.92 | 108.54 | 2 | 9 | 350.52 | 418.07 | 6 | 41 | 336.92 | 261.54 |
| | 10 | 134.85 | 144.18 | | 42 | 136.48 | 170.05 | | 10 | 334.85 | 392.18 | | 42 | 320.48 | 423.05 |
| | 11 | 147.56 | 163.23 | | 43 | 157.95 | 115.47 | | 11 | 336.56 | - | | 43 | 415.95 | 319.47 |
| | 12 | 118.24 | 166.85 | | 44 | 136.81 | 143.71 | | 12 | 345.24 | 421.85 | | 44 | 317.81 | 345.71 |
| | 13 | 125.32 | 153.65 | | 45 | 157.97 | 146.69 | | 13 | 322.32 | 383.65 | | 45 | 417.97 | - |
| | 14 | 142.97 | 147.59 | | 46 | 136.32 | 112.53 | | 14 | 410.97 | 411.59 | | 46 | - | 299.53 |
| | 15 | 126.18 | 225.14 | | 47 | 134.63 | 163.46 | | 15 | - | 339.14 | | 47 | 334.63 | 418.46 |
| | 16 | 110.38 | 174.29 | | 48 | 132.55 | 157.79 | | 16 | 266.38 | 428.29 | | 48 | 355.55 | 423.79 |
| 3 | 17 | 128.83 | 154.43 | 7 | 49 | 139.44 | 109.44 | 3 | 17 | 325.83 | 437.43 | 7 | 49 | 377.44 | 286.44 |
| | 18 | 139.98 | 170.07 | | 50 | 132.65 | 135.49 | | 18 | 357.98 | 433.07 | | 50 | 326.65 | 357.49 |
| | 19 | 127.81 | 155.82 | | 51 | 144.46 | 136.33 | | 19 | 380.81 | - | | 51 | 380.46 | 390.33 |
| | 20 | 124.60 | 161.22 | | 52 | 124.40 | 93.46 | | 20 | - | 421.22 | | 52 | 355.40 | 339.46 |
| | 21 | 132.42 | 158.49 | | 53 | 116.63 | 144.33 | | 21 | 402.42 | 464.49 | | 53 | 277.63 | 383.33 |
| | 22 | 98.11 | 108.44 | | 54 | 119.20 | 144.21 | | 22 | 324.11 | 342.44 | | 54 | 311.20 | 415.21 |
| | 23 | 112.43 | 134.58 | | 55 | 131.73 | 133.75 | | 23 | 311.43 | 365.58 | | 55 | - | - |
| | 24 | 133.08 | 144.01 | | 56 | 127.45 | 141.23 | | 24 | 369.08 | 405.01 | | 56 | 339.45 | 359.23 |
| 4 | 25 | 139.19 | 140.39 | 8 | 57 | 125.33 | 128.82 | 4 | 25 | 337.19 | 390.39 | 8 | 57 | 292.33 | 362.82 |
| | 26 | 73.30 | 167.03 | | 58 | 98.13 | 123.36 | | 26 | - | 447.03 | | 58 | 292.13 | 311.36 |
| | 27 | 134.85 | 167.35 | | 59 | 122.42 | 110.32 | | 27 | 349.85 | - | | 59 | 366.42 | 290.32 |
| | 28 | 120.80 | 183.79 | | 60 | 134.83 | 151.52 | | 28 | 287.80 | 482.79 | | 60 | 370.83 | 390.52 |
| | 29 | 138.19 | 116.03 | | 61 | 147.87 | 104.97 | | 29 | 289.19 | 345.03 | | 61 | 393.87 | 270.97 |
| | 30 | 121.70 | 140.79 | | 62 | 128.22 | 148.97 | | 30 | 325.70 | 392.79 | | 62 | - | 349.97 |
| | 31 | 117.11 | 169.52 | | 63 | 116.65 | 123.65 | | 31 | - | 430.52 | | 63 | 330.65 | 329.65 |
| | 32 | 99.99 | 165.08 | | 64 | 152.71 | 139.64 | | 32 | 240.99 | 430.08 | | 64 | 363.71 | - |

[1] Tratamientos: 1: aves sometidas al modelo de desafío A; 2: aves sometidas al modelo de desafío B

## ANEXO 26. Experimento presentado en el Anexo 1 – ganancia de peso individual, gramos (2 de 4)

| De 0 a 21 días de edad | | | | | | | | De 0 a 28 días de edad | | | | | | | |
|---|---|---|---|---|---|---|---|---|---|---|---|---|---|---|---|
| Jaula | N° | Tratamientos [1] | | Jaula | N° | Tratamientos [1] | | Jaula | N° | Tratamientos [1] | | Jaula | N° | Tratamientos [1] | |
| | | 1 | 2 | | | 1 | 2 | | | 1 | 2 | | | 1 | 2 |
| 1 | 1 | 833.39 | - | 5 | 33 | 488.52 | 904.14 | 1 | 1 | 1353.39 | - | 5 | 33 | - | 1385.14 |
| | 2 | 639.08 | - | | 34 | - | 930.98 | | 2 | 1039.08 | - | | 34 | - | 1461.98 |
| | 3 | 791.50 | - | | 35 | 680.16 | - | | 3 | 1345.50 | - | | 35 | 1000.16 | - |
| | 4 | - | 627.72 | | 36 | 548.83 | 815.30 | | 4 | - | - | | 36 | 761.83 | 1286.30 |
| | 5 | - | 854.90 | | 37 | 759.52 | 598.33 | | 5 | - | 1195.90 | | 37 | 1219.52 | - |
| | 6 | 551.01 | 762.38 | | 38 | 872.62 | 751.59 | | 6 | - | 1185.38 | | 38 | 1372.62 | 1137.59 |
| | 7 | 740.83 | 965.44 | | 39 | - | - | | 7 | 1240.83 | 1366.44 | | 39 | - | - |
| | 8 | 720.95 | 938.60 | | 40 | - | 869.85 | | 8 | 1110.95 | 1409.60 | | 40 | - | 1300.85 |
| 2 | 9 | 608.52 | - | 6 | 41 | 672.92 | - | 2 | 9 | 868.52 | - | 6 | 41 | 982.92 | - |
| | 10 | 742.85 | 863.18 | | 42 | 509.48 | 911.05 | | 10 | 1182.85 | 1354.18 | | 42 | - | 1282.05 |
| | 11 | 680.56 | - | | 43 | 769.95 | 701.47 | | 11 | - | - | | 43 | 1249.95 | 1080.47 |
| | 12 | - | 809.85 | | 44 | 575.81 | 683.71 | | 12 | - | 1280.85 | | 44 | 881.81 | - |
| | 13 | 722.32 | 696.65 | | 45 | 877.97 | - | | 13 | - | - | | 45 | 1487.97 | - |
| | 14 | 853.97 | 878.59 | | 46 | - | 670.53 | | 14 | 1149.97 | 1419.59 | | 46 | - | 991.53 |
| | 15 | - | 708.14 | | 47 | - | 872.46 | | 15 | - | 1075.14 | | 47 | - | 1273.46 |
| | 16 | 508.38 | 835.29 | | 48 | 700.55 | 901.79 | | 16 | - | - | | 48 | 1080.55 | 1252.79 |
| 3 | 17 | 646.83 | 963.43 | 7 | 49 | 670.44 | - | 3 | 17 | 1027.83 | 1584.43 | 7 | 49 | - | - |
| | 18 | 686.98 | 842.07 | | 50 | - | 667.49 | | 18 | 1066.98 | 1343.07 | | 50 | - | - |
| | 19 | 823.81 | - | | 51 | 796.46 | 841.33 | | 19 | 1373.81 | - | | 51 | 1346.46 | 1262.33 |
| | 20 | - | 832.22 | | 52 | 782.40 | 762.46 | | 20 | - | 1213.22 | | 52 | 1322.40 | 1197.46 |
| | 21 | 803.42 | 952.49 | | 53 | 607.63 | 647.33 | | 21 | 1263.42 | 1513.49 | | 53 | 995.63 | 988.33 |
| | 22 | - | - | | 54 | 637.20 | 748.21 | | 22 | - | - | | 54 | 1087.20 | 1169.21 |
| | 23 | 569.43 | - | | 55 | - | - | | 23 | - | - | | 55 | - | - |
| | 24 | 712.08 | 741.01 | | 56 | 698.45 | 721.23 | | 24 | 1112.08 | - | | 56 | 1088.45 | 1072.23 |
| 4 | 25 | 577.19 | - | 8 | 57 | - | 878.82 | 4 | 25 | - | - | 8 | 57 | - | 1289.82 |
| | 26 | - | 844.03 | | 58 | 635.13 | 718.36 | | 26 | - | 1245.03 | | 58 | 1035.13 | 975.36 |
| | 27 | 800.85 | - | | 59 | 762.42 | 726.32 | | 27 | 1280.85 | - | | 59 | 1232.42 | 1077.32 |
| | 28 | 587.80 | 1014.79 | | 60 | 750.83 | - | | 28 | - | 1545.79 | | 60 | 1130.83 | - |
| | 29 | 611.19 | 745.03 | | 61 | 723.87 | 546.97 | | 29 | 1001.19 | 1195.03 | | 61 | 971.87 | - |
| | 30 | 689.70 | 768.79 | | 62 | - | 796.97 | | 30 | 994.70 | - | | 62 | - | 1087.97 |
| | 31 | - | 818.52 | | 63 | 629.65 | 746.65 | | 31 | - | 1169.52 | | 63 | 1029.65 | 1097.65 |
| | 32 | - | 828.08 | | 64 | 669.71 | - | | 32 | - | 1129.08 | | 64 | - | - |

[1] Tratamientos: 1: aves sometidas al modelo de desafío A; 2: aves sometidas al modelo de desafío B

## ANEXO 27. Experimento presentado en el Anexo 1 – ganancia de peso individual, gramos (3 de 4)

| Semana 2 | | | | | | | | Semana 3 | | | | | | | |
|---|---|---|---|---|---|---|---|---|---|---|---|---|---|---|---|
| Jaula | N° | Tratamientos [1] | | Jaula | N° | Tratamientos [1] | | Jaula | N° | Tratamientos [1] | | Jaula | N° | Tratamientos [1] | |
| | | 1 | 2 | | | 1 | 2 | | | 1 | 2 | | | 1 | 2 |
| 1 | 1 | 192.00 | 220.00 | 5 | 33 | 85.00 | 268.00 | 1 | 1 | 480.00 | - | 5 | 33 | 188.00 | 464.00 |
| | 2 | 165.00 | 254.00 | | 34 | - | 266.00 | | 2 | 359.00 | - | | 34 | - | 523.00 |
| | 3 | 206.00 | - | | 35 | 206.00 | 233.00 | | 3 | 452.00 | - | | 35 | 311.00 | - |
| | 4 | 130.00 | 167.00 | | 36 | 160.00 | 254.00 | | 4 | - | 318.00 | | 36 | 290.00 | 414.00 |
| | 5 | - | 243.00 | | 37 | 222.00 | 196.00 | | 5 | - | 461.00 | | 37 | 393.00 | 267.00 |
| | 6 | 174.00 | 248.00 | | 38 | 229.00 | 230.00 | | 6 | 269.00 | 370.00 | | 38 | 495.00 | 385.00 |
| | 7 | 139.00 | 287.00 | | 39 | 150.00 | - | | 7 | 454.00 | 503.00 | | 39 | - | - |
| | 8 | 175.00 | 289.00 | | 40 | 191.00 | 282.00 | | 8 | 423.00 | 494.00 | | 40 | - | 433.00 |
| 2 | 9 | 213.00 | 248.00 | 6 | 41 | 199.00 | 153.00 | 2 | 9 | 258.00 | - | 6 | 41 | 336.00 | - |
| | 10 | 200.00 | 248.00 | | 42 | 184.00 | 253.00 | | 10 | 408.00 | 471.00 | | 42 | 189.00 | 488.00 |
| | 11 | 189.00 | - | | 43 | 258.00 | 204.00 | | 11 | 344.00 | - | | 43 | 354.00 | 382.00 |
| | 12 | 227.00 | 255.00 | | 44 | 181.00 | 202.00 | | 12 | - | 388.00 | | 44 | 258.00 | 338.00 |
| | 13 | 197.00 | 230.00 | | 45 | 260.00 | - | | 13 | 400.00 | 313.00 | | 45 | 460.00 | - |
| | 14 | 268.00 | 264.00 | | 46 | - | 187.00 | | 14 | 443.00 | 467.00 | | 46 | - | 371.00 |
| | 15 | - | 114.00 | | 47 | 200.00 | 255.00 | | 15 | - | 369.00 | | 47 | - | 454.00 |
| | 16 | 156.00 | 254.00 | | 48 | 223.00 | 266.00 | | 16 | 242.00 | 407.00 | | 48 | 345.00 | 478.00 |
| 3 | 17 | 197.00 | 283.00 | 7 | 49 | 238.00 | 177.00 | 3 | 17 | 321.00 | 526.00 | 7 | 49 | 293.00 | - |
| | 18 | 218.00 | 263.00 | | 50 | 194.00 | 222.00 | | 18 | 329.00 | 409.00 | | 50 | - | 310.00 |
| | 19 | 253.00 | - | | 51 | 236.00 | 254.00 | | 19 | 443.00 | - | | 51 | 416.00 | 451.00 |
| | 20 | - | 260.00 | | 52 | 231.00 | 246.00 | | 20 | - | 411.00 | | 52 | 427.00 | 423.00 |
| | 21 | 270.00 | 306.00 | | 53 | 161.00 | 239.00 | | 21 | 401.00 | 488.00 | | 53 | 330.00 | 264.00 |
| | 22 | 226.00 | 234.00 | | 54 | 192.00 | 271.00 | | 22 | - | - | | 54 | 326.00 | 333.00 |
| | 23 | 199.00 | 231.00 | | 55 | - | - | | 23 | 258.00 | - | | 55 | - | - |
| | 24 | 236.00 | 261.00 | | 56 | 212.00 | 218.00 | | 24 | 343.00 | 336.00 | | 56 | 359.00 | 362.00 |
| 4 | 25 | 198.00 | 250.00 | 8 | 57 | 167.00 | 234.00 | 4 | 25 | 240.00 | - | 8 | 57 | - | 516.00 |
| | 26 | - | 280.00 | | 58 | 194.00 | 188.00 | | 26 | - | 397.00 | | 58 | 343.00 | 407.00 |
| | 27 | 215.00 | - | | 59 | 244.00 | 180.00 | | 27 | 451.00 | - | | 59 | 396.00 | 436.00 |
| | 28 | 167.00 | 299.00 | | 60 | 236.00 | 239.00 | | 28 | 300.00 | 532.00 | | 60 | 380.00 | - |
| | 29 | 151.00 | 229.00 | | 61 | 246.00 | 166.00 | | 29 | 322.00 | 400.00 | | 61 | 330.00 | 276.00 |
| | 30 | 204.00 | 252.00 | | 62 | - | 201.00 | | 30 | 364.00 | 376.00 | | 62 | - | 447.00 |
| | 31 | - | 261.00 | | 63 | 214.00 | 206.00 | | 31 | - | 388.00 | | 63 | 299.00 | 417.00 |
| | 32 | 141.00 | 265.00 | | 64 | 211.00 | - | | 32 | - | 398.00 | | 64 | 306.00 | - |

[1] Tratamientos: 1: aves sometidas al modelo de desafío A; 2: aves sometidas al modelo de desafío B

## ANEXO 28. Experimento presentado en el Anexo 1 – ganancia de peso individual, gramos (4 de 4)

| Semana 4 | | | | | | | | | | | | | | | |
|---|---|---|---|---|---|---|---|---|---|---|---|---|---|---|---|
| Jaula | N° | Tratamientos [1] | | Jaula | N° | Tratamientos [1] | | Jaula | N° | Tratamientos [1] | | Jaula | N° | Tratamientos [1] | |
| | | 1 | 2 | | | 1 | 2 | | | 1 | 2 | | | 1 | 2 |
| 1 | 1 | 520.00 | - | 3 | 17 | 381.00 | 621.00 | 5 | 33 | - | 481.00 | 7 | 49 | - | - |
| | 2 | 400.00 | - | | 18 | 380.00 | 501.00 | | 34 | - | 531.00 | | 50 | - | - |
| | 3 | 554.00 | - | | 19 | 550.00 | - | | 35 | 320.00 | - | | 51 | 550.00 | 421.00 |
| | 4 | - | - | | 20 | - | 381.00 | | 36 | 213.00 | 471.00 | | 52 | 540.00 | 435.00 |
| | 5 | - | 341.00 | | 21 | 460.00 | 561.00 | | 37 | 460.00 | - | | 53 | 388.00 | 341.00 |
| | 6 | - | 423.00 | | 22 | - | - | | 38 | 500.00 | 386.00 | | 54 | 450.00 | 421.00 |
| | 7 | 500.00 | 401.00 | | 23 | - | - | | 39 | - | - | | 55 | - | - |
| | 8 | 390.00 | 471.00 | | 24 | 400.00 | - | | 40 | - | 431.00 | | 56 | 390.00 | 351.00 |
| 2 | 9 | 260.00 | - | 4 | 25 | - | - | 6 | 41 | 310.00 | - | 8 | 57 | - | 411.00 |
| | 10 | 440.00 | 491.00 | | 26 | - | 401.00 | | 42 | - | 371.00 | | 58 | 400.00 | 257.00 |
| | 11 | - | - | | 27 | 480.00 | - | | 43 | 480.00 | 379.00 | | 59 | 470.00 | 351.00 |
| | 12 | - | 471.00 | | 28 | - | 531.00 | | 44 | 306.00 | - | | 60 | 380.00 | - |
| | 13 | - | - | | 29 | 390.00 | 450.00 | | 45 | 610.00 | - | | 61 | 248.00 | - |
| | 14 | 296.00 | 541.00 | | 30 | 305.00 | - | | 46 | - | 321.00 | | 62 | - | 291.00 |
| | 15 | - | 367.00 | | 32 | - | 351.00 | | 47 | - | 401.00 | | 63 | 400.00 | 351.00 |
| | 16 | - | - | | 32 | - | 301.00 | | 48 | 380.00 | 351.00 | | 64 | - | - |

[1] Tratamientos: 1: aves sometidas al modelo de desafío A; 2: aves sometidas al modelo de desafío B

## ANEXO 29. Experimento presentado en el Anexo 1 - hallazgos en las necropsias*

| Lesiones | Tratamiento 1 | | | | Tratamiento 2 | | | |
|---|---|---|---|---|---|---|---|---|
| | Día 7 | Día 14 | Día 21 | Día 28 | Día 7 | Día 14 | Día 21 | Día 28 |
| Riñones pálidos | 5 | 3 | 4 | 4 | 4 | 2 | 4 | 4 |
| Riñones congestionados | 0 | 0 | 0 | 0 | 0 | 0 | 0 | 1 |
| Riñones inflamados | 1 | 1 | 4 | 5 | 0 | 0 | 1 | 7 |
| Riñones con hemorragias | 1 | 1 | 0 | 0 | 0 | 0 | 0 | 0 |
| Presencia de uratos en uréteres | 2 | 1 | 2 | 0 | 0 | 0 | 1 | 0 |
| Timos congestionados | 0 | 2 | 0 | 0 | 0 | 0 | 0 | 0 |
| Timos con petequias | 0 | 1 | 0 | 0 | 0 | 0 | 0 | 0 |
| Atrofia del timo | 0 | 0 | 0 | 0 | 0 | 0 | 0 | 0 |
| Involución anormal de la bursa | 0 | 0 | 0 | 0 | 0 | 0 | 0 | 0 |
| Edema testicular | 0 | 2 | 1 | 3 | 0 | 0 | 0 | 1 |
| Ascitis | 0 | 1 | 0 | 2 | 0 | 0 | 0 | 0 |
| Hidropericardio | 3 | 2 | 3 | 5 | 2 | 2 | 3 | 4 |
| Rotura de cabeza de fémur [1] | 1 | 1 | 1 | 0 | 2 | 1 | 1 | 0 |
| Retención de cartílago de fémur | 0 | 0 | 1 | 0 | 0 | 0 | 1 | 0 |
| Contenido anaranjado en el íleon | 0 | 0 | 2 | 0 | 0 | 0 | 1 | 2 |
| Tonsilas cecales activas | 0 | 0 | 0 | 3 | 0 | 0 | 0 | 3 |
| Hígado pálido | 0 | 1 | 0 | 0 | 0 | 1 | 0 | 0 |
| Hemorragias hepáticas subcapsulares | 0 | 0 | 0 | 1 | 0 | 1 | 0 | 0 |
| Vesícula biliar agrandada | 0 | 0 | 0 | 0 | 0 | 0 | 0 | 0 |
| Hemorragias musculares | 0 | 0 | 0 | 0 | 0 | 0 | 0 | 0 |
| Hemorragias subcutáneas | 0 | 0 | 0 | 0 | 0 | 0 | 0 | 0 |
| Lesiones en pico y cavidad oral | 0 | 0 | 0 | 0 | 0 | 0 | 0 | 0 |
| Emplume deficiente o helicóptero | 0 | 0 | 0 | 0 | 0 | 0 | 0 | 0 |
| Proventriculitis | 0 | 0 | 0 | 0 | 0 | 0 | 0 | 0 |

Tratamientos 1 y 2: aves sometidas a los modelos de desafío A o B, respectivamente.
(*) Los valores corresponden al número de aves que presentaron el hallazgo por cada 8 aves muestreadas.
[1] Las fracturas se observaron después de intentar dislocar la cabeza femoral del acetábulo.

**ANEXO 30. Experimento presentado en el Anexo 1 – peso vivo, gramos**

| Jaula | Día 0 de edad | | Día 7 de edad | | Día 14 de edad | | Día 21 de edad | | Día 28 de edad | |
|---|---|---|---|---|---|---|---|---|---|---|
| | Tratamientos [1] | | Tratamientos [1] | | Tratamientos [1] | | Tratamientos [1] | | Tratamientos [1] | |
| | 1 | 2 | 1 | 2 | 1 | 2 | 1 | 2 | 1 | 2 |
| 1 | 49.75 | 46.25 | 179.88 | 193.25 | 349.57 | 437.14 | 762.33 | 877.60 | 1266.80 | 1337.50 |
| 2 | 49.00 | 47.88 | 179.38 | 216.00 | 386.86 | 447.00 | 734.50 | 846.50 | 1112.00 | 1328.50 |
| 3 | 46.72 | 47.62 | 171.38 | 196.00 | 400.00 | 458.29 | 754.33 | 915.00 | 1216.20 | 1462.00 |
| 4 | 48.48 | 47.63 | 166.63 | 203.88 | 353.83 | 464.86 | 702.40 | 884.00 | 1141.67 | 1305.80 |
| 5 | 47.31 | 48.72 | 182.88 | 193.63 | 361.86 | 440.71 | 719.20 | 860.67 | 1138.25 | 1363.00 |
| 6 | 47.42 | 47.97 | 188.75 | 187.75 | 404.71 | 404.14 | 732.67 | 838.33 | 1185.20 | 1223.60 |
| 7 | 47.38 | 49.35 | 176.88 | 179.13 | 385.86 | 411.43 | 746.33 | 780.17 | 1215.60 | 1186.80 |
| 8 | 49.86 | 48.09 | 178.13 | 177.00 | 393.86 | 377.71 | 745.33 | 784.67 | 1129.60 | 1155.20 |

[1] Tratamientos: 1: aves sometidas al modelo de desafío A; 2: aves sometidas al modelo de desafío B

**ANEXO 31. Experimento presentado en el Anexo 1 – ganancia de peso, gramos**

| Jaula | 0 a 7 días | | 0 a 14 días | | 0 a 21 días | | 0 a 28 días | | Semana 2 | | Semana 3 | | Semana 4 | |
|---|---|---|---|---|---|---|---|---|---|---|---|---|---|---|
| | Tratamientos [1] | | Tratamientos [1] | | Tratamientos [1] | | Tratamientos [1] | | Tratamientos [1] | | Tratamientos [1] | | Tratamientos [1] | |
| | 1 | 2 | 1 | 2 | 1 | 2 | 1 | 2 | 1 | 2 | 1 | 2 | 1 | 2 |
| 1 | 130.13 | 147.00 | 299.99 | 390.82 | 712.79 | 829.81 | 1217.95 | 1289.33 | 168.71 | 244.00 | 406.17 | 429.20 | 472.80 | 409.00 |
| 2 | 130.38 | 168.13 | 338.12 | 399.25 | 686.10 | 798.62 | 1067.11 | 1282.44 | 207.14 | 230.43 | 349.17 | 402.50 | 332.00 | 467.50 |
| 3 | 124.66 | 148.38 | 353.09 | 409.89 | 707.09 | 866.24 | 1168.82 | 1413.55 | 228.43 | 262.57 | 349.17 | 434.00 | 434.20 | 516.00 |
| 4 | 118.14 | 156.25 | 305.12 | 416.95 | 653.35 | 836.54 | 1092.25 | 1256.89 | 179.33 | 262.29 | 335.40 | 415.17 | 391.67 | 406.80 |
| 5 | 135.56 | 144.91 | 313.96 | 391.82 | 669.93 | 811.70 | 1088.53 | 1314.37 | 177.57 | 247.00 | 335.40 | 414.33 | 373.25 | 460.00 |
| 6 | 141.33 | 139.78 | 357.04 | 355.94 | 684.45 | 790.17 | 1136.64 | 1176.06 | 215.00 | 217.14 | 323.67 | 418.50 | 417.20 | 364.60 |
| 7 | 129.50 | 129.78 | 338.32 | 361.64 | 698.76 | 731.34 | 1168.03 | 1137.91 | 209.14 | 232.43 | 358.50 | 357.17 | 463.60 | 393.80 |
| 8 | 128.27 | 128.91 | 344.28 | 329.37 | 695.27 | 735.68 | 1079.98 | 1105.62 | 216.00 | 202.00 | 342.33 | 416.50 | 379.60 | 332.20 |

[1] Tratamientos: 1: aves sometidas al modelo de desafío A; 2: aves sometidas al modelo de desafío B

**ANEXO 32. Experimento presentado en el Anexo 1 – consumo de alimento, gramos/pollo**

| Jaula | 0 a 7 días | | 0 a 14 días | | 0 a 21 días | | 0 a 28 días | | Semana 2 | | Semana 3 | | Semana 4 | |
|---|---|---|---|---|---|---|---|---|---|---|---|---|---|---|
| | Tratamientos [1] | | Tratamientos [1] | | Tratamientos [1] | | Tratamientos [1] | | Tratamientos [1] | | Tratamientos [1] | | Tratamientos [1] | |
| | 1 | 2 | 1 | 2 | 1 | 2 | 1 | 2 | 1 | 2 | 1 | 2 | 1 | 2 |
| 1 | 194 | 193 | 513 | 580 | 954 | 1157 | 1712 | 1827 | 319 | 388 | 442 | 576 | 758 | 670 |
| 2 | 198 | 197 | 511 | 550 | 944 | 1120 | 1476 | 1930 | 314 | 353 | 433 | 570 | 532 | 810 |
| 3 | 198 | 197 | 578 | 575 | 1051 | 1161 | 1795 | 1867 | 380 | 378 | 473 | 586 | 744 | 706 |
| 4 | 198 | 197 | 562 | 596 | 910 | 1210 | 1287 | 1984 | 364 | 399 | 348 | 614 | 377 | 774 |
| 5 | 197 | 198 | 497 | 557 | 871 | 1135 | 1385 | 1977 | 299 | 358 | 375 | 578 | 514 | 842 |
| 6 | 180 | 198 | 497 | 499 | 965 | 1071 | 1609 | 1919 | 318 | 301 | 467 | 572 | 644 | 848 |
| 7 | 197 | 199 | 548 | 587 | 1022 | 1124 | 1696 | 1840 | 351 | 389 | 474 | 537 | 674 | 716 |
| 8 | 194 | 197 | 553 | 500 | 992 | 1063 | 1542 | 1835 | 359 | 303 | 440 | 563 | 550 | 772 |

[1] Tratamientos: 1: aves sometidas al modelo de desafío A; 2: aves sometidas al modelo de desafío B

**ANEXO 33. Experimento presentado en el Anexo 1 – conversión alimentaria y mortalidad**

| Jaula | Conversión alimentaria | | | | | | | | | | | | | | Mortalidad, % | |
|---|---|---|---|---|---|---|---|---|---|---|---|---|---|---|---|---|
| | 0 a 7 días | | 0 a 14 días | | 0 a 21 días | | 0 a 28 días | | Semana 2 | | Semana 3 | | Semana 4 | | 0 a 28 días | |
| | Tratamientos [1] | | Tratamientos [1] | | Tratamientos [1] | | Tratamientos [1] | | Tratamientos [1] | | Tratamientos [1] | | Tratamientos [1] | | Tratamientos [1] | |
| | 1 | 2 | 1 | 2 | 1 | 2 | 1 | 2 | 1 | 2 | 1 | 2 | 1 | 2 | | |
| 1 | 1.49 | 1.31 | 1.71 | 1.48 | 1.34 | 1.39 | 1.41 | 1.42 | 1.89 | 1.59 | 1.09 | 1.34 | 1.60 | 1.64 | 0.00 | 12.50 |
| 2 | 1.51 | 1.17 | 1.51 | 1.38 | 1.38 | 1.40 | 1.38 | 1.51 | 1.51 | 1.53 | 1.24 | 1.42 | 1.60 | 1.73 | 25.00 | 12.50 |
| 3 | 1.59 | 1.32 | 1.64 | 1.40 | 1.49 | 1.34 | 1.54 | 1.32 | 1.66 | 1.44 | 1.36 | 1.35 | 1.71 | 1.37 | 0.00 | 12.50 |
| 4 | 1.67 | 1.26 | 1.84 | 1.43 | 1.39 | 1.45 | 1.18 | 1.58 | 2.03 | 1.52 | 1.04 | 1.48 | 0.96 | 1.90 | 25.00 | 0.00 |
| 5 | 1.46 | 1.37 | 1.58 | 1.42 | 1.30 | 1.40 | 1.27 | 1.50 | 1.69 | 1.45 | 1.12 | 1.39 | 1.38 | 1.83 | 12.50 | 0.00 |
| 6 | 1.27 | 1.42 | 1.39 | 1.40 | 1.41 | 1.35 | 1.42 | 1.63 | 1.48 | 1.38 | 1.44 | 1.37 | 1.54 | 2.33 | 0.00 | 0.00 |
| 7 | 1.52 | 1.53 | 1.62 | 1.62 | 1.46 | 1.54 | 1.45 | 1.62 | 1.68 | 1.67 | 1.32 | 1.50 | 1.45 | 1.82 | 0.00 | 0.00 |
| 8 | 1.51 | 1.53 | 1.61 | 1.52 | 1.43 | 1.45 | 1.43 | 1.66 | 1.66 | 1.50 | 1.28 | 1.35 | 1.45 | 2.32 | 0.00 | 0.00 |

[1] Tratamientos: 1: aves sometidas al modelo de desafío A; 2: aves sometidas al modelo de desafío

**ANEXO 34. Exp. presentado en Anexo 1 - heces (1 de 2)**

| Rep. | Semana 1 | | Semana 2 | | Semana 3 | | Semana 4 | |
|---|---|---|---|---|---|---|---|---|
| | Tratamientos [1] | | | | | | | |
| | 1 | 2 | 1 | 2 | 1 | 2 | 1 | 2 |
| | **Heces normales, %** | | | | | | | |
| 1 | 22.22 | 33.33 | 29.41 | 38.10 | 29.41 | 23.81 | 30.00 | 23.08 |
| 2 | 30.00 | 35.29 | 20.00 | 33.33 | 18.75 | 30.00 | 23.08 | 22.22 |
| 3 | 28.57 | 27.27 | 28.57 | 40.00 | 23.08 | 22.22 | 33.33 | 27.27 |
| 4 | 16.67 | 25.00 | 25.00 | 31.25 | 27.27 | 18.75 | 28.57 | 30.00 |
| 5 | 25.00 | 30.00 | 23.08 | 29.41 | 30.00 | 27.27 | 36.84 | 14.29 |
| 6 | 25.00 | 26.32 | 17.39 | 31.58 | 27.78 | 21.43 | 23.53 | 21.05 |
| 7 | 27.27 | 25.00 | 26.32 | 38.10 | 18.75 | 20.00 | 30.43 | 28.57 |
| 8 | 25.00 | 30.43 | 28.57 | 40.00 | 26.09 | 31.82 | 33.33 | 26.09 |
| | **Heces acuosas, %** | | | | | | | |
| 1 | 77.78 | 66.67 | 70.59 | 61.90 | 70.59 | 76.19 | 70.00 | 76.92 |
| 2 | 70.00 | 64.71 | 80.00 | 66.67 | 81.25 | 70.00 | 76.92 | 77.78 |
| 3 | 71.43 | 72.73 | 71.43 | 60.00 | 76.92 | 77.78 | 66.67 | 72.73 |
| 4 | 83.33 | 75.00 | 75.00 | 68.75 | 72.73 | 81.25 | 71.43 | 70.00 |
| 5 | 75.00 | 70.00 | 76.92 | 70.59 | 70.00 | 72.73 | 63.16 | 85.71 |
| 6 | 75.00 | 73.68 | 82.61 | 68.42 | 72.22 | 78.57 | 76.47 | 78.95 |
| 7 | 72.73 | 75.00 | 73.68 | 61.90 | 81.25 | 80.00 | 69.57 | 71.43 |
| 8 | 75.00 | 69.57 | 71.43 | 60.00 | 73.91 | 68.18 | 66.67 | 73.91 |
| | **Heces con alimento sin digerir, %** | | | | | | | |
| 1 | 66.67 | 60.00 | 52.94 | 52.38 | 52.94 | 61.90 | 40.00 | 61.54 |
| 2 | 70.00 | 64.71 | 60.00 | 44.44 | 56.25 | 65.00 | 38.46 | 55.56 |
| 3 | 71.43 | 63.64 | 66.67 | 55.00 | 53.85 | 61.11 | 38.89 | 54.55 |
| 4 | 66.67 | 56.25 | 60.00 | 50.00 | 50.00 | 62.50 | 42.86 | 50.00 |
| 5 | 66.67 | 65.00 | 61.54 | 47.06 | 45.00 | 68.18 | 36.84 | 57.14 |
| 6 | 62.50 | 52.63 | 60.87 | 47.37 | 50.00 | 64.29 | 47.06 | 52.63 |
| 7 | 54.55 | 62.50 | 57.89 | 52.38 | 50.00 | 73.33 | 39.13 | 57.14 |
| 8 | 66.67 | 60.87 | 57.14 | 50.00 | 47.83 | 63.64 | 38.10 | 52.17 |

[1] Tratamientos: 1: modelo de desafío A; 2: modelo de desafío B

**ANEXO 35. Exp. presentado en Anexo 1 - heces (2 de 2)**

| Rep. | Semana 1 | | Semana 2 | | Semana 3 | | Semana 4 | |
|---|---|---|---|---|---|---|---|---|
| | Tratamientos [1] | | | | | | | |
| | 1 | 2 | 1 | 2 | 1 | 2 | 1 | 2 |
| | **Heces con descamaciones mucosas, %** | | | | | | | |
| 1 | 0.00 | 0.00 | 0.00 | 0.00 | 0.00 | 4.76 | 5.00 | 7.69 |
| 2 | 0.00 | 0.00 | 0.00 | 0.00 | 0.00 | 5.00 | 7.69 | 5.56 |
| 3 | 0.00 | 0.00 | 4.76 | 0.00 | 0.00 | 5.56 | 5.56 | 9.09 |
| 4 | 0.00 | 0.00 | 0.00 | 0.00 | 4.55 | 0.00 | 4.76 | 5.00 |
| 5 | 0.00 | 0.00 | 0.00 | 0.00 | 5.00 | 4.55 | 5.26 | 7.14 |
| 6 | 0.00 | 0.00 | 0.00 | 5.26 | 5.56 | 0.00 | 5.88 | 5.26 |
| 7 | 0.00 | 0.00 | 0.00 | 0.00 | 0.00 | 0.00 | 4.35 | 9.52 |
| 8 | 0.00 | 0.00 | 0.00 | 0.00 | 4.35 | 4.55 | 4.76 | 8.70 |
| | **Heces hemorrágicas, %** | | | | | | | |
| 1 | 0.00 | 0.00 | 0.00 | 0.00 | 0.00 | 0.00 | 0.00 | 0.00 |
| 2 | 0.00 | 0.00 | 0.00 | 0.00 | 0.00 | 0.00 | 0.00 | 0.00 |
| 3 | 0.00 | 0.00 | 0.00 | 0.00 | 0.00 | 0.00 | 0.00 | 0.00 |
| 4 | 0.00 | 0.00 | 0.00 | 0.00 | 0.00 | 0.00 | 0.00 | 0.00 |
| 5 | 0.00 | 0.00 | 0.00 | 0.00 | 0.00 | 0.00 | 0.00 | 0.00 |
| 6 | 0.00 | 0.00 | 0.00 | 0.00 | 0.00 | 0.00 | 0.00 | 0.00 |
| 7 | 0.00 | 0.00 | 0.00 | 0.00 | 0.00 | 0.00 | 0.00 | 0.00 |
| 8 | 0.00 | 0.00 | 0.00 | 0.00 | 0.00 | 0.00 | 0.00 | 0.00 |
| | **Índice Dimar, %** | | | | | | | |
| 1 | 36.11 | 31.67 | 30.88 | 28.57 | 30.88 | 35.71 | 28.75 | 36.54 |
| 2 | 35.00 | 32.35 | 35.00 | 27.78 | 34.38 | 35.00 | 30.77 | 34.72 |
| 3 | 35.71 | 34.09 | 35.71 | 28.75 | 32.69 | 36.11 | 27.78 | 34.09 |
| 4 | 37.50 | 32.81 | 33.75 | 29.69 | 31.82 | 35.94 | 29.76 | 31.25 |
| 5 | 35.42 | 33.75 | 34.62 | 29.41 | 30.00 | 36.36 | 26.32 | 37.50 |
| 6 | 34.38 | 31.58 | 35.87 | 30.26 | 31.94 | 35.71 | 32.35 | 34.21 |
| 7 | 31.82 | 34.38 | 32.89 | 28.57 | 32.81 | 38.33 | 28.26 | 34.52 |
| 8 | 35.42 | 32.61 | 32.14 | 27.50 | 31.52 | 34.09 | 27.38 | 33.70 |

[1] Tratamientos: 1: modelo de desafío A; 2: modelo de desafío B

**ANEXO 36. Exp. presentado en Anexo 1 - daño intestinal (1 de 4)**

| Rep. | Semana 1 | | Semana 2 | | Semana 3 | | Semana 4 | |
|---|---|---|---|---|---|---|---|---|
| | Tratamientos [1] | | | | | | | |
| | 1 | 2 | 1 | 2 | 1 | 2 | 1 | 2 |
| | **Enteritis** | | | | | | | |
| 1 | 0 | 0 | 0 | 0 | 0 | 1 | 1 | 1 |
| 2 | 0 | 0 | 0 | 0 | 1 | 1 | 1 | 0 |
| 3 | 0 | 0 | 1 | 0 | 1 | 1 | 1 | 1 |
| 4 | 0 | 0 | 0 | 0 | 0 | 0 | 0 | 1 |
| 5 | 0 | 0 | 0 | 0 | 0 | 1 | 1 | 0 |
| 6 | 0 | 0 | 1 | 0 | 1 | 0 | 0 | 1 |
| 7 | 0 | 1 | 0 | 0 | 1 | 1 | 1 | 0 |
| 8 | 1 | 0 | 0 | 1 | 1 | 0 | 1 | 1 |
| | **Presencia de lesiones intestinales** | | | | | | | |
| 1 | 0 | 0 | 0 | 0 | 1 | 1 | 0 | 0 |
| 2 | 0 | 0 | 0 | 0 | 0 | 0 | 1 | 1 |
| 3 | 0 | 0 | 0 | 0 | 0 | 1 | 1 | 1 |
| 4 | 0 | 0 | 0 | 0 | 0 | 1 | 0 | 1 |
| 5 | 0 | 0 | 0 | 0 | 0 | 1 | 1 | 1 |
| 6 | 0 | 0 | 0 | 0 | 1 | 1 | 1 | 1 |
| 7 | 0 | 0 | 0 | 0 | 1 | 1 | 1 | 1 |
| 8 | 0 | 0 | 0 | 0 | 0 | 0 | 1 | 1 |
| | **Densidad de las lesiones** | | | | | | | |
| 1 | 0 | 0 | 0 | 0 | 0 | 10 | 15 | 15 |
| 2 | 0 | 0 | 0 | 0 | 0 | 1 | 0 | 15 |
| 3 | 0 | 0 | 0 | 0 | 0 | 5 | 12 | 12 |
| 4 | 0 | 0 | 0 | 0 | 0 | 0 | 0 | 20 |
| 5 | 0 | 0 | 0 | 0 | 0 | 0 | 5 | 15 |
| 6 | 0 | 0 | 0 | 0 | 0 | 12 | 0 | 18 |
| 7 | 0 | 0 | 0 | 0 | 4 | 20 | 0 | 0 |
| 8 | 0 | 0 | 0 | 0 | 0 | 4 | 0 | 20 |

[1] Tratamientos: 1: modelo de desafío A; 2: modelo de desafío B

**ANEXO 37. Exp. presentado en Anexo 1 - daño intestinal (2 de 4)**

| Rep. | Semana 1 | | Semana 2 | | Semana 3 | | Semana 4 | |
|---|---|---|---|---|---|---|---|---|
| | Tratamientos [1] | | | | | | | |
| | 1 | 2 | 1 | 2 | 1 | 2 | 1 | 2 |
| | **Score de lesiones - *Clostridium perfringens*** | | | | | | | |
| 1 | 0 | 0 | 0 | 0 | 0 | 0 | 0 | 0 |
| 2 | 0 | 0 | 0 | 0 | 0 | 0 | 1 | 0 |
| 3 | 0 | 0 | 0 | 0 | 0 | 0 | 1 | 0 |
| 4 | 0 | 0 | 0 | 0 | 0 | 1 | 0 | 0 |
| 5 | 0 | 0 | 0 | 0 | 0 | 1 | 1 | 1 |
| 6 | 0 | 0 | 0 | 0 | 1 | 0 | 1 | 1 |
| 7 | 0 | 0 | 0 | 0 | 1 | 1 | 1 | 1 |
| 8 | 0 | 0 | 0 | 0 | 0 | 0 | 1 | 1 |
| | **Score de lesiones por *Eimeria acervulina*** | | | | | | | |
| 1 | 0 | 0 | 0 | 0 | 0 | 2 | 0 | 0 |
| 2 | 0 | 0 | 0 | 0 | 0 | 0 | 0 | 2 |
| 3 | 0 | 0 | 0 | 0 | 0 | 0 | 3 | 1 |
| 4 | 0 | 0 | 0 | 0 | 0 | 3 | 0 | 3 |
| 5 | 0 | 0 | 0 | 0 | 0 | 2 | 1 | 1 |
| 6 | 0 | 0 | 0 | 0 | 1 | 2 | 1 | 2 |
| 7 | 0 | 0 | 0 | 0 | 0 | 2 | 0 | 1 |
| 8 | 0 | 0 | 0 | 0 | 0 | 0 | 0 | 2 |
| | **Score de lesiones por *Eimeria maxima*** | | | | | | | |
| 1 | 0 | 0 | 0 | 0 | 2 | 0 | 0 | 0 |
| 2 | 0 | 0 | 0 | 0 | 0 | 0 | 0 | 0 |
| 3 | 0 | 0 | 0 | 0 | 0 | 1 | 2 | 0 |
| 4 | 0 | 0 | 0 | 0 | 0 | 0 | 0 | 0 |
| 5 | 0 | 0 | 0 | 0 | 0 | 0 | 0 | 0 |
| 6 | 0 | 0 | 0 | 0 | 2 | 2 | 0 | 0 |
| 7 | 0 | 0 | 0 | 0 | 0 | 0 | 0 | 0 |
| 8 | 0 | 0 | 0 | 0 | 0 | 0 | 0 | 0 |

[1] Tratamientos: 1: modelo de desafío A; 2: modelo de desafío B

**ANEXO 38. Exp. presentado en Anexo 1 - daño intestinal (3 de 4)**

| Rep. | Semana 1 | | Semana 2 | | Semana 3 | | Semana 4 | |
|---|---|---|---|---|---|---|---|---|
| | Tratamientos [1] | | | | | | | |
| | 1 | 2 | 1 | 2 | 1 | 2 | 1 | 2 |
| | Aves positivas a *Eimeria acervulina* (macroscópico) | | | | | | | |
| 1 | 0 | 0 | 0 | 0 | 0 | 1 | 0 | 0 |
| 2 | 0 | 0 | 0 | 0 | 0 | 0 | 0 | 1 |
| 3 | 0 | 0 | 0 | 0 | 0 | 0 | 1 | 1 |
| 4 | 0 | 0 | 0 | 0 | 0 | 1 | 0 | 1 |
| 5 | 0 | 0 | 0 | 0 | 0 | 1 | 1 | 1 |
| 6 | 0 | 0 | 0 | 0 | 1 | 1 | 1 | 1 |
| 7 | 0 | 0 | 0 | 0 | 0 | 1 | 0 | 1 |
| 8 | 0 | 0 | 0 | 0 | 0 | 0 | 0 | 1 |
| | Aves positivas a *Eimeria máxima* (macroscópico) | | | | | | | |
| 1 | 0 | 0 | 0 | 0 | 1 | 0 | 0 | 0 |
| 2 | 0 | 0 | 0 | 0 | 0 | 0 | 0 | 0 |
| 3 | 0 | 0 | 0 | 0 | 0 | 1 | 1 | 0 |
| 4 | 0 | 0 | 0 | 0 | 0 | 0 | 0 | 0 |
| 5 | 0 | 0 | 0 | 0 | 0 | 0 | 0 | 0 |
| 6 | 0 | 0 | 0 | 0 | 1 | 1 | 0 | 0 |
| 7 | 0 | 0 | 0 | 0 | 0 | 0 | 0 | 0 |
| 8 | 0 | 0 | 0 | 0 | 0 | 0 | 0 | 0 |
| | Aves positivas a coccidia (macroscópico) | | | | | | | |
| 1 | 0 | 0 | 0 | 0 | 1 | 1 | 0 | 0 |
| 2 | 0 | 0 | 0 | 0 | 0 | 0 | 0 | 1 |
| 3 | 0 | 0 | 0 | 0 | 0 | 1 | 1 | 1 |
| 4 | 0 | 0 | 0 | 0 | 0 | 1 | 0 | 1 |
| 5 | 0 | 0 | 0 | 0 | 0 | 1 | 1 | 1 |
| 6 | 0 | 0 | 0 | 0 | 1 | 1 | 1 | 1 |
| 7 | 0 | 0 | 0 | 0 | 0 | 1 | 0 | 1 |
| 8 | 0 | 0 | 0 | 0 | 0 | 0 | 0 | 1 |

[1] Tratamientos: 1: modelo de desafío A; 2: modelo de desafío B

**ANEXO 39. Exp. presentado en Anexo 1 - daño intestinal (4 de 4)**

| Rep. | Semana 1 | | Semana 2 | | Semana 3 | | Semana 4 | |
|---|---|---|---|---|---|---|---|---|
| | Tratamientos [1] | | | | | | | |
| | 1 | 2 | 1 | 2 | 1 | 2 | 1 | 2 |
| | Score microscópico de coccidia | | | | | | | |
| 1 | 0 | 0 | 0 | 0 | 0 | 0 | 3 | 2 |
| 2 | 0 | 0 | 0 | 0 | 0 | 0 | 2 | 1 |
| 3 | 0 | 0 | 0 | 0 | 0 | 0 | 3 | 3 |
| 4 | 0 | 0 | 0 | 0 | 0 | 1 | 2 | 3 |
| 5 | 0 | 0 | 0 | 0 | 0 | 1 | 2 | 3 |
| 6 | 0 | 0 | 0 | 0 | 1 | 2 | 3 | 2 |
| 7 | 0 | 0 | 0 | 0 | 2 | 2 | 2 | 3 |
| 8 | 0 | 0 | 0 | 0 | 0 | 1 | 0 | 3 |
| | Aves positivas a coccidia (microscópico) | | | | | | | |
| 1 | NO | NO | NO | NO | NO | NO | SI | SI |
| 2 | NO | NO | NO | NO | NO | NO | SI | SI |
| 3 | NO | NO | NO | NO | NO | NO | SI | SI |
| 4 | NO | NO | NO | NO | NO | SI | SI | SI |
| 5 | NO | NO | NO | NO | NO | SI | SI | SI |
| 6 | NO | NO | NO | NO | SI | SI | SI | SI |
| 7 | NO | NO | NO | NO | SI | SI | SI | SI |
| 8 | NO | NO | NO | NO | NO | SI | NO | SI |

[1] Tratamientos: 1: modelo de desafío A; 2: modelo de desafío B

**ANEXO 40. Exp. presentado en Anexo 1 - órganos linfoides (1 de 3)**

| Rep. | Semana 1 | | Semana 2 | | Semana 3 | | Semana 4 | |
|---|---|---|---|---|---|---|---|---|
| | Tratamientos [1] | | | | | | | |
| | 1 | 2 | 1 | 2 | 1 | 2 | 1 | 2 |
| | **Diámetro de la bursa (medido con bursómetro)** | | | | | | | |
| 1 | 2 | 4 | 4 | 4 | 6 | 6 | 6 | 8 |
| 2 | 3 | 4 | 4 | 5 | 6 | 6 | 7 | 8 |
| 3 | 3 | 3 | 4 | 5 | 5 | 7 | 7 | 6 |
| 4 | 3 | 3 | 4 | 4 | 5 | 6 | 7 | 7 |
| 5 | 3 | 3 | 4 | 5 | 5 | 5 | 6 | 8 |
| 6 | 3 | 4 | 5 | 5 | 5 | 7 | 6 | 7 |
| 7 | 3 | 3 | 6 | 4 | 6 | 6 | 5 | 7 |
| 8 | 4 | 4 | 5 | 4 | 5 | 5 | 5 | 7 |
| | **Índice morfométrico de la bursa** | | | | | | | |
| 1 | 0.9827 | 3.1500 | 2.0178 | 1.5226 | 2.5497 | 1.4870 | 1.6284 | 2.4735 |
| 2 | 1.4607 | 2.1429 | 1.7464 | 2.4266 | 3.1072 | 3.3828 | 2.2575 | 3.5280 |
| 3 | 2.1053 | 2.6733 | 1.5957 | 2.7897 | 2.9642 | 3.5294 | 3.1175 | 1.7333 |
| 4 | 1.7791 | 1.7647 | 1.8482 | 2.1206 | 2.3259 | 1.9240 | 3.0335 | 2.3552 |
| 5 | 1.7816 | 2.0833 | 2.2849 | 2.6201 | 3.4328 | 1.5537 | 3.2226 | 3.6000 |
| 6 | 1.8539 | 2.4138 | 2.6053 | 4.7200 | 2.7158 | 3.7517 | 2.4514 | 3.3826 |
| 7 | 2.2989 | 2.3196 | 3.5635 | 2.2938 | 2.8134 | 2.6452 | 1.4981 | 2.6987 |
| 8 | 2.5389 | 2.4064 | 3.1250 | 1.8684 | 1.4958 | 1.9325 | 1.3458 | 3.5172 |
| | **Índice morfométrico del bazo** | | | | | | | |
| 1 | 1.0405 | 0.8000 | 0.7122 | 0.8848 | 2.1026 | 1.0095 | 1.0043 | 1.9743 |
| 2 | 0.7865 | 1.2381 | 1.0048 | 0.7045 | 1.4111 | 1.1135 | 1.2375 | 0.8224 |
| 3 | 1.2281 | 0.8416 | 1.2766 | 1.2661 | 2.0358 | 1.2000 | 0.8141 | 1.4741 |
| 4 | 0.8589 | 1.3122 | 1.1551 | 0.8108 | 1.8671 | 1.0369 | 1.2823 | 1.8059 |
| 5 | 0.8621 | 0.8854 | 1.0484 | 0.9389 | 2.0896 | 1.7373 | 0.9225 | 0.8784 |
| 6 | 0.9551 | 3.7931 | 1.6316 | 1.3867 | 1.4928 | 1.4686 | 1.5443 | 1.2315 |
| 7 | 0.9770 | 1.1340 | 1.7127 | 1.2371 | 0.7382 | 0.8258 | 0.4485 | 1.7091 |
| 8 | 0.6736 | 0.5348 | 1.0511 | 1.2527 | 0.8726 | 1.0583 | 0.5894 | 1.3924 |

[1] Tratamientos: 1: modelo de desafío A; 2: modelo de desafío B

**ANEXO 41. Exp. presentado en Anexo 1 - órganos linfoides (2 de 3)**

| Rep. | Semana 1 | | Semana 2 | | Semana 3 | | Semana 4 | |
|---|---|---|---|---|---|---|---|---|
| | Tratamientos [1] | | | | | | | |
| | 1 | 2 | 1 | 2 | 1 | 2 | 1 | 2 |
| | **Índice morfométrico del timo** | | | | | | | |
| 1 | 5.6069 | 4.6500 | 5.9644 | 6.3374 | 4.4536 | 6.8349 | 4.1822 | 5.4009 |
| 2 | 5.7865 | 5.0000 | 8.7799 | 8.9824 | 8.5346 | 5.0371 | 3.6204 | 6.7434 |
| 3 | 5.0877 | 5.6436 | 7.6330 | 4.8712 | 6.0098 | 6.1529 | 4.2368 | 6.0370 |
| 4 | 4.6626 | 6.9231 | 4.9505 | 5.3430 | 2.7848 | 5.3111 | 7.3301 | 3.3710 |
| 5 | 4.4828 | 6.3021 | 7.0430 | 4.5197 | 4.5821 | 4.6893 | 4.7232 | 7.1529 |
| 6 | 5.7865 | 7.1429 | 5.9737 | 8.4000 | 3.5432 | 6.3151 | 4.9460 | 5.7964 |
| 7 | 8.5632 | 4.7938 | 6.6298 | 5.8247 | 7.4652 | 6.2452 | 3.9122 | 6.0270 |
| 8 | 5.2332 | 6.1497 | 6.6761 | 5.3928 | 5.8033 | 4.7853 | 3.8310 | 5.3797 |
| | **Relación bursa / bazo** | | | | | | | |
| 1 | 0.9444 | 3.9375 | 2.8333 | 1.7209 | 1.2126 | 1.4730 | 1.6214 | 1.2529 |
| 2 | 1.8571 | 1.7308 | 1.7381 | 3.4444 | 2.2019 | 3.0381 | 1.8243 | 4.2900 |
| 3 | 1.7143 | 3.1765 | 1.2500 | 2.2034 | 1.4560 | 2.9412 | 3.8295 | 1.1759 |
| 4 | 2.0714 | 1.3448 | 1.6000 | 2.6154 | 1.2458 | 1.8556 | 2.3657 | 1.3042 |
| 5 | 2.0667 | 2.3529 | 2.1795 | 2.7907 | 1.6429 | 0.8943 | 3.4933 | 4.0982 |
| 6 | 1.9412 | 0.6364 | 1.5968 | 3.4038 | 1.8193 | 2.5545 | 1.5874 | 2.7467 |
| 7 | 2.3529 | 2.0455 | 2.0806 | 1.8542 | 3.8113 | 3.2031 | 3.3404 | 1.5789 |
| 8 | 3.7692 | 4.5000 | 2.9730 | 1.4915 | 1.7143 | 1.8261 | 2.2833 | 2.5260 |
| | **Relación bursa / timo** | | | | | | | |
| 1 | 0.1753 | 0.6774 | 0.3383 | 0.2403 | 0.5725 | 0.2176 | 0.3894 | 0.4580 |
| 2 | 0.2524 | 0.4286 | 0.1989 | 0.2702 | 0.3641 | 0.6716 | 0.6236 | 0.5232 |
| 3 | 0.4138 | 0.4737 | 0.2091 | 0.5727 | 0.4932 | 0.5736 | 0.7358 | 0.2871 |
| 4 | 0.3816 | 0.2549 | 0.3733 | 0.3969 | 0.8352 | 0.3623 | 0.4138 | 0.6987 |
| 5 | 0.3974 | 0.3306 | 0.3244 | 0.5797 | 0.7492 | 0.3313 | 0.6823 | 0.5033 |
| 6 | 0.3204 | 0.3379 | 0.4361 | 0.5619 | 0.7665 | 0.5941 | 0.4956 | 0.5836 |
| 7 | 0.2685 | 0.4839 | 0.5375 | 0.3938 | 0.3769 | 0.4236 | 0.3829 | 0.4478 |
| 8 | 0.4851 | 0.3913 | 0.4681 | 0.3465 | 0.2578 | 0.4038 | 0.3513 | 0.6538 |

[1] Tratamientos: 1: modelo de desafío A; 2: modelo de desafío B

**ANEXO 42. Exp. presentado en Anexo 1 - órganos linfoides (3 de 3)**

| Rep. | Semana 1 | | Semana 2 | | Semana 3 | | Semana 4 | |
|---|---|---|---|---|---|---|---|---|
| | Tratamientos [1] | | | | | | | |
| | 1 | 2 | 1 | 2 | 1 | 2 | 1 | 2 |
| | **Relación timo / bazo** | | | | | | | |
| 1 | 5.3889 | 5.8125 | 8.3750 | 7.1628 | 2.1181 | 6.7703 | 4.1643 | 2.7356 |
| 2 | 7.3571 | 4.0385 | 8.7381 | 12.7500 | 6.0481 | 4.5238 | 2.9257 | 8.2000 |
| 3 | 4.1429 | 6.7059 | 5.9792 | 3.8475 | 2.9520 | 5.1275 | 5.2045 | 4.0955 |
| 4 | 5.4286 | 5.2759 | 4.2857 | 6.5897 | 1.4915 | 5.1222 | 5.7164 | 1.8667 |
| 5 | 5.2000 | 7.1176 | 6.7179 | 4.8140 | 2.1929 | 2.6992 | 5.1200 | 8.1429 |
| 6 | 6.0588 | 1.8831 | 3.6613 | 6.0577 | 2.3735 | 4.3000 | 3.2028 | 4.7067 |
| 7 | 8.7647 | 4.2273 | 3.8710 | 4.7083 | 10.1132 | 7.5625 | 8.7234 | 3.5263 |
| 8 | 7.7692 | 11.5000 | 6.3514 | 4.3051 | 6.6508 | 4.5217 | 6.5000 | 3.8636 |

[1] Tratamientos: 1: modelo de desafío A; 2: modelo de desafío B

**ANEXO 43. Exp. presentado en Anexo 1 - carcasa y vísceras (1 de 3)**

| Rep. | Semana 1 | | Semana 2 | | Semana 3 | | Semana 4 | |
|---|---|---|---|---|---|---|---|---|
| | Tratamientos [1] | | | | | | | |
| | 1 | 2 | 1 | 2 | 1 | 2 | 1 | 2 |
| | **Rendimiento de carcasa, %** | | | | | | | |
| 1 | 66.47 | 65.50 | 64.69 | 70.99 | 70.20 | 74.08 | 67.58 | 70.35 |
| 2 | 61.80 | 64.76 | 70.10 | 69.86 | 71.23 | 73.81 | 72.24 | 71.79 |
| 3 | 64.33 | 68.32 | 78.72 | 70.60 | 71.66 | 72.12 | 71.05 | 72.67 |
| 4 | 64.42 | 66.06 | 70.63 | 72.77 | 74.68 | 71.31 | 71.00 | 71.48 |
| 5 | 63.22 | 64.58 | 73.92 | 69.21 | 72.39 | 68.36 | 70.36 | 73.10 |
| 6 | 64.05 | 67.98 | 68.68 | 68.80 | 71.76 | 74.77 | 73.11 | 74.06 |
| 7 | 62.07 | 67.53 | 72.10 | 68.81 | 72.70 | 73.94 | 61.83 | 70.01 |
| 8 | 64.77 | 61.50 | 69.60 | 52.61 | 72.85 | 73.93 | 67.19 | 72.42 |
| | **Rendimiento de pechuga, %** | | | | | | | |
| 1 | 12.61 | 14.22 | 14.31 | 18.75 | 19.87 | 19.65 | 19.08 | 19.59 |
| 2 | 12.97 | 13.31 | 16.41 | 18.07 | 19.54 | 21.95 | 22.07 | 19.90 |
| 3 | 12.03 | 13.79 | 16.31 | 16.95 | 17.60 | 20.94 | 21.09 | 24.30 |
| 4 | 12.85 | 13.14 | 16.20 | 19.94 | 19.62 | 20.60 | 18.66 | 21.97 |
| 5 | 11.35 | 13.23 | 17.86 | 18.27 | 19.29 | 19.38 | 18.33 | 20.31 |
| 6 | 12.40 | 15.23 | 17.74 | 15.06 | 18.88 | 20.56 | 20.63 | 21.02 |
| 7 | 12.75 | 13.74 | 17.81 | 18.20 | 17.83 | 20.77 | 17.27 | 20.54 |
| 8 | 11.61 | 12.38 | 18.72 | 20.92 | 17.59 | 20.32 | 20.73 | 20.80 |
| | **Porcentaje de pechuga, %** | | | | | | | |
| 1 | 18.97 | 21.70 | 22.12 | 26.42 | 28.30 | 26.52 | 28.24 | 27.85 |
| 2 | 20.99 | 20.55 | 23.41 | 25.86 | 27.43 | 29.74 | 30.56 | 27.72 |
| 3 | 18.70 | 20.19 | 20.72 | 24.02 | 24.56 | 29.04 | 29.69 | 33.44 |
| 4 | 19.95 | 19.88 | 22.94 | 27.40 | 26.27 | 28.89 | 26.28 | 30.74 |
| 5 | 17.95 | 20.49 | 24.16 | 26.40 | 26.64 | 28.36 | 26.05 | 27.79 |
| 6 | 19.36 | 22.41 | 25.83 | 21.88 | 26.32 | 27.50 | 28.21 | 28.38 |
| 7 | 20.55 | 20.34 | 24.71 | 26.45 | 24.52 | 28.10 | 27.93 | 29.34 |
| 8 | 17.92 | 20.13 | 26.90 | 39.77 | 24.14 | 27.49 | 30.85 | 28.71 |

[1] Tratamientos: 1: modelo de desafío A; 2: modelo de desafío B

## ANEXO 44. Exp. presentado en Anexo 1 - carcasa y vísceras (2 de 3)

| Rep. | Semana 1 | | Semana 2 | | Semana 3 | | Semana 4 | |
|---|---|---|---|---|---|---|---|---|
| | Tratamientos [1] | | | | | | | |
| | 1 | 2 | 1 | 2 | 1 | 2 | 1 | 2 |
| | **Relación carcasa / pechuga** | | | | | | | |
| 1 | 5.27 | 4.61 | 4.52 | 3.79 | 3.53 | 3.77 | 3.54 | 3.59 |
| 2 | 4.76 | 4.87 | 4.27 | 3.87 | 3.65 | 3.36 | 3.27 | 3.61 |
| 3 | 5.35 | 4.95 | 4.83 | 4.16 | 4.07 | 3.44 | 3.37 | 2.99 |
| 4 | 5.01 | 5.03 | 4.36 | 3.65 | 3.81 | 3.46 | 3.81 | 3.25 |
| 5 | 5.57 | 4.88 | 4.14 | 3.79 | 3.75 | 3.53 | 3.84 | 3.60 |
| 6 | 5.17 | 4.46 | 3.87 | 4.57 | 3.80 | 3.64 | 3.54 | 3.52 |
| 7 | 4.87 | 4.92 | 4.05 | 3.78 | 4.08 | 3.56 | 3.58 | 3.41 |
| 8 | 5.58 | 4.97 | 3.72 | 2.51 | 4.14 | 3.64 | 3.24 | 3.48 |
| | **Peso relativo del intestino, %** | | | | | | | |
| 1 | 8.29 | 9.54 | 9.07 | 8.49 | 5.58 | 4.92 | 5.97 | 7.05 |
| 2 | 9.52 | 9.01 | 9.22 | 7.10 | 7.06 | 5.79 | 4.90 | 6.13 |
| 3 | 9.69 | 8.63 | 8.30 | 7.46 | 6.70 | 5.81 | 7.05 | 4.82 |
| 4 | 9.71 | 9.61 | 7.31 | 6.39 | 7.10 | 6.10 | 6.17 | 7.53 |
| 5 | 9.37 | 8.90 | 7.67 | 7.03 | 6.21 | 6.89 | 8.19 | 6.34 |
| 6 | 8.15 | 9.30 | 6.98 | 7.91 | 5.83 | 5.96 | 6.04 | 4.79 |
| 7 | 12.86 | 10.88 | 7.44 | 7.47 | 6.55 | 4.92 | 5.49 | 6.45 |
| 8 | 10.16 | 11.19 | 5.28 | 6.04 | 5.63 | 4.13 | 4.63 | 6.10 |
| | **Peso relativo del hígado, %** | | | | | | | |
| 1 | 4.14 | 3.29 | 4.57 | 3.03 | 3.38 | 2.56 | 2.60 | 2.90 |
| 2 | 4.96 | 3.21 | 3.62 | 3.46 | 3.45 | 2.61 | 3.00 | 1.95 |
| 3 | 3.87 | 3.91 | 3.54 | 3.15 | 3.31 | 2.47 | 3.42 | 2.57 |
| 4 | 4.55 | 4.30 | 3.05 | 2.50 | 4.26 | 3.14 | 2.75 | 2.95 |
| 5 | 3.61 | 4.01 | 3.42 | 3.05 | 3.30 | 3.94 | 3.32 | 2.06 |
| 6 | 4.59 | 4.09 | 4.19 | 4.70 | 3.53 | 3.18 | 3.30 | 2.85 |
| 7 | 3.84 | 4.39 | 3.52 | 3.24 | 3.64 | 2.45 | 3.74 | 2.92 |
| 8 | 3.63 | 4.50 | 3.82 | 3.24 | 2.82 | 2.95 | 2.91 | 2.91 |

[1] Tratamientos: 1: modelo de desafío A; 2: modelo de desafío B

## ANEXO 45. Exp. presentado en Anexo 1 - carcasa y vísceras (3 de 3)

| Rep. | Semana 1 | | Semana 2 | | Semana 3 | | Semana 4 | |
|---|---|---|---|---|---|---|---|---|
| | Tratamientos [1] | | | | | | | |
| | 1 | 2 | 1 | 2 | 1 | 2 | 1 | 2 |
| | **Peso relativo del páncreas, %** | | | | | | | |
| 1 | 0.636 | 0.375 | 0.525 | 0.329 | 0.328 | 0.392 | 0.293 | 0.325 |
| 2 | 0.427 | 0.452 | 0.426 | 0.341 | 0.354 | 0.329 | 0.287 | 0.299 |
| 3 | 0.567 | 0.351 | 0.404 | 0.348 | 0.379 | 0.304 | 0.301 | 0.329 |
| 4 | 0.387 | 0.475 | 0.396 | 0.364 | 0.369 | 0.336 | 0.233 | 0.178 |
| 5 | 0.466 | 0.510 | 0.333 | 0.402 | 0.319 | 0.421 | 0.339 | 0.317 |
| 6 | 0.438 | 0.527 | 0.363 | 0.408 | 0.304 | 0.348 | 0.297 | 0.283 |
| 7 | 0.511 | 0.392 | 0.445 | 0.477 | 0.340 | 0.306 | 0.252 | 0.294 |
| 8 | 0.415 | 0.332 | 0.509 | 0.285 | 0.277 | 0.387 | 0.254 | 0.291 |
| | **Relación páncreas / hígado** | | | | | | | |
| 1 | 0.154 | 0.114 | 0.115 | 0.109 | 0.097 | 0.153 | 0.113 | 0.112 |
| 2 | 0.086 | 0.141 | 0.118 | 0.098 | 0.103 | 0.126 | 0.096 | 0.153 |
| 3 | 0.147 | 0.090 | 0.114 | 0.110 | 0.115 | 0.123 | 0.088 | 0.128 |
| 4 | 0.085 | 0.110 | 0.130 | 0.145 | 0.087 | 0.107 | 0.085 | 0.060 |
| 5 | 0.129 | 0.127 | 0.097 | 0.132 | 0.097 | 0.107 | 0.102 | 0.154 |
| 6 | 0.095 | 0.129 | 0.087 | 0.087 | 0.086 | 0.110 | 0.090 | 0.099 |
| 7 | 0.133 | 0.089 | 0.126 | 0.147 | 0.093 | 0.125 | 0.067 | 0.101 |
| 8 | 0.114 | 0.074 | 0.133 | 0.088 | 0.098 | 0.131 | 0.087 | 0.100 |

[1] Tratamientos: 1: modelo de desafío A; 2: modelo de desafío B

## ANEXO 46. Experimento 4 – comportamiento productivo

| Rep. | Peso día 0 | | Peso día 28 | | Ganancia de peso | | Consumo de alimento | | Conversión alimentaria | |
|---|---|---|---|---|---|---|---|---|---|---|
| | Tratamientos [1] | | Tratamientos [1] | | Tratamientos [1] | | Tratamientos [1] | | Tratamientos [1] | |
| | 1 | 2 | 1 | 2 | 1 | 2 | 1 | 2 | 1 | 2 |
| 1 | 46.25 | 48.93 | 1337.5 | 1457.3 | 1289.33 | 1407.06 | 1827.01 | 1817.68 | 1.4170 | 1.2918 |
| 2 | 47.88 | 47.95 | 1328.5 | 1569.8 | 1282.44 | 1521.92 | 1930.48 | 1772.12 | 1.5053 | 1.1644 |
| 3 | 47.62 | 46.91 | 1462.0 | 1451.0 | 1413.55 | 1403.90 | 1867.04 | 1920.89 | 1.3208 | 1.3683 |
| 4 | 47.63 | 50.82 | 1305.8 | 1480.6 | 1256.89 | 1429.09 | 1983.92 | 1955.55 | 1.5784 | 1.3684 |
| 5 | 48.72 | 49.09 | 1363.0 | 1436.2 | 1314.37 | 1387.41 | 1976.64 | 1974.27 | 1.5039 | 1.4230 |
| 6 | 47.97 | 47.68 | 1223.6 | 1442.8 | 1176.06 | 1394.27 | 1918.57 | 1880.01 | 1.6314 | 1.3484 |

[1] Tratamientos: 1: dieta basal; 2: dieta basal + 500 ppm de Orevitol®

## ANEXO 47. Experimento 4 – estado antioxidante

| Jaula | Superóxido dismutasa | | Caroteno sérico | |
|---|---|---|---|---|
| | Tratamientos (1) | | Tratamientos (1) | |
| | 1 (2) | 2 (3) | 1 (2) | 2 (3) |
| 1 | 67 | 85 | 2.0 | 2.0 |
| 2 | 72 | 90 | 2.0 | 2.0 |
| 3 | 61 | 74 | 2.0 | 2.0 |
| 4 | 62 | 69 | 2.0 | 2.0 |
| 5 | 67 | 78 | 2.0 | 2.0 |
| 6 | 70 | 64 | 2.0 | 2.0 |

(1) Tratamientos: 1: dieta basal; 2: dieta basal + 500 ppm de Orevitol®
(2) Informes 011889573, 011889576, 011889577, 011889578, 011889580, 011889581 ROE
(3) Informes 011889567, 011889570, 011889572, 011889574, 011889575, 011889579 ROE

**ANEXO 48. Experimento 5 - evaluación de heces y órganos linfoides**

| Rep. | Día 21 | | | Día 28 | | | Día 21 | | | Día 28 | | | Día 28 | | |
|---|---|---|---|---|---|---|---|---|---|---|---|---|---|---|---|
| | Tratamientos[1] | | | | | | Tratamientos[1] | | | | | | Tratamientos[1] | | |
| | 1 | 2 | 3 | 1 | 2 | 3 | 1 | 2 | 3 | 1 | 2 | 3 | 1 | 2 | 3 |
| | Heces normales, % | | | | | | Heces con descamaciones mucosas, % | | | | | | Diámetro de la bursa (bursómetro) | | |
| 1 | 23.81 | 50.00 | 50.00 | 23.08 | 56.25 | 50.00 | 4.76 | 0.00 | 4.55 | 7.69 | 0.00 | 0.00 | 8 | 7 | 7 |
| 2 | 30.00 | 50.00 | 44.44 | 22.22 | 60.00 | 40.00 | 5.00 | 0.00 | 0.00 | 5.56 | 5.00 | 0.00 | 8 | 6 | 7 |
| 3 | 22.22 | 52.17 | 50.00 | 27.27 | 55.00 | 50.00 | 5.56 | 4.35 | 5.00 | 9.09 | 5.00 | 5.00 | 6 | 6 | 7 |
| 4 | 18.75 | 47.37 | 50.00 | 30.00 | 61.11 | 40.91 | 0.00 | 0.00 | 0.00 | 5.00 | 0.00 | 4.55 | 7 | 6 | 8 |
| 5 | 27.27 | 55.00 | 50.00 | 14.29 | 60.00 | 42.86 | 4.55 | 0.00 | 0.00 | 7.14 | 0.00 | 0.00 | 8 | 6 | 7 |
| 6 | 21.43 | 58.82 | 52.63 | 21.05 | 52.63 | 47.37 | 0.00 | 0.00 | 0.00 | 5.26 | 0.00 | 5.26 | 7 | 6 | 8 |
| 7 | 20.00 | 53.33 | 50.00 | 28.57 | 61.90 | 42.86 | 0.00 | 0.00 | 0.00 | 9.52 | 4.76 | 4.76 | 7 | 7 | 6 |
| 8 | 31.82 | 52.38 | 45.00 | 26.09 | 60.00 | 43.75 | 4.55 | 4.76 | 5.00 | 8.70 | 0.00 | 0.00 | 7 | 6 | 7 |
| | Heces acuosas, % | | | | | | Heces hemorrágicas, % | | | | | | Índice morfométrico de la bursa | | |
| 1 | 76.19 | 50.00 | 50.00 | 76.92 | 43.75 | 50.00 | 0.00 | 0.00 | 0.00 | 0.00 | 0.00 | 0.00 | 2.4735 | 2.4221 | 2.8714 |
| 2 | 70.00 | 50.00 | 55.56 | 77.78 | 40.00 | 60.00 | 0.00 | 0.00 | 0.00 | 0.00 | 0.00 | 0.00 | 3.5280 | 1.7699 | 1.7545 |
| 3 | 77.78 | 47.83 | 50.00 | 72.73 | 45.00 | 50.00 | 0.00 | 0.00 | 0.00 | 0.00 | 0.00 | 0.00 | 1.7333 | 2.0935 | 2.6621 |
| 4 | 81.25 | 52.63 | 50.00 | 70.00 | 38.89 | 59.09 | 0.00 | 0.00 | 0.00 | 0.00 | 0.00 | 0.00 | 2.3552 | 1.4135 | 3.3652 |
| 5 | 72.73 | 45.00 | 50.00 | 85.71 | 40.00 | 57.14 | 0.00 | 0.00 | 0.00 | 0.00 | 0.00 | 0.00 | 3.6000 | 1.5625 | 2.3272 |
| 6 | 78.57 | 41.18 | 47.37 | 78.95 | 47.37 | 52.63 | 0.00 | 0.00 | 0.00 | 0.00 | 0.00 | 0.00 | 3.3826 | 1.8487 | 3.7376 |
| 7 | 80.00 | 46.67 | 50.00 | 71.43 | 38.10 | 57.14 | 0.00 | 0.00 | 0.00 | 0.00 | 0.00 | 0.00 | 2.6987 | 2.7415 | 2.1570 |
| 8 | 68.18 | 47.62 | 55.00 | 73.91 | 40.00 | 56.25 | 0.00 | 0.00 | 0.00 | 0.00 | 0.00 | 0.00 | 3.5172 | 1.9146 | 3.2045 |
| | Heces con alimento sin digerir, % | | | | | | Índice Dimar | | | | | | Relación bursa / timo | | |
| 1 | 61.90 | 38.89 | 36.36 | 61.54 | 25.00 | 38.89 | 35.71 | 22.22 | 22.73 | 36.54 | 17.19 | 22.22 | 0.4580 | 0.4272 | 0.2967 |
| 2 | 65.00 | 37.50 | 44.44 | 55.56 | 25.00 | 33.33 | 35.00 | 21.88 | 25.00 | 34.72 | 17.50 | 23.33 | 0.5232 | 0.2667 | 0.4682 |
| 3 | 61.11 | 34.78 | 40.00 | 54.55 | 25.00 | 35.00 | 36.11 | 21.74 | 23.75 | 34.09 | 18.75 | 22.50 | 0.2871 | 0.6258 | 0.4327 |
| 4 | 62.50 | 31.58 | 43.75 | 50.00 | 22.22 | 36.36 | 35.94 | 21.05 | 23.44 | 31.25 | 15.28 | 25.00 | 0.6987 | 0.3170 | 0.7799 |
| 5 | 68.18 | 40.00 | 42.86 | 57.14 | 26.67 | 35.71 | 36.36 | 21.25 | 23.21 | 37.50 | 16.67 | 23.21 | 0.5033 | 0.3346 | 0.6133 |
| 6 | 64.29 | 35.29 | 42.11 | 52.63 | 26.32 | 31.58 | 35.71 | 19.12 | 22.37 | 34.21 | 18.42 | 22.37 | 0.5836 | 0.3194 | 0.6205 |
| 7 | 73.33 | 33.33 | 37.50 | 57.14 | 23.81 | 33.33 | 38.33 | 20.00 | 21.88 | 34.52 | 16.67 | 23.81 | 0.4478 | 0.3967 | 0.4888 |
| 8 | 63.64 | 33.33 | 40.00 | 52.17 | 25.00 | 37.50 | 34.09 | 21.43 | 25.00 | 33.70 | 16.25 | 23.44 | 0.6538 | 0.4180 | 0.8000 |

[1] Tratamientos: 1: aves desafiadas control; 2: aves desafiadas + 500 ppm de Orevitol®; 3: aves desafiadas + 150 ppm de neomicina.

## ANEXO 49. Experimento 5 – pesos y ganancias individuales, gramos

**Peso vivo - día 1 de edad (g)**

| Jaula | N° | Tratamientos [1] | | | Jaula | N° | Tratamientos [1] | | |
|---|---|---|---|---|---|---|---|---|---|
| | | 1 | 2 | 3 | | | 1 | 2 | 3 |
| 1 | 1 | 44.55 | 46.81 | 46.01 | 5 | 33 | 46.86 | 45.86 | 47.80 |
| | 2 | 40.73 | 49.91 | 48.18 | | 34 | 50.02 | 42.65 | 47.06 |
| | 3 | 45.77 | 46.40 | 45.09 | | 35 | 48.48 | 53.19 | 48.07 |
| | 4 | 46.28 | 51.42 | 44.60 | | 36 | 45.70 | 48.83 | 48.90 |
| | 5 | 46.10 | 46.07 | 43.19 | | 37 | 50.67 | 49.52 | 50.15 |
| | 6 | 48.62 | 50.82 | 47.99 | | 38 | 49.41 | 51.22 | 49.37 |
| | 7 | 45.56 | 44.49 | 45.32 | | 39 | 47.43 | 56.57 | 44.76 |
| | 8 | 52.40 | 55.53 | 48.17 | | 40 | 51.15 | 44.87 | 49.64 |
| 2 | 9 | 46.93 | 50.93 | 51.42 | 6 | 41 | 48.46 | 49.46 | 50.40 |
| | 10 | 47.82 | 43.46 | 48.28 | | 42 | 49.95 | 50.49 | 42.51 |
| | 11 | 48.77 | 46.70 | 46.59 | | 43 | 49.53 | 43.86 | 49.78 |
| | 12 | 41.15 | 44.27 | 47.65 | | 44 | 51.29 | 54.20 | 54.14 |
| | 13 | 54.35 | 54.73 | 44.58 | | 45 | 46.31 | 47.60 | 47.19 |
| | 14 | 42.41 | 46.59 | 49.92 | | 46 | 40.47 | 46.55 | 50.52 |
| | 15 | 52.86 | 48.67 | 43.80 | | 47 | 48.54 | 42.81 | 46.12 |
| | 16 | 48.71 | 48.26 | 44.82 | | 48 | 49.21 | 46.49 | 49.50 |
| 3 | 17 | 47.57 | 48.56 | 46.72 | 7 | 49 | 55.56 | 51.33 | 45.36 |
| | 18 | 48.93 | 49.23 | 53.61 | | 50 | 48.51 | 41.08 | 50.97 |
| | 19 | 42.18 | 45.46 | 47.10 | | 51 | 49.67 | 52.41 | 42.46 |
| | 20 | 48.78 | 46.39 | 49.62 | | 52 | 48.54 | 53.88 | 46.58 |
| | 21 | 48.51 | 44.13 | 50.92 | | 53 | 43.67 | 46.69 | 50.83 |
| | 22 | 48.56 | 47.62 | 52.61 | | 54 | 52.79 | 45.85 | 47.00 |
| | 23 | 46.42 | 47.15 | 45.93 | | 55 | 46.25 | 50.01 | 47.53 |
| | 24 | 49.99 | 46.72 | 52.76 | | 56 | 49.77 | 50.57 | 50.98 |
| 4 | 25 | 50.61 | 56.46 | 47.60 | 8 | 57 | 62.18 | 45.93 | 50.03 |
| | 26 | 46.97 | 53.63 | 45.91 | | 58 | 42.64 | 47.11 | 46.34 |
| | 27 | 45.65 | 49.29 | 47.95 | | 59 | 44.68 | 45.62 | 56.94 |
| | 28 | 46.21 | 50.54 | 49.58 | | 60 | 44.48 | 45.89 | 46.32 |
| | 29 | 45.97 | 51.17 | 40.40 | | 61 | 46.03 | 47.87 | 40.92 |
| | 30 | 40.21 | 47.86 | 56.72 | | 62 | 54.03 | 51.72 | 49.71 |
| | 31 | 52.48 | 46.12 | 53.18 | | 63 | 44.35 | 48.39 | 42.98 |
| | 32 | 52.92 | 51.51 | 43.72 | | 64 | 46.36 | 41.91 | 54.54 |

**Peso vivo - día 28 de edad (g)**

| Jaula | N° | Tratamientos [1] | | | Jaula | N° | Tratamientos [1] | | |
|---|---|---|---|---|---|---|---|---|---|
| | | 1 | 2 | 3 | | | 1 | 2 | 3 |
| 1 | 1 | ** | ** | 1513 | 5 | 33 | 1432 | 1257 | 1553 |
| | 2 | ** | 1601 | ** | | 34 | 1512 | ** | ** |
| | 3 | ** | ** | 1433 | | 35 | ** | 1381 | ** |
| | 4 | ** | ** | 1343 | | 36 | 1332 | 1471 | 1443 |
| | 5 | 1242 | ** | ** | | 37 | ** | ** | 1125 |
| | 6 | 1234 | 1441 | 1257 | | 38 | 1187 | 1531 | 1303 |
| | 7 | 1412 | 1511 | ** | | 39 | ** | ** | 1303 |
| | 8 | 1462 | 1276* | ** | | 40 | 1352 | 1541 | ** |
| 2 | 9 | ** | 1571 | 1353 | 6 | 41 | ** | ** | 1473 |
| | 10 | 1402 | 1761 | 1503 | | 42 | 1332 | 1580 | ** |
| | 11 | ** | ** | 1533 | | 43 | 1130 | 1311 | 1203 |
| | 12 | 1322 | ** | ** | | 44 | ** | 1471 | 1453 |
| | 13 | ** | ** | 1163 | | 45 | ** | 1501 | ** |
| | 14 | 1462 | ** | 1329 | | 46 | 1032 | ** | 1255 |
| | 15 | 1128 | 1456* | ** | | 47 | 1322 | ** | 1303 |
| | 16 | ** | 1491 | ** | | 48 | 1302 | 1351 | ** |
| 3 | 17 | 1632 | 1231 | 1363 | 7 | 49 | ** | 1401 | 1343 |
| | 18 | 1392 | ** | ** | | 50 | ** | ** | 1423 |
| | 19 | ** | 1521* | 1393 | | 51 | 1312 | 1661 | ** |
| | 20 | 1262 | ** | ** | | 52 | 1246 | ** | 1381 |
| | 21 | 1562 | ** | 1623 | | 53 | 1032 | 1311 | ** |
| | 22 | ** | 1441 | 1553 | | 54 | 1222 | ** | ** |
| | 23 | ** | 1521 | ** | | 55 | ** | 1571 | 1493 |
| | 24 | ** | 1541 | 1373 | | 56 | 1122 | 1383 | 1473 |
| 4 | 25 | ** | 1581 | 1473 | 8 | 57 | 1352 | 1132 | 1693 |
| | 26 | 1292 | ** | 1293 | | 58 | 1018 | ** | 1313 |
| | 27 | ** | ** | ** | | 59 | 1122 | 1391 | 1543 |
| | 28 | 1592 | 1301 | ** | | 60 | ** | 1581 | ** |
| | 29 | 1241 | 1631 | 1403 | | 61 | ** | 1301 | 1063 |
| | 30 | ** | 1461 | 1303 | | 62 | 1142 | 1461 | ** |
| | 31 | 1222 | ** | ** | | 63 | 1142 | ** | ** |
| | 32 | 1182 | 1429* | 1073 | | 64 | ** | ** | 1331 |

**Ganancia de peso (g/pollo)**

| Jaula | N° | Tratamientos [1] | | | Jaula | N° | Tratamientos [1] | | |
|---|---|---|---|---|---|---|---|---|---|
| | | 1 | 2 | 3 | | | 1 | 2 | 3 |
| 1 | 1 | - | - | 1467 | 5 | 33 | 1385 | 1211 | 1505 |
| | 2 | - | 1551 | - | | 34 | 1462 | - | - |
| | 3 | - | - | 1388 | | 35 | - | 1328 | - |
| | 4 | - | - | 1298 | | 36 | 1286 | 1422 | 1394 |
| | 5 | 1196 | - | - | | 37 | - | - | 1075 |
| | 6 | 1185 | 1390 | 1209 | | 38 | 1138 | 1480 | 1254 |
| | 7 | 1366 | 1467 | - | | 39 | - | - | 1258 |
| | 8 | 1410 | 1220 | - | | 40 | 1301 | 1496 | - |
| 2 | 9 | - | 1520 | 1302 | 6 | 41 | - | - | 1423 |
| | 10 | 1354 | 1718 | 1455 | | 42 | 1282 | 1530 | - |
| | 11 | - | - | 1486 | | 43 | 1080 | 1267 | 1153 |
| | 12 | 1281 | - | - | | 44 | - | 1417 | 1399 |
| | 13 | - | - | 1118 | | 45 | - | 1453 | - |
| | 14 | 1420 | - | 1279 | | 46 | 992 | - | 1204 |
| | 15 | 1075 | 1407 | - | | 47 | 1273 | - | 1257 |
| | 16 | - | 1443 | - | | 48 | 1253 | 1305 | - |
| 3 | 17 | 1584 | 1182 | 1316 | 7 | 49 | - | 1350 | 1298 |
| | 18 | 1343 | - | - | | 50 | - | - | 1372 |
| | 19 | - | 1476 | 1346 | | 51 | 1262 | 1609 | - |
| | 20 | 1213 | - | - | | 52 | 1197 | - | 1334 |
| | 21 | 1513 | - | 1572 | | 53 | 988 | 1264 | - |
| | 22 | - | 1393 | 1500 | | 54 | 1169 | - | - |
| | 23 | - | 1474 | - | | 55 | - | 1521 | 1445 |
| | 24 | - | 1494 | 1320 | | 56 | 1072 | 1332 | 1422 |
| 4 | 25 | - | 1525 | 1425 | 8 | 57 | 1290 | 1086 | 1643 |
| | 26 | 1245 | - | 1247 | | 58 | 975 | - | 1267 |
| | 27 | - | - | - | | 59 | 1077 | 1345 | 1486 |
| | 28 | 1546 | 1250 | - | | 60 | - | 1535 | - |
| | 29 | 1195 | 1580 | 1363 | | 61 | - | 1253 | 1022 |
| | 30 | - | 1413 | 1246 | | 62 | 1088 | 1409 | - |
| | 31 | 1170 | - | - | | 63 | 1098 | - | - |
| | 32 | 1129 | 1377 | 1029 | | 64 | - | - | 1276 |

[1] Tratamientos: 1: aves desafiadas control; 2: aves desafiadas + 500 ppm de Orevitol®; 3: aves desafiadas + 150 ppm de neomicina.

* Aves seleccionadas como muestras o muertas en la semana inmediata siguiente. ** Aves que fueron muestreadas o murieron antes de completar el periodo

## ANEXO 50. Experimento 5 – comportamiento productivo

| Rep. | Peso vivo – día 0 de edad (g) | | | Peso vivo – día 28 de edad (g) | | | Ganancia de peso, g/pollo | | |
|---|---|---|---|---|---|---|---|---|---|
| | Tratamientos [1] | | | Tratamientos [1] | | | Tratamientos [1] | | |
| | 1 | 2 | 3 | 1 | 2 | 3 | 1 | 2 | 3 |
| 1 | 46.25 | 48.93 | 46.07 | 1337.50 | 1457.25 | 1386.50 | 1289.33 | 1407.06 | 1340.58 |
| 2 | 47.88 | 47.95 | 47.13 | 1328.50 | 1569.75 | 1376.20 | 1282.44 | 1521.92 | 1328.04 |
| 3 | 47.62 | 46.91 | 49.91 | 1462.00 | 1451.00 | 1461.00 | 1413.55 | 1403.90 | 1410.98 |
| 4 | 47.63 | 50.82 | 48.13 | 1305.80 | 1480.60 | 1309.00 | 1256.89 | 1429.09 | 1262.13 |
| 5 | 48.72 | 49.09 | 48.22 | 1363.00 | 1436.20 | 1345.40 | 1314.37 | 1387.41 | 1297.20 |
| 6 | 47.97 | 47.68 | 48.77 | 1223.60 | 1442.80 | 1337.40 | 1176.06 | 1394.27 | 1287.21 |
| 7 | 49.35 | 48.98 | 47.71 | 1186.80 | 1465.40 | 1422.60 | 1137.91 | 1415.20 | 1374.32 |
| 8 | 48.09 | 46.81 | 48.47 | 1155.20 | 1373.20 | 1388.60 | 1105.62 | 1325.79 | 1338.85 |

| Rep. | Consumo de alimento, g/pollo | | | Conversión alimentaria | | | Mortalidad, % | | |
|---|---|---|---|---|---|---|---|---|---|
| | Tratamientos [1] | | | Tratamientos [1] | | | Tratamientos [1] | | |
| | 1 | 2 | 3 | 1 | 2 | 3 | 1 | 2 | 1 |
| 1 | 1827 | 1818 | 1669 | 1.42 | 1.29 | 1.25 | 12.50 | 12.50 | 12.50 |
| 2 | 1930 | 1772 | 1930 | 1.51 | 1.16 | 1.45 | 12.50 | 12.50 | 0.00 |
| 3 | 1867 | 1921 | 1943 | 1.32 | 1.37 | 1.38 | 12.50 | 0.00 | 0.00 |
| 4 | 1984 | 1956 | 1875 | 1.58 | 1.37 | 1.49 | 0.00 | 0.00 | 0.00 |
| 5 | 1977 | 1974 | 1944 | 1.50 | 1.42 | 1.50 | 0.00 | 0.00 | 0.00 |
| 6 | 1919 | 1880 | 1997 | 1.63 | 1.35 | 1.55 | 0.00 | 0.00 | 0.00 |
| 7 | 1840 | 1916 | 2014 | 1.62 | 1.35 | 1.47 | 0.00 | 0.00 | 0.00 |
| 8 | 1835 | 1770 | 2033 | 1.66 | 1.33 | 1.52 | 0.00 | 0.00 | 0.00 |

[1] Tratamientos: 1: aves desafiadas control; 2: aves desafiadas + 500 ppm de Orevitol®; 3: aves desafiadas + 150 ppm de neomicina.

## ANEXO 51. Experimento 6 – características las heces

| Variable | Tratamiento [1] | | | | | | | |
|---|---|---|---|---|---|---|---|---|
| | 1 | | | | 2 | | | |
| | Repetición | | | | Repetición | | | |
| | 1 | 2 | 3 | 4 | 1 | 2 | 3 | 4 |
| 5 días después de iniciado el tratamiento | | | | | | | | |
| Heces normales, % | 37.2 | 50.0 | 27.5 | 23.7 | 42.0 | 77.0 | 63.9 | 64.7 |
| Heces acuosas, % | 52.1 | 44.7 | 62.7 | 64.9 | 46.3 | 16.8 | 15.9 | 24.0 |
| Heces con alimento sin digerir, % | 62.8 | 50.0 | 72.5 | 76.3 | 58.0 | 23.0 | 36.1 | 35.3 |
| Heces con descamación mucosa, % | 2.3 | 6.3 | 0.0 | 0.0 | 1.2 | 1.1 | 1.7 | 0.0 |
| Heces hemorrágicas, % | 0.0 | 0.0 | 0.0 | 0.0 | 0.0 | 0.0 | 0.0 | 0.0 |
| Índice Dimar | 31.98 | 26.58 | 36.25 | 38.15 | 29.30 | 11.78 | 18.48 | 17.65 |
| Disbacteriosis | + | + | + | + | + | - | - | - |
| 7 días después de iniciado el tratamiento | | | | | | | | |
| Heces normales, % | 53.5 | 75.0 | 25.0 | 36.9 | 85.4 | 66.4 | 86.1 | 76.6 |
| Heces acuosas, % | 36.2 | 20.6 | 64.1 | 55.4 | 12.1 | 26.5 | 12.9 | 18.0 |
| Heces con alimento sin digerir, % | 46.5 | 25.0 | 75.0 | 63.1 | 14.6 | 33.6 | 13.9 | 23.4 |
| Heces con descamación mucosa, % | 0.0 | 0.0 | 0.0 | 0.0 | 0.0 | 0.0 | 0.0 | 0.0 |
| Heces hemorrágicas, % | 0.0 | 0.0 | 0.0 | 0.0 | 0.0 | 0.0 | 0.0 | 0.0 |
| Índice Dimar | 23.25 | 16.67 | 36.59 | 36.35 | 12.90 | 15.19 | 8.93 | 13.28 |
| Disbacteriosis | + | - | + | + | - | - | - | - |

[1] 1: control sin tratamiento; 2: aves tratadas (Orevitol® 300 ml/$m^3$ por cinco días).

## ANEXO 52. Experimento 6 – comportamiento productivo

| Variable | Tratamiento (1) | | | | | | | |
|---|---|---|---|---|---|---|---|---|
| | 1 | | | | 2 | | | |
| | Repetición | | | | Repetición | | | |
| | 1 | 2 | 3 | 4 (2) | 1 | 2 | 3 | 4 (2) |
| Edad, días | 33 | 33 | 23 | 23 | 33 | 33 | 23 | 23 |
| Sexo | M | H | M | H | M | H | M | H |
| Peso vivo día 0 (3), kg | 1.597 | 1.450 | 0.520 | 0.536 | 1.597 | 1.450 | 0.520 | 0.536 |
| Peso vivo 5° día (3), kg | 2.068 | 1.765 | 0.797 | 0.849 | 2.011 | 1.842 | 0.826 | 0.849 |
| Peso vivo 7° día (3), kg | 2.288 | 1.974 | 1.012 | 0.997 | 2.373 | 2.051 | 1.043 | 0.982 |
| Diferencias en peso al 7° día, % | - | - | - | - | 3.7 | 3.9 | 3.2 | -1.6 |
| Peso vivo 10° día (4), kg | 2.617 | 2.288 | 1.333 | 1.220 | 2.915 | 2.366 | 1.369 | 1.182 |
| Ganancia de peso de 0 a 7 días, g | 691 | 524 | 492 | - | 776 | 601 | 523 | - |

(1) Tratamientos: 1: aves control sin tratamiento; 2: aves tratadas con 300 ml de Orevitol® / $m^3$ de agua de bebida por cinco días.

(2) Se aplicó las pruebas de Dixon (Dean and Dixon, 1951) y Grubbs (Grubbs, 1969) a las diferencias en peso al 7° día y se verificó un valor anómalo (P<0.05) en la repetición 4. Por ello se obvió dicha repetición, en cuanto a comportamiento productivo, para el cálculo de los valores promedio.

(3) 0, 5 y 7 días transcurridos desde el inicio del tratamiento.

(4) El peso vivo en el 10° día se calculó por regresión lineal a partir de los pesos previos de cada unidad experimental.

## ANEXO 53. Experimento 7 - peso y ganancias de peso individuales (gramos)

| Peso vivo - día 0 de edad (g) | | | | | | | | Peso vivo - día 21 de edad (g) | | | | | | | |
|---|---|---|---|---|---|---|---|---|---|---|---|---|---|---|---|
| Jaula | N° | Tratamientos [1] | | Jaula | N° | Tratamientos [1] | | Jaula | N° | Tratamientos [1] | | Jaula | N° | Tratamientos [1] | |
| | | 1 | 2 | | | 1 | 2 | | | 1 | 2 | | | 1 | 2 |
| 1 | 1 | 44.55 | 46.81 | 5 | 33 | 46.86 | 45.86 | 1 | 1 | - | - | 5 | 33 | 951 | 797 |
| | 2 | 40.73 | 49.91 | | 34 | 50.02 | 42.65 | | 2 | - | 967 | | 34 | 981 | 737 |
| | 3 | 45.77 | 46.40 | | 35 | 48.48 | 53.19 | | 3 | - | - | | 35 | - | 907 |
| | 4 | 46.28 | 51.42 | | 36 | 45.70 | 48.83 | | 4 | 674 | - | | 36 | 861 | 1037 |
| | 5 | 46.10 | 46.07 | | 37 | 50.67 | 49.52 | | 5 | 901 | 658 | | 37 | 649 | - |
| | 6 | 48.62 | 50.82 | | 38 | 49.41 | 51.22 | | 6 | 811 | 907 | | 38 | 801 | 967 |
| | 7 | 45.56 | 44.49 | | 39 | 47.43 | 56.57 | | 7 | 1011 | 997 | | 39 | - | - |
| | 8 | 52.40 | 55.53 | | 40 | 51.15 | 44.87 | | 8 | 991 | 877 | | 40 | 921 | 1047 |
| 2 | 9 | 46.93 | 50.93 | 6 | 41 | 48.46 | 49.46 | 2 | 9 | - | 977 | 6 | 41 | - | - |
| | 10 | 47.82 | 43.46 | | 42 | 49.95 | 50.49 | | 10 | 911 | 1117 | | 42 | 961 | 1027 |
| | 11 | 48.77 | 46.70 | | 43 | 49.53 | 43.86 | | 11 | - | - | | 43 | 751 | 787 |
| | 12 | 41.15 | 44.27 | | 44 | 51.29 | 54.20 | | 12 | 851 | 880 | | 44 | 735 | 1077 |
| | 13 | 54.35 | 54.73 | | 45 | 46.31 | 47.60 | | 13 | 751 | - | | 45 | - | 1047 |
| | 14 | 42.41 | 46.59 | | 46 | 40.47 | 46.55 | | 14 | 921 | - | | 46 | 711 | - |
| | 15 | 52.86 | 48.67 | | 47 | 48.54 | 42.81 | | 15 | 761 | 927 | | 47 | 921 | 867 |
| | 16 | 48.71 | 48.26 | | 48 | 49.21 | 46.49 | | 16 | 884 | 917 | | 48 | 951 | 897 |
| 3 | 17 | 47.57 | 48.56 | 7 | 49 | 55.56 | 51.33 | 3 | 17 | 1011 | 847 | 7 | 49 | - | 927 |
| | 18 | 48.93 | 49.23 | | 50 | 48.51 | 41.08 | | 18 | 891 | - | | 50 | 716 | - |
| | 19 | 42.18 | 45.46 | | 51 | 49.67 | 52.41 | | 19 | - | 967 | | 51 | 891 | 1087 |
| | 20 | 48.78 | 46.39 | | 52 | 48.54 | 53.88 | | 20 | 881 | - | | 52 | 811 | 865 |
| | 21 | 48.51 | 44.13 | | 53 | 43.67 | 46.69 | | 21 | 1001 | 793 | | 53 | 691 | 867 |
| | 22 | 48.56 | 47.62 | | 54 | 52.79 | 45.85 | | 22 | - | 997 | | 54 | 801 | - |
| | 23 | 46.42 | 47.15 | | 55 | 46.25 | 50.01 | | 23 | - | 1037 | | 55 | - | 1067 |
| | 24 | 49.99 | 46.72 | | 56 | 49.77 | 50.57 | | 24 | 791 | 1037 | | 56 | 771 | 907 |
| 4 | 25 | 50.61 | 56.46 | 8 | 57 | 62.18 | 45.93 | 4 | 25 | - | 1027 | 8 | 57 | 941 | 817 |
| | 26 | 46.97 | 53.63 | | 58 | 42.64 | 47.11 | | 26 | 891 | 682 | | 58 | 761 | 641 |
| | 27 | 45.65 | 49.29 | | 59 | 44.68 | 45.62 | | 27 | - | - | | 59 | 771 | 897 |
| | 28 | 46.21 | 50.54 | | 60 | 44.48 | 45.89 | | 28 | 1061 | 877 | | 60 | - | 1017 |
| | 29 | 45.97 | 51.17 | | 61 | 46.03 | 47.87 | | 29 | 791 | 1107 | | 61 | 593 | 847 |
| | 30 | 40.21 | 47.86 | | 62 | 54.03 | 51.72 | | 30 | 809 | 967 | | 62 | 851 | 967 |
| | 31 | 52.48 | 46.12 | | 63 | 44.35 | 48.39 | | 31 | 871 | - | | 63 | 791 | - |
| | 32 | 52.92 | 51.51 | | 64 | 46.36 | 41.91 | | 32 | 881 | 937 | | 64 | - | - |

[1] Tratamientos: 1: aves desafiadas control; 2: aves desafiadas + 500 ppm de Orevitol®.

\- Los valores faltantes corresponden a aves que fueron muestreadas o murieron antes de completar el periodo

**ANEXO 54. Experimento 7 – resultados generales**

| Jaula | Peso inicial (g) | | Peso final (g) | | Ganancia de peso (g/ave) | | Consumo de alimento (g/ave) | | Conversión alimentaria | | Positivo a coccidia micro. | | Score micro. de coccidia | | Pigmentación | | N° ooquistes / g cama | | | | | | | |
|---|---|---|---|---|---|---|---|---|---|---|---|---|---|---|---|---|---|---|---|---|---|---|---|---|
| | T1[1] | T2[1] | T1[1] | T2[1] | T1[1] | T2[1] | T1[1] | T2[1] | T1[1] | T2[1] | T1[1] | T2[1] | T1[1] | T2[1] | T1[1] | T2[1] | T1[1] | | | | T2[1] | | | |
| 1 | 46.25 | 48.93 | 877.60 | 881.20 | 831.35 | 832.27 | 1157 | 1155 | 1.32 | 1.31 | 0 | 0 | 0 | 0 | 1 | 3 | 1675 | 1809 | 2010 | 1876 | 1072 | 1273 | 1273 | 1206 |
| 2 | 47.88 | 47.95 | 846.50 | 963.60 | 798.62 | 915.65 | 1120 | 1132 | 1.32 | 1.17 | 0 | 0 | 0 | 0 | 3 | 1 | 1407 | 1608 | 1675 | 1809 | 1139 | 1407 | 1340 | 1273 |
| 3 | 47.62 | 46.91 | 915.00 | 946.33 | 867.38 | 899.42 | 1161 | 1125 | 1.27 | 1.19 | 0 | 0 | 0 | 0 | 2 | 2 | 1541 | 1876 | 2144 | 1809 | 1273 | 1474 | 1407 | 1474 |
| 4 | 47.63 | 50.82 | 884.00 | 932.83 | 836.37 | 882.01 | 1210 | 1162 | 1.37 | 1.25 | 1 | 0 | 1 | 0 | 3 | 2 | 1474 | 1608 | 1742 | 1541 | 1206 | 1407 | 1340 | 1474 |
| 5 | 48.72 | 49.09 | 860.67 | 915.33 | 811.95 | 866.24 | 1135 | 1182 | 1.32 | 1.29 | 1 | 1 | 1 | 1 | 1 | 3 | 1407 | 1608 | 1474 | 1541 | 1273 | 1608 | 1541 | 1541 |
| 6 | 47.97 | 47.68 | 838.33 | 950.33 | 790.36 | 902.65 | 1071 | 1090 | 1.28 | 1.15 | 1 | 0 | 2 | 0 | 2 | 4 | 1407 | 1541 | 1675 | 1474 | 1005 | 1273 | 1407 | 1206 |
| 7 | 49.35 | 48.98 | 780.17 | 953.33 | 730.82 | 904.35 | 1124 | 1120 | 1.44 | 1.17 | 1 | 0 | 2 | 0 | 2 | 2 | 1541 | 1675 | 1474 | 1541 | 1340 | 1474 | 1474 | 1541 |
| 8 | 48.09 | 46.81 | 784.67 | 864.33 | 736.58 | 817.52 | 1063 | 1050 | 1.35 | 1.21 | 1 | 0 | 1 | 0 | 3 | 3 | 1407 | 1608 | 1474 | 1675 | 1273 | 1407 | 1474 | 1608 |

| Jaula | Mortalidad (%) | | Score de lesiones de *E. acervulina* (macro) | | Score de lesiones de *E. máxima* (macro) | | Diámetro de bursa | | Rbu | | Rba | | Rti | | Bu-Ba | | Bu-Ti | | Ti-Ba | |
|---|---|---|---|---|---|---|---|---|---|---|---|---|---|---|---|---|---|---|---|---|
| | T1[1] | T2[1] | T1[1] | T2[1] | T1[1] | T2[1] | T1[1] | T2[1] | T1[1] | T2[1] | T1[1] | T2[1] | T1[1] | T2[1] | T1[1] | T2[1] | T1[1] | T2[1] | T1[1] | T2[1] |
| 1 | 12.5 | 12.5 | 2 | 1 | 0 | 0 | 6 | 5 | 1.4870 | 2.9797 | 1.0095 | 0.5616 | 6.8349 | 8.0811 | 1.4730 | 5.3056 | 0.2176 | 0.3687 | 6.7703 | 14.3888 |
| 2 | 0.0 | 12.5 | 0 | 2 | 0 | 0 | 6 | 6 | 3.3828 | 2.7231 | 1.1135 | 2.0742 | 5.0371 | 5.8517 | 3.0381 | 1.3129 | 0.6716 | 0.4653 | 4.5238 | 2.8212 |
| 3 | 12.5 | 0.0 | 0 | 2 | 1 | 0 | 7 | 5 | 3.5294 | 2.0490 | 1.2000 | 0.8892 | 6.1529 | 4.9613 | 2.9412 | 2.3044 | 0.5736 | 0.4130 | 5.1275 | 5.5797 |
| 4 | 0.0 | 0.0 | 3 | 0 | 0 | 0 | 6 | 5 | 1.9240 | 1.4135 | 1.0369 | 2.6015 | 5.3111 | 2.6015 | 1.8556 | 0.5434 | 0.3623 | 0.5434 | 5.1222 | 1.0000 |
| 5 | 0.0 | 0.0 | 2 | 1 | 0 | 0 | 5 | 5 | 1.5537 | 2.2639 | 1.7373 | 0.9861 | 4.6893 | 4.3889 | 0.8943 | 2.2958 | 0.3313 | 0.5158 | 2.6992 | 4.4507 |
| 6 | 0.0 | 0.0 | 2 | 1 | 2 | 0 | 7 | 6 | 3.7517 | 2.5647 | 1.4686 | 1.0824 | 6.3151 | 5.4824 | 2.5546 | 2.3696 | 0.5941 | 0.4678 | 4.3000 | 5.0652 |
| 7 | 0.0 | 0.0 | 2 | 1 | 0 | 1 | 6 | 5 | 2.6452 | 1.6509 | 0.8258 | 0.8373 | 6.2452 | 6.7335 | 3.2031 | 1.9718 | 0.4236 | 0.2452 | 7.5625 | 8.0423 |
| 8 | 0.0 | 0.0 | 0 | 0 | 0 | 0 | 5 | 5 | 1.9325 | 2.1314 | 1.0583 | 0.9135 | 4.7853 | 5.1763 | 1.8261 | 2.3333 | 0.4038 | 0.4118 | 4.5217 | 5.6667 |

[1] Tratamientos: T1: aves desafiadas control; T2: aves desafiadas + 500 ppm de Orevitol®.

## ANEXO 55. Experimento 8 - peso y ganancias de peso individuales (gramos)

**Peso vivo - día 0 de edad (g)**

| Jaula | N° | Tratamientos [1] | | Jaula | N° | Tratamientos [1] | |
|---|---|---|---|---|---|---|---|
| | | 1 | 2 | | | 1 | 2 |
| 1 | 1 | 44.55 | 46.81 | 5 | 33 | 46.86 | 45.86 |
| | 2 | 40.73 | 49.91 | | 34 | 50.02 | 42.65 |
| | 3 | 45.77 | 46.40 | | 35 | 48.48 | 53.19 |
| | 4 | 46.28 | 51.42 | | 36 | 45.70 | 48.83 |
| | 5 | 46.10 | 46.07 | | 37 | 50.67 | 49.52 |
| | 6 | 48.62 | 50.82 | | 38 | 49.41 | 51.22 |
| | 7 | 45.56 | 44.49 | | 39 | 47.43 | 56.57 |
| | 8 | 52.40 | 55.53 | | 40 | 51.15 | 44.87 |
| 2 | 9 | 46.93 | 50.93 | 6 | 41 | 48.46 | 49.46 |
| | 10 | 47.82 | 43.46 | | 42 | 49.95 | 50.49 |
| | 11 | 48.77 | 46.70 | | 43 | 49.53 | 43.86 |
| | 12 | 41.15 | 44.27 | | 44 | 51.29 | 54.20 |
| | 13 | 54.35 | 54.73 | | 45 | 46.31 | 47.60 |
| | 14 | 42.41 | 46.59 | | 46 | 40.47 | 46.55 |
| | 15 | 52.86 | 48.67 | | 47 | 48.54 | 42.81 |
| | 16 | 48.71 | 48.26 | | 48 | 49.21 | 46.49 |
| 3 | 17 | 47.57 | 48.56 | 7 | 49 | 55.56 | 51.33 |
| | 18 | 48.93 | 49.23 | | 50 | 48.51 | 41.08 |
| | 19 | 42.18 | 45.46 | | 51 | 49.67 | 52.41 |
| | 20 | 48.78 | 46.39 | | 52 | 48.54 | 53.88 |
| | 21 | 48.51 | 44.13 | | 53 | 43.67 | 46.69 |
| | 22 | 48.56 | 47.62 | | 54 | 52.79 | 45.85 |
| | 23 | 46.42 | 47.15 | | 55 | 46.25 | 50.01 |
| | 24 | 49.99 | 46.72 | | 56 | 49.77 | 50.57 |
| 4 | 25 | 50.61 | 56.46 | 8 | 57 | 62.18 | 45.93 |
| | 26 | 46.97 | 53.63 | | 58 | 42.64 | 47.11 |
| | 27 | 45.65 | 49.29 | | 59 | 44.68 | 45.62 |
| | 28 | 46.21 | 50.54 | | 60 | 44.48 | 45.89 |
| | 29 | 45.97 | 51.17 | | 61 | 46.03 | 47.87 |
| | 30 | 40.21 | 47.86 | | 62 | 54.03 | 51.72 |
| | 31 | 52.48 | 46.12 | | 63 | 44.35 | 48.39 |
| | 32 | 52.92 | 51.51 | | 64 | 46.36 | 41.91 |

**Peso vivo - día 21 de edad (g)**

| Jaula | N° | Tratamientos [1] | | Jaula | N° | Tratamientos [1] | |
|---|---|---|---|---|---|---|---|
| | | 1 | 2 | | | 1 | 2 |
| 1 | 1 | - | - | 5 | 33 | 951 | 797 |
| | 2 | - | 967 | | 34 | 981 | 737 |
| | 3 | - | - | | 35 | - | 907 |
| | 4 | 674 | - | | 36 | 861 | 1037 |
| | 5 | 901 | 658 | | 37 | 649 | - |
| | 6 | 811 | 907 | | 38 | 801 | 967 |
| | 7 | 1011 | 997 | | 39 | - | - |
| | 8 | 991 | 877 | | 40 | 921 | 1047 |
| 2 | 9 | - | 977 | 6 | 41 | - | - |
| | 10 | 911 | 1117 | | 42 | 961 | 1027 |
| | 11 | - | - | | 43 | 751 | 787 |
| | 12 | 851 | 880 | | 44 | 735 | 1077 |
| | 13 | 751 | - | | 45 | - | 1047 |
| | 14 | 921 | - | | 46 | 711 | - |
| | 15 | 761 | 927 | | 47 | 921 | 867 |
| | 16 | 884 | 917 | | 48 | 951 | 897 |
| 3 | 17 | 1011 | 847 | 7 | 49 | - | 927 |
| | 18 | 891 | - | | 50 | 716 | - |
| | 19 | - | 967 | | 51 | 891 | 1087 |
| | 20 | 881 | - | | 52 | 811 | 865 |
| | 21 | 1001 | 793 | | 53 | 691 | 867 |
| | 22 | - | 997 | | 54 | 801 | - |
| | 23 | - | 1037 | | 55 | - | 1067 |
| | 24 | 791 | 1037 | | 56 | 771 | 907 |
| 4 | 25 | - | 1027 | 8 | 57 | 941 | 817 |
| | 26 | 891 | 682 | | 58 | 761 | 641 |
| | 27 | - | - | | 59 | 771 | 897 |
| | 28 | 1061 | 877 | | 60 | - | 1017 |
| | 29 | 791 | 1107 | | 61 | 593 | 847 |
| | 30 | 809 | 967 | | 62 | 851 | 967 |
| | 31 | 871 | - | | 63 | 791 | - |
| | 32 | 881 | 937 | | 64 | - | - |

**Ganancia de peso (g/pollo/día)**

| Jaula | N° | Tratamientos [1] | | Jaula | N° | Tratamientos [1] | |
|---|---|---|---|---|---|---|---|
| | | 1 | 2 | | | 1 | 2 |
| 1 | 1 | - | - | 5 | 33 | 43.05 | 35.76 |
| | 2 | - | 43.67 | | 34 | 44.33 | 33.05 |
| | 3 | - | - | | 35 | - | 40.67 |
| | 4 | 29.90 | - | | 36 | 38.81 | 47.05 |
| | 5 | 40.71 | 29.14 | | 37 | 28.48 | - |
| | 6 | 36.29 | 40.76 | | 38 | 35.81 | 43.62 |
| | 7 | 45.95 | 45.38 | | 39 | - | - |
| | 8 | 44.71 | 39.10 | | 40 | 41.43 | 47.71 |
| 2 | 9 | - | 44.10 | 6 | 41 | - | - |
| | 10 | 41.10 | 51.14 | | 42 | 43.38 | 46.52 |
| | 11 | - | - | | 43 | 33.38 | 35.38 |
| | 12 | 38.57 | 39.81 | | 44 | 32.57 | 48.71 |
| | 13 | 33.19 | - | | 45 | - | 47.57 |
| | 14 | 41.86 | - | | 46 | 31.95 | - |
| | 15 | 33.71 | 41.81 | | 47 | 41.52 | 39.24 |
| | 16 | 39.76 | 41.38 | | 48 | 42.95 | 40.52 |
| 3 | 17 | 45.86 | 38.00 | 7 | 49 | - | 41.71 |
| | 18 | 40.10 | - | | 50 | 31.76 | - |
| | 19 | - | 43.90 | | 51 | 40.05 | 49.29 |
| | 20 | 39.62 | - | | 52 | 36.29 | 38.62 |
| | 21 | 45.33 | 35.67 | | 53 | 30.81 | 39.05 |
| | 22 | - | 45.19 | | 54 | 35.62 | - |
| | 23 | - | 47.14 | | 55 | - | 48.43 |
| | 24 | 35.29 | 47.14 | | 56 | 34.33 | 40.76 |
| 4 | 25 | - | 46.24 | 8 | 57 | 41.86 | 36.71 |
| | 26 | 40.19 | 29.90 | | 58 | 34.19 | 28.29 |
| | 27 | - | - | | 59 | 34.57 | 40.52 |
| | 28 | 48.33 | 39.33 | | 60 | - | 46.24 |
| | 29 | 35.48 | 50.29 | | 61 | 26.05 | 38.05 |
| | 30 | 36.62 | 43.76 | | 62 | 37.95 | 43.57 |
| | 31 | 39.00 | - | | 63 | 35.57 | - |
| | 32 | 39.43 | 42.14 | | 64 | - | - |

[1] Tratamientos: 1: dieta basal; 2: dieta basal + 500 ppm de Orevitol®.

\- Los valores faltantes corresponden a aves que fueron muestreadas o murieron antes de completar el periodo

**ANEXO 56. Experimento 8 – digestibilidad, balance nitrogenado y eficiencia de nutrientes**

| Jaula | Peso vivo Inicial (g) | | Peso vivo día14 de edad (g) | | Peso vivo día 21 de edad (g) | | Ganancia de peso de 0 a 21 días (g/ave/d) | | Ganancia de peso de 14 a 21 días (g) | | Consumo de alimento de 0 a 21 d (a/ave/d) | | Consumo alimento 3° semana (g) | | Conversión alimentaria de 0 a 21 d | | Digestibilidad de la proteína (%) | |
|---|---|---|---|---|---|---|---|---|---|---|---|---|---|---|---|---|---|---|
| | T1 (1) | T2 (1) | T1 (1) | T2 (1) | T1 (1) | T2 (1) | T1 (1) | T2 (1) | T1 (1) | T2 (1) | T1 (1) | T2 (1) | T1 (1) | T2 (1) | T1 (1) | T2 (1) | T1 (1) | T2 (1) |
| 1 | 46.25 | 48.93 | 448 | 469 | 877.60 | 881.20 | 39.51 | 39.61 | 429 | 413 | 55.10 | 55.00 | 27.43 | 25.71 | 1.39 | 1.39 | 77.65 | 71.15 |
| 2 | 47.88 | 47.95 | 444 | 511 | 846.50 | 963.60 | 38.03 | 43.64 | 403 | 453 | 53.33 | 53.90 | 27.14 | 25.95 | 1.40 | 1.24 | 60.02 | 73.93 |
| 3 | 47.62 | 46.91 | 481 | 503 | 915.00 | 946.33 | 41.25 | 42.84 | 434 | 443 | 55.29 | 53.57 | 27.90 | 25.43 | 1.34 | 1.25 | 65.48 | 56.53* |
| 4 | 47.63 | 50.82 | 469 | 523 | 884.00 | 932.83 | 39.84 | 41.95 | 415 | 410 | 57.62 | 55.33 | 29.24 | 25.81 | 1.45 | 1.32 | 58.82 | 70.56 |
| 5 | 48.72 | 49.09 | 446 | 505 | 860.67 | 915.33 | 38.65 | 41.31 | 414 | 410 | 54.05 | 56.29 | 27.52 | 26.95 | 1.40 | 1.36 | 71.49 | 74.48 |
| 6 | 47.97 | 47.68 | 420 | 489 | 838.33 | 950.33 | 37.63 | 42.99 | 419 | 461 | 51.00 | 51.90 | 27.24 | 27.67 | 1.35 | 1.21 | 52.07 | 73.75 |
| 7 | 49.35 | 48.98 | 423 | 491 | 780.17 | 953.33 | 34.83 | 42.98 | 357 | 463 | 53.52 | 53.33 | 25.57 | 25.86 | 1.54 | 1.24 | 61.55 | 73.26 |
| 8 | 48.09 | 46.81 | 368 | 464 | 784.67 | 864.33 | 35.03 | 38.90 | 417 | 401 | 50.62 | 50.00 | 26.81 | 25.48 | 1.45 | 1.28 | 64.57 | 79.68 |

| Jaula | Digestibilidad de la materia seca (%) | | Ingesta de nitrógeno (g/ave/día) | | Excreción de nitrógeno (g/ave/día) | | Retención aparente de nitrógeno (g/ave/día) | | Balance nitrogenado ($g/kg^{0.75}$/día) | | Balance nitrogenado (g/kg/día) | | Relación de eficiencia energética (%) | | Relación de eficiencia proteica (g/g) | |
|---|---|---|---|---|---|---|---|---|---|---|---|---|---|---|---|---|
| | T1 (1) | T2 (1) | T1 (1) | T2 (1) | T1 (1) | T2 (1) | T1 (1) | T2 (1) | T1 (1) | T2 (1) | T1 (1) | T2 (1) | T1 (1) | T2 (1) | T1 (1) | T2 (1) |
| 1 | 79.52 | 75.58 | 2.588 | 2.427 | 0.579 | 0.700 | 2.010 | 1.726 | 2.763 | 2.339 | 3.031 | 2.558 | 24.686 | 25.350 | 3.7884 | 3.8902 |
| 2 | 66.17 | 76.61 | 2.561 | 2.449 | 1.024 | 0.638 | 1.537 | 1.811 | 2.155 | 2.296 | 2.382 | 2.455 | 23.434 | 27.550 | 3.5962 | 4.2278 |
| 3 | 72.01 | 58.99 * | 2.633 | 2.400 | 0.909 | 1.043* | 1.724 | 1.356* | 2.279 | 1.743* | 2.470 | 1.872* | 24.548 | 27.497 | 3.7671 | 4.2197 |
| 4 | 67.53 | 77.78 | 2.759 | 2.436 | 1.136 | 0.717 | 1.623 | 1.719 | 2.195 | 2.197 | 2.399 | 2.361 | 22.403 | 25.073 | 3.4379 | 3.8477 |
| 5 | 77.94 | 75.35 | 2.597 | 2.543 | 0.740 | 0.649 | 1.857 | 1.894 | 2.580 | 2.469 | 2.841 | 2.668 | 23.741 | 24.010 | 3.6433 | 3.6845 |
| 6 | 60.49 | 79.99 | 2.570 | 2.611 | 1.232 | 0.685 | 1.338 | 1.926 | 1.915 | 2.489 | 2.128 | 2.676 | 24.280 | 26.300 | 3.7259 | 4.0359 |
| 7 | 65.47 | 76.55 | 2.413 | 2.440 | 0.928 | 0.652 | 1.485 | 1.788 | 2.193 | 2.305 | 2.469 | 2.476 | 22.035 | 28.262 | 3.3815 | 4.3371 |
| 8 | 69.02 | 80.69 | 2.530 | 2.404 | 0.896 | 0.488 | 1.634 | 1.916 | 2.501 | 2.627 | 2.834 | 2.885 | 24.550 | 24.844 | 3.7674 | 3.8125 |

(1) Tratamientos: T1: dieta basal; T2: dieta basal + 500 ppm de Orevitol®.

* Por medio de las pruebas de Dixon (Dean and Dixon, 1951) y Grubbs (Grubbs, 1969) se determinó que estos valores son anómalos (P<0.05), por ello se obvió dicha repetición para el cálculo de los valores promedio del tratamiento 2 y se consideró como unidad perdida. Debido a que el balance nitrogenado se calcula a partir de la excreción de nitrógeno, esta misma repetición se obvió también para el cálculo de dicha variable

**ANEXO 57. Índices de correlación de las características de los órganos linfoides en pollos de carne**

| Variables asociadas | Índices de correlación | | |
|---|---|---|---|
| | Hernández (1998) | Ulloa *et al* (1999) [1] | Perozo-Marín *et al* (2004) [2] |
| Peso vivo – peso del timo | 0.90 | 0.87 | 0.99 |
| Peso vivo – peso del bazo | 0.94 | 0.89 | 0.99 |
| Peso vivo – peso de la bursa | 0.84 | 0.92 | 0.64 |
| Peso del timo – peso del bazo | 0.88 | 0.77 | 0.98 |
| Peso de la bursa – peso del timo | 0.81 | 0.87 | 0.62 |
| Peso de la bursa – peso del bazo | 0.80 | 0.83 | 0.61 |
| Peso de la bursa – diámetro de la bursa | - | - | 0.93 |

[1] Las aves presentaron retraso en el crecimiento debido a contaminación fúngica del alimento y a un probable caso de Anemia del Pollo.

[2] Evaluación realizada en una zona geográfica endémica a la Enfermedad de Gumboro